\pdfoutput=1

\documentclass[aps,rmp,twocolumn,superscriptaddress,nofootinbib]{revtex4-1}

\usepackage[breaklinks]{hyperref}
\usepackage[all]{hypcap}

\usepackage{placeins}

\usepackage{lineno}  

\usepackage{graphicx}  
\usepackage{dcolumn}  
\renewcommand{\arraystretch}{1.1}

\usepackage{xspace}

\usepackage{amsmath}

\usepackage{ifthen} 

\newboolean{pdflatex}
\setboolean{pdflatex}{true} 
\newboolean{articletitles}
\setboolean{articletitles}{true} 
\usepackage{amssymb}
\usepackage{amsfonts}
\usepackage{upgreek}

\newcommand*\patchAmsMathEnvironmentForLineno[1]{%
\expandafter\let\csname old#1\expandafter\endcsname\csname #1\endcsname
\expandafter\let\csname oldend#1\expandafter\endcsname\csname
end#1\endcsname
 \renewenvironment{#1}%
   {\linenomath\csname old#1\endcsname}%
   {\csname oldend#1\endcsname\endlinenomath}%
}
\newcommand*\patchBothAmsMathEnvironmentsForLineno[1]{%
  \patchAmsMathEnvironmentForLineno{#1}%
  \patchAmsMathEnvironmentForLineno{#1*}%
}
\AtBeginDocument{%
\patchBothAmsMathEnvironmentsForLineno{equation}%
\patchBothAmsMathEnvironmentsForLineno{align}%
\patchBothAmsMathEnvironmentsForLineno{flalign}%
\patchBothAmsMathEnvironmentsForLineno{alignat}%
\patchBothAmsMathEnvironmentsForLineno{gather}%
\patchBothAmsMathEnvironmentsForLineno{multline}%
\patchBothAmsMathEnvironmentsForLineno{eqnarray}%
}

\newcommand{\ee}{e^{+}e^{-}}

\newcommand{\jp}{J/\psi}

\newcommand{\psip}{\psi^{\prime}}

\newcommand{\mumu}{\mu^{+}\mu^{-}}
\newcommand{\pipi}{\pi^{+}\pi^{-}}

\newcommand{\pim}{\pi^{-}}
\newcommand{\pip}{\pi^{+}}

\newcommand{\piz}{\pi^{0}}

\newcommand{\qqbar}{q\bar{q}}
\newcommand{\ccbar}{c\bar{c}}

\newcommand{\rt}{\rightarrow}

\newcommand{\jpsi}{J/\psi}

\newcommand{\bbbar}{b\bar{b}}
\newcommand{\QQbar}{Q\bar{Q}}

\newcommand{\tev}{\ifthenelse{\boolean{inbibliography}}{\ensuremath{~T\kern -0.05em eV}}{\ensuremath{\mathrm{\,Te\kern -0.1em V}}}}
\newcommand{\gev}{\ensuremath{\mathrm{\,Ge\kern -0.1em V}}}
\newcommand{\mev}{\ensuremath{\mathrm{\,Me\kern -0.1em V}}}
\newcommand{\kev}{\ensuremath{\mathrm{\,ke\kern -0.1em V}}}
\newcommand{\ev}{\ensuremath{\mathrm{\,e\kern -0.1em V}}}
\newcommand{\gevc}{\ensuremath{{\mathrm{\,Ge\kern -0.1em V\!/}c}}}
\newcommand{\mevc}{\ensuremath{{\mathrm{\,Me\kern -0.1em V\!/}c}}}
\newcommand{\gevcc}{\ensuremath{{\mathrm{\,Ge\kern -0.1em V\!/}c^2}}}
\newcommand{\gevgevcccc}{\ensuremath{{\mathrm{\,Ge\kern -0.1em V^2\!/}c^4}}}
\newcommand{\mevcc}{\ensuremath{{\mathrm{\,Me\kern -0.1em V\!/}c^2}}}

\newcommand\babar{BaBar}
\newcommand\DZero{D0}

\newcommand\Bs{\ensuremath{B_s^0}}
\newcommand\Bsjp{\ensuremath{B_s^0\to\jpsi\phi}}
\newcommand\Bsdecay{\ensuremath{\Bs\to\mu^{\mp}D_s^{\pm}X}}

\newcommand\ecm{\ensuremath{E_{\rm cm}}}
\def\ecmx#1{\ensuremath{\ecm(#1)}}

\usepackage{mciteplus}

\def\myclearpage{}

\newboolean{prl}
\setboolean{prl}{true} 

\newboolean{arxiv}
\setboolean{arxiv}{true} 

\begin{document}

\title{ \quad\\[0.5cm] Non-Standard Heavy Mesons and Baryons, an Experimental Review}

\author{Stephen Lars Olsen}\affiliation{Center for Underground Physics, Institute for Basic Science,
Daejeon 305-811 Korea}

\author{Tomasz Skwarnicki}\affiliation{Department of Physics, Syracuse University, 
Syracuse, NY 13244 U.S.A.}

\author{Daria Zieminska}\affiliation{Department of Physics, Indiana University, 
Bloomington, IN 47405-71055 U.S.A.}

\date{August 13, 2017}

\begin{abstract}
Quantum Chromodynamics (QCD), the generally accepted theory for
the strong interactions, describes the interactions between quarks and gluons.
The strongly interacting particles that are seen in nature are hadrons, which are
composites of quarks and gluons.  Since QCD is a strongly coupled theory at
distance scales that are characteristic of observable hadrons, there are no
rigorous, first-principle methods to derive the spectrum and properties of the
hadrons from the QCD Lagrangian, except for Lattice QCD simulations that
are not yet able to cope with all aspects of complex and short-lived states.
Instead, a variety of ``QCD inspired'' phenomenological models have been proposed.  
Common features of these models
are predictions for the existence of hadrons with substructures that are more
complex than the standard  quark-antiquark mesons and the three quark baryons
of the original quark model that provides a concise description of most of the
low-mass hadrons. Recently, an assortment of candidates for non-standard
multi-quark mesons, meson-gluon hybrids and pentaquark baryons 
that contain heavy (charm or bottom) quarks have been discovered.  Here
we review the experimental evidence for these states and make some general
comparisons of their measured properties with standard quark-model expectations
and predictions of various models for non-standard hadrons.
We conclude that the spectroscopy of all but simplest hadrons is not yet understood.
\ifthenelse{\boolean{arxiv}}{
\begin{center}
Submitted to {\it Reviews of Modern Physics}.
\end{center}
}{
\begin{center}
\quad
\end{center}
}
\end{abstract}
\maketitle
{\renewcommand{\thefootnote}{\fnsymbol{footnote}}}
\setcounter{footnote}{0}

\tableofcontents

\section{Introduction}

The major breakthrough in our understanding of the spectrum of subatomic hadrons was the
nearly simultaneous realization in 1964 by Gell-Mann~\cite{Gellmann:1964nj} and Zweig~\cite{Zweig:1981pd}
that hadrons could be succinctly described as composites of fractionally charged fermions with
baryon number $B=1/3$, called ``quarks'' by Gell-Mann and ``aces'' by Zweig. The original
quark model had three different flavored quarks: $q=u^{+2/3},d^{-1/3},s^{-1/3}$ (now called the
light flavors)\footnote{The $u$ and $d$ quark form an isospin doublet: $u$ with $I_3=1/2$
and $d$ with $I_3=-1/2$. The $s$-quark has a non-zero additive flavor quantum number called
strangeness; for historical reasons the $s$ quark has negative strangeness ${\mathcal S}=-1$
and the $\bar{s}$ quark has positive strangeness ${\mathcal S}=+1$.} 
and their $B=-1/3$ antiparticles $\bar{q}=\bar{u}^{-2/3},\bar{d}^{+1/3},\bar{s}^{+1/3}$.
The most economical quark combinations for producing $B=0$ mesons and $B=1$ baryons,
are $q\bar{q}$ and $qqq$,\footnote{For simplicity of notation, 
flavor indices are suppressed.  In combinations, such as $qqq$ and $q\bar{q}$, it is implicitly
assumed that each $q$ can have any one of the three light quark flavors.}  respectively, and these
combinations reproduce the pseudoscalar and vector meson octets and the spin-1/2 and spin-3/2 baryon
octet and decuplet that were known at that time.  Nevertheless, both authors noted in their
original papers that more complex structures with integer charges and $B=0$ or $B=1$ could exist,
such as $qq\bar{q}\bar{q}$ ``tetraquark'' mesons and $qqq\bar{q}q$ ``pentaquark'' baryons.
However, no candidates for these more complicated configurations were known at the time.

\subsection{Color charges, gluons and QCD}
The original quark model implied pretty grievous violations of the
Pauli Exclusion Principle. For example, the quark model identifies the $J=3/2$ $\Omega^-$ baryon as
a state that contains three $s$ quarks that are all in a relative S-wave and with parallel spins;
{\it i.e.}, the three $s$ quarks occupy the same quantum state, in violation of Pauli's principle. This
inspired a suggestion by Greenberg~\cite{Greenberg:1964pe} that quarks were not fermions but, instead,
``parafermions'' of order three, with an additional, hidden quantum number that made them distinct. In this
picture, the three $s$ quarks in the $\Omega^-$ have different values of this hidden quantum
number and are, therefore, non-identical particles.    

\begin{figure}[htb]
  \includegraphics[width=0.48\textwidth]{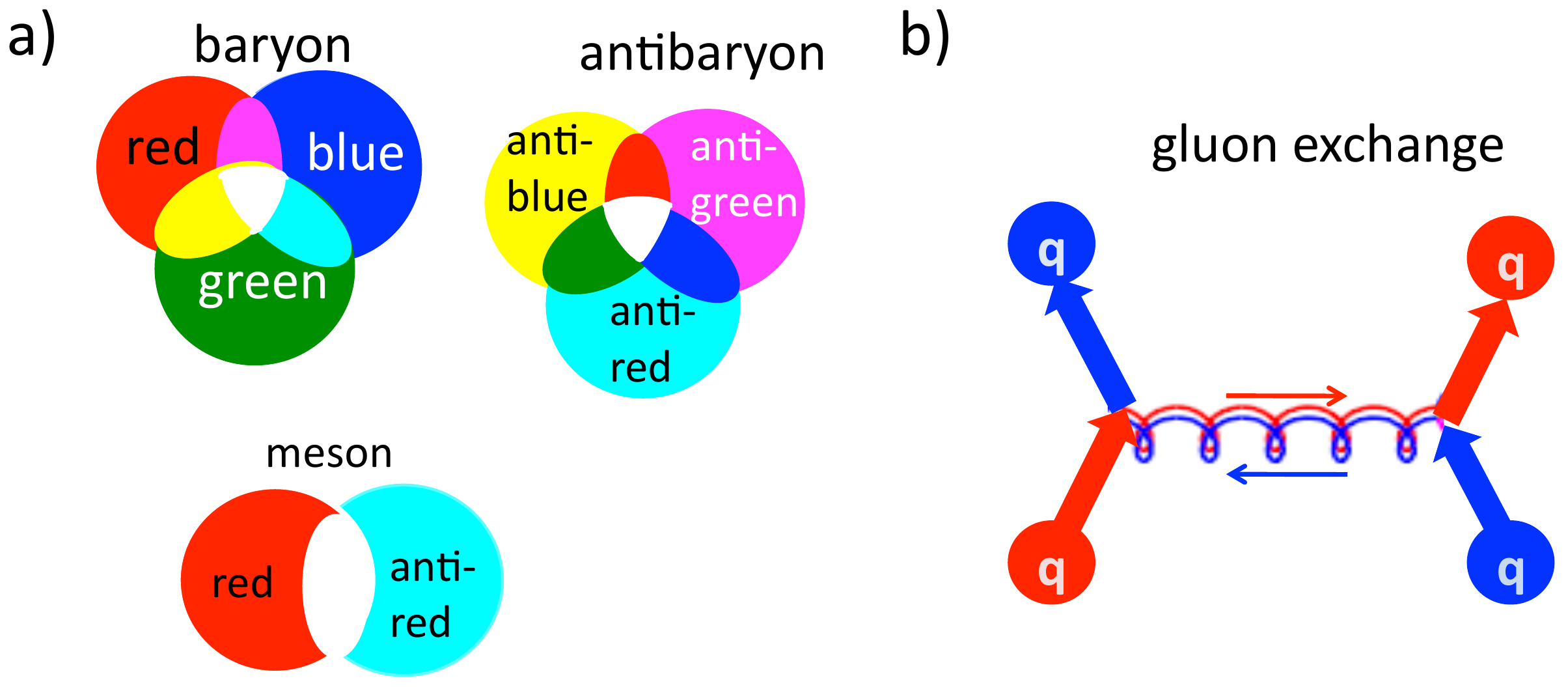}
\caption{\footnotesize {\bf a)} The color makeup of baryons, antibaryons and a meson.
{\bf b)} Single gluon exchange between two quarks. Gluons have two color indices that
can be viewed as two color charges that propagate in opposite directions.
}
\label{fig:colored-qcd}
\end{figure}

In the following year, Han and Nambu~\cite{Han:1965pf} proposed a model in which each of the quarks
are $SU(3)$ triplets in flavor-space (and with integer electric charges) with strong-interaction ``charges''
that are a triplet in another $SU(3)$ space.  They identified Greenberg's hidden quantum numbers with
three different varieties of strong charges, $q\rt q_i$, $i=1,2,3$, and associated the
observable hadrons as singlets in the space of this additional $SU(3)$ symmetry group. This can be done
with three-quark combinations in which each quark has a different strong charge 
(${\rm baryons}=\epsilon_{ijk}q_i q_j q_k$) or quark-antiquark combinations, where the quark's strong charge and
the antiquark's strong anticharge are the same type (${\rm mesons}=\delta^i_j q_i \bar{q}^j$).  Because of the uncanny
correspondence between these prescriptions with the rules for human color perception, where white can be
produced either by triplets of three primary colors or by color plus complementary-color pairs,
the strong-interaction charges were soon dubbed ``color'' charges: red, green and blue, with anticharges
that are the corresponding complementary colors: cyan, magenta and yellow.  The color neutral combinations
that form baryons, antibaryons and mesons are illustrated in Fig.~\ref{fig:colored-qcd}a.

Measurements of the total cross section for $e^+e^-\rt$~hadrons were consistent with the existence of the three color
degrees of freedom~\cite{Litke:1973np}.  The generalization of the Han-Nambu idea to a gauge theory with quarks
of fractional electric charge was done in 1973~\cite{Bardeen:1972xk} and is called Quantum Chromodynamics or
QCD. This is now the generally accepted theory for the strong interactions. 

\subsubsection{Asymptotic freedom and confinement}
 
\begin{figure}[htb]
  \includegraphics[width=0.50\textwidth]{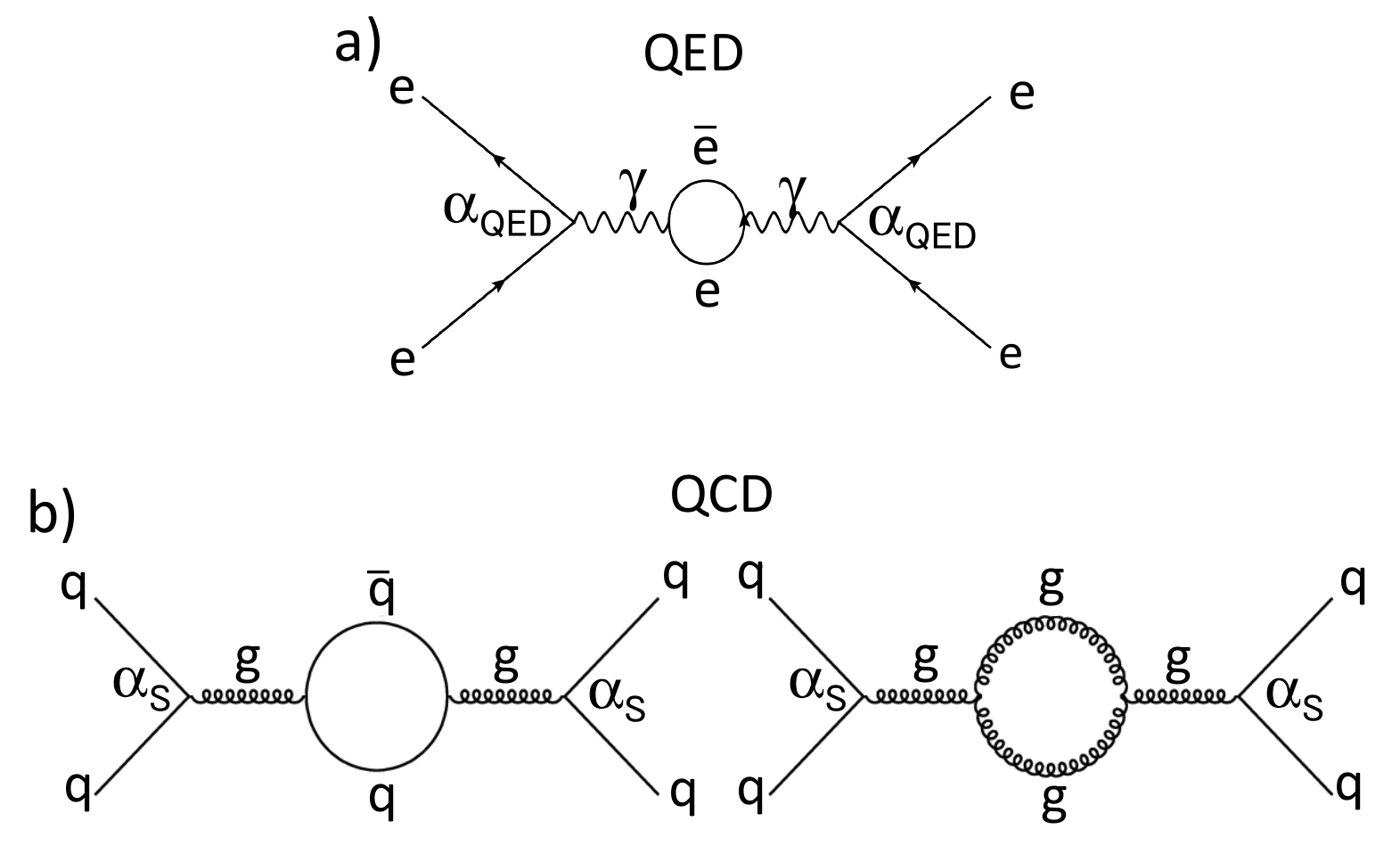}
\caption{\footnotesize 
{\bf a)}  The lowest-order QED vacuum polarization diagram for electron-electron scattering.
{\bf b)}  The lowest-order QCD vacuum polarization diagrams for quark-quark scattering.
}
\label{fig:vac-pol}
\end{figure}

In QCD, the color force is mediated by eight massless vector particles called gluons, which are the
the generalization of the photon in QED.  Unlike QED in which the photons are electrically
neutral and do not interact with each other, the gluons of QCD have color charges and, thus, interact
with each other.  Figure~\ref{fig:colored-qcd}b shows a single gluon exchange between two colored quarks.
In QED, the vacuum polarization diagram, shown in Fig.~\ref{fig:vac-pol}(a), results in a modification
of the QED coupling strength $\alpha_{\rm QED}$ that makes it decrease with increasing distance.  For distance
scales of order 1~meter, $\alpha_{\rm QED}\simeq 1/137$; at a distance scale of 0.002~fm, comparable to the Compton
wavelength of the $Z^0$ weak vector boson, $\alpha_{\rm QED}\simeq 1/125$.  

\begin{figure}[htb]
  \includegraphics[height=0.35\textwidth,width=0.5\textwidth]{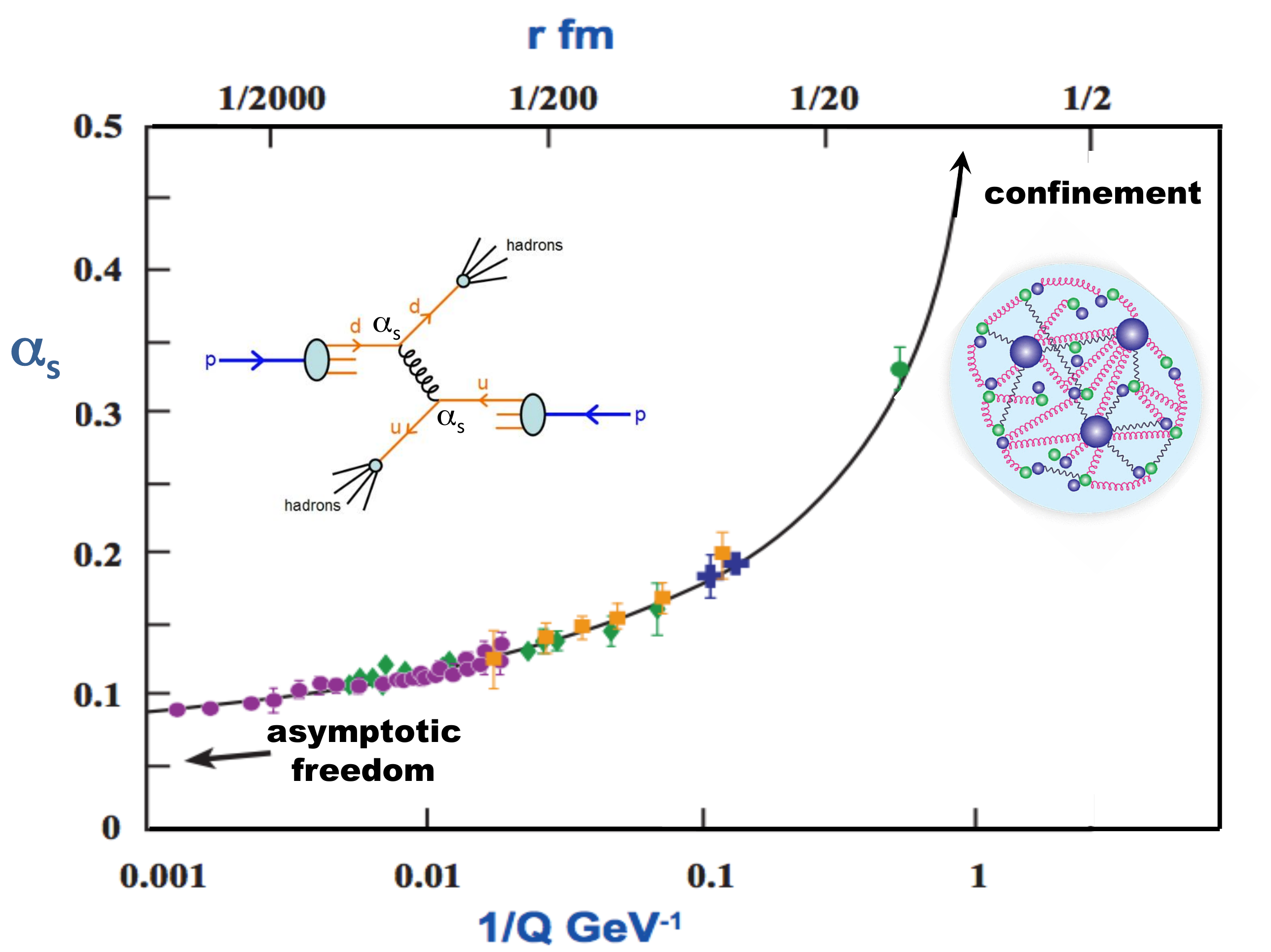}
\caption{\footnotesize The behavior of the QCD coupling strength $\alpha_s$ as a function
of the inverse momentum transfer $1/Q$ or, equivalently, the quark separation distance $r$.
Descriptions of the data points and the associated references are provided in ref.~\cite{Olive:2016xmw}.
}
\label{fig:alphas}
\end{figure}

In QCD, the gluon-gluon interaction includes additional vacuum polarization diagrams that have virtual gluon
loops as shown in Fig.~\ref{fig:vac-pol}(b).  These gluon loops  modify the QCD coupling strength $\alpha_s$
in a way that is opposite to that of its QED counterpart:  they cause $\alpha_s$ to {\em decrease} at short
distances and {\em increase} at long distances~\cite{Gross:1973id,Politzer:1973fx} as
illustrated in Fig.~\ref{fig:alphas}.  The relatively small value of the coupling strength at short distances:
$\alpha_s=0.1185\pm0.0006$ at $r\simeq 0.002$~fm, results in what is called ``asymptotic freedom,''
and facilitates the use of perturbation expansions to make reliable (albeit difficult) first-principle
calculations for short-distance, high-momentum-transfer processes such as those studied in the high-$p_T$
detectors at CERN's Large Hadron Collider (LHC).  In contrast, for distance scales of that approach 
$r\sim 1$~fm,  which are characteristic of the sizes of hadrons, $\alpha_s \sim {\mathcal O}(1)$ and perturbation expansions
do not converge.  This increase in the coupling strength for large quark separations is the source of
``confinement,'' {\it i.e.}, the reason that isolated colored particles, be they quarks or gluons, are never
seen.  The only strongly interacting particles that can exist in isolation are color-charge-neutral ({\em i.e.}
white) hadrons.  

\subsection{The QCD dilemma}

In QCD, the component of the Standard Model (SM) of elementary processes that deals with the strong
interaction, the elementary particles are the color-charged quarks and gluons.  However, a consequence of
confinement is that these particles are never seen in experiments.  Although the QCD Lagrangian is expected,
in principle, to completely describe the spectrum of hadrons and all of their properties, there is no
rigorous first-principle translation of this into any useful mathematical expressions.

The quark and gluon composition of hadrons can be hopelessly complex, as illustrated in the inset
on the right side of Fig.~\ref{fig:alphas}.  For distance scales on the order 1~fm, the typical size of a
hadron, $\alpha_s\sim 1$ and the pattern illustrated in the figure is just one of an infinite number of
possible quark-gluon configurations that are only subject to the constraints that they have appropriate
quantum numbers and are color neutral.  In fact, while the traditional three quarks form baryons and
quark-antiquark pairs form mesons prescription works well for the meson octets and the baryon octet \& decuplet
that were known at the time quarks were first introduced, it fails in a number of other areas. Soon after the quark model
was proposed, it was realized that these simple rules failed to provide a satisfactory explanation for the
properties of the lowest-mass scalar-meson octet~\cite{Jaffe:1976ig} and were unable to provide a simple explanation for
the positive parity of the lowest-lying excitation of the proton, the $J^P=1/2^+$ $N^*(1440)$ (the ``Roper
resonance'')~\cite{AlvarezRuso:2010xr} or the mass of the lowest-lying excitation of the $\Lambda$ hyperon, the
$J^P=1/2^-$ $\Lambda(1405)$~\cite{Close:1980ab}.

A fundamental process that can be computed with perturbative QCD is quark-quark elastic scattering at
high-momentum transfer.  This shows up in high energy $pp$ collider experiments as events with two high
transverse momentum jets of hadrons that are nearly back-to-back in azimuth. 
The theoretical description of this process
is based on calculations of the diagram shown in the inset on the left side
of Fig.~\ref{fig:alphas}.  Here, in lieu of a beam or target of isolated quarks, the
beam and target particles are quarks contained inside the colliding protons. 
The momentum distribution
of quarks inside the proton is governed by long-distance QCD and approximated by universal parton
distribution functions that are taken from fits to data from hadron collider
measurements at lower center of mass (c.m.) energies, deep-inelastic lepton-proton scattering experiments, etc.  
The fundamental
QCD $qq\rt qq$ process at the core of the diagram has been computed up to ${\mathcal O}(\alpha_s^3)$, but the
properties of the final-state quarks cannot be directly measured and, instead, have to be inferred from the
jets of hadrons that they produce; for this, empirical ``fragmentation functions'' are employed.  Thus, even
processes that are amenable to perturbative QCD calculations involve significant long-distance QCD effects
both in the initial and final state.

This nearly total disconnect between the hadrons that we observe in experiments and the quarks and gluons 
that appear in the theory is a problem of huge proportions in particle physics.\footnote{As Frank Wilczek,
put it in a recent interview~\cite{wilczek16}: ``We have something called a standard model, but its
foundations are kind of scandalous. We have not known how to define an important part of it
mathematically rigorously,...''}  This is what we refer to as the ``QCD Dilemma.''  In addition
to the intellectual dissatisfaction with a theory that is not directly applicable to the particles
that are used and detected in experiments, there is also a practical problem in that many SM tests and
searches for New Physics (NP) involve strongly interacting hadrons in the initial and/or final states
of the associated measurements.  Even experiments that do not use initial or final states that contain
hadrons are still subject to their effects from virtual quantum fluctuations.  As a result, the sensitivities
of many NP search experiments are ultimately limited by hadron-related theoretical uncertainties. Because of
this, as the experimental sensitivities of NP searches improve, commensurate improvements in our understanding
of long-distance QCD become more and more important.
A good example of the latter are hadronic contributions to the
predicted value of the muon anomalous magnetic moment \cite{Jegerlehner:2009ry}.

\subsection{Searches for light ``non-standard'' hadrons}
\label{sec:intro_light}

A possible way experiments may be able to contribute to these improvements is by identifying patterns in
hadron physics that may help guide the development of improved theoretical mdoels.  One peculiar pattern, and a 
long-standing puzzle, has been the lack of any evidence for light-flavored hadrons with substructures
that are more complex than the three-quark baryons and quark-antiquark mesons of the original quark model.
During the fifty years that have ensued since the birth of the quark idea, numerous experiments have searched for
pentaquark baryons and light-flavored mesons with $J^{PC}$ quantum numbers that are not accessible in $q\bar{q}$
systems.  Although during the same time period a large number of additional $qqq$ baryon and $q\bar{q}$ meson
resonances were found, no unambiguous examples of light hadrons with non-standard structures have emerged.

In particular, from the very earliest days of the quark model, $K^+ p$ and $K^+d$ cross section data were
scoured for evidence of resonances with positive strangeness (${\mathcal S} =+1$) quantum numbers in either
the isospin $I=1$ or $I=0$ channels that would necessarily contain an $\bar{s}$ quark in a minimal
$qud\bar{s}u$ ($q=u$~or~$d$) five-quark (``pentaquark'') array.\footnote{Since the $s$ quark has ${\mathcal S}=-1$,
conventional three quark baryons that contain one or more $s$ quarks have negative strangeness; the $K^+$
meson contains an $\bar{s}$ quark and has ${\mathcal S}=+1$.}   Candidates for baryon states with positive strangeness,
two with $I=0$, dubbed the $Z_0(1780)$ and $Z_0(1865)$, and three with $I=1$, the
$Z_1(1900)$, $Z_1(2150)$ and $Z_1(2500)$, appeared in the 1976 Particle data Group (PDG) tables ~\cite{pdg76},
but were absent by the time of the 1994~\cite{pdg94} and subsequent versions.  History repeated itself in 2003,
when an experiment studying  $\gamma n\rt K^+K^-n$ reactions using a beam of energy-tagged $\gamma$ rays
impinging on a $^{12}C$ target, reported a ``sharp baryon resonance peak'' in the $K^+n$ invariant mass
distribution with a mass and width, $M=1.54\pm 0.01$~GeV and $\Gamma<25$~MeV~\cite{Nakano:2003qx}, that closely
matched the 1.53~GeV mass and 15~MeV width that was predicted for an ${\mathcal S}=+1$ pentaquark in a 1997
paper by Diakonov {\it et al.}~\cite{Diakonov:1997mm}.   The observation of this peak, which
was called the $\Theta^+$, started a great flurry of activity that produced a number of conflicting
experimental results. This ended three years later when results from some definitive experiments became
available.  Based on these, the PDG 2006 report~\cite{pdg06} declared that
``The conclusion that pentaquarks in general, and the $\Theta^+$, in particular, do not exist, appears
compelling.'' Instructive reviews of this recent pentaquark episode and references to the many related
experimental reports are provided in refs.~\cite{Schumacher:2005wu,Dzierba:2004db,Hicks:2012zz}. 

Searches for non-standard mesons have mostly concentrated on looking for meson resonances with ``exotic''
quantum numbers, {\em i.e.} $J^{PC}$ values that {\em cannot} be formed from a fermion-antifermion pair, namely
$0^{--}$ \& $0^{+-}$, $1^{-+}$, $2^{+-}$, etc. A number of experiments have reported evidence for resonance-like
behavior with $J^{PC}=1^{-+}$, but their interpretations as true resonances remain a subject of some dispute. The
situation is summarized in refs.~\cite{Meyer:2010ku} and \cite{Meyer:2015eta}.

On the other hand, the scalar mesons with masses below 1 GeV,  $f_0(500)$, $K_0^*(800)$, 
$a_0(980)$, and $f_0(980)$, which have non-exotic $J^{PC}=0^{++}$ quantum numbers that can be accessed
by a spin-singlet ($S=0$) $q\bar{q}$ pair in a P-wave, have frequently been cited as candidates for
multiquark states (see, for example ref.~\cite{Achasov:2008ut}). In $q\bar{q}$ systems, the lowest-lying
$J=0$ P-wave $q \bar q$  states are expected to have masses that are above 1.2 GeV, close to those of the
$J^P=1^+$ and  $J^P=2^+$ P-wave mesons, such as the $a_1(1260)$ and $a_2(1320)$ resonances.  In fact, an
octet of $0^{++}$ states with the expected masses ({\em i.e.} the $a_0(1450),$ $K_0^*(1430)$, etc.) has been
identified, and this makes the lighter scalar octet supernumerary. An especially puzzling feature of the low
mass scalars is their mass hierarchy, which is inverted with respect to what would be expected from the quark
model: the strange state, $K_0^*(800)=\bar d s$, is lighter than the $I=1$ $a_0(980)$, which is nominally comprised
of $q\bar q$ pairs ( $q= u$~or~$d$), and the  $f_0(500)$, which, in the standard $q\bar{q}$ meson scheme, would
be an $s\bar s$ state, is the lightest member of the octet. This is contrary to the well established quark model
feature of other mesons and baryons, where states with more $s$ quarks are heavier. This peculiar features led
to speculation that that the lightest $0^{++}$ mesons are comprised of some kind of four quark
configuration~\cite{Jaffe:1976ig,Weinstein:1982gc,Jaffe:2004ph}.
The isosinglet scalar mesons with masses above 1 GeV, $f_0(1370)$, $f_0(1500)$ and $f_0(1710)$ are also supernumerary
since at most two can be attributed to the $1^3{\rm P}_0$ $q \bar{q}$ ($q=u$, $d$, $s$) states. 
They fall into the region of the lightest predicted 
glueball, {\em i.e.} a meson comprised only of gluons, with no quarks. However, these
can mix with conventional $q\bar q$ states, thereby
making a clear cut experimental identification of a glue-glue bound state component difficult. 
Detailed discussion of glueballs and other light exotic hadron candidates can be found, {\em e.g.},   
in ref.~\cite{Olive:2016xmw}.\footnote{See review notes on \lq\lq Quark Model'', \lq\lq Non-$q\bar{q}$ Mesons'', 
\lq\lq Note on Scalar Mesons Below 2 GeV'', and \lq\lq Pole Structure of the $\Lambda(1405)$ Region''.}

\subsection{Heavy quarks and the quarkonium spectra}

During the decade that immediately followed the introduction of the notion of fractionally charged quarks, their
actual existence was met with considerable skepticism.  Although fractionally charged particles produced in high energy
particle collisions would have very distinct experimental signatures and should be relatively easy to observe,
numerous searches at accelerators and in cosmic rays all reported negative results. The conservation of electric charge
ensures that at least one of the fractionally charged quark types should be stable, in which case there could be a
fractionally charged component of ordinary matter. Searches in minerals, deep sea water, meteorites, moon rocks, etc. all
failed to find any sign of this.  Reviews of this interesting era of quark search experiments are provided in
refs.~\cite{Jones:1976xy} and~\cite{Lyons:1980ad}.

\begin{figure}[htb]
  \includegraphics[width=0.48\textwidth]{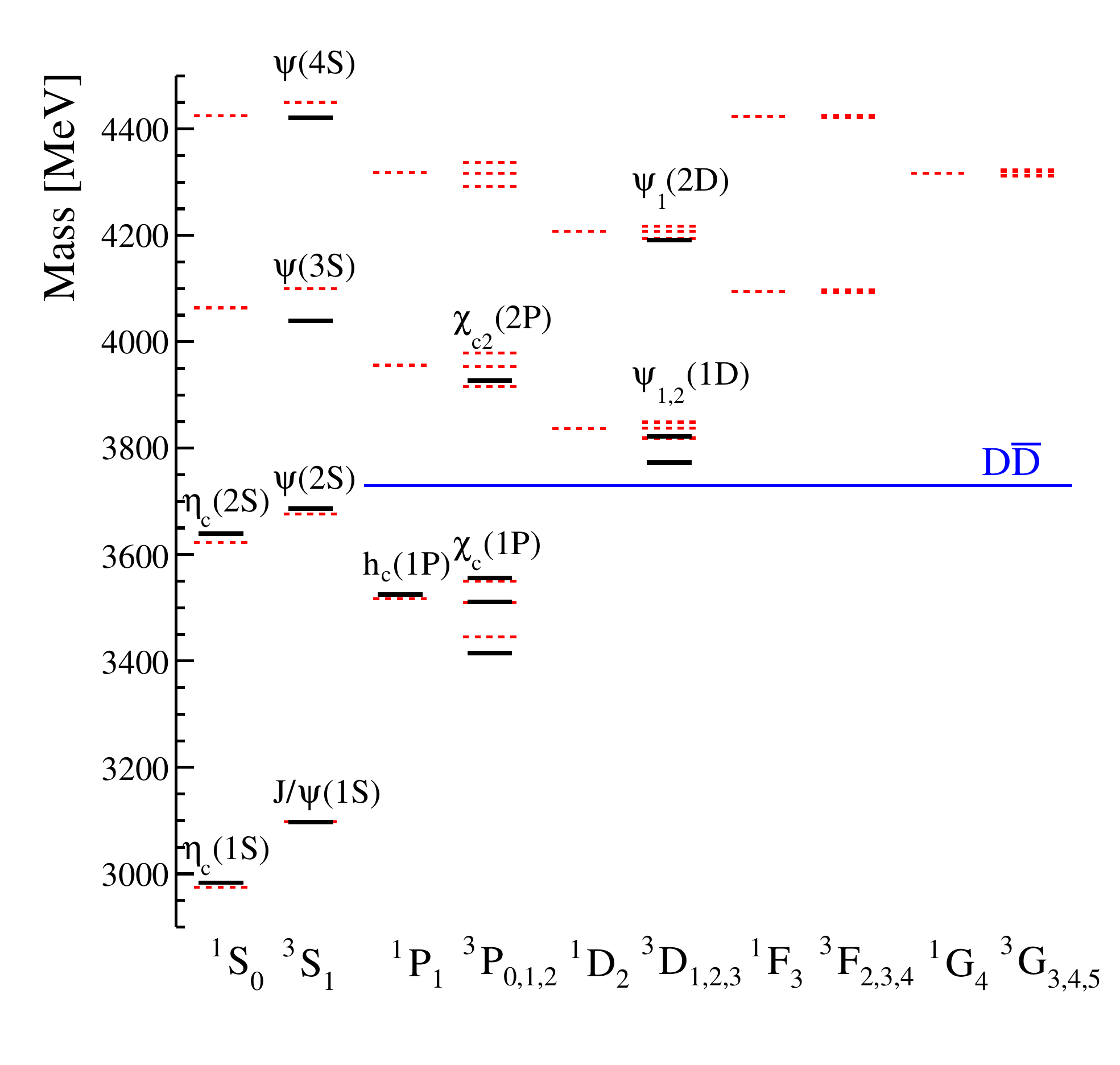}
\caption{\footnotesize The current status of the charmonium spectrum. The dashed (red) lines indicate the expected
states and their masses based on recent calculations~\cite{Barnes:2005pb} based on the Godfrey-Isgur
relativized potential model~\cite{Godfrey:1985xj}.   The solid (black) lines indicates the experimentally established
charmonium states, with masses and spin-parity ($J^{PC}$) quantum number assignments taken from ref.~\cite{Olive:2016xmw},
and labeled by their spectroscopic designations.
The open-flavor threshold is also indicated (blue line).
}
\label{fig:charmonium}
\end{figure}

Thus, while the usefulness of the quark idea as an effective classifier of the spectrum of hadronic particles and for
describing the results of deep-inelastic electron-nucleon scattering experiments was without question, their existence
as real physical entities, as opposed to a useful mathematical mnemonic aid, was strongly debated. However, this debate
was put to rest in 1974--75 with the discovery of the $J/\psi$~\cite{Aubert:1974js,Augustin:1974xw}, $\psip$~\cite{Abrams:1974yy}
and $\chi_{c0,1,2}$~\cite{Feldman:1975bq} mesons\footnote{The $\psip$ and $\chi_{cJ}$ are commonly used names 
for the spin-triplet $\psi (2{\rm S})$, and $\chi_{cJ}(1{\rm P})$ charmonium states.} 
with masses between 3~and~4~GeV. These new resonance states, all
of which are strikingly narrow,  were accurately described by Appelquist and Politzer as bound states of a
$c$- and a $\bar{c}$ quark~\cite{Appelquist:1974zd}, where $c$ denotes the charge$=+2/3$ ``charmed quark''  
with charm flavor ${\mathcal C}=+1$. The assortment of possible $\ccbar$ mesons are collectively known as
charmonium. The large $c$ quark mass ($m_c\simeq 1.3$~GeV) ensures that the $c$ quark motion in bound $\ccbar$
systems is nearly non-relativistic and the spectrum of observed states can be reasonably well described by the
Schr\"{o}dinger equation with a potential that is Coulombic at short distances (in accord with the notion
of asymptotic freedom) and joined to a linearly increasing ``confining'' term at large
distances~\cite{Eichten:1978tg,Necco:2001xg}. 
The charmonium spectrum of states, which have a one-to-one
correspondence to the allowed atomic levels in the hydrogen atom, is indicated in Fig.~\ref{fig:charmonium}.
All of the states below the $M=2m_D$~($=3730$~MeV) open-charm threshold\footnote{Here $m_D$ is the mass of the
$D^0$, the lightest ``open-charm'' meson with quark content $c\bar{u}$ and charm quantum number ${\mathcal C}=1$.}
have been experimentally identified and found to have
have masses and other properties that are in good agreement with potential model expectations. The simplicity
and dramatic success of the charmonium model resulted in a rapid and almost universal acceptance in the particle 
physics community that quarks are real, physical entities.
A systematic theoretical framework for implementing corrections to the static potential 
approach was later developed in form of NRQCD\cite{Brambilla:1999xf,Brambilla:2014jmp}.

\begin{figure}[htb]
  \includegraphics[width=0.48\textwidth]{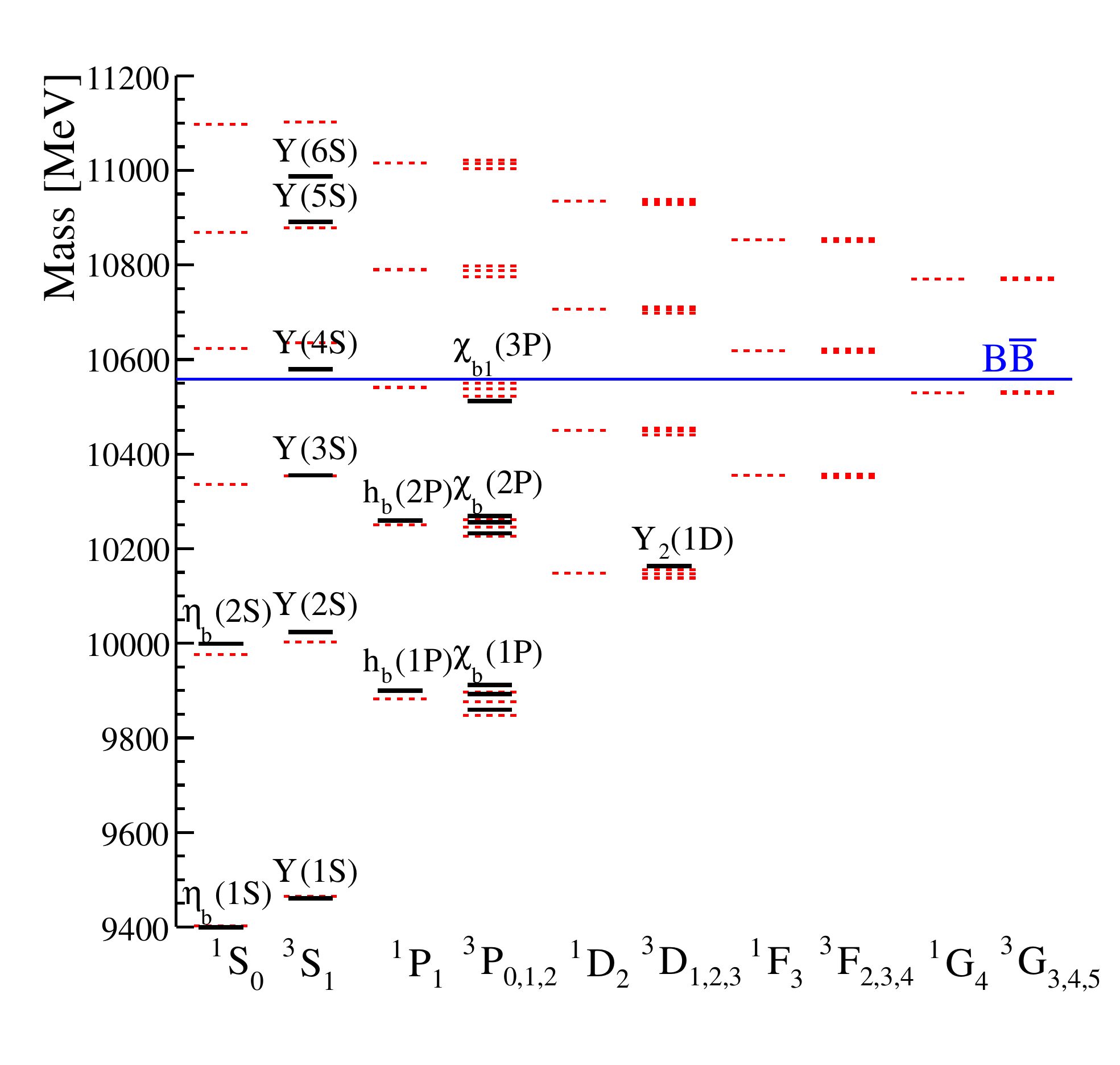}
\caption{\footnotesize 
The current status of the bottomonium spectrum. The dashed lines indicate the expected
states and their masses based on recent calculations~\cite{Godfrey:2015dia} based on the Godfrey-Isgur
relativized potential model~\cite{Godfrey:1985xj}.   The solid lines indicates the experimentally established
bottomonium states, with masses and spin-parity ($J^{PC}$) quantum number assignments taken from ref.~\cite{Olive:2016xmw},
and labeled by their spectroscopic designations. 
The open-flavor threshold is also indicated (blue line).
}
\label{fig:bottomonium}
\end{figure}

\subsubsection{The $b$ quark and the spectrum of bottomonium mesons}
Three years later, in 1977, a similar family of narrow meson resonances (the $\Upsilon$, $\Upsilon'$ \& $\Upsilon''$)
was discovered in the 9.4~to~10.4~GeV mass region~\cite{Herb:1977ek,Ueno:1978vr,Andrews:1980ha}.  
These states were identified as $\bbbar$ bound states, where $b$ designates the ${\rm charge} =-1/3$ ``bottom,'' or
``beauty'' quark with beauty quantum number ${\mathcal B}=-1$, and are now called the bottomonium mesons. It
was found that the application the same potential that was used for charmonium, with a $b$ quark mass of
$m_b\simeq 4.2$~GeV, could produce a reasonable description of the bottomonium system
(see Fig.~\ref{fig:bottomonium}). In this case there are more states below the $M=2m_B$~($=10.56$~GeV) open-bottom
threshold,\footnote{To conform to the nomenclature
of ${\rm charge} =-1/3$ $s$ quark system, $B$ mesons contain a $\bar{b}$ quark and have ``beauty flavor''
${\mathcal B}=+1$ while $\bar{B}$ mesons contain a $b$ quark and have ${\mathcal B}=-1$; {\it i.e.} 
$B=\bar{b}q$ and $\bar{B}=b\bar{q}$, where $q=u$~or~$d$} and most of these have been identified and found to have
masses and other properties that are in good agreement with potential model expectations.  The $c$ and $b$ quark 
are known as ``heavy quarks'' and often denoted as $Q$ ($Q=c$~or~$b$); likewise charmonium and bottomonium mesons  
are collectively referred to as ``quarkonium'' mesons and denoted as $\QQbar$.

\subsubsection{Non-standard quarkonium-like mesons and quarkonium pentaquarks}

The large heavy-quark masses strongly suppress the production of $\QQbar$ pairs from the vacuum during the
quark-to-hadron fragmentation process. Thus, if a $Q$ and a $\bar{Q}$ quark are found among the decay products
of a previously unseen meson resonance, they must have been present as constituents of the meson itself.  If the $Q$ and $\bar{Q}$ quarks
are the parent meson's only constituents, it must be a $\QQbar$ quarkonium state and, thus, have properties that
match those of one of the as yet unassigned allowed quarkonium levels. If it does not fit into
one of the available levels, it must have a substructure that is more complex than just $\QQbar$ and, thus,
qualify as a non-standard hadron. The limited number of unassigned charmonium states with masses below 4.5~GeV
and the theoretical expectation that most of the unassigned charmonium states will have relatively low and
non-overlapping widths, make the identification of non-standard charmonium-like mesons less ambiguous than is
the case for light-quark hadrons.  For similar reasons, a baryon resonance that decays to a final state containing
a $Q$  and a $\bar{Q}$ quark must contain a $\QQbar$ pair among its constituents and, thus, have a valence
configuration that contains at least five quarks.

In contrast to experiments in the light-quark sector, recent searches for non-standard hadrons containing
heavy quark pairs, {\it i.e.}~hadrons that contain a $\ccbar$ quark pair, have uncovered a number of intriguing
states including the $Z(4430)^{\pm}$, which is electrically charged and smoking-gun evidence\footnote{Since the
$Z(4430)^{\pm}$  decays to final state that contains a $\psip$, it must have constituent $c$- and
$\bar{c}$-quarks plus additional light quarks to account for its non-zero electric charge.} for a four-quark
meson that decays to $\psip \pi^{\pm}$~\cite{Choi:2007wga,Aaij:2014jqa}, and two strong candidates for pentaquark
states, the $P_c(4380)$ and $P_c(4450)$ that both decay to $\jpsi p$~\cite{Aaij:2015tga}.  In addition to these,
about twenty other candidate non-standard hadron states containing $\ccbar$ quarks have been found and studied by
the BESIII experiment at the BEPCII  $\tau$-charm factory in Beijing, the Belle and BaBar experiments at the
KEKB and PEPII $B$-factories, the CDF and D0 experiments at the Tevatron, and the LHCb, ATLAS and CMS experiments at the
LHC. In addition, two non-standard bottomonium-like meson candidates were seen by Belle and a candidate for a
mixed-flavor $\bar{b}s \bar{d} u$ was reported by $D0$~\cite{D0:2016mwd}. The non-standard hadron candidates and
some of their properties are listed in Table~\ref{tab:Q-Qbar}, where the charmed pentaquark candidates are labeled
as $P_c$, the charged ($I=1$) states as $Z$, the $J^{PC}=1^{--}$ states as $Y$, and all the rest as $X$.


\begin{table*}[p]
\footnotesize
\caption{ Recently discovered non-standard hadron candidates with hidden charm or beauty. 
The masses $M$ and widths $\Gamma$ are
averages of measurements with uncertainties added in quadrature, except for
$X(4140)$, $X(4274)$ ($Z^+(4200)$), where 
ref.~\cite{Aaij:2016iza,Aaij:2016nsc} 
(ref.~\cite{Chilikin:2014bkk}) values are listed.
See sec.~\ref{sec:x4140etc} (sec.~\ref{sec:z4430}) for more detailed 
discussion. 
The errors on the average values include scale factors in case of 
tensions between individual measurements \cite{Olive:2016xmw}.
We do not quote a mass or width for the $Y(4260)$ structure, since the latest 
precision data have revealed its double-peak composition \cite{Ablikim:2016qzw}, 
with the main component listed under $Y(4220)$ and 
its high-mass shoulder under $Y(4360)$. 
The results from single-peak fits to the $Y(4260)$ structure are not included when 
determining the $Y(4220)$ parameters.   
For $X(3872)$, only $\pipi\jpsi $ decays are used in the mass average.
Ellipses (...) indicate an inclusive reaction.
Question marks indicate informed guesses at $J^{PC}$ values or no information.
For charged states, $C$ refers to the neutral isospin partner.} 
\setlength{\tabcolsep}{0.21pc}
\label{tab:Q-Qbar}
\begin{center}
\renewcommand{\arraystretch}{1.2}
\begin{tabular}{lcccll}
\hline\hline
\rule[10pt]{-1mm}{0mm}
 State & $M$~(MeV) & $\Gamma$~(MeV) & $J^{PC}$ & Process~(decay mode) & 
     Experiment \\ 
\hline
\rule[10pt]{-1mm}{0mm}
${X(3872)}$& $3871.69\pm0.17$ & $<1.2$ &
    $1^{++}$
    & $B \to K + (J/\psi\, \pi^+\pi^-)$ &
    Belle~\cite{Choi:2003ue,Choi:2011fc}, \babar~\cite{Aubert:2004ns},    
    LHCb~\cite{Aaij:2013zoa,Aaij:2015eva}  \\ 
& & & & $p\bar p \to (J/\psi\, \pi^+\pi^-)+ ...$ &
    CDF~\cite{Acosta:2003zx,Abulencia:2005zc,Aaltonen:2009vj}, \DZero~\cite{Abazov:2004kp}   \\ 
& & &   & $B \to K + (J/\psi\, \pi^+\pi^-\pi^0)$ &
    Belle~\cite{Abe:2005ax},    \babar~\cite{delAmoSanchez:2010jr}  \\ 
& & & & $B \to K + (D^0 \bar D^0 \pi^0)$ &
    Belle~\cite{Gokhroo:2006bt,Adachi:2008sua},    \babar~\cite{Aubert:2007rva} \\ 
& & & & $B \to K + (J/\psi\, \gamma)$ &
    \babar~\cite{delAmoSanchez:2010jr},
    Belle~\cite{Bhardwaj:2011dj}, 
    LHCb~\cite{Aaij:2011sn} \\ 
& & & & $B \to K + (\psip \, \gamma)$ &
    \babar~\cite{Aubert:2008ae},
    Belle~\cite{Bhardwaj:2011dj}, 
    LHCb~\cite{Aaij:2014ala} \\ 
& & & &  $pp \to (J/\psi\, \pi^+\pi^-)+ ...$   & 
    LHCb~\cite{Aaij:2011sn},   CMS~\cite{Chatrchyan:2013cld}, ATLAS~\cite{Aaboud:2016vzw}  \\
& & & &  $\ee \to \gamma + (J/\psi\, \pi^+\pi^-)+ ...$   & 
    BESIII~\cite{Ablikim:2013dyn}      \\ 
$X(3915)$ & $3918.4\pm1.9$ & $20\pm5$ & $0^{++}$ &
    $B \to K + (J/\psi\, \omega)$ &
    Belle~\cite{Abe:2004zs},
    \babar~\cite{Aubert:2007vj,delAmoSanchez:2010jr} \\ 
     & & & & $e^+e^- \to e^+e^- + (J/\psi\, \omega)$ &
    Belle~\cite{Uehara:2009tx},
    \babar~\cite{Lees:2012xs} \\ 
$X(3940)$ & $3942^{+9}_{-8}$ & $37^{+27}_{-17}$ & $0^{-+}(?)$ &
     $e^+e^- \to J/\psi + (D^* \bar D)$ &
     Belle~\cite{Abe:2007sya} \\ 
     &&&& $e^+e^- \to J/\psi + (...)$ &
     Belle~\cite{Abe:2007jna}\\ 
$X(4140)$ & $4146.5^{+6.4}_{-5.3}$ & $83^{+27}_{-25}$ & $1^{++}$ &
     $B \to K + (J/\psi\, \phi)$ &
     CDF~\cite{Aaltonen:2009tz}, CMS~\cite{Chatrchyan:2013dma}, D0~\cite{Abazov:2013xda}, LHCb~\cite{Aaij:2016iza,Aaij:2016nsc} \\ 
          &                         &                             &
     & $p\bar{p}\to(J/\psi\phi) + ...$ &   D0~\cite{Abazov:2015sxa} \\
$X(4160)$ & $4156^{+29}_{-25} $ & $139^{+113}_{-\ 65}$ & $0^{-+}(?)$ &
     $e^+e^- \to J/\psi + (D^* \bar{D}^*)$ &
     Belle~\cite{Abe:2007sya} \\ 
$Y(4260)$ & \multicolumn{2}{c}{see $Y(4220)$ entry} & $1^{--}$      
          & $e^+e^- \to \gamma + (J/\psi\, \pi^+\pi^-)$ &
            \babar~\cite{Aubert:2005rm,Lees:2012cn},
           CLEO~\cite{He:2006kg}, Belle~\cite{Yuan:2007sj,Liu:2013dau}  \\ 
$Y(4220)$ & $4222 \pm 3$ & $48\pm7$ & $1^{--}$ &
       $e^+e^-\to (J/\psi\, \pi^+\pi^-)$ &  BESIII~\cite{Ablikim:2016qzw}\\ 
& & & & $e^+e^-\to (h_c\, \pi^+\pi^-)$ &  BESIII~\cite{BESIII:2016adj}\\ 
& & & & $e^+e^-\to (\chi_{c0}\, \omega)$ &  BESIII~\cite{Ablikim:2014qwy} \\
& & & & $e^+e^-\to (J/\psi\, \eta)$ &  BESIII~\cite{Ablikim:2015xhk}\\ 
& & & & $e^+e^-\to (\gamma\, X(3872))$ &  BESIII~\cite{Ablikim:2013dyn}\\ 
& & & & $e^+e^-\to (\pim\, Z_c^+(3900))$ &  BESIII~\cite{Ablikim:2013mio}, Belle~\cite{Liu:2013dau}\\ 
& & & & $e^+e^-\to (\pim\, Z_c^+(4020))$ &  BESIII~\cite{Ablikim:2013wzq}\\ 
\rule[10pt]{-1mm}{0mm}
$X(4274)$ & $4273 ^{+19}_{-\ 9}$ & $56^{+14}_{-16}$ & $1^{++}$ &
     $B\to K + (J/\psi\, \phi )$ &
     CDF~\cite{Aaltonen:2011at}, CMS~\cite{Chatrchyan:2013dma}, LHCb~\cite{Aaij:2016iza,Aaij:2016nsc} \\ 
$X(4350)$ & $4350.6^{+4.6}_{-5.1}$ & $13.3^{+18.4}_{-10.0}$ & (0/2)$^{++}$ &
     $e^+e^-\to e^+e^- + (J/\psi\, \phi)$ &
     Belle~\cite{Shen:2009vs}  \\ 
$Y(4360)$ & $4341\pm 8$ & $102\pm9$ & $1^{--}$ &
     $e^+e^-\to\gamma + (\psip \, \pi^+\pi^-)$ &
     \babar~\cite{Aubert:2007zz,Lees:2012pv},
     Belle~\cite{Wang:2007ea,Wang:2014hta} \\ 
     & & & & $e^+e^-\to (J/\psi\, \pi^+\pi^-)$ &  BESIII~\cite{Ablikim:2016qzw}\\ 
$Y(4390)$ & $4392\pm6$ & $140\pm16$ & $1^{--}$ &
     $e^+e^-\to (h_c \, \pi^+\pi^-)$ &
     BESIII~\cite{BESIII:2016adj}\\ 
$X(4500)$ & $4506 ^{+16}_{-19}$  & $92^{+30}_{-21}$ & $0^{++}$ &
     $B \to K + (J/\psi\, \phi)$ &
      LHCb~\cite{Aaij:2016iza,Aaij:2016nsc} \\ 
$X(4700)$ & $4704^{+17}_{-26}$  & $120^{+52}_{-45}$ & $0^{++}$ &
     $B \to K + (J/\psi\, \phi)$ &
      LHCb~\cite{Aaij:2016iza,Aaij:2016nsc} \\ 
$Y(4660)$ & $4643\pm9$ & $72\pm11$ & $1^{--}$ &
     $e^+e^-\to\gamma + (\psip \, \pi^+\pi^-)$ &
     Belle~\cite{Wang:2007ea,Wang:2014hta}, \babar~\cite{Aubert:2007zz,Lees:2012pv} \\ 
& & & &     $e^+e^-\to\gamma + (\Lambda_c^+ \Lambda_c^-)$ &
     Belle~\cite{Pakhlova:2008vn}   \\ 
\hline
${Z_c^{+,0}(3900)}$ & $3886.6\pm 2.4$ & $28.1\pm 2.6$ & $1^{+-}$ &
     $e^+e^- \to \pi^{-,0} + (J/\psi\, \pi^{+,0})$ &
     BESIII~\cite{Ablikim:2013mio,Ablikim:2015tbp},~Belle~\cite{Liu:2013dau}  \\ 
 &  &  &  &
     $e^+e^- \to \pi^{-,0} + (D\bar{D}^{*})^{+,0}$ &
     BESIII~\cite{Ablikim:2013xfr,Ablikim:2015gda}  \\ 
${Z_c^{+,0}(4020)}$ & $4024.1\pm 1.9$ & $13\pm 5$ & $1^{+-}$(?) &
     $e^+e- \to \pi^{-,0} + (h_c\, \pi^{+,0})$ &
     BESIII~\cite{Ablikim:2013wzq,Ablikim:2014dxl}  \\ 
 &  &  &  &
     $e^+e^- \to \pi^{-,0} + (D^*\bar{D}^{*})^{+,0}$ &
     BESIII~\cite{Ablikim:2013emm,Ablikim:2015vvn} \\ 
$Z^+(4050)$ & $4051^{+24}_{-43}$ & $82^{+51}_{-55}$ & $?^{?+}$&
     $ B \to K + (\chi_{c1}\, \pi^+)$ &
     Belle~\cite{Mizuk:2008me},
     \babar~\cite{Lees:2011ik} \\ 
$Z^+(4200)$ & $4196^{+35}_{-32}$ & $370^{+\ 99}_{-149}$ & $1^{+}$&
     $ B \to K + (\jpsi \, \pi^+)$ &
     Belle~\cite{Chilikin:2014bkk} \\ 
 &  &  &  &
     $ B \to K + (\psip \pi^+)$ &
     LHCb~\cite{Aaij:2014jqa}  \\ 
$Z^+(4250)$ & $4248^{+185}_{-\ 45}$ &
     177$^{+321}_{-\ 72}$ &  $?^{?+}$ &
     $ B \to K + (\chi_{c1}\, \pi^+)$ &
     Belle~\cite{Mizuk:2008me},
     \babar~\cite{Lees:2011ik} \\ 
$Z^+(4430)$ & $4477\pm20$ & $181 \pm 31$ & $1^{+}$&
     $B \to K + (\psip \, \pi^+)$ &
     Belle~\cite{Choi:2007wga,Mizuk:2009da,Chilikin:2013tch},
     LHCb~\cite{Aaij:2014jqa,Aaij:2015zxa} \\ 
 &  &  & &
     $B \to K + (J\psi\, \pi^+)$ &
     Belle~\cite{Chilikin:2014bkk}  \\ 
\hline
$P_c^+(4380)$ & $4380 \pm 30$ & $205 \pm 88$ & $(\frac{3}{2}/\frac{5}{2})^\mp$ &
     $\Lambda_b^{+} \to K + (J/\psi \, p)$ &
     LHCb~\cite{Aaij:2015tga} \\ 
$P_c^+(4450)$ & $4450\pm 3$ & $39 \pm 20$ & $(\frac{5}{2}/\frac{3}{2})^\pm$ &
     $\Lambda_b^{+} \to K + (J/\psi \, p)$ &
     LHCb~\cite{Aaij:2015tga} \\ 
\hline\hline
$Y_b(10860)$ & $10891.1^{+3.4}_{-3.8}$ & $53.7^{+7.2}_{-7.8}$ & $1^{--}$ &
      $e^+e^- \to (\Upsilon(nS)\, 
\pi^+\pi^-)$ &
      Belle~\cite{Abe:2007tk,Santel:2015qga} \\ 
\hline
$Z_{b}^{+,0}(10610)$ & $10607.2\pm2.0$ & $18.4\pm2.4$ & $1^{+-}$ &
       $Y_b(10860) \to \pi^{-,0}+ (\Upsilon(nS)\,\pi^{+,0})$  &
      Belle~\cite{Belle:2011aa,Garmash:2014dhx,Krokovny:2013mgx} \\ 
 &  &  &  &
      $Y_b(10860) \to \pi^- + (h_b(nP)\,\pi^+)$  & 
       Belle~\cite{Belle:2011aa} \\ 
 &  &  &  &
      $Y_b(10860) \to \pi^- + (B\bar{B}^*)^+$  & 
       Belle~\cite{Garmash:2015rfd}  \\ 
$Z_{b}^+(10650)$ & $10652.2\pm1.5$ & $11.5\pm2.2$ & $1^{+-}$ &
       $Y_b(10860) \to\pi^- + (\Upsilon(nS)\,\pi^+)$  &
      Belle~\cite{Belle:2011aa,Garmash:2014dhx}  \\ 
 &  &  &  &
       $Y_b(10860) \to \pi^- + (h_b(nP)\,\pi^+)$  & 
       Belle~\cite{Belle:2011aa} \\ 
 &  &  &  &
      $Y_b(10860) \to \pi^- + (B^*\bar{B}^*)^+$  & 
       Belle~\cite{Garmash:2015rfd} \\ 

\hline\hline
\end{tabular}
\end{center}
\end{table*}

In this review, we summarize the results from this huge amount of experimental activity and discuss how these findings
reflect on theoretical ideas concerning long-distance QCD. The emphasis is on the experimental evidence, for recent
reviews that have more focus on theoretical issues, 
see refs.~\cite{Lebed:2016hpi,Esposito:2016noz,Guo:2017jvc,Ali:2017jda}.

\subsubsection{Comments on units, terminology and notation}

In this report we use ``natural units'' where $\hbar = c =1$; energy, momentum and mass are expressed in
units of either MeV or GeV.  In the case of MeV, the units of both length ([L]) and time ([T]) are 1~MeV$^{-1}$.
These can be related to conventional units by: [L]= $\hbar c/(1\ $MeV) = 197~fm and 
[T]=[L]/$c$=$\hbar/(1\ {\rm MeV})=6.58\times 10^{-22}$~s.  Also, when an experimental number is quoted, we
usually list the quadrature sum of the statistical and systematic errors. 

The spectra of $\ccbar$ charmonium and $\bbbar$ bottomonium mesons are shown above
in Figs.~\ref{fig:charmonium} and~\ref{fig:bottomonium}, respectively, where their $J^{PC}$ quantum numbers
and commonly used names are listed.   Sometimes it is convenient to describe these states using spectroscopic
notation: $n_r^{2S+1}{\rm L}_J$, where $n_r$ is the $\QQbar$ radial quantum number, 
$S=0$~or~1 is their combined spin, L=S,~P,~D... denotes their relative orbital
angular momentum and $J$ is the total angular momentum. Thus, for example, the 
$J/\psi$ and $\psip$ states shown in Fig.~\ref{fig:charmonium} are
the $1^{3}{\rm S}_{1}$ and $2^{3}{\rm S}_{1}$ $\ccbar$ states, respectively, while the $\chi_{c0}$,
$\chi_{c1}$ and $\chi_{c2}$ are the $1^{3}{\rm P}_{0,1,2}$ triplet states.

The charmonium (bottomonium)
states contain $\ccbar$ ($\bbbar$) pairs and, thus, have a zero net charm (beauty) quantum number;
these are sometimes referred to as hidden-charm (hidden-beauty) states.  Particles with a single
charmed (bottom) quark are referred to as open-charm (open-bottom) states.   Properties of the 
lowest-lying open-charm and open-bottom mesons and baryons mentioned in this report are listed
in Table~\ref{open-flavor-mesons}.  

\begin{table}[hbtp]
  \caption{\footnotesize
Properties of the lowest-lying open-charm and open-bottom particles.   Here $I:I_3$ denote
the total and third component of the isospin and ${\mathcal S}$, ${\mathcal C}$ and ${\mathcal B}$ are
the strangeness, charm and beauty quantum numbers.
}
  \label{open-flavor-mesons}\footnotesize
  \begin{center}
     \begin{tabular}{l|cccccccc}
        \hline\hline

particle   & quark    &~$J^{P}$~&~$I:I_3$~~&~~${\mathcal S}$~&~${\mathcal C}$~&~${\mathcal B}$~&  $M$     &   $c\tau$   \\ 
           & content  &        &       &               &                &                & (MeV)~    &  ~($\mu$m) \\\hline
$D^+$      &$c\bar{d}$& $0^-$  & $1/2:~~1/2$&      $0$       &      $1$       &      $0$       & $1869.6$ &   $312$    \\
$D^0$      &$c\bar{u}$& $0^-$  & $1/2:-1/2$&      $0$       &      $1$       &      $0$       & $1864.8$ &   $123$    \\
$D^{*+}$   &$c\bar{d}$& $1^-$  & $1/2:~~1/2$&      $0$        &     $1$       &      $0$        &$2010.3$  &   $\sim 0$ \\
$D^{*0}$   &$c\bar{u}$& $1^-$  & $1/2:-1/2$&      $0$        &     $1$       &      $0$        & $2007.0$ &   $\sim 0$ \\\hline
$D^+_s$    &$c\bar{s}$& $0^-$  & $0:0$   &      $1$       &     $1$       &      $0$        & $1968.3$  &   $150$    \\\hline
$\Lambda^+_c$& $cud$  &$(1/2)^+$& $0:0$   &      $0$        &     $1$       &      $0$        & $2286.5$  &   $60$     \\\hline
$\Sigma^{++}_c$& $cuu$  &$(1/2)^+$& $1:~~1$   &      $0$        &     $1$       &      $0$     & $2454.0$  & $\sim 0$  \\
$\Sigma^{+}_c$& $cud$  &$(1/2)^+$& $1:~~0$   &      $0$        &     $1$       &      $0$     & $2452.9$  & $\sim 0$    \\
$\Sigma^{0}_c$& $cdd$  &$(1/2)^+$& $1:-1$   &      $0$        &     $1$       &      $0$     & $2453.8$  & $\sim 0$    \\\hline\hline
$\bar{B}^0$&$b\bar{d}$& $0^-$  & $1/2:~~1/2$ &    $0$        &    $0$       &      $-1$        & $5279.6$  &   $455$   \\
$B^-$      &$b\bar{u}$& $0^-$  & $1/2:-1/2$ &    $0$        &    $0$       &      $-1$        & $5279.3$  &   $491$   \\
$\bar{B}^{*0}$&$b\bar{d}$& $1^-$ & $1/2:~~1/2$ &   $0$       &    $0$       &      $-1$        & $5325.2$  &  $\sim 0$ \\            
$B^{*-}$   &$b\bar{u}$& $1^-$    & $1/2:-1/2$ &     $0$      &    $0$       &      $-1$        & $5325.2$  & $\sim 0$  \\\hline 
$\bar{B}^{0}_s$ &$b\bar{s}$& $0^-$ & $0:0$ &     $1$      &    $0$       &      $-1$        & $5366.8$  &   $453$   \\\hline
$\Lambda_b$& $bud$  &$(1/2)^+$   & $0:0$   &    $0$       &    $0$       &      $-1$        & $5619.5$  &   $435$   \\\hline\hline 
  \end{tabular}
  \end{center}
\end{table}

Limits on the electric dipole moment of the neutron confirm that QCD is
matter-antimatter symmetric to a high degree of confidence~\cite{Afach:2015sja}.
In addition, the experimental environments of the measurements discussed in this
report are also mostly matter-antimatter symmetric.  Thus, the data samples that are
used for these measurements usually include charge-conjugate reactions.  For
example, a measurement of a $\pi D\bar{D}^*$ system will use a combined set of
$\pip D\bar{D}^*$, $\pip D^*\bar{D}$, $\pim D\bar{D}^*$, and $\pim D^*\bar{D}$
events. In this report, for simplicity and readability we abbreviate this to
$\pip D\bar{D}^*$ with the implicit assumption that charge-conjugate 
combinations are included.  For similar reasons, when we discuss
meson-antimeson molecule-like possibilities, we abbreviate combinations like
$(D\bar{D}^*\pm \bar{D}D^*)/\sqrt{2}$ to simply $D\bar{D}^*$.

\section{Models for non-standard hadrons}

In the absence of any rigorous analytical method for making first-principle calculations of the
spectrum of non-standard hadrons, simplified models that are motivated by the color structure and
other general features of QCD have been developed.  The current best hope for a rigorous,
first-principle treatment of at least some of the issues discussed here is lattice QCD, which is
discussed later in this section.

The color structure of QCD suggests the existence of three types of non-standard hadronic particles.
These include: multiquark hadrons (tetraquark mesons and pentaquark baryons) formed from
tightly bound colored diquarks; hybrid mesons and baryons comprised of color-singlet combinations
of quarks and one or more ``valence'' gluons; and glueball mesons that are comprised only of gluons (with
no quarks). Other possible forms of multiquark states are meson-meson and/or meson-baryon
molecule-like systems that are bound (or nearly bound) via Yukawa-like nuclear forces, and
bound states comprised of quarkonium cores surrounded by clouds of light quarks and gluons.

\subsection{QCD-color-motivated models}

\subsubsection{QCD diquarks}

\begin{figure}[htb]
  \includegraphics[width=0.45\textwidth]{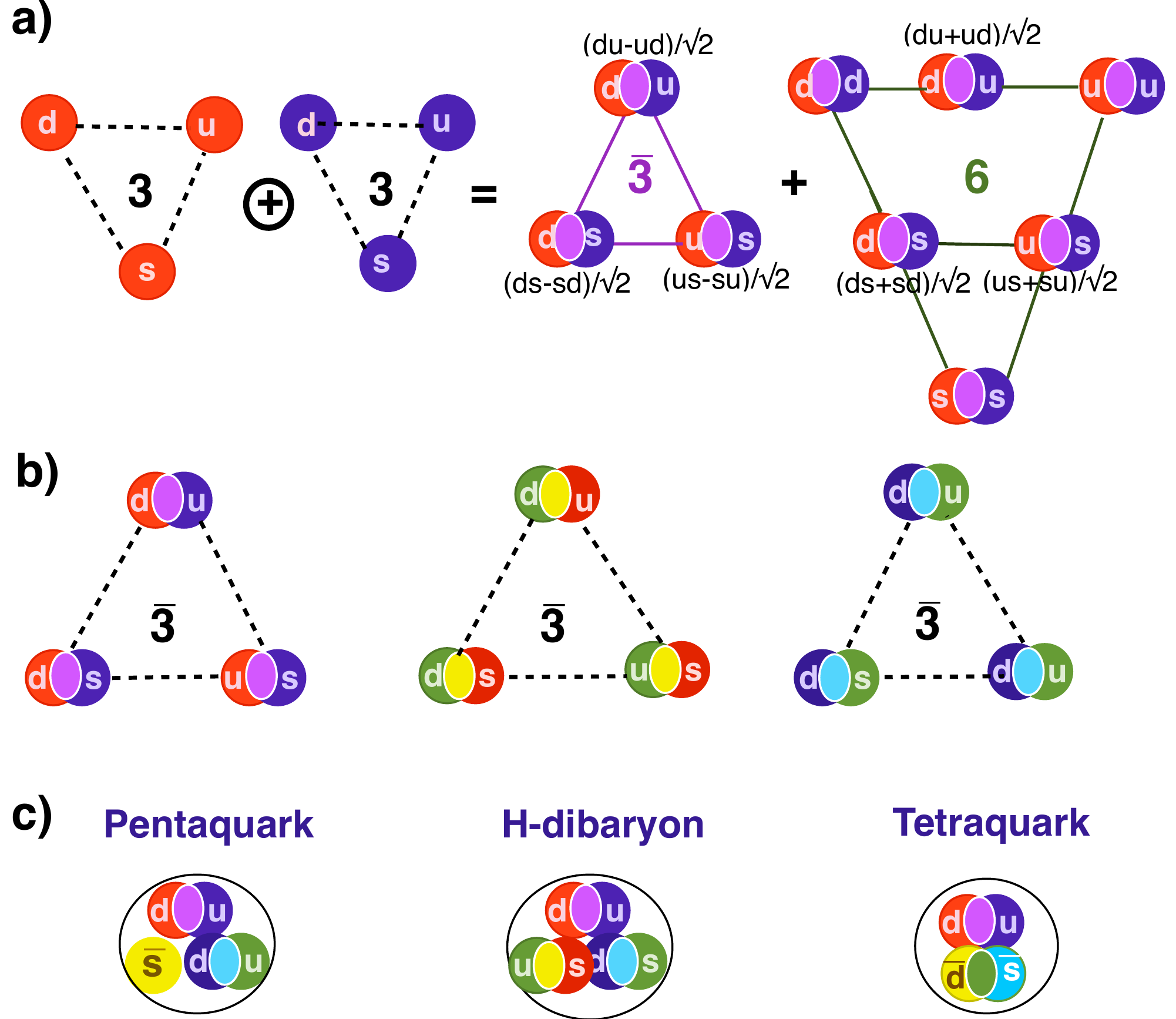}
\caption{\footnotesize {\bf a)} Combining a red and blue quark triplet produces a
magenta (antigreen) antitriplet and sextet.  The antitriplet is antisymmetric in color and flavor
while the sextet is symmetric in both quantities. {\bf b)}  The three anticolored diquark antitriplets.
{\bf c)}  Some of the multiquark, color-singlet states that can be formed from quarks,
antiquarks, diquarks and diantiquarks.}
\label{fig:diquarks}
\end{figure}

It is well known that the combination of a $q=u,d,s$ light-quark triplet with a
$\bar{q}=\bar{u},\bar{d},\bar{s}$ antiquark antitriplet gives the familiar meson nonets (an octet plus 
a singlet) of flavor-$SU(3)$.  Using similar considerations based on QCD~\cite{Jaffe:1976ig}, a red and a blue
quark triplet can be combined to form a magenta (antigreen) antitriplet of $qq'$ ``diquarks'' that 
is antisymmetric in both color and flavor, and a magenta flavor-symmetric sextet, as illustrated in
Fig.~\ref{fig:diquarks}a.  The Pauli principle restricts the spin state of antitriplet quarks to
$S=0$ and that of the sextet quarks to $S=1$.  Since the single-gluon-exchange color force between the
quarks in an $S=0$ antitriplet diquark is attractive, Jaffe designated these as ``good'' diquarks and
those in an $S=1$ sextet, where the short-range force is repulsive,  as ``bad'' diquarks~\cite{Jaffe:2004ph}.
From the nucleon and $\Delta^0$-baryon mass difference he estimated the difference in binding between light ``bad''
and ``good'' diquarks to be $\sim\frac{2}{3}(m_{\Delta}-m_N)\sim 200$~MeV.

Likewise, green-red and blue-green diquarks form yellow (anti-blue)
and cyan (anti-red) antitriplets as shown in Fig.~\ref{fig:diquarks}b.  Thus, in color space,
a ``good'' diquark antitriplet looks like an antiquark triplet with baryon number $B=+2/3$ and spin=0. 

Since these diquarks are not color-singlets, they cannot exist as free particles but, instead, they should be
able to combine with other colored objects in a manner similar to antiquark antitriplets, thereby forming
multiquark color-singlet states with a more complex substructure than the $\qqbar$ mesons and $qqq$ baryons of
the original quark model. Jaffe proposed that the puzzles associated with the low-mass $0^{++}$ mesons, discussed
above in Section~\ref{sec:intro_light}, could be explained by identifying them as four-quark combinations
of a diquark and a diantiquark.  In this scheme, the $a_0(980)$ isotriplet mesons are formed from
[$qs$]-[$\bar{q}\bar{s}$] ($q=u$~or~$d$) configurations and their large mass relative to other octet members
is due to the two $s$ quarks among its constituents~\cite{Jaffe:1976ih,Maiani:2004uc,Hooft:2008we}.  In addition to the light scalar mesons,
diquarks and/or diantiquarks could be constituents of other octets of tetraquark mesons, as well as pentaquark
baryons and six-quark $H$-dibaryons, as illustrated in Fig.~\ref{fig:diquarks}c.

These considerations are expanded to include heavy-light diquarks ($Qq$) and diantiquarks
($\bar{Q}\bar{q}$) in refs.~\cite{Maiani:2004vq} and~\cite{Terasaki:2004yx}.  The $Qq$ ($\bar{Q}\bar{q}$)
combinations are color-$SU(3)$ antitriplets (triplets) and flavor-$SU(3)$ triplets (antitriplets).
In this case, since the spin-spin force between the quarks is reduced by a factor of $m_q/m_Q$ and the
mass difference between ``bad'' and ``good'' diquarks is reduced.  As  a result $S=1$ $Qq$ diquarks are not so
``bad'' and both $S=0$ and $S=1$ diquarks could be expected to play important roles in hadron
spectroscopy~\cite{Manohar:2000dt}.  More detailed discussions of diquark models are provided in 
refs.~\cite{Esposito:2014rxa,Esposito:2016noz}

\subsubsection{QCD hybrids}

The linear confining term in the color-force potential produces a force between a meson's constituent 
quark and antiquark that is constant with increasing separation.  As a result, unlike the electric
field lines between opposite charges in QED, which spread out in space, the color field-lines are configured
in a tightly confined ``flux tube'' that runs between the $q$ and the $\bar{q}$~\cite{Isgur:1983wj}.

In their lowest mass configurations, the flux-tube is in a ground state with angular momentum
quantum numbers $L=0$ and $S=0$, and only the relative orbital angular momentum of the quarks
and their net spin determine the quantum numbers of a state; the gluonic degrees of freedom
do not play any role.  As a result, the $J^{PC}$ quantum numbers of these ground-state
or ``conventional'' mesons, where $\vec{J}=\vec{L}+\vec{S}$, $P=(-1)^{L+1}$ and $C=(-1)^{L+S}$ are 
restricted to values that can be accessed by a quark-antiquark pair: $J^{PC}=0^{++}$, $0^{-+}$,
$1^{++}$, $1^{+-}$, $1^{--}$, $2^{++}$, $2^{-+}$, $2^{--}$...; other quantum number combinations, namely
$J^{PC}=0^{--}$, $0^{+-}$, $1^{-+}$, $2^{+-}$..., are inaccessible and are called ``exotic.''  However, if the
flux tube is in an excited state, its orbital angular momentum and/or spin can
be non-zero, and contribute $L$ and $S$ values that are consistent with one or more gluons.
In this case they contribute to the overall quantum numbers of the state, producing mesons
with exotic quantum number assignments~\cite{Horn:1977rq}.  Since gluons have zero isospin,
quarkonium hybrids, {\it i.e.} $Q\bar{Q}$-$g$ states, are necessarily isospin singlets.

Models for the decays of hybrids find that decays to identical mesons are strongly suppressed,
while decays to two different mesons where one is a $\qqbar$ in an S-wave and the other a
$\qqbar$ in a P-wave are enhanced~\cite{Isgur:1985vy,Page:1998gz}. The predicted widths for $\pi\pi$ or
$K\bar{K}$ final states for light quark hybrids are small, as are the $D\bar{D}$ and $B\bar{B}$ decay
widths for quarkonium hybrids. In contrast, light hybrid decays to $a_1\pi$, $b_1\pi$ and $K_1(1400)\bar{K}$
decays, where $a_1$, $b_1$ and $K_1$ are axial-vector mesons, in which the $\qqbar$ pair is in a
relative P-wave, are expected to be strong.  Likewise, quarkonium hybrids are expected to have 
strong decay widths for $D^{**}\bar{D}^{(*)}$ and $B^{**}\bar{B}^{(*)}$ final states, where $D^{**}$ and $B^{**}$
denote open charm $(c\bar{q})$ and beauty $(b\bar{q})$ ($q=u,d$) P-wave states, respectively.

A recent review of hybrid mesons by  Meyer and Swanson~\cite{Meyer:2015eta} points out limitations
 in this naive but useful 30-year old picture
and provides references to current computations based on the lattice gauge theory.

\subsection{Other models}

\subsubsection{Hadronic molecules}

\begin{figure}[htbp]
\includegraphics[width=0.48\textwidth]{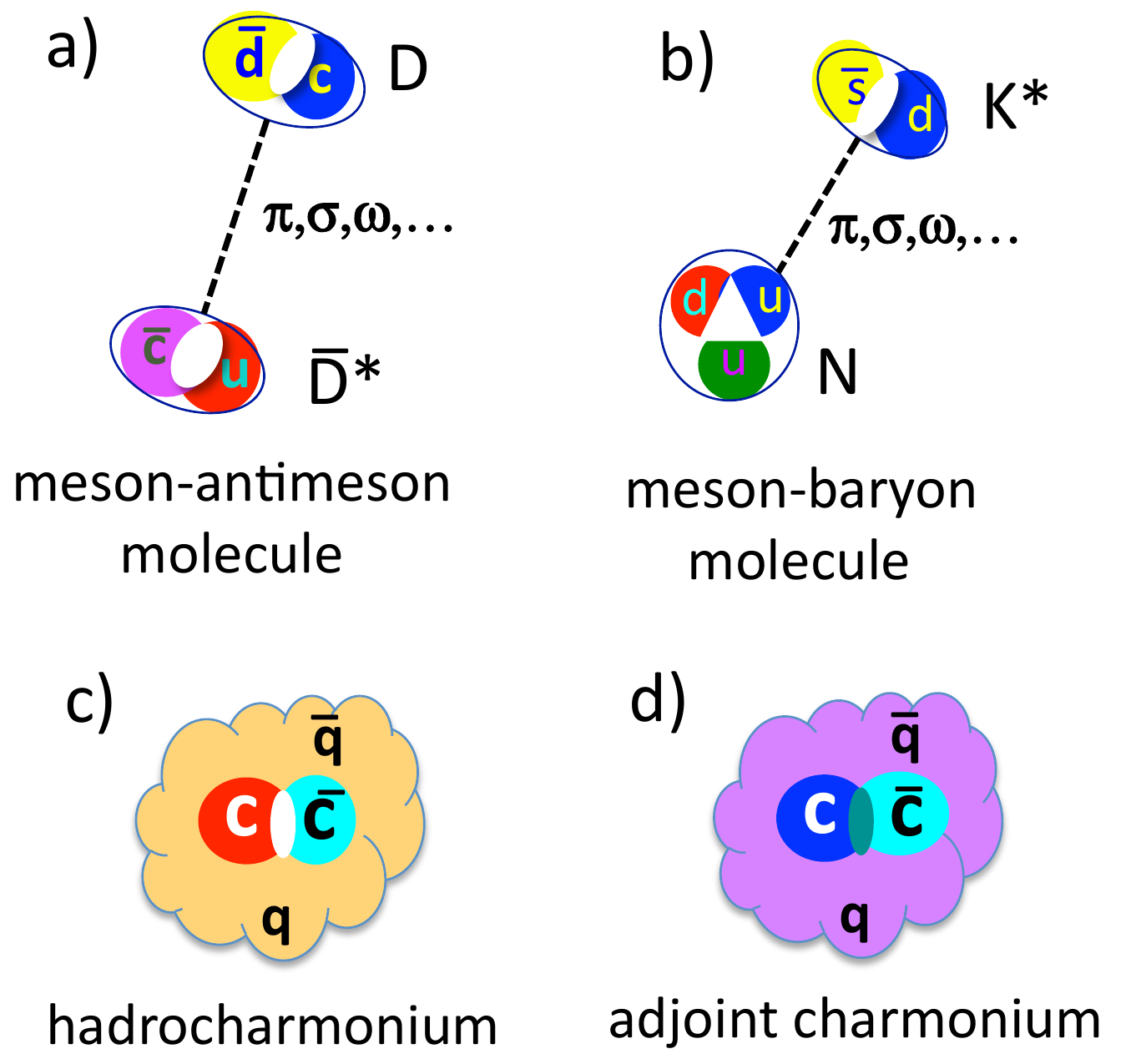}
\caption{\footnotesize Illustrations of a {\bf a)} meson-meson and a {\bf b)} meson-baryon molecular-like
structure bound by Yukawa-type meson exchange forces.  {\bf c)} A sketch of the hadrocharmonium configuration
of multiquark states.  Here a color-singlet $Q\bar{Q}$ core state interacts with a surrounding `blob'' of
gluons and light quarks via QCD versions of Van der Waals type forces.  {\bf d)} In adjoint charmonium states,
a color-octet $Q\bar{Q}$ pair interacts with surrounding gluons and light quarks via color forces.}
\label{fig:molecule-fig}
\end{figure}

The idea that Yukawa-type meson exchange forces could produce deuteron-like bound states of ordinary,
color-singlet hadrons, as illustrated in Fig.~\ref{fig:molecule-fig}a~and~\ref{fig:molecule-fig}b,
has been around for a long time~\cite{Voloshin:1976ap,Bander:1975fb,DeRujula:1976zlg,Manohar:1992nd}.
These ``molecular'' states are expected to have masses that are near the constituent particles' mass
threshold, and to have spin-parity
($J^{PC}$) quantum numbers that correspond to an S-wave combination of the constituent particles.  For the
deuteron, single pion exchange is the most important contributor to its binding. T\"ornqvist studied the
possibility for forming deuteron-like $B\bar{B}^*$ and $B^*\bar{B}^*$ states, which he called ``deusons,''
using a single pion exchange potential and concluded that such states ``certainly must
exist''~\cite{Tornqvist:1993ng}; he also predicted that if some small additional attraction was provided
by shorter range exchanges, bound $D\bar{D}^*$ and $D^*\bar{D}^*$ systems would also exist.

Since three-pseudoscalar couplings like $D\bar{D}\pi$ and $B\bar{B}\pi$ are forbidden by rotation plus parity
invariance, single-pion exchange forces do not contribute to $D\bar{D}$ or $B\bar{B}$ binding and, thus,
molecule-like structures in these systems are not expected to occur. 

In molecule-like states formed from pairs of
open-charm or open-beauty mesons that are primarily bound  by single $\pi$-meson exchange, the heavy $Q$
and $\bar{Q}$ quarks are typically well separated in space with very little overlap. This suggests that
``fall-apart'' decay modes to pairs of open-flavor mesons would be dominant, while decays to final states in
which the $Q$ and $\bar{Q}$ quarks coalesce to form a hidden-flavor quarkonium state would be rather strongly
suppressed.  More detailed discussions of molecular models are provided in
refs.~\cite{Swanson:2006st,Polosa:2015tra,Lebed:2016hpi,Guo:2017jvc}.

\subsubsection{Hadrocharmonium}
For conventional charmonium states with masses above the open-charm ({\it i.e.}, $D\bar{D}^{(*)}$) threshold,
the branching fractions for fall-apart decays to pairs of open-charm mesons are measured to be two or three
orders-of-magnitude higher than decays to hidden-charm final states.\footnote{For example, 
${\mathcal B}(\psi(3770)\rt D\bar{D})=(93^{+8}_{-9})\%$  while
${\mathcal B}(\psi(3770)\rt\pipi\jp)=(0.193\pm 0.028)\%$~\cite{Olive:2016xmw}.}  This is not the case
for many of the non-standard hadrons discussed here, where hidden quarkonium modes are a common discovery channel
with branching fractions that are lower that open-flavor fall-apart modes, but only by factors of ten or less. 
The hadrocharmonium model was proposed by Dubynskiy and Voloshin~\cite{Dubynskiy:2008mq} in order to account for this
property.  In this model,  a compact color-singlet $Q\bar{Q}$ charmonium core state is embedded in a spatially extended
``blob'' of light hadronic matter.  These two components interact via QCD versions of the Van der Waals force. They
find that the mutual forces in this configuration are strong enough to form bound states if the light hadronic
matter is a highly excited resonant state.  In this model, decays to the hidden charmonium core state are enhanced
to a level where they are competitive with those for fall-apart modes \cite{Dubynskiy:2008di}.  
Allowing for a sizable branching fraction into open charm modes 
requires a careful tuning of the model parameters.

\subsubsection{Born-Oppenheimer model}
An ``all of the above'' approach that incorporates all of the configurations discussed above, plus the
adjoint charmonium configuration illustrated in Fig.~\ref{fig:molecule-fig}d, which is like hadrocharmonium
except with an allowance for the possibility that the $Q\bar{Q}$ core state has non-zero color, has been advocated
by Braaten, Langmack and Smith~\cite{Braaten:2013boa,Braaten:2014qka}.
This model is modeled on the Born-Oppenheimer approximation that is used in atomic and molecular physics
to treat the binding of atoms into molecules. In this approach, the slow-moving
atomic nuclei are replaced by the heavy quarks and the potential that describes the interaction of
the positive nuclear charges and the surrounding negative electron clouds are replaced by lattice-QCD
computed gluon-induced potentials~\cite{Juge:1999ie}.  A first application of this approach was used
to predict the masses of the lowest lying charmonium hybrid and tetraquark mesons.

\subsubsection{Kinematically induced resonance-like mass peaks}
\label{sec:kinematicpeaks}
While the classic signal for the presence of an unstable hadron resonance is a peak in the invariant mass
distribution of its decay products, not all mass-spectrum peaks are genuine hadron states. Some can be
produced by near-threshold kinematic effects.  These include threshold ``cusps,'' and anomalous triangle
singularities.

\begin{figure}[htbp]
\includegraphics[width=0.45\textwidth]{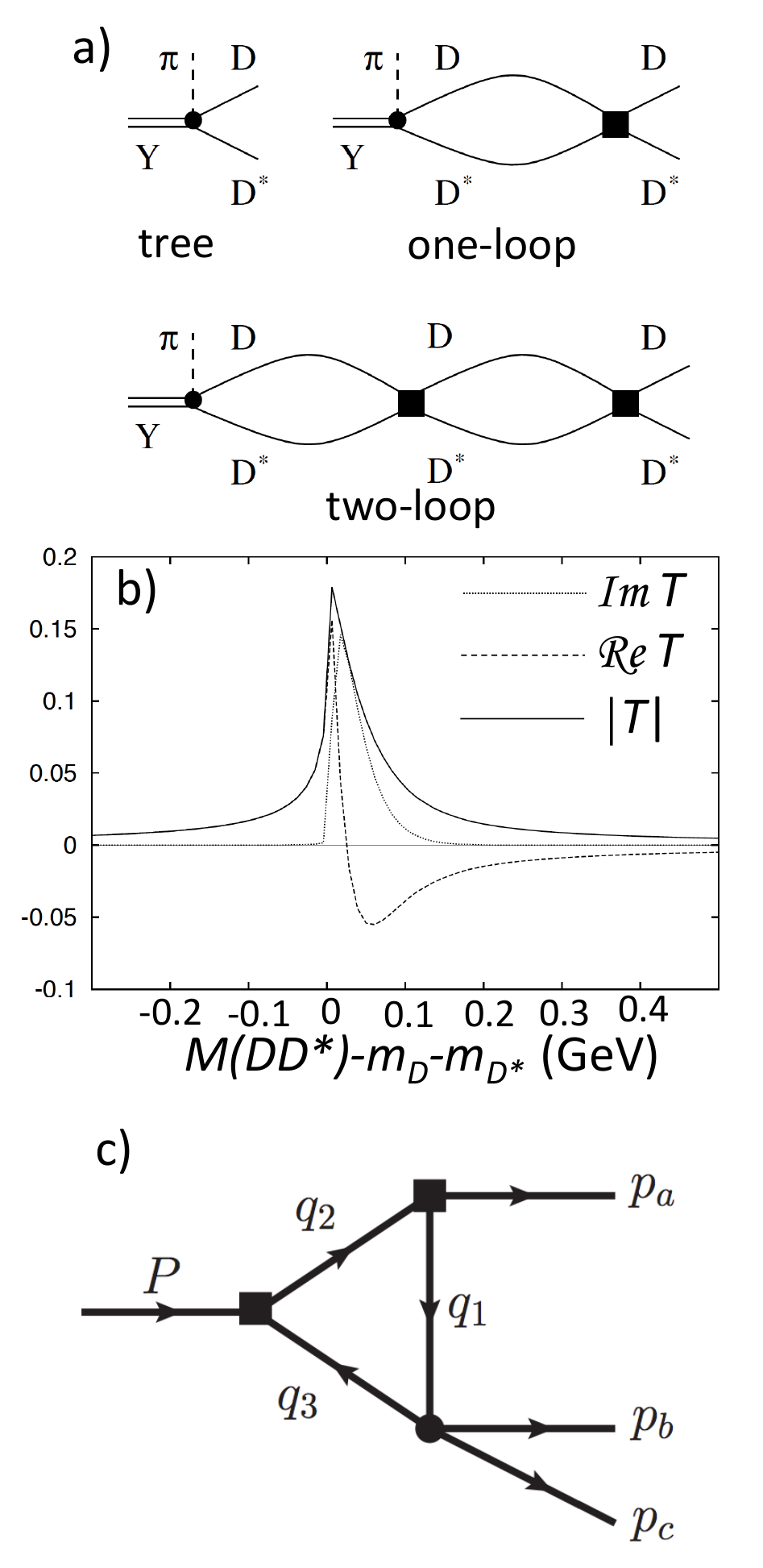}
\caption{\footnotesize {\bf a)} Low order diagrams that describe a three-body decay process
$Y\rt\pi D\bar{D}^*$ scattering near the $\sqrt{s}=m_D + m_{D^*}$ threshold (taken from ref.~\cite{Swanson:2014tra}).
The dotted curve is $Im T$; the dashed curve is $Re T$ and the solid curve is $|T|$.
{\bf b)} A three-body decay with an internal triangle.  This diagram is singular when the three 
virtual particles that form the triangle are all simultaneously on the mass shell. 
 \label{fig:cusp-diagrams}
}
\end{figure}

\underline{Threshold cusps:} Figure~\ref{fig:cusp-diagrams}a shows the three lowest-order diagrams for three-body
decays $Y\rt\pi D\bar{D}^*$. Consider the one-loop diagram, where the $D$ and $\bar{D}^*$ elastically rescatter.
If the two particles are in an S-wave, the imaginary part of the scattering amplitude is zero for
$M(D\bar{D}^*)<(m_D+m_{\bar{D}^*})$ and abruptly rises at threshold 
as~\cite{Bugg:2011jr,Blitz:2015nra,Swanson:2015bsa}:
\begin{equation}
Im T(s)\propto g^2\rho(s),
\label{eq:gq}
\end{equation}
where $g$ is the coupling constant and $\rho(s)$ is the phase-space factor.  In order to
eventually terminate the growth of $\rho(s)$ before $Im T$ increases to an unphysically large value, 
it has to be attenuated by a hadronic form factor, $F(s)$, in which case
\begin{equation}
\rho(s)=\frac{2k}{\sqrt{s}}F(s),
\end{equation} 
where $k$ is the momentum of one of the particles in the two-particle rest frame. 
For $T(s)$ to be analytic, it must have a real part of the form
\begin{equation} 
Re T(s) = \frac{1}{\pi}P \int_{s_{\rm thresh}}^{\infty} \frac{ds' g^2(s')\rho(s')}{s'-s},
\end{equation}
where $P$ denotes the principal value integral and $s_{\rm thresh}$ is the mass-squared at threshold. 
The resulting $|T|$ has a very sharp, cusp-like structure that peaks slightly above the $M=\sqrt{s_{\rm thresh}}$ 
threshold; this peak originates from kinematics and has nothing to do with any resonant structure in
the $D\bar{D}^*$ two-body system.  The $M(D\bar{D}^*)$ behavior of $|T|$ and its real and imaginary
parts is shown in Fig.~\ref{fig:cusp-diagrams}b.  
It is argued in ref.~\cite{Guo:2014iya} that genuine cusp effects are small, 
and that the cusp-based models \cite{Bugg:2011jr,Blitz:2015nra,Swanson:2015bsa}, 
which described the significant near-threshold peaks in the experimental data, 
enhanced the cusp effect by introduction of ad-hoc, non-analytic form factors in the coupling constant,
$g\to g(s)$ in Eq.~\ref{eq:gq}, which invalidates the approach.

\underline{Anomalous triangle singularity:} In three-body decays, diagrams that contain internal triangles, as illustrated in
Fig.~\ref{fig:cusp-diagrams}c, may contribute.  In 1959, Landau showed that
this diagram becomes singular when the three virtual particles that form the triangle
are all simultaneously on the mass shell~\cite{Landau:1959fi}. This is called the anomalous
triangle singularity (ATS). 
In kinematic regions where the conditions for this
singularity are satisfied \cite{Coleman:1965xm}, 
resonance-like peaking structures that have nothing to do with
true particle resonances can be produced.  
It has been shown that, if the rescattering is purely elastic, 
the effect of the triangle singularity integrates
to zero in the Dalitz plot projections \cite{Schmid:1967ojm}.
However, in case of many coupled channels, this theorem applies
to the sum of intensities of all of them, thus the Dalitz plot 
projections to individual channels 
can produce mass peaks \cite{Szczepaniak:2015hya}. 
Properties of the ATS and methods for
distinguishing ATS-induced mass peaks from genuine resonances are discussed in
refs.~\cite{Liu:2015taa,Szczepaniak:2015eza,Pilloni:2016obd}.

\subsection{Lattice QCD}
\label{sec:latticeQCD}

In QCD, information about the mass of a  hadron $H$ is encoded in the correlation function $C_H(t)$ of the hadron
creation operator ${\mathcal O}_{H}$ evaluated at different times: 
$C_H(t) = \langle \Omega | \mathcal{O}_H^\dagger(t) \mathcal{O}_H(0) | \Omega \rangle$
(see {\em e.g.} ref.~\cite{Weber:2013eba}).  
Here the state $H$ is created from the vacuum $|\Omega\rangle$ at time $t=0$
and  propagates until time $t$ when it is annihilated.  The operator $\mathcal{O}_H(t)$ is a suitable composition of
quark and gluon field operators, {\it e.g.}, for a pion, which is a pseudoscalar $\bar u  d$ state, it is 
$\mathcal{O}_\pi(t) \ \ = \ \ \int d^3r \, \bar{u}(\mathbf{r},t) \gamma_5 d(\mathbf{r},t)$,
where $d(\mathbf{r},t)$ and $\bar{u}(\mathbf{r},t)$ are the $d$- and $\bar{u}$-quark creation operators.
The integration extends over all possible  spatial configurations of the quark
and gluon
fields.
To avoid the oscillating behavior of the correlator in real time, the integration is
performed in the Euclidean space-time where the time coordinate is imaginary.
The hadron mass is determined from the
correlation function's asymptotic exponential behavior  $C_H(t) \propto \exp(-m_Ht)$.  

The path integral is impossible to solve analytically.
A major conceptual breakthrough occurred in 1974 when Wilson proposed~\cite{Wilson:1974sk} that long-distance QCD could be digitized by
transcribing the relevant integrals to a lattice of discrete space-time points, where quark fields placed at
lattice sites interact with each other by interconnected gluon links.  The resulting equations are solved
numerically by using Monte Carlo techniques to generate random samples of all possible
configurations.  

The difficulty with this approach is that realistic lattice QCD (LQCD) computations require extreme computational
resources, much beyond those that were available when Wilson first proposed his ideas. Ideally, the lattice should
be several fermi in extent in order to fully contain a hadron while the lattice spacing must be small enough to minimize
discretization errors.  With a  lattice of 32 sites in each of the four dimensions, there are $32^4 \approx 10^6$
lattice sites.  With quark fields restricted to just two quark flavors, $u$ and $d$, each with a real and 
imaginary part, with three colors and four spin components, plus 32 gluon degrees of freedom (8~color$\times$4~spin),
the dimension of the integral is  $32^4\times (2\times 24+32)\approx 10^8$.

Happily, Wilson's dream is now becoming a reality thanks to the relentless, Moore's law-like advance in high-performance
computing.   Systems capable of providing hundreds of teraflop/s-yrs by exploiting
large-scale parallel programming techniques with calculations running cooperatively across thousands of processors are now
available~\cite{Blum:2013mhx}.\footnote{One teraflop/s-yr is defined as the number of floating-point operations performed in
a year by a computer that sustains one trillion operations per second.}  This, coupled with major advances in LQCD algorithms
(see ref.~\cite{Orginos:2015tha} for a recent review), has resulted in a number important recent results related to hadron masses.

\begin{figure}[htb]
\begin{center}
\includegraphics[width=0.48\textwidth]{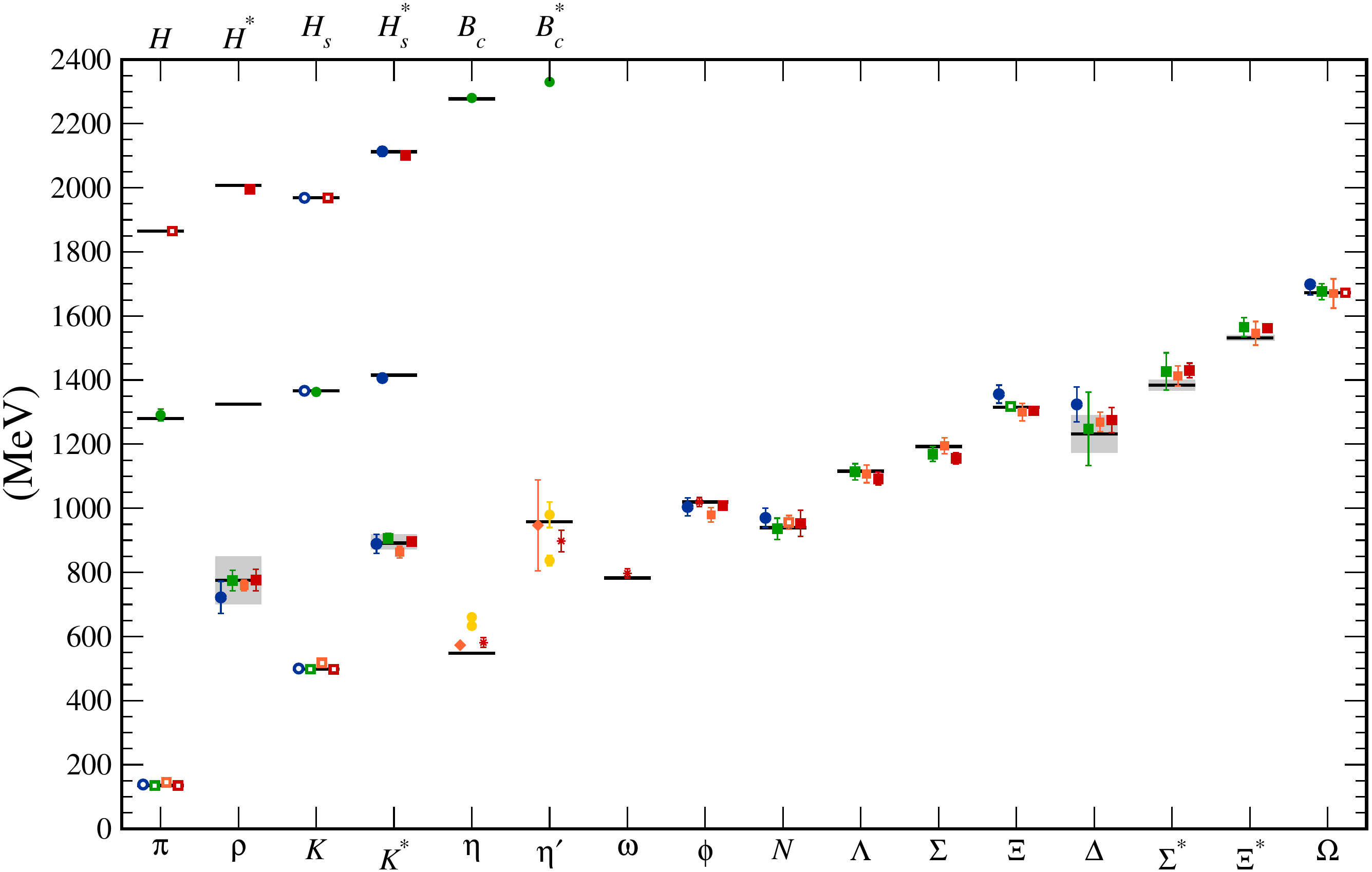}
\end{center}
\caption{\footnotesize The LQCD hadron spectrum 
from   MILC~\cite{Aubin:2004wf,Bazavov:2009bb},
PACS-CS~\cite{Aoki:2008sm}, BMW~\cite{Durr:2008zz}, QCDSF~\cite{Bietenholz:2011qq},
RBC \& UKQCD~\cite{Christ:2010dd},
Hadron Spectrum~\cite{Dudek:2011tt}, 
UKQCD~\cite{Gregory:2011sg}, Fermilab-MILC~\cite{Bernard:2010fr},
HPQCD~\cite{Gregory:2010gm}, and
Mohler \& Woloshyn~\cite{Mohler:2011ke}.
The $b$-flavored meson masses are offset by $-4000$~MeV.
 Horizontal bars (gray boxes) denote experimentally measured masses (widths).
(Figure from ref.~\cite{Kronfeld:2012uk}.)}
\label{fig:QCDSFspectrum}
\end{figure}

For example, the QCDSF Collaboration~\cite{Bietenholz:2011qq}  reported a lattice calculation of the masses
of hadrons composed of $u$, $d$, and $s$ quarks, ranging from the $\eta$ meson to the $\Omega ^-$ baryon using only the charged
pion and kaon masses and a combination of the $p$, $\Sigma$, and $\Xi$ masses as inputs; the only tuneable parameters are the  quark masses and the coupling constant $\alpha_s$.
Recent Results for  mesons and baryons   
are shown in Fig.~\ref{fig:QCDSFspectrum}, where there is a good agreement 
with the established values.

\begin{figure}[htbp]
\centering
\includegraphics*[width=0.48\textwidth]{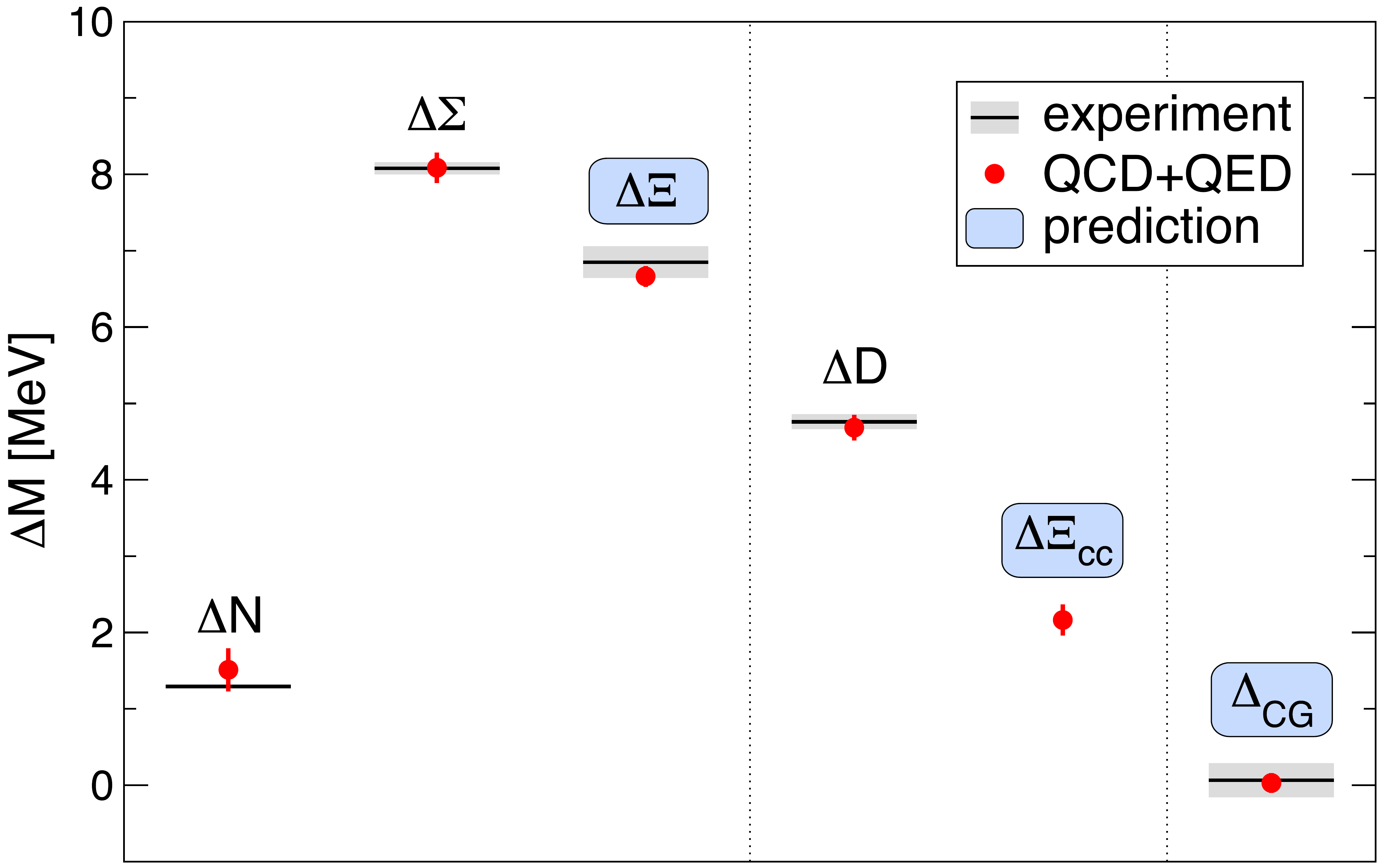}
\caption{ \footnotesize
Results of the lattice computations of $\Delta N = m_{n}-m_{p}$, $\Delta\Sigma = m_{\Sigma^-}- m_{\Sigma^+}$,
$\Delta \Xi = m_{\Xi^-}-m_{\Xi^0}$, $\Delta D= m_{D^+}-m_{D^0}$ and $\Delta \Xi_{cc} = m_{\Xi_{cc}^{++}}-m_{\Xi_{cc}^{+}}$ 
isospin mass splittings, and a test of the Coleman-Glashow
relation~\cite{Coleman:1961jn} $\Delta_{\rm CG}\equiv \Delta M_{\rm N}-\Delta M_{\Sigma}-\Delta M_{\Xi}=0$  
from ref.~\cite{Borsanyi:2014jba}.   The horizontal lines are the experimental values
and the grey shaded regions represent the experimental error.
The computed precision for the quantities with labels in  blue shaded boxes is better than that of
current measurements.  
}
\label{fig:qced_splittings}
\end{figure}

To date, because of computing-power constraints, most LQCD computations ignore isospin violations and set the $u$-~and~$d-$quark
masses equal.  However, precision lattice results on QCD-generated isospin violations are now being realized. 
Borsanyi {\em et al.}~\cite{Borsanyi:2014jba} have reported a lattice-based, ab-initio computation of the
(1.293~MeV) neutron-proton mass difference that results from the competition between electromagnetic and QCD-induced
isospin-breaking effects\footnote{
The calculation reported in ref.~\cite{Borsanyi:2014jba}
finds a QCD contribution to $m_{n}-m_{p}$ that is $2.52\pm 0.49$ times larger than that from the (opposite-sign)
electromagnetic effect.  The magnitude of this QCD contribution has huge existential significance; an increase or
decrease by as little as $\sim 20$\% would have dire consequences on Nature's ability to support life
(see ref.~\cite{Wilczek:2015exa}).}  with an accuracy of $300$~keV.
They also determined mass splittings in the $\Sigma$, $\Xi$, $D$ and $\Xi_{cc}$\footnote{The $\Xi_{cc}$ is a candidate
for a  doubly charmed $ccq$ baryon with mass $M=3820\pm 1.0$~MeV that was reported by the SELEX
experiment~\cite{Mattson:2002vu,Ocherashvili:2004hi}  but was not confirmed by other
experiments~\cite{Chistov:2006zj,Aubert:2006qw,Aaij:2013voa}. 
The LHCb group recently reported a $12\sigma$ signal for a $\Xi_{cc}$ candidate at a lower mass of $3621.4\pm0.8$ MeV \cite{Aaij:2017ueg}.} 
isospin multiplets with precision that is better, in
some cases, than that of the currently available experimental measurements, as shown in Fig.~\ref{fig:qced_splittings}. 

\noindent

The spectrum of mesons carrying one charmed quark, or a charmed-anticharmed pair, has been recently
computed on the lattice by Cichy {\em et al.}~\cite{Cichy:2016bci}.
To tune the valence quark masses the authors used
experimental values of the masses of electrically neutral and charged $\pi$, $K$,
and $D$ mesons. 
Using a variety of quark-antiquark meson creation operators the authors
 were able to determine the masses of the lowest-lying
1S and 1P charmonium states with levels of precision that are in the range $0.2 \sim 0.8$ percent.
Cichy {\em et al.}~\cite{Cichy:2016bci}   also successfuly verified  the  masses of several
charm mesons with the exception of  the $D_{s0}^*(2317)$ and $D_{s1}(2460)$ (see sec.~\ref{sec:charmheavylight} below), 
which have masses close to  two-meson thresholds and, thus, require more advanced  
techniques \cite{Leskovec:2015naf}, as discussed in sec.~\ref{sec:charmheavylight}.

Determining the highly-excited resonance spectra has recently become possible thanks
to a technique proposed by Luscher~\cite{Luscher:1990ux}.
The Hadron Spectrum Collaboration~\cite{Liu:2012ze} did a comprehensive study of the spectrum of excited charmonium
mesons with masses up to 4.5~GeV that included possible $\ccbar$-gluon hybrid states.  
They find the
lightest $\ccbar$-gluon hybrids are a $0^{-+}$ pseudoscalar with $M\simeq 4195$~GeV; a $1^{-+}$ ``exotic''
with $M\simeq 4215$~MeV and a $1^{--}$ vector with $M\simeq 4285$~MeV. One of the non-standard mesons discussed in
this report is the $Y(4260)$ vector state that is considered by some authors to be a promising candidate for a
$1^{--}$ hybrid state. (This is discussed below in Sect.~\ref{sec:y4260}). 
It will be interesting to see what happens to the LQCD-computed mass value when calculations 
extended to three-particle resonances~\cite{Hansen:2014eka,Hansen:2015zga,Briceno:2016ffu}    become feasible in the next decades.

Recently, first attempts to address questions related to possible quarkonium-like, four-quark mesons
have been reported.  For example, in order to get insight into  the structure of the $X(3872)$, a candidate
non-standard meson with mass very near the $D\bar{D}^*$  threshold ({\it i.e.}, $M(X(3872)\simeq m_D + m_{D^*}$)
and discussed in detail in Sect.~\ref{sec:x3872}, Padmanath {\it et al.}~\cite{Padmanath:2015era} made a LQCD
study that included standard $\ccbar$ charmonium, $D\bar{D}^*$ meson-meson, and  diquark-diantiquark operators,
22 in total, that allowed the particle to be a superposition of all three configurations. Their result indicated
the presence of a $c\bar c$ and a meson-meson ($D\bar D^*$) component, but with no sign of a diquark-diantiquark
component,  leading the authors to the conclusion that  a QCD tetraquark interpretation of the  $X(3872)$ was disfavored. 
Bicudo {\em et al.}~\cite{Bicudo:2015kna} 
considered the possible existence of bound states in the $\bar b \bar b u d$ four-quark systems with a net
beauty flavor of ${\mathcal B}=2$. Their first exploratory simulations found signs of a state with a binding energy of
$-90^{+43}_{-36}$ MeV, {\it i.e.}, about 2$\sigma$ from zero.
This was followed up by more precise calculations - see ref.~\cite{Francis:2016hui} and references within.
For a review of the searches of resonances with LQCD see ref.~\cite{Briceno:2017max}.

Lattice QCD efforts in the area of non-standard hadron spectroscopy, while still in their infancy, are
very encouraging.  With the order-of-magnitude increase in the available computing power expected during the
next decade (see, {\em e.g.},~ref.~\cite{Blum:2013mhx}) and continued advances in the sophistication of the
algorithms that will be used to extract physics information from the improved configurations, LQCD seems to be
on the verge of becoming a very powerful tool for deriving a more profound theoretical understanding of the
recently discovered states and for providing important guidance for future experiments.   
 
\myclearpage

\section{Heavy Flavor Experiments}
\label{sec:experiments}

The results reviewed here come from experiments that operate at vastly different energies.
At the low energy extreme is the BESIII experiment at the Institute of High Energy Physics
in Beijing that operates at the BEPCII $\ee$ collider and can access c.m.~energies between
2~and~4.6~GeV. At the high energy extreme are the ATLAS, CMS and LHCb experiments operating
at the LHC $pp$ collider at CERN, with c.m.~energies that are more than three orders of
magnitude higher, {\em i.e.}~7~to~13~TeV.  In between are the now defunct BaBar and Belle
$\ee$ $B$-factory experiments and, earlier, the CLEO experiment, that all ran at c.m.~energies
near 10~GeV, and the CDF and D0 experiments at the 1.96~TeV Tevatron $p\bar{p}$ collider.

The low energy $\ee$ experiments have the advantage of clean experimental environments and
the ability to exploit energy-momentum-conservation constraints to help extract signals
from complex final states, including those containing $\gamma$-rays and $\pi^0$-mesons.  However,
the relevant production cross sections are at the few nanobarn level and, even with the high
luminosities achieved by BEPCII and the $B$-factories, event rates are low.  In contrast, the
high energy experiments at hadron colliders 
enjoy charm particle production cross sections of a few millibarns 
and beauty particle cross sections of the order of a hundred microbarns, so large event samples
can be accumulated.  The charmed and beauty particles are usually highly boosted, thereby producing
decay vertices that are well separated from the production point and experimentally quite distinct.
This makes it possible to  isolate very clean samples of events, but only for all-charged-particle
final states. Since detected  $\gamma$-rays and $\pi^0$-mesons originating from a displaced vertex do not
have associated trajectories, they cannot be distinguished from $\gamma$-rays and $\pi^0$-mesons
that originate from the primary high-energy $pp$ (or $p\bar{p}$) interaction point and, thus, the
reconstruction of final states containing neutral particles suffer from severe combinatorial backgrounds.

A common feature of all the contributing experiments is that they were motivated and designed
to do something other than heavy hadron spectroscopy.  The original goals and highest priorities for
BESIII were precision measurements of charmed quark decays and studies of light mesons
and baryons produced in $J/\psi$ decays; the $B$-factory and LHCb experiments were aimed
at studies of weak interaction processes in the decays of particles containing
a $b$-quark; the Tevatron experiment's main jobs were the discovery of the top-quark and
precision studies of the $Z$ and $W^{\pm}$ weak vector bosons; the CMS and ATLAS experiments
discovered the Higgs' boson and have done numerous searches for new physics particles.
The discoveries of heavy non-standard hadrons have, for the most part, been unexpected but, in
many cases, have generated levels of interest that rival those of the high priority
topics.

\subsection{Experiments at $e^+e^-$ colliders}
\label{sec:exp_ee}

Many of the early contributions to this research were from the
BaBar~\cite{Aubert:2001tu} and Belle~\cite{Abashian:2000cg} experiments that
operated at the PEPII~\cite{pepii:1994} and KEKB~\cite{Kurokawa:2001nw} $B$-factories,
respectively, between 1999 and 2010. These facilities accumulated data
at and near $E_{\rm cm}=10.58$~GeV 
to study matter-antimatter asymmetries ($CP$ violations)
in the decays of $B$ mesons and to validate the Kobayashi-Maskawa mechanism for $CP$
violation~\cite{Kobayashi:1973fv}.

\begin{figure}[htb]
  \includegraphics[width=0.5\textwidth]{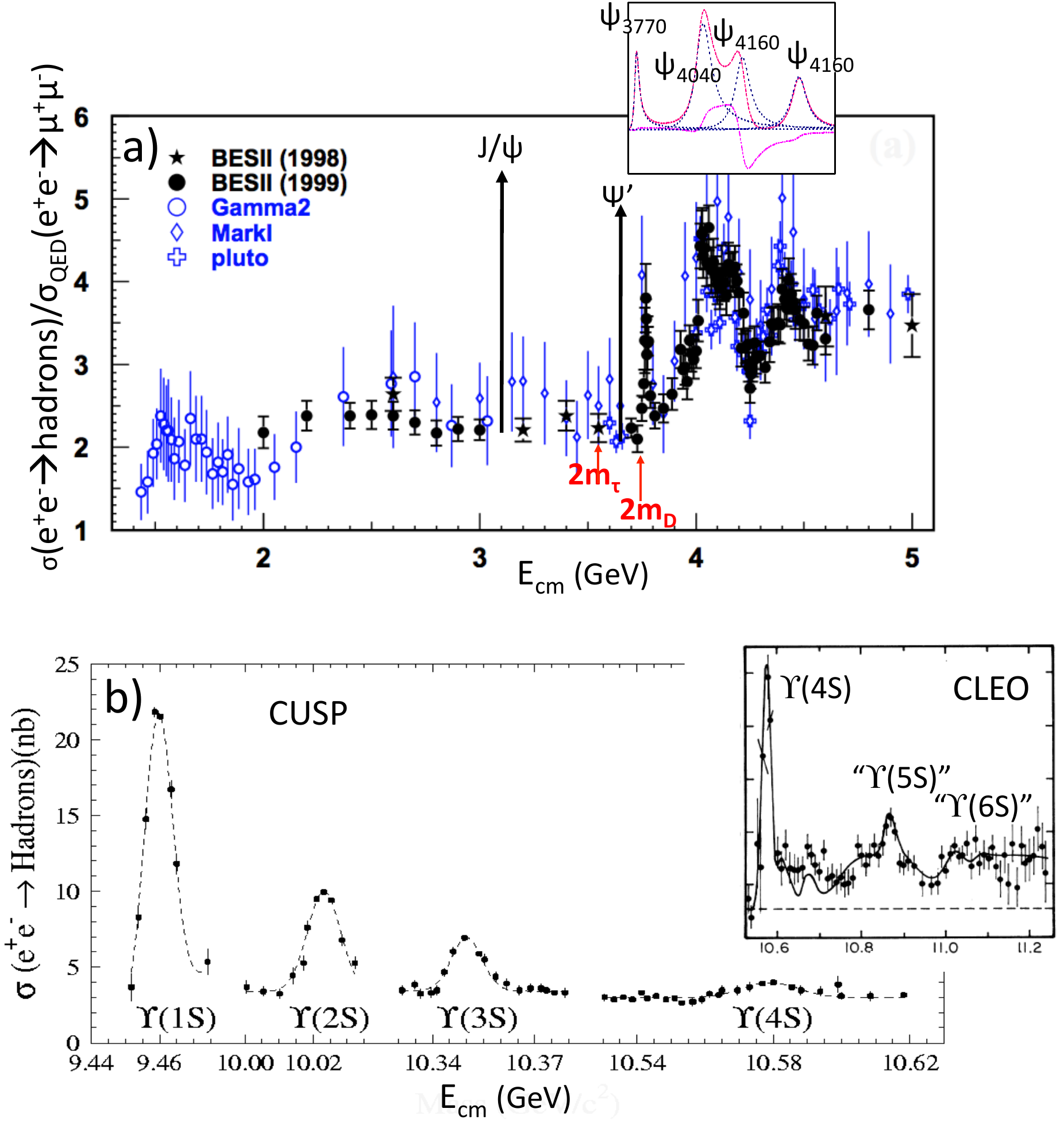}
\caption{\footnotesize 
{\bf a)} Total cross section measurements for $\ee$ annihilation into hadronic final states
in units of the QED cross section $\sigma_{\rm QED}(\ee\rt\mumu)=86.8\ {\rm nb}/s({\rm GeV}^2)$
from refs.~\cite{Criegee:1981qx,Augustin:1975yq,Bacci:1980zs,Bai:1999pk,Bai:2001ct}.
The insert shows the results of a fit to the measurements in the 3.6~to~4.6~GeV energy interval that
identifies the $1^{--}$ charmonium states in this region~\cite{Ablikim:2007gd}.
{\bf b)} Measurements of $\sigma(\ee\rt$hadrons) in the 9.45~to~10.62~GeV energy region from
CUSP~\cite{Rice:1982br}.  The inset shows measurements between 10.55~and~11.5~GeV from
CLEO~\cite{Besson:1984bd}.
 }
\label{fig:R-charm}
\end{figure}

The total cross section for $\ee\rt$~hadrons at energies that were accessed by
the $B$-factory colliders is shown in Fig.~\ref{fig:R-charm}b. There are three
narrow peaks, called the $\Upsilon(1{\rm S}),$ $\Upsilon(2{\rm S})$ and $\Upsilon(3{\rm S})$,
at c.m.~energies of 9.46, 10.02 and 10.36~GeV, respectively. These are the three lowest
triplet S-wave bottomonium ($\bbbar$) states.  Since they are below the
$E_{\rm cm}=2m_B=10.56$~GeV open-bottom threshold and, thus, not able to decay into a $B$
and a $\bar{B}$ meson pair, their primary decay channel is via $b$-quark $\bar{b}$-quark
annihilation into gluons. Since this is a suppressed process,\footnote{The suppression of strong
interaction processes in which no quark lines connect the initial to the final state is known
as the {\it Okubo-Zweig-Iizuka (OZI) rule}~\cite{Okubo:1963fa,Zweig:1981pd,Iizuka:1966fk}.}
the three below-threshold bottomonium states have natural widths that are less than 100~keV
and much smaller than the colliders' c.m.~energy spreads, which are typically $\sim 6$~MeV.
The fourth peak, the $\Upsilon(4{\rm S})$, with a peak mass of $10.58$~GeV, can decay to $B\bar{B}$
meson pairs and has a natural width of $\Gamma(\Upsilon(4{\rm S}))\simeq 20$~MeV, and is distinctly
broader than the c.m.~energy spread. Most of the data taking by both $B$-factory experiments
occurred at a c.m.~energy corresponding to the peak mass of the $\Upsilon(4{\rm S})$ resonance.  

The BEPCII $e^+e^-$ collider~\cite{Gu:2003ck} in Beijing and
its associated BESIII experiment~\cite{Ablikim:2009aa} started operation in 2008 and covers
the c.m.~energy range between 2.0~and~4.6~GeV, which includes the thresholds for producing
$\tau^+\tau^-$ lepton and $\ccbar$ quark pairs.  The $\ccbar$ system has two narrow
$J^{PC}=1^{--}$ resonances below the $E_{\rm cm}=2m_D=3.73$~GeV open-charm threshold: the $\jpsi$
and the $\psip$ (the latter is often denoted as $\psi(2S)$ or $\psi(3686)$). 
The first $1^{--}$ resonance above that threshold is the
$\psi(3770)$, which decays almost exclusively into $D\bar{D}$ final states. Figure~\ref{fig:R-charm}a
shows $\sigma(e^+e^-\rt$~hadrons) for energies that are accessible at the BEPCII collider. 

The BaBar~\cite{Aubert:2001tu}, Belle~\cite{Abashian:2000cg} and BESIII~\cite{Ablikim:2009aa}
detectors are similar in overall structure to the CLEO-II detector~\cite{Kubota:1991ww}.  Each
is a nearly 4$\pi$ solid-angle magnetic spectrometer surrounding the $\ee$ interaction point that
contains an assortment of detection systems.  All of these experiments use a large cylindrical
gas-filled drift chamber for charged-particle trajectory measurements. Surrounding these are
particle identification (PID) devices for distinguishing charged pions, kaons and protons followed
by an array of CsI(Tl) crystals that serves as an electromagnetic calorimeter for detecting
$\gamma$-rays and $\pi^0$-mesons and identifying electrons. These systems are all situated 
inside a large superconducting solenoidal magnet coil with an external iron
magnetic-flux return yoke that is instrumented to identify muons and detect $K_L$ mesons. In addition,
BaBar and Belle had elaborate, high-precision vertex measuring systems comprised of silicon-strip
detector arrays that immediately surrounded the interaction point and were essential for their
$CP$-violation measurements.  

\begin{figure}[htb]
  \includegraphics[width=0.45\textwidth]{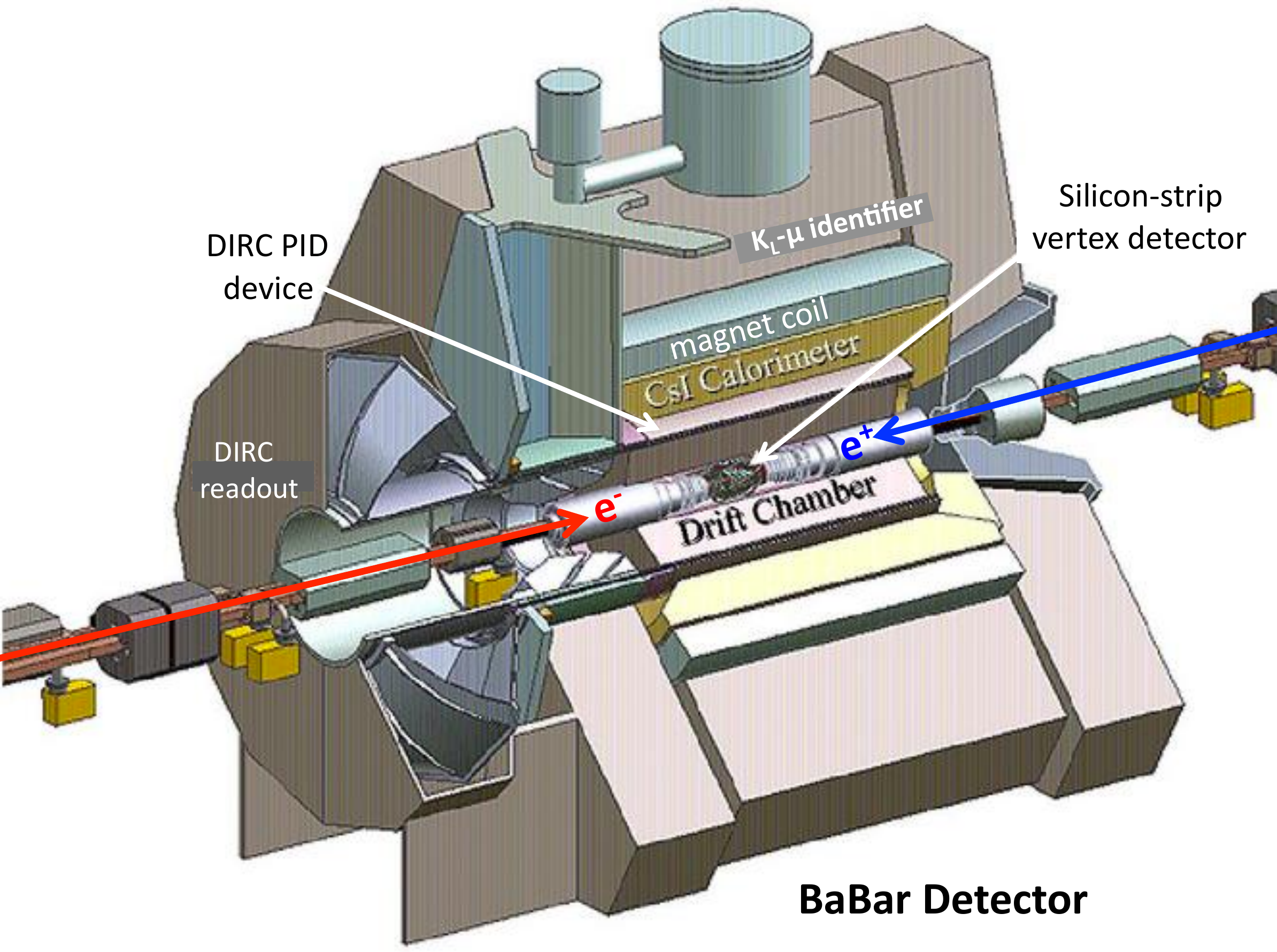}
\caption{\footnotesize 
A schematic view of the BaBar detector. Here the bulk of the charged particle tracking information is
provided by a cylindrical gas drift chamber.  Particle identification (PID) is provided by the
reconstruction of Cerenkov light cones produced in a barrel-like array of quartz bars in a device
named the ``DIRC.''  The electron and positron beams traverse the detector in a vacuum chamber
that is surrounded by magnetic bending and focusing devices. The beams collide near the center of the
cylindrical vertex detector.}
\label{fig:BaBar-iso}
\end{figure}

While similar in overall structure and capabilities, these detectors differ
in many of their specific details, especially in the PID systems, as described in the cited references.
As an example, Fig.~\ref{fig:BaBar-iso} shows an isometric drawing of the BaBar detector. 

\subsubsection{The $B$-factory experiments}

\begin{figure}[htb]
  \includegraphics[height=0.3\textwidth,width=0.5\textwidth]{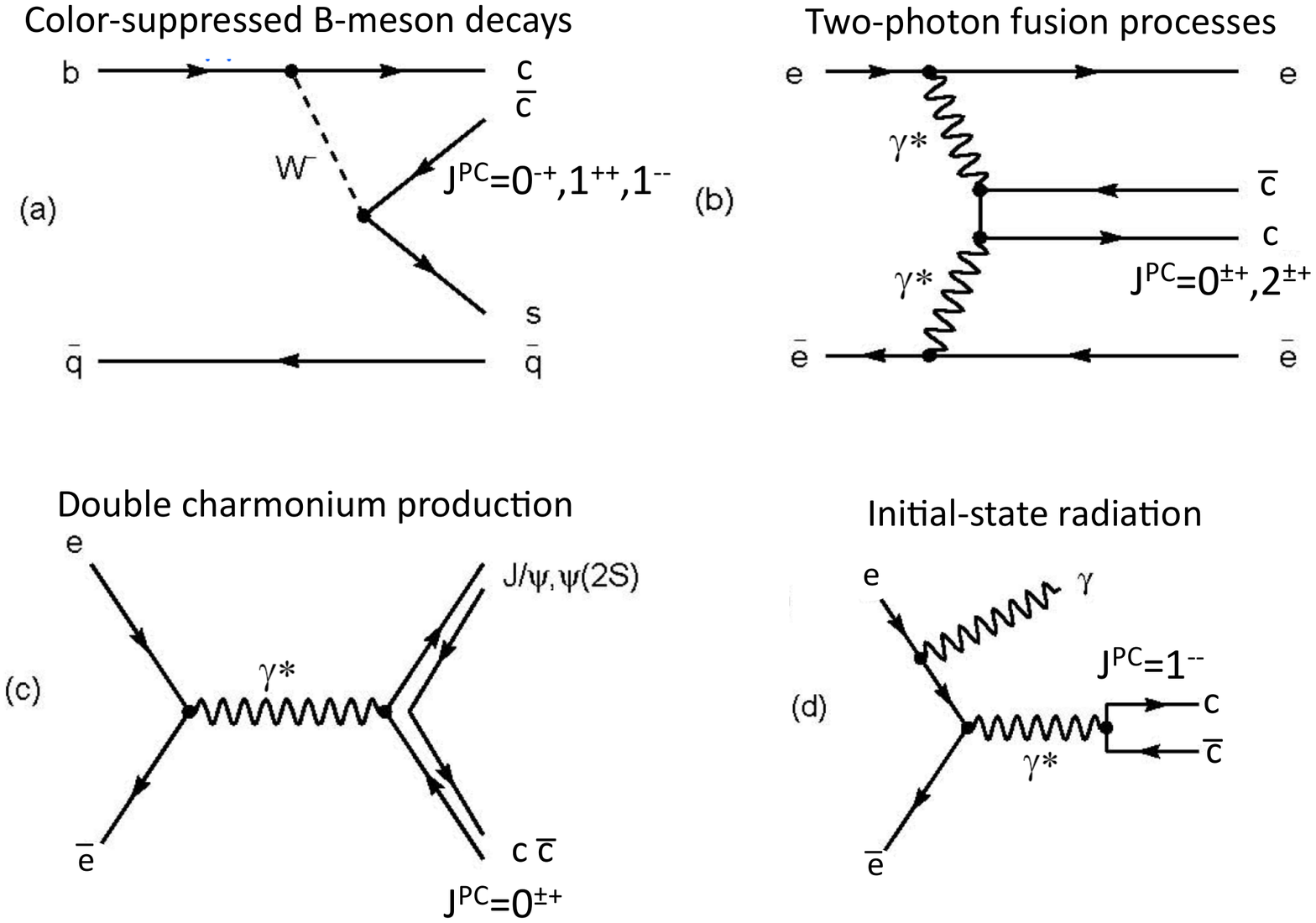}
\caption{\footnotesize Processes that produce $\ccbar$ pairs in $\ee$ collisions near $E_{\rm cm}=10.6$~GeV:
{\bf a)} $B\rt K(\ccbar)$ decays;  {\bf b)} two-photon fusion processes; 
{\bf c)} $\ee$ annihilation into $\ccbar\ccbar$; and
{\bf d)} initial state radiation.
}
\label{fig:ccbar-prod}
\end{figure}

As it is clear from the cross section plot in Fig.~\ref{fig:R-charm}b, $\ee$ annihilations
with c.m.~energies between 9.4~and~10.8~GeV, the energy range that was accessible to BaBar and Belle, are
good sources of particles containing $b$-quarks and $\bbbar$ quark pairs. In addition, $\ee$ collisions
in this energy range also produce relatively large numbers of particles containing $c$-quarks and $\ccbar$ quark
pairs. The processes involved are illustrated in Fig.~\ref{fig:ccbar-prod} and described in the following.

\underline{Charmed  quark production in $B$-meson decays:}
$\bar{B}$ mesons have a $b\bar{q}$ quark content ($\bar{q}=\bar{u}$ or $\bar{d}$), a mass of 5.28~GeV,
and a lifetime of approximately 1.5~ps. In $E_{\rm cm}=10.58$~GeV $\ee$ collisions they are produced
in $B\bar{B}$ pairs that are nearly at rest and with no accompanying particles. The primary decay
mechanism is the weak interaction transition $b\rt c$ with the emission of a virtual $W^-$ boson.
In about 15\% of these decays, the $W^-$ materializes as a $\bar{c}$- and $s$-quark.
Figure~\ref{fig:ccbar-prod}a illustrates this process for the cases where the $s$-quark combines
with the spectator $\bar{q}$ to form a $K$ meson. In these events the system recoiling against
the $K$ meson contains a $\ccbar$ quark pair and, to the extent that the original $\bar{q}$ quark is a
passive spectator to the decay process (the ``factorization approximation''~\cite{Beneke:1999br}),
the $\ccbar$ pair has $J^{PC}$ quantum numbers of $0^{-+}$, $1^{--}$ and $1^{++}$, 
which finds support in the experimental results \cite{Aubert:2005vi}.  
Sizeable corrections to the factorization approximation may occur. 

Since the $B$ mesons are produced in pairs
with no accompanying particles, the c.m.~energy of each meson is $E_{\rm cm}/2$, which is precisely
known from the operating conditions of the collider. This provides two powerful and weakly correlated 
experimental signatures for identifying events of interest:
\begin{equation}
\Delta E_{\rm cm}\equiv\ E_{\rm cm}/2 - \sum_i E_i^* =0
\end{equation}
and
\begin{equation}
M_{\rm bc}\equiv \sqrt{(E_{\rm cm}/2)^2 - |\sum_i \vec{p_i}^*|^2}=m_B, 
\end{equation}  
where $E_i^*$ and $\vec{p_i}^*$ are the c.m.~energy and three-momentum of the $i^{\rm th}$ decay product
of the $B$ meson candidate.  The experimental resolutions for $\Delta E_{\rm cm}$ and $M_{\rm bc}$ depend upon the
decay mode that is under consideration; typical rms values are 10$\sim$15~MeV for $\Delta E_{\rm cm}$ and
2.5$\sim$3~MeV for $M_{\rm bc}$. 
For a more detailed discussion of these variables 
and an improved definition of $M_{\rm bc}$ for asymmetric $e^+e^-$ 
colliders, see sec.~7.1.1.2 of ref.~\cite{Bevan:2014iga}.

\begin{figure}[htb]
  \includegraphics[width=0.48\textwidth]{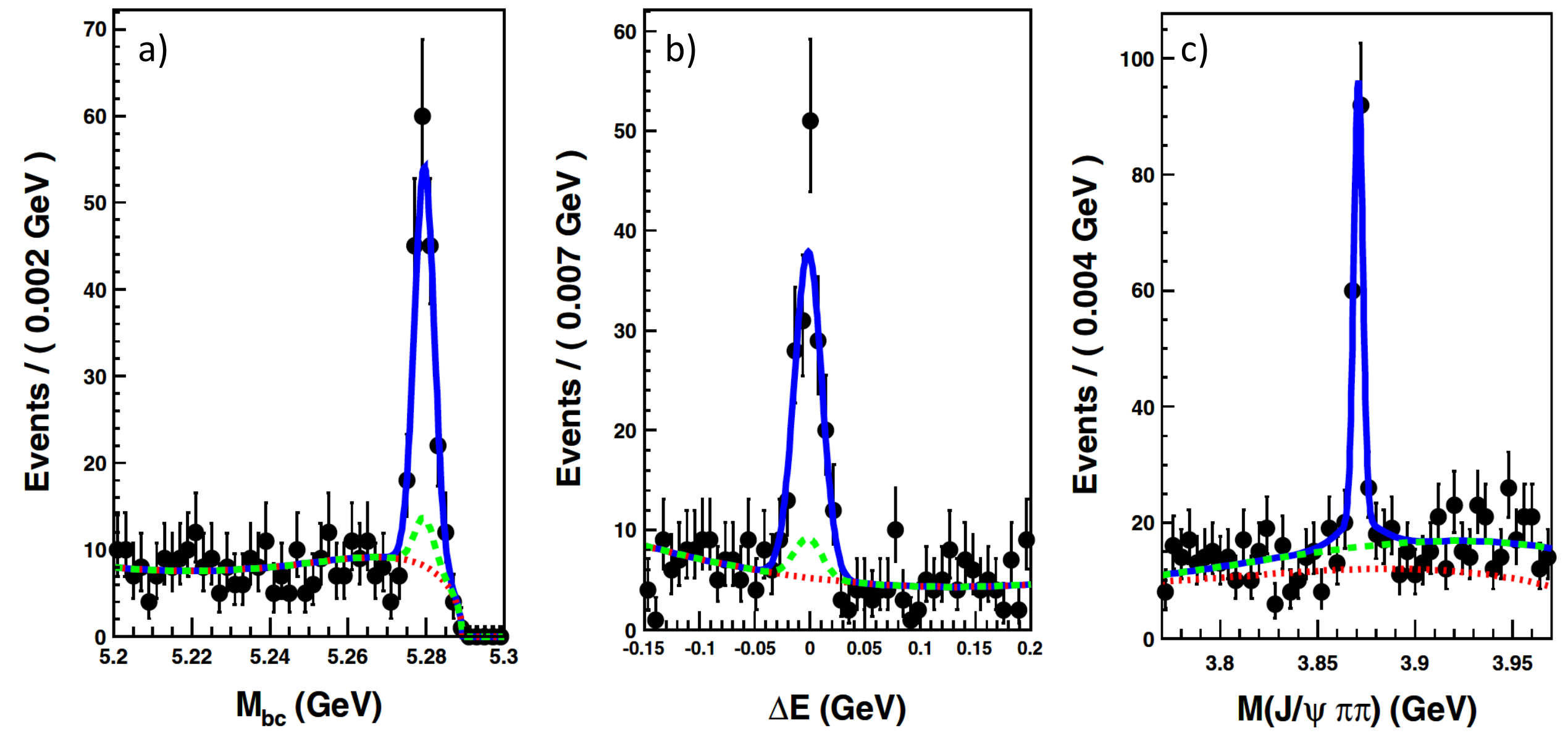}
\caption{\footnotesize 
The {\bf a)} $M_{\rm bc}$, {\bf b)} $\Delta E_{\rm cm}$ and {\bf c)} $M(\pipi\jpsi$ distributions for $B^-\rt K^-\pipi\jpsi$
decays (from ref.~\cite{Choi:2011fc}). The narrow $\pipi\jpsi$ mass peak in {\bf c)} is the $X(3872)$ signal.  
}
\label{fig:x3872-prd}
\end{figure}

The $X(3872)$, the first of the non-standard $XYZ$ meson candidates to be seen, was discovered by Belle as
a narrow peak in the $\pipi\jpsi$ invariant mass distribution for $B\rt K\pipi\jpsi$ decays~\cite{Choi:2003ue}.
Figure~\ref{fig:x3872-prd} shows the $M_{\rm bc}$, $\Delta E_{\rm cm}$  and $M(\pipi\jpsi)$ distributions for
$B^-\rt K^- X(3872)$; $X(3872)\rt\pipi\jpsi$ event candidates, where the events in each plot are selected from
the signal peak regions of the other two distributions. The background under the signal peaks is mainly 
combinatorial, {\em i.e.}~where one of the tracks assigned to the reconstructed $B$ meson is, in fact, a decay
product of the accompanying $\bar{B}$.  The $X(3872)$ is discussed in detail below in Section~\ref{sec:x3872}.
 
\begin{figure*}[htbp]
  \includegraphics[width=\textwidth]{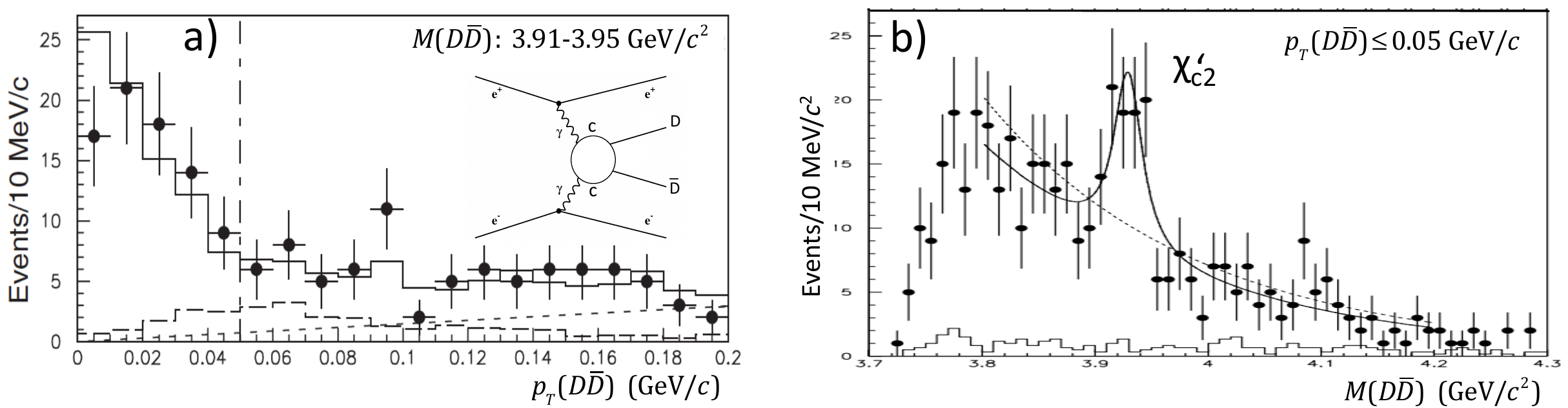}
\caption{\footnotesize 
{\bf a)} The $p_T$ distribution for $\gamma\gamma\rt D\bar{D}$ event
candidates with $3.91<M(D\bar{D})<3.95$~GeV. {\bf b)} The $M(D\bar{D})$ distribution for event
candidates with $p_T <0.05$~GeV (from ref.~\cite{Uehara:2005qd}). 
}
\label{fig:chic2prime}
\end{figure*}

\underline{Two-photon fusion processes:}
In the two-photon fusion process, both the incoming $e^-$ and $e^+$ radiate photons that subsequently interact
via the diagram shown in Fig.~\ref{fig:ccbar-prod}b. In this process, the quark-antiquark pair is produced
with a probability that is proportional to $e_q^4$, where $e_q$ is the quark charge; since $e_c=2/3$, this favors
$\ccbar$ production.   This process is dominated by events where both of the incoming photons
are nearly real ($q^2\simeq 0$), in which case both the $e^+$ and $e^-$ scatter at very small angles and are
undetectable. For these ``untagged'' events the net transverse momentum $p_T$ of the $\ccbar$ system's decay
products is very small, and this provides an important experimental signature. The allowed $J^{PC}$ quantum numbers
for $\ccbar$ systems produced via this process are restricted to $0^{\pm +}$ and $2^{\pm +}$. The conventional
$\chi_{c2}'$ ($2^3{\rm P}_2$) charmonium state was first seen by Belle~\cite{Uehara:2005qd} as the distinct peak near
3930~MeV in the $D\bar{D}$ invariant mass distribution in selected, low $p_T$ $\gamma\gamma\rt D\bar{D}$
candidate events shown in Fig.~\ref{fig:chic2prime}b.
 
\begin{figure*}[htbp]
  \includegraphics[width=\textwidth]{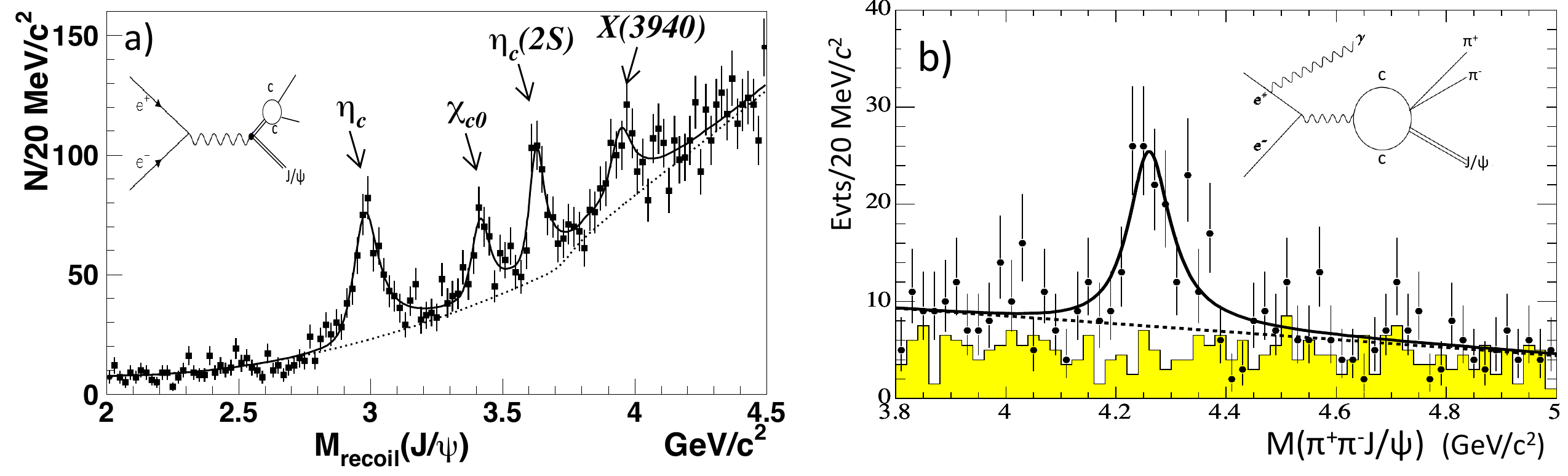}
\caption{\footnotesize 
{\bf a) }The distribution of masses recoiling from the $\jpsi$ in inclusive $\ee\rt \jpsi X$ events
from ref.~\cite{Abe:2007jna}.
{\bf b)} The $\pipi\jpsi$ invariant mass distribution for $\ee\rt \gamma_{\rm isr}\pipi\jpsi$ events
from ref.~\cite{Aubert:2005rm}
}
\label{fig:x3940-y4260}

\end{figure*}

\underline{Double charmonium production:}
Prior to the operation of the $B$ factories, computations based on a QCD-motivated effective field theory,
NRQCD, predicted that prompt inclusive $\jpsi$ production in continuum $\ee$ annihilation events at
$E_{\rm cm}\approx 10$~GeV would be dominated by $\ee\rt\jpsi g$ and $\ee\rt \jpsi gg$ processes, where $g$
denotes a gluon, and that $\ee\rt\jpsi \ccbar$ processes would account for no more than 10\% of the inclusive
$\jpsi$ production rate~\cite{Berezhnoy:2003hz}.  One of the surprises from the $B$ factories was the Belle
observation that the $\ee\rt \jpsi\ccbar$ process is, in fact, the dominant mechanism for prompt $\jpsi$ production,
accounting for approximately 60\% of the total rate for these events~\cite{Pakhlov:2009nj}. Charge conjugation
invariance requires that the $\ccbar$ system recoiling against the $\jpsi$ have even charge conjugation
parity.\footnote{Two charmonium states with the same $C$-parity can be produced in two-photon annihilation
processes, but these are suppressed relative to single-photon annihilation by a factor of
$(\alpha_{\rm QED}/\alpha_s)^2$~\cite{Bodwin:2002fk}.} 
Figure~\ref{fig:x3940-y4260}a shows Belle's measured distribution of masses recoiling from the $\jpsi$ in inclusive
$\ee\rt\jpsi+X$ production, $M_{\rm recoil}=\sqrt{(E_{\rm cm}-E_{\jpsi}^*)^2-|\vec{p}_{\jpsi}^{~*}|^2}$, where $E_{\jpsi}^*$
and $\vec{p}_{\jpsi}^{~*}$ are the c.m.~energy and three-momentum of the $\jpsi$. In the figure, clear peaks corresponding
to the $\eta_c$, $\chi_{c0}$ and $\eta_c^{\prime}$, the well established $1^1{\rm S}_0$, $1^3{\rm P}_0$ and $2^1{\rm S}_0$
charmonium  states, respectively, are evident~\cite{Abe:2007jna}. In addition, there is a distinct peak near
3940~MeV that cannot be identified with any known charmonium state; Belle called this peak the $X(3940)$.  

The established charmonium states seen in Fig.~\ref{fig:x3940-y4260}a all have even charge conjugation
and zero angular momentum ($J=0$).  The absence of signals for the $\chi_{c1}$ and $\chi_{c2}$, which are
in the same mass range and have even $C$, provides some circumstantial evidence that the
$\ee\rt J/\psi + (\ccbar)$ production process favors $(\ccbar)$ systems with $J=0$. 

\underline{Initial state radiation:}
In energy $\ee$ colliders, the initial state $e^+$ or $e^-$ occasionally radiates a high energy $\gamma$ 
and the $e^+$ and $e^-$ subsequently annihilate at a correspondingly reduced c.m.~energy: 
$E_{\rm cm}^{\prime}=E_{\rm cm}\sqrt{1-x}$, where $x=2E_{\gamma}/E_{\rm cm}$ is the fraction of the radiating
beam particle's c.m.~energy that is carried off by the photon. $B$-factory experiments typically operate
at $E_{\rm cm}=10.58$~GeV and, when an $E_{\gamma}\simeq 4.5$~GeV $\gamma$ is radiated, the $\ee$ annihilation 
occurs at a c.m.~energy of $E_{\rm cm}^{\prime}\simeq 4$~GeV, which is in the $\ccbar$ threshold region. As
a result, $\ee$ collisions at $E_{\rm cm}=10.58$~GeV can directly produce $1^{--}$ $\ccbar$ states in
association with a single high-energy $\gamma$. Advantages of measurements that exploit this
{\it initial state radiation} process (isr) are that they can be made parasitically with other measurements
and a broad range of reduced energies can be accessed at the same time. Although isr is a higher-order QED
process and, thus, suppressed, the very high luminosities provided by the $B$-factory colliders have made
it a valuable research tool.   Figure~\ref{fig:x3940-y4260}b shows the invariant mass distribution from
BaBar for $\pipi\jpsi$ events produced in association with a high energy $\gamma$~\cite{Aubert:2005rm}. In
this measurement, the isr $\gamma$ was not detected and, instead, its existence was established by selecting
events with a {\it missing mass}, $M_{\rm miss}$, consistent with zero,  where
$M_{\rm miss}\equiv \sqrt{(E_{\rm cm}-\sum_i E_i^*)^2 -|\sum_i\vec{p_i}^*|^2}$ and $E_i^*$ and $\vec{p_i}^*$ are the 
c.m.~energy and three-momentum of the $i^{\rm th}$ detected particle. The $\pipi\jpsi$ mass distribution in
Fig.~\ref{fig:x3940-y4260}b is dominated by an unexpected distinct peak near 4.26~GeV that BaBar called
the $Y(4260)$. Its production from a single virtual photon ensures that the $J^{PC}$ quantum numbers of the
$Y(4260)$ must be the same as those of the photon, {\it i.e.}~$1^{--}$. There are no unassigned $1^{--}$
charmonium states near 4260~MeV and, so, the $Y(4260)$ cannot be a conventional $\ccbar$ meson.
The $Y(4260)$ is discussed in detail below in Section~\ref{sec:y4260}.

\subsubsection{The BESIII experiment in the $\tau$-charm threshold region}
\label{sec:besIII}

For $J^{PC}=1^{--}$ states such as the $Y(4260)$, the BESIII experiment has the advantage that they can be
directly produced in $\ee$ annihilation, {\it e.g.,} $\ee\rt Y(4260) \rt\pipi \jpsi$, with no isr photons
and their associated luminosity penalty.   

By working with exclusive events near threshold, and exploiting the possibility of applying energy and momentum
kinematic constraints that improve resolution and signal to noise, the BESIII experiment is uniquely able to
isolate events with complex final states. This provides opportunities that are not available to $B$-factory and
hadron collider experiments that are mostly restricted to studies of processes that include a final-state $\jpsi$
or $\psip$ that decays to a pair of leptons.  This is because dilepton events are experimentally distinct, simple
to reconstruct, and have low combinatorial backgrounds. In contrast, the BESIII experiment routinely isolates
clean signals of charmonium states that only decay to complex multihadron final states and are plagued by huge
combinatorial backgrounds in $B$-factory and hadron-collider environments. For example, BESIII has made studies
of reactions that have an $\eta_c$ and/or an $h_c$ in the final state, by selecting and reconstructing complex,
multihadron final states.\footnote{The $\eta_c$ is the $S=0$ hyperfine partner of the $\jpsi$ and the $h_c$ is
the $S=0$ partner of the ($\chi_{c0},\chi_{c1},\chi_{c2}$) $S=1$, P-wave triplet of $\ccbar$ states. Although
the existence of the $h_c$ was predicted in 1974, it remained undiscovered until thirty years later, when it
was found in $\psip\rt\piz h_c$ decays by the CLEOc experiment~\cite{Rosner:2005ry}.} 

\begin{figure*}[htb]
\includegraphics[width=\textwidth]{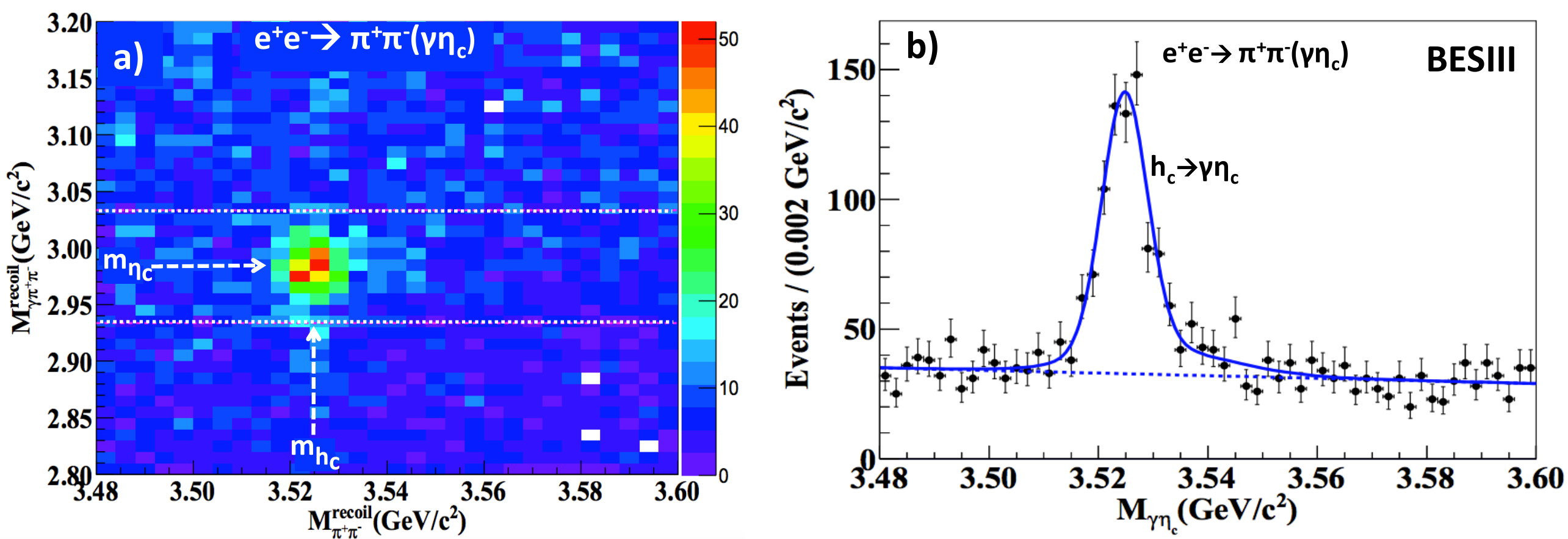}
\caption{\footnotesize 
{\bf a) } A scatter-plot of $M^{\rm recoil}_{\gamma\pipi}$ (vertical) {\em vs.} $M^{\rm recoil}_{\pipi}$
for $\ee\rt\pipi\gamma\eta_c$ candidate events from ref.~\cite{BESIII:2016adj}. The cluster of events
is the signal for the $\ee\rt\pipi h_c$; $h_c\rt\gamma\eta_c$ reaction chain. {\bf b)} The projection
of events with $\gamma\pipi$ recoil mass near $m_{\eta_c}$ ({\em i.e.}, between the horizontal dotted
lines in panel~a) on the $M^{\rm recoil}_{\pipi}(=M(\gamma \eta_c))$ axis. The peak near 3525~MeV is 
the $h_c\rt\gamma\eta_c$ decay signal.
}
\label{fig:bes3_hc-vs-etac}
\end{figure*}

This capability is illustrated by a recent BESIII study of the $\ee\rt\pipi h_c$ process at c.m.~energies in
the vicinity of the $Y(4260)$ peak, in which they detected the $h_c$ via its decay to $h_c\rt \gamma\eta_c$,
which is its dominant decay mode with a branching fraction of $51\pm 6$\%~\cite{Olive:2016xmw}. To accomplish
this, they selected  $\ee\rt\pipi\gamma \eta_c$ events, where the candidate $\eta_c$ mesons were reconstructed
in one of 16 different exclusive hadronic decay channels, and applied conservation of energy and momentum
constraints.  Figure~\ref{fig:bes3_hc-vs-etac}a shows a scatter-plot of the mass recoiling against the $\gamma\pipi$
particle combination (vertical) {\em vs.} the mass recoiling against the $\pipi$ pair (horizontal) for selected
events, where a distinct cluster of events near $M^{\rm recoil}_{\gamma\pipi}\simeq m_{\eta_c}$ and
$M^{\rm recoil}_{\pipi}\simeq m_{h_c}$, is evident~\cite{BESIII:2016adj}.  Figure~\ref{fig:bes3_hc-vs-etac}b shows the
projection of events with $M^{\rm recoil}_{\gamma\pipi}$ in the $\eta_c$ mass region onto the $M^{\rm recoil}_{\pipi}$ axis
-- which is equivalent to $M(\gamma\eta_c)$ -- where there is a large $h_c\rt\gamma\eta_c$ signal at
$m_{h_c}=3525$~MeV on a relatively small background. With this clean $\ee\rt\pipi h_c$ event sample, the
BESIII experiment discovered a narrow enhancement in the $\pi^{\pm}h_c$ invariant mass distribution that is a
candidate for an electrically charged charmonium-like four-quark state, called the
$Z_c(4020)$~\cite{Ablikim:2013wzq}, as discussed below in Section~\ref{sec:z3900zb}.  Reconstruction of events
with an $h_c$ in the final state has never been accomplished in $B$-factory or hadron-collider experiments.

\subsubsection{Future prospects}

The BESIII experiment is scheduled to continue to run until 2025, with some minor upgrades to the detector and the
BEPCII collider. There are serious discussions in China~\cite{zhao:2016zg} and in Russia~\cite{Bondar:2013cja} about
the possibility of a new facility in the charm quark threshold energy region with a luminosity of order
$10^{35}$cm$^{-2}$s$^{-1}$, a two-order-of-magnitude increase over that of BEPCII.
  
SuperKEKB is a major upgrade upgrade to the KEKB facility with the ambitious goal of a factor of forty increase
in instantaneous luminosity (to $8\cdot 10^{35}$~cm$^{-2}$s$^{-1}$) that will start operation in late 2017 with a targeted
total integrated luminosity of 50,000~fb$^{-1}$ (50 inverse attobarns) by about 2025~\cite{Ohnishi:2013fma}. BelleII
is an upgraded version of the Belle detector that is being constructed by a large international collaboration to
exploit the physics opportunities provided by SuperKEKB~\cite{Aushev:2010bq}. While the main emphasis of the
BelleII program is on searches for new, beyond the Standard Model physics processes, a high-sensitivity program of
studies of hadron spectroscopy is planned (see, {\em e.g.,} ref.~\cite{Bondar:2016hva}).

\myclearpage

\subsection{High-p$_T$  detectors at high-energy hadron colliders}
\label{sec:exp_gpd}

\begin{figure}[htb]
\includegraphics[width=0.45\textwidth]{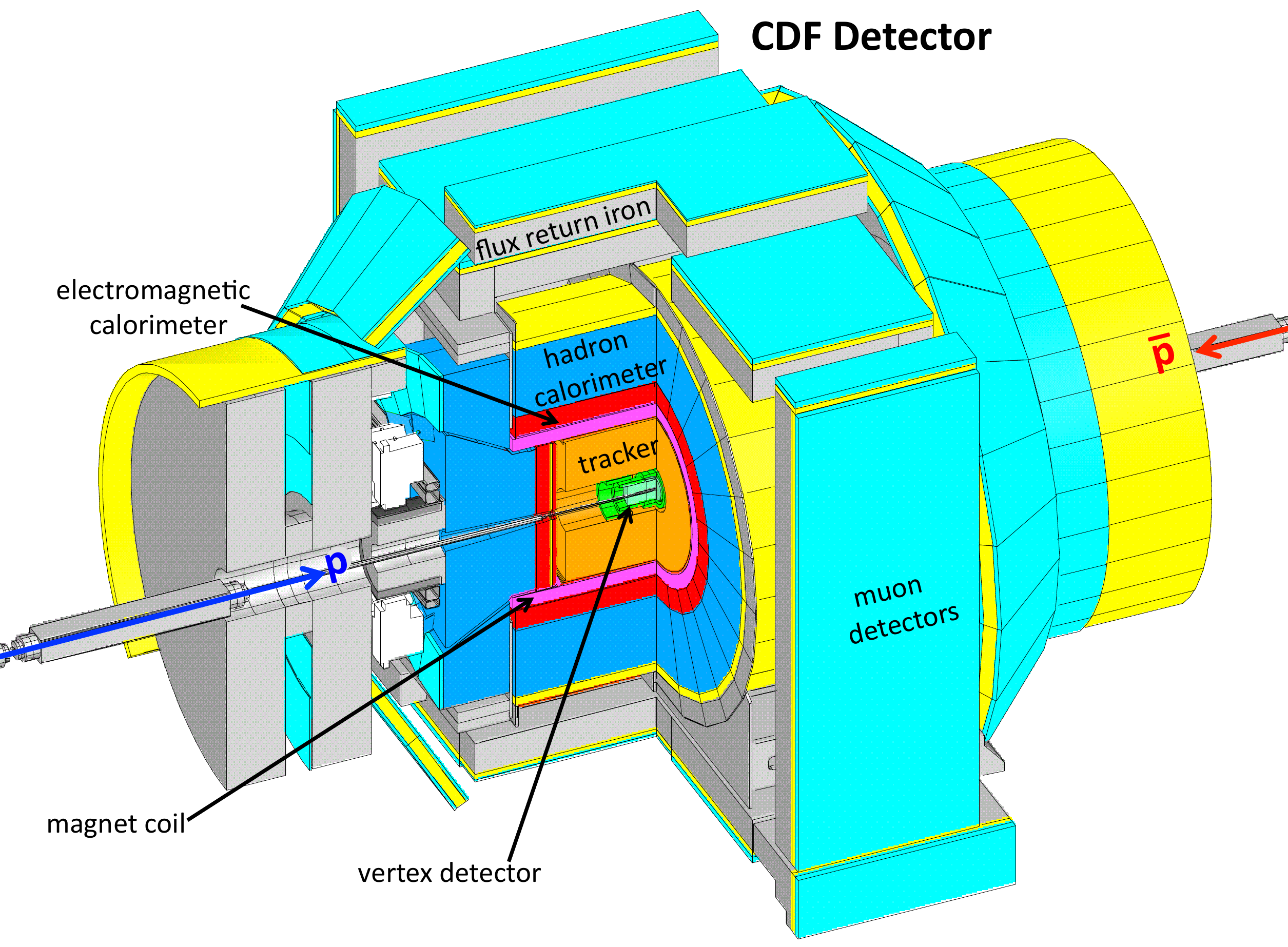}
\caption{\footnotesize A schematic view of the CDF detector.  Here the tracking device is a cylindrical gas drift chamber.
The electromagnetic calorimeters (in red) are comprised of alternating layers of plastic scintillators and lead sheets; the hadron
calorimeters (in blue) are similar structures with lead replaced by steel.
}
\label{fig:CDF-iso}
\end{figure}

Between 1985 and 2011, the high-energy frontier was centered on the CDF and D0 experiments at the Fermilab Tevatron 
that studied particles produced in proton-antiproton collisions at a c.m. energy of 1.96~TeV. This frontier moved to CERN
in 2010 at the LHC, when the ATLAS and CMS experiments started initial operations with proton-proton collisions at 7~TeV.
These experiments were designed to study processes at high momentum transfer using large, multi-story magnetic detectors
surrounding the regions where the beams were brought into collision. The detector designs, while different in almost all
details,  follow a common concept. In the innermost region, closest to the collision region, vertex detection is provided
by silicon microstrip and pixel devices that measure trajectory coordinates with about 10 micron precision. The vertex detectors
are surrounded by charged-particle tracking devices to give more complete trajectory information.  Massive calorimeters
surround the tracking systems and detect $\gamma$-rays, identify electrons, and measure the energies of hadronic particles.
The only particles that escape the calorimeters are muons, which are identified in external muon detectors of varying degree of
complexity, and neutrinos.   As an example, a schematic view of the CDF detector is provided in Fig.~\ref{fig:CDF-iso}. 
 
\subsubsection{The Tevatron experiments}

The highest priority of the Tevatron program was the discovery of
the top quark, the last undiscovered quark of the Standard Model.
Its observation   was simultaneously announced by the CDF~\cite{Abe:1995hr} and D0~\cite{Abachi:1995iq} 
collaborations in March 1995.  The Tevatron experiments  have also contributed to the spectroscopy
of heavy hadrons. CDF~\cite{Acosta:2003zx} and D0~\cite{Abazov:2004kp} were the first experiments to
confirm Belle's observation of the $X(3872)$.  These experiments made the important observations that
characteristics of $X(3872)$ production in hadron collisions were very similar to those of the $\psip$, including
strong signals for prompt production via QCD processes.  In addition, CDF performed the most precise measurement to date
of the $X(3872)$ mass \cite{Aaltonen:2009vj}.

\begin{figure*}[htbp]
\includegraphics[width=0.8\textwidth]{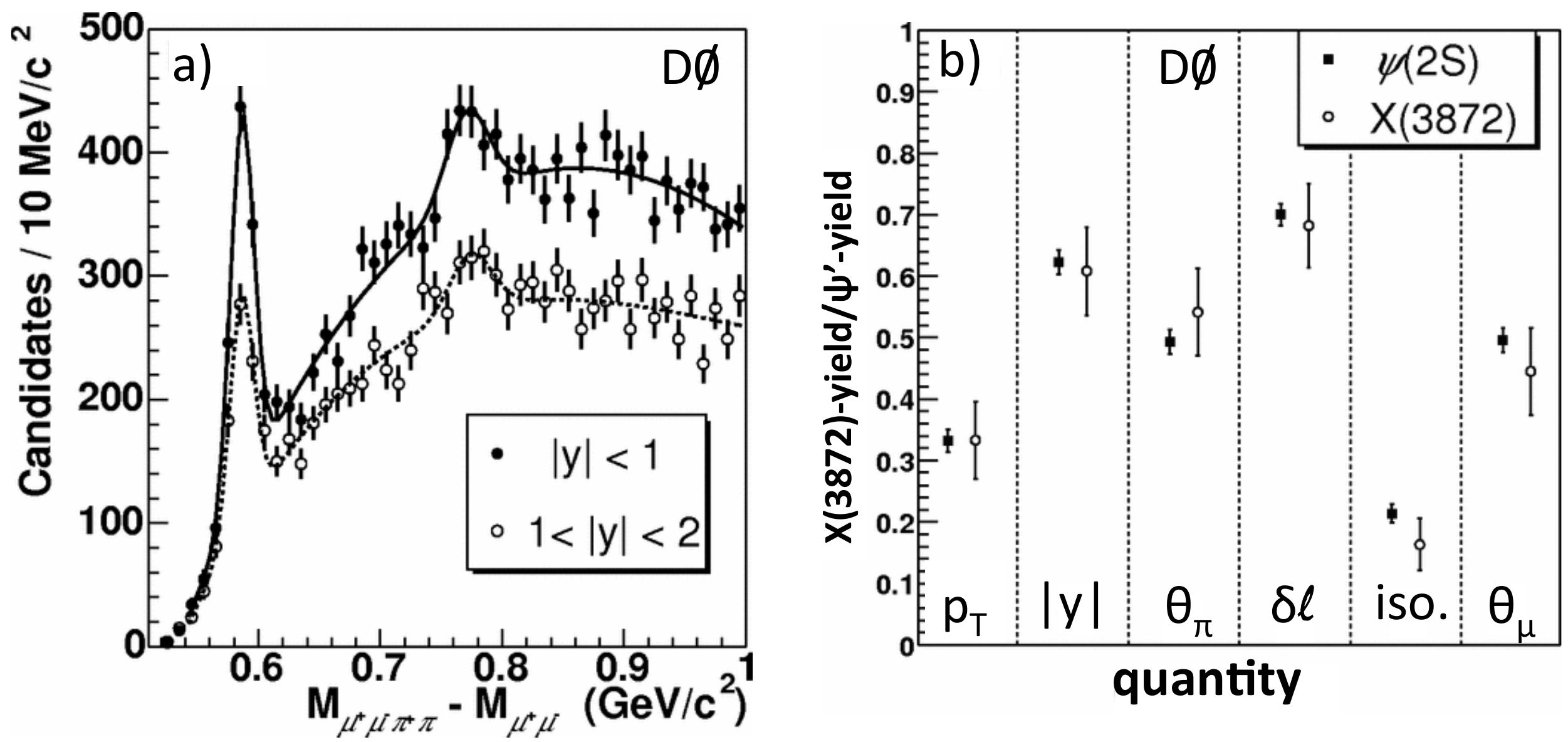}
\caption{\footnotesize {\bf a)} $\Delta M=M(\pipi\mumu)-M(\mumu)$ distributions for central events (solid points) and end-cap events
(open circles) from the D0 experiment~\cite{Abazov:2004kp}.  The similarity between the relative signal yields in the $X(3872)$ and
$\psip$ peaks indicate that the production mechanism for the two states has similar dependence on pseudorapidity ($|y|$).  {\bf b)}
Similar comparisons are shown for other quantities: $p_T(\pipi\jpsi)$; $\cos\theta_{\pi}$;
proper decay length $\delta\ell$; jet isolation parameter; and $\cos\theta_{\mu}$
(for details, see ref.~\cite{Abazov:2004kp}).}
\label{fig:D0-x3872}
\end{figure*}

In Fig.~\ref{fig:D0-x3872}a, the $\Delta M=M(\pipi\mumu)-M(\mumu)$ distribution for events in the central
part of the D0 detector (with pseudorapidity\footnote{Pseudorapidity is defined as $y=-\ln[\tan(\theta /2)]$, 
where $\theta$ is the polar angle.} $|y|<1$) are shown as solid circles;  that for events in the endcap
regions ($1<|y|<2)$ are shown as open circles~\cite{Abazov:2004kp}.   The ratio of the number of central events in the
$X(3872)$ peak (near $\Delta M=775$~MeV) to that in the $\psip$ peak (near $\Delta M=589$~MeV) is $0.43\pm 0.08$
and in good agreement with the corresponding ratio for endcap events, namely $0.45\pm 0.11$.  The D0 experiment made similar
comparisons of $X(3872)$  and $\psip$ production rates for a number of other quantities and found good agreement in all cases,
as shown in Fig.~\ref{fig:D0-x3872}b. This is a strong indication the $X(3872)$ and $\psip$ production mechanisms are the same.
This is discussed further below in Section~\ref{sec:x3872}.

\begin{figure*}[htb]
\includegraphics[width=0.8\textwidth]{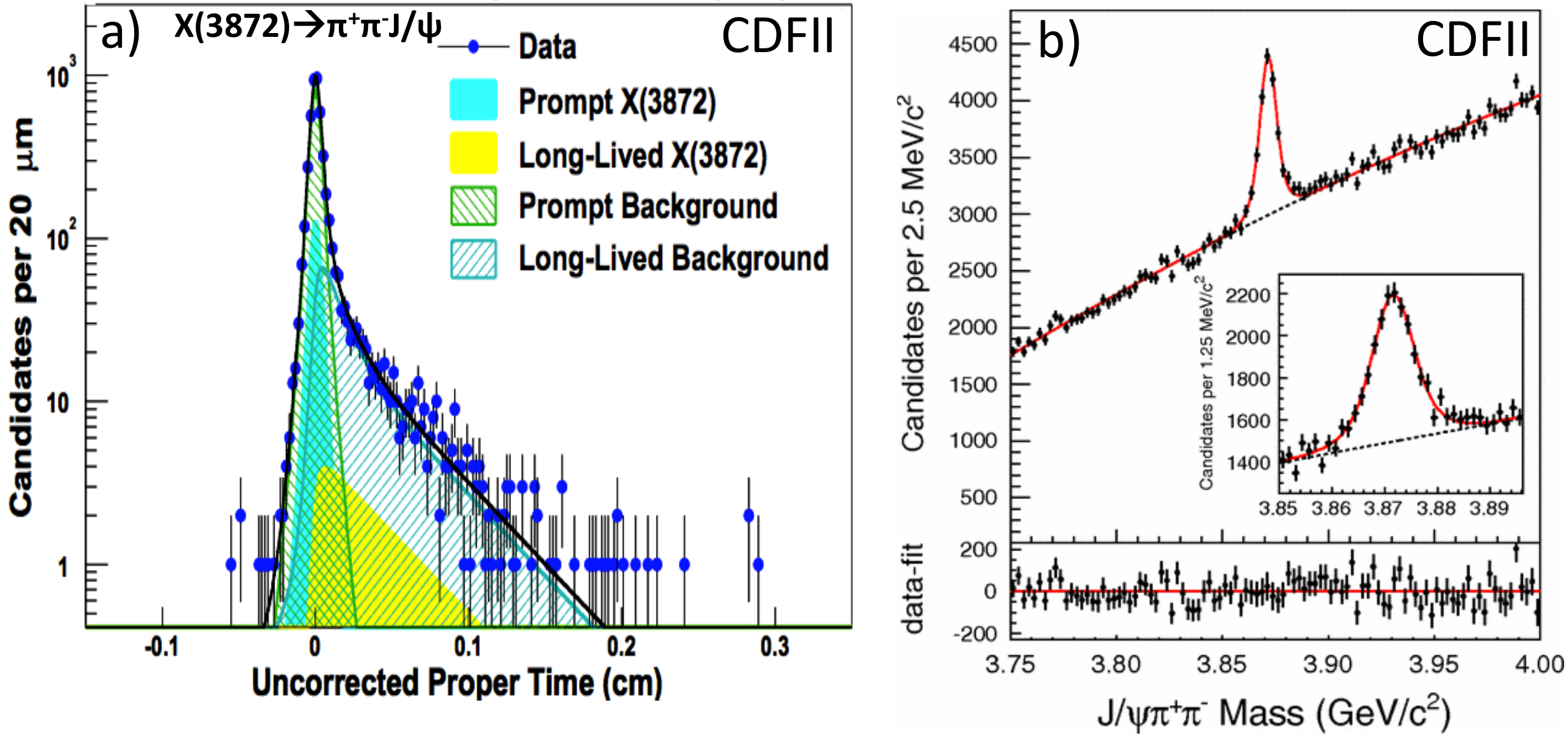}
\caption{\footnotesize {\bf a)} The proper time distribution of $\pipi\jpsi$ vertices for events in the region of the $X(3872)$
peak from ref.~\cite{Bauer:2004bc}.  The main source of $X(3872)$ are prompt $p\bar{p}$ interactions (turquoise peak); 
production via open-bottom-particle decays (yellow exponential) is only $16\pm 5$\% of the total.
{\bf b)} The data points in the upper panel show the invariant mass distribution for $X(3872)\rt\pipi\jpsi$ candidates in the CDF
detector~\cite{Aaltonen:2009vj}; the solid curve shows the result of an unbinned likelihood fit and the dashed curve shows the background
component.  The lower panel shows the fit residuals.
}
\label{fig:CDF-x3872}
\end{figure*}

In the proper time distribution for $X(3872)\rt\pipi\jpsi$ vertex positions, shown in Fig.~\ref{fig:CDF-x3872}a,
CDF found that most of the $X(3872)$ production in $p\bar{p}$ collisions was due to prompt QCD processes; only
$16\pm 5$\% of the signal is from displaced vertices that are characteristic of open-bottom-particle decays~\cite{Bauer:2004bc}. 
The displaced-vertex fraction for $\psip$ production is somewhat larger, $28\pm 1$\%, but comparable.  In 2009, with an
order-of-magnitude larger data set, CDF reported the most precise single mass measurement to date,
$M(X(3872))=3871.61\pm 0.16\ {\rm (stat)} \pm 0.19\ {\rm (syst)}$~MeV, based on a fit to the approximately 6K event
$X(3872)\rt\pipi\jpsi$ invariant mass peak shown in Fig.~\ref{fig:CDF-x3872}b~\cite{Aaltonen:2009vj}.  This mass value is
indistinguishable from the $D^0\bar{D}^{*0}$ threshold mass, $(m_{D^0}+m_{D^{*0}})=3871.68\pm 0.10$~MeV~\cite{Olive:2016xmw}, with
high precision, and this is one of the most striking properties of the $X(3872)$.

\subsubsection{The LHC  experiments}

\begin{figure*}[htbp]
\includegraphics[width=0.8\textwidth]{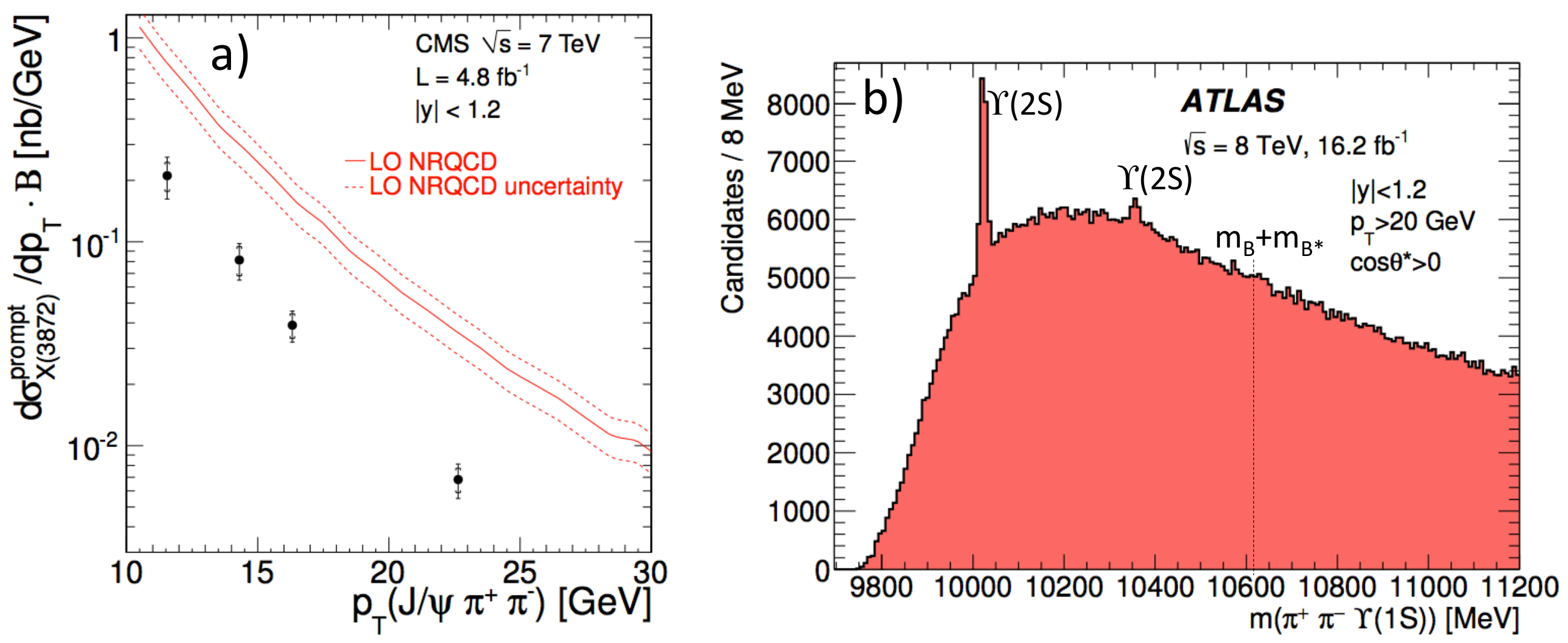}
\caption{\footnotesize {\bf a)} The $p_T$ dependence of the prompt $X(3872)$ differential production
cross section measured by CMS~\cite{Chatrchyan:2013cld}.  The red curves show a theoretical prediction
that is based on an NRQCD calculation.
{\bf b)}  The inclusive $M(\pipi\Upsilon(1S))$ invariant mass distribution measured by ATLAS~\cite{Aad:2014ama}.
Aside from peaks corresponding to  $\Upsilon(2S)$ and $(\Upsilon(3S))$ decays to $\pipi\Upsilon(1S)$,
no additional structures are evident.  The location of the $B\bar{B}^*$ mass threshold is indicated.
}
\label{fig:LHC-x3872}
\end{figure*}

The new-generation of high-$p_T$ hadron collider experiments, 
ATLAS~\cite{Aad:2008zzm} and CMS~\cite{Chatrchyan:2008aa}, 
are producing complementary results on the production of exotic hadrons.
For example, CMS has measured the $p_T$ dependence of the prompt $X(3872)$
production cross section~\cite{Chatrchyan:2013cld} and found it to be about
a factor of four below an NRQCD-based theoretical prediction~\cite{Artoisenet:2009wk}; these
results are shown in Fig.~\ref{fig:LHC-x3872}a.  CMS~\cite{Chatrchyan:2013mea} and
ATLAS~\cite{Aad:2014ama} have also performed measurements of the inclusive $\pipi\Upsilon(1S)$
mass distribution at a c.m. energy of 8~TeV in  search 
for a bottomonium-like counterparts of the $X(3872)$ or $Y(4260)$.
No dramatic signals were found, but the sensitivity of these searches was not very high.
Cross-section times branching-fraction upper limits for new state production that are $\sim$6.5\% of
the $\Upsilon(2S)\rt\pipi\Upsilon(1S)$ production were established~\cite{Chatrchyan:2013cld}; this is
about the same as the CMS measured ratio for $X(3872)\rt\pipi\jpsi$ and $\psip\rt\pipi\jpsi$ production: 
$6.6\pm 0.7$\%~\cite{Chatrchyan:2013cld}.
The inclusive $M(\pipi\Upsilon(1S))$ distribution measured at ATLAS is shown in Fig.~\ref{fig:LHC-x3872}b. 

\myclearpage

\subsection{LHCb - forward detector at LHC}
\label{sec:exp_lhcb}

\begin{figure}[htbp]
\includegraphics*[width=0.5\textwidth]{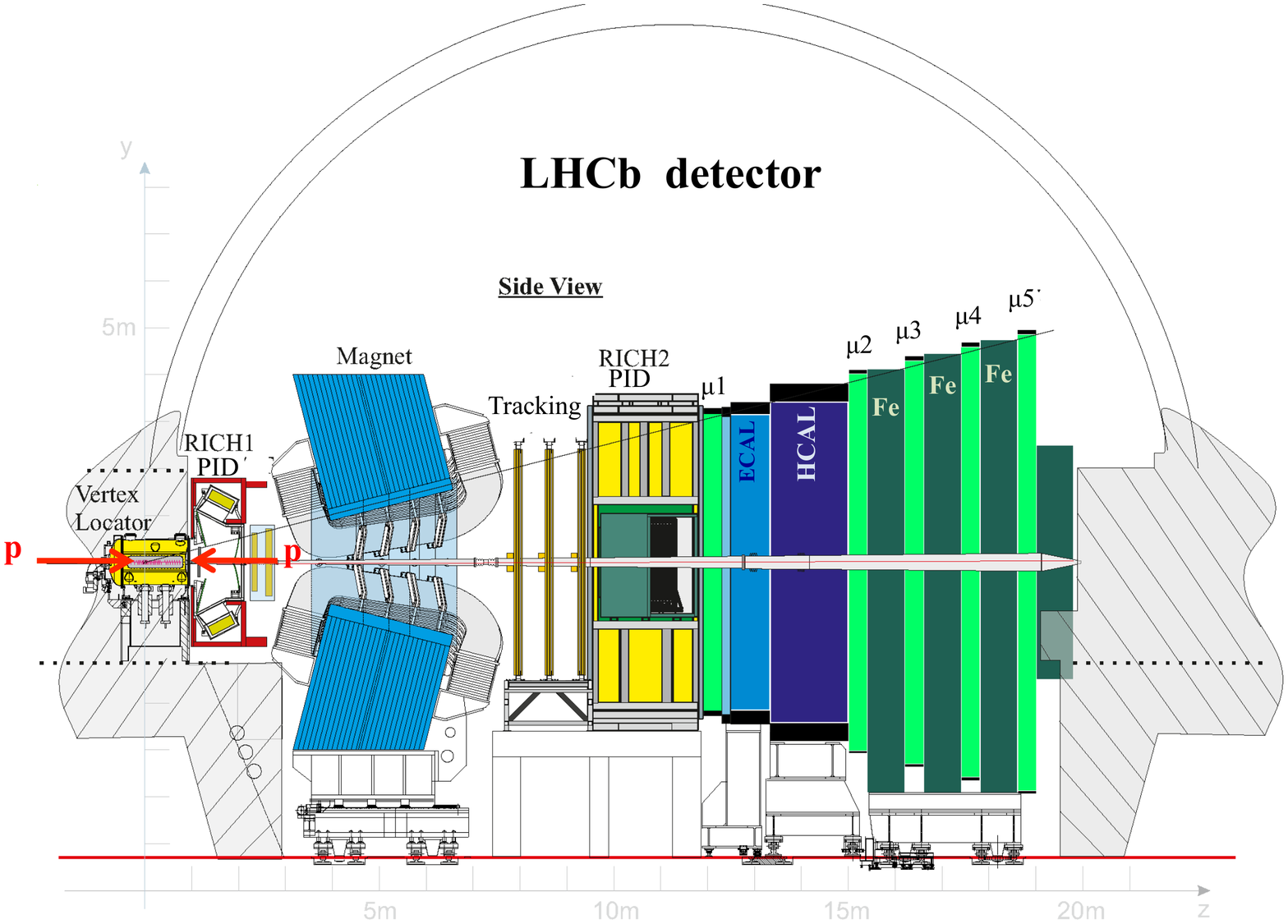}
\caption{\footnotesize
  The LHCb detector.  Vertex detection and tracking is provided by an array of silicon-strip
detectors that surround the $pp$ interaction point, a large-area plane of silicon-strip detectors located upstream,
and three planes of silicon-strips and straw drift tubes located downstream of a 4~Tm bending magnet. Charged
particles are identified by two ring-imaging Cherenkov (RICH) detectors. 
}
  \label{fig:lhcbdet}
\end{figure}

The LHCb experiment is the first hadron-collider experiment that is dedicated to heavy flavor physics.
Its detector~\cite{Alves:2008zz,LHCb-DP-2014-002}, shown in Fig.~\ref{fig:lhcbdet}, is a single-arm forward
spectrometer that captures heavy-quark production cross-sections that are comparable to those in the
high-$p_T$, central detectors at LHC, but concentrated in a compact solid angle
near the forward direction ($0.8^o<\theta <15.4^o$). 
Because of this concentration, a much smaller number of electronic channels are required and, thus, there is
a smaller data record for each event. As a result, the LHCb data acquisition system can record events at a higher
frequency than the high-$p_T$ detectors; the LHCb recorded data at a 5~kHz rate in Run-I (2011-12) and 12.5~kHz
in Run-II (2015-present), which are about a factor of five higher than the corresponding rates for the ATLAS and
CMS detectors.   Furthermore, in contrast with the central detectors, most of the trigger bandwidth is dedicated
to heavy flavor physics and includes dimuon events with low transverse-momentum thresholds and purely hadronic
events  that have secondary decay vertices that are well separated from the $pp$ collision points.  The price of
these capabilities is a limit on the tolerable instantaneous luminosity of $4\cdot10^{32}$ cm$^{-2}$s$^{-1}$,
which is  almost two orders of magnitude lower than the maximum values that the LHC is capable of delivering.
Therefore, integrated luminosities of LHCb data sets are smaller than that of CMS and ATLAS data sets.
This makes CMS and ATLAS experiment competitive in detection of some $B$ decays to simple final states with muon pairs.  
The LHCb detector is equipped with two ring-imaging Cherenkov (RICH) detectors that provide good suppression of pion
backgrounds for final states that include kaons and protons. 
The central detectors at LHC and Tevatron lack efficient hadron identification and, thus, have to cope with high
backgrounds in such channels.  The Tevatron experiments also had the disadvantage of the lower cross-sections.

The production cross-sections for heavy flavors at the LHC 
are three orders of magnitude larger than those at the $e^+e^-$ $B$ factories.
Even after correcting for the smaller reconstruction efficiencies and shorter accumulated beam time, 
the Run-I  LHCb data samples of $B$ decays to $\jpsi$ and light hadrons are an order of magnitude larger than those
accumulated during the ten-year operating lifetimes of the $B$ factories.
The signal purity is even slightly better than in BaBar or Belle, thanks to the long visible lifetimes of
the lightest open-bottom-flavored hadrons. 
The identification of tracks produced from a displaced vertex reduces combinatorial backgrounds associated with 
additional particles produced in the primary $pp$ collisions as well as those from the decays of the companion
bottom-flavored hadron. The large signal sample enabled the LHCb group to make the first determination of $X(3872)$'s
$J^{PC}$ quantum numbers~\cite{Aaij:2013zoa,Aaij:2015eva}; confirm the $Z(4430)^+$ structure ~\cite{Aaij:2014jqa,Aaij:2015zxa};
and demonstrate its consistency with a Breit-Wigner (BW) resonance hypothesis by means of an Argand diagram \cite{Aaij:2014jqa}.
Recently, LHCb has performed the first amplitude analysis  of $B^+\to\jpsi\phi K^+$ decays that made the first determination of
the quantum numbers of $X(4140)$ and established the existence of three other $\jpsi\phi$ mass peaks: the $X(4274)$, $X(4500)$
and $X(4700)$ \cite{Aaij:2016iza,Aaij:2016nsc}.  A big advantage of collecting data at a hadronic collider is the simultaneous
accumulation of large  $B$, $B_s$, $B_c$ and $\Lambda_b^0$ data sets, as opposed to $B$ factories, where $B_s$ samples 
require dedicated data runs, and $B_c$ mesons and $\Lambda_b^0$ baryons are not accessible.  The large sample of $\Lambda_b$
events in LHCb's Run~I data sample led to the discovery of pentaquark-like $\jpsi p$ mass structures $P_c(4450)^+$ and
$P_c(4380)^+$  in $\Lambda_b^0\to\jpsi p K^-$ decays~\cite{Aaij:2015tga}.

\begin{figure*}[htbp]
\includegraphics*[width=0.8\textwidth]{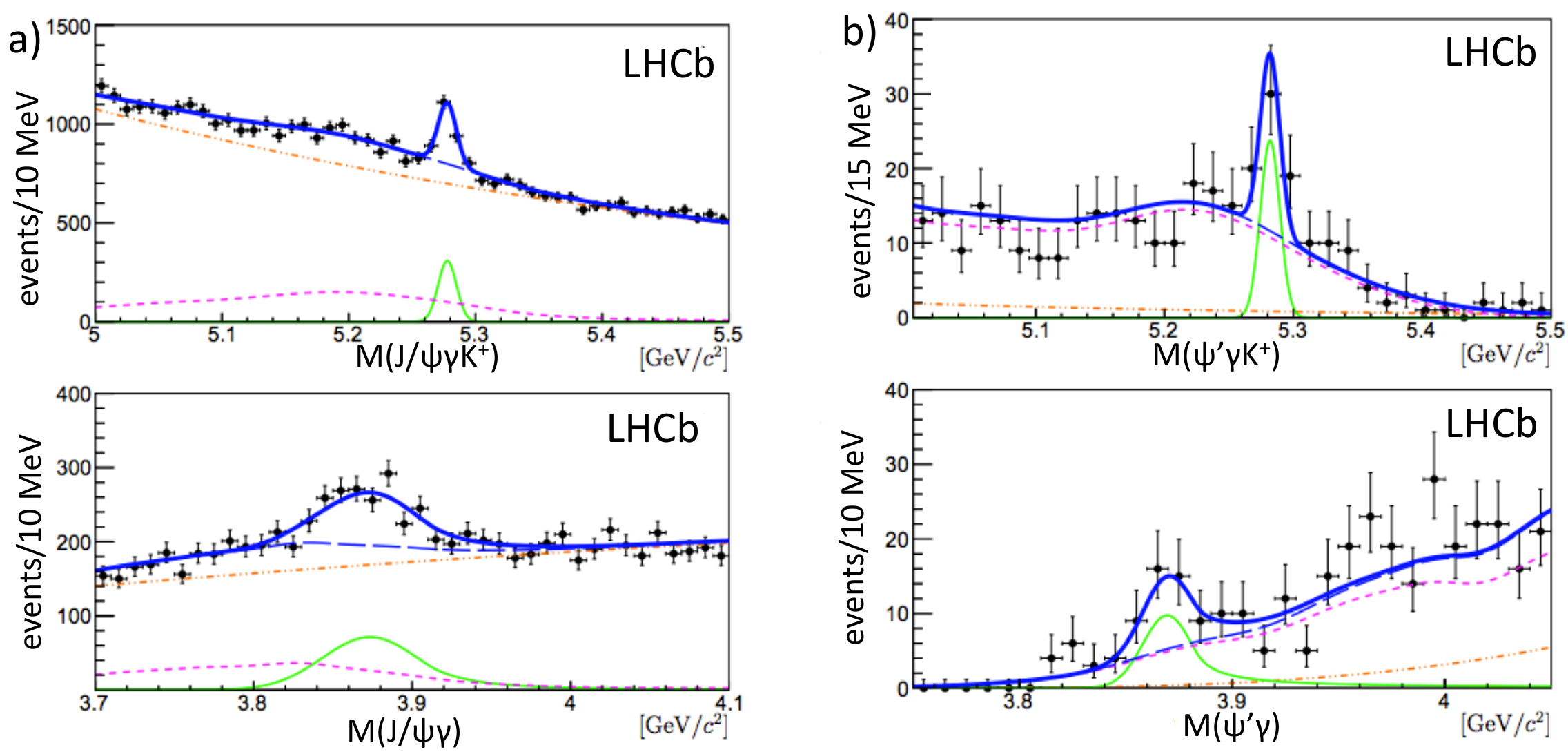}
\caption{\footnotesize
{\bf a)} The upper panel shows the $M(\jpsi\gamma K^+)$ distribution for $B\rt X(3872) K^+$; $X(3872)\rt\jpsi\gamma$
candidate events with $M(\jpsi\gamma)$ within $\pm 3\sigma$ of $3872$~MeV, where $\sigma$ is the $\jpsi\gamma$ mass resolution.
The lower panel shows the $M(\jpsi\gamma)$ distribution for events with $M(\jpsi\gamma K^+)$ within $3\sigma$
of $m_{B^+}=5.28$~GeV.
{\bf b)}  Corresponding plots for the $B\rt X(3872) K^+$; $X(3872)\rt \psip\gamma$ candidate events.
The curves are projections of fits described in ref.~\cite{Aaij:2014ala}.
}
  \label{fig:LHCb_gammapsip}
\end{figure*}

LHCb is equipped with an electromagnetic calorimeter. However, $\gamma$-ray and $\pi^0$ reconstruction
efficiencies are much lower than in the $e^+e^-$ experiments and, because of the lack of vertex information,
combinatorial backgrounds for photons are large.  Nevertheless, exploration of channels with one $\gamma$-ray,
or $\pi^0$ or $\eta$ is possible.  For example,  Fig.~\ref{fig:LHCb_gammapsip} shows LHCb signals for
$\gamma\jpsi$ and $\gamma\psip$ decays of the $X(3872)$.
These data were used to provide the most precise measurement 
to data of the branching fraction ratio~\cite{Aaij:2014ala}
\begin{equation}
\frac{{\mathcal B}(X(3872)\rt\gamma\psip)}{{\mathcal B}(X(3872)\rt\gamma\jpsi)}=2.46\pm 0.70,
\label{eq:gpsip_over_gjpsi}
\end{equation}
which is an important quantity for distinguishing between different theoretical interpretations of the $X(3872)$.

All of the LHCb results on hadron spectroscopy that have been published to date have been based  on analyses of Run-I data;
{\em i.e.} with integrated luminosities as high as 3~fb$^{-1}$ at $\ecmx{pp}=7-8$~TeV. The on-going Run-II is expected to conclude
in 2018 with an additional 8~fb$^{-1}$ of integrated luminosity collected mostly at $\ecmx{pp}=13$~TeV, where the heavy-flavor
production cross-sections are about a factor of two higher~\cite{Aaij:2015bpa}.
A major upgrade of the LHCb detector is currently in preparation~\cite{Bediaga:1443882}, 
which will allow collecting data at $2\cdot10^{33}$~cm$^{-2}$s$^{-1}$ starting around 2021.
About 50 fb$^{-1}$ of integrated luminosity is expected by 2030. 
A second major upgrade is under consideration, which would allow data taking with an instantaneous luminosity of
$2\cdot10^{34}$ cm$^{-2}$s$^{-1}$ with the goal of a final data sample corresponding to 300 fb$^{-1}$ by the end
of LHC operations.

\myclearpage

\section{Heavy-light exotic hadron candidates}
\label{sec:heavy_light}

\subsection{Charm}
\label{sec:charmheavylight}

The  modern quark model~\cite{Godfrey:1985xj,Godfrey:2015dva}
predicts a rich spectrum of heavy-light mesons containing a heavy quark, $Q$, and an anti-light
quark, $q=u,d,s$. In contrast to quarkonium, where the states are best characterized
by the $\QQbar$ spin $S$ and the relative orbital angular momentum $L$, the heavy-light $Q\bar{q}$
mesons are expected to be best described by $\vec{j}_q$, the orbital angular momentum $\vec{L}$ plus
the spin of the light quark $\vec{s}_q$, since the heavy quark
spin interactions are suppressed by its large mass.

For S-wave $Q\bar{q}$ systems, $j_q=1/2$ and there are two meson states, one with $J^P=0^-$ and
the other with $J^P=1^-$, corresponding to antiparallel and parallel $\vec{j}_q$ and $\vec{s}_Q$
configurations.  For P-wave systems, $j_q$ can be either $1/2$ or $3/2$, and two doublets of
meson states are expected: one for $j_q=1/2$ that contains a $J^P=0^+$ and $1^+$ meson, and the
other for $j_q=3/2$ with a $J^P=1^+$ and a $2^+$ meson.

In the charm quark sector ($Q=c$) the non-strange ($q=u,d$) S-wave states are the
well established pseudo-scalar $D$ and vector $D^*$ isospin doublets. The corresponding 
strange $(q=s$) mesons are the $D_s^+$ and $D_s^{*+}$ isospin singlets.  The ``hyperfine'' mass
splitting, $m_{1^-}-m_{0^-}$, for the $D^*$-$D$ and $D_s^*$-$D_s$ are nearly equal; for both
systems it is about 140~MeV \cite{Olive:2016xmw}.  

The P-wave $D$ mesons, which are hard to produce and tend to be wide and overlapping, are difficult
to study experimentally.  For example, in the
$q=u,d$ system, the $j_q=1/2$, $J^P=0^+$ $D_0(2400)$-meson and the $j_q=1/2$, $J^P=1^+$ $D_1(2430)$ meson
have $\sim 300$~MeV natural widths and their properties have only recently been
established~\cite{Link:2003bd,Abe:2003zm,Aubert:2009wg}.  
In the $j_s=3/2$ $c\bar{s}$ system, the very narrow ($\Gamma=0.92\pm 0.05$~MeV) $D_{s1}(2536)$-meson
and the relatively narrow ($\Gamma=17\pm 4$~MeV) $D_{s2}^*(2573)$ meson were well established in the
1990s but, prior to the operation of $B$ factories, the $j_s=1/2$, $0^+$ and $1^+$ states had still
not been identified.  According to the quark model, these states were expected to have masses above the
$m_{D^{(*)}}+m_{K}$ threshold and it was suspected that strong decays to $DK$ and $D^*K$
final states made them wide, and difficult to see.

One of the  biggest surprises from the $B$ factories was the BaBar discovery of a very narrow
state decaying to $D_s\pi^0$ with mass near 2317~MeV produced in inclusive $\ee\rt D_s^+\pi^0+X$
interactions. 
CLEO quickly confirmed the BaBar $D_{s0}^*(2317)$ discovery and reported
the observation of a second narrow state decaying to $D_s^*\pi^0$ with mass near 2460~MeV that is now called the
$D_{s1}(2460)$~\cite{Besson:2003cp}.  The Belle experiment established the production of both states in
exclusive $B\rt\bar{D}D_{sJ}$ decays, observed the $D_{s1}(2460)\rt D_s^*\gamma$ decay mode, and established
the spin-parity of the $D_{s1}(2460)$ to be $J^P=1^+$~\cite{Krokovny:2003zq}.

The latest world-average results for their masses and 95\% CL upper limits on their natural
widths are~\cite{Olive:2016xmw}:
\begin{eqnarray}
D_{s0}^*(2317):~~M=2317.7 \pm 0.6~{\rm MeV}    &~&  \Gamma<3.8~{\rm MeV} \nonumber \\
D_{s1}(2460):~~M=2459.5\pm 0.6~{\rm MeV}       &~&  \Gamma<3.5~{\rm MeV}. \nonumber \\
\quad &~& \quad
\end{eqnarray}
Since the $J^P=1^+$ quantum numbers of the $D_{s1}(2460)$ match those expected for the P-wave $j_s=1/2$
 $1^+$ state, and the mass $D_{s1}$-$D_{s0}^*$ mass splitting, $141.8\pm 0.8$~MeV, closely matches the
hyperfine splitting measured in the S-wave systems, a natural interpretation of these two states is
that they are the ``missing'' P-wave $j_s=1/2$ $0^+$ and $1^+$ $c\bar{s}$ states.  Since the $D_{s0}^*$
($D_{s1}$) mass is below the $DK$ $(D^* K)$ mass threshold, it can only decay via isospin-violating 
processes, which accounts for its small width.\footnote{A $0^+$ assignment for the $D_{s0}^*$ is supported by
the absence of any evidence for $D_{s0}^*\rt D_s\gamma$ decays.}

However, their  masses, which are  much  lower than quark-model expectations for  P-wave $c \bar{s}$
states,\footnote{A recent quark calculation of the masses of the $j_q=1/2$ P-wave $c\bar{s}$ mesons finds
2484~MeV for the $0^+$ and 2549~MeV for the $1^+$ states~\cite{Godfrey:2015dva}.}
are a puzzle. This is illustrated in Table~\ref{tab:Dsj}, where masses of $c\bar{d}$ mesons are compared
with those of the corresponding $c\bar{s}$ mesons. The $c\bar{s}$-$c\bar{d}$ mass differences for the
S-wave $0^-$ and $1^-$ mesons are both very close to 100~MeV, as is also the case for the $1^+$ and $2^+$
$j_q=3/2$ P-wave mesons.  In the quark model, the corresponding $c\bar{d}$ and $c\bar{s}$ mesons have the
same configurations, and the $\approx 100$~MeV mass difference reflects the $s$~and~$d$ quark
mass difference.

\begin{table}[htbp]
\caption{\footnotesize
Comparison of the masses of the low-lying $D_s$ ($c\bar{s}$) and charged $D$ ($c\bar{d}$)
meson states. Here the $D_{s0}^*(2317)$ and the $D_{s1}(2460)$ are identified with the
P-wave, $0^+$ $j_q=1/2$ and $1^+$ $j_q=3/2$ mesons, respectively.  The masses are taken from
ref.~\cite{Olive:2016xmw}.
}
\label{tab:Dsj}
\begin{center}
\begin{tabular}{lc|c|cc|c}
  $L$   &   $j_{q}$   &  $J^P$    &  $m(c\bar{d})$    &  $m(c\bar{s})$  &   $m_{c\bar{s}}$-$m_{c\bar{d}}$ \\
        &            &           &     (MeV)         &     (MeV)       &      (MeV)           \\
\hline 
\hline 
 $0$    &  $1/2$     &   $0^-$   & $1869.6\pm 0.1$   & $1968.3\pm 0.1$  &     $98.7\pm 0.1 $         \\
        &            &   $1^-$   & $2010.3\pm 0.1$ & $2112.1\pm 0.4$  &     $101.7\pm 0.4 $        \\
\hline 
\hline 
 $1$    &  $1/2$     &   $0^+$   & $2318\pm 29$      & $2317.7\pm 0.6$  &     $0.3\pm 29$            \\
        &            &   $1^+$   & $2430\pm 36$      & $2449.5\pm 0.1$ &      $30\pm 36$            \\
        &  $3/2$     &   $1^+$   & $2423.2\pm 2.4$   & $2335.1\pm 0.6$  &     $111.9\pm 2.4$          \\
        &            &   $2^+$   & $2464.3\pm 1.6$   & $2571.9\pm 0.8$  &     $107.6\pm 1.8$         \\
\hline 
\hline 
\end{tabular}
\end{center}
\end{table}

This is not the case for the $D_{s0}^*(2317)$ and $D_{s1}(2460)$ mesons, if they are taken
to be the $j_q=1/2$ and $j_q=3/2$ P-wave mesons, in which case the masses of the $c\bar{s}$ and $c\bar{d}$ systems
are nearly equal, in spite of the $s$- and $d$-quark mass difference.
  This puzzling behavior led to speculation
that these states may contain four-quark components, either in a QCD tetraquark arrangement or as
molecule-like structure formed from $D$ and $K$ mesons.

In the tetraquark 
picture~\cite{Cheng:2003kg,Terasaki:2003qa,
Maiani:2004vq,Bracco:2005kt},  
the  $D^*_{s0}(2317)$ could be a $[cq][\bar s \bar q']$ state, in which case
a rich  spectrum of similar states is expected to exist. In particular there should
be electrically neutral and doubly charged partners. A BaBar study of $D_s^+\pi^{\pm}$ systems
produced in $\ee$ annihilations found no states in the $D_{s0}^*(2317)$ mass region in either
channel~\cite{Aubert:2006bk} and Belle reported upper limits on the production of
neutral or doubly charged partner states in $B\rt \bar{D} D_s^+\pi^{\pm}$ that are an order
of magnitude below isospin-based predictions, making the tetraquark option
implausible~\cite{Choi:2015lpc}. 

The $D_{s0}^*$ and $D_{s1}$ lie about 40~MeV below the $DK$ and $D^*K$ mass thresholds, 
respectively, suggesting that they might be $DK^{(*)}$ molecules~\cite{Barnes:2003dj},
or mixtures of a $DK^{(*)}$ molecule and a conventional $c\bar{q}$ meson~\cite{Browder:2003fk}.
The idea that the  $D^*_{s0}(2317)$ and $D_{s1}(2460)$ may be  $DK^{(*)}$ bound states was studied
with lattice QCD by Leskovec {\em et al.}~\cite{Leskovec:2015naf}.  The authors examined 
the $J^P=0^+$ and $J^P=1^+$ states that are produced when $DK$ and $D^*K$ scattering operators
are included with the standard $c\overline s$ meson operators. The analysis established the
existence of below-threshold poles with binding energies consistent with  $D_{s0}^*(2317)$ and
$D_{s1}(2460)$.

\subsection{Bottom}

\begin{figure*}[htbp]
\begin{center}
\quad\\[-3cm]
 \includegraphics[width=0.48\textwidth]{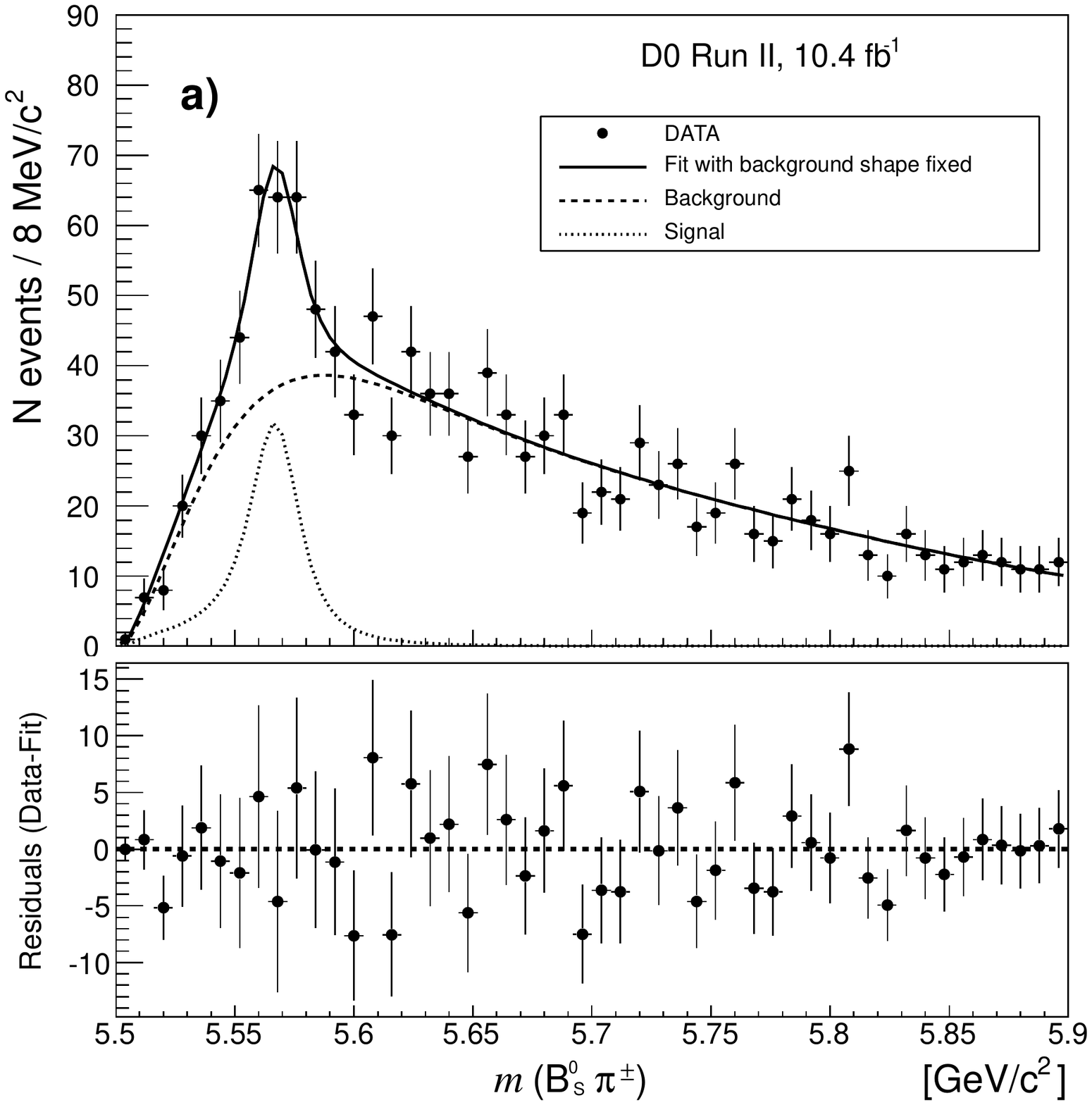} \,
 \includegraphics[width=0.48\textwidth]{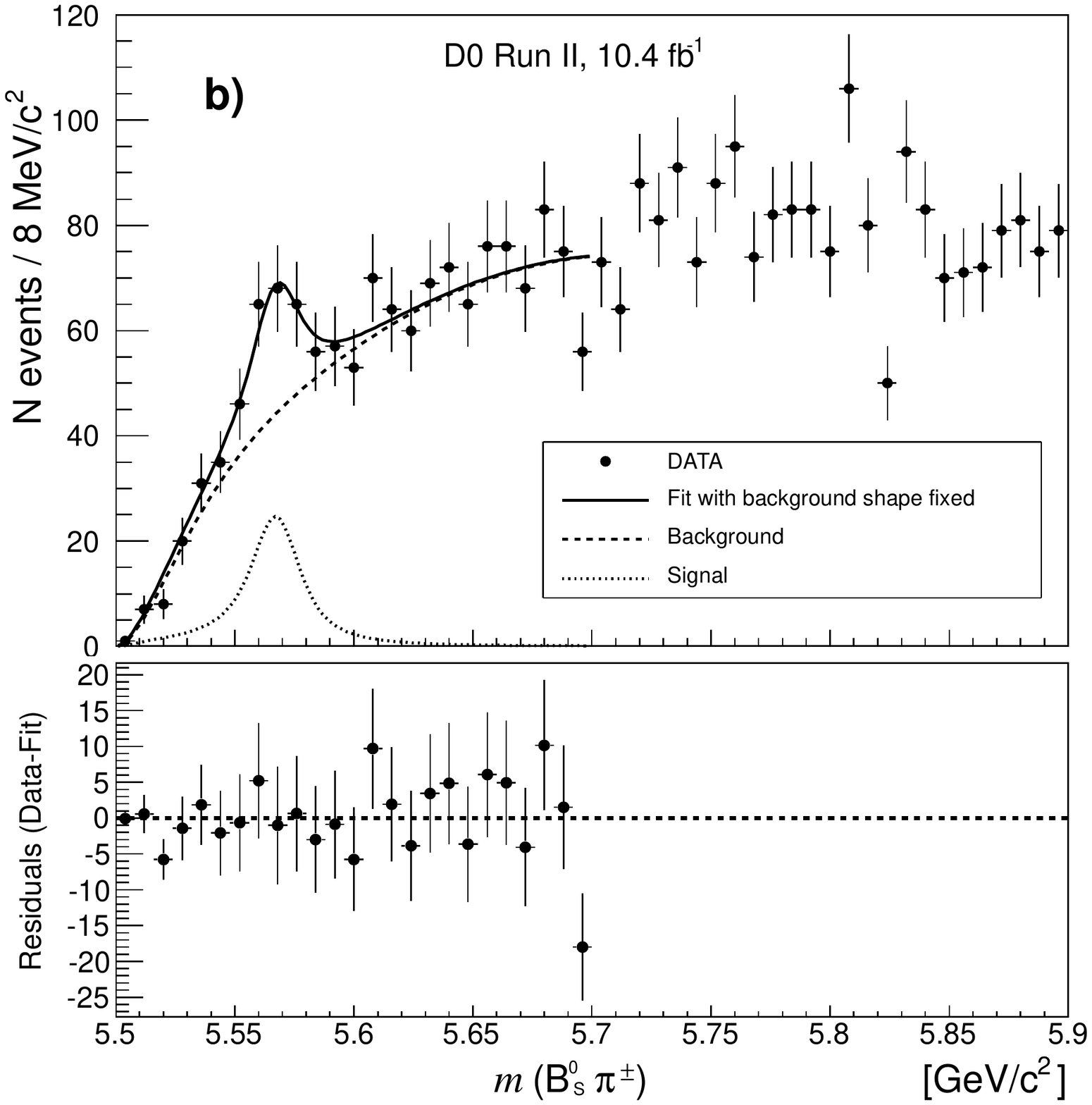}    
\caption{\footnotesize The $M(B_s^0 \pi^{\pm})$ distribution
together with the background distribution (dashed curve) and fit results (solid curve)
for events {\bf a)} with  and {\bf b)} without the application of the
$\Bs$-$\pi^{\pm}$ opening angle requirement (from ref.~\cite{D0:2016mwd}).
}
\label{fig:mbspidr}
\end{center}
\end{figure*}

Recently the D0 collaboration reported  evidence for  a possible four-quark state 
that decays to $\Bs\pi^\pm$, where the $\Bs$ decays via \Bsjp\ \cite{D0:2016mwd}. 
In the analysis, 5,500 reconstructed \Bsjp\ decay candidates were paired with
a charged particle that was assumed to be a pion.  Multiple entries for a single
event, which occur when more than one pion candidate passes the track selection criteria,
were suppressed by limiting the angular separation of the $\Bs$ candidate and the
charged track.\footnote{The angular separation requirement was
$\Delta R< \sqrt{\Delta\eta^2 + \Delta\phi^2}$, where $\eta=\ln[\tan\theta/2]$ is the
``pseudorapidity'' and $\phi$ is the azimuthal angle.}
The resultant $\Bs\pi^\pm$ invariant mass spectrum, shown in Fig.~\ref{fig:mbspidr}a,  
has a narrow structure that is approximately 60~MeV above the $m_{B_s}+m_{\pi}$ threshold.

The solid curve in  Fig.~\ref{fig:mbspidr}a shows the results of a fit that uses a
resolution-broadened  BW line shape to represent a signal (dotted curve) plus an
incoherent background modeled by an empirical shape that was determined from $\Bs$-mass
sideband events (for non-$\Bs$ background) and MC-simulated inclusive $\Bs$ production
events (for combinatoric backgrounds associated with real $\Bs$ mesons). The fitted mass and
width of the peak, which the D0 group called the $X(5568)$, are 
$M=5567.8 \pm  2.9  {\rm \thinspace (stat)} ^{+0.9}_{-1.9}  {\rm \thinspace (syst)}$~MeV
and 
$\Gamma=21.9 \pm 6.4  {\rm \thinspace (stat)}   ^{+5.0}_{-2.5} {\rm \thinspace (syst)} $~MeV.
The signal significance, including look-elsewhere and systematic-uncertainty effects,
is 5.1$\sigma$.  The $X(5568)$ signal is also evident in the $\Bs\pi^{\pm}$ spectrum
without the pion direction restriction, shown in Fig.~\ref{fig:mbspidr}b, albeit
with reduced significance. The ratio of the number of $B_s$ mesons that originate from 
$X(5568)$ decays to all $B_s$ mesons, is determined for the D0 acceptance to be
$\rho_X^{D0}=(8.1\pm 2.4)\%$.

In an alternative approach,  the authors extracted  a  $B_s^0 \pi^{\pm}$ signal by performing
fits of the number of $B_s^0$ events in the  $J/\psi \phi$ mass distribution in 20-MeV intervals 
of  $M(B_s^0\pi^{\pm})$. The results of that fit  confirm that
the observed signal is due to events with genuine $B_s^0$ mesons and eliminates the possibility
that some non-$B_s^0$ process may mimic the signal.

\begin{figure}[bthp]
\begin{center}
\quad\\[-6.5cm]
\quad \includegraphics[width=0.5\textwidth]{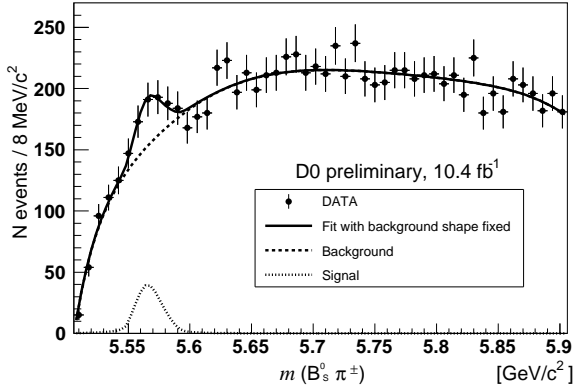} \quad
\caption{\footnotesize 
 The invariant mass distribution  $M(\Bs\pi^\pm)$ for the $B_s^0 \rightarrow D_s \mu X$
event sample with the results of the fit superimposed (from ref.~\cite{bspisl}).}
\label{fig:mbspisl}
\end{center}
\end{figure}

To confirm the production of
the  $X(5568)$ with an independent sample, D0 studied  events in which the $B_s^0$ meson decayed 
semileptonically~\cite{bspisl}. Decays \Bsdecay, where $X$ denotes a neutrino possibly accompanied by mesons, are much more
abundant than the exclusive decay \Bsjp.  The mass resolution is worse due to the presence of the undetected
neutrino, but it is possible to select events where this effect is minimized by requiring that the $D_s$ meson
and the muon account for a large fraction of the $B_s^0$ momentum.  The data show an excess over the simulated
background at the mass expected from events produced by the decay $X(5568) \to \Bs\pi^\pm$ with $\Bsjp$,
as seen in Fig.~\ref{fig:mbspisl}, thereby providing a confirmation of the results of ref.~\cite{D0:2016mwd}. 
The combined significance of the signal in the two channels
is 5.7 standard deviations.

\begin{figure*}[bthp]
  \quad\hskip-1.3cm 
  \includegraphics*[width=0.5\textwidth]{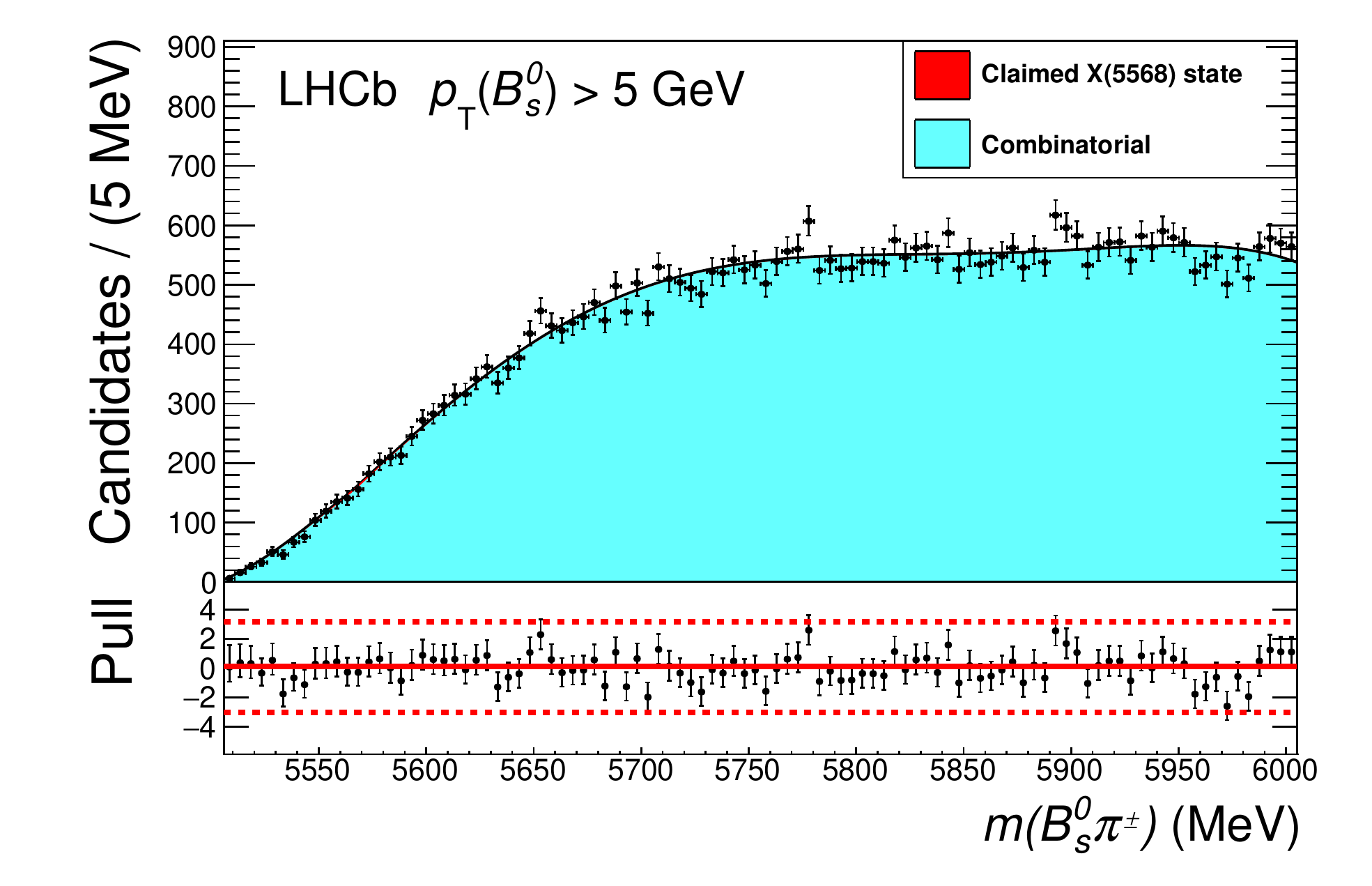}
  \quad\hskip-0.5cm 
  \hbox{ \includegraphics*[width=0.46\textwidth]{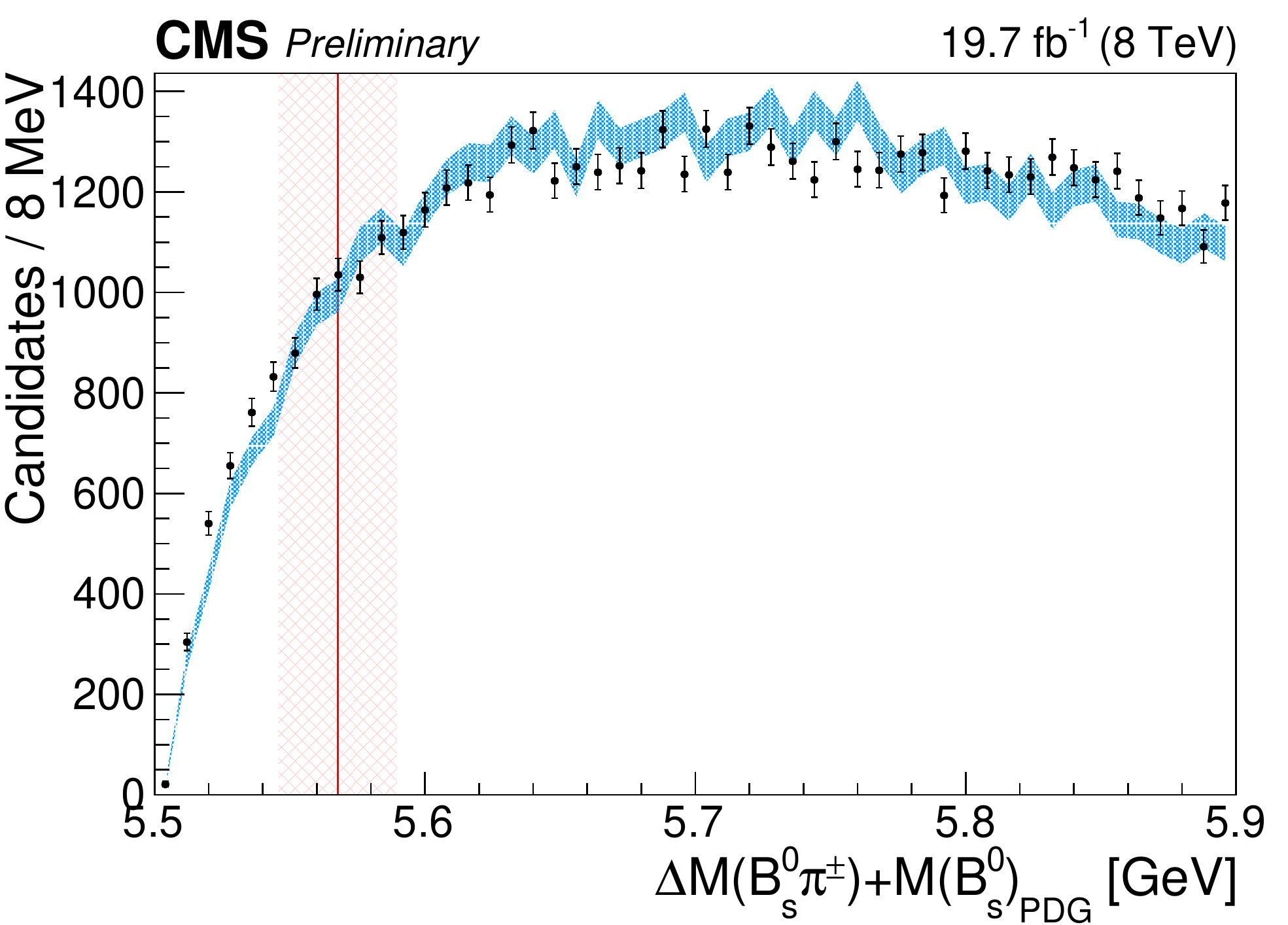} 
  \quad\hskip-0.5cm } 
\caption{
  The $\Bs\pi^{\pm}$ distributions (points with error bars) 
  observed by the LHCb \cite{Aaij:2016iev} {\em (left)} 
  and by the CMS \cite{CMS:2016fvl} {\em (right)}.
  The LHCb plot shows the result of a fit an $X(5568)$ signal (not visible) included 
  on top of the combinatorial background. 
  The continuous blue band in the CMS plot is the distribution observed for the $\Bs$-mass sideband data.
  The vertical red band illustrates the $M_X\pm\Gamma_X$ region of the X(5568) state claimed by D0. 
  \label{fig:lhcbcms_bspi}
}
\end{figure*}

The LHCb group searched for $X(5568)$ production in $pp$ collisions at $\ecm=7-8$ TeV \cite{Aaij:2016iev}.
With 44,000 $\Bsjp$  and 65,000 $\Bs\to D_s^-\pi^+$ reconstructed events and a superior signal-to-background ratio, 
no peaking structure in the $\Bs\pi^{\pm}$ invariant mass distribution is observed in the $X(5568)$ mass region (see
the left panel of Fig.~\ref{fig:lhcbcms_bspi}). Upper limits on $\rho_X^{\rm LHCb}$ of $<1.2\%$~($<2.4\%$) for $p_{\rm T}(X)>5$ GeV~($>10$ GeV)
are established at the 95\%\ C.L. In addition, the CMS group found no sign of an $X(5568)$ peak in a preliminary analysis of
a 48,000 event sample of reconstructed $\Bsjp$ decays, and reported an upper limit of $\rho_X^{\rm CMS}<3.9\%$ at the
95\%\ C.L.~\cite{CMS:2016fvl} (see the right panel of Fig.~\ref{fig:lhcbcms_bspi}).

No satisfactory theoretical description of the $X(5568)$ structure has yet 
been proposed~\cite{Burns:2016gvy,Guo:2016nhb,Esposito:2016itg,Yang:2016sws}.
If confirmed, this state would be comprised of two quarks and two antiquarks of four different flavors: $b, s, u, d$.
Such a state  might be a  tightly  bound $B_d^0$-$K^\pm$ molecule or a $[bd]$-$[\bar{s}\bar{u}]$ tetraquark. However, the
low mass of the $X(5568)$, about 200 MeV below both the $B_d^0 K^\pm$ threshold and the three-quark ($bsu$) $\Xi_b$ baryon,
disfavors both of these interpretations.  A Lattice QCD study of the $B_s \pi$ scattering~\cite{Lang:2016jpk}
does not support the existence a $J^P=0^+$ state with the $X(5568)$ parameters.

\myclearpage

\section{Neutral exotic hadron candidates}
\label{sec:neutral}

\subsection{$X(3872)$}
\label{sec:x3872}
\begin{figure*}[htb]
\includegraphics[height=0.36\textwidth,width=0.45\textwidth]{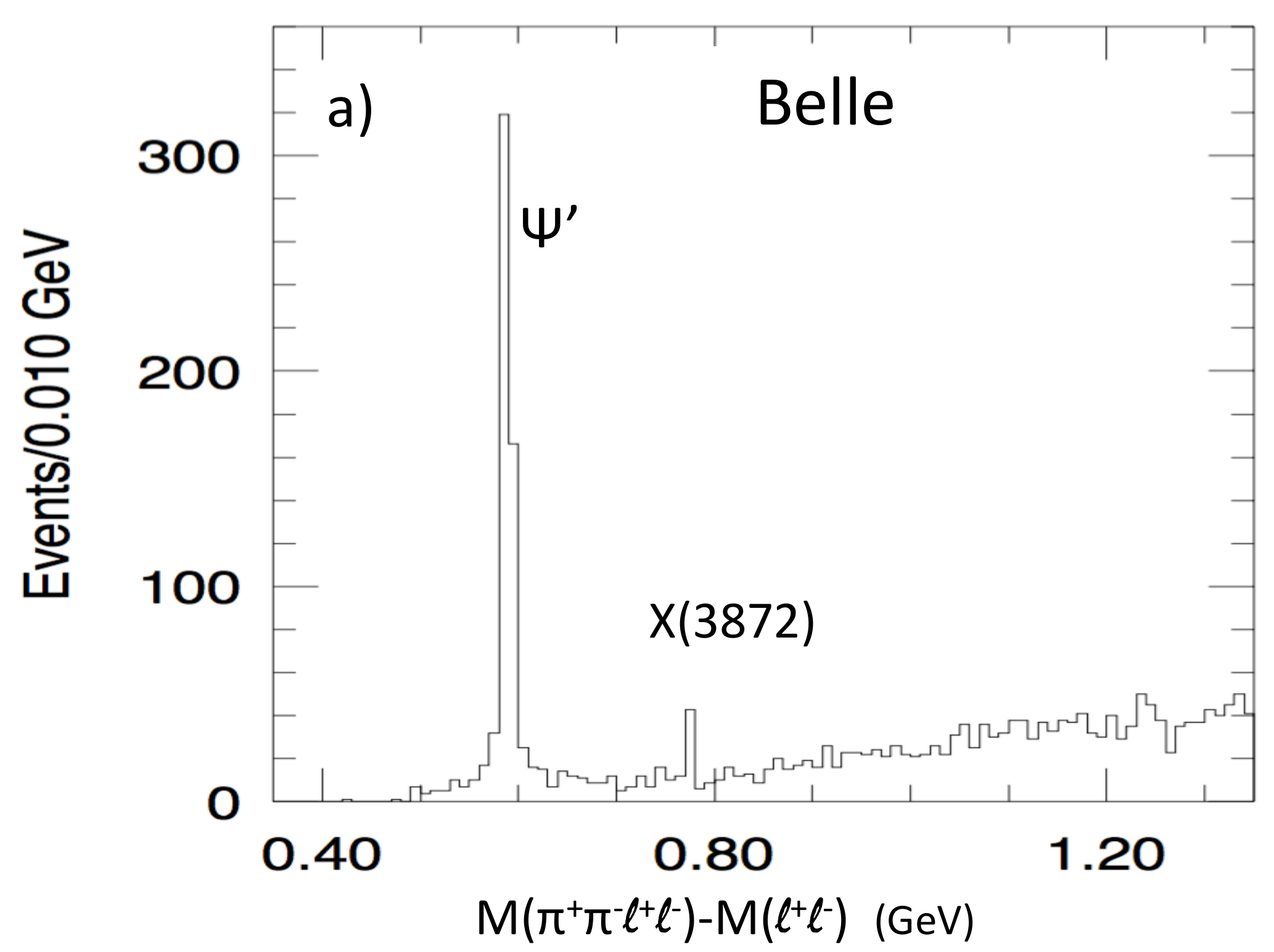}
\includegraphics[height=0.36\textwidth,width=0.45\textwidth]{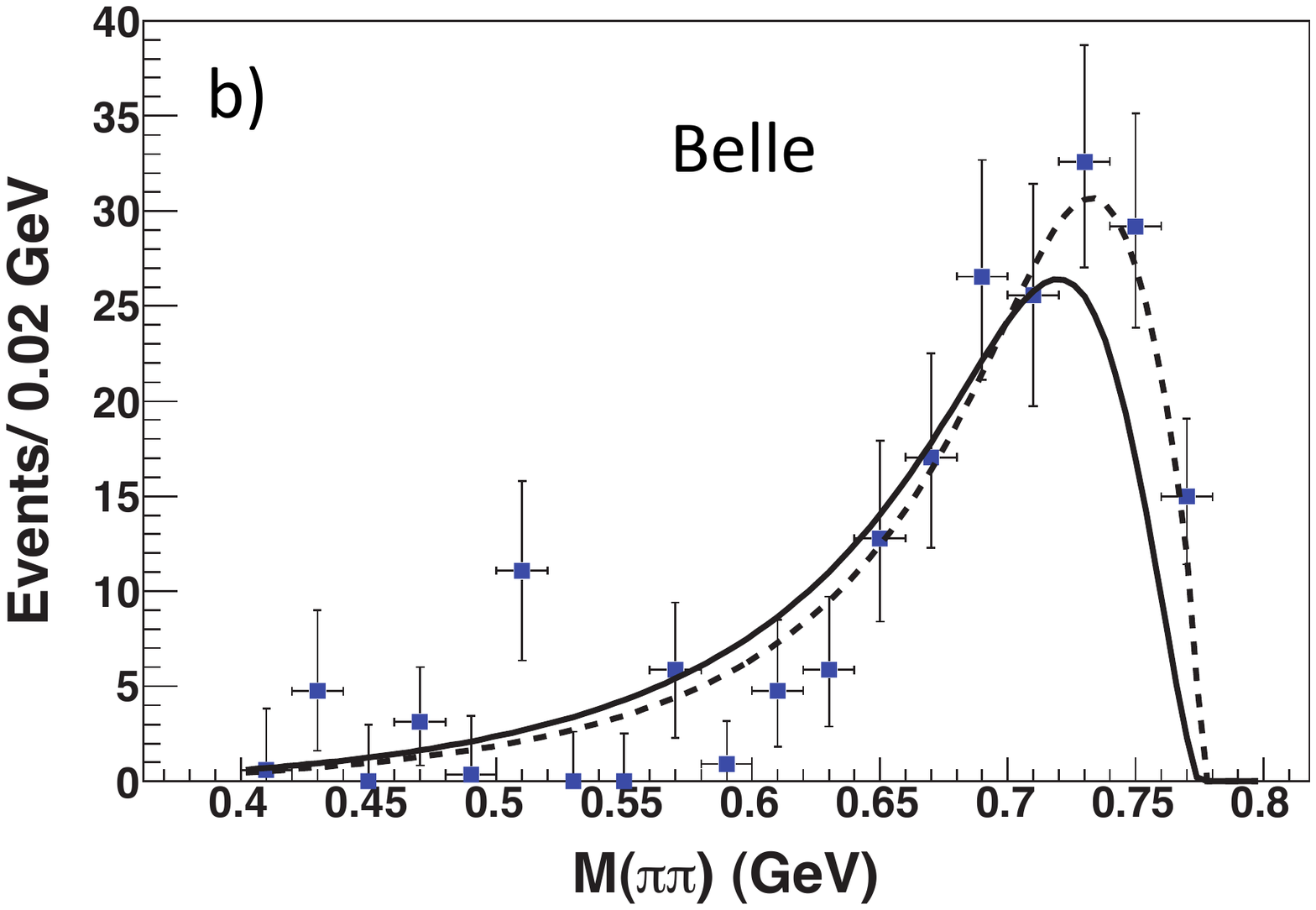}
\caption{\footnotesize {\bf a)} The $M(\pipi\ell^+\ell^-)-M(\ell^+\ell^-)$ distribution for
$B\rt K\pipi\ell^+\ell^-$ events with $|M(\ell^+\ell^-)-m_{\jpsi}|<20$~MeV~\cite{Choi:2003ue}. 
{\bf b)}
The $\pipi$ invariant mass distribution for $X(3872)\rt\pipi\jp$ events in Belle.
The curves shows results of fits to a $\rho\rt\pipi$ line shape including $\rho$-$\omega$
interference~\cite{Choi:2011fc}.  The dashed (solid) curve is for even (odd) $X(3872)$ parity
}
\label{fig:x3872}
\end{figure*}

The first quarkonium-like candidate for a non-standard hadron to be seen was the $X(3872)$, which was found by
Belle~\cite{Choi:2003ue} as an unexpected narrow peak in the $\pipi\jp$ invariant mass distribution in
$B\rt K\pipi\jp$ decays shown in Fig.~\ref{fig:x3872}a. It is experimentally well established, having been
seen and studied by a number of
experiments~\cite{Acosta:2003zx,Abazov:2004kp,Aubert:2004ns,Aaij:2011sn,Chatrchyan:2013cld,Ablikim:2013dyn,Aaboud:2016vzw}.  
Its most intriguing feature is its mass: the 2016 PDG world-average value is $M(X(3872)) = 3871.69\pm 0.17$~MeV, which,
at current levels of precision, is indistinguishable from the $D^0\bar{D}^{*0}$ mass threshold
$m_{D^0} + m_{D^{*0}}=3871.68\pm 0.10$~MeV~\cite{Olive:2016xmw};  the difference is
$\delta m_{00} \equiv (m_{D^0}+m_{D^{*0}})-M(X(3872)) = -0.01 \pm 0.20$~MeV.  Whether this close proximity of the $X(3872)$
to the $D^0\bar{D}^{*0}$ mass threshold is a coincidence or a feature of hadron dynamics is an issue that has attracted
considerable interest.  The $X(3872)$ is also quite narrow; Belle has reported a 90\% C.L. upper limit on its total width
of $\Gamma <1.2$~MeV~\cite{Choi:2011fc}.   
In addition to the production in $B\rt X(3872) K$ decays, the $X(3872)$ state was
also observed in $B\rt X(3872) K\pi$ decays \cite{Bala:2015wep}. 

The radiative decay process $X(3872)\rt\gamma\jp$ has been measured by BaBar~\cite{Aubert:2006aj} and 
Belle~\cite{Bhardwaj:2011dj} to have a branching fraction that is $0.24\pm 0.05$ that for the $\pipi\jp$ mode.
This, plus BaBar~\cite{Aubert:2008ae} and LHCb~\cite{Aaij:2014ala} reports of strong evidence for
$X(3872)\rt\gamma\psip$ decays (see Fig.~\ref{fig:LHCb_gammapsip}b), establish the charge conjugation
parity of the $X(3872)$ as even ($C=+$), in which case the $\pipi$ system in the $X\rt\pipi\jpsi$ decay process
must be from $\rho^0\rt\pipi$ decay. This is consistent with $\pipi$ line-shape measurements done by
CDF~\cite{Abulencia:2005zc}, Belle~\cite{Choi:2011fc}, and CMS~\cite{Chatrchyan:2013cld}.
The $\pipi$ line shape measured by Belle is shown in Fig.~\ref{fig:x3872}b.

\begin{figure*}[htb]
  \includegraphics[width=\textwidth]{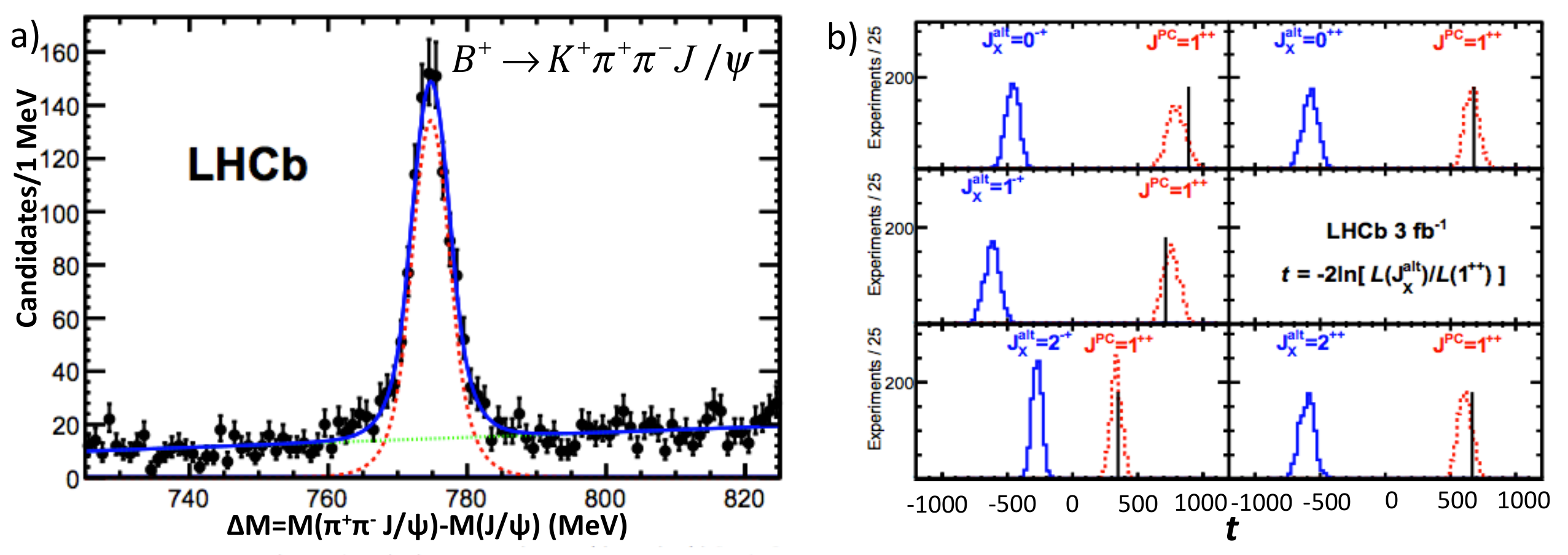}
\caption{\footnotesize {\bf a)} The $M(\pipi\jpsi)-m_{\jpsi}$ distribution for
$B^+ \rt K^+ \pipi\jpsi$ events from LHCb~\cite{Aaij:2015eva}. 
{\bf b)}  Distributions of  $t\equiv -2\ln[{\mathcal L}^{\rm alt}/{\mathcal L}^{++}]$  for simulated experiments
for alternative $J^{PC}$ hypotheses (blue histograms) and $1^{--}$ (red dashed histograms).  The vertical
lines show the values of $t$ determined from the data.  Similar results for $J^{PC}=3^{\pm +}$ and $4^{\pm +}$
are not shown.}
\label{fig:lhcb_Jpc}
\end{figure*}

In their 3~fb$^{-1}$ Run-I data sample, the LHCb experiment detected a $1011\pm 38$~signal events for the decay
chain: $B^+\rt K^+ X(3872)$;  $X(3872)\rt\pipi\jpsi$; $\jpsi\rt\mumu$ on a small background as shown in
Fig.~\ref{fig:lhcb_Jpc}a. In an amplitude analysis based on the angular correlations among the five final-state
particles in these events, the LHCb group found that the $J^{PC}=1^{++}$ quantum number hypothesis had the highest
likelihood value~\cite{Aaij:2015eva,Aaij:2013zoa}. They evaluated the significance of the $1^{++}$ assignment using
the likelihood ratio $t\equiv -2\ln[{\mathcal L}^{\rm alt}/{\mathcal L}^{++}]$ as a test variable, where ${\mathcal L}^{++}$
is the likelihood for the $1^{++}$ hypothesis and ${\mathcal L}^{\rm alt}$ is that for an alternative. (With 
this definition, positive values of $t$ favor $1^{++}$.)  The solid-blue (dashed-red) histograms in Fig.~\ref{fig:lhcb_Jpc}b
show $t$ value distributions for ensembles of Monte Carlo (MC) experiments generated with alternative and $1^{++}$ $J^{PC}$
values,\footnote{The experiments are generated with numbers of signal and background events that fluctuate around those
in the experimental data according to the observed statistical errors.}  with the $t$ value determined for the real data
indicated by vertical black lines.  The experimental results favor the $1^{++}$ hypotheses over all alternative even-$C$
$J^{PC}$ assignments with $J\le 4$ by a wide margin; in all comparisons the statistical significance of the $1^{++}$
assignment, determined from the distributions for the ensembles of MC experiments, is more than $16\sigma$. 

The only available $1^{++}$ standard charmonium level that is expected to have a mass near 3872~MeV is the
$2^{3}{\rm P}_{1}$ $\ccbar$ state, commonly known as the $\chi_{c1}({\rm 2P})$ or $\chi_{c1}^{\prime}$.  However,
for a number of reasons, the assignment of the $X(3872)$ as the $\chi_{c1}^{\prime}$ charmonium state has been
deemed to be ``improbable''~\cite{Eichten:2005ga}.  Among these are the $X(3872)$ mass and width values, and
the apparent isospin violation in its discovery decay channel $X(3872)\rt\rho\jpsi$. 

\underline{\it Mass:}~~
The $\chi_{c2}^{\prime}$, the $J=2$ spin-multiplet partner of the $\chi_{c1}^{\prime}$, was identified by
Belle in 2006 as a distinct peak in the $\gamma\gamma\rt D\bar{D}$ cross section at $3927\pm 3$~MeV (see
Fig.~\ref{fig:chic2prime}b), with an angular distribution that is characteristic of a D-wave $D\bar{D}$
meson system and a production rate that is consistent with charmonium model expectations for the
$2^{3}{\rm P}_2$~$\ccbar$ state~\cite{Uehara:2005qd}.  BaBar confirmed this observation in 2010~\cite{Aubert:2010ab}
and found properties that are consistent with those reported by Belle.  There is general agreement in the quarkonium
community that the identification of the Belle peak as the $\chi_{c2}^{\prime}$ is reliable~\cite{Brambilla:2010cs}. If,
with this assignment for the $\chi_{c2}^{\prime}$, the $X(3872)$ is identified as the $\chi_{c1}^{\prime}$, the
$\chi_{c2}^{\prime}$-$\chi_{c1}^{\prime}$ mass splitting would be $\delta M_{2-1}(2{\rm P})=56\pm 3$~MeV and larger than the
measured splitting for the $1{\rm P}$ states: $\delta M_{2-1}(1{\rm P})=46.5\pm 0.1$~MeV.  This conflicts with $\ccbar$
potential model expectations that $\delta M_{2-1}(n_r {\rm P})$ decreases with increasing radial quantum
$n_r$~\cite{Godfrey:1985xj,Barnes:2005pb}.  For $\ccbar$ states above the threshold for decays into open-charmed
$D\bar{D}$ and $D\bar{D}^*$ mesons, potential model calculations should be modified to include the effects of
intermediate on-mass-shell open-charmed-meson loops.  These effects have been estimated by three group using three
different approaches~\cite{Eichten:2004uh,Li:2009zu,Wang:2014voa}; all three of these analyses predict that
$\delta M_{2-1}(2{\rm P})$ decreases to values that are even lower than potential model expectations. 

\underline{\it Width:}~~
The Belle group's upper limit $\Gamma(X(3872))<1.2$~MeV is only slightly higher than the measured width of the
the 1P $\chi_{c1}$ state, $\Gamma(\chi_{c1})=0.84\pm 0.04$~MeV~\cite{Olive:2016xmw}.  However, since the $X(3872)$
has a number of additional allowed decay channels that are not accessible to the $\chi_{c1}$, including, for example,
the $X(3872)\rt\rho\jpsi$ discovery mode and the order of magnitude stronger $D^0\bar{D}^{*0}$ mode that is discussed
below, these are expected to be reflected in a substantially larger total width if the $X(3872)$ were, in fact, the
$\chi_{c1}^{\prime}$~\cite{Eichten:2004uh}.

\underline{\it Isospin Violation:}~~
Since standard charmonium states contain no constituent $u$- or $d$-quarks, they necessarily have zero isospin. On the
other hand, since the $\rho$-meson is an isovector, the $\rho\jpsi$ decay final state has isospin $I=1$, and the
$\chi_{c1}^{\prime}\rt \rho\jp$ decay process violates isospin and should be strongly suppressed, and an unlikely discovery
mode for a charmonium state~\cite{Eichten:2005ga}. 

These reasons, plus the close correspondence between its mass and the $m_{D^0}+m_{D^{*0}}$ threshold, led to considerable
speculation that the substructure of the $X(3872)$ is more complex than that of a simple $\ccbar$ charmonium
state~\cite{Tornqvist:2003na}.

An interesting question about the $X(3872)$ is the value of its isospin. Explicit evidence for strong isospin violation
in $X(3872)$ decays came from observations by both Belle~\cite{Abe:2005ax} and BaBar~\cite{delAmoSanchez:2010jr} of the
$X(3872)\rt\omega\jpsi$ decay mode with a branching fraction that is nearly equal to that for $\rho\jpsi$; the PDG
average is ${\mathcal B}(X(3872)\rt\omega\jpsi)/{\mathcal B}(X(3872)\rt\pipi\jpsi)=0.8\pm 0.3$~\cite{Olive:2016xmw}.
Since $M_{X(3872)}-m_{\jpsi} \simeq 775$~MeV, the upper kinematic boundary for the mass of the $\pipi$ system is right at the
peak mass of the $\rho$ resonance and $\sim 7$~MeV below $m_{\omega}$. Thus, while the decay $X(3872)\rt \rho\jp$ is
kinematically allowed to proceed through nearly the entire low-mass side of the $\rho$ resonance, $X(3872)\rt \omega\jpsi$
can only proceed via a small fraction of the low-mass tail of the $\omega$ peak. These considerations imply a kinematic
suppression of the amplitude for $X(3872)\rt\omega\jp$ decays relative to the $\rho\jp$ channel by a factor of about $\sim $4, in which
case the near equality of the $\rho\jp$ and $\omega\jp$ decay rates implies that an $I=0$ assignment is
favored~\cite{Suzuki:2005ha}, but a sizable isovector component in the X(3872) wavefunction is still allowed.
If the $X(3872)$ had $I=1$, it would have charged partners. Searches for
narrow, charged partners of the $X(3872)$ decaying into $\rho^{\pm}\jpsi$ by BaBar~\cite{Aubert:2004zr} and
Belle~\cite{Choi:2011fc} set branching ratio limits that are well below expectations based on isospin conservation. These
results suggest that the $X(3872)$ is mostly an isospin singlet and that the $\rho\jpsi$ decay mode violates isospin
symmetry. 

\begin{figure*}[htb]
\includegraphics[width=0.45\textwidth]{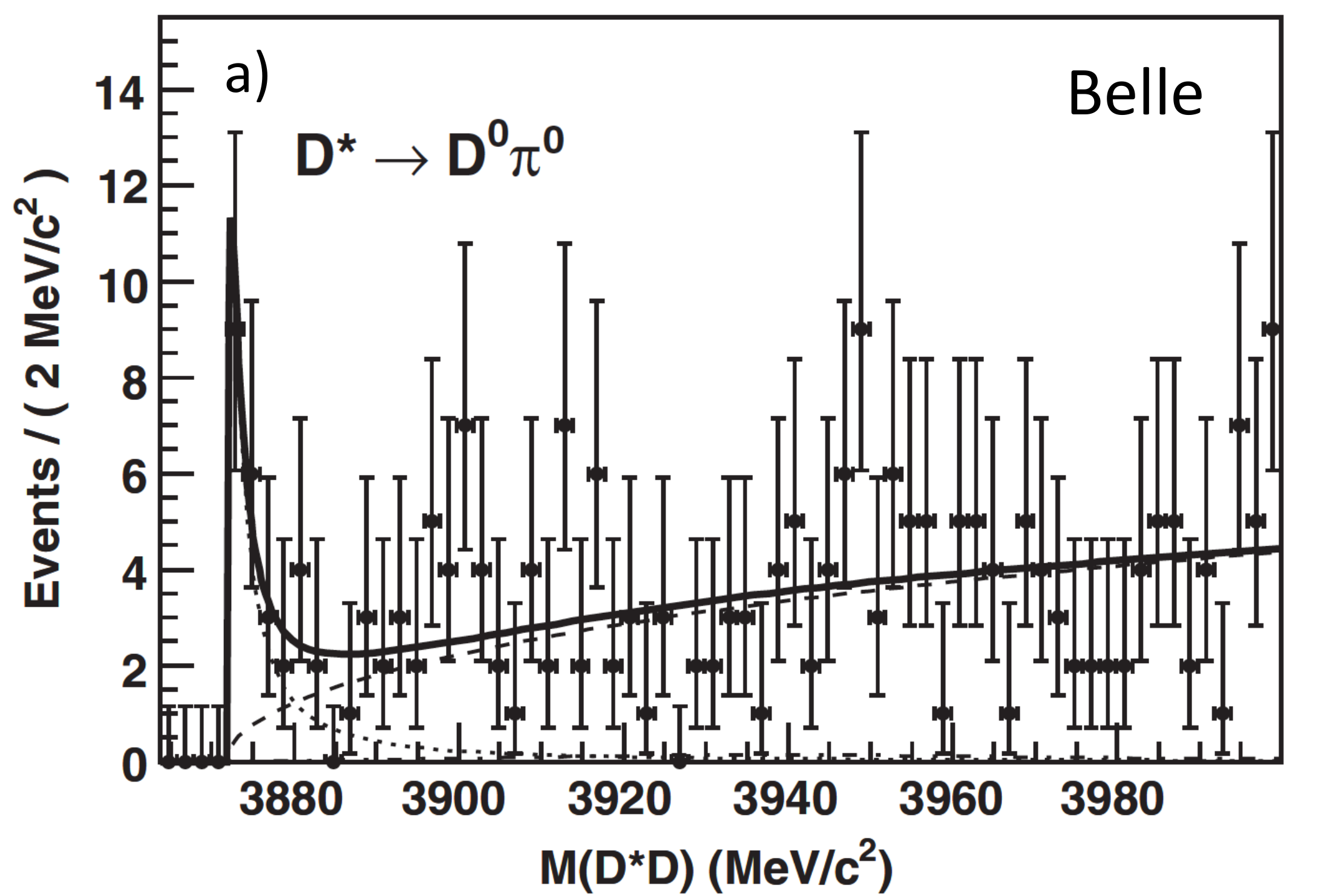}
\includegraphics[width=0.45\textwidth]{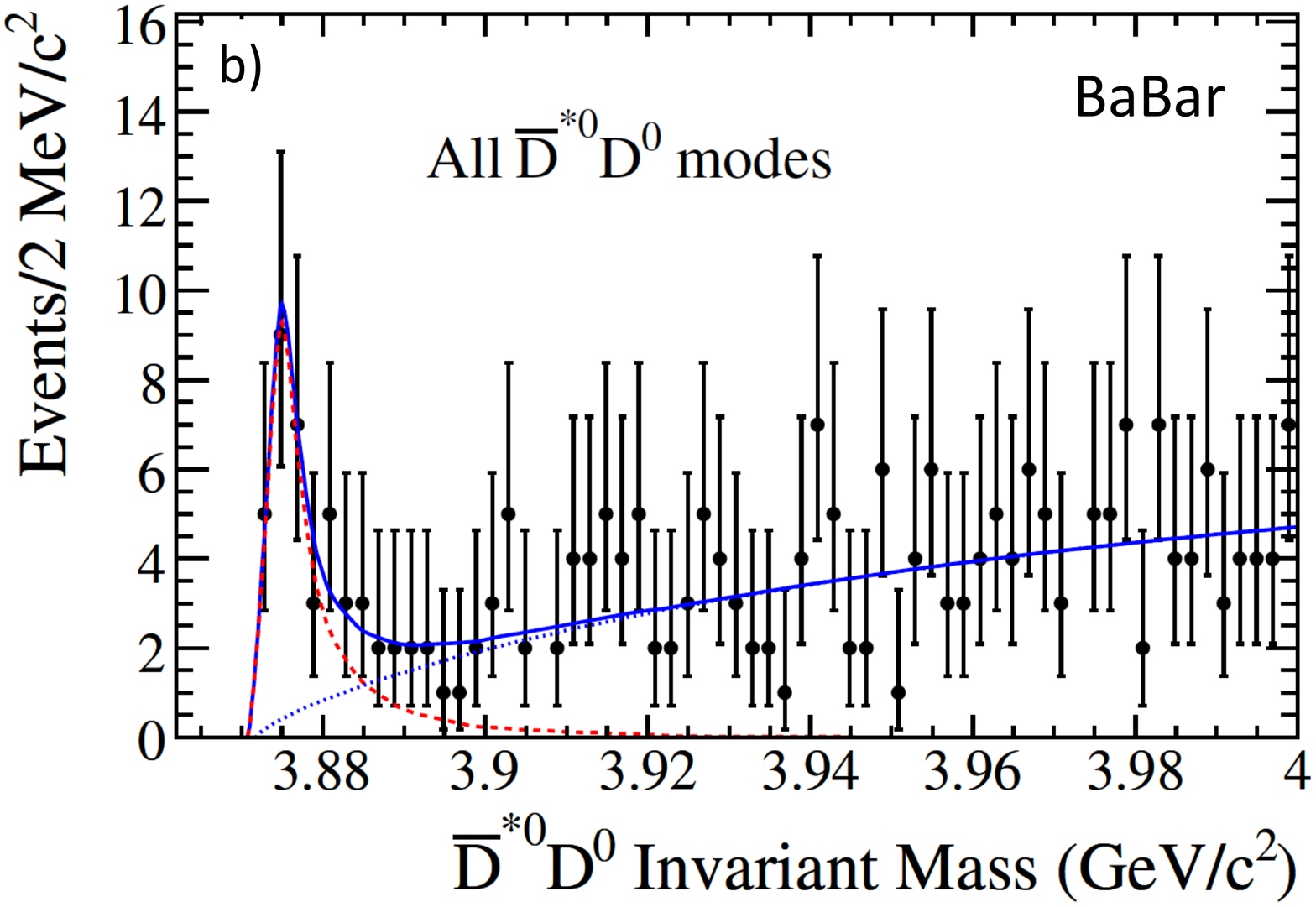}
\caption{\footnotesize $M(D^0\bar{D}^{*0})$ distributions from $B\rt K D^0\bar{D}^{*0}$ from
{\bf a)} Belle~\cite{Aushev:2008sua} and {\bf b)} BaBar~\cite{Aubert:2007rva}.
The peaks near threshold are the signals
$X(3872)\rt D^0\bar{D}^{*0}$ decays.
}
\label{fig:x3872-ddstr}
\end{figure*}

The $X(3872)\rt D^0\bar{D}^{*0}$ decay mode has been observed by both Belle~\cite{Aushev:2008sua} and
BaBar~\cite{Aubert:2007rva} with a  measured branching fraction that is $9.9\pm 2.3$~times that for the $\pipi\jp$ 
channel (see Figs.~\ref{fig:x3872-ddstr}a and b). The $J^{PC}=1^{++}$ quantum number assignment implies that the $X(3872)$
couples to a $D^0\bar{D}^{*0}$ pair in an S- and/or D-wave and, because the $D^0\bar{D}^{*0}$ system is right at
threshold, the S-wave can be expected to be dominant. In this case some very general and universal theorems
apply~\cite{Braaten:2007dw,Braaten:2004rn,Coito:2012vf,Polosa:2015tra}. One consequence of these theorems is that, independently of its
dynamical origin, the $X(3872)$ should exist for a significant fraction of the time as a $D^0\bar{D}^{*0}$ molecule-like
state (either bound or virtual) with a size comparable to its scattering length: $a_{00}=\hbar/\sqrt{\mu |\delta m_{00}}|$,
where $\mu_{00}$ is the $D^0\bar{D}^{*0}$ reduced mass. The limited experimentally allowed range for non-zero $\delta m_{00}$
values given above implies that the mean $D^0$-$\bar{D}^{*0}$ separation has to be huge: $a_{00}\ge 7$~fm.  Although the
$X(3872)$ mass is well below the $D^{+}D^{*-}$ mass threshold, it is expected to exist for a smaller fraction of the time as
a $D^+D^{*-}$ molecule-like state. The extent of the $D^{+}D^{*-}$ configuration, for which
$\delta m_{+-}=(m_{D^+}-m_{D^{*-}})-M(X(3872))= 8.2$~MeV and $a_{+-}\simeq 2$~fm, is much different.  The very different
properties of the $D^{0}\bar{D}^{*0}$ and $D^{+}D^{*-}$ configurations ensure that the $X(3872)$ isospin is not precisely
defined\footnote{Since $|I=1;I_{3}=0\rangle=[|D^0\bar{D}^{*0}\rangle+|D^{+}D^{*-}\rangle]/\sqrt{2}$
and $|I=0;I_{3}=0\rangle=[|D^0\bar{D}^{*0}\rangle-|D^{+}D^{*-}\rangle]/\sqrt{2}$, a well defined $I=1$ or $I=0$ state
implies equal $D^0\bar{D}^{*0}$ and $D^+D^{*-}$content.} as was first pointed out in ref.~\cite{Tornqvist:2004qy}.

One diagnostic of the nature of $X(3872)$ is the relative strength of the $\gamma\psip$ and $\gamma\jpsi$ decay
modes~\cite{Swanson:2004pp}. The preference for $\gamma\psip$ over $\gamma\jp$, as indicated in
Eq.~\ref{eq:gpsip_over_gjpsi}, is in accord with expectations for a $1^{++}$ charmonium, where the $\chi_{c1}^{\prime}$
and the $\psip$ have the same radial wave-function and the $\chi_{1}^{\prime}\rt \gamma\psip$ E1 transition is favored
over that for $\gamma\jp$, which are ``hindered''~\cite{Barnes:2005pb} by the mis-match between the orthogonal initial-
and final-state radial wave functions.\footnote{In this case the $2{\rm P}$-$1{\rm S}$ overlap integral is only non-zero
because the final-state $1{\rm S}$ radial wave-function is boosted relative to that for the initial-state $1{\rm P}$ state.}
In contrast, in models in which the $X(3872)$ is a pure molecular state, the $\gamma\psip$ decay channel is strongly
suppressed relative to that for $\gamma\jpsi$~\cite{Swanson:2003tb,Dong:2009uf}.

\begin{figure}[htb]
  \includegraphics[width=0.48\textwidth]{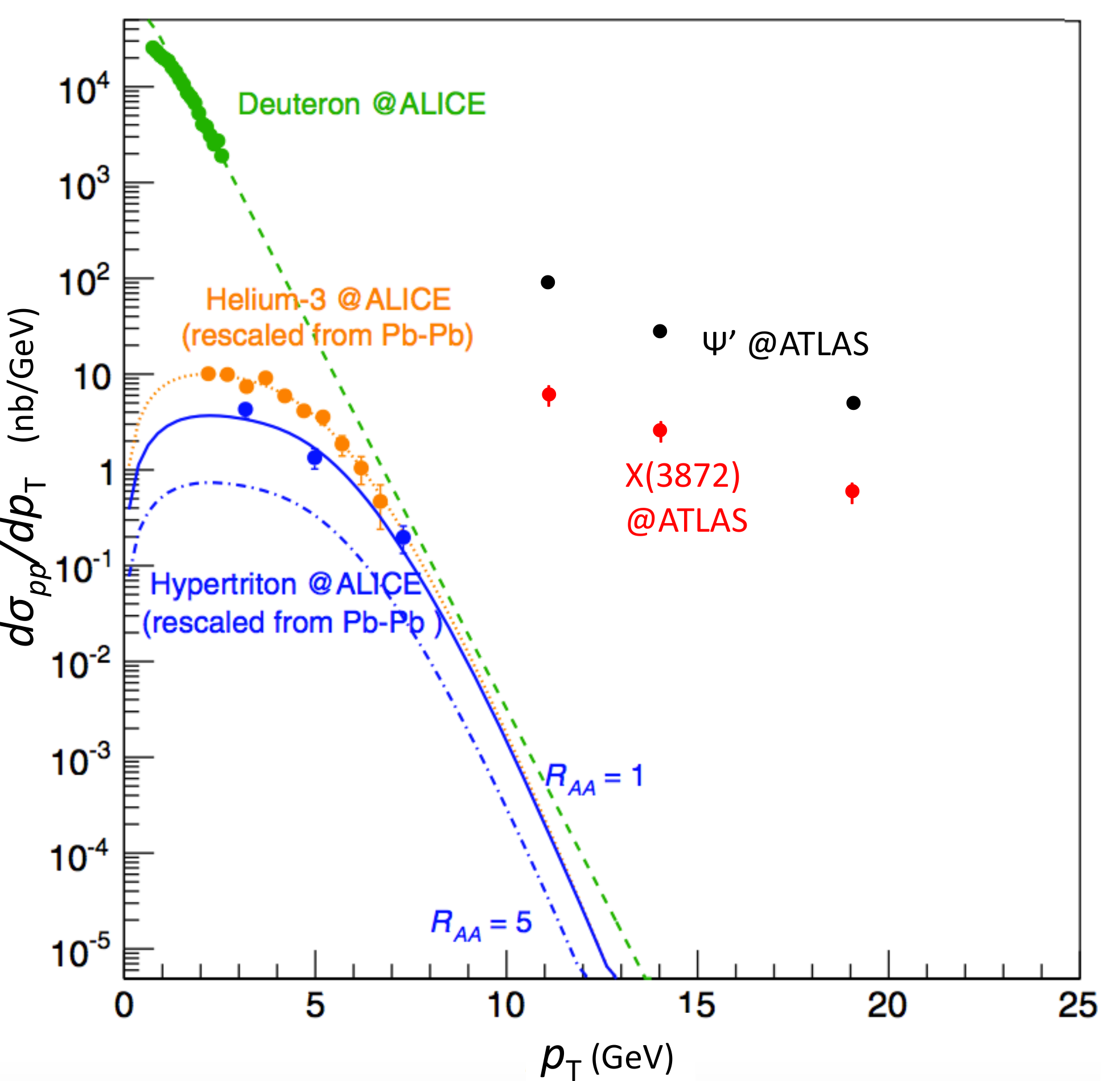}
\hspace{\fill}
\caption{\footnotesize 
Differential cross sections for particle production {\em vs.} $p_T$ at the LHC.  The solid red (black) circles
are ATLAS measurements of prompt $X(3872)$ ($\psip$) production in $\ecm=8$~TeV pp collisions~\cite{Aaboud:2016vzw}.
Results for deuteron (green) $^3He$ (orange), and $^3_{\Lambda}H$ (blue) are extrapolations of ALICE Pb-Pb measurements
at $\ecmx{NN}=2.76$~TeV to $\ecmx{pp}=7$~TeV using a Glauber model~\cite{Esposito:2015fsa}. The associated
curves are the results of fits of blast-wave model~\cite{Schnedermann:1993ws} expectations in the absence of any
corrections for multi-nucleon collective effects.  The blue dash-dot curve are the extrapolated results for $^3_{\Lambda}H$
when collective effects are included. }
\label{fig:atlas_prompt-alice-He3}
\end{figure}

Another  diagnostic that has been proposed is the nature of its prompt production in high-energy hadron
collisions~\cite{Bignamini:2009sk}. As discussed in Section~\ref{sec:exp_gpd} above, the hadron collider experiments
see strong signals for prompt $X(3872)$ production in $\ecm = 1.96$~TeV $p\bar{p}$ collisions at the
Tevatron~\cite{CDFprompt:2004,Abazov:2004kp}, and in $pp$ collisions at $\ecm=7$~TeV~\cite{Chatrchyan:2013cld} and
8~TeV~\cite{Aaboud:2016vzw} collisions at the LHC. In each of these experiments, the measured properties of $X(3872)$
production are quite similar to those for the $\psip$ aside from an overall scale factor of about one-tenth, as illustrated
in Fig.~\ref{fig:atlas_prompt-alice-He3}, where ATLAS measurements of the transverse momentum ($p_T$) dependence of prompt
$\psip$ and $X(3872)$ production are shown as solid black and red circles, respectively~\cite{Aaboud:2016vzw}.

If the $X(3872)$ is a composite $D\bar{D}^*$ molecule-like object, as suggested by the closeness of its mass to the 
$m_{D^0}+m_{D^{*0}}$ threshold, one would expect that its production properties in prompt, high-energy hadron collisions
would be less like those of the $\psip$ and more like those of known composite objects, like light nuclei or hypernuclei.
In the absence of any direct measurements of light nuclei and hypernuclei production in 7$\sim$8~TeV $pp$ collisions, the
authors of ref.~\cite{Esposito:2015fsa} extrapolated measurements from the ALICE experiment~\cite{Adam:2015yta,Adam:2015vda}
of inclusive deuteron, $^3$He and hypertriton $^3_{\Lambda}H$ production cross sections in Pb-Pb collisions (with nucleon-nucleon
c.m.~energies of $\ecmx{NN}=2.76$~TeV), to $pp$ collisions at 7~TeV by means of a Glauber-model calculation. These are
included in Fig.~\ref{fig:atlas_prompt-alice-He3} where the associated curves are results of fits to the commonly used
blast-wave-model function for particle production in high energy heavy ion collisions~\cite{Schnedermann:1993ws}. 

In the Glauber model, the nucleons inside heavy ions interact independently, and multi-nucleon, collective effects are
ignored. The blue dash-dot curve shows the ref.~\cite{Esposito:2015fsa} estimate for how the hypertriton extrapolation
and fit would change if large collective effects were included. The extrapolations from Pb-Pb measurements to $pp$
collisions and the blast-wave model are very approximate and likely to be wrong by large factors. However, the differences
between these extrapolations and the measured $X(3872)$ $p_T$-dependence are many orders-of-magnitude too large to be
accounted for by refinements in the models. 

\begin{figure*}[htb]
\includegraphics[width=0.45\textwidth]{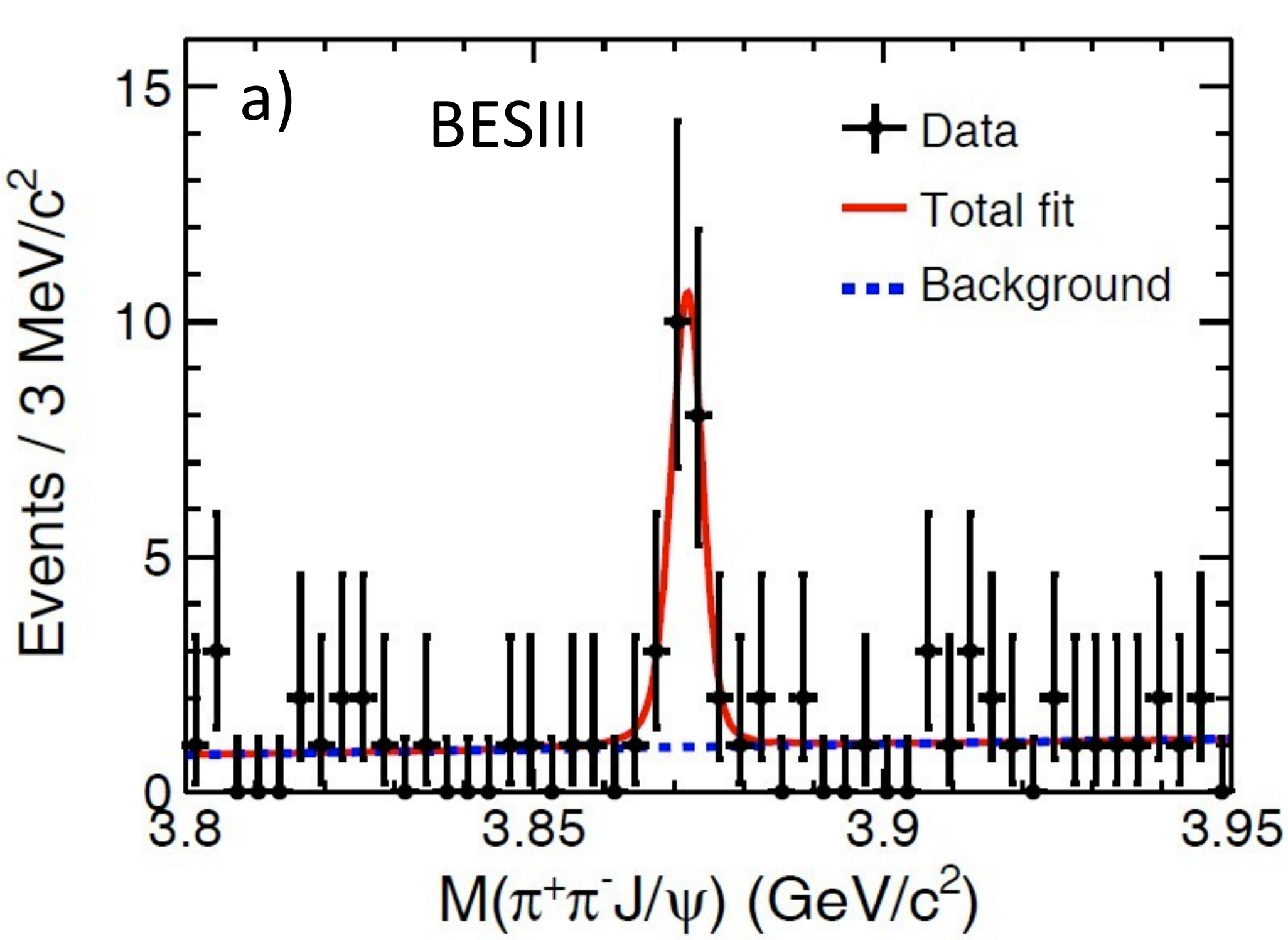}
\includegraphics[width=0.45\textwidth]{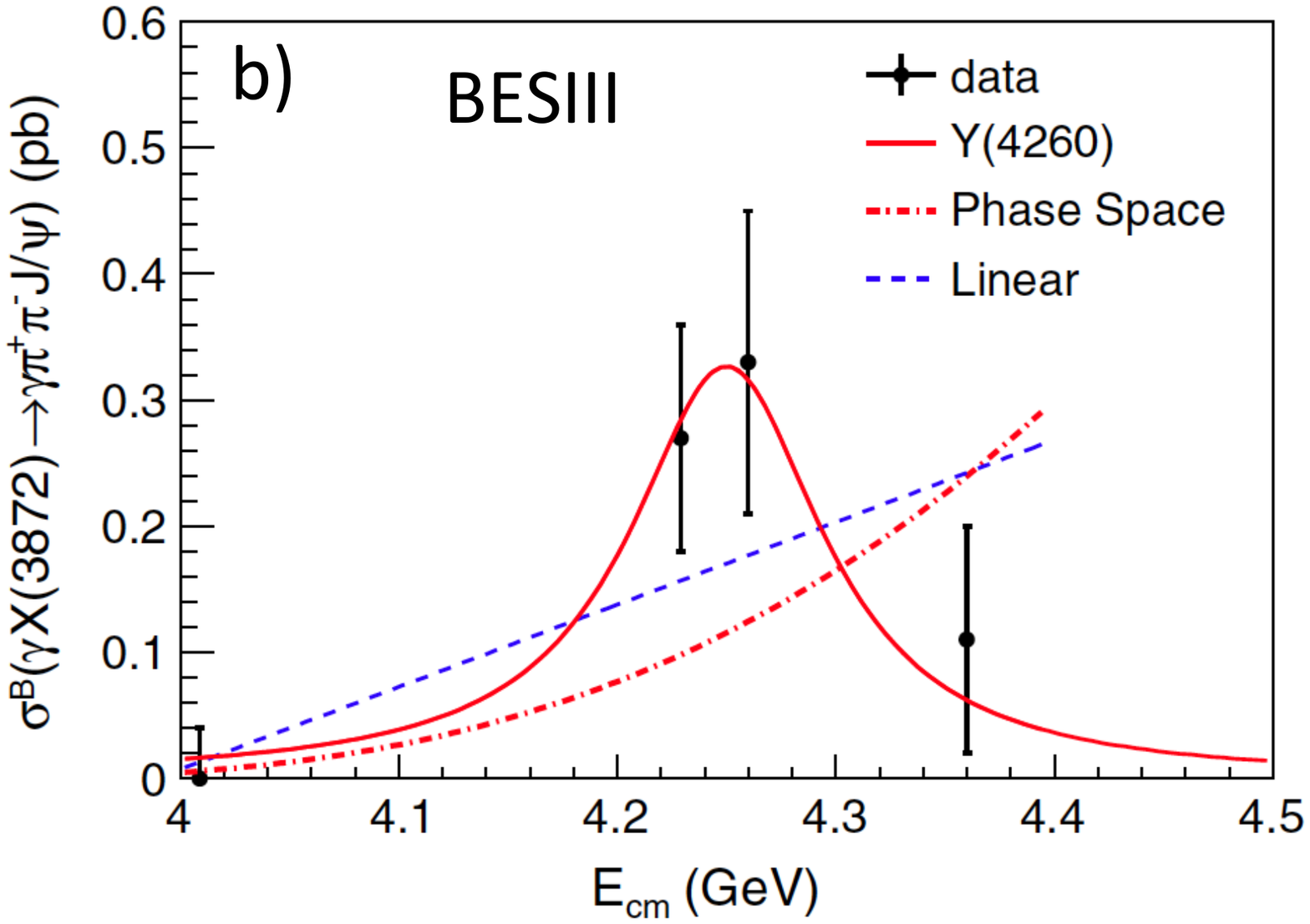}
\caption{\footnotesize {\bf a)} 
The data points show the BESIII experiment's $M(\pipi\jpsi)$ distribution for $\ee\rt\gamma\pipi\jpsi$
events at energies near the $Y(4260)$ resonance~\cite{Ablikim:2013dyn}.  The fitted peak has a mass and width of
$M=3871.9\pm 0.7$~MeV and $\Gamma = 0.0^{+1.7}_{-0.0}$~MeV ($<2.4$~MeV), which are in good
agreement with the PDG world average values for the $X(3872)$.
{\bf b)} The energy dependence of the BESIII
$\sigma(\ee\rt\gamma X(3872))\times {\mathcal B}(X(3872)\rt\pipi\jpsi)$ measurement.
The solid curve is the $Y(4260)$ line shape fitted
to the data; the dashed curves show phase-space and linear production model expectations. 
}
\label{fig:y_2_gx}
\end{figure*}

The BESIII experiment recently reported  $X(3872)$ production in the process $\ee\rt\gamma \pipi\jp$ at
c.m.~energies in the region of the $Y(4260)$ charmonium-like resonance peak~\cite{Ablikim:2013dyn}. The $X(3872)$ was
detected via its $\pipi\jpsi$ decay channel; a $\pipi\jpsi$ invariant mass distribution summed over the data
at four energy points is shown in Fig.~\ref{fig:y_2_gx}a, where a $6.3\sigma$ peak at the mass
of the $X(3872)$ is evident.  Figure~\ref{fig:y_2_gx}b shows the energy dependence of the $X(3872)$
production rate where there is some indication that the observed signal is associated with the
$Y(4260)$. Assuming that $Y(4260)\rt\gamma X(3872)$ decays are the source of this signal, and
using the PDG lower limit ${\mathcal B}(X(3872)\rt\pipi\jpsi)>0.026$~\cite{Olive:2016xmw}, BESIII determines
\begin{equation}
\frac{{\mathcal B}(Y(4260)\rt\gamma X(3872))}{{\mathcal B}(Y(4260)\rt \pipi\jpsi)}>0.05,
\end{equation}
which is substantial and suggests that there is some commonality in the nature of the $Y(4260)$, $X(3872)$
and $Z_c(3900)$.\footnote{The $Y(4260)$ meson is discussed below in Sect.~\ref{sec:y4260} and the $Z_c(3900)$
is discussed in Sect.~\ref{sec:charged}}

\subsection{$X(3915)$}
\label{sec:x3915}

\begin{figure}[htb]
\includegraphics[width=0.35\textwidth]{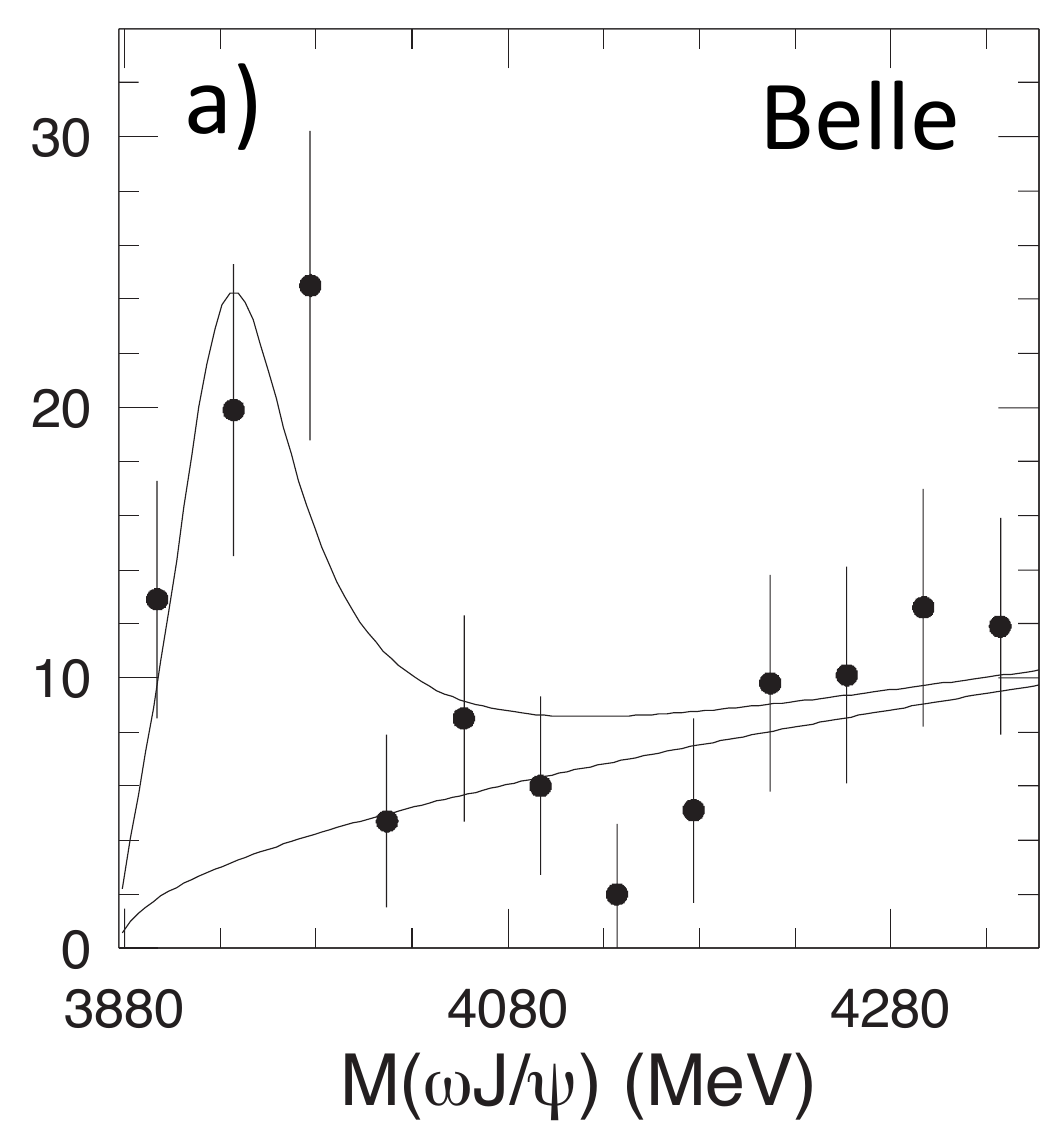}
\quad\\
\includegraphics[width=0.4\textwidth]{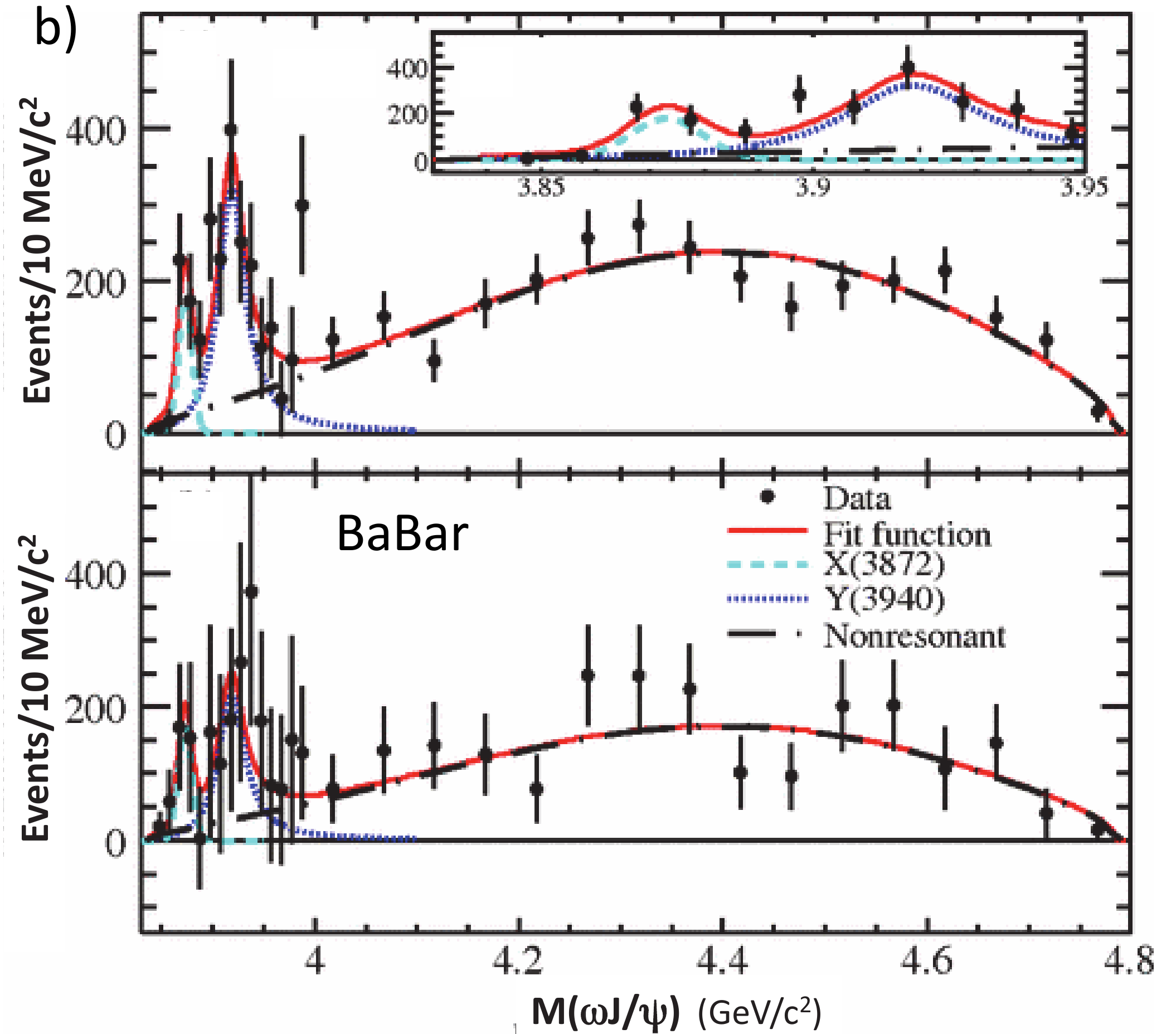}
\caption{\footnotesize $X(3915)\rt\omega\jpsi$ signals in $B\rt K\omega\jpsi$ decays from {\bf a)}
Belle (Fig.~2a from ref.~\cite{Abe:2004zs}) and {\bf b)} BaBar (Fig.~2 from
ref.~\cite{delAmoSanchez:2010jr}). In the latter, the upper panel shows results for
$B^+\rt K^+\omega\jpsi$  and the lower panel shows those for $B^0\rt K_S \omega\jpsi$.
The inset in the upper panel shows an expanded view of the low end of the $\omega\jpsi$ mass scale,
where the smaller, low-mass peak is due the $X(3872)\rt\omega\jpsi$ and the larger, higher mass peak
is the $X(3915)\rt\omega\jpsi$ signal.  
}
\label{fig:b_2_x3915}
\end{figure}

After finding the $X(3872)$ in $B\rt K\rho\jpsi$ decays,  Belle studied $B\rt K\omega\jpsi$  decay, where,
in a data sample containing 275M $B\bar{B}$ meson pairs, they observed a prominent, near-threshold enhancement
in the $\omega\jpsi$ invariant mass distribution shown in Fig.~\ref{fig:b_2_x3915}a~\cite{Abe:2004zs}. 
Belle fitted this enhancement with an S-wave BW resonance shape and found a mass and width
for this peak, which they originally dubbed the $Y(3940)$, of $M=3943\pm 17$~MeV and $\Gamma = 87\pm 24$~MeV. The Belle
result was confirmed by BaBar with a data sample containing 383M $B\bar{B}$ meson pairs~\cite{Aubert:2007vj}
and, later, with BaBar's final, 467M $B\bar{B}$-meson-pair data sample~\cite{delAmoSanchez:2010jr}. BaBar's
fits to the data yielded lower mass and width values: $M=3919\pm 4$~MeV and $\Gamma = 31 \pm 11$~MeV.  With
their full data sample, BaBar was able to resolve an $X(3872)\rt\omega\jpsi$ contribution to the enhancement
(see the inset in the upper panel of Fig.~\ref{fig:b_2_x3915}b).  The weighted average of the Belle and
BaBar results are $M=3920\pm 4$~MeV and $\Gamma=41\pm 10 $~MeV.

\vspace{0.1in}

\begin{figure*}[htb]
\includegraphics[height=0.27\textwidth,width=0.45\textwidth]{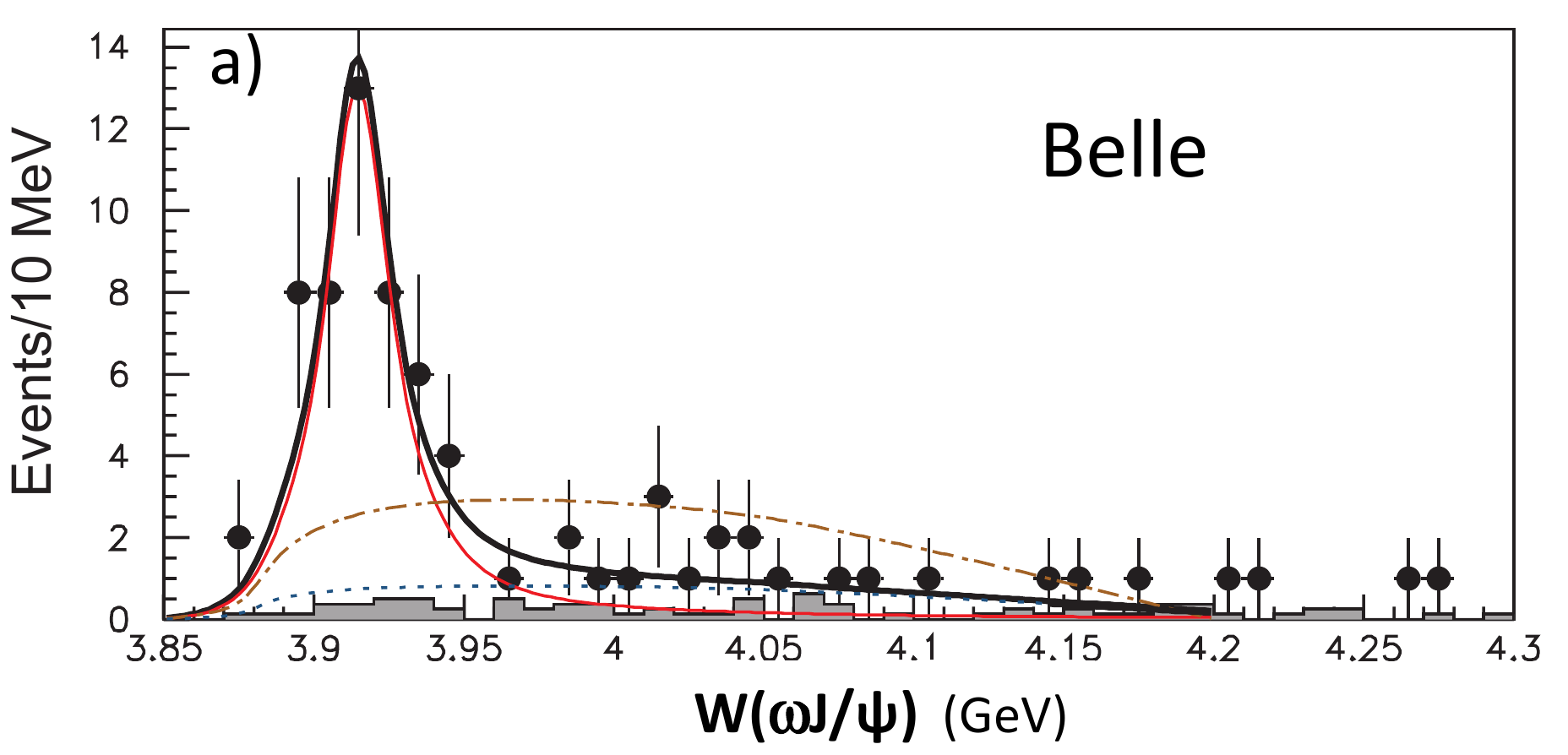}
\includegraphics[height=0.27\textwidth,,width=0.45\textwidth]{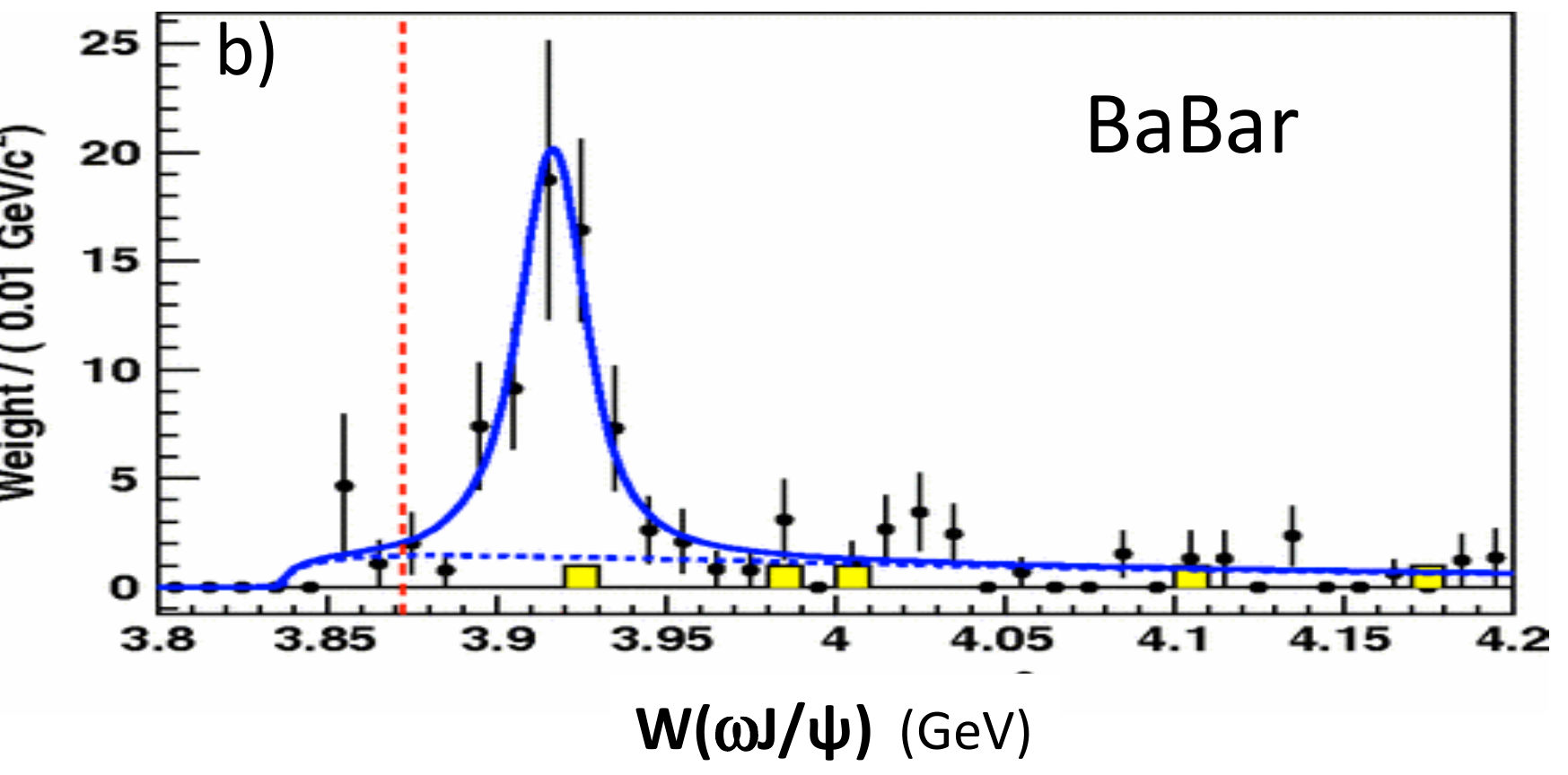}
\caption{\footnotesize $X(3915)\rt\omega\jpsi$ signals in $\gamma\gamma\rt\omega\jpsi$
fusion reactions from {\bf a)} Belle~\cite{Uehara:2009tx}and {\bf b)} BaBar~\cite{Lees:2012xs}. 
The bold solid curves in each figure are results of fits with a BW resonance shape to represent
the signal and a smooth function of $p^*$ to represent the background, where $p^*$ is the $\jpsi$
momentum in the $\gamma\gamma$ c.m.~system. The dash-dot curve in {\rm a)} is the result of a fit
with no BW resonance term; the dashed vertical line in {\rm b)} indicates the location of $W=3872$~MeV.
The shaded histograms in both plots show the non-$\jpsi$ background estimated from events in the
$\jpsi$ mass sidebands. 
}
\label{fig:gg_2_x3915}
\end{figure*}

An $\omega\jpsi$ mass peak with similar mass and width on a very small background was reported
by Belle in the two-photon process $\gamma\gamma\rt\omega\jpsi$, and shown in
Fig.~\ref{fig:gg_2_x3915}a~\cite{Uehara:2009tx}.   The BaBar group subsequently observed a very
similar peak~\cite{Lees:2012xs} in  the same process with mass and width values that were in good
agreement with those reported by Belle. The weighted average of the Belle and BaBar measurements
are $M=3917.4\pm 2.4$ MeV and $\Gamma=14\pm 6$~MeV. The close agreement between the masses
determined for the $Y(3940)\rt \omega\jpsi$  peak in $B\rt K\omega\jpsi$ decays and the
$X(3915)\rt\omega\jpsi$ signal seen in $\gamma\gamma\rt\omega\jpsi$ production, and the similar
values of the widths suggest that these are two different production mechanisms for the same state.
In the following we assume this to be the case and refer to this state as the $X(3915)$. The PDG
tables \cite{Olive:2016xmw} list the results from both channels as a single entry with average mass and width values of
\begin{eqnarray}
     M(X(3915)) &=& 3918.4 \pm 1.9 \ {\rm MeV}\nonumber \\
\Gamma(X(3915)) &=&   20.0 \pm 5.0 \ {\rm MeV}.
\label{eqn:x3915-mass-width}
\end{eqnarray}   

\vspace{0.1in}

\subsubsection{Is the $X(3915)$ the $\chi_{c0}$ charmonium state?}
BaBar performed a spin-parity analysis with their $\gamma\gamma\rt\omega\jpsi$ events that favored
a $J^{PC}=0^{++}$ quantum number assignment. Based on this result, they identified the $X(3915)$ as
a candidate for the $2^{3}{\rm P}_{0}$ charmonium state, commonly known as the  $\chi_{c0}^{\prime}$, and
the PDG classified the $X(3915)$ as the $\chi_{c0}^{\prime}$ in the 2014 Meson Summary Tables~\cite{pdg14}.
However, although BaBar's preferred $J^{PC}=0^{++}$ values match expectations for the $\chi_{c0}^{\prime}$, 
other properties of the $X(3915)$ make it a poor candidate for the $2^{3}{\rm P}_{0}$ charmonium
state~\cite{Guo:2012tv,Wang:2014voa,Olsen:2014maa}.  The mass is too high; the $\chi_{c2}^{\prime}$-$X(3915)$
mass splitting, $\delta M_{2-0}(2P)=8.8\pm 3.2$~MeV is only 6\% of the $1{\rm P}$ splitting:
$\delta M_{2-0}(1P)=141.5 \pm 0.3$~MeV, in strong contradiction with theoretical
expectations~\cite{Eichten:2004uh,Li:2009zu,Wang:2014voa}.  Another peculiarity of the
$X(3915)=\chi_{c0}^{\prime}$ assignment is the lack of any experimental evidence for $X(3915)\rt  D\bar{D}$
decays and the apparent strength of the $X(3915)\rt\omega\jpsi$ {\em discovery} mode, which conflicts with
expectations that $\chi_{c0}\rt D\bar{D}$ would be a strongly favored {\em fall-apart} mode and
$\chi_{c0}^{\prime}\rt\omega\jpsi$ an OZI-rule-violating process that is expected to be strongly
suppressed~\cite{Okubo:1963fa,Zweig:1981pd,Iizuka:1966fk}.

The authors of ref.~\cite{Zhou:2015uva} added to the controversy by pointing out that the BaBar 
spin-parity analysis that ruled out the $J^{PC}=2^{++}$ hypothesis assumed the dominance of the
helicity-2 amplitude over that for helicity-0. Their reanalysis of the BaBar angular distributions
showed that when a helicity-0 amplitude is included, a $J^{PC}=2^{++}$ assignment cannot be ruled out,
and make an argument that identifies the $X(3915)$ as the $\chi_{c2}^{\prime}$ charmonium state.
However, their argument for the $X(3915)=\chi_{c2}^{\prime}$ assignment ignores the consequences of
$X(3915)$ production in $B\rt K\omega\jpsi$ decays. For example, since the {\it total} branching
fraction ${\mathcal B}(B^+\rt K^+\chi_{c2}(1{\rm P}))=(1.1 \pm 0.4)\times 10^{-5}$ is smaller than
the {\it product} of branching fractions 
\begin{equation}
{\mathcal B}(B^+\rt K^+ X(3915))\times {\mathcal B}(X\rt \omega\jpsi)=3.0^{+0.9}_{-0.7}\times 10^{-5},
\label{eqn:x3915-2-omega-jpsi}
\end{equation} 
the $X(3915)=\chi_{c2}^{\prime}$ assignment would imply a $B^+\rt K^+ \chi_{c2}^{\prime}$ partial decay width
that is substantially larger than that for $B^+\rt K^+\chi_{c2}$, which is contrary to models
for $B$-meson decays to charmonium states, where these widths are expected to be proportional to the
square of the $\ccbar$ wave function at the origin, which decreases with increasing radial quantum
number~\cite{Bodwin:1992qr}.  

On the other hand, the $X(3915)\rt\omega\jpsi$ signals seen in $B$ decays and in $\gamma\gamma$
production may be unrelated. More data and separate spin-parity determinations for the $\omega\jpsi$
systems produced in $B\rt K\omega\jpsi$ and $\gamma\gamma\rt\omega\jpsi$ processes and with fewer
assumptions are needed. At present, the situation remains confused, as evidenced by the 2016 edition
of the PDG report, which no longer identifies this as the $\chi_{c0}^{\prime}$ and
has reverted to calling this state the $X(3915)$~\cite{Olive:2016xmw}.
Recently Belle reported the observation of an alternative $\chi_{c0}^{\prime}$ candidate, the $X^*(3860)$,
with none of the problems associated with the $X(3915)=\chi_{c0}^{\prime}$ assignment~\cite{Chilikin:2017evr}.
This is discussed below in Section~\ref{sec:x3940}.

\subsubsection{Is the $X(3915)$ a $\ccbar s\bar{s}$ four-quark state?}

The above-mentioned provisos notwithstanding, the most parsimonious interpretation of existing data is to
assume that the $\omega\jpsi$ peaks seen in $B$ decays and $\gamma\gamma$ production are due to the same state.
In that case, the most likely $J^{PC}$ assignment is $0^{++}$, and the mass, strength of the $\omega\jpsi$
decay channel, and absence of any evidence for a significant $D\bar{D}$ decay mode rule against its
identification as a $\ccbar$ charmonium state. The mass is 18.2~MeV below the $2m_{D_s}$ threshold, and this
suggests that it may contain a significant $\ccbar s\bar{s}$ component, either in a $D_s^+ D_s^-$
molecule-like configuration~\cite{Li:2015iga}, a $[\bar{c}\bar{s}][cs]$ tetraquark~\cite{Lebed:2016yvr}
or a mixture of the two. In any of these pictures, the $D\bar{D}$ decay mode would strictly violate the
OZI-rule, while the $\omega$ meson's small, but non-negligible $s\bar{s}$ content \cite{Benayoun:1999fv}
would partially mitigate the
$\omega\jpsi$ mode's violation of the rule and an $\omega\jpsi$ decay width that is comparable or greater
than that for $D\bar{D}$ would not be {\it a priori} ruled out.  For a $\ccbar s\bar{s}$ combination
configured either as a molecule-like or a tetraquark arrangement, the decay mode least effected by OZI
suppression would be $X(3915)\rt\eta\eta_c$ and this could be expected to be a dominant decay mode. However,
Belle searched for this mode, saw no significant signal, 
and established a (90\% CL) product branching fraction upper limit~\cite{Vinokurova:2015txd}:
\begin{equation}
{\mathcal B}(B^+\rt K^+ X(3915))\times {\mathcal B}(X\rt \eta\eta_c)< 4.7 \times 10^{-5}.
\label{eqn:x3915-2-eta-etac}
\end{equation} 
Since this limit is not very stringent, it is difficult to draw a definite conclusion from it. A comparison
of it with the $\omega\jpsi$ measurement given in Eq.~\ref{eqn:x3915-2-omega-jpsi} indicates that, in spite
of Belle's null result, the partial decay width for $X(3915)\rt\eta\eta_c$ could still 
be larger than that for $\omega\jpsi$ by as much
as a factor of $\simeq 2$.  The expectation that the $\eta\eta_c$ partial width
should be large is only qualitative and our limited level of understanding of these processes precludes the
ability of making a reliable quantitative estimate of just how large it should be. Because of this, the
absence of an $\eta\eta_c$ mode would probably only be fatal to the $\ccbar s\bar{s}$ quark assignment if its partial
width was shown to be definitely much smaller than that for $\omega\jpsi $
decays.\footnote{This issue is discussed in ref.~\cite{Li:2015iga}.}

\subsubsection{Discussion}
 
Since it is relatively narrow and is seen as clear signals in both $B$-meson decays and $\gamma\gamma$
fusion reactions, the $X(3915)$ is one of the most intriguing of the $XYZ$ exotic meson candidates.
However, significant progress in our understanding of its underlying nature will probably not be
forthcoming until larger data samples are available in future experiments such as
BelleII~\cite{Abe:2010gxa}.  With the order-of-magnitude larger event samples that are expected for
BelleII, we can expect definitive $J^{PC}$ determinations and measurements of, or more stringent limits
on, the strengths of the $D\bar{D}$ and $\eta\eta_c$ decay channels for both the $B$-meson-decay and
$\gamma\gamma$-fusion production modes.
The LHCb experiment has demonstrated ability to detect $\omega$ mesons in $B$ decays \cite{LHCb:2012cw}
and should also be able to probe the $X(3915)$ quantum numbers.


\subsection{$Y(4260)$ and other $J^{PC}=1^{--}$ states}
\label{sec:y4260}

\begin{figure}[htb]
  \includegraphics[height=0.2\textwidth,width=0.5\textwidth]{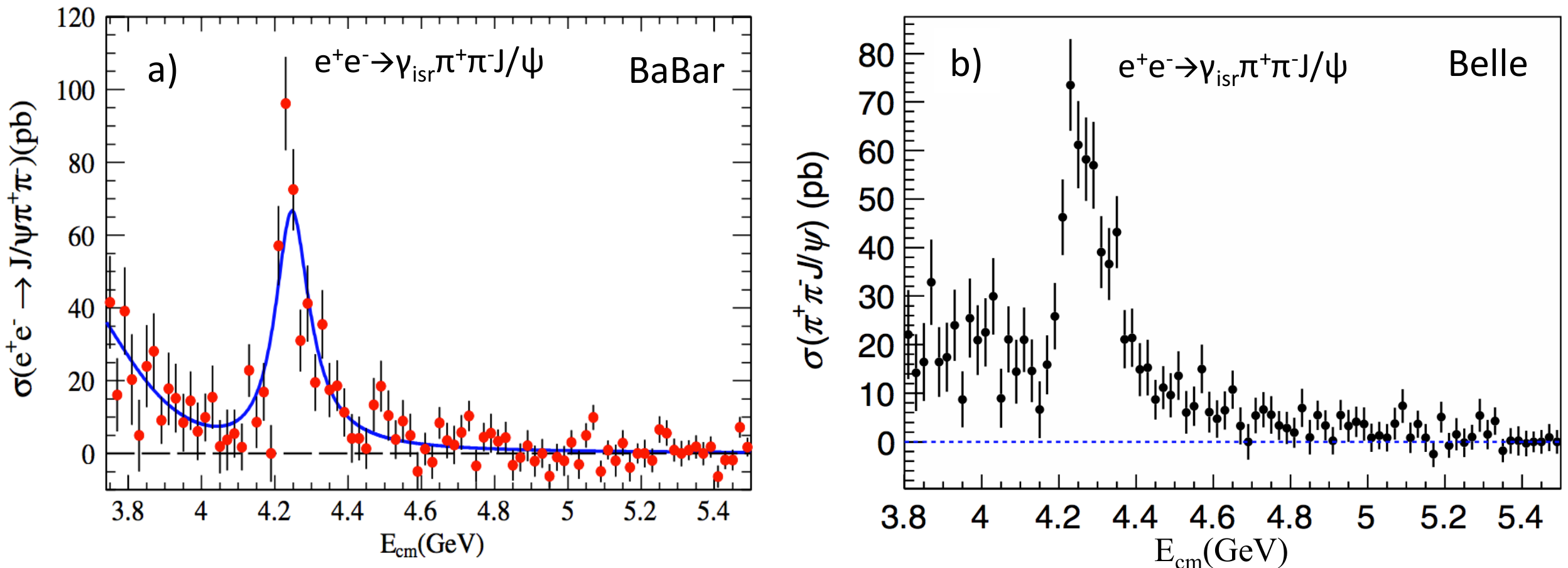}
\caption{\footnotesize {\bf a)} The data points show the Born cross sections for $\ee\rt\pipi\jpsi$,
measured via the initial-state radiation process $\ee\rt\gamma_{\rm isr}\pipi\jpsi$ by BaBar~\cite{Lees:2012cn}.
The curve show results of a fit that used a single BW resonance to represent the $Y(4260)$ resonance plus a
linear background term. 
{\bf b)} Belle measurements of the same cross sections~\cite{Liu:2013dau}.
}
\label{fig:babar-belle_y4260}
\end{figure}

After the discovery of the $X(3872)$ in $\pipi\jpsi$ decays and before its $J^{PC}=1^{++}$ quantum number
assignment was established, The BaBar group considered the possibility that it might be a $1^{--}$ vector
state and searched its direct production in the initial-state-radiation process
$\ee\rt\gamma_{\rm isr}\pipi\jpsi$~\cite{Aubert:2005eg}. They did not see an $X(3872)$ signal and were able
to conclude that the $X(3872)$'s $J^{PC}$ quantum numbers were not $1^{--}$. They did see, however, an
unexpected strong accumulation of events with $\pipi \jpsi$ invariant masses that peaked near
4.26~GeV, shown in Fig.~\ref{fig:x3940-y4260}b of Section~\ref{sec:exp_ee}, that they called the
$Y(4260)$~\cite{Aubert:2005rm}.  Subsequent BaBar measurements~\cite{Lees:2012cn} of the ``Born'' cross
sections\footnote {``Born'' cross sections are cross sections that correspond to the lowest-order Feynman
diagram and are determined by ``radiatively  correcting'' observed cross sections for higher-order QED
effects such as initial-state-radiation and vacuum polarization.}
for $\ee\rt\pipi\jpsi$ at c.m.~energies near the $Y(4260)$ mass peak, using their full data set, are shown
in Fig.~\ref{fig:babar-belle_y4260}a. This peak was quickly confirmed by CLEO~\cite{Coan:2006rv} and
Belle~\cite{Yuan:2007sj}. The most recent Belle measurements~\cite{Liu:2013dau} of $\sigma(\ee\rt\pipi\jpsi)$
in the $Y(4260)$ region, based on their full data set, are shown in Fig.~\ref{fig:babar-belle_y4260}b, where
the similarity with the BaBar measurements is apparent. The weighted average of the mass and width values
determined by BaBar, CLEO and Belle from fits of a single BW resonance line shape to the $Y(4260)$ peak in
their data are~\cite{Olive:2016xmw}:
\begin{eqnarray} 
M(Y(4260))       &=& 4251 \pm 9~{\rm MeV}  \nonumber \\
\Gamma(Y(4260))  &=& 120 \pm 12~{\rm MeV}.
\label{eqn:y4260-mass}
\end{eqnarray}
The excess of events 
near 4 GeV in their $\pi^+\pi^-\jpsi$ cross section measurements 
was attributed by Belle to an additional possible resonance that they called the $Y(4008)$ \cite{Yuan:2007sj,Olive:2016xmw},
but a similar excess was not observed by BaBar \cite{Lees:2012cn} and was
not confirmed by recent BESIII results \cite{Ablikim:2016qzw}.

\vspace{0.1in}

\begin{figure}[htb]
  \includegraphics[width=0.48\textwidth]{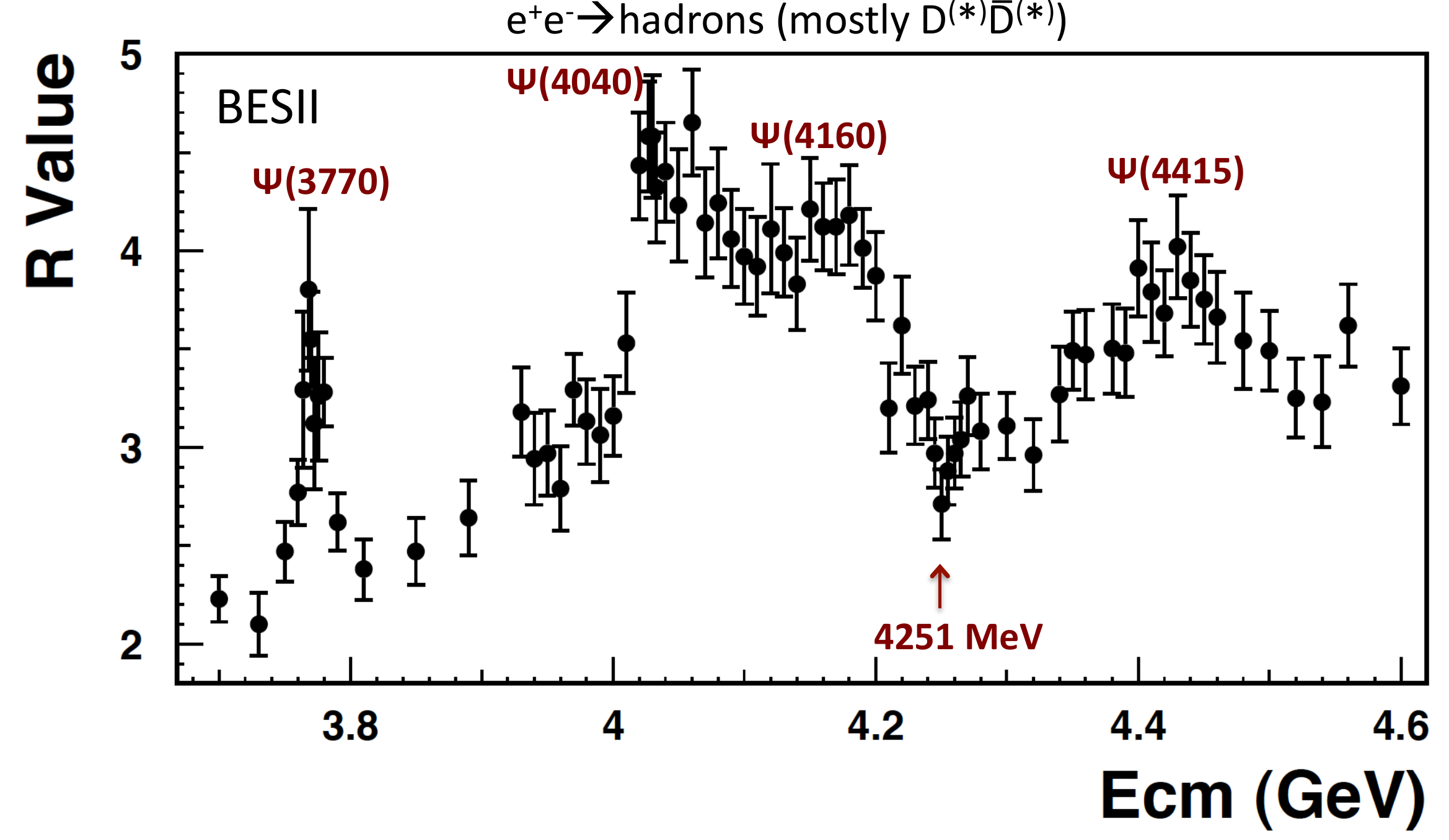}
\caption{\footnotesize 
Measurements of the ratio $R=\sigma_{\rm tot}(\ee\rt {\rm hadrons})/\sigma_{\rm QED}(\ee\rt\mumu)$, where
$\sigma_{\rm QED}(\ee\rt\mumu)=86.85\ {\rm nb}/s$ ($s$ in GeV$^2$), from ref.~\cite{Bai:2001ct}.  The
structures above $R\simeq 2$ are attributed to the indicated $1^{--}$ charmonium mesons decaying to
$D\bar{D}$, $D\bar{D}^*$ open-charmed or $D^*\bar{D}^*$ final states.  The expected position for a
$Y(4260)$ signal, indicated by an arrow, is located at a local minimum in the measured cross section. 
  }
\label{fig:bes2_R}
\end{figure}

The production mode of the $Y(4260)$ ensures that its $J^{PC}$ quantum numbers are the same as those
of the photon, {\em i.e.} $1^{--}$. Its discovery decay mode, $Y(4260)\rt\pipi\jpsi$, provides strong
evidence that its constituents contain a $\ccbar$ quark pair. However, all of the $1^{--}$ $\ccbar$
charmonium levels with mass below 4500~MeV have already been assigned to well established $1^{--}$
resonances that are seen in the total cross section for $\ee\rt {\rm hadrons}$ between 2.6~and
4.6~GeV~\cite{Bai:1999pk,Bai:2001ct} (see the inset in Fig.~\ref{fig:R-charm}a).
In addition, even though its mass is well above all of the $D^{(*)}\bar{D}^{(*)}$ open-charmed-meson
mass thresholds, there is no evidence for its decay to pairs of open-charmed mesons in the inclusive
$\ee\rt$~hadrons total cross section. BESII measurements of $\sigma_{\rm tot}(\ee\rt {\rm hadrons})$ at
c.m.~energies between 3.7~and~4.6~GeV, shown in Fig.~\ref{fig:bes2_R}, exhibit considerable structure
that is primarily due to the production and decay to pairs of open-charmed mesons of the established
$1^{--}$ $\psi(3770)$, $\psi(4040)$, $\psi(4160)$ and $\psi(4415)$ charmonium states. The strong signals
for these states in $\sigma_{\rm tot}(\ee\rt {\rm hadrons})$ plus their absence in the $M(\pipi\jpsi)$
invariant mass distributions shown in Figs.~\ref{fig:babar-belle_y4260}a~and~\ref{fig:babar-belle_y4260}b
reflect the expected strong dominance of fall-apart decays to open-charmed-meson pairs over
OZI-rule-suppressed decays to hidden-charm final states that is characteristic of
above-open-charm-threshold charmonium states. In contrast, the absence of any sign of $Y(4260)$ decays
to charmed mesons in $\sigma_{\rm tot}(\ee\rt {\rm hadrons})$ plus its strong signal in the $\pipi\jpsi$
decay channel is opposite to expectations for charmonium. As a result, there has been considerable
theoretical speculation that the $Y(4260)$ might be some kind of a multi-quark meson or a $\ccbar$-gluon
hybrid state~\cite{Brambilla:2010cs,Maiani:2014aja,Guo:2014zva}. 

\vspace{0.1in}

\begin{figure}[htb]
\begin{minipage}[t]{84mm}
  \includegraphics[height=0.40\textwidth,width=0.95\textwidth]{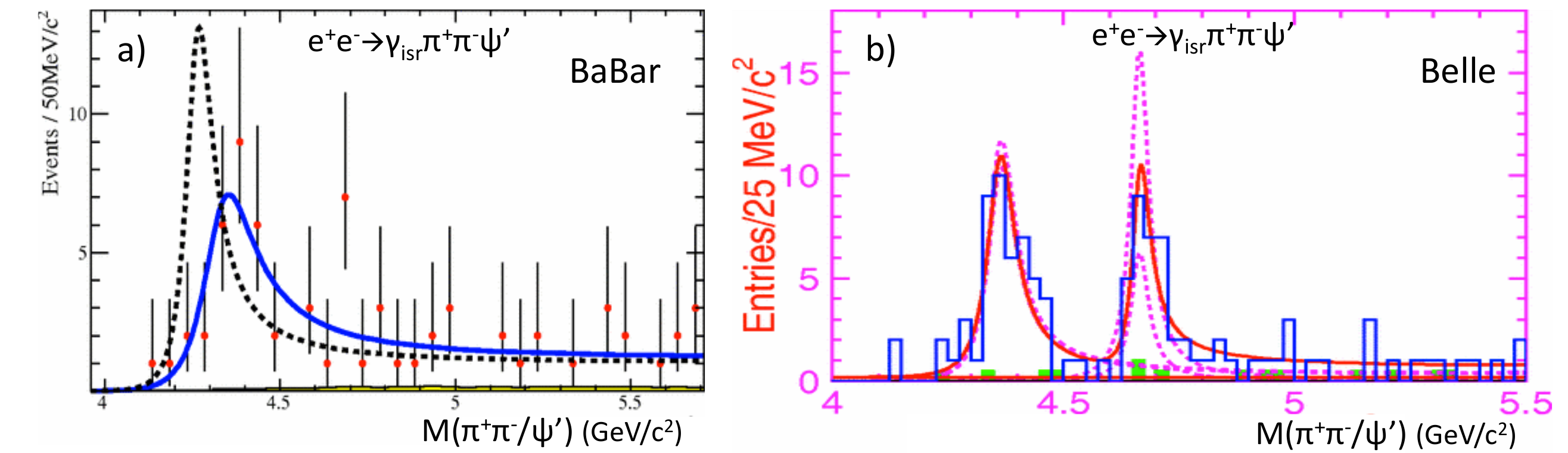}
\end{minipage}\hspace{\fill}
\caption{\footnotesize {\bf a)} The data points show the $\pipi\psip$ invariant mass distribution
for $\ee\rt\gamma_{\rm isr}\pipi\psip$ events in BaBar~\cite{Aubert:2007zz}.  The solid curve shows
results of a fit that used a single BW resonance to represent the signal plus a linear background
term.  The dashed curve shows the results of a fit with mass and width constrained to the $Y(4260)$
values. {\bf b)} The blue histogram shows the corresponding results from Belle~\cite{Wang:2007ea}.
The solid curve shows the results of a fit that uses two interfering BW amplitudes, one with mass
near 4360~MeV and the other near 4660~MeV. The dashed curves show the individual resonance
contributions from two equally good fits that have different interference phases.}
\label{fig:babar-belle_y4360}
\end{figure}

A BaBar search for $Y(4260)\rt \pipi\psip$ decays in $\ee\rt\pipi\psip $ events resulted in the
$\pipi\psip$ invariant mass distribution shown in Fig.~\ref{fig:babar-belle_y4360}a, where there is
a strong peaking of events near 4320~MeV on a nearly negligible background~\cite{Aubert:2007zz}.
This peak is not compatible with the measured mass and width of the $Y(4260)$, as indicated by the dashed
curve in the figure.  A subsequent study of the same reaction with a larger data sample by Belle confirmed
the BaBar observation, albeit at a somewhat higher mass near 4360~MeV as shown in
Fig.~\ref{fig:babar-belle_y4360}b~\cite{Wang:2007ea}. 
The current PDG values for the mass and width of this peak, called the $Y(4360)$, are~\cite{Olive:2016xmw}:
\begin{eqnarray}
M(Y(4360))        &=& 4346\pm  6\ {\rm MeV}  \nonumber \\
\Gamma(Y(4360))   &=&  102\pm 12\ {\rm MeV}.
\label{eqn:y4360-mass-width}
\end{eqnarray}

\vspace{0.1in}

In addition, Belle observed a second distinct $\pipi\psip$ invariant mass peak near 4660~MeV 
that is evident in Fig.~\ref{fig:babar-belle_y4360}b, an observation that was confirmed by
BaBar~\cite{Lees:2012pv}.  In addition,  Belle also reported a peak with similar mass and width in
the $\Lambda_c^{+} \Lambda_c^{-} $ invariant mass in $\ee\rt\gamma_{\rm isr}\Lambda_c^{+}\Lambda_c^{-}$
events~\cite{Pakhlova:2008vn}.  The PDG average of the Belle and BaBar mass and width measurements
of this second peak, called the $Y(4660)$, are~\cite{Olive:2016xmw}:
\begin{eqnarray}
M(Y(4660))        &=& 4643\pm  9\ {\rm MeV}  \nonumber \\
\Gamma(Y(4660))   &=&   72\pm 11\ {\rm MeV}.
\label{eqn:y4660-mass-width}
\end{eqnarray}

\subsubsection{BESIII as a ``$Y(4260)$-Factory''}
The BaBar and Belle results on the $Y(4260)$, $Y(4360)$ and $Y(4660)$ all relied on production of
these states via the isr process illustrated in Fig.~\ref{fig:ccbar-prod}d and discussed in
Section~\ref{sec:exp_ee}. This process has the advantage of sampling many $\ee$ c.m.~energies at once,
but is limited by a severe, order $\alpha_{\rm QED}$, luminosity penalty associated with the radiation of
a hard photon.  For detailed studies of these states, the BESIII experiment has the advantage of operating
at and near c.m.~energies corresponding to the $Y(4260)$ and $Y(4360)$ peaks, thereby functioning as a
``$Y$-factory.''  In this mode of operation, large event samples can be accumulated near the peaks of these
resonances.  On the other hand, lineshape measurements of the resonance parameters and the separation
of resonance signals from underlying non-resonant backgrounds require time-consuming energy-by-energy scans. 

\begin{figure}[htb]
\begin{minipage}[t]{84mm}
  \includegraphics[width=1.0\textwidth]{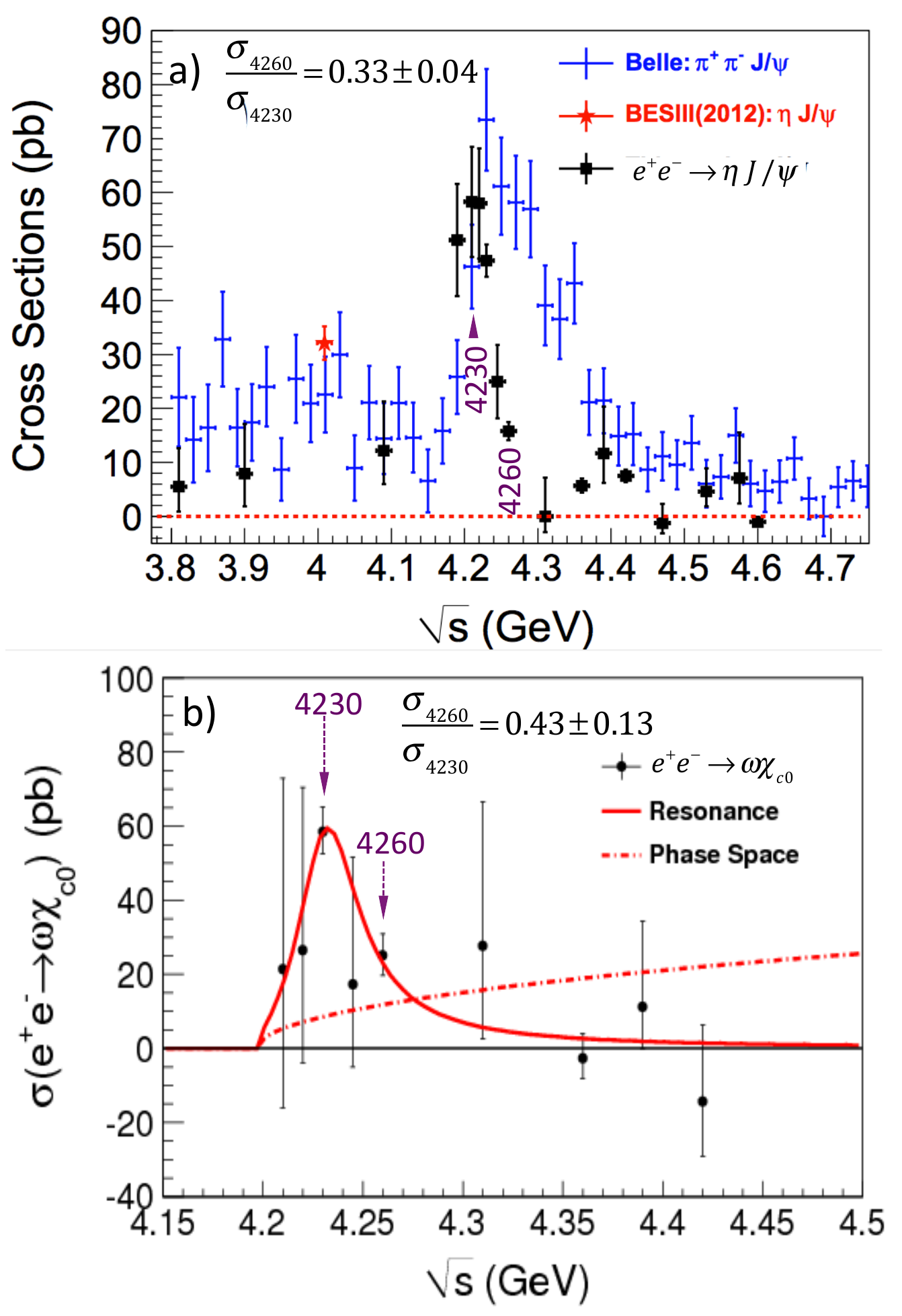}
\end{minipage}\hspace{\fill}
\caption{\footnotesize {\bf a)} Cross section measurements for $\ee\rt \eta\jpsi$ from BESIII are
shown as black points~\cite{Ablikim:2015xhk}.  For comparison, Belle isr measurements of the $\ee\rt\pipi\jpsi$
cross section over the same energy range are shown as blue crosses~\cite{Liu:2013dau}. The red star indicates
a previous BESIII $\eta\jpsi$ cross section measurement near the peak of the $\psi(4040)$ charmonium
state~\cite{Ablikim:2012ht}.  {\bf b)} The data points show BESIII measurements of the cross section for
$\ee\rt\omega\chi_{c0}$~\cite{Ablikim:2014qwy}.  The solid curve shows the result of a fit of a 
threshold-constrained BW resonance shape to the data. The dash-dot curve indicates what a phase-space-only
distribution would be like. 
}
\label{fig:eta-jpsi_omega-chic0}
\end{figure}

BESIII's first data-taking run in this energy range accumulated a 525~pb$^{-1}$ data sample at 4260~MeV in
which they found $1477\pm 43$ $\pipi\jpsi$ events that included $307\pm 48$ events of the type
$\ee\rt \pi^{\mp}Z_c(3900)^{\pm}$; $Z_c(3900)^{\pm}\rt\pi^{\pm}\jpsi$, where the $Z_c(3900)^{\pm}$ is a relatively
narrow, resonance-like structure with non-zero electric charge that is discussed in some detail below in
Section~\ref{sec:charged}. BESIII subsequently did a scan of measurements around the $Y(4260)$ and $Y(4360)$
peaks, including relatively high statistics points at $\ecm=4230$~MeV, $4260$~MeV and
$4360$~MeV.\footnote{The three ``high luminosity'' data samples have integrated luminosities of 1047~pb$^{-1}$
at 4230~MeV, 827~pb$^{=-1}$ at 4260~MeV; and 540~pb$^{-1}$ at 4360~MeV. The other points have luminosities of
about 50~pb$^{-1}$.}  With these data, BESIII measured the cross sections for $\ee\rt\eta\jpsi$ shown in
Fig.~\ref{fig:eta-jpsi_omega-chic0}a~\cite{Ablikim:2015xhk} and $\ee\rt\omega\chi_{c0}$ shown in
Fig.~\ref{fig:eta-jpsi_omega-chic0}b~\cite{Ablikim:2014qwy}. 

\vspace{0.1in}

Figure~\ref{fig:eta-jpsi_omega-chic0}a includes a comparison of BESIII's $\ee\rt\eta\jpsi$ cross sections with
Belle isr results for $\sigma(\ee\rt\pipi\jpsi)$~\cite{Liu:2013dau}, where it is evident that the peak seen in
the $\eta\jpsi$ channel is much narrower than Belle's $Y(4260)\rt\pipi\jpsi$ peak. The $\omega\chi_{c0}$ cross
section (Fig.~\ref{fig:eta-jpsi_omega-chic0}b) shows a behavior that is similar as that for $\eta\jpsi$.  The
red curve in this figure is the result of a fit of a threshold-constrained BW resonance to the $\omega\chi_{c0}$
data points,  which returns mass and width values, $M_{\omega\chi_{c0}}=4230 \pm 10$~MeV and
$\Gamma_{\omega\chi_{c0}}=38 \pm 12$~MeV that are a poor match to the PDG values for the $Y(4260)$ given in
Eq.\ref{eqn:y4260-mass}. The absence of any constraining $\eta\jpsi$ data points on the lower side of the peak,
{\it i.e.},\ between $M(\eta\jpsi)=4100$~and~4200~MeV, precluded BESIII from doing a meaningful fit for a
$\eta\jpsi$ line shape.  Instead they characterized the shapes of the $\eta\jpsi$ and $\omega\chi_{c0}$ peaks by
the ratio of their cross sections at the high-statistics $\ecm=4230$~and~$4260$~MeV data points:
$R^{4260}_{4230}(f)=\frac{\sigma^{4260}(\ee\rt f)}{\sigma^{4230}(\ee\rt f)}$, where they find good agreement:
$R^{4260}_{4230}(\eta\jpsi)=0.33\pm 0.04$ and $R^{4260}_{4230}(\omega\chi_{c0})=0.43\pm 0.13$. 

\begin{figure}[htb]
\begin{minipage}[t]{84mm}
  \includegraphics[width=1.0\textwidth]{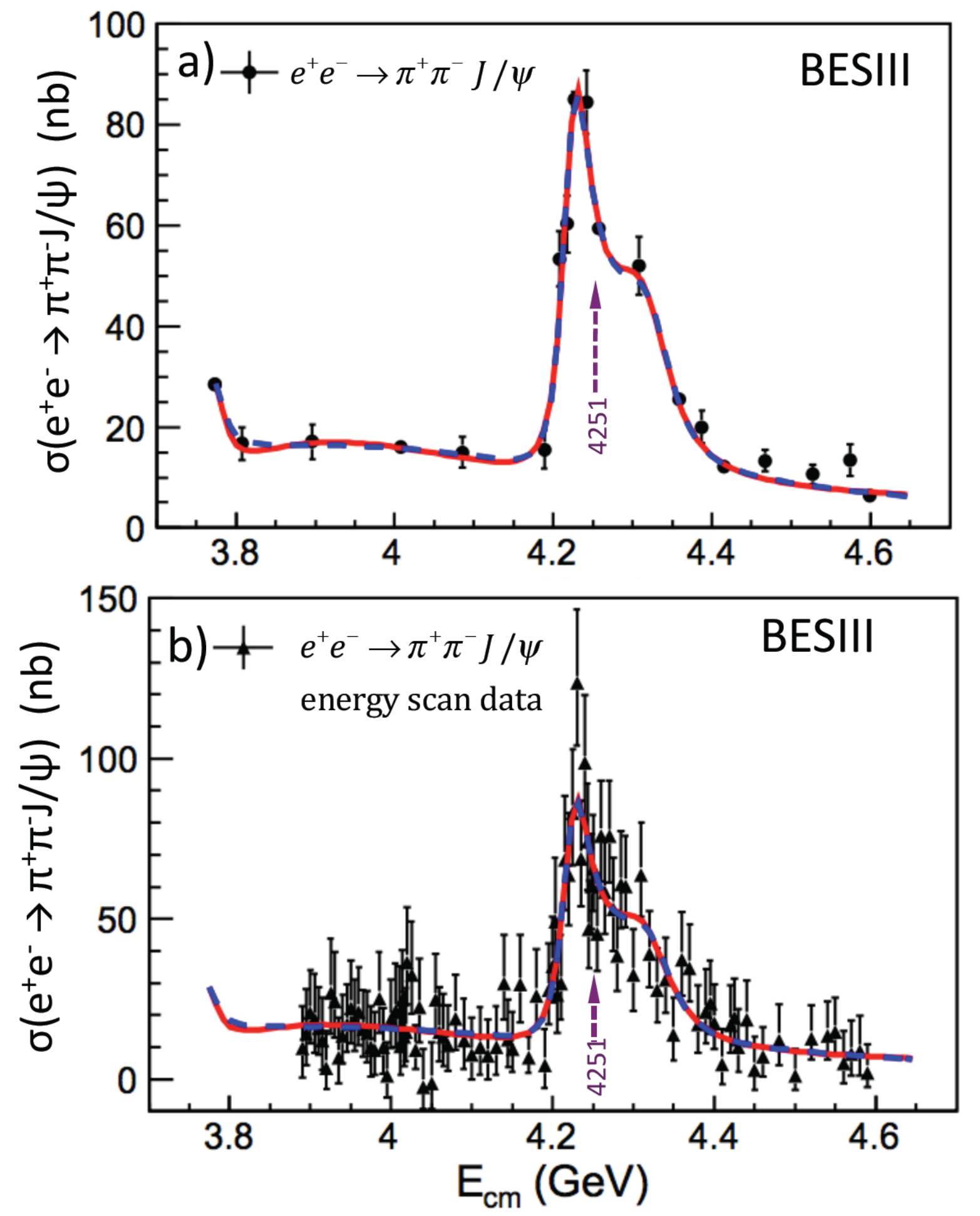}
\end{minipage}\hspace{\fill}
\caption{\footnotesize BESIII measurements~\cite{Ablikim:2016qzw} of the cross section for
$\ee\rt\pipi\jpsi$ for {\bf a)} the ``high luminosity'' scan data and {\bf b)} the ``low luminosity'' scan
data.  Dashed arrows in both plots indicate the ``Y(4260)'' mass value from the PDG-2016 average of results
based on single-BW resonance fits to pre-2016 measurements given in Eq.~\ref{eqn:y4260-mass}.
}
\label{fig:bes3_pipijpsi-scan}
\end{figure}

The evident incompatibility of the narrow structures in the $\ee\rt\eta\jpsi$ and $\ee\rt\omega\chi_{c0}$ cross
sections with the broad $Y(4260)\rt\pipi\jpsi$ peak prompted BESIII to map out the $\ecm$ energy region
in the vicinity of the $Y(4260)$ with two additional, independent data sets~\cite{Ablikim:2016qzw}. One consists
of 19 ``high luminosity'' data runs with at least 40~pb$^{-1}$/point between $\ecm=3773$~MeV and 4599~MeV.
The other consists of 60 ``low luminosity''  energy-scan data runs with $7-9$~pb$^{-1}$/point between
$\ecm=3882$~MeV and 4567~MeV. Figures~\ref{fig:bes3_pipijpsi-scan}a~and~b show $\ee\rt\pipi\jpsi$ cross
section measurements from the high- and low-luminosity data scans, respectively, where it is evident that the
line shape of the ``$Y(4260)$'' peak is not well described by a single BW resonance function.  The curves in
the figures show the results from fits to the data in both plots with two interfering BW resonance amplitudes
that provides mass and width values of
\begin{eqnarray}
M_1=4222 \pm 4\ {\rm MeV}  ~ &~~& \Gamma_1=44 \pm 5\ {\rm MeV} \nonumber \\
M_2=4320 \pm 13\ {\rm MeV}   &~~& \Gamma_2=101^{+27}_{-22} \ {\rm MeV}, 
\label{eqn:two-BW-masses}
\end{eqnarray}  
where the statistical and (smaller) systematic errors are added in quadrature. The simplest interpretation of
these results is that the first peak is the $Y(4260)$, which has a significantly lower mass and narrower width
than the $B$-factory-measured values that are given above in Eq.~\ref{eqn:y4260-mass}, and that the second peak
is the due to a $\pipi\jpsi$ decay mode of the $Y(4360)$ resonance, with slightly lower mass and narrower width
values than those determined from the $\pipi\psip$ decay mode listed in Eq.~\ref{eqn:y4360-mass-width}.

\vspace{0.1in}

\begin{figure}[htb]
\begin{minipage}[t]{84mm}
  \includegraphics[width=\textwidth]{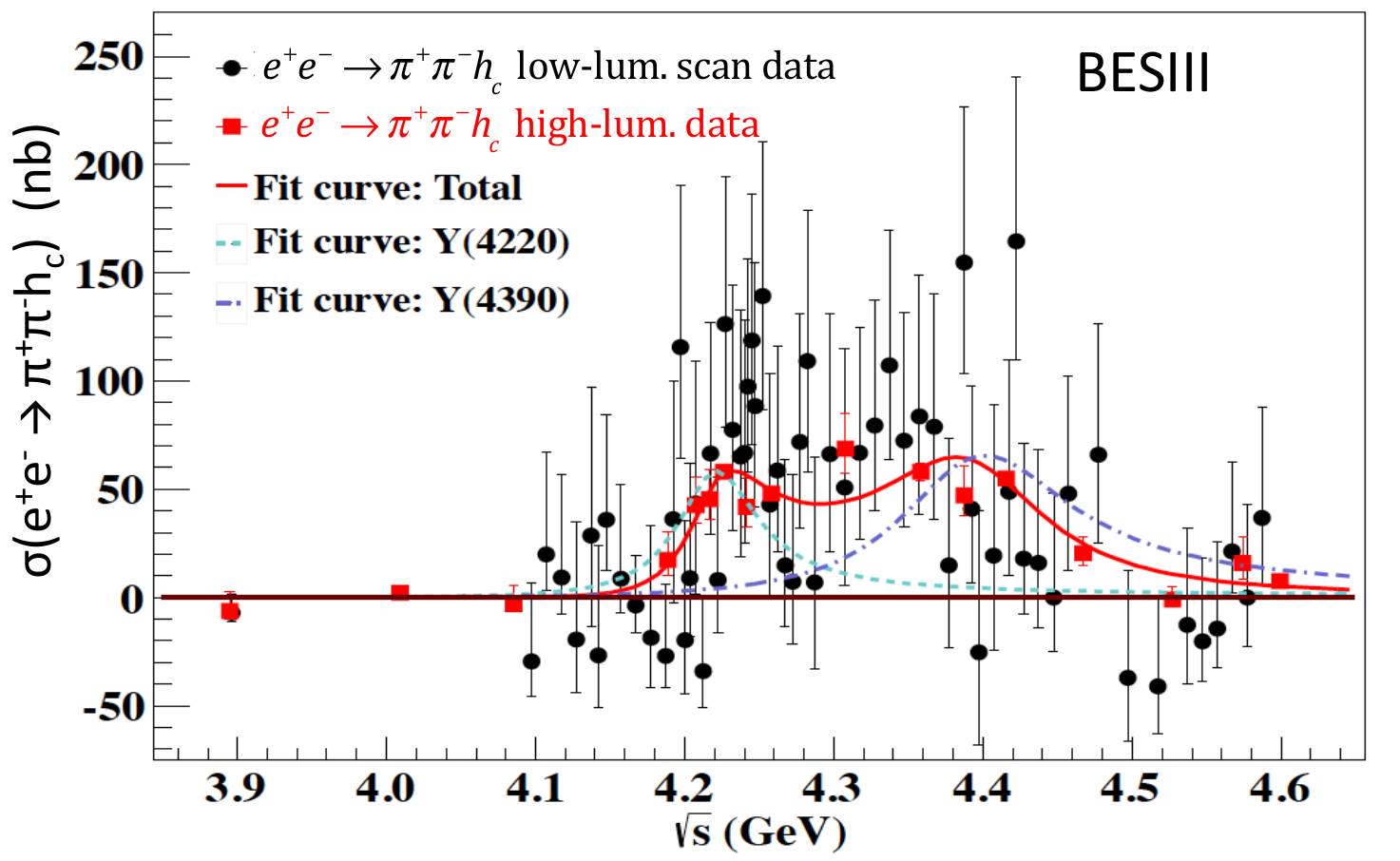}
\end{minipage}\hspace{\fill}
\caption{\footnotesize 
The data points show BESIII measurements of the cross section for $\ee\rt\pipi h_c$, where the $h_c$ was
detected via its $h_c\rt\gamma\eta_c$ decay mode with the $\eta_c$ reconstructed in one of 16 exclusive
multihadron decay channels~\cite{BESIII:2016adj}.  The solid black dots are from the low-luminosity-scan
and the solid red points are the high-luminosity-scan points. The solid red curve shows the results of a
fit to the data with a coherent sum of two interfering BW amplitudes discussed in the text. 
}
\label{fig:bes3_pipihc-scan}
\end{figure}

BESIII measurements of the energy dependence of the cross section for  $\ee\rt\pipi h_c$  with the same
two data sets~\cite{BESIII:2016adj} are shown in Fig.~\ref{fig:bes3_pipihc-scan}. The solid red curve in the
figure shows the results of a fit to the measurements with a coherent sum of two BW amplitudes.
The parameters determined from the fit are:
\begin{eqnarray}
M_1=4218 \pm 4\ {\rm MeV}  ~ &~~& \Gamma_1=66 \pm  9\ {\rm MeV} \nonumber \\
M_2=4392 \pm 6\ {\rm MeV}   &~~& \Gamma_2=140 \pm 16 \ {\rm MeV},
\label{eqn:two-BW-masses-bis}
\end{eqnarray}  
where the statistical and (smaller) systematic errors are added in quadrature.
The lower mass BW term, shown in the figure as a dashed green line, has a fitted mass and width that is
consistent with the $M\simeq 4220$~MeV peak seen in $\eta\jpsi$, $\omega\chi_{c0}$ and $\pipi\jpsi$.
No evidence for the higher mass $\pipi h_c$ peak is seen in the $\eta\jpsi$ or $\omega\chi_{c0}$ channels
and its measured parameters are inconsistent with those of the $Y(4360)$, for both the $\pipi\jpsi$ and
$\pipi\psip$ channels.

\vspace{0.1in}

\subsubsection{Discussion}
The $Y(4260)$ and the other, higher-mass $1^{--}$ states have attracted considerable attention; the
BaBar~\cite{Aubert:2005rm,Aubert:2007zz} and Belle~\cite{Wang:2007ea} papers reporting their discoveries
rank among these experiments most highly cited papers. Most of the theoretical discussions to date have
been focused on $Y(4260)$ mass and width parameters that were determined from single BW fits to isr line
shapes shown in Fig.~\ref{fig:babar-belle_y4260}. However, the recent measurements of the $\ee\rt \eta\jpsi$,
$\omega\chi_{c0}$ and $\pipi h_c$ cross sections, shown in Figs.~\ref{fig:eta-jpsi_omega-chic0} 
and~\ref{fig:bes3_pipihc-scan}, respectively, and precise results for $\ee\rt\pipi\jpsi$ shown in
Fig.~\ref{fig:bes3_pipijpsi-scan}, demonstrate that the single-resonance assumption that was used to determine
the mass and width values given in Eq.~\ref{eqn:y4260-mass} was too na\"{i}ve and the values that were derived
are not reliable.  
The older results based on single-peak fits to $\pipi\jpsi$ mass distribution ought to be ignored and
the $Y(4260)$ label retired. Averaging the mass and width determinations in 
$\pipi\jpsi$, $\pipi h_c$ and $\omega\chi_{c0}$ channels, we obtain:
\begin{eqnarray}
M(Y(4220))        &=& 4222\pm  3\ {\rm MeV}  \nonumber \\
\Gamma(Y(4220))   &=&   48\pm  7\ {\rm MeV},
\label{eqn:y4220-mass-width}
\end{eqnarray}
where the error on the width was scaled up to account for mild disagreements between 
the different channels.

Some theoretical papers interpreted the $Y(4260)$ as a bound state
of a $D$ meson and a $\bar{D}_1(2420)$, a $J^P=1^+$ P-wave excitation of the $D$ meson with mass
2421~MeV and width $\Gamma=27.4$~MeV~\cite{Ding:2008gr,Wang:2013cya}. For the Eq.~\ref{eqn:y4260-mass} mass
value for the $Y(4260)$, this implied a $D\bar{D}_1$ binding energy of $\simeq 35$~MeV, which is somewhat larger
than typical values for nuclear systems that are bound by Yukawa meson-exchange forces. With the lower,
Eq.~\ref{eqn:y4220-mass-width} value for the $Y(4220)$ mass, the implied $D\bar{D}_1$ binding energy nearly doubles
to 66~MeV, which suggests that the $D\bar{D}_1$ molecule interpretation should be reevaluated. Other authors
have suggested that the $Y(4260)$ might be a $\ccbar$-gluon hybrid meson~\cite{Zhu:2005hp,Close:2005iz,Kou:2005gt}.
A lattice QCD calculation (with pion mass $\sim 400$MeV) finds a candidate for a $1^{--}$ hybrid state at a
mass of $4285\pm 14$~MeV that the authors suggested as 
a possible interpretation for the $Y(4260)$ \cite{Liu:2012ze}.
A large radiative width of the $Y(4260)$ would be at odds 
with the hybrid interpretation, and this calls for improved measurements of the   
$e^+e^-\to\gamma X(3872)$ \cite{Ablikim:2013dyn} cross-section in the relevant mass range, 
to correlate it better with the observed $Y$ structures and to extract 
their absolute radiative branching ratios.

\myclearpage

\subsection{$X(4140)$ and other $\jpsi\phi$ structures}
\label{sec:x4140etc}
\def\bujphik{B^+\to\jpsi \phi K^+}

\begin{figure*}[htb]
\includegraphics[width=0.8\textwidth]{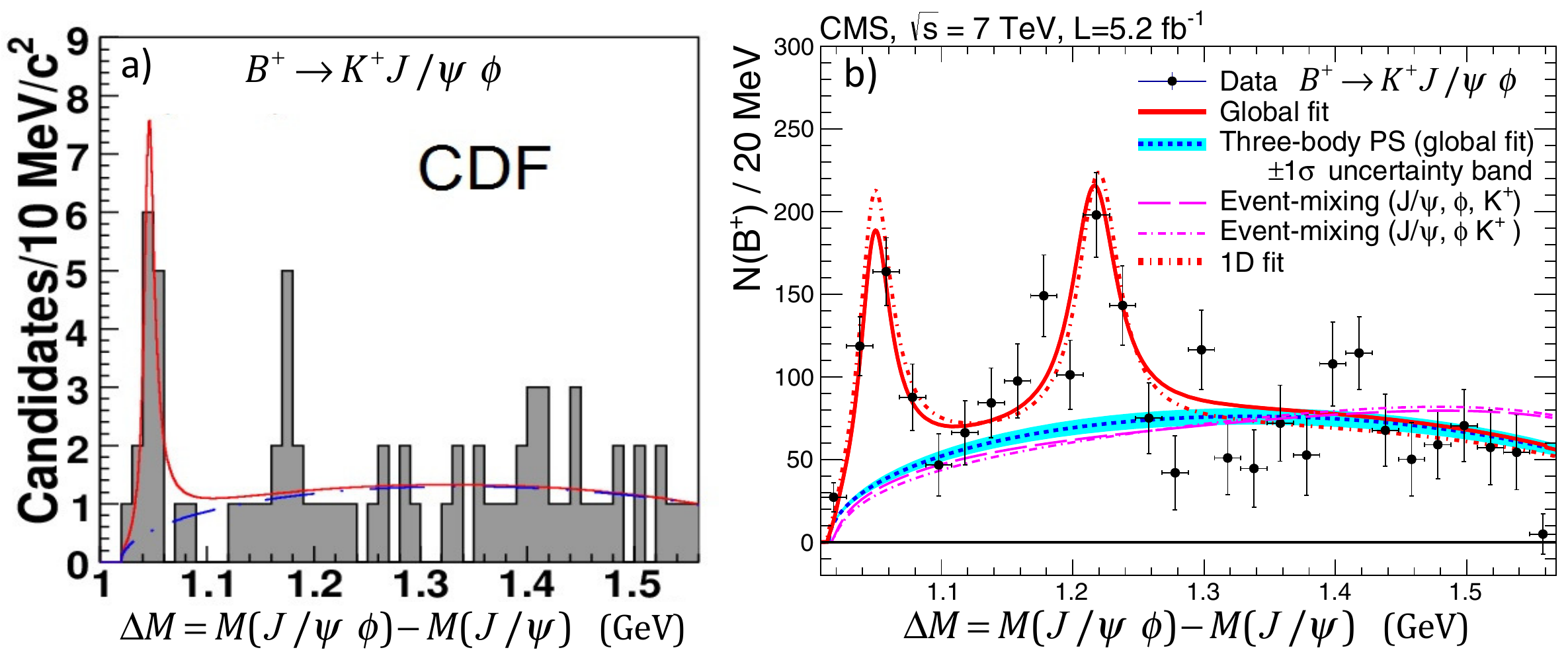}
\caption{\footnotesize
{\bf a)} The $\Delta M=M(\jpsi\phi) - M(\jpsi)$ distribution for a 58 event sample of candidate  
$B^+\to\jpsi\phi K^+$ decays from the CDF experiment \protect\cite{Aaltonen:2009tz}.
The histogram shows the data and the red curve shows the result of a fit with a BW signal shape and a
three-body phase-space term to represent the non-resonant background. 
{\bf b)} The corresponding plot for a 2.5K event sample of candidate $B^+$ decays from the CMS
experiment \protect\cite{Chatrchyan:2013dma}.  Here the fit includes two BW signal shapes, one for the
$X(4140)$ and the other for the enhancement near $\Delta M\simeq 1.22$~GeV. }
\label{fig:CDFCMS4140}
\end{figure*}

Studies of mass structures in $\jpsi\phi$ have a vivid and controversial history that involves a number
of experiments, as summarized in Tables~\ref{tab:x4140} and \ref{tab:x4274plus}. The relative ease of
triggering on $\jpsi\to\mu^+\mu^-$ decays, supplemented with the distictively narrow $\phi\to K^+K^-$ mass peak,
provides a relatively clean signature, even in hadron collider experiments with no hadron identification
capabilities. The history started in 2008, when the CDF collaboration presented $3.8\sigma$ evidence for a
near-threshold $\jpsi\phi$ mass peak in $\bujphik$ decays, shown in Fig.~\ref{fig:CDFCMS4140}a, with
mass $M=4143 \pm 3$~MeV and width $\Gamma=11.7\,^{+9.1}_{-6.2}$~MeV, that is called the
$X(4140)$~\cite{Aaltonen:2009tz}.\footnote{In the literature, this is sometimes referred to as the $Y(4140)$.}

A conventional $\ccbar$ charmonium state with this mass would be able to decay to a variety of
open-charmed-meson pair final states via allowed fall-apart decays and have an expected width that is
much higher than CDF's measured value for the $X(4140)$.  Moreover, the observed $\jpsi\phi$ decay mode
would be OZI-suppressed for charmonium state decays and expected to have an undetectably small branching
fraction. Because of these conflicts with charmonium-model-based expectations, the CDF observation
triggered considerable interest.  It was suggested that the $X(4140)$ structure could be a molecular state 
\cite{Liu:2009ei,Branz:2009yt,Albuquerque:2009ak,Ding:2009vd,Zhang:2009st,Liu:2009pu,Wang:2009ry,Molina:2009ct,Chen:2015fdn,Karliner:2016ith},
a tetraquark state \cite{Stancu:2009ka,Drenska:2009cd,Wang:2015pea,Anisovich:2015caa,Lebed:2016yvr}, 
a hybrid state \cite{Mahajan:2009pj,Wang:2009ue} 
or a rescattering effect \cite{Liu:2009iw,Swanson:2014tra}.

\begin{figure*}[htb]
\includegraphics[width=0.8\textwidth]{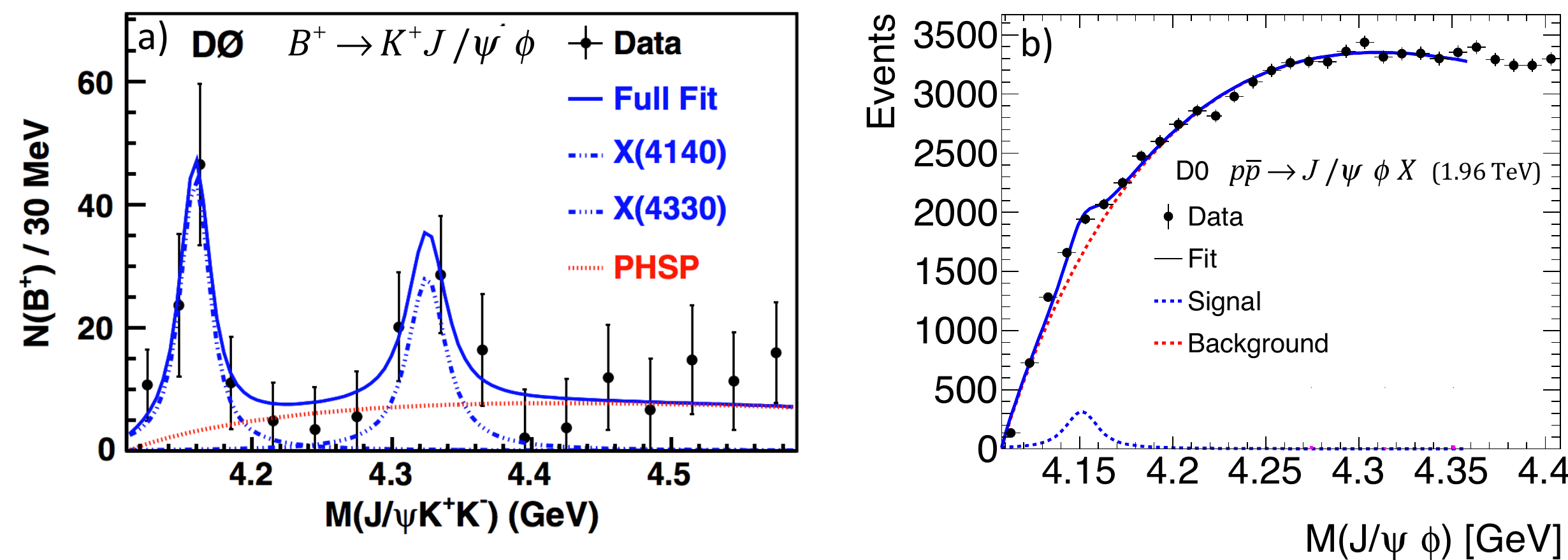}
\caption{\footnotesize 
{\bf a)} The $M(\jpsi\phi)$ distribution for a 215~event sample of candidate $\bujphik$ decays from the D0
experiment~\cite{Abazov:2013xda}.
The solid blue curve is the result of a fit with two BW signal shapes and a  three-body phase-space term to
represent the non-resonant background.
{\bf b)} 
The distribution of invariant masses for prompt $\jpsi\phi$ combinations produced in inclusive
 $p\bar{p}\rt\jpsi\phi X$ reactions from the D0 experiment~\cite{Abazov:2015sxa}.
 }
\label{fig:D04140}
\end{figure*}

A analysis of $\bujphik$ decays by the LHCb collaboration, based on a fraction of their Run-I data 
sample~\cite{Aaij:2012pz}, found no evidence for a narrow $X(4140)$-like peak, and set an upper limit on its
production that was in $2.4\sigma$ tension with the CDF results~\cite{Aaltonen:2011at}.  
Belle~\cite{Brodzicka:2010zz,ChengPing:2009vu}~(unpublished) and BaBar~\cite{Lees:2014lra} searches for a
narrow $X(4140)$ state did not confirm its presence, but the limits that they set were not in serious conflict
with the CDF measurements.  In 2014, an $X(4140)\rt\jpsi\phi$-like signal with mass and width values consistent
with the CDF results and a statistical significance of $5\sigma$, shown in Fig.~\ref{fig:CDFCMS4140}b, was
reported in $\bujphik$ decays by the CMS collaboration~\cite{Chatrchyan:2013dma}.  Also in 2014, D0 reported the
$M(\jpsi\phi)$ distribution shown in Fig.~\ref{fig:D04140}a, where there is $3\sigma$ evidence for an narrow
$X(4140)$-like structure, but with a mass, $4159\pm 8$~MeV, that was about two standard deviations higher than the
CDF value~\cite{Abazov:2013xda}. In addition, D0 reported a $4.7\sigma$ signal for prompt $X(4140)$ production in
$\ecm=1.96$~TeV $p\bar{p}$ collisions as shown in Fig.~\ref{fig:D04140}b~\cite{Abazov:2015sxa}. The BESIII
collaboration did not find evidence for $X(4140)\to\jpsi\phi$ in $e^+e^-\to\gamma X(4140)$ and set upper limits on
its production cross-sections at c.m.~energies of $4.23$, $4.26$ and $4.36$ GeV ]~\cite{Ablikim:2014atq}.

In an unpublished update to their $\bujphik$ analysis~\cite{Aaltonen:2011at}, 
the CDF collaboration presented $3.1\sigma$ evidence for a second relatively
narrow $\jpsi\phi$ mass peak near $4274\pm 8$ MeV, an observation that has also
received considerable attention in the literature \cite{He:2011ed,Finazzo:2011he}.
There are signs of a second $\jpsi\phi$ mass peak in the CMS distribution shown in
Fig.~\ref{fig:CDFCMS4140}b, but at a mass of $4314\pm 5$~MeV, which is $3.2$ standard
deviations higher than the CDF value; no statistical significance of this structure
is reported \cite{Chatrchyan:2013dma}.  There is some hint of a second peak near 4330~MeV
in the D0 $\jpsi\phi$ mass distribution shown in Fig.~\ref{fig:D04140}a, but with a small, 
$\sim 1.7\sigma$ significance.  The Belle collaboration saw $3.2\sigma$ evidence for a 
narrow $\jpsi\phi$ peak at $4351\,\pm 5$~MeV  in two-photon collisions, which implies
$J^{PC}=0^{++}$ or $2^{++}$, and found no evidence for $X(4140)$ in the same analysis~\cite{Shen:2009vs}.

\subsubsection{The 6-dimensional LHCb Amplitude Analysis}
All the analyses mentioned above had limited data sets and were based on simple
$\jpsi\phi$ mass  fits, with a Breit-Wigner shape to represent the signal and an incoherent
background described by an ad-hoc functional shape (usually a three-body $\bujphik$
phase-space distribution).  While the $M(\phi K)$ distribution in this decay process has been
observed to be featureless, several resonant contributions from $K^*\rt K\phi$ excitations are
expected. The first amplitude analysis of $\bujphik$ decays that was capable of separately
resolving possible $K^{*+}\to\phi K^+$ and $X\to\jpsi\phi$ resonances was recently reported 
by the LHCb collaboration.  This was based on a nearly background-free sample of $\bujphik$
decays that was larger than that for any of the previous analyses~\cite{Aaij:2016iza,Aaij:2016nsc}.
In this analysis, it was found that the data across the full, 6-dimensional (6D) phase-space
of invariant masses and decay angles spanned by the five final-state particles could not be
described by a model that contains only excited kaon states that decay into $\phi K$; an
acceptable description of the data was only obtained when four coherent
$X\to\jpsi\phi$ peaking structures were included. The $K^{*+}$ amplitude model determined from the analysis
that included the four $\jpsi\phi$ resonant structures is consistent with expectations based
on the quark model and previous experimental $K^{*+}\rt\phi K^+$ resonance results.

\begin{figure}[bthp]
  \includegraphics*[width=0.5\textwidth]{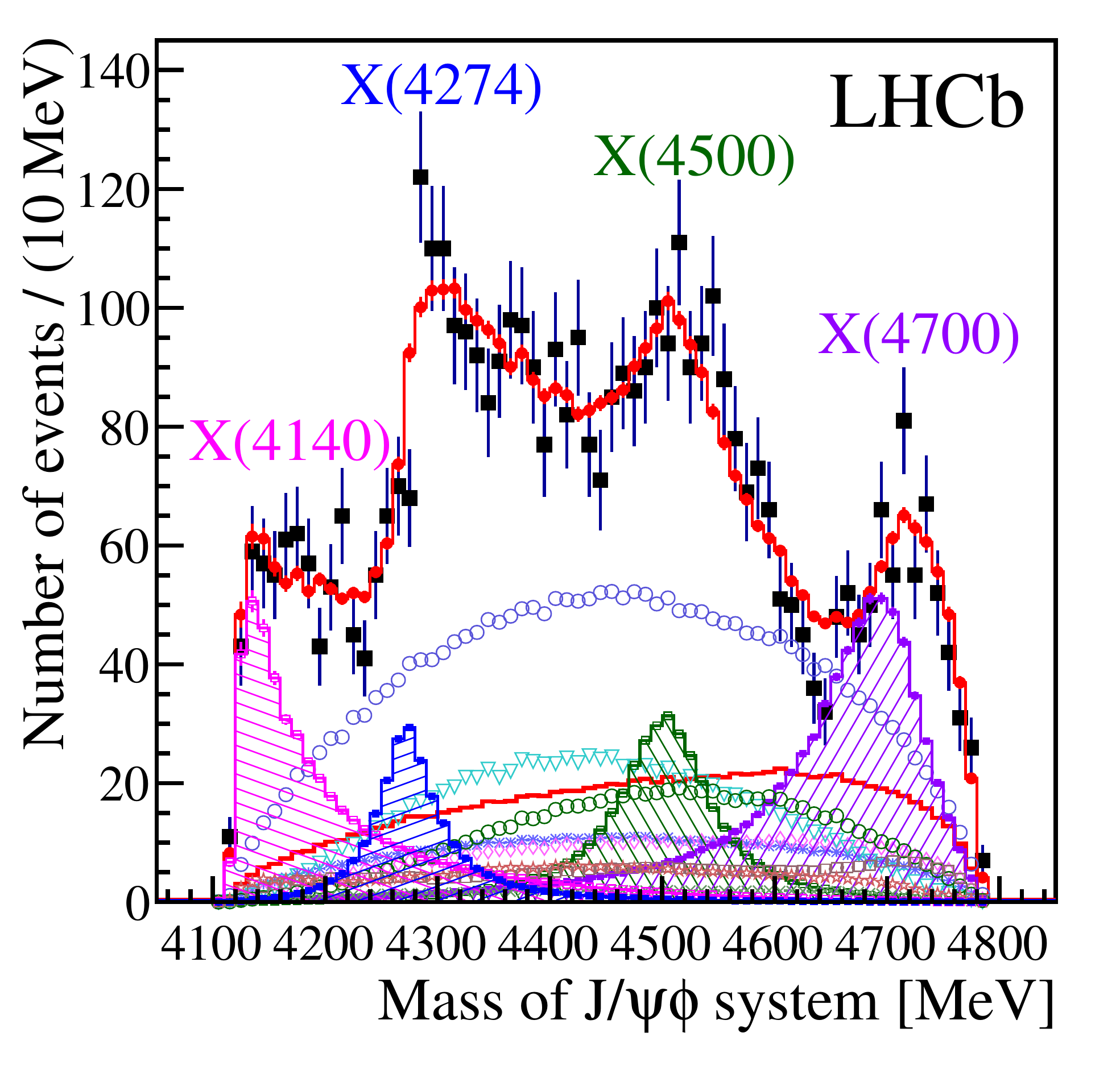}
\caption{
The points with error bars show the distribution of $\jpsi\phi$ invariant masses in the LHCb experiment's
4.3K event sample of candidate $B^+\to\jpsi\phi K^+$ decays \protect\cite{Aaij:2016iza,Aaij:2016nsc}.
The non-$B$-decay background estimate is shown by the lower red histogram. Projections of the 6-dimensional
amplitude fit, with the four $X\to\jpsi\phi$ resonance terms shown as hatched histograms plus
contributions from a $X\to \jpsi\phi$ non-resonant amplitude (blue circles) and $K^{**}\to\phi K$ excitations,
are shown by the solid-line red histogram with error bars.
\label{fig:LHCbJpsiPhi}
}
\end{figure}

Figure~\ref{fig:LHCbJpsiPhi} shows the $\jpsi\phi$ invariant mass distribution from the 4.3K
reconstructed $\bujphik$ decays in the LHCb Run-II data sample with the projected results from the 6D fit
superimposed as a red histogram with error bars.  There is no narrow $\jpsi\phi$ mass peak just above
the kinematic threshold as first reported by CDF. Instead, a broad enhancement with a
mass, $M=4146.5\pm 4.5^{+4.6}_{-2.8}$~MeV, that is consistent with the $X(4140)$ values from CDF and CMS,
but with a width, $\Gamma=83\pm21\,^{+21}_{-14}$~MeV, 
that is substantially broader than the CDF value\footnote{This should be considered ``tension'', 
rather than disagreement since the CDF and LHCb results differ by $2.7\sigma$.}, 
is observed with high ($8.4\sigma$) significance. 
The $J^{PC}$ quantum numbers of this structure are determined
from the LHCb fit to be $1^{++}$; other hypotheses are ruled out with a significance of $5.7\sigma$ or more.
The $1^{++}$ quantum number assignment has an important impact on possible interpretations for the $X(4140)$,
in particular, it rules out the $0^{++}$ or $2^{++}$ $D_s^{*+}D_s^{*-}$ molecular models proposed in
refs.~\cite{Liu:2009ei,Branz:2009yt,Albuquerque:2009ak,Ding:2009vd,Zhang:2009st,Molina:2009ct,Chen:2015fdn}. 
It was suggested that below-$\jpsi\phi$-threshold $D_s^{\pm}D_{s}^{*\mp}$ kinematically 
induced cusp~\cite{Swanson:2014tra,Karliner:2016ith} may be responsible for 
the observed $X(4140)$ structure (Appendix D in ref.~\cite{Aaij:2016nsc}), 
though the cusp model used in this analysis is theoretically controversial as discussed in 
sec.~\ref{sec:kinematicpeaks}.

The PDG's 2017 update to ref.~\cite{Olive:2016xmw} lists an average of all published measurements 
\cite{Aaltonen:2009tz,Chatrchyan:2013dma,Abazov:2013xda,Abazov:2015sxa}
of the $X(4140)$ parameters as its mass, $4146.8\pm2.5$ MeV, 
and width, $19^{+8}_{-7}$ MeV. 
The evolution of the $Z(4430)^+$ mass and width determination \cite{Choi:2007wga,Mizuk:2009da,Chilikin:2013tch},
discussed below in sec.~\ref{sec:z4430disc},  
provides the valuable lesson that a one-dimensional fit to a mass distribution
of a resonance peak, together with an ad hoc assumption about the background shape and its 
incoherence, is prone to yield biased mass and width results and underestimated systematic errors.
Therefore, in Table~\ref{tab:Q-Qbar} we list mass and width values that are based only on the results from 
the full amplitude 
analysis \cite{Aaij:2016iza}, since this is the only one that
resolved various background contributions and added them coherently to the signal
amplitude.

The analysis also established the existence of the $X(4274)$ structure with $M=4273.3\pm 8.3^{+17.2}_{-3.6}$~MeV
at the $6\sigma$ significance level and with quantum numbers that were determined to be
$1^{++}$ at the $5.8\sigma$ level. No proposed molecular bound-state or cusp model can account for these $X(4274)$
$J^{PC}$ values. A hybrid charmonium state in this mass region would have $J^{PC}=1^{-+}$~\cite{Mahajan:2009pj,Wang:2009ue}. 
Most models that interpret the $X(4140)$ as a  tetraquark state predicted that the $J^{PC}$ values of the next
higher-mass state to be different 
from $1^{++}$~\cite{Drenska:2009cd,Wang:2015pea,Anisovich:2015caa,Lebed:2016yvr,Maiani:2016wlq}.  
An exception is a tetraquark model implemented by Stancu~\cite{Stancu:2009ka} that not only correctly assigned
$1^{++}$ to the $X(4140)$,  but also predicted a second $1^{++}$ state at a mass that is not much higher than that
of the $X(4274)$. 
A lattice QCD calculation with
diquark operators found no evidence for a $1^{++}$ tetraquark below $4.2$ GeV~\cite{Padmanath:2015era}.
However, given than not all dynamical effects were simulated, this calculation probably does not rule them out.   

In addition, the LHCb analysis, which was the first high-sensitivity investigation of the high $\jpsi\phi$ mass region, uncovered
three significant $0^{++}$ contributions: a $0^{++}$ nonresonant term plus two, previously unseen $0^{++}$ resonances,
the $X(4500)$ (with $6.1\sigma$ significance) and the $X(4700)$ (with $5.6\sigma$ significance).  The $0^{++}$ quantum
numbers of these states are established with significances of more than $4\sigma$.  Wang {\it et al.}~\cite{Wang:2009ry}
predicted a virtual $D_s^{*+}D_s^{*-}$ state at $4.48\pm0.17$ GeV.

None of the $\jpsi\phi$ structures observed in $B$ decays are consistent with the
state seen in two-photon collisions by the Belle collaboration \cite{Shen:2009vs}.

\setboolean{prl}{true} 
\begin{table*}[bhtp]
\caption{Results related to the $X(4140)\to\jpsi\phi$ mass peak, 
first observed in $B^+\to\jpsi\phi K^+$ decays.
The first (second) significance quoted for ref.~\cite{Abazov:2015sxa} 
is for the non-prompt (prompt) production components 
(the mass and width were determined from the non-prompt sample). 
The last column gives a fraction of $B^+\to\jpsi\phi K^+$ rate 
attributed to the $X(4140)$ structure, however, the CDF and  
D0 results were normalized to a $B^+\to\jpsi\phi K^+$ rate
that excluded the high $\jpsi\phi$ mass range.
}
\label{tab:x4140}
\hbox{
\hbox{\ifthenelse{\boolean{prl}}{}{\quad\hskip-2.5cm}
\hbox{
\ifthenelse{\boolean{prl}}{}{\begin{footnotesize}}
\renewcommand{\arraystretch}{1.2}
\def\1#1{\multicolumn{1}{c}{#1}}
\def\2{}
\def\3{\ifthenelse{\boolean{prl}}{}{\!\!\!}}
\def\pms{\ifthenelse{\boolean{prl}}{\pm}{\!\pm\!}}
\begin{tabular}{ccrllcl}
\hline
Year & Experiment  & \1{$B\to\jpsi\phi K$} & \multicolumn{4}{c}{$X(4140)$ peak} \\
     & luminosity  & \1{yield}        & \1{Mass [MeV]} & \1{Width [MeV]} & 
\1{\ifthenelse{\boolean{prl}}{Significance}{Sign.}} & \1{Fraction \%} \\
\hline\hline
2008 & CDF 2.7 fb$^{-1}$  \cite{Aaltonen:2009tz} & 
$58\pm10$ \2 &  
$4143.0\pms2.9\pms1.2$ \ifthenelse{\boolean{prl}}{\quad}{\3} &
$11.7\,^{+8.3}_{-5.0}\pms3.7$ &
$3.8\sigma$ & \\
{\it 2009} & {\it Belle} \cite{Brodzicka:2010zz} &
$\mathit{325\pm21}$ \2 &
$\mathit{4143.0}$ {\it fixed} & 
$\mathit{11.7}$ {\it fixed} &  
$\mathit{1.9\sigma}$ & \\
{\it 2011} & {\it CDF 6.0 fb$^{-1}$}  \cite{Aaltonen:2011at} & 
$\mathit{115\pm12}$ \2 &
$\mathit{4143.4\,^{+2.9}_{-3.0}\pms0.6}$ &
$\mathit{15.3\!^{+10.4}_{-\phantom{0}6.1}\!\pms\!2.5}$ & 
$\mathit{5.0\sigma}$ &
$\mathit{14.9\pms3.9\pms2.4}$ \\
2011 & LHCb 0.37 fb$^{-1}$  \cite{Aaij:2012pz} &
$346\pm20$ \2 &
$4143.4$ fixed & 
$15.3$ fixed   &
$1.4\sigma$    &
$<7$ @~90\%CL \\
2013 &
CMS 5.2 fb$^{-1}$  \cite{Chatrchyan:2013dma} &
$2480\pm160$ &
$4148.0\pms2.4\pms6.3$ &
$28\phantom{.0}\,^{+15\phantom{.0}}_{-11\phantom{.0}}\pm19$ & 
$5.0\sigma$ &
$10\phantom{.0}\pms3$ (stat.)  \\
2013 & D0 10.4 fb$^{-1}$  \cite{Abazov:2013xda} &
$215\pm37$ \2 &
$4159.0\pms4.3\pms6.6$ &
$19.9\pms12.6\,^{+1.0}_{-8.0}$ &
$3.0\sigma$ &
$21\phantom{.0}\pms8\phantom{.0}\pms4$ \\
2014 & BaBar  \cite{Lees:2014lra} &
$189\pm14$ \2 &
$4143.4$ fixed & 
$15.3$ fixed  &
$1.6\sigma$ &
$<13.3$ @~90\%CL \\
2016 & LHCb 3.0 fb$^{-1}$  \cite{Aaij:2016iza} &
$4289\pm151$ \2 &
$4146.5\pms4.5\,^{+4.6}_{-2.8}$ & 
$83\phantom{.0}\pms21\phantom{.0}\,^{+21}_{-14}$  &
$8.4\sigma$    &
$13.0\pms3.2\,^{+4.8}_{-2.0}$ \\
\hline
2015 & D0 10.4 fb$^{-1}$  \cite{Abazov:2015sxa} &
\1{\3$p\bar{p}\to\jpsi\phi...$\3} &
$4152.5\pms1.7\,^{+6.2}_{-5.4}$ &
$16.3\pms5.6\pms11.4$ &
\3$5.7\sigma$ ($4.7\sigma$){\hskip-1cm\quad} &
 \\
\hline
\end{tabular}
\ifthenelse{\boolean{prl}}{}{\end{footnotesize}}
}
}
}
\end{table*} 

\begin{table*}[bhtp]
\caption{Results related to $\jpsi\phi$ mass structures 
heavier than the $X(4140)$ peak.   
The unpublished results are shown in italics.
The last column gives a fraction of the total $B^+\to\jpsi\phi K^+$ rate 
attributed to the given structure. 
}
\label{tab:x4274plus}
\hbox{
\hbox{\ifthenelse{\boolean{prl}}{}{\quad\hskip-2.2cm}
\hbox{
\ifthenelse{\boolean{prl}}{}{\begin{footnotesize}}
\def\1#1{\multicolumn{1}{c}{#1}}
\def\2{}
\def\3{\ifthenelse{\boolean{prl}}{}{\!\!\!}}
\def\pms{\ifthenelse{\boolean{prl}}{\pm}{\!\pm\!}}
\renewcommand{\arraystretch}{1.2}
\begin{tabular}{ccrllcl}
\hline
Year & Experiment  & \1{$B\to\jpsi\phi K$} & \multicolumn{3}{c}{$X(4274-4351$) peaks(s)} \\
     & luminosity  & \1{yield}        & \1{Mass [MeV]} & \1{Width [MeV]} & 
\1{\ifthenelse{\boolean{prl}}{Significance}{Sign.}}  & \1{Fraction [\%]} \\
\hline\hline
{\it 2011} & {\it CDF 6.0 fb$^{-1}$}  \cite{Aaltonen:2011at} & 
$\mathit{115\pm12}$ \2 &
$\mathit{4274.4\,^{+8.4}_{-6.7}\pms1.9}$ &
$\mathit{32.3{^{+21.9}_{-15.3}}\!\pms7.6}$ &
$\mathit{3.1\sigma}$ & \\
2011 & LHCb 0.37 fb$^{-1}$  \cite{Aaij:2012pz} &
$346\pm20$ \2 &
$4274.4$ fixed &
$32.3$ fixed  &
  &
$<\phantom{0}8$ @~90\%CL \\
2013 &
CMS 5.2 fb$^{-1}$  \cite{Chatrchyan:2013dma} &
$2480\pm160$ &
$4313.8\pms5.3\pms7.3$ \ifthenelse{\boolean{prl}}{\quad}{\3} &
$38\phantom{.0}\,^{+30\phantom{.0}}_{-15\phantom{.0}}\pms16$ &
  & \\
2013 & D0 10.4 fb$^{-1}$  \cite{Abazov:2013xda} &
$215\pm37$ \2 &
$4328.5\pms12.0$ &
$30\phantom{.0}$ fixed &
  & \\
2014 & BaBar \cite{Lees:2014lra} &
$189\pm14$ \2 &
$4274.4$ fixed &
$32.3$ fixed  &
$1.2\sigma$  &
$<18.1$ @~90\%CL \\
2016 & LHCb 3.0 fb$^{-1}$  \cite{Aaij:2016iza} &
$4289\pm151$ \2 &
$4273.3\pms8.3\,^{+17.2}_{-\phantom{1}3.6}$ & 
$56\phantom{.0}\pms11\phantom{.0}\,^{+\phantom{1}8}_{-11}$  &
$6.0\sigma$    &
$7.1\pms2.5\,^{+3.5}_{-2.4}$ \\
     &                                         &
                &
$4506\phantom{.3}\pms11\phantom{.0}\,^{+12}_{-15}$ &
$92\phantom{.0}\pms21\phantom{.0}\,^{+21}_{-20}$  &
$6.1\sigma$    &
$6.6\pms2.4\,^{+3.5}_{-2.3}$ \\
     &                                         &
                &
$4704\phantom{.3}\pms10\phantom{.0}\,^{+14}_{-24}$ &
$120\pms31\phantom{.0}\,^{+42}_{-33}$  &
$5.6\sigma$    &
$12\phantom{.0}\pms5\phantom{.0}\,^{+9\phantom{.0}}_{-5\phantom{.0}}$ \\
\hline
2010 & Belle \cite{Shen:2009vs} &
\1{\3$\gamma\gamma\to\jpsi\phi$\3} &
$4350.6\,^{+4.6}_{-5.1}\pms0.7$ &
$13\,^{+18}_{-\phantom{0}9}\pms4$ & 
$3.2\sigma$ & \\
\hline
\end{tabular}
\ifthenelse{\boolean{prl}}{}{\end{footnotesize}}
}
}
}
\end{table*} 

\subsubsection{Charmonium Assignments for the $\jpsi \phi$ states?}

The main reason that the $X(4140)$ attracted a lot of interest was the narrow width reported by the
early measurements.  However, the widths determined from the LHCb analysis are larger, ranging between
56~and~120~MeV, depending on the $\jpsi\phi$ peak, and these cannot {\em a priori} be considered to be
too narrow to be charmonium states.  The $X(4140)$ and $X(4274)$ both have quantum numbers that match
the $\chi_{c1}(3{\rm P})$ state, and their masses are in the range of potential model predictions 
for this state~\cite{Chen:2016iua,Li:2009ad,Barnes:2005pb,Godfrey:1985xj,Lu:2016cwr,Ortega:2016hde}.

The dominant decay modes are expected to be to  $D\bar{D}^{*}$, $D^{*}\bar{D}^{*}$ and
$D_s\bar{D}_s^{*}$, final states with total width predictions that range from low values near
30~MeV~\cite{Ortega:2016hde,Barnes:2005pb} to values of 58~MeV~\cite{Chen:2016iua}.  
Given the considerable theoretical uncertainties of these predictions, either the $X(4140)$
or the $X(4274)$ can be considered as a candidate for the $\chi_{c1}(3{\rm P})$ state.  
The $X(4500)$ and the $X(4700)$ have been suggested as
candidates for the $\chi_{c0}(4{\rm P})$ and $\chi_{c0}(5{\rm P})$ states,
since they lie in the predicted mass and width ranges for these states~\cite{Lu:2016cwr,Ortega:2016hde}. 
These higher charmonium states would have a large number of 
allowed decay modes to open charm mesons but, unfortunately, there are no published measurements
of mass spectra in the relevant mass ranges for $B\to D^{(*)}_{(s)}\bar{D}^{*}_{(s)} K$ decays. 

\begin{figure}[bthp]
  \includegraphics*[width=0.5\textwidth]{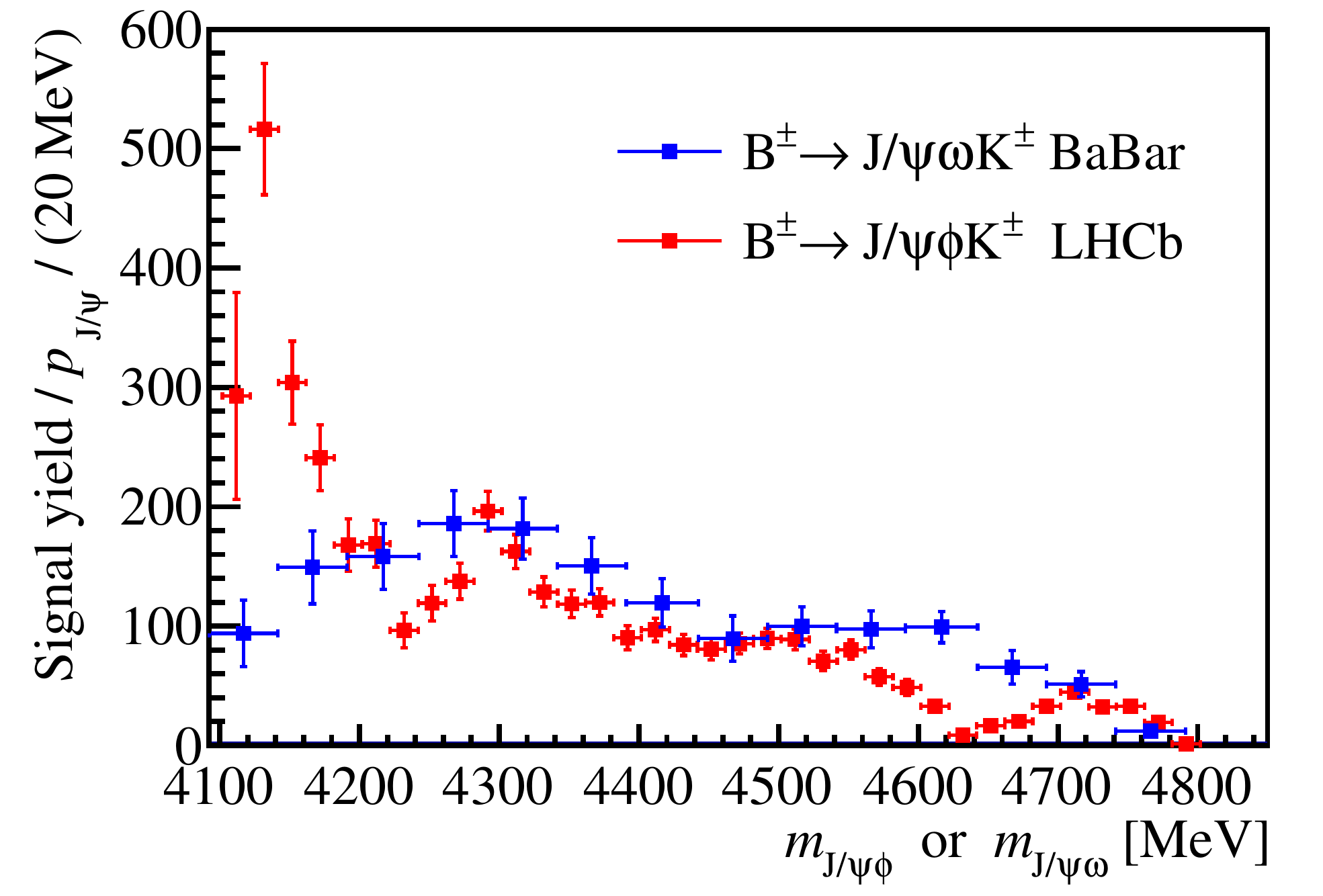}
\caption{
The efficiency-corrected and background-subtracted $\jpsi\phi$ invariant 
mass spectrum for $B^+\to\jpsi\phi K^+$ decays (red points) from LHCb~\cite{Aaij:2016iza,Aaij:2016nsc} and
the $\jpsi\omega$ mass spectrum for $B^+\rt\jpsi\phi K^+$ decays (blue points) from BaBar~\cite{Lees:2014lra},
where the signal yields (in arbitrary units) have been divided by the $\jpsi$ momentum in the $X$ rest frame
to account for phase-space differences.  The two distributions are normalized to have equal areas for masses
above the $\jpsi\phi$ threshold.
\label{fig:JpsiPhivsJpsiOmega}
}
\end{figure}

Na\"ively, one expects the couplings of the $\chi_{cJ}(nP)$ states to $\jpsi\phi$ and to $\jpsi\omega$   
to be very similar, with the rate for the latter being enhanced by the larger phase-space
that is available for the lighter states and the relative ease of producing light $u\bar{u}$
or $d\bar{d}$ quark pairs that comprise the $\omega$ from the vacuum compared to that for more
massive $s\bar{s}$ pairs that comprise the $\phi$. 
The $\jpsi\omega$ mass spectrum in $B\to\jpsi\omega K$ decays measured by the BaBar
collaboration~\cite{Lees:2014lra} does not show any structures resembling the $\jpsi\phi$ mass peaks,
as illustrated in Fig.~\ref{fig:JpsiPhivsJpsiOmega}, which argues against a charmonium interpretation
for any state among the $X(4140)$, $X(4274)$, $X(4500)$ and $X(4700)$.

\myclearpage

\subsection{$X^{*}(3860)$, $X(3940)$ and $X(4160)$}
\label{sec:x3940}

As discussed above in Section~\ref{sec:exp_ee}, the $X(3940)$ was first seen by
Belle~\cite{Abe:2007jna} as an unexpected peak in the distribution of masses 
($M_{\rm recoil}(\jpsi)$) recoiling against a $\jp$ in inclusive $\ee\rt\jp X$ annihilation at
$\ecm\simeq 10.6$~GeV shown in Fig.~\ref{fig:x3940-y4260}a. 
In this figure, there are four distinct peaks: the lower three are due to the
exclusive  processes $\ee\rt\jp\eta_c$, $\ee\rt\jp\chi_{c0}$ and $\ee\rt\jp\eta_c^{\prime}$.
The fourth peak, near 3940~MeV, cannot be associated with any known or expected charmonium
state and has been named the $X(3940)$.  The curve shows results of a fit that includes
four BW line shapes, three for the established $\eta_c$, $\chi_{c0}$ and $\eta_c(2{\rm S})$
charmonium states plus a fourth one to accommodate the unexpected peak near 3940~MeV.  From 
the fit, the mass of the fourth state was found to be $M=3943\pm 6$~MeV, and a limit on
the total width of $\Gamma \le 52$~MeV was established.

Belle did subsequent studies of the exclusive processes $\ee\rt\jpsi D^{(*)}\bar{D}^{(*)}$ decays in the
same energy region, where, to compensate for the low detection efficiency for $D$ and $D^*$ mesons, a
partial reconstruction technique was used that required the reconstruction of the $\jpsi$ and only
one $D$ or $D^*$ meson, and the presence of the undetected $\bar{D}$ or $\bar{D}^*$ was inferred from 
energy-momentum conservation.  With this technique, Belle found a strong signal for
$X(3940)\rt D\bar{D}^*$~\cite{Abe:2007sya} plus two other states: the
$X^*(3860)\rt D\bar{D}$~\cite{Chilikin:2017evr} and $Y(4160)\rt D^*\bar{D}^*$~\cite{Abe:2007sya}.  
Although these three states have not been confirmed by any other experiment, the significance of
the Belle observations in all three cases is above the $5\sigma$ level and we briefly discuss
them here. 

\subsubsection{$X^*(3860)\rt D\bar{D}$; an alternative  $\chi_{c0}(2{\rm P})$ candidate?}
\begin{figure*}[htb]
  \includegraphics[width=0.8\textwidth]{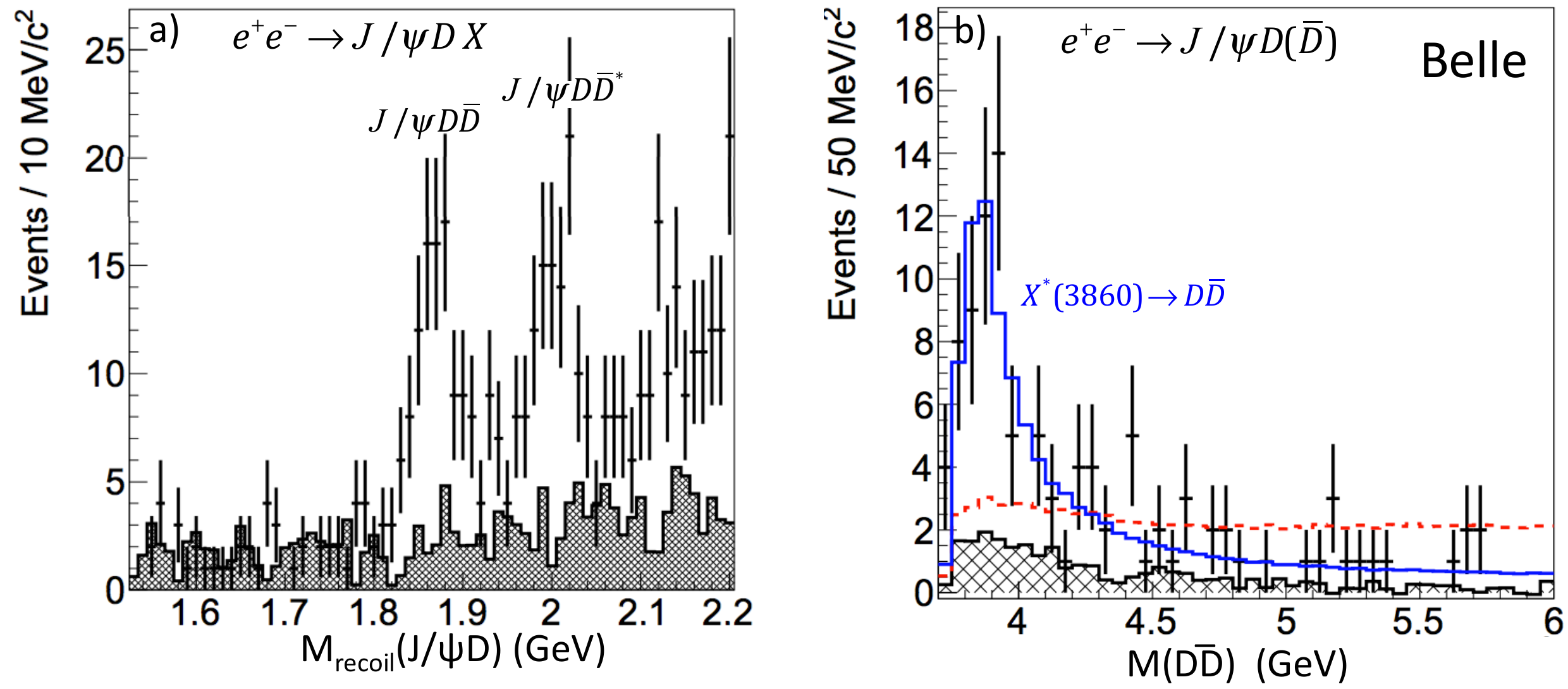}
\caption{\footnotesize {\bf a)}  
The distribution of masses recoiling from a reconstructed $\jpsi$ and $D$-meson in
$\ee\rt \jpsi D X$ annihilation at and near $\ecm=10.58$~GeV.  The two peaks correspond
to exclusive $\ee\rt\jpsi D\bar{D}$ and $\jpsi D\bar{D}^*$ events. The shaded histogram indicates
the background level determined from the $\jpsi$ and $D$ mass sidebands.
{\bf b)} The $D\bar{D}$ invariant mass distribution for the exclusive $\ee\rt \jpsi D\bar{D}$
events.  The solid blue histogram is the result of a fit with a BW resonant amplitude plus a
coherent non-resonant background. The red dash histogram is the fit result when only a
non-resonant amplitude is included~\cite{Chilikin:2017evr}.
}
\label{fig:belle_x3860}
\end{figure*}

~~~
Figure~\ref{fig:belle_x3860}a shows the distribution of masses recoiling
against a detected $\jpsi$ and $D$ meson in $\ee\rt \jpsi D+X$ annihilation events collected in
Belle at c.m.~energies at and near 10.58~GeV. There two peaks are apparent, one centered at
$M_{\rm recoil}(\jpsi D)= m_{D}$, corresponding to exclusive $\ee\rt\jpsi D\bar{D}$ events and the
other centered at $m_{D^*}$, corresponding to exclusive $\ee\rt\jpsi D\bar{D}^*$
events~\cite{Chilikin:2017evr}. Figure~\ref{fig:belle_x3860}b shows the distribution of $D\bar{D}$
pairs in the exclusive $\jpsi D\bar{D}$ event sample, where there is a
strong peaking at small masses.  Fits to the data with a variety of nonresonant-model
amplitudes were unable to describe the data over the four-dimensional phase-space spanned by the
final state particles.  For each choice of nonresonant amplitude, an additional, coherent BW amplitude
was needed. The mass and width of the BW resonance determined from the best fit to the data (shown in
Fig.~\ref{fig:belle_x3860}b as a solid blue histogram) are $M=3862^{+26~+40}_{-32~-13}$~MeV and
$\Gamma=201^{+154~+88}_{~-67~-82}$~MeV. 
The $J^{PC}=0^{++}$ quantum number hypothesis gives the best fit to the data and was favored over
the $2^{++}$ hypothesis by $2.5\sigma$.  The mass, $0^{++}$ quantum numbers and strong
$D\bar{D}$ decay mode of the $X^*(3860)$ all match well to expectations for the $\chi_{c0}(2{\rm P})$
charmonium state, making it a superior candidate for this assignment than the $X(3915)$.

\subsubsection{$X(3940)\rt D\bar{D}^*$}
\begin{figure}[htb]
  \includegraphics[width=0.48\textwidth]{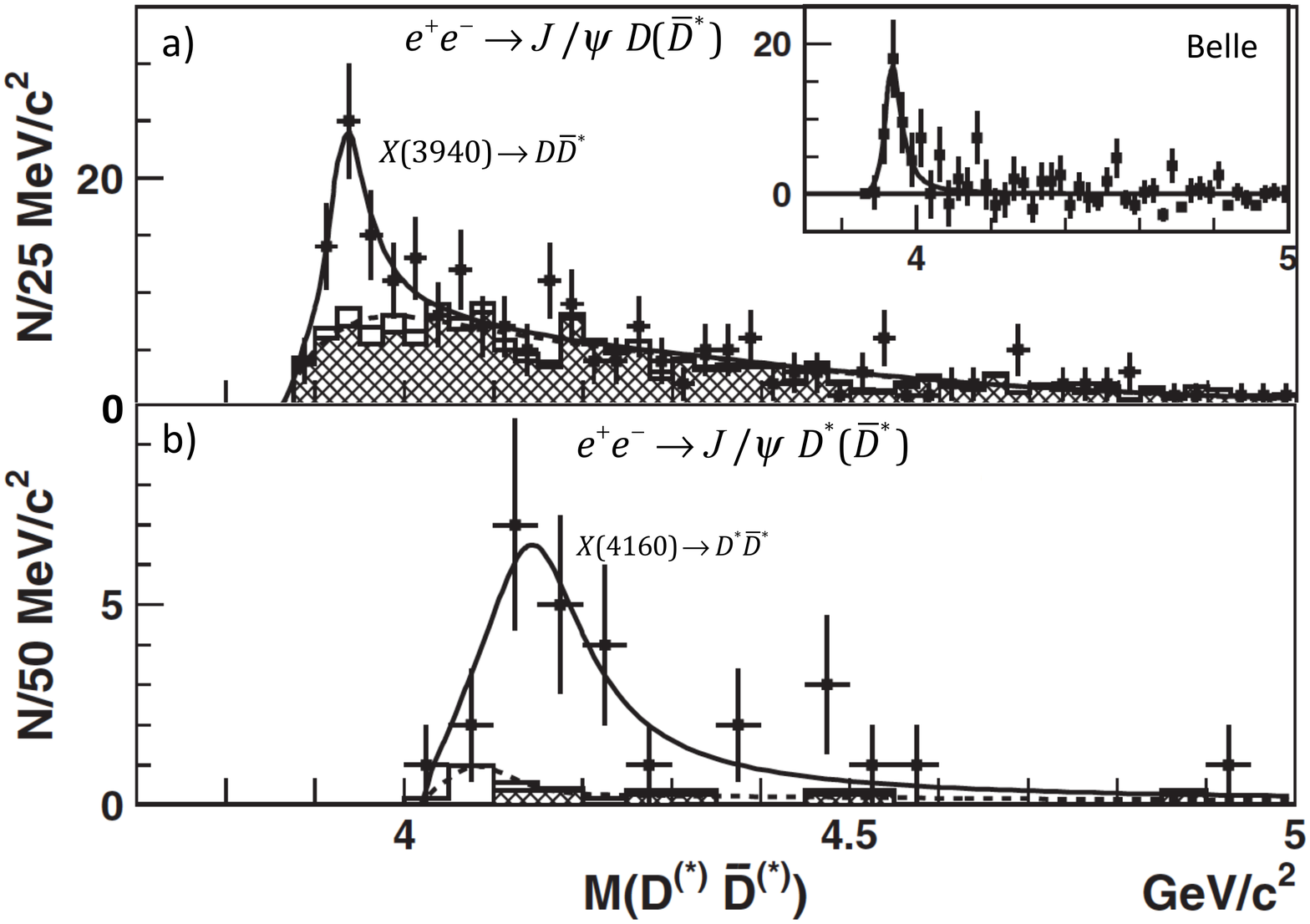}
\caption{\footnotesize {\bf a)}  
The $M(D\bar{D}^*)$ distribution for $\ee\rt\jpsi D\bar{D}^*$ events where the $\jpsi$ and $D$-meson
are reconstructed and the four-momentum of the undetected $\bar{D}^*$-meson is inferred
from energy-momentum conservation.  The hatched histogram is the non-$\jpsi$ and/or non-$D$ meson
background determined from the $\jpsi$ and $D$-meson mass sidebands.  The curve shows the results of
the fit to the  $X(3940)$ resonance described in the text. 
{\bf b)}The $M(D^*\bar{D}^*$ distribution for exclusive $\ee\rt\jpsi D^*\bar{D}^*$ events.  The curve
is the result of the fit for the $X(4160)$. 
}
\label{fig:belle_x3940-x4160-1}
\end{figure}

~~~
Figure~\ref{fig:belle_x3940-x4160-1}a shows the the $D\bar{D}^*$ invariant mass distribution for
$\ee\rt\jpsi D X$ annihilation events where  $M_{\rm recoil}(\jpsi D)$ is in the $\bar{D}^*$ 
peak~\cite{Abe:2007sya}.  Here the hatched histogram is the non-$\jpsi$ and/or non-$D$-meson
background determined from $\jpsi$ and $D$-meson mass sideband events.  The inset shows the
background-subtracted $M(D \bar{D}^*)$ distribution, which is dominated by a near-threshold peak. A
fit of a BW resonance to these data, shown as a solid curve in the figure, returns a mass and width
of $M=3942\pm 9$~MeV and $\Gamma=37^{+27}_{-17}$~MeV, values that are consistent with results for the 
$X(3940)$ determined from the inclusive $\ee\rt\jpsi X$~missing mass distribution~\cite{Abe:2007jna}.

\subsubsection{$X(4160)\rt D^*\bar{D}^*$}~~~
Figure~\ref{fig:belle_x3940-x4160-1}b shows the the $D^*\bar{D}^*$ invariant mass distribution for
$\ee\rt\jpsi D^* X$ annihilation events where  $M_{\rm recoil}(\jpsi D^*)$ is in the $\bar{D}^*$
mass region. Here the mass-sideband-estimated non-$\jpsi$ and/or non-$D^*$-meson backgrounds are
very small. The curve shows the result of a fit to a single BW resonance term plus a
phase-space-like background.  The fitted mass and width for this peak, which is called the $X(4160)$,
is $M=4156\pm 27$~MeV and $\Gamma= 139^{+113}_{-\ 65}$~MeV~\cite{Abe:2007sya}.

\subsubsection{Discussion}
Neither the $X(3940)$ nor the $X(4160)$ show up in the $D\bar{D}$ invariant mass distribution for
exclusive $\ee\rt\jpsi D\bar{D}$ at the same energies.  Also, as mentioned above, the absence of
signals for any of the known non-zero spin charmonium states in the inclusive spectrum of
Fig.~\ref{fig:x3940-y4260}a provides circumstantial evidence for $J=0$ assignments for the
$X(3940)$ and $X(4160)$.  If the $X(3940)$ has $J=0$, its $D\bar{D}^*$ decay mode ensures that
its $J^{PC}$ quantum numbers are $0^{-+}$. If the $X(4160)$ has $J=0$ the absence of any sign of 
$X(4160)\rt D\bar{D}$ decay supports a $0^{-+}$ assignment for this state as well. In both cases,
the measured masses are far below expectations for the only available unassigned $0^{-+}$ charmonium
levels: the $\eta_c(3S)$ and $\eta_c(4S)$.  Since there are no strong reasons to doubt the generally
accepted identifications of the $\psi(4040)$ peak seen in the inclusive cross section for
$\ee\rt${\rm hadrons} as the $\psi(3S)$ and the $\psi(4415)$ peak as the $\psi(4S)$~\cite{Bai:2001ct},
these assignments would imply hyperfine $n_r^3{\rm S}-n_r^1{\rm S}$ mass splittings that increase from the measured
value of $47.2\pm 1.2$~MeV for $n_r=2$~\cite{Olive:2016xmw}, to $\sim 100$~MeV for $n_r=3$ and $\sim 250$~MeV
for $n_r=4$~\cite{Chao:2007it}.  This pattern conflicts with expectations from potential models, where hyperfine
splittings are proportional to the square of the $\ccbar$ radial wave-function at $r=0$ and
decrease with increasing $n_r$~\cite{Godfrey:1985xj}, and is the main reason that the $X(3940)$ and
$X(4160)$ are considered candidates for non-standard charmonium-like hadrons.

\myclearpage

\section{Charged non-standard hadron candidates}
\label{sec:charged}

Distinguishing neutral candidates for non-standard mesons that decay into quarkonia states, from 
excitations of conventional $\QQbar$ states is a complex task that can be fraught with ambiguities. 
In contrast, charged quarkonium-like candidates are explicitly non-standard and the only outstanding 
issues concern the nature of their internal dynamics. Candidates for both charmonium-like and 
bottomonium-like charged states are discussed in the section.

\subsection{$Z(4430)^+$ and similar structures in $B$ decays}
\label{sec:z4430}

\subsubsection{The $Z(4430)^+\rt\psip\pi^+$ in $B\rt\psip\pip K$ decays}
The first established candidate for a charged charmonium-like state dates back to 2007, when the
$Z(4430)^+$ was observed by the Belle collaboration as a peak in the the invariant mass of the
$\psip\pip$ system in $\bar{B}\to\psip\pip K$ decays~\cite{Choi:2007wga} ($K=K^0_s$ or $K^-$).
The BaBar experiment, with a data sample containing a similar number of $B\rt \psip\pip K$ decay
events, did not find strong evidence for a $Z(4430)^+$ signal that could not be attributed to
reflections from various kaon excitations decaying to $K\pi^+$ that were analyzed in a
model-independent way~\cite{Aubert:2008aa}.  However, the BaBar results were not sensitive enough to
directly contradict the Belle observation.  Subsequently, the Belle collaboration reanalyzed their
data with an amplitude model that combined coherent $K\pi^+$ and $\psip\pi^+$ resonant contributions
to fit to the data distribution across a two-dimensional ($M^2(K\pi)$~{\em vs.}~$M^2(\psip\pi)$) Dalitz
plane~\cite{Mizuk:2009da}. This was later refined to include two additional kinematic variables that
were angles that describe $\psip\to\ell^+\ell^-$ decays~\cite{Chilikin:2013tch}. Both analyses 
reaffirmed Belle's original claim for a significant $Z(4430)^+$ signal, albeit with a substantially
larger mass and total width than the values given in the initial Belle report, which was based on a na\"ive fit to the
$\psip\pi^+$ mass distribution. The latter, four-dimensional amplitude analyses favored a $J^P=1^+$
spin-parity of the $Z(4430)^+$ over other possible $J^P$ assignments at the $3.4\sigma$ level. 

\begin{figure}[bthp]
  \includegraphics*[width=0.45\textwidth]{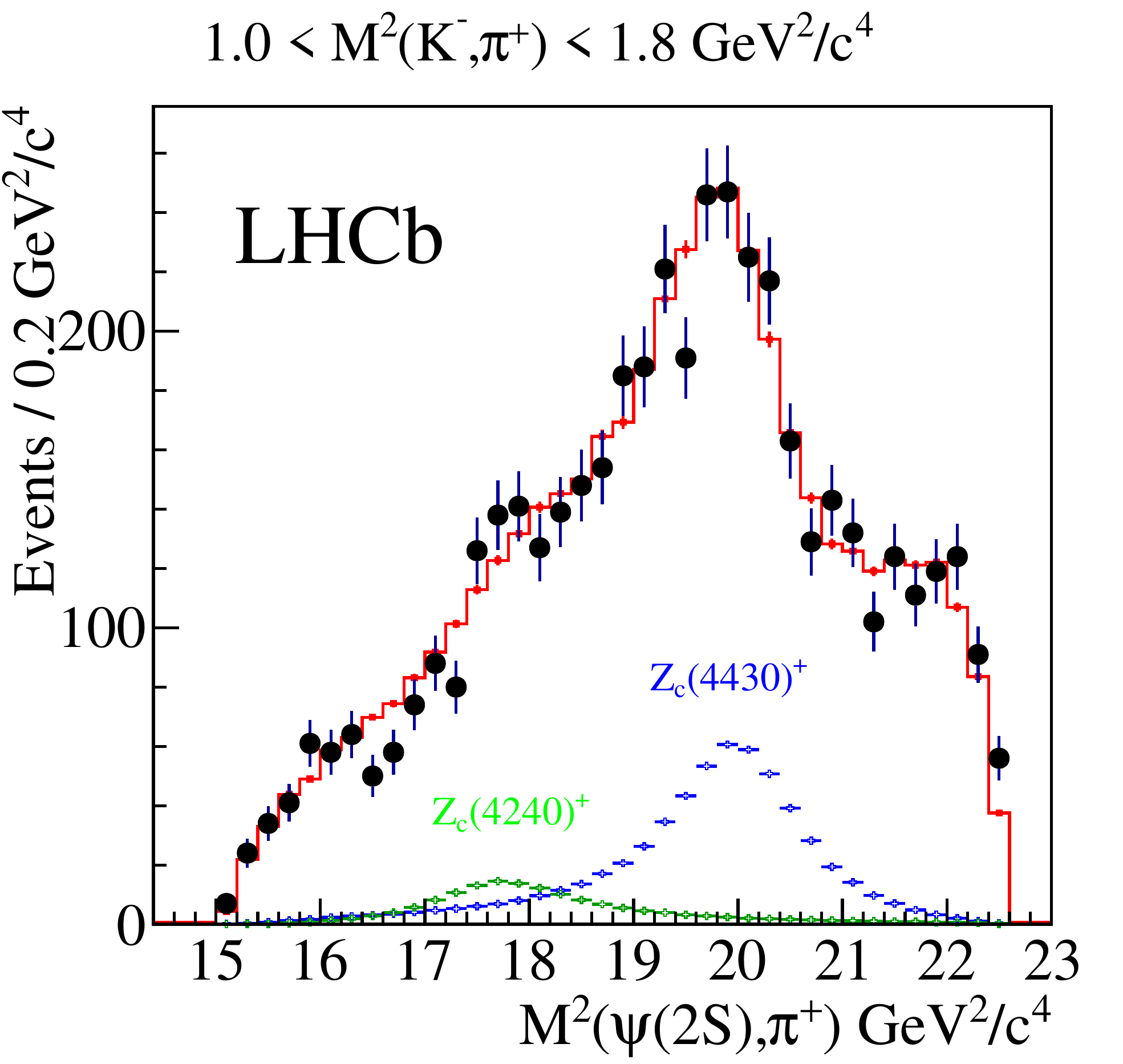}
\quad\\[-0.5cm]
  \hbox{ \includegraphics*[width=0.60\textwidth]{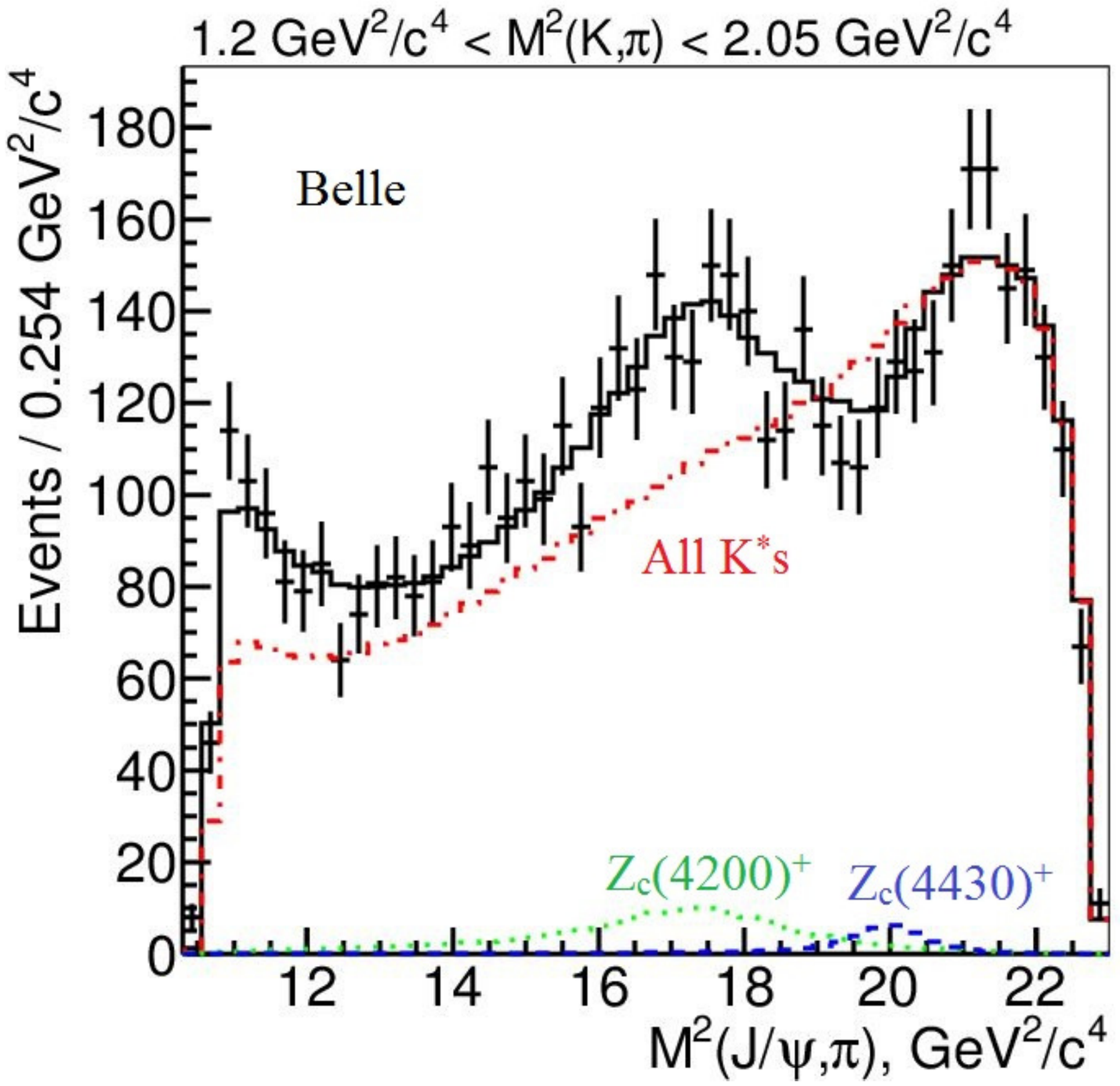} \quad\hskip-3.6cm}
\vskip-0.6cm
\caption{
The points with errors show distributions of {\it (upper)} $M^2(\psip\pi^+)$ 
in $\bar{B}^0\to\psip\pi^+K^-$ decays from LHCb \protect\cite{Aaij:2014jqa} 
and {\it (lower)} $M^2(\jpsi\pi^+)$ in $\bar{B}^0\rt \jpsi \pip K^-$ from
Belle \protect\cite{Chilikin:2014bkk}.
Here only events with $K^-\pi^+$ invariant masses that are between the 
$K^*(892)^0$ and $K^*_2(1430)^0$ resonances are included in order to
suppress contributions from the $K\pi$ channel.  Projections of four-dimensional
amplitude fits that include coherent contributions from kaon excitations and two $Z^+$ 
terms are superimposed as solid-line histograms. The
individual $Z^+$ terms are shown as blue and green points; the dashed red
curve in the Belle plot shows the projection of all the $K^-\pi^+$ terms combined.
  \label{fig:BtoZc}
}
\end{figure}

The existence of the $Z(4430)^+$ structure was independently confirmed in 2014 (with $13.9\sigma$ significance)
by the LHCb experiment~\cite{Aaij:2014jqa}, 
which was based on a four-dimensional analysis of a $\bar{B}^0\rt\psip\pip K^-$ event sample that was an order of
magnitude larger than those used in the Belle and BaBar experiments (see Fig.~\ref{fig:BtoZc}(upper)). The LHCb
amplitude analysis yielded results that were consistent with the Belle determination, including the confirmation
of the $J^P=1^+$ assignment, but in this case at the $9.7\sigma$ level.  The average of the Belle and LHCb mass
and width values are~\cite{Olive:2016xmw}
\begin{eqnarray}
M(Z(4430))      &=& 4478^{+15}_{-81}~{\rm MeV} \nonumber \\
\Gamma(Z(4430)) &=& 181 \pm 31~{\rm MeV}.
\label{eqn:z4430-m-w}
\end{eqnarray}

\begin{figure}[bthp]
  \quad\hskip-0.8cm 
  \includegraphics*[width=0.25\textwidth]{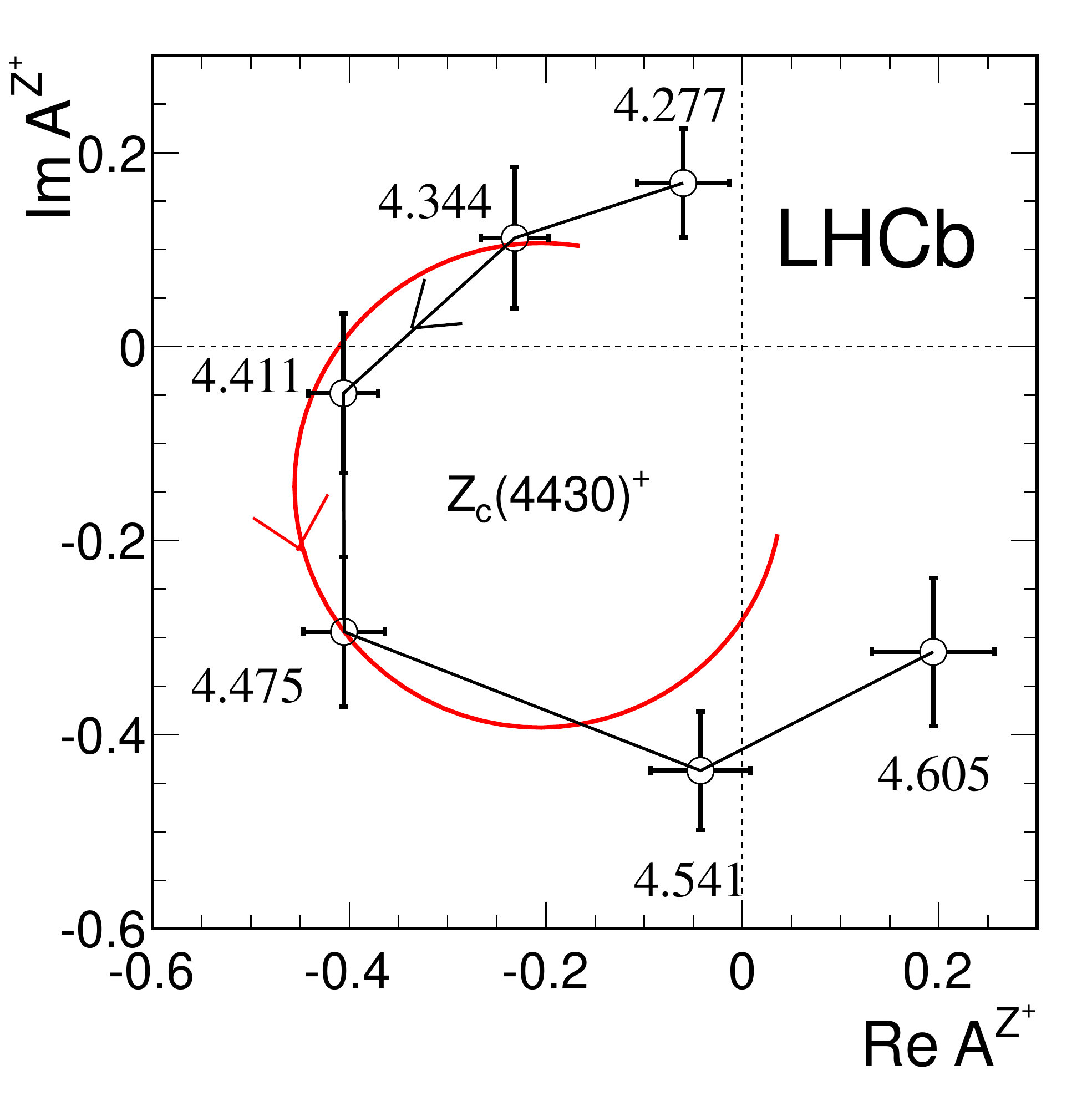}
  \quad\hskip-0.9cm 
  \hbox{ \includegraphics*[width=0.335\textwidth]{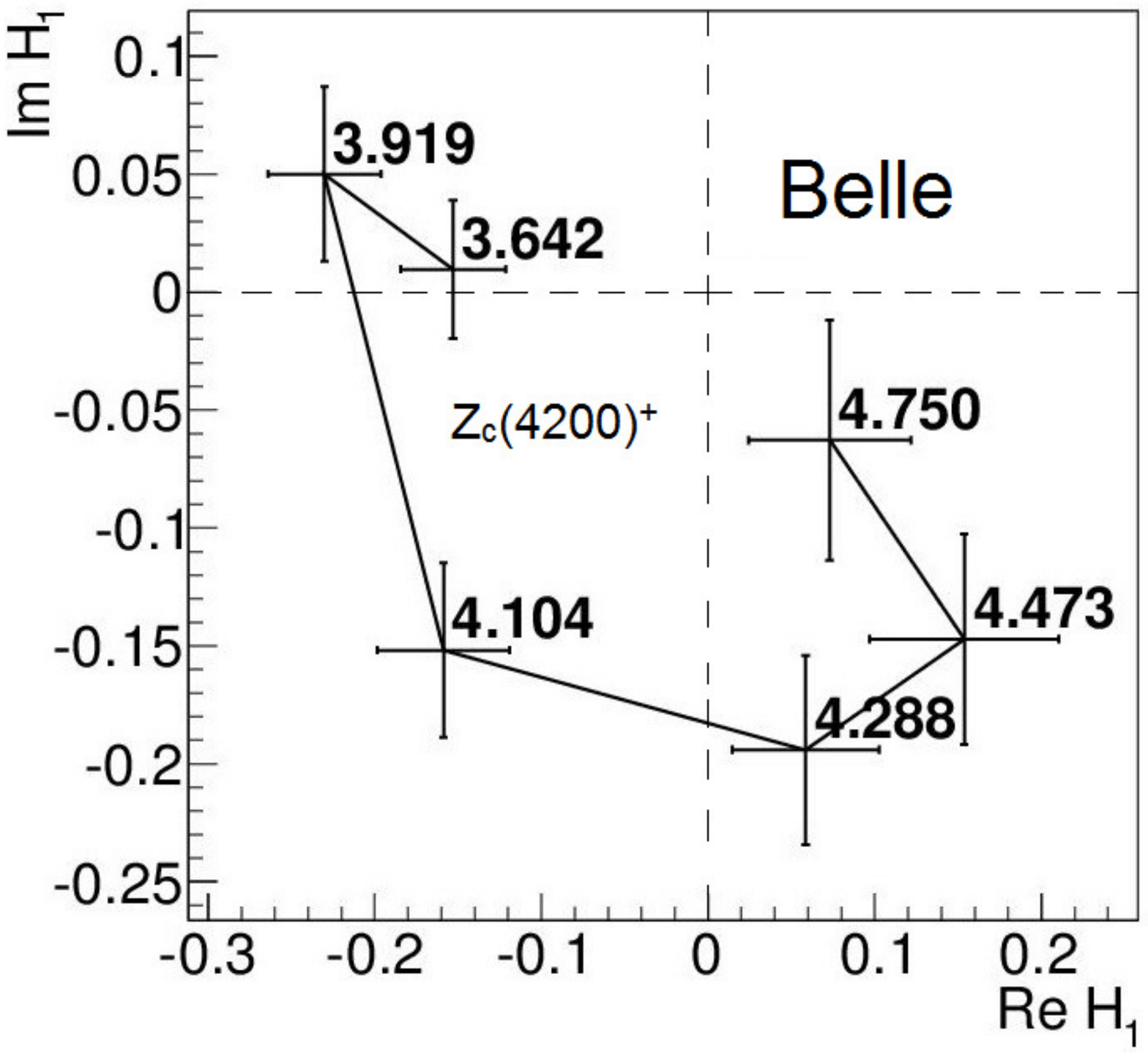} 
  \quad\hskip-2.0cm } 
\caption{
Fitted values of the real and imaginary parts 
of the amplitude for {\it (left)} LHCb's $Z(4430)^+\rt\psip\pip$ signal \protect\cite{Aaij:2014jqa} 
and {\it (right)} Belle's $Z(4200)^+\rt\jpsi\pip$ signal \protect\cite{Chilikin:2014bkk},
for six $M^2(\psi,\pi^+)$ bins of equal width that span the resonance.
The solid red curve in the LHCb plot shows the pattern expected for a Breit-Wigner resonance amplitude.
Units are arbitrary. 
  \label{fig:ArgandZ}
}
\end{figure}

The large event sample enabled the LHCb
group to measure the $1^+$ $Z(4430)^+$ amplitude's dependence on $\psip\pi^+$ mass independently of any
assumptions about the resonance line shape. The resulting ``Argand diagram'' of the real {\em vs.} imaginary parts
of the $1^+$ amplitude, shown in Fig.~\ref{fig:ArgandZ}(left), shows a nearly circular, counter-clockwise motion
with an abrupt change in the amplitude phase at the peak of its magnitude that is characteristic of a BW resonance
amplitude.  This diagram rules out an interpretation of the $Z(4430)$ peak as being due to the effects of a
rescattering process proposed by Pakhlov and Uglov~\cite{Pakhlov:2014qva} that predicted a clock-wise phase motion.
Since other rescattering mechanisms or coupled-channel cusps could produce a counter-clockwise phase motion similar
to that of a BW resonance, higher statistics studies of the Argand diagram are needed to probe the
amplitude-dependence on mass at a level of detail that is fine enough to distinguish a resonance pole from other types
of meson-meson interactions. 

\begin{figure}[bthp]
\includegraphics*[width=0.5\textwidth]{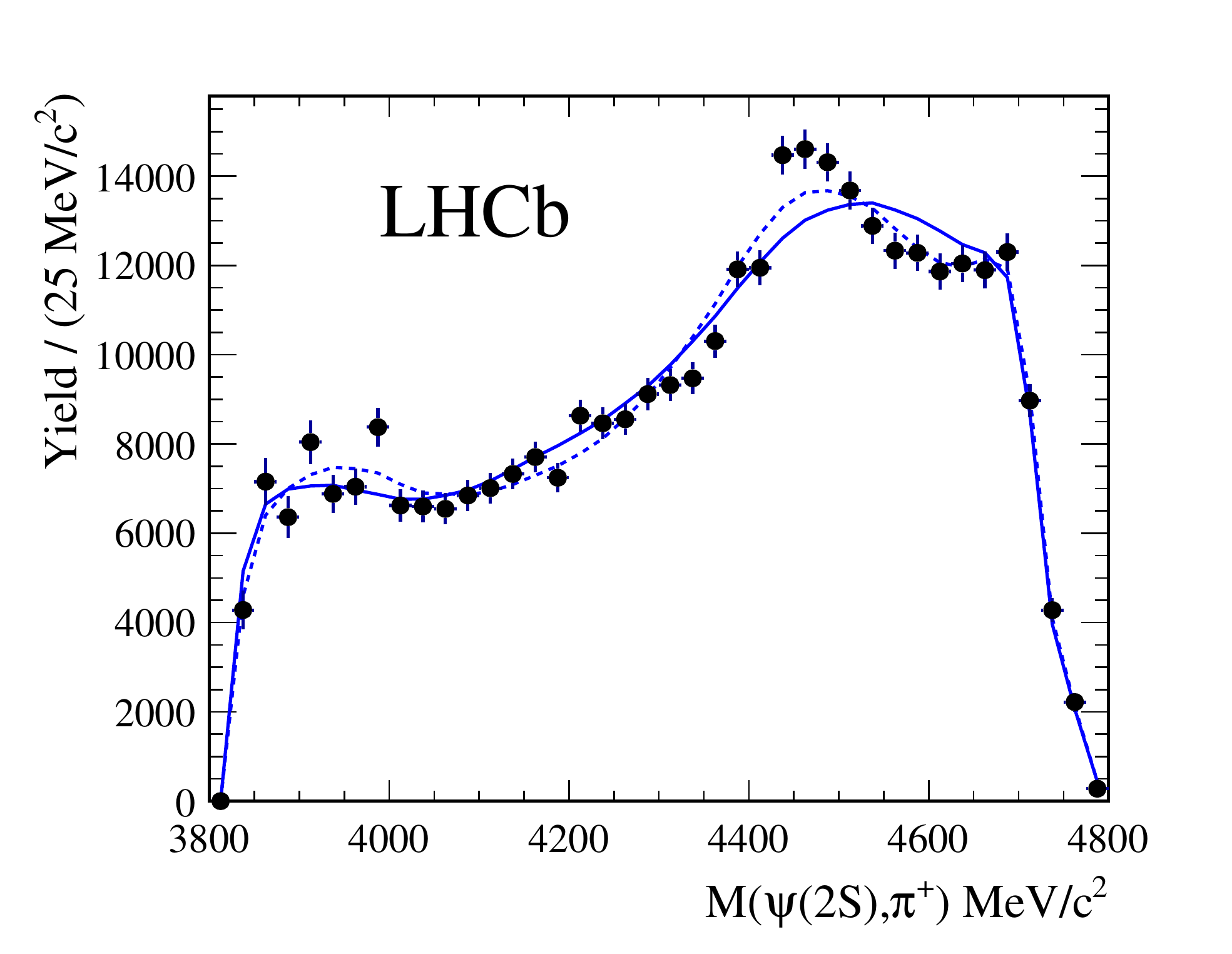}
\caption{
The dots with error bars show background-subtracted and efficiency-corrected $M(\psip\pip)$ distribution
for $\bar{B}^0\to\psip\pip K^-$ events in LHCb \protect\cite{Aaij:2015zxa}.  The solid (dashed) blue lines
correspond to contributions from the $K^-\pip$ channel for a  maximum-allowed angular momentum
of $J_{\rm max}=2$ ($J_{\rm max}=3$). Since the lightest known $J=3$ resonance, the $K_3^*(1780)$, is
already beyond the kinematically allowed $K\pi$ mass limit for $\bar{B}\rt\psip\pi K$ decay,
no plausible contribution from the $K\pi$ channel can account for the shape of the $M(\psip\pi)$
distribution in the $Z(4430)^+$ mass region. 
  \label{fig:miz4430}
}
\end{figure}

The LHCb collaboration also performed an analysis of their $\bar{B}^0\to\psip\pip K^-$ events 
using a $K^-\pi^+$ model-independent approach~\cite{Aaij:2015zxa} 
similar to the one that was earlier performed by the BaBar collaboration~\cite{Aubert:2008aa}. 
This approach yielded conclusive results that demonstrate the
requirement for non-$K\pi$ contributions to $\bar{B}^0\to\psip\pip K^-$ decays at the $8\sigma$
level, as shown in Fig.~\ref{fig:miz4430}.  While this approach demonstrates the need for contributions
from a non-$K\pi$ source in the $Z(4430)^+$ mass region, it does not provide any independent way to
extract any of the characteristics of these contributions; for this, an amplitude analysis approach
is necessary.  

\subsubsection{The $Z(4200)^+\rt\jpsi\pip$ in $\bar{B}^0\rt\jpsi\pip K^-$ decays}
The Belle collaboration performed an amplitude analysis of $\bar{B}^0\to\jpsi \pi^+K^-$ 
decays \cite{Chilikin:2014bkk} and found that in this channel too, the data could not be well
described solely with contributions from the $K\pi$ channel.  A satisfactory fit was obtained only
after contributions from two $\jpsi\pi^+$ resonances were included: one corresponding to a very broad
$1^+$ $Z(4200)^+$ state with width $\Gamma=370\pm70^{+\phantom{0}70}_{-132}$~MeV, 
mass $4196\,^{+31}_{-29}\,^{+17}_{-13}$ MeV, 
and a significance of $6.2\sigma$, and the second corresponding to $\jpsi\pi$
decay mode of the $Z(4430)$.  
The analysis showed that the $Z(4200)$
interferes destructively with the $Z(4430)^+\rt\jpsi\pip$ amplitude, producing a dip in the
$M^2(\jpsi\pip)$ distribution near the $Z(4430)$ peak mass, as shown in Fig.~\ref{fig:BtoZc}(lower).
In the analysis, the mass and width of the $Z(4430)\rt\jpsi\pip$ BW amplitude were fixed
at the values determined from the $\psip\pip$ analysis. The statistical significance (not
including systematic errors) of the $Z(4430)\rt\jpsi\pip$ amplitude was determined to be
$5.1\sigma$, and the magnitude of the $Z(4430)\rt\jpsi\pi$ term
was found to be much smaller than that for $Z(4430)\rt\psip\pi$ in spite of the larger available
phase-space. The $Z(4200)^+$ state awaits independent confirmation, although some indication for it
in $\psi(2S)\pi^+$ decays may have been seen in the LHCb $B\rt\psip\pi K$ analysis~\cite{Aaij:2014jqa},
where they reported evidence for a state in this mass region with either $0^-$  or $1^+$ 
quantum numbers that is shown by the green points in 
Fig.~\ref{fig:BtoZc}(left).\footnote{Using a $0^-$ hypothesis, the LHCb obtained a mass of 
$4239\pm18^{+45}_{-10}$ MeV and a width of $220\pm47^{+108}_{-\phantom{0}74}$ MeV, which are 
consistent with the Belle results but cannot be averaged with them since they are obtained 
using different $J^P$ assignments. For the $1^+$ hypothesis, LHCb only reported a width, 
 $660\pm150$ MeV, and that without a systematic uncertainty.}

The Belle collaboration presented Argand diagrams for the $J^P=1^+$~$\jpsi\pip$ amplitude for two
helicity amplitudes. One of them, shown in Fig.~\ref{fig:ArgandZ}(right), displays a 
nearly circular phase motion
that is consistent with expectations for a BW resonance amplitude.

\begin{figure}[bthp]
\hbox{ \includegraphics*[width=0.6\textwidth]{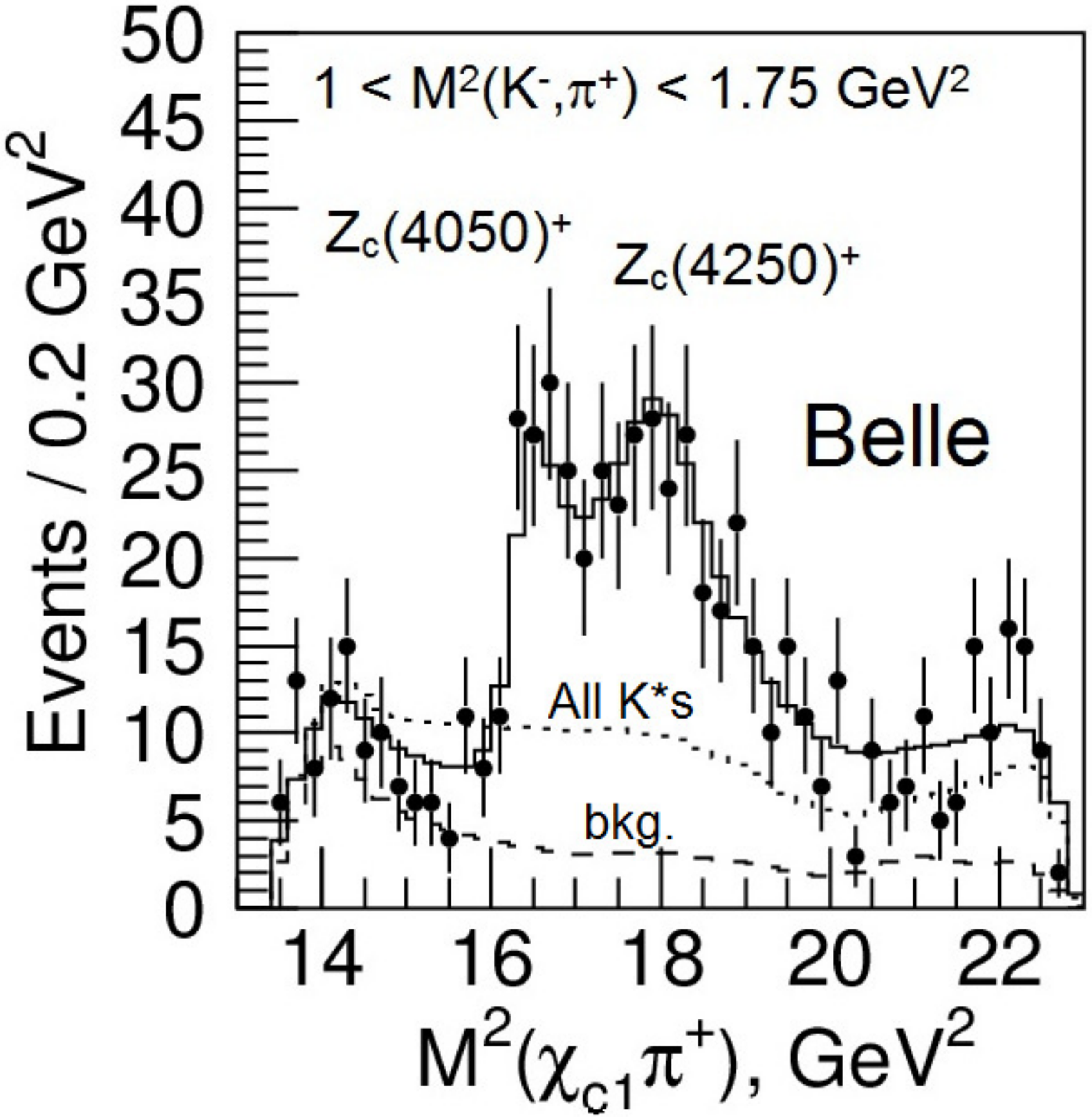} \quad\hskip-1cm }
\caption{
The points with error bars show the $M^2(\chi_{c1}\pip)$ distribution for
in $\bar{B}^0\to\chi_{c1}\pip K^-$ decays detected in Belle \protect\cite{Mizuk:2008me}.
The projection of a two-dimensional amplitude fit that included kaon excitations and
two $Z^+$ terms are superimposed as a solid-line histogram.  The dotted line corresponds
to all $K^-\pi^+$ terms combined and the non-$B$ and/or non-$\chi_{c1}$ background contribution is
shown by the dashed line.
\label{fig:chicpi}
}
\end{figure}

\subsubsection{Charged $\chi_{c1}\pip$ resonances in $\bar{B}^0\rt\chi_{c1}\pip K^-$ decays}
The Belle collaboration also reported evidence for charged $\chi_{c1}\pip$
resonances in a two-dimensional ($M^2(K\pi)$~{\em vs.}~$M^2(\chi_{c1}\pi)$) Dalitz plot
analysis of $\bar{B}^0\to\chi_{c1}\pip K^-$ decays.   The data could not be fitted with resonances in
the $K\pi$ channel only, and was best described when two $\chi_{c1}\pip$ resonances, the $Z(4050)^+$
and $Z(4250)^+$, were included, as shown in Fig.~\ref{fig:chicpi}~\cite{Mizuk:2008me}.
BaBar saw an enhancement in the same mass region, but could account for it with reflections from
$K\pi^+$ resonances analyzed with the model-independent method  described above~\cite{Lees:2011ik};
their results neither confirmed nor contradicted the Belle results.  These two candidate
$\chi_{c1}\pi^+$ resonant states still await independent confirmation and a complete amplitude analysis
that spans all 6 dimensions of the decay phase-space and determines their quantum numbers. 

\subsubsection{Discussion}
\label{sec:z4430disc}
The neutral $X(3872)$, $X(3915)$ and $Y(4260)$ charmonium-like states discussed in Section~\ref{sec:neutral}
showed up as distinct, relatively narrow peaks on a small background. In these cases, fitting the peaks with
simple BW line shapes and ignoring the effects of possible signal interference with a coherent components
of the non-resonant backgrounds were reasonable approximations.  However, for the charged charmonium states 
discussed in this subsection, this approximation is no longer valid. Here, since the states are broad and
contributions from coherent non-resonant processes are substantial, interference effects distort the signal
line shapes to such an extent that they no longer resemble that of a standard BW resonance peak.

\begin{figure}[bthp]
\hbox{ \includegraphics*[width=0.45\textwidth]{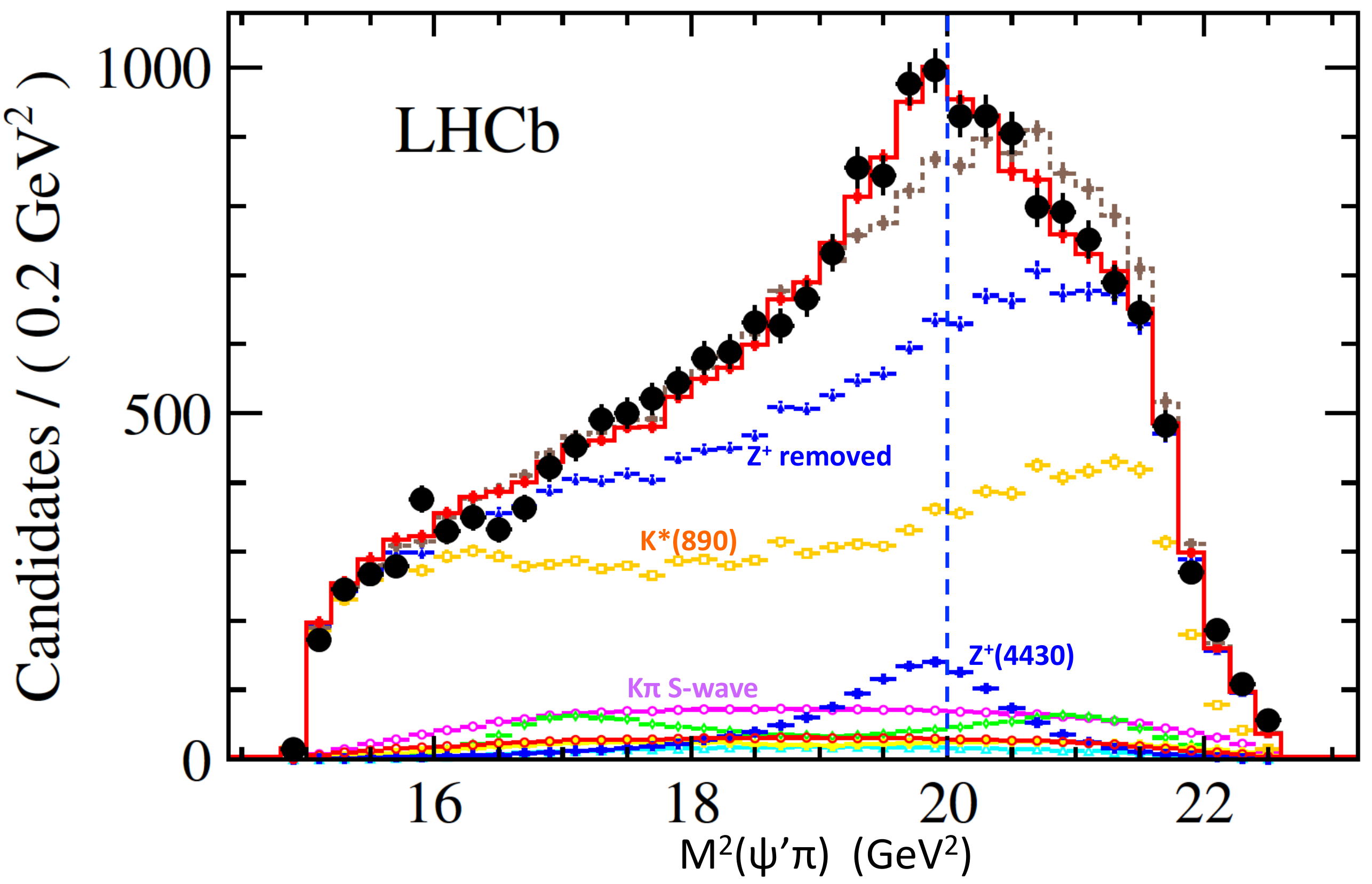} \quad\hskip-1cm }
\caption{
The black points show the $M^2(\psip\pip)$ distribution for all of the LHCb $\bar{B}^0\to\psip\pip K^-$
events \protect\cite{Aaij:2014jqa}. The solid-red histogram is the
projection of the four-dimensional fit that includes the $Z(4430)^+$ amplitude and the
dashed-brown histogram shows the best fit that was found with no $\psip\pip$ resonances.
Contributions from individual fit components are shown, with the dominant ones labeled.
The upper blue points show the final fit results with the $Z^+$ terms removed. The vertical dashed-blue
line indicates the fitted $Z^+$ resonance mass value.
\label{fig:LHCb_z4430_fit}
}
\end{figure}
 
This is illustrated for the case of the LHCb group's $Z(4430)^+\rt\psip\pip$ analysis~\cite{Aaij:2014jqa}
in Fig.~\ref{fig:LHCb_z4430_fit}, where the black data points show the $M^2(\psip\pip)$ distribution for
all of their selected $\bar{B}^0\rt\psip\pip K^-$ events (with no $K\pi$ mass selection). The solid red
histogram shows the results of the final LHCb four-dimensional fit that included a $Z(4430)^+$ amplitude
and the upper blue points show that fit's results with the all terms involving the $Z(4430)^+\rt\psip\pip$
removed. The difference between the red histogram and the upper blue points is the total $Z^+$ contribution
to the fit, including effects of interference with amplitudes in the $K\pi$ channel. The shape of the total 
$Z^+$ contribution is much different than that of the $Z^+$ term alone, shown as the lower blue points, because
of strong interference effects that are constructive on the low mass side of the $Z^+$ resonance and switch to
destructive at higher masses, reflecting the abrupt phase change that occurs at the peak of a BW resonance
amplitude. The Belle group's original $Z(4430)^+$ results were based on a na\"ive BW line-shape fit to the
visible peak, which only corresponds to the lower lobe of the actual pattern. As a result, they reported
low values for the mass and width~\cite{Choi:2007wga}.

While the presence of coherent non-resonant processes complicates the extraction of resonance signals, it
provides the possibility of measuring the signal amplitude's phase motion across the resonance (see
Fig.~\ref{fig:ArgandZ}), thereby providing valuable information that would otherwise be inaccessible.

\myclearpage

\subsection{Charged $Z_b^+$ and $Z_c^+$ states produced in $\ee$ processes}
\label{sec:z3900zb}

\subsubsection{The $Z_b$ charged bottomonium-like mesons}

\begin{figure} 
\includegraphics[width=0.48\textwidth]{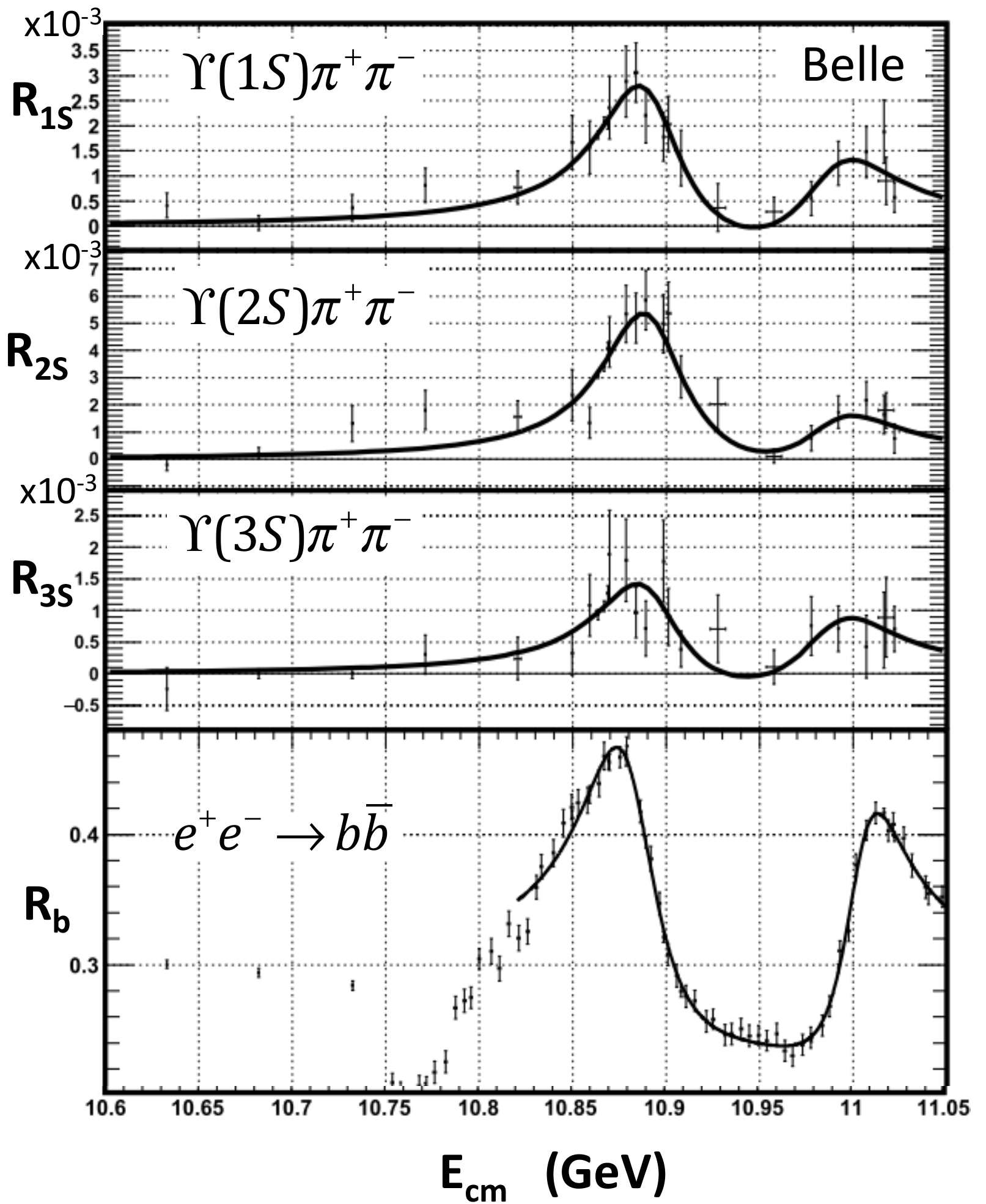}
\caption{\footnotesize 
Cross sections for $\ee\rt \pipi\Upsilon (n_r{\rm S})$ ($n_r=1,2,3$) ({\it upper}) and 
$\ee\rt b\bar{b}$ {\it (lower)} in units of the Born QED cross section for $\ee\rt\mumu$
($\sigma_{\rm QED}(\ee\rt\mumu)=(4\pi\alpha^2)/(3 \ecm^2)$) in the vicinity of the
$\Upsilon(5{\rm S})$ resonance (from ref.~\cite{Santel:2015qga}). }
\label{fig:Rbb-Rpipiups}
\end{figure}

The large $Y(4260)\rt\pipi\jp$ signal discovered in the charmonium
mass region by BaBar motivated a Belle search for similar behavior
in the bottomonium system~\cite{Hou:2006it}. This uncovered anomalously large
$\pipi\Upsilon(n_r{\rm S})$ $(n_r=1,2,3)$ production rates that peak around
$\ecmx{\ee}=10.89$~GeV as shown in the upper three panels of
Fig.~\ref{fig:Rbb-Rpipiups}~\cite{Santel:2015qga}. This peak energy is close to 
a peak in the $\ee\rt b\bar{b}$ cross section near $\ecm\simeq 10.87$~GeV, 
shown in the lower panel of Fig.~\ref{fig:Rbb-Rpipiups}, that is usually associated
with the conventional $\Upsilon(5{\rm S})$ bottomonium meson. 

If the peaks in the $\pipi\Upsilon(n_r{\rm S})$
cross sections are attributed to $\Upsilon(5{\rm S})$ decays, it implies
$\Upsilon(5{\rm S})\rt \pipi\Upsilon(n_r{\rm S})$ ($n_r=1,2,3$) partial widths that are 
two orders of magnitude larger than theoretical predictions~\cite{Abe:2007tk}, and
the measured value of the $\Upsilon(4{\rm S})$ width \cite{Olive:2016xmw}.\footnote{For example, the
PDG average value of $\Upsilon(4{\rm S})$ branching fraction to $\pipi\Upsilon(1{\rm S})$ measurements is
${\mathcal B}(\Upsilon(4{\rm S})\rt\pipi\Upsilon(1{\rm S})=(8.1\pm 0.6)\times 10^{-5}$~\cite{Olive:2016xmw}.
In contrast, the Belle measurement for the peak near 10.86~GeV is more than 50~times larger, 
${\mathcal B}(\Upsilon(10860)\rt\pipi\Upsilon(1{\rm S}))=(5.3\pm 0.6)\times 10^{-3}$~\cite{Chen:2008xia}.}
This suggests that {\it either} the peak in the $\ee$ annihilation cross
section near $\ecm= 10.87$~GeV that has long been identified as the $\Upsilon(5{\rm S})$
$\bbbar$ bottomonium state~\cite{Besson:1984bd} is not a standard $\bbbar$ meson but,
instead, a $b$-quark-sector equivalent of the $Y(4260)$~\cite{Ali:2009pi}, {\it or} there is an
overlap of the conventional $\Upsilon(5{\rm S})$ with a nearby $b$-quark-sector equivalent of the $Y(4260)$,
{\it or} the $\Upsilon(5{\rm S})$ experiences some dynamical effects that have little or no influence on
the $\Upsilon(4{\rm S})$. We follow the PDG and refer to this peak as the $\Upsilon(10860)$.

\begin{figure} 
\includegraphics[width=0.48\textwidth]{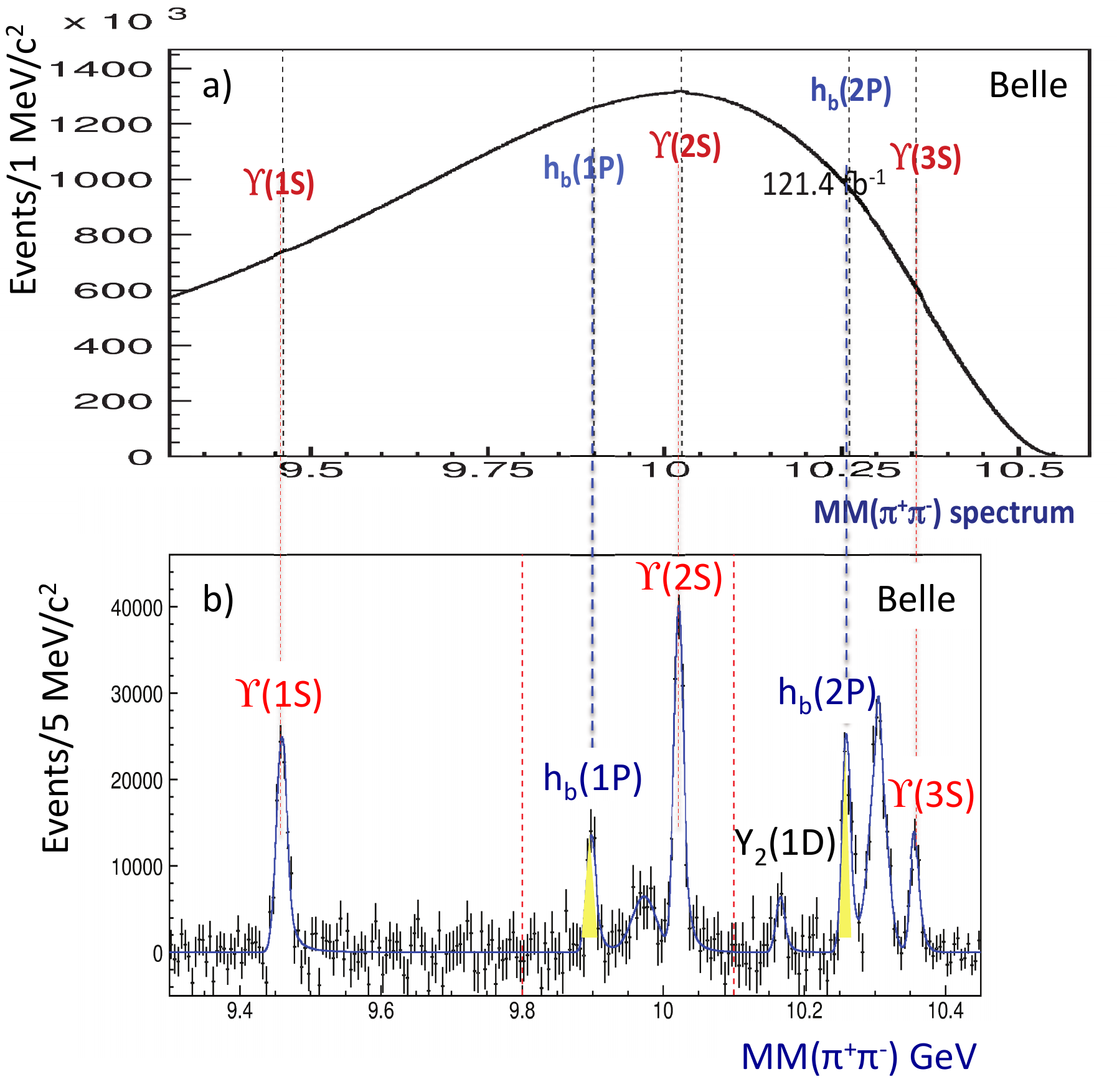}
\caption{\footnotesize 
{\bf a)} Distribution of masses
recoiling against $\pipi$ pairs at c.m.~energies near 
$10.87$~GeV and
({\bf b)} residuals from piece-wise fits to the data with
smooth polynomials  (from ref.~\cite{Adachi:2011ji}).~The $h_b(1{\rm P})$
and $h_b(2{\rm P})$ peaks, shaded in yellow, were the first observations of these two states.}
\label{fig:belle_pipi-recoil}
\end{figure}

Belle accumulated a large sample of data at and near the energy of the $\Upsilon(10860)$
mass peak ($\ecmx{\ee}=10.866$~GeV) in order to investigate the source of this
anomaly, 121.4~fb$^{-1}$ in total.  Figure~\ref{fig:belle_pipi-recoil}a shows
the distribution of masses recoiling against all of the $\pipi$ pairs in these
events~\cite{Adachi:2011ji}. The combinatoric
background is huge -- there are typically $10^6$ entries in each
1~MeV bin -- and the statistical errors are tiny ($\sim 0.1\%$).~The
data were fit piece-wise with sixth-order polynomials and the
residuals from the fits are shown in the lower panel of Fig.~\ref{fig:belle_pipi-recoil}b,
where, in  addition to peaks at the $\Upsilon(1{\rm S})$, $\Upsilon(2{\rm S})$,
$\Upsilon(3{\rm S})$ masses and some expected reflections, there are
unambiguous signals for the $h_b(1{\rm P})$ and $h_b(2{\rm P})$, the $1^1{\rm P}_1$
and $2^1{\rm P}_1$ bottomonium states. This was the first observation of
these two elusive levels~\cite{Adachi:2011ji}. One puzzle is that the $\pipi
h_b(m_r{\rm P})$, ($m_r=1,2$) final states are produced at rates that are
nearly the same as those for $\pipi\Upsilon(n_r{\rm S})$ ($n_r=1,2,3$), even
though the $\Upsilon(5{\rm S})\rt\pipi h_b$ transition requires a heavy-quark spin flip
that is expected to result in a strong suppression ~\cite{Bondar:2011ev}.

\begin{figure*} 
\includegraphics[width=0.8\textwidth]{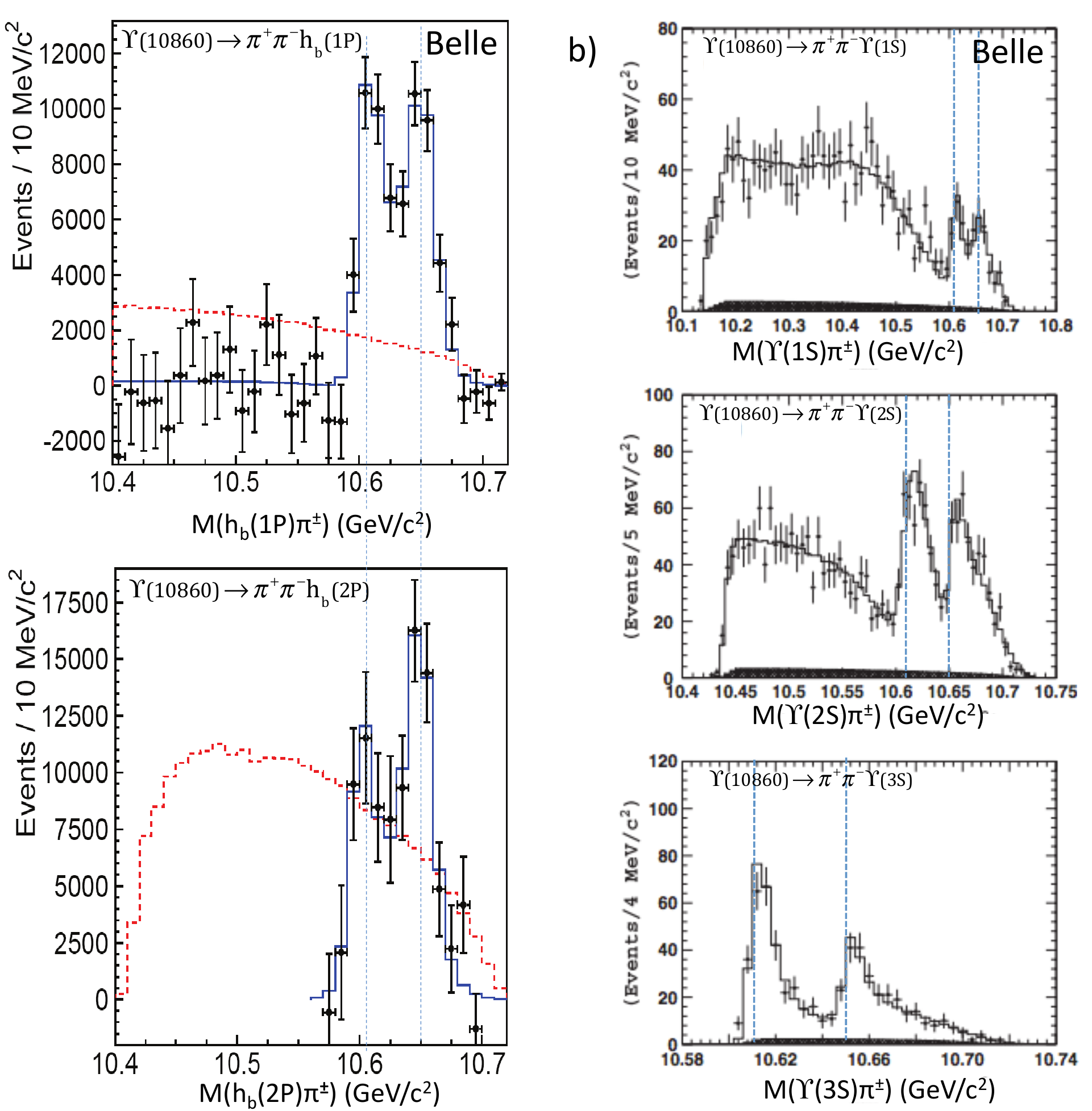}
\caption{\footnotesize 
{\bf a)} Invariant mass
distributions for $h_b(1{\rm P})\pip$ ({\it upper}) and $h_b(2{\rm P})\pip$
({\it lower}) from $\ee\rt\pipi h_b(n_r{\rm P})$ events.
~{\bf b)} Invariant
mass distributions for $\Upsilon(1{\rm S})\pip$ ({\it upper}), $\Upsilon(2{\rm S})\pip$
({\it center}) and $\Upsilon(3{\rm S})\pip$ ({\it lower}) in
$\ee\rt\pipi\Upsilon(n_r{\rm S})$ events.  The vertical dashed lines in each panel
indicate the mass values given in Eq.~\ref{eqn:zb_mass_and_width}. 
The figures are from ref.~\cite{Belle:2011aa}.}
\label{fig:belle_mhbpi-upspi}
\end{figure*}

Figure~\ref{fig:belle_mhbpi-upspi}a show the $\pipi h_b$ yields {\em vs.} the maximum $h_b
\pi^{\pm}$ invariant mass for $h_b=h_b(1{\rm P})$ (upper) and $h_b=h_b(2{\rm P})$ (lower), where it can be
seen that essentially all of the $\pipi h_b$ events are associated with the production of an
$h_b\pi$ system with an $M(h_b\pi)$ value near either $10\,610$~MeV or $10\,650$~MeV~\cite{Belle:2011aa}.
Studies of fully reconstructed $\pipi \Upsilon(n_r{\rm S})$, ($n_r=1,2,3)$
$\Upsilon(n_r{\rm S})\rt\ell^+\ell^-$ events in the same data sample found
$\Upsilon(n_r{\rm S})\pi$ mass peaks at the same masses in the $M_{\rm max}(\Upsilon(n_r{\rm S})\pi)$
distributions for all three narrow $\Upsilon(n_r{\rm S})$ states; these are shown
in the three panels of Fig.~\ref{fig:belle_mhbpi-upspi}b.  Here the fractions of
$\pipi\Upsilon(n_r{\rm S})$ events in the two peaks are substantial -- $\sim6\%$ for the
$\Upsilon(1{\rm S}) $, $\sim 22\%$ for the $\Upsilon(2{\rm S})$ and $\sim 43\%$
for the $\Upsilon(3{\rm S})$ -- but, unlike the case for the $\pipi
h_b(m_r{\rm P})$ channels, they only account for a fraction of the anomalous
$\pipi\Upsilon(n_r{\rm S})$ event yield~\cite{Garmash:2014dhx}.  Thus, the production and decays
of the two $Z_b$ states can explain some, but not all of the anomalously large
$\Upsilon(5{\rm S})\rt \pipi\Upsilon(1,2,3{\rm S})$ decay rates.  The fitted values of the peak
masses (indicated by the vertical dashed lines in each panel of Fig.~\ref{fig:belle_mhbpi-upspi})
and widths in all five channels are consistent with each
other; the weighted average mass and width values of the two
peaks, dubbed the $Z_b(10610)$ and $Z_b(10650)$, are
\begin{eqnarray}
Z_b(10610):~~M_1 &=& 10\,607\pm 2~{\rm MeV} \nonumber \\
        \Gamma_1 &=& 18.4\pm 2.4~{\rm MeV} \nonumber \\
Z_b(10650):~~M_2 &=& 10\,652\pm 2~{\rm MeV} \nonumber \\
        \Gamma_2 &=& 11.5\pm 2.2~{\rm MeV},  \label{eqn:zb_mass_and_width}
\end{eqnarray}
\noindent
respectively. A Belle study of $\piz\piz\Upsilon(n_r{\rm S})$,
$(n_r=1,2,3)$ found a $6.5\sigma$ signal for the neutral
$Z_b(10610)^0$ isospin partner state with a mass $M(Z_b(10610)^0)=
10\,609\pm 6$~MeV and a production rate that is consistent with
isospin-based expectations~\cite{Krokovny:2013mgx}.

The $Z_b(10610)$ mass is only $2.6\pm 2.2$~MeV above the
$m_B+m_{B^*}$ mass threshold and the $Z_b(10650)$ mass is only
$2.0\pm 1.6$~MeV above $2m_{B^*}$.  Dalitz-plot analyses of the
$\pipi\Upsilon(n_r{\rm S})$ final states establish $J^P=1^+$ quantum number
assignments for both states~\cite{Garmash:2014dhx}. The close proximity of the
$Z_b(10610)$ and $Z_b(10650)$ to the $B\bar{B}^*$ and $B^*\bar{B}^*$ thresholds, respectively,
and the $J^P=1^+$ quantum number assignment suggests that they may be virtual S-wave
molecule-like states~\cite{Bondar:2011ev}.

\begin{figure} 
\includegraphics[width=0.48\textwidth]{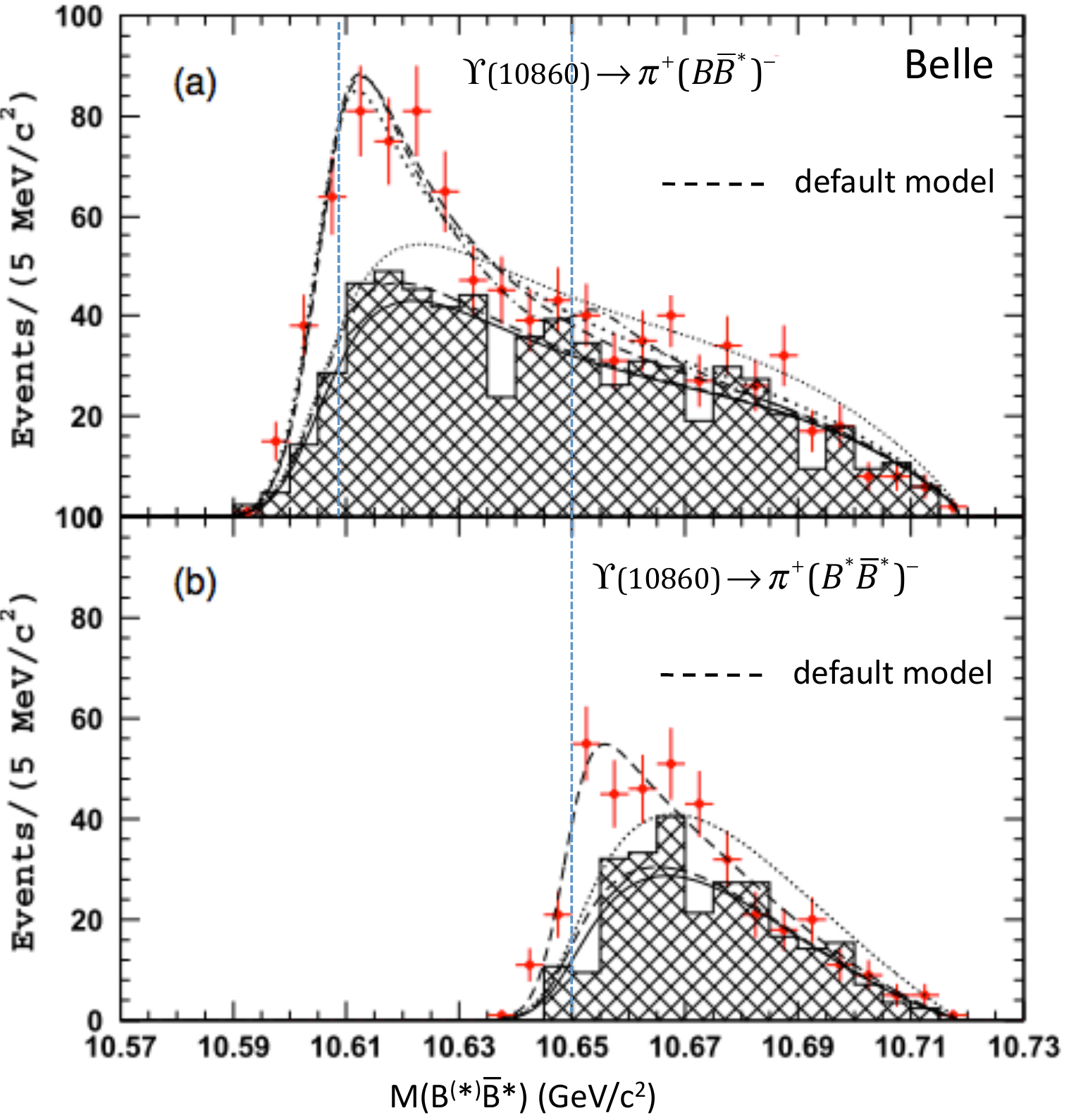}
\caption{\footnotesize 
{\bf a)} The $M(B\bar{B}^*)$ distribution for $\pip B\bar{B}^*$ events
and {\bf b)} the $M(B^*\bar{B}^*)$ distribution for $\pip B^*\bar{B}^*$ events.
The short-dashed curves show results of fits with only a $Z_b(10610)\rt B\bar{B}^*$
contribution to {\bf a)} and a $Z_b(10650)\rt B^*\bar{B}^*$ contribution the {\bf b)}. 
The other curves and the hatched (background) histogram are described in the text.   
The figures are from ref.~\cite{Garmash:2015rfd}.}
\label{fig:belle_Zb_2_BB}
\end{figure}

The $B^{(*)}\bar{B}^{*}$ molecule picture is supported by a Belle
study of $\Upsilon(10860)\rt \pi B^{(*)}\bar{B}^*$ final states in the same data
sample~\cite{Garmash:2015rfd}, where the the pion and one $B$-meson is reconstructed and
the presence of the accompanying $\bar{B}$ and the distinction between $\pi B\bar{B}^*$
and $\pi B^*\bar{B}^*$ are inferred from energy-momentum conservation; the $B\bar{B}^*$
and $B^*\bar{B}^*$ invariant masses are inferred from the pion momentum.
The data points in Figures~\ref{fig:belle_Zb_2_BB}a and 
\ref{fig:belle_Zb_2_BB}b show the $B\bar{B}^*$ and $B^*\bar{B}^*$
invariant mass distributions, respectively, where the background, mostly from continuum
$\ee\rt\ccbar$ events and estimated from events where the pion charge and the flavor of the
detected $B$ meson do not match, is shown as hatched histograms. 

The $M(B\bar{B}^*)$ distribution (Fig.~\ref{fig:belle_Zb_2_BB}a) has a distinct peak near the
mass of the $Z_b(10610)$ and the $M(B^*\bar{B}^*)$ distribution (Fig.~\ref{fig:belle_Zb_2_BB}b) 
peaks at the $Z_b(10650)$ mass.  Fits to the data with various combinations of $Z_b$ BW amplitudes,
both with and without a coherent non-resonant phase-space term, are shown as curves in the figures.
In these fits, the masses and widths of the BW amplitudes are fixed at the values given
in Eq.~\ref{eqn:zb_mass_and_width}.  The default fit, shown as short-dashed curves, uses only a 
$Z_b(10610)\rt B\bar{B}^*$ amplitude for the $B\bar{B}^*$ (Fig.~\ref{fig:belle_Zb_2_BB}a) fit and a
$Z_b(10650)\rt B^*\bar{B}^*$ amplitude for the $B^*\bar{B}^*$ (Fig.~\ref{fig:belle_Zb_2_BB}b) fit and
gives an adequate description of the data. Other variations: phase-space only (dotted); single BW
amplitudes plus phase-space (dashed-dot) and  two BW amplitudes plus phase-space (long dashed) do
not make any significant improvements.

From the default fit,  the branching fraction values ${\mathcal
B}(Z_b(10610)\rt B^+\bar{B}^{*0}+\bar{B}^0B^{*+}) =
(86\pm 3)\%  $ and ${\mathcal B}(Z_b(10610)\rt
B^{*+}\bar{B}^{*0}) = (74\pm 6)\%  $ are inferred.~The
$B^{(*)}\bar{B}^{*}$ {\em fall-apart} modes are stronger than the sum
total of the $\pip\Upsilon(n_r{\rm S})$ and $\pip h(m_r{\rm P})$ modes, but only by
factors of $\sim 6$ for the $Z_b(10610)$ and $\sim 3$ for the
$Z_b(10650)$.~The measured branching fraction for $Z_b(10650)\rt
B\bar{B}^{*}$ is consistent with zero.~This pattern, where
$B\bar{B}^*$ decays dominate for the $Z_b(10610)$ and $B^*\bar{B}^*$
decays are dominant for the $Z_b(10650)$ are consistent with
expectations for molecule-like structures~\cite{Karliner:2015ina},
which were proposed even before 
these states were observed \cite{Liu:2008fh,Liu:2008tn}.
A tetraquark interpretation of these states \cite{Ali:2014dva} 
was also made before their discovery \cite{Karliner:2008rc}.

\subsubsection{The $Z_c$ charged charmonium-like mesons}

As discussed in the previous Section, the discovery in the $c$-quark sector of the
unexpected  $Y(4260)\rt \pipi\jpsi$ signal in the initial-state-radiation process
$\ee\rt\gamma_{\rm{isr}}\pipi \jpsi$ at $\ecm\simeq 10.6$~GeV motivated Belle
to look for possible related anomalies in $\ee\rt\pipi\Upsilon(n_r{\rm S})$ ($n_r=1,2,3$)
reactions near $\ecm = 10.86$~GeV. This resulted in the discovery of
anomalously high rates transitions of what was thought to be the $\Upsilon(5{\rm S})$
charmonium state to $\pipi\Upsilon(1,2,3{\rm S})$, as shown in the top three panels
of Fig.~\ref{fig:Rbb-Rpipiups}~\cite{Chen:2008xia}.  Further investigations 
of these anomalies led to the discovery of the $Z_b(10610)$ and $Z_b(10650)$
charged bottomonium-like states. These discoveries in the $b$-quark sector
prompted the BESIII group to take data with the BEPCII collider operating at
$\ecm=4.26$~GeV, to see if there were  $c$-quark sector equivalents of
the $Z_b$ states produced in the decays of the $Y(4260)$.

\begin{figure} 
\includegraphics[width=0.48\textwidth]{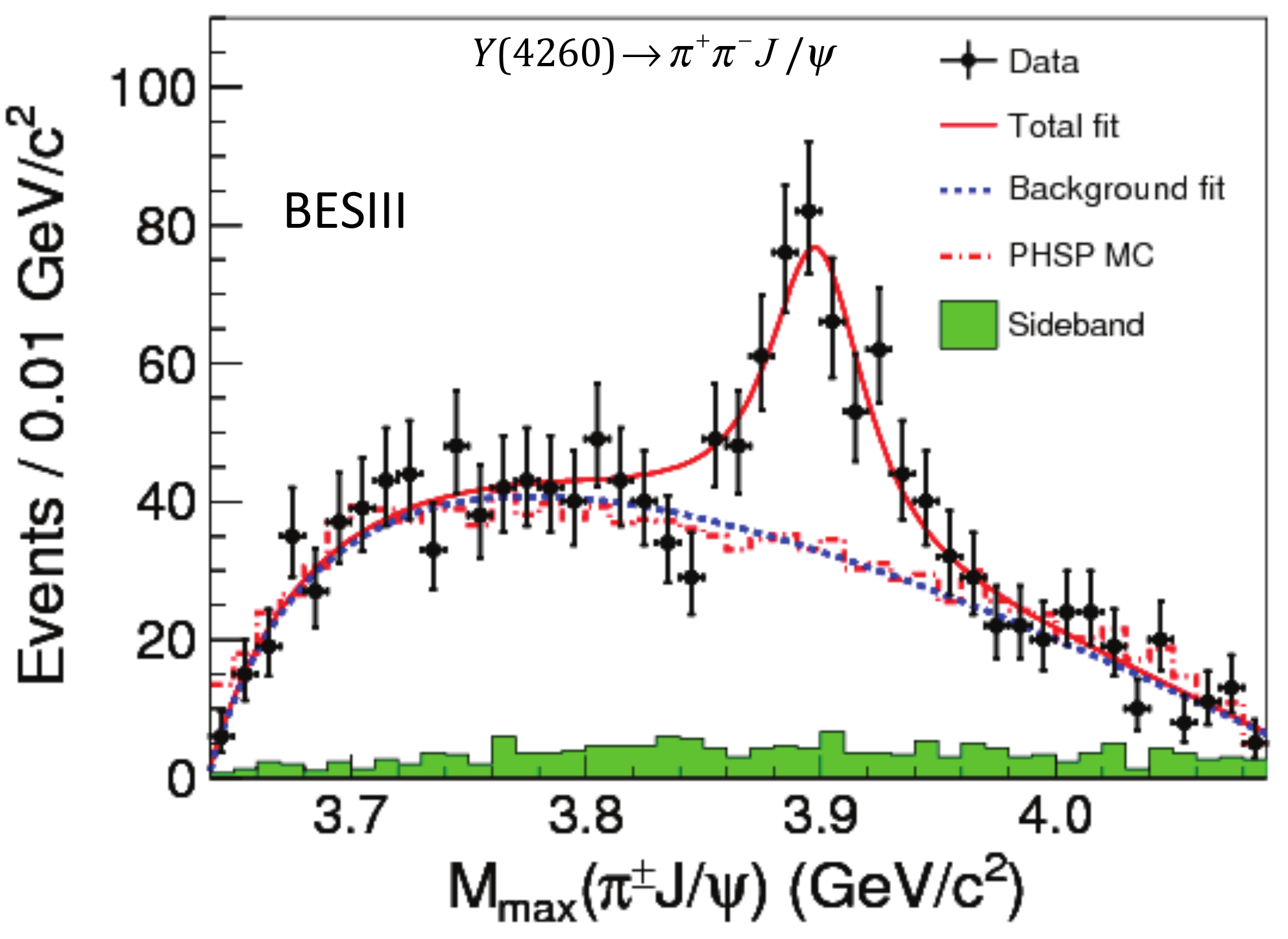}
\caption{\footnotesize 
Distribution of the larger of the two $\pi^{\pm}\jpsi$ masses in $\ee\rt\pipi\jpsi$
events collected in the BESIII detector at $\ecm=4.260$~GeV. The histogram shows the
level of the non-$\jpsi$ background, which is determined from $\jpsi$ mass sideband events.
 The figure is from
ref.~\cite{Ablikim:2013mio}.}
\label{fig:bes3_zc3900}
\end{figure}

\underline{\it The $Z_c(3900):$}~~Figure~\ref{fig:bes3_zc3900} shows the distribution of
the largest of the $\pi^{+}\jpsi$ and $\pi^{-}\jpsi/$ invariant mass combinations in
$Y(4260)\rt\pipi\jpsi$ events in a 525~pb$^{-1}$ BESIII data sample accumulated at
$\ecm=4.260$~GeV~\cite{Ablikim:2013mio}.  Here a distinct peak, called the
$Z_c(3900)$, is evident near $3900$~MeV. A fit using a BW amplitude to represent the
$\pi^{\pm}\jp$ mass peak and an incoherent phase-space-like function to represent the
non-resonant background gives a mass and width of $M(Z_c(3900))=3899.0\pm 6.1$~MeV and
${\Gamma}(Z_c(3900))= 46 \pm 22$ ~MeV, which is $\sim 24$~MeV above the 
$m_{D^{*+}} + m_{\bar{D}^{0}}$ (or $m_{D^+} + m_{\bar{D}^{*0}}$) threshold.  The $Z_c(3900)$ was
observed by Belle in isr data at about the same time~\cite{Liu:2013dau}.

\begin{figure*} 
\includegraphics[width=\textwidth]{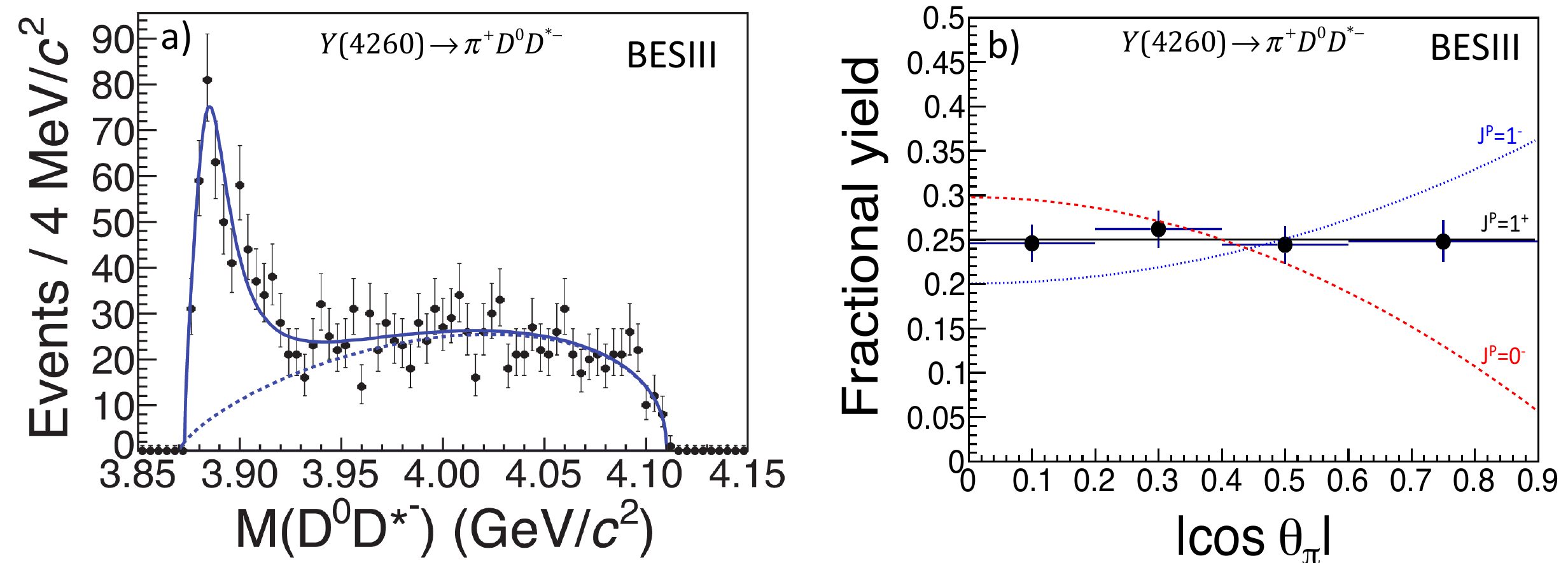}
\caption{\footnotesize 
{\bf a)} The $D^0 D^{*-}$ invariant mass distribution in $\ee\rt\pi^+ D^0 D^{*-}$ events collected
in the BESIII detector at $\ecm=4.260$~GeV~\cite{Ablikim:2013xfr}. The solid curve shows the result of 
a fit to the data points with a threshold-modified BW amplitude plus an incoherent phase-space-like background
(dashed curve). 
{\bf b)} The efficiency corrected $Z_c(3885)$ production angle distribution compared to expectations for
different $J^P$ quantum number assignments.
 The figures are from ref.~\cite{Ablikim:2013xfr}.}
\label{fig:bes3_zc3885-ddstr}
\end{figure*}

A subsequent BESIII study of the $D^0D^{*-}$ systems produced in $\ee\rt \pi^+ D^0 D^{*-}$
final states in the same data sample~\cite{Ablikim:2013xfr} found the very strong near-threshold
peak in the $D^0D^{*-}$ and invariant mass distribution shown in Fig.~\ref{fig:bes3_zc3885-ddstr}a. The
solid curve in the figure shows the results of a fit to the data with a threshold-modified BW amplitude to
represent the peak and an incoherent phase-space-like function to represent the background. The same
analysis found a similar peak in the $D^-D^{*0}$ invariant mass distribution in $\ee\rt\pi^+ D^-D^{*0}$
events. The masses and widths from the two channels are consistent and their average values 
are $M= 3883.9\pm 4.5$~MeV and ${\Gamma}=24.8\pm 12$~MeV,

Since the mass is $\simeq 2\sigma$ lower than the $Z_c(3900)$ mass reported in
ref.~\cite{Ablikim:2013mio}, BESIII cautiously named this $D\bar{D}^*$ state the $Z_c(3885)$.
In the mass determinations of both the $Z_c(3885)$ and $Z_c(3900)$, effects of possible
interference with a coherent component of the non-resonant background are ignored, an approximation
that can bias mass measurements by amounts comparable to the resonance widths, and this effect could
account for the different mass values.  Thus, we consider it highly likely that the
$Z_c(3885)$ is the $Z_c(3900)$ in a different decay channel.~If this
is the case, the partial width for $Z_c(3900)\rt D\bar{D}^*$ decays
is $6.2\pm 2.9$ times larger than that for $\jp\pip$, which is small
compared to open-charm {\em vs.} hidden-charm decay-width ratios for
established charmonium states above the open-charm threshold, such
as the $\psi(3770)$ and $\psi(4040)$, where corresponding ratios are
measured to be more than an order-of-magnitude larger~\cite{Olive:2016xmw}.
On the other hand, this ratio is similar to the properties of 
the $X(3872)$ and $Z_b$ states. 

The strong $Z_c(3885)\rt D\bar{D}^*$ signal enabled the BESIII group to
determine its $J^P$ quantum numbers from the dependence of its
production on $\theta_{\pi}$, the bachelor pion production angle
relative to the beam direction in the $\ee$ cm system.~For
$J^P=0^-$, $d N/d|\cos\theta_{\pi} |$ should go as
$\sin^2\theta_{\pi}$; for $1^-$ it should follow
$1+\cos^2\theta_{\pi}$ and for $1^+$ it should be flat ($0^+$ is
forbidden by Parity).  Figure~\ref{fig:bes3_zc3885-ddstr}b shows the efficiency-corrected
$Z_c(3885)$ signal yield as a function of $|\cos\theta_{\pi}|$,
together with expectations for $J^P=0^+$ (dashed red), $1^-$ (dotted
blue) and $J^P=1^+$.  The $J^P=1^+$ assignment is clearly preferred
and the $0^-$ and $1^-$ assignments are ruled out with high
confidence.

BESIII also reported neutral counterparts of the $Z_c(3900)$ in the $\pi^0\jpsi$ channel
in $\ee\rt\pi^0\pi^0 \jpsi$ events~\cite{Ablikim:2015tbp}, and the $D^+D^{*-}$ and $D^0\bar{D}^{*0}$
channels in $\ee\rt\pi^0(D\bar{D}^*)^0$ events~\cite{Ablikim:2015gda},
with mass and width values that are in good agreement with the charged $Z_c(3900)$ state
measurements.  The relative signal yields in the charged and neutral channels are
consistent with expectations based on isospin conservation.

\begin{figure*} 
\includegraphics[width=\textwidth]{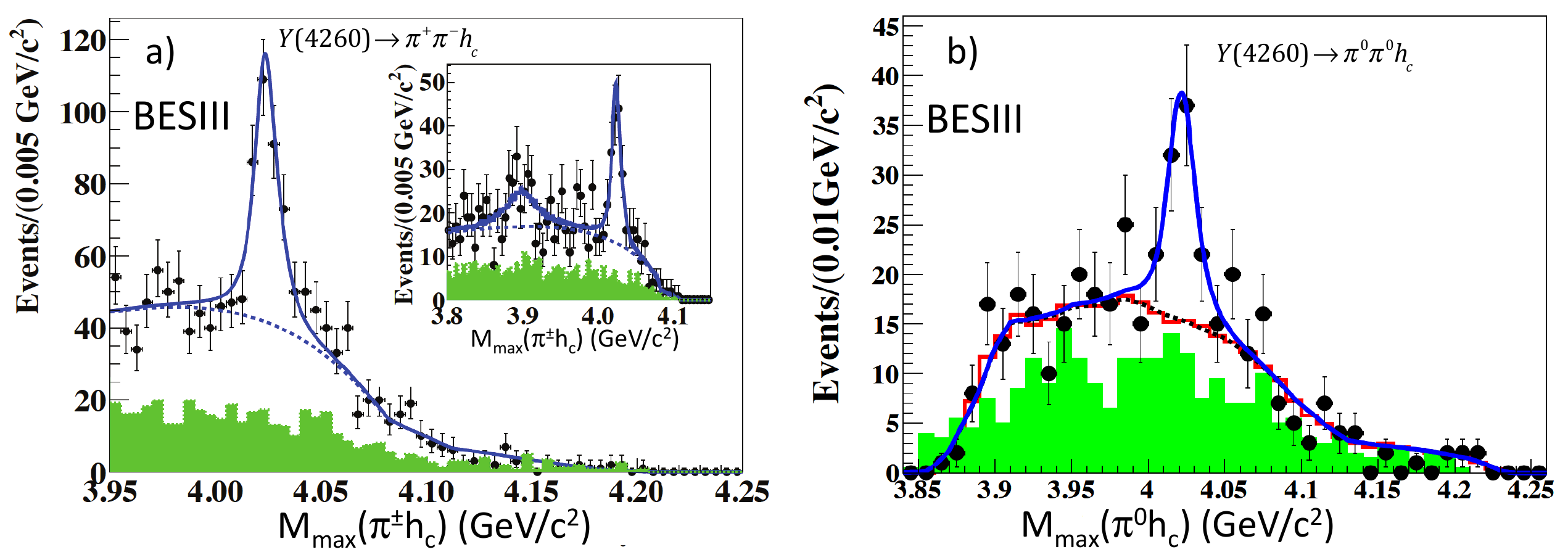}
\caption{\footnotesize 
{\bf a)} The distribution of the larger of the two $\pi^{\pm}h_c$ masses in $\ee\rt\pipi h_c$
events collected in the BESIII detector at $\ecm=4.260$ and 4.360~GeV. 
The inset shows a larger $M(\pi h_c)$ that includes the $Z_c(3900)$ mass region.
{\bf b)} The $M_{\rm max}(\pi^0 h_c)$ distribution for $\ee\rt \pi^0\pi^0 h_c$ events.
The histograms show the level of the non-$h_c$ background determined from $h_c$ mass sideband events.
The curves are described in the text.
The figures are from refs.~\cite{Ablikim:2013wzq} and~\cite{Ablikim:2014dxl}, respectively.}
\label{fig:bes3_zc4020_pihc}
\end{figure*}

\underline{\it The $Z_c(4020):$}~~ With data accumulated at the
peaks of the $Y(4260)$, $Y(4360)$ and nearby energies, BESIII made a
study of $\pipi h_c(1{\rm P})$ final states~\cite{Ablikim:2013wzq}. Exclusive
$h_c(1{\rm P})$ decays were detected via the $h_c\rt\gamma\eta_c$ transition,
where the $\eta_c$ was reconstructed in 16 exclusive hadronic decay
modes.  With these data, BESIII observed a distinct peak near 4020~MeV
in the $M_{\rm max}(\pi^{\pm}h_c)$ distribution that is shown in
Fig.~\ref{fig:bes3_zc4020_pihc}a.  A fit to this peak, which the BESIII group
called the $Z_c(4020)^+$, with a signal BW amplitude (assuming $J^P=1^+$)
plus a smooth background, returns a $\sim 9\sigma$ significance signal
with a mass of $M(Z_c(4020))=4022.9\pm 2.8$~MeV -- about $ 5$~MeV
above $m_{D^{*+}}+m_{\bar{D}^{*0}}$ -- and a width of $\Gamma(Z_c(4020))=7.9\pm 3.7$~MeV.

The inset in Fig.~\ref{fig:bes3_zc4020_pihc}a shows the result of including a
$Z_c(3900)^+\rt \pip h_c$ term in the fit. In this case, a marginal
$\sim 2\sigma$ signal for $Z_c(3900)^+\rt \pip h_c$ is seen to the
left of the $Z_c(4020)$ peak.~This translates into  an upper limit
on the branching fraction for $Z_c(3900)^+\rt\pip h_c$ decay that is less
than that for $Z_c(3900)^+\rt\pip\jpsi$ by a factor of five.

BESIII also observed the neutral isospin partner of the
$Z_c(4020)$~\cite{Ablikim:2014dxl}. The $M_{\rm max}(\piz h_c )$
distribution for $\ee\rt\pi^0\pi^0 h_c$ events in the same data set
used for the $Z_c(4020)^{\pm}$, shown in Fig.~\ref{fig:bes3_zc4020_pihc}b,
looks qualitatively like the $M_{\rm max}(\pip h_c )$ distribution with a
distinct peak near $4020$~MeV.~A fit to the data that includes a BW
term with a width fixed at the value measured for the $Z_c(4020)^+$
and floating mass returns a mass of $4023.9\pm 4.4$~MeV; this and
the signal yield are in good agreement with expectations based on
isospin symmetry.

\begin{figure*} 
\includegraphics[width=0.8\textwidth]{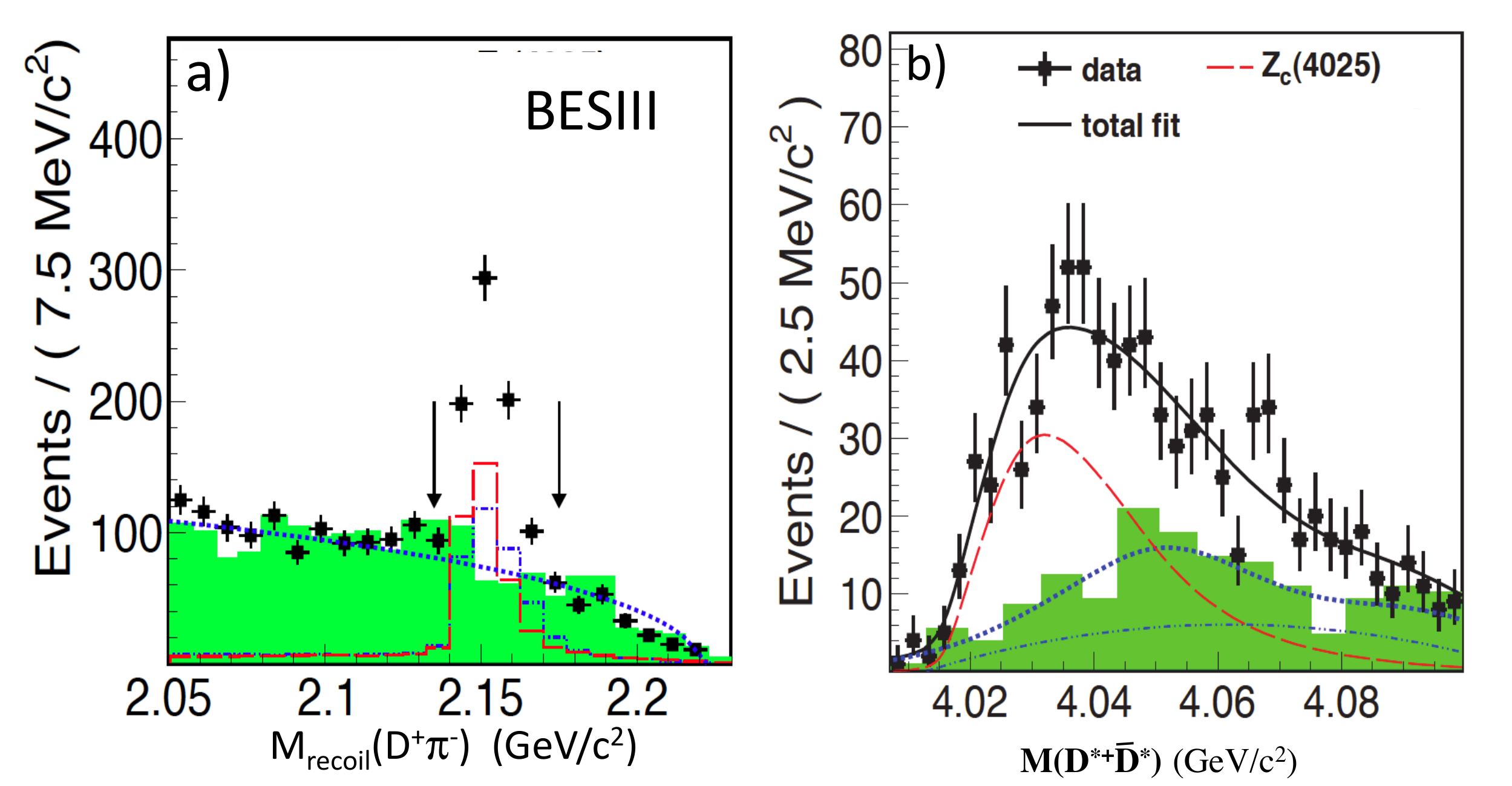}
\caption{\footnotesize  
{\bf a)} Distribution of masses recoiling against the reconstructed  
$\pim$ and $D^+$ in $\ee\rt\pim (D^*\bar{D}^*)^+$ events collected in the BESIII
detector at $\ecm=4.260$~GeV. 
The peak near 2.15~GeV corresponds to $\ee\rt\pim (D^{*+}\bar{D}^{*0})^+$ signal events.
{\bf b)} The $M(D^{*+}\bar{D}^{*0})$ distribution inferred from the $\pim$ momentum for
signal events.  The histograms show the level of the combinatoric background determined from
events where the reconstructed pion and $D$ meson have the same electric charge.
The curves are described in the text.
The figures are from ref.~\cite{Ablikim:2013emm}.
}
\label{fig:bes3_zc4020_dstrdstr}
\end{figure*}

A study of $\ee\rt D^{*+}\bar{D}^{*0}\pim$ events in the
$\ecm=4.260$~GeV data sample using a partial reconstruction
technique that only required the detection of the bachelor $\pim$,
the $D^+$ from the $D^{*+}\rt \piz D^+$ decay and one $\piz$, either
from the $D^{*+}$ or the $\bar{D}^{*0}$ decay, to isolate the process and
measure the $D^{*+}\bar{D}^{*0}$ invariant mass~\cite{Ablikim:2013emm}.
The signal for real $D^{*+}\bar{D}^{*0}\pim$ final states is
the distinct peak near 2.15~GeV in the distribution of masses 
recoiling from the reconstructed $D^+$ and $\pim$, shown in
Fig.~\ref{fig:bes3_zc4020_dstrdstr}a.  The measured
$D^*\bar{D}^*$ invariant mass distribution for events in the 2.15~GeV peak
and inferred from the $\pim$ momentum is shown as the data
points in Fig.~\ref{fig:bes3_zc4020_dstrdstr}b, where a strong near-threshold
peaking behavior that cannot be described by a phase-space-like
distribution, shown as a dash-dot blue curve, or by combinatoric
background, which is determined from wrong-sign (WS) events in the
data ({\em i.e.}~events where the pion and charged $D$ meson
have the same sign) that are shown as the shaded histogram. The
solid black curve shows the results of a fit to the data points that
includes an efficiency weighted S-wave BW function (long dashes),
the combinatoric background shape (short dashes) scaled to measured non-$D^{*+}\bar{D}^{*0}\pim$
background level under the signal peak in Fig.~\ref{fig:bes3_zc4020_dstrdstr}a,
and a phase-space term (dash-dot). The fit returns a $13\sigma$ signal with mass and
width $M=4026.3\pm 4.5 $~MeV and $\Gamma=24.8\pm 9.5$~MeV,  values that agree within
errors to those measured for the $Z_c(4020)^+\rt\pi^+ h_c $ channel. 
Although BESIII cautiously calls this $(D^*\bar{D}^*)^+$ signal the $Z_c(4025)$, we
consider this to likely be another decay mode of the $Z_c(4020)$.

A neutral $D^*\bar{D}^*$ state with a mass and width that are consistent with the 
$Z_c(4025)^+$ was seen by BESIII in $\ee\rt\pi^0(D^*\bar{D}^*)^0$ events~\cite{Ablikim:2015vvn}.

The $Z_c(4020)\rt D^*\bar{D}^*$ and $\pim h_c$ signal yields reported in
refs.~\cite{Ablikim:2013emm} and~\cite{Ablikim:2013wzq}, respectively, 
imply a partial width for $Z_c(4020)\rt D^*\bar{D}^*$ that is larger than that
for $Z_c(4020)\rt\pi h_c$, but only by a factor of $12\pm 5$, not by the
large factors that are characteristic of conventional charmonium.  In addition,
there is no sign of $Z_c(4020)\rt D\bar{D}^*$ in Fig.~\ref{fig:bes3_zc3885-ddstr}a, where
there is a $\sim 500$ event $Z_c(3885)\rt D^0 D^{*-}$ signal.  A recent BESIII study of
$Z_c(3900)\rt D\bar{D}^*$ decays~\cite{Ablikim:2015swa} set a 90\% confidence level upper
limit ${\mathcal B}(Z_c(4020)\rt D\bar{D}^*)<0.13\times {\mathcal B}(Z_c(3885)\rt D\bar{D}^*)$.
This absence of any evident signal for $Z_c(4020)\rt D\bar{D}^*$ suggests that the
$Z_c(4020)\rt D\bar{D}^*$ partial width is considerably smaller than that for
$Z_c(4020)\rt D^*\bar{D}^*$, which mirrors the behavior of the two $Z_b$ states and is
suggestive of some relation to the $Z_c(4020)$'s proximity to the $2m_{D^*}$ threshold.

Many similarities between the $Z_c$ and $Z_b$ states discovered in $e^+e^-$ 
annihilations to $\pi\pi$ plus a heavy quarkonium states can be taken as a reflection
of Heavy Quark Symmetry. 
However, there are also differences between them that await an explanation.
While the $Z_b(10610)$ and $Z_b(10650)$
both decay to $\pi\Upsilon(n_r{\rm S})$ and $\pi h_b$,
the $Z_c(4020)$ is not observed in $\pi\jpsi$ and
the $Z_c(3900)$ is not observed  in $\pi h_c$.\footnote{There is also possibly a different resonant $\pi\psip$ substructure 
in $e^+e^-\to\pipi\psip$ \cite{Ablikim:2017oaf}.}
The latter is particularly difficult to accommodate in a purely molecular interpretation, in
which both $Z_c$ states should copiously decay to $\pi h_c$ \cite{Esposito:2014hsa}.

\FloatBarrier

\myclearpage

\section{Pentaquark candidates}
\label{sec:pentaquarks}

At the birth of the quark model, 
Gell-Mann \cite{Gellmann:1964nj} and Zweig \cite{Zweig:1981pd}  both 
suggested the possibility of particles built from more than the minimal quark content,
including pentaquark baryons $qqqq\bar q$. 
Experimental searches for pentaquarks comprised 
of light flavors have a long and often controversial history that is briefly 
summarized in sec.~\ref{sec:intro_light}.
No undisputed candidates have been found in over fifty years of searches, 
although unusual properties of some ordinary baryons, such as, {\em e.g.}~the $\Lambda(1405)$, 
are often attributed to mixing of $qqq$ and $qqqq\bar q$ systems.

\begin{figure}[bthp]
        \includegraphics*[width=0.5\textwidth]{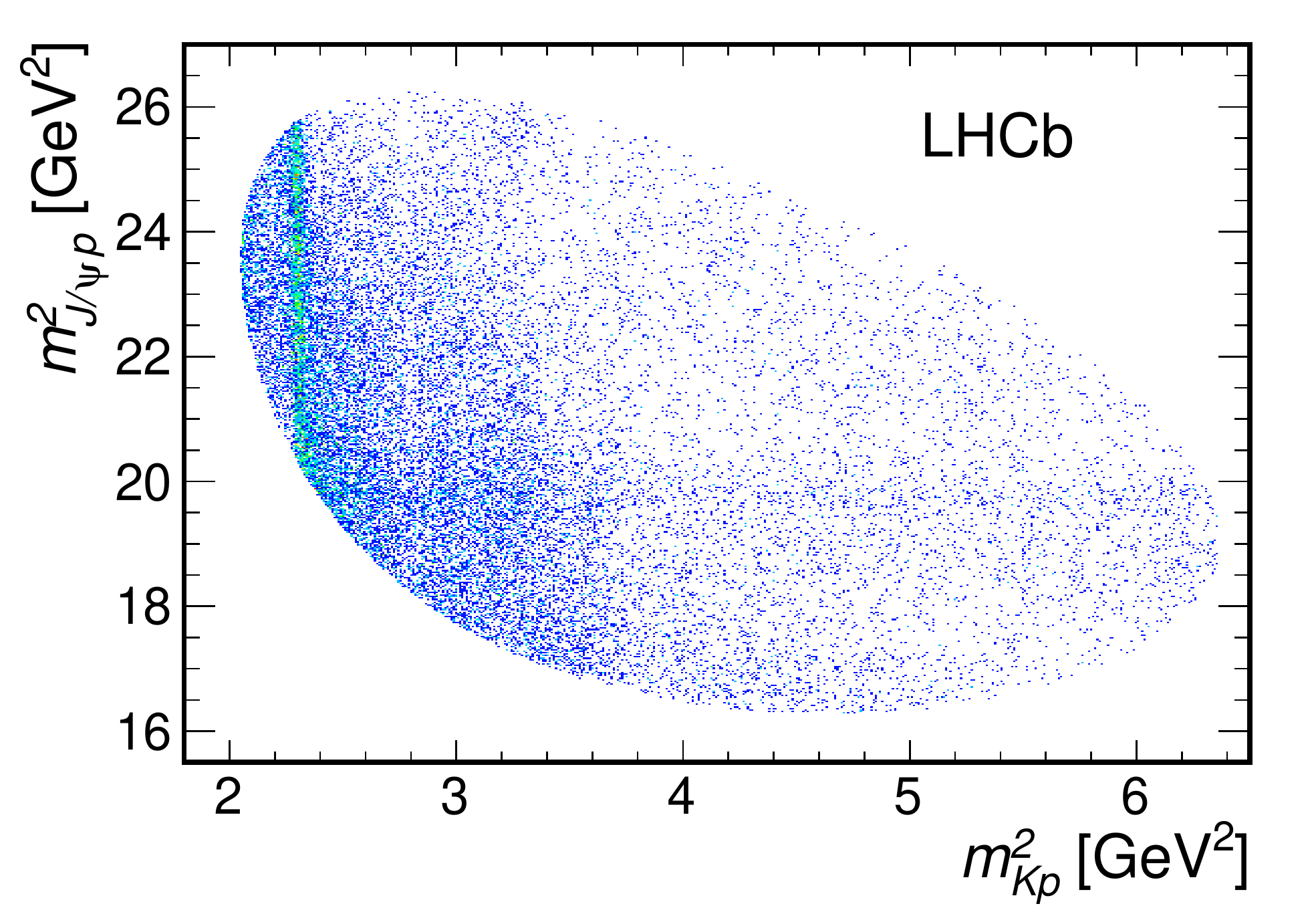}
  \vskip-0.1cm\caption{\small 
  Dalitz plot distribution for $\Lambda_b^0\to\jpsi p K^-$ decays as observed 
  by LHCb \cite{Aaij:2015tga}. 
  \label{fig:dalitzLb2jpsipK}
  }
\end{figure}

\begin{figure*}[bthp]
  \quad\hskip-3.6cm 
     \includegraphics*[width=0.48\textwidth]{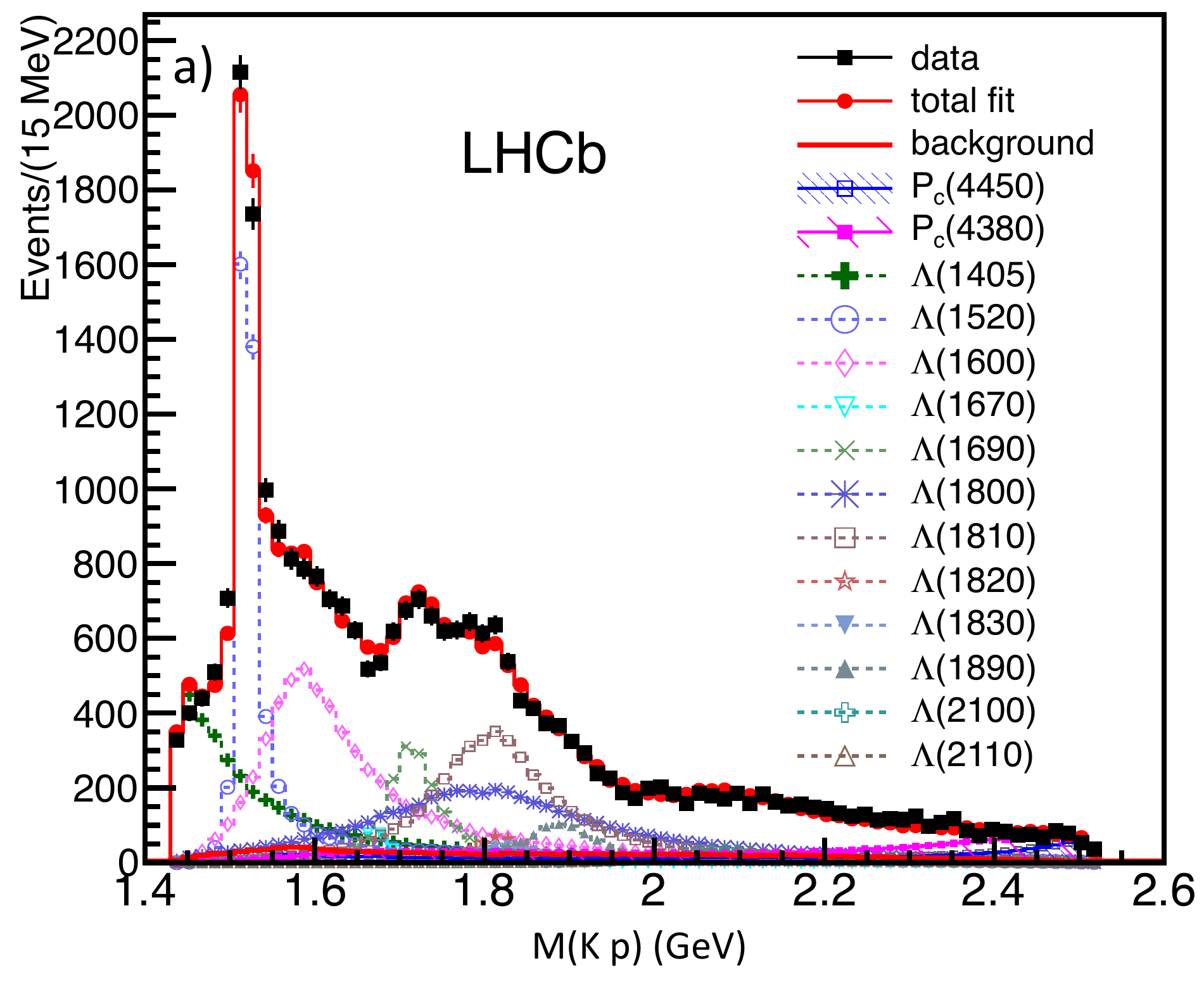}
  \quad\hskip-0.4cm 
    \hbox{ \includegraphics*[width=0.48\textwidth]{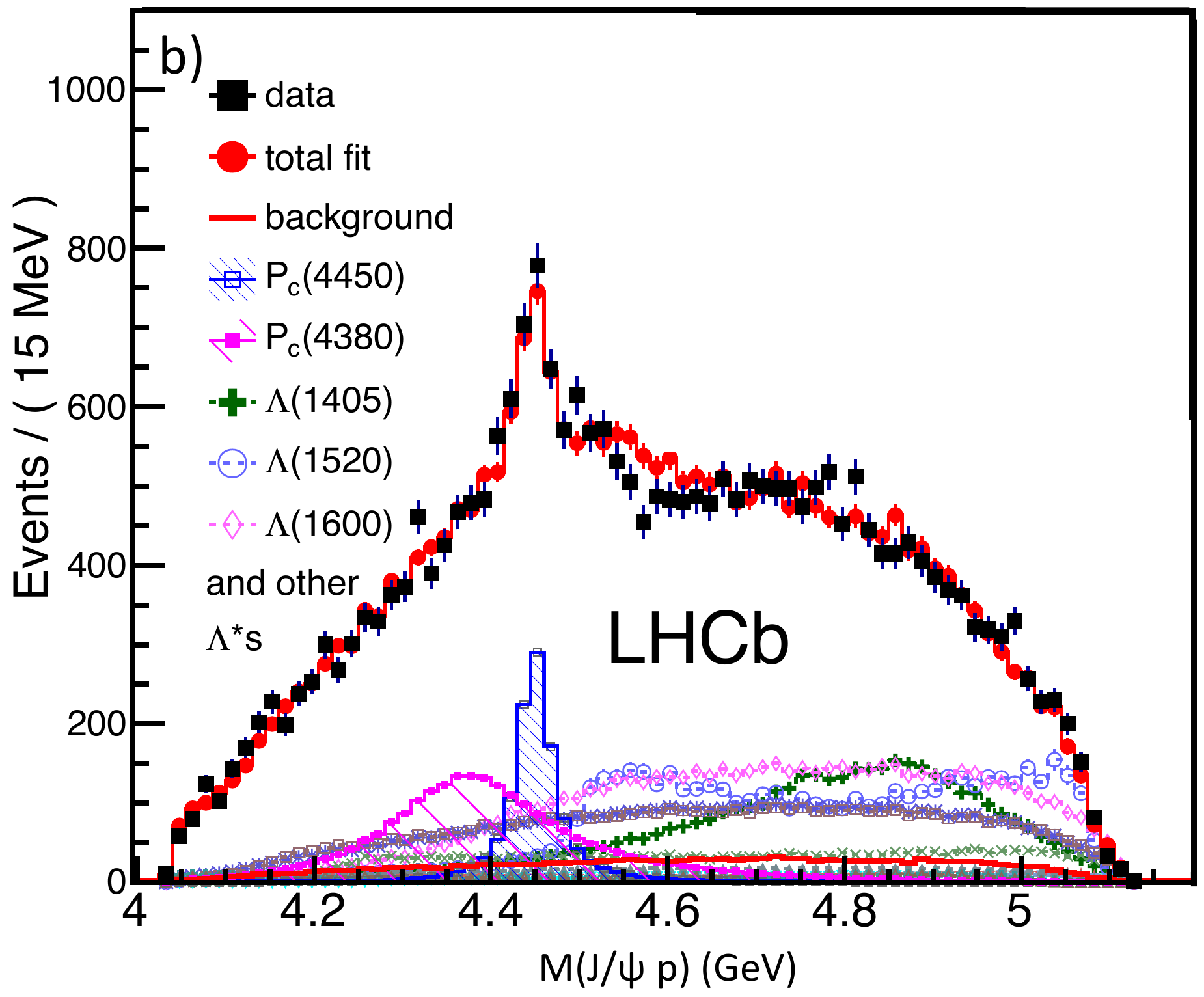}
    \quad\hskip-4.0cm } 
  \vskip-0.2cm\caption{\small 
  Projections of the amplitude fit with $P_c(4380)^+$ and $P_c(4450)^+$ states included (red histogram)
  onto the  {\bf a)} $M(K p)$ and {\bf b)} $M(\jpsi p)$ distributions (black squares with error bars)
  in the LHCb group's $\Lambda_b^0\to\jpsi p K^-$ event sample~\cite{Aaij:2015tga}.  Contributions from
  individual components of the fit are indicated as different color/line-style histograms as labeled.    
  \label{fig:lhcb_PcFit_masses}
  }
\end{figure*}

In 2015, convincing evidence for pentaquark-like structures with a minimal quark content 
of $uudc\bar c$  was reported by LHCb in a study of 
$\Lambda_b^0\to\jpsi p K^-$ ($\jpsi\to\mu^+\mu^-$) decays~\cite{Aaij:2015tga}.
In addition to contributions from many conventional $\Lambda^*\rt K^-p$ baryon
resonances (with a quark content of $uds$), the data contain a narrow peak in the
$\jpsi p$ mass distribution that is evident as a distinct horizontal band in the
$M^2(\jpsi p)$~{\em vs.}~$M^2(K^- p)$ Dalitz plot shown in Fig.~\ref{fig:dalitzLb2jpsipK}),
and the distribution of $\jpsi p$ invariant masses shown in
Fig.~\ref{fig:lhcb_PcFit_masses}b.
 
In order to clarify the nature of this band, an amplitude analysis 
was performed that was modeled after
the four-dimensional analysis of $\bar{B}^0\to\psip \pi^+ K^-$ ($\psip\to\mu^+\mu^-$) that
the LHCb group used to study the $Z(4430)^+\to\psip\pi^+$ charmonium-like state as described
in Section~\ref{sec:z4430}. Although the properties of the initial state particles are quite
different -- the spin~1/2 $\Lambda_b$ {\em vs.}~the spin~0 $B$-meson -- the final states for the
two processes are very similar, with $\pi^+$ being replaced by $p$. The signal statistics,
$26\,000\pm170$, and the background level, $5.4\%$, are also very comparable.  A
quasi-two-body amplitude model was used that was based
on an isobar approximation (i.e.\ summing up Breit-Wigner amplitudes) with the dynamics of the
contributing decay processes parameterized by a helicity formalism.  The amplitude fit spanned
a kinematically complete,  six-dimensional space of independent kinematic variables, 
including invariant masses $M(Kp)$ and $M(\jpsi p)$, decay helicity 
angles ($\theta$) of $\Lambda_b$, $\jpsi$, $\Lambda^*$ or pentaquark candidate
$P_c^+\to\jpsi p$, and angles between the decay planes.\footnote{The decay 
helicity angle is the angle between one of the decay products and the
boost direction in the rest frame of the parent particle.} 
For the $K^-p$ channel,
fourteen reasonably well established $\Lambda^*$ resonances were included  
with masses and widths set to their values given in the 2014 PDG tables~\cite{pdg14}, 
and varied within their uncertainties when evaluating systematic errors. 
Their helicity couplings (between 1 and 6 complex numbers per resonance) were 
determined from the fit to the data.
It was found that these $\Lambda^*$ contributions taken by themselves fail to describe the data.
It was necessary to add two non-standard $P_c^+\to\jpsi p$ pentaquark contributions to the
matrix element (10 free parameters per resonance)
before the narrow structure seen in $M(\jpsi p)$ could be reasonably well reproduced,
as illustrated in Fig.~\ref{fig:lhcb_PcFit_masses}.

The lower mass state, $P_c(4380)^+$, 
has a fitted mass of $4380\pm8\pm29$ \mev, width of
$205\pm18\pm86$ \mev, a fit fraction of $8.4\pm0.7\pm4.2$ \%\ and a significance of $9\sigma$.
The higher mass state, $P_c(4450)^+$, 
has a fitted mass of $4449.8\pm1.7\pm2.5$ \mev, a much narrower width of
$39\pm5\pm19$ \mev, 
a fit fraction of $4.1\pm0.5\pm1.1$ \%\ 
and a significance of $12\sigma$.

\begin{figure}[bthp]
    \includegraphics*[width=0.5\textwidth]{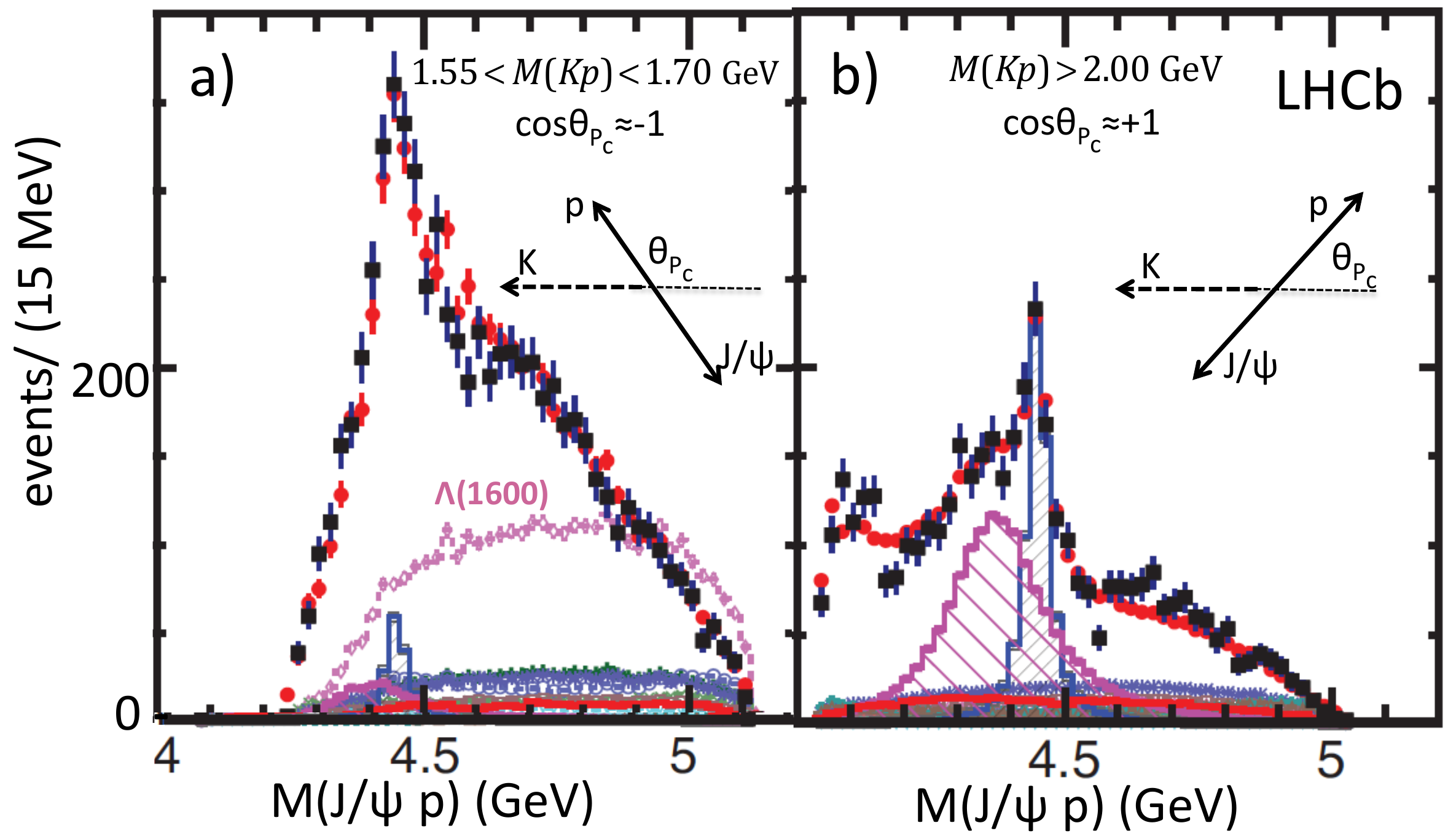}
 \caption{\footnotesize
  Projections of the amplitude fit with $P_c(4380)^+$ and $P_c(4450)^+$ states included (red
  histogram) onto the $M(\jpsi p)$ invariant mass distributions from the LHCb group's
  $\Lambda_b^0\to\jpsi p K^-$ events (black squares with error bars) with 
  {\bf a)} $1.55\le M(K p)\le 1.70$~GeV and {\bf b)} $M(K p)>2.0$~GeV~\cite{Aaij:2015tga}.
  The individual fit components are shown with the same color/line-style designations that are
  used in Fig.~\ref{fig:lhcb_PcFit_masses}.  
  \label{fig:Pc-parity}
  }
\end{figure}

The need for a second $P_c^+$ state is visually more apparent in Fig.~\ref{fig:Pc-parity}b that
shows the $M(\jpsi p)$ projections for large $Kp$ invariant masses ($M(Kp)\ge 2.0$~GeV), where
contributions from $\Lambda^*$ resonances are the smallest.  Even though contributions from the
two $P_c^+$ states are most visible in this region, they interfere destructively in this part of
the Dalitz plane, as is evident in the figure.  In contrast, in the $M(\jpsi p)$ projection at
the other extreme of the Dalitz plane, at low $Kp$ mass values ($1.55\le M(Kp)\le 1.70$~GeV) shown
in Fig.~\ref{fig:Pc-parity}a, the interference between the two $P_c^+$ states is
constructive.  High $M(Kp)$ values correspond to $\cos\theta_{P_c}$ values near $+1$, while low 
$M(Kp)$ values correspond to $\cos\theta_{P_c}\approx-1$.  The observed pattern, with interference
that is constructive near $\cos\theta_{P_c}\approx -1$ and destructive near $\cos\theta_{P_c}\approx +1$,
can only occur between odd and even partial waves, thereby indicating that the two $P_c^+$ states
must have opposite parities. A similar interference pattern is observed in the $\cos\theta_{\Lambda^*}$
distribution (Fig.~7 in ref.~\cite{Aaij:2015tga}) that reflects the presence of
parity-doublets in the $\Lambda^*$ spectrum.  Unfortunately, the spins of the two $P_c^+$ states
were not uniquely determined.  
Within the statistical and systematic ambiguities, $(3/2,5/2)$ and
$(5/2,3/2)$ combinations with either $(-,+)$ or $(+,-)$ parities, are not well resolved. All other
combinations are disfavored.
There is a strong
dependence of the data preference for the $P_c^+$ spins on the $\Lambda^*$ model
used in the amplitude fit (see sec.~13.1 in ref.~\cite{Jurik:2016bdm}), and this
calls for some caution in the interpretation of the observed $P_c^+$ states until their quantum
numbers are more firmly determined.

\begin{figure}[htbp]
        \includegraphics*[width=0.5\textwidth]{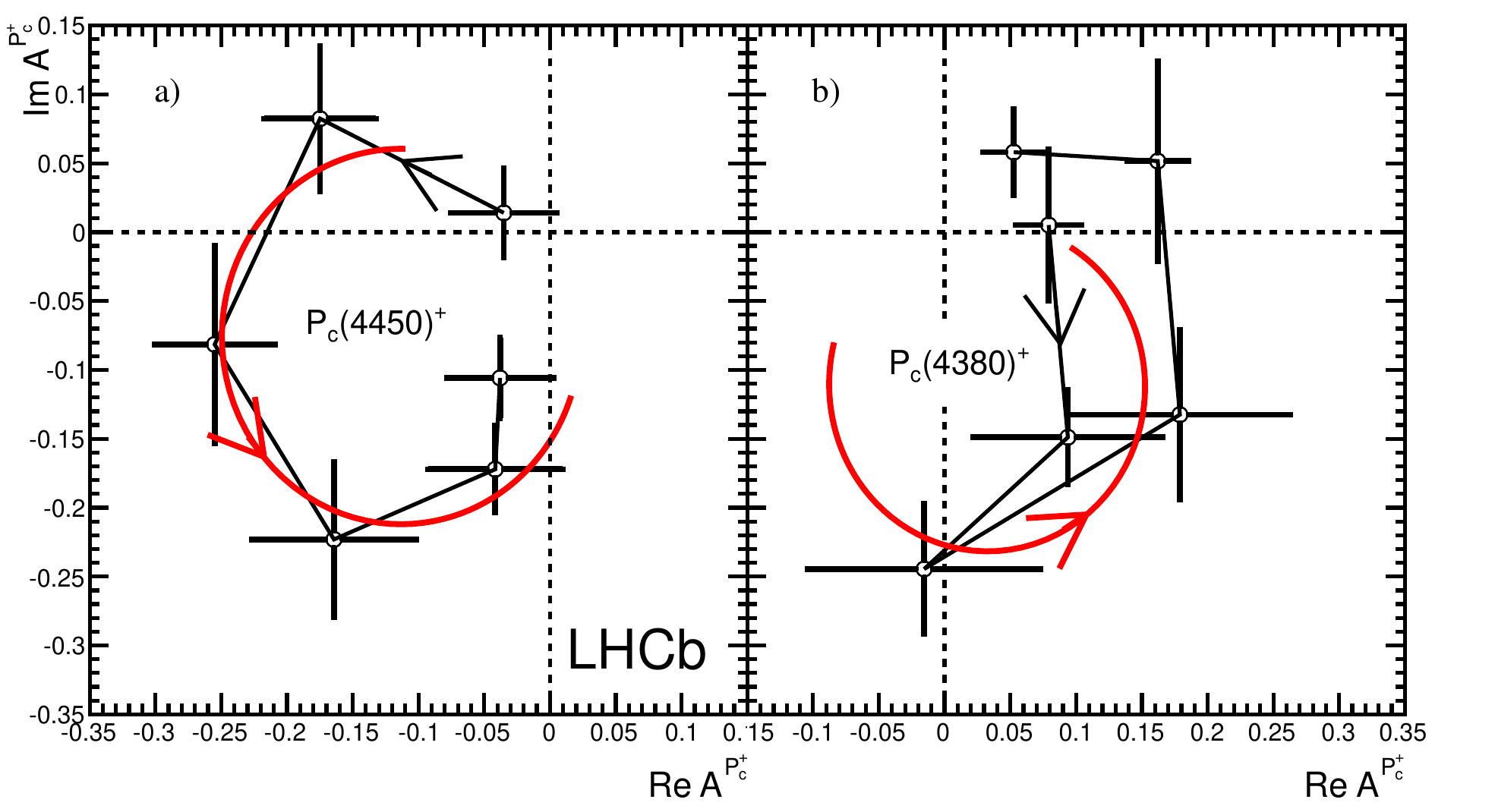}
  \vskip-0.1cm\caption{\footnotesize
Fitted values of the real and imaginary parts of the amplitudes 
of the $P_c(4450)^+$ (left) and $P_c(4380)^+$ (right) states for $\Lambda_b^0\to\jpsi p K^-$
shown in the Argand diagrams
as connected points with the error bars 
(masses increase counterclockwise) \cite{Aaij:2015tga}. 
The solid red curves are the predictions
from the Breit-Wigner formula, with resonance masses and widths
set to the nominal fit results and scaled to the displayed points.
  \label{fig:Pcargand}
  }
\end{figure}

Argand diagrams for the two $P_c^+$ states are shown in Fig.~\ref{fig:Pcargand}.
These were obtained by replacing the Breit-Wigner amplitude for one of the $P_c^+$ states at a time 
by a combination of independent complex amplitudes at six equidistant points in the $\pm\Gamma_0$ range
(interpolated in mass for continuity) 
that were fit to the data simultaneously with the other parameters of the full matrix element model.  
While the narrower $P_c(4450)^+$ state shows the expected resonant behavior, 
the diagram for the $P_c(4380)^+$ deviates somewhat from BW expectations.   However,
the statistical errors are large, especially for the broader $P_c(4380)^+$ state.

The inclusion of additional $\Lambda^*$ states beyond the well established ones,
of $\Sigma^*$ excitations (expected to be suppressed) 
and of non-resonant contributions with a constant amplitude, 
did not remove the need for two pentaquark states in the model to describe the data.
On the other hand, $\Lambda^*$ spectroscopy is a complex subject, 
from both experimental and theoretical points of view. 
This was demonstrated by a recent reanalysis of $\bar{K}N$ 
scattering data~\cite{Fernandez-Ramirez:2015tfa}   
in which the $\Lambda(1800)$ state, which was previously considered to be \lq\lq well established,''
is not seen, and where evidence for a few previously unidentified states is included. 
In fact, all theoretical models for $\Lambda^*$ 
baryons~\cite{SPECFG, SPECCI, SPECLMP, SPECMPS, SPECSF, SPECELMS} 
predict a much larger number of higher mass excitations than is established experimentally.
The high density of predicted states, which are generally expected to have large widths, 
makes it difficult to identify individual states experimentally. Non-resonant contributions
with a non-trivial $K^-p$ mass dependence may also occur. Therefore, LHCb also inspected their
data with an approach that is nearly independent of way the $K^-p$
contributions are modeled~\cite{Aaij:2016phn}.
\begin{figure}[hbtp]
  \begin{center}
  \includegraphics*[width=0.5\textwidth]{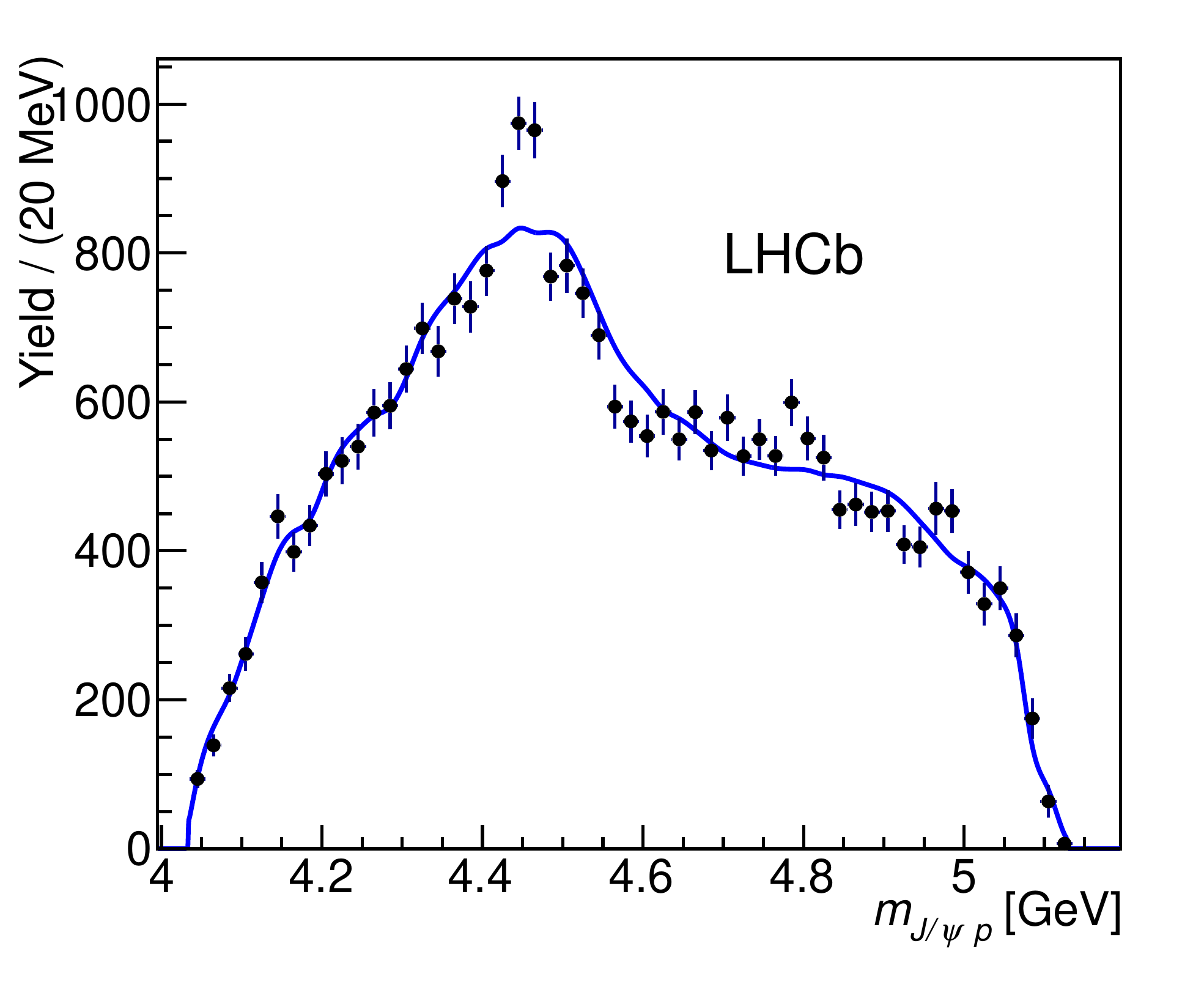}
  \end{center}
  \vskip-0.5cm\caption{\footnotesize
    The efficiency-corrected and background-subtracted distribution 
    of $M(\jpsi p)$ for the LHCb data, shown as black points with error bars, is 
    compared with the best fit obtained using reflections based on the observed $K^-p$ mass
    distribution and moments of the $K^-p$ helicity angle patterns, shown as the solid blue
    curve~\cite{Aaij:2016phn}.  The possibility that the $M(\jpsi p)$ distribution can be 
    accommodated by any plausible reflections from the $K^-p$ system is ruled out at the 
    $>9\sigma$ level.
   \label{fig:lhcb_mjpsip}
  }
   \vskip-0.2cm
\end{figure}
\noindent
A representation of the Dalitz plane distribution 
was constructed using the measured $M(Kp)$ distribution 
and Legendre polynomial moments of the cosine of the $\Lambda^*$ helicity angle determined from
the data as a function of $M(Kp)$. The maximal rank of the moments generated by the $K^-p$
contributions alone cannot be higher than twice their largest total angular momentum.
Since high-spin $\Lambda^*$ states cannot significantly contribute at low $M(Kp)$ values,
high rank moments were excluded from the representation 
(see Figs.~1 and 3 in ref.~\cite{Aaij:2016phn}).
When projected onto the $M(\jpsi p)$ axis of the Dalitz plane,  
this representation cannot describe the data
as shown in Fig.~\ref{fig:lhcb_mjpsip}.
The disagreement was quantified to be at least at the 9$\sigma$ level, and demonstrates
that the hypothesis that $K^-p$ contributions alone can generate the observed
$m_{\jpsi p}$ mass structure can be rejected with very high confidence
without any assumptions about the number of $K^-p$ contributions, 
their resonant or non-resonant character, or their mass shapes or interference phases.
This establishes the presence of contributions from either non-standard hadron channels or 
from rescattering effects of conventional ones. However, this approach says nothing about
their characterization.

The LHCb collaboration also inspected Cabibbo-suppressed decays 
$\Lambda_b^0\to\jpsi p \pi^-$ for signs of the $P_c(4380)^+$ and $P_c(4450)^+$
states~\cite{Aaij:2016ymb}.  The reconstructed $\Lambda_b^0$ signal yield in this channel
is more than an order of magnitude smaller and has a background fraction that is worse by
a factor of three than the Cabibbo-favored mode, thereby precluding the ability to perform
an unconstrained search for $\jpsi p$ states. Instead, the $P_c^+$ parameters were fixed to
the values measured  in the $\Lambda_b^0\to\jpsi p K^-$ channel, and varied within their
uncertainties (including $J^P$ ambiguities) in systematic studies.  Only the production
helicity couplings were allowed to be different (4 free parameters per state).
The possible presence of the $Z_c(4200)^-\to\jpsi\pi^-$ resonance, observed by Belle in
$B^0\to\jpsi\pi^- K^+$ decays, further complicates the amplitude analysis, adding 10 free
parameters even after its mass and width had been fixed to the measured values.
Up to 14 known $N^*\to p\pi^-$ resonances were included in the fit.  As a result, even
after neglecting contributions from higher orbital angular momenta in the $N^*$ decays,
there were as many as 106 free parameters in the six-dimensional fit to the relevant masses
and helicity angles.

\begin{figure*}[bthp]
  \quad\hskip-3.6cm 
     \includegraphics*[width=0.5\textwidth]{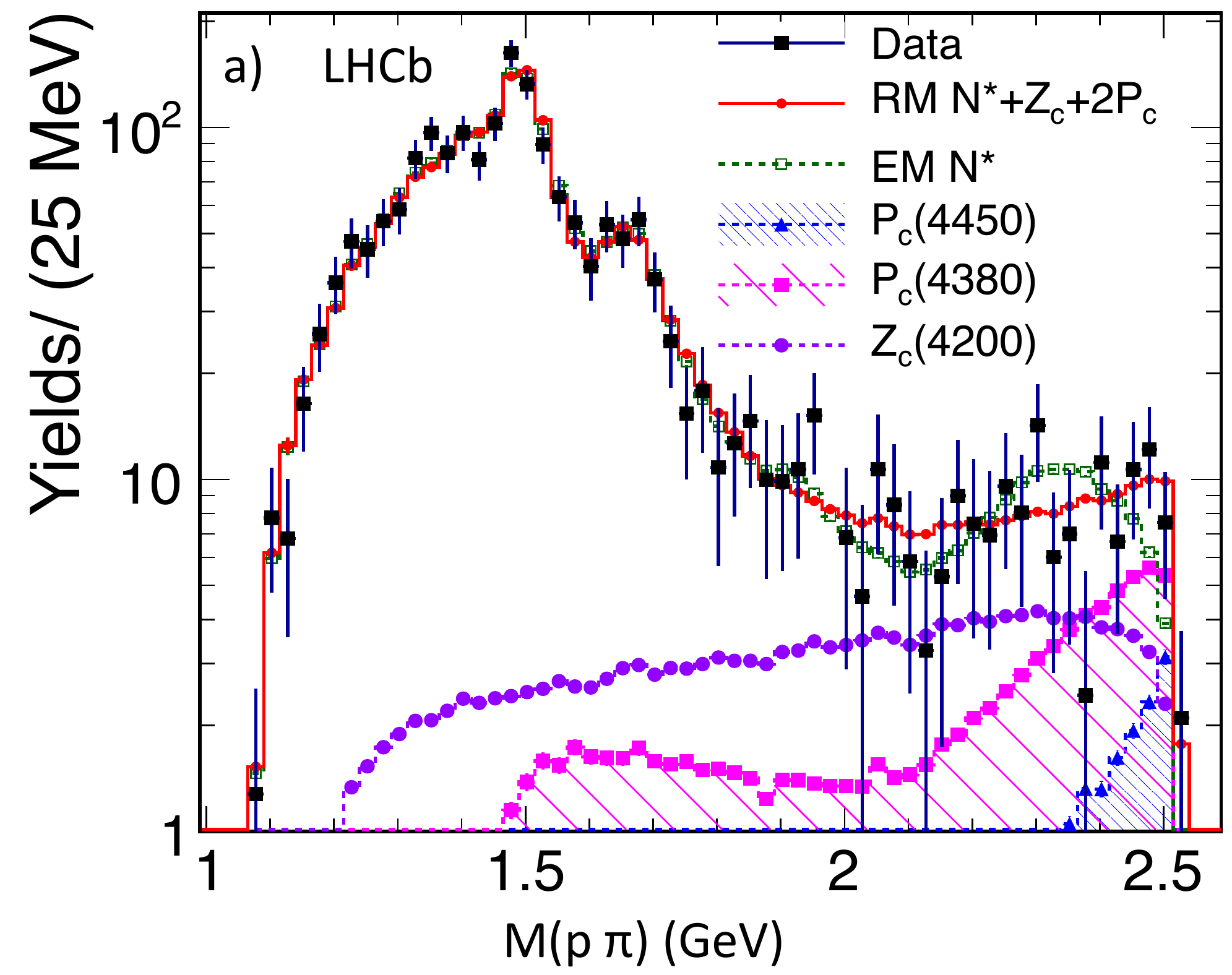}
  \quad\hskip-0.4cm 
    \hbox{ \includegraphics*[width=0.5\textwidth]{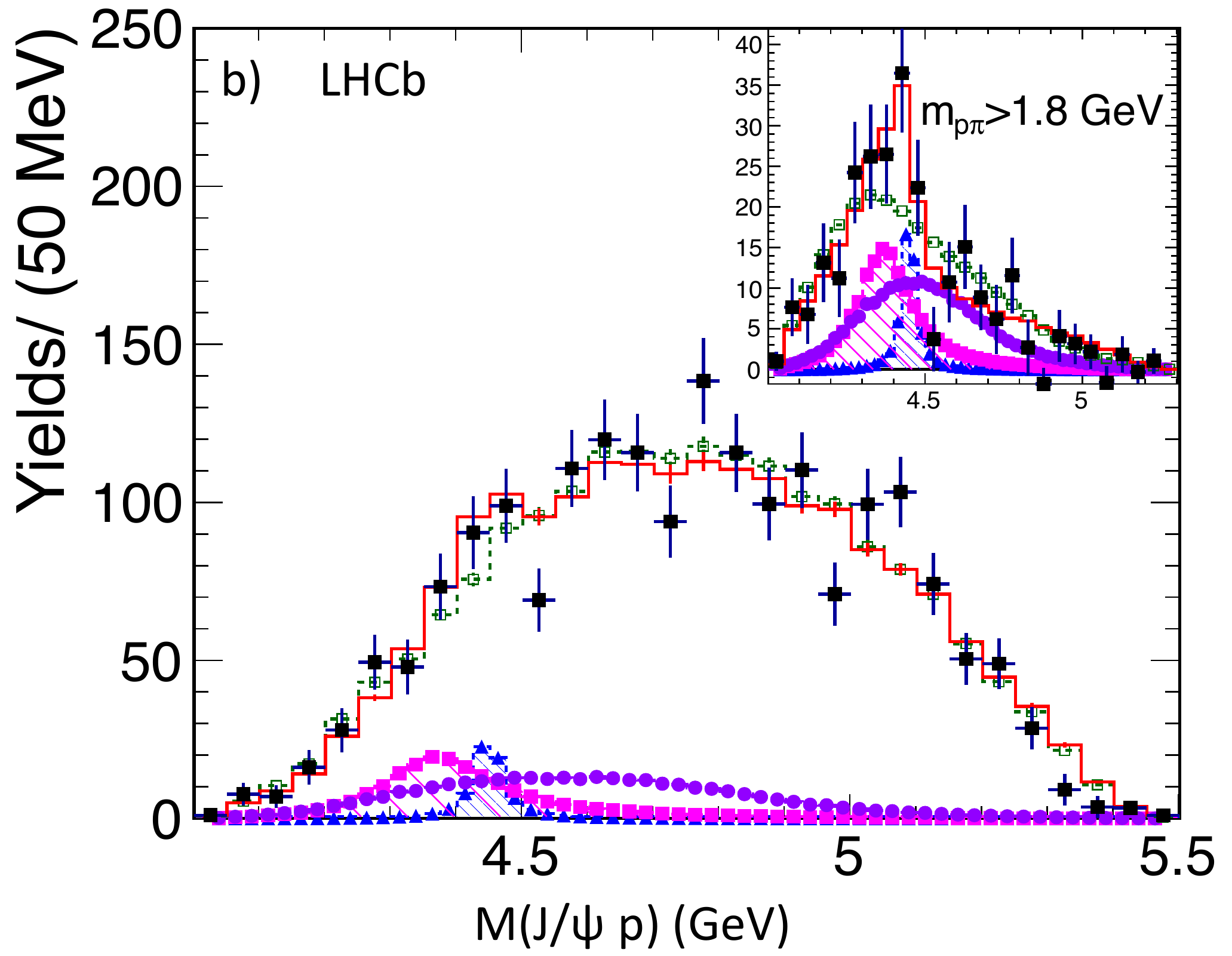}
    \quad\hskip-4.0cm } 
  \vskip-0.2cm\caption{\footnotesize Comparisons of  projections of the amplitude fits (red histogram) 
  onto {\bf a)} $M(p\pi)$ and {\bf b)} $M(\jpsi p)$ data distributions
  for $\Lambda_b^0\to\jpsi p \pi^-$ decays in LHCb~\cite{Aaij:2016ymb}.
  The dashed green histogram shows the results of a fit with no non-standard hadrons in  the $p\pi$  
  and $\jpsi p$ channels.  The inset in b) shows the $M(\jpsi p)$ fit-data comparison for events with
  $M(p\pi)>1.8$~GeV, where
  there is some hint of a $P_c(4450)^+$ signal.
  \label{fig:lhcb_PcFit_pimasses}
  }
\end{figure*}

The analysis yielded $3.1\sigma$ evidence for the summed presence of non-standard 
($Z_c(4200)^-$, $P_c(4380)^+$ and $P_c(4450)^+$) hadron contributions.  The $M(p\pi)$ and
$M(\jpsi p)$ projections of the fit are compared with the experimental data in 
Figs.~\ref{fig:lhcb_PcFit_pimasses}a~and~\ref{fig:lhcb_PcFit_pimasses}b, respectively.
The inset in Fig.~\ref{fig:lhcb_PcFit_pimasses}b shows the $M(\jpsi p)$ projection for
events with $M(p\pi)>1.8$~GeV, where there is some, but not very significant, indication of
a $P_c(4450)^+\rt\jpsi p$ signal.  However, ambiguities between $P_c^+$ and $Z_c^-$ terms
eliminate any ability to establish the presence of any individual non-standard hadron
contribution.  As a result, these results failed to confirm any of these states, and
more data are needed for more conclusive results.

\myclearpage

\section{Summary and Future Prospects}
\label{sec:summary}

Here we do not attempt to provide a detailed review of the successes and failures of
the large variety of theoretical models that have been proposed as explanations of the
non-standard hadrons discussed in this report.  For this we refer the reader to three 
recent and comprehensive discussions of these issues by experts who have somewhat different
viewpoints~\cite{Chen:2016qju,Lebed:2016hpi,Esposito:2016noz}.  Instead we provide our
experimentally oriented sense of where this field is at, and where it is heading.
  
\begin{figure*}[htbp]
  \includegraphics[width=0.8\textwidth]{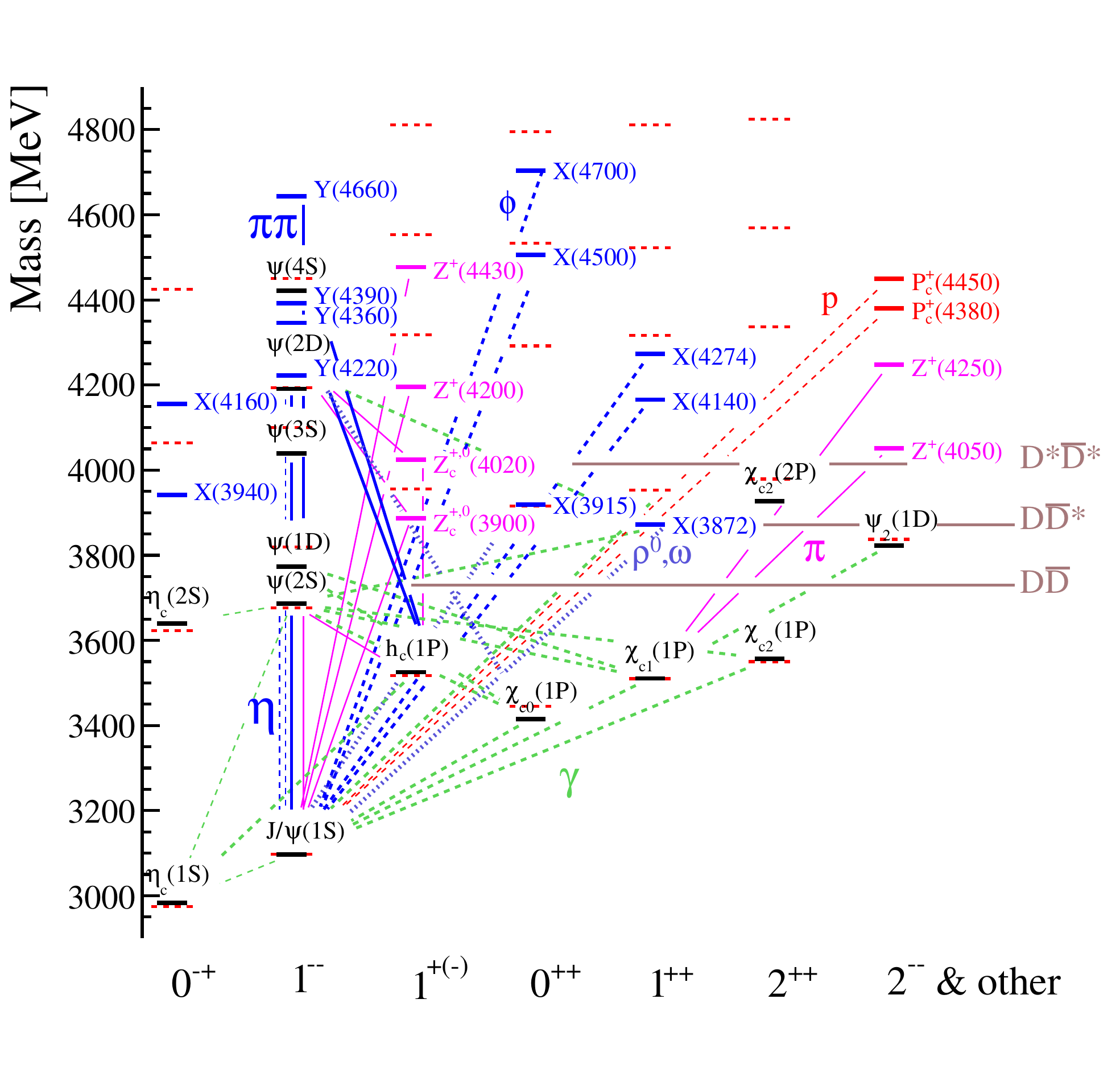}
\caption{\footnotesize The current status of the charmonium-like spectrum. The dashed (red) 
horizontal lines indicate the expected
states and their masses based on recent calculations~\cite{Barnes:2005pb} 
based on the Godfrey-Isgur
relativized potential model~\cite{Godfrey:1985xj}, 
supplemented by the calculations in ref.~\cite{Lu:2016cwr}
for high radial excitations of the P-wave states.
The solid (black) horizontal lines indicate the experimentally established
charmonium states, with masses and spin-parity ($J^{PC}$) quantum number assignments taken
from ref.~\cite{Olive:2016xmw}, and labeled by their spectroscopic assignment. 
The open-flavor decay channel thresholds are shown with longer solid (brown) horizontal lines.
The candidates for exotic charmonium-like states are also shown with shorter solid (blue or magenta) 
horizontal lines with labels reflecting their most commonly used names. 
All states are organized according to their quantum numbers given on horizontal axis. 
The last column includes states with unknown quantum numbers, the two pentaquark candidates and the
lightest charmonium $2^{--}$ state.  
The lines connecting the known states indicate known photon or hadron transitions between them:
dashed-green are $\gamma$ transitions; (thick E1, thin M1), solid-magenta are $\pi$;
thin (thick) dashed-blue are $\eta$ ($\phi$); dashed-red are $p$; dotted-blue are $\rho^0$ or $\omega$;  and solid-blue other $\pi\pi$ 
transitions, respectively.
}
\label{fig:charmoniumExotic}
\end{figure*}

Experimentally, it is remarkable that the number of 
charmonium-like states above the $D\bar{D}$ threshold that have unexpected properties, 
like narrow total widths and/or relatively large decay rates to hidden-charm states in spite 
of the existence of easily accessible decay channels with open charm, is about double
the number of known charmonium states in this mass range that conform to expected behavioral
patterns~(see Fig.~\ref{fig:charmoniumExotic}).
Clearly, the simple $Q\bar{Q}$ model of charmonium, which works so well for states below the
open-flavor threshold, fails at higher masses, and some new degrees of freedom have become relevant.

\begin{figure*}[htbp]
  \includegraphics[width=0.8\textwidth]{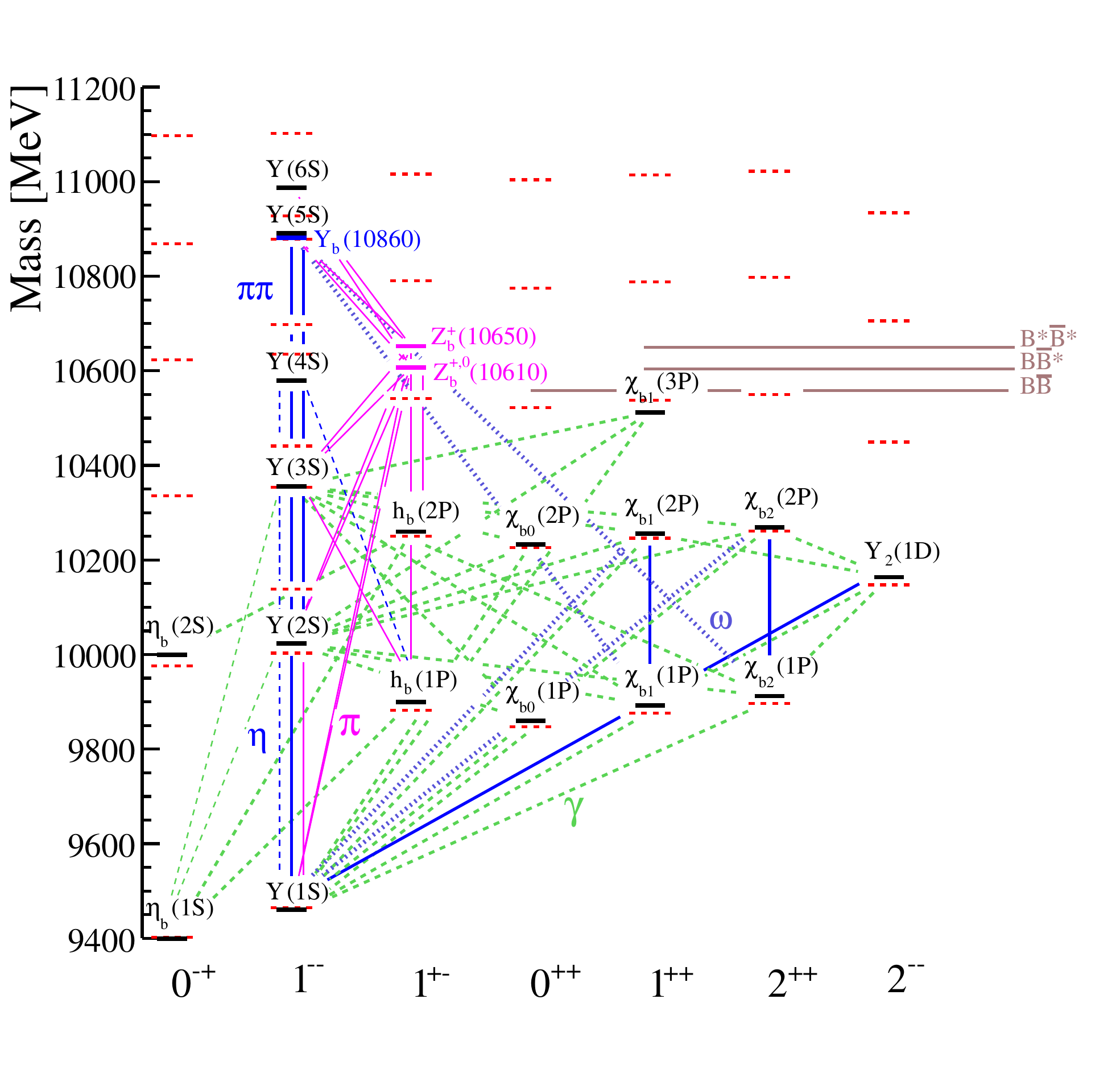}
\caption{\footnotesize The current status of the bottomonium-like spectrum. 
The current status of the bottomonium-like spectrum. The dashed (red) lines indicate the expected
states and their masses based on recent calculations~\cite{Godfrey:2015dia} based on the Godfrey-Isgur
relativized potential model~\cite{Godfrey:1985xj}. 
The solid (black) horizontal lines indicate the experimentally established
charmonium states, with masses and spin-parity ($J^{PC}$) quantum number assignments taken
from ref.~\cite{Olive:2016xmw}, and labeled by their spectroscopic assignment. 
The open-flavor decay channel thresholds are shown with longer solid (brown) horizontal lines.
The candidates for exotic bottomonium-like states are also shown with shorter solid (blue or magenta) 
horizontal lines with labels reflecting their most commonly used names. 
The known photon and hadron transitions are also indicated
(see the caption of Fig.~\ref{fig:charmoniumExotic}).}
\label{fig:bottomoniumExotic}
\end{figure*}

The high-mass bottomonium system is not as experimentally well explored as that
for charmonium (Fig.~\ref{fig:bottomoniumExotic}).
However, when explored with the same technique, {\em i.e.}~$e^+e^-$ energy scans, 
it revealed anomalous states with apparently similar characteristics as those
observed in the charmonium system, as expected from heavy-quark symmetry. 
 
Finding and understanding these new degrees of freedom presents a great challenge
to current theoretical models of hadronic structures.

\subsection{Theory}

\subsubsection{Molecules}

We, as sentient beings, owe our existence to residual strong forces between nucleons, {\em i.e.}~between individually
confined, color-singlet groups of three quarks in baryons. The resulting nuclei support the rich diversity
of atoms in nature. In the context of QCD, this binding resembles the binding of atoms into molecules by residual
electromagnetic forces and, thus, nuclei can be pictured as baryon-baryon molecules.
Nuclear physics provides good models for the ``molecular'' forces between nucleons, and 
it is natural to expect that meson-meson and meson-baryon combinations may also experience 
similar forces. But, since nuclear models are not based on direct derivations from the 
fundamental theory of strong interactions (QCD), it is not {\em a priori} known how strong the forces 
in these other, non-nucleon-nucleon, systems are, or if bound molecule-like meson-meson or meson-baryon
combinations actually exist.

\label{sec:summ-moecules}
The 2003 Belle paper~\cite{Choi:2003ue} that reported the discovery of the $X(3872)\rt\pipi\jpsi$
emphasized two intriguing experimental features. One was the close proximity of the
$X(3872)$ mass and the $D^0\bar{D}^{*0}$ mass threshold; at that time, the measurement precision
of the $X(3872)$ mass was $\pm 0.8$~MeV and that of the PDG-2002 value world average for
$m_{D^0}+m_{D^{*0}}$ was $\pm 1$~MeV~\cite{Hagiwara:2002fs} and 
$\delta m_{00}=(m_{D^0}+m_{D^{*0}})-M(X(3872))=-0.9\pm 1.3$~MeV.  The second intriguing feature was
the concentration of $\pipi$ invariant masses near the $\rho$ meson mass, that was a strong
indication that the decay violated isospin symmetry in a substantial way~\cite{Choi:2003ue}.

Within a few weeks after the Belle results were made public, papers were posted by T\"ornqvist~\cite{Tornqvist:2003na},
and Close and Page~\cite{Close:2003sg} that pointed out that these mass and isospin-breaking
properties were characteristic of expectations for a $D\bar{D}^*$ molecular state. In fact, a $J^{PC}=1^{++}$,
$D\bar{D}^*$ bound state with mass near $3870$ MeV had been predicted (and named!) in a 1994 paper by
T\"ornqvist~\cite{Tornqvist:1993ng}; inspired by its similarity to the deuteron, T\"ornqvist called
the state a ``deuson.''  As a result, at that time, experimenters and theorists expected that a
thorough understanding of the underlying nature of the $X(3872)$ would be a straightforward exercise and
that they could look forward to exploring a rich spectroscopy of related deuson states, both in the
charm quark and bottom quark sectors.

However, this optimism turned out to be short-lived. As discussed above, in Section~\ref{sec:x3872},
the CDF and D0 groups found the $X(3872)$ was produced promptly in $\ecm=1.96$~TeV $p\bar{p}$
annihilations with production cross sections and other characteristics that are similar to those for
prompt $\psip$ production~\cite{CDFprompt:2004,Abazov:2004kp}, while detailed computations for
a loosely bound $D\bar{D}^*$ composite showed that such similarities were highly
unlikely~\cite{Bignamini:2009sk}.  Also, in the deuson picture, the $X(3872)$ is primarily a
$D^0\bar{D}^{*0}$ bound state. Searches for other near-threshold $D\bar{D}^*$ combinations, such
as mostly $D^+D^{*-}$, or $D^0D^{*-}$ states, with the same $J^{PC}=1^{++}$ quantum numbers,
came up empty~\cite{Aubert:2004zr,Choi:2011fc}.  Another problem with the deuson idea is the large
rate for  $X(3872)\rt\gamma\psi(2S)$ reported by BaBar~\cite{Aubert:2008ae} and LHCb~\cite{Aaij:2014ala}
(see Eq.~\ref{eq:gpsip_over_gjpsi}), which is expected for the $2^3{\rm P}_1$ charmonium state but counter to
predictions for molecules~\cite{Swanson:2003tb,Dong:2009uf}.

At present, and as pointed out in ref.~\cite{Braaten:2007dw}, it seems likely that the
$X(3872)$ is a quantum mechanical mixture of a tightly bound $\ccbar$ core in $2^3{\rm P}_1$ 
configuration and a molecule-like $D\bar{D}^{*}$ combination.  This idea was
verified by detailed computations in refs.~\cite{Coito:2012vf} and~\cite{Takeuchi:2014rsa};
the latter found that the bulk of the $D\bar{D}^*$ binding comes from the coupling between
the $\ccbar$ core and the $D\bar{D}^*$ components, and not much comes from the mutual attraction
between the $D$ and the $\bar{D}^*$, which is the key feature of deuson models. 
In this picture for the $X(3872)$, its prompt production in high energy hadron collisions and radiative
decays to the $\psip$ proceed via the $\ccbar$ core component of the $X(3872)$ and, thus, have
characteristics that are similar to those of the $\chi_{c1}'$.

The bulk of the huge theoretical literature on non-standard hadrons
is on molecular models~\cite{Pakvasa:2003ea,Braaten:2004rw,Voloshin:2003nt,Fleming:2007rp,Zhu:2007wz,Gamermann:2009fv,Molina:2009ct,Zhang:2009bv,Sun:2012sy,Wang:2013cya,Polosa:2015tra,Karliner:2016ith}.
In these, binding is provided by pion- and other light-meson-exchange forces. Since the
binding provided by these forces is not expected to be very large, molecular states are expected to be near the masses
of their constituent hadrons and have appropriate S-wave $J^{PC}$ quantum numbers. This is the case
for the $Z_b(10610)$ and the $Z_b(10650)$, which are within a few~MeV of the $B\bar{B}^*$ and
$B^*\bar{B}^*$ thresholds, respectively, and applies reasonably well to the $Z_c(3900)$ and $Z_c(4020)$, which
are $\simeq 10$~MeV above the $D\bar{D}^*$ and $D^*\bar{D}^*$ thresholds, respectively.  However, the
interpretation of these states as molecules is controversial. Peaks at masses that are slightly above
threshold are dangerously similar to expectations for kinematically induced
cusps~\cite{Bugg:2011jr,Blitz:2015nra,Swanson:2015bsa} (see Fig.~\ref{fig:cusp-diagrams}b and related
text).  Anomalous triangle singularities are another mechanism that can produce above-threshold peaks
that are not related to a physical resonance~\cite{Chen:2013coa}.  Moreover, unlike the $X(3872)$, no
evidence for these states have been found in lattice QCD
calculations~\cite{Prelovsek:2013xba,Prelovsek:2014swa,Lee:2014bea,Ikeda:2016zwx}. On the other hand, detailed
studies of the BESIII's $Z_c(3900)\rt\jpsi\pi$ and $D\bar{D}^*$ signals~\cite{Guo:2014iya}
and Belle's corresponding $Z_b$ signals~\cite{Guo:2016bjq,Pilloni:2016obd,Albaladejo:2015lob} show that the observed peaks
can be identified as virtual states with associated poles in the complex scattering
$t$-matrices.

The $J^{P}=1^{+}$ $Z(4430)$ (now with a mass near $4478$~MeV) has been proposed as a radial excitation
of the $Z_c(3900)$, comprised of a molecule-like $D\bar{D}^*(2S)$
configuration~\cite{Barnes:2014csa,Ma:2014zua}, where the $D^*(2S)$ is the radial excitation of
the $D^*$. Although the existence of the $D^*(2S)$ has not been firmly established, the BaBar group reported a strong
candidate for this state in the $D^+\pi^-$ and $D^{*+}\pi^-$ invariant mass distributions in inclusive
$\ee\rt D^{(*)+}\pi^-\ X$ reactions at $\ecm=10.58$~GeV~\cite{delAmoSanchez:2010vq};  LHCb subsequently
reported observations of $D^{(*)}\pi$ invariant mass structures with peak and width values similar to BaBar's
that were produced in high energy $pp$ collisions~\cite{Aaij:2013sza}.
The averages of the BaBar and LHCb mass and width measurements for this state,
which is called the $D^*_J(2600)$, are $2622 \pm 12$~MeV and $104\pm 20$~MeV,
respectively~\cite{Olive:2016xmw}. 
If we assume $D^*(2S)=D^*_J(2600)$,  the $D\bar{D}^*(2S)$
``threshold'' is at $\simeq 4490$~MeV, and $\simeq 12$~MeV  above the $Z(4430)$ mass.   This association
with a radially excited $D^*$ meson may account for the observed preference for the $Z(4430)$ to decay to
$\psi(2S)\pi$ over $\jpsi\pi$ (recall that $\psip$=$\psi(2S)$). 
The large $D^*_J(2600)$ width could also explain the large $Z(4430)$ width, although for 
such a broad constituent, it is not clear whether the molecular formalism still applies.
Also, unlike the $Z(4430)$ state, the $Z_c(3900)$ state is not produced in $B\to Z\,K$ decays \cite{Chilikin:2014bkk},
which casts doubt on any model in which these two states have essentially the same internal structure, differing
only by a radial excitation. 

Molecules are not likely explanations for most of the other hidden-charm non-standard mesons. For example,
when the $Y(4260)$ was first reported by BaBar with a mass of $4259^{+8}_{-10}$~MeV, its interpretation
as a $D\bar{D}_1(2420)$ molecule with a binding energy of $\simeq 25$~MeV might have been
plausible~\cite{Wang:2013cya}.  But recent high-statistics measurements by BESIII have shown that
single resonance fits to the $Y(4260)$ were na\"ive and the peak is, in fact, best fit by two
resonances (see Fig.~\ref{fig:bes3_pipijpsi-scan}) with masses $4220\pm 4$~MeV and
$4320\pm 13$~MeV~\cite{Wang:2013cya}.  This implies that the required $D\bar{D}_1(2420)$ binding
energy for the lower-mass peak would be much higher, at $\simeq 65$~MeV, while the higher-mass state
would be unbound by $\simeq 35$~MeV,  making a molecular interpretation for either component of this
peak implausible.

The only relevant two-particle threshold with $J^{PC}=0^{++}$ S-wave quantum numbers that is near the
$X(3915)$ meson is the $D_s^+ D_s^-$ threshold at $3937$~MeV.  However, since three-pseudoscalar-meson
couplings are forbidden by parity plus rotation invariance, $\pi$-meson exchange forces, which provide
the bulk of the binding for the deuteron, are not applicable.  Thus, there is no plausible deuson-like
model that can account for the nature of the $X(3915)$.  Likewise, before the LHCb group established the
$J^{PC}$ of the $X(4140)$ resonance as $1^{++}$, there were a number of suggestions that it was an S-wave
$D^{*+}_sD^{*-}_s$ molecule ($2m_{D^{*-}_s}=4224$~MeV).  But an S-wave $D^{*+}_sD^{*-}_s$ system can only
have $J^{PC}=0^{++}$ or $2^{++}$ (since $C=(-1)^{L+S}$) \cite{Bondar:2016hva}.
The LHCb measurement \cite{Aaij:2016iza} rules out this possibility and, in addition, they
reported three additional candidates for $c\bar{c}s\bar{s}$ states: the $X(4274)$, $X(4500)$ and 
$X(4700)$.  None of these are close to the thresholds of any S-wave combinations of $D_s$ excitations that
would reproduce their measured quantum numbers.  

Some authors have suggested that the narrow $P_c(4450)$ pentaquark might be an S-wave,
$J^{P}=3/2^-$ molecule comprised of a $\Sigma_c$ baryon and a $\bar{D}^*$ meson, with a binding energy
of $\simeq  13$~MeV~\cite{Karliner:2015ina,Roca:2015dva}. This, taken alone, might be a reasonable suggestion.  
However, if the $P_c(4450)$ is $3/2^-,$ the LHCb data indicate that the $P_c(4380)$ should be $5/2^+$~\cite{Aaij:2015tga}. 
The lightest meson-baryon combinations that can produce spin $5/2$ in an S-wave  are $p\chi_{c2}$ (for
which $\pi$-exchange forces are not allowed) and $\Sigma_c^*\bar{D}^*$, with mass thresholds that are
115~and~145 MeV above the $P_c(4380)$ mass, respectively.  
Moreover, the $\sim 200$~MeV width of the lower-mass $P_c(4380)$ state makes 
it pretty short-lived  for a molecular state with $c$ and $\bar{c}$ spatially separated
into two different confining volumes, resulting in a small overlap of its wave function with the $J/\psi$ 
state that is produced in its decay~\cite{Karliner:2016via}. 
These thresholds are also too high to provide a plausible molecular explanation for the case where the
$J^P=5/2^{\pm}$ assignment is associated with the heavier and narrower $P_c(4450)$ peak. This is not only
the case for a bound molecule interpretation, 
but also for cusp or triangle anomaly mechanisms, which require S-wave interactions to make
significant contributions~\cite{Bayar:2016ftu}. 

Thus, a reasonable conclusion is that while molecule models have relevance for some of the observed
candidates, they are not the whole story.

\subsubsection{Diquarks}
\label{sec:summ-diquarks}

QCD, which explains the existence of $q\bar{q}$ mesons and $qqq$ baryons, also predicts that 
the short-distance color force between two quarks in an $S=0$ diquark antitriplet state
(see Fig.~\ref{fig:diquarks}b) is attractive and one-half the strength of the attraction
between a quark and antiquark in a standard meson.  
Therefore, diquarks and diantiquarks are expected
to play a strong role in shaping the spectroscopy of multi-quark
hadrons, when pairs of quarks exist inside the confinement volume~\cite{Jaffe:1976ig}.  

In 2004, Maiani {\em et.~al}~\cite{Maiani:2004vq}  proposed that the $X(3872)$ is formed from
a symmetric combination of an $S=0$ and $S=1$ diquark and diantiquark in a relative S-wave:
$[cq]_{S=0}[\bar{c}\bar{q}]_{S=1} + [cq]_{S=1}[\bar{c}\bar{q}]_{S=0}$, with $q=u$~or~$d$. This implies the
existence of two nearly degenerate neutral states, $X_h$ and $X_l$, that are mixtures of $X_u=[cu][\bar{c}\bar{u}]$
and $X_d=[cd][\bar{c}\bar{d}]$, with a mass difference:
\begin{eqnarray}
M(X_h)-M(X_l)&=&2(m_d-m_u)/\cos 2\theta  \nonumber \\
             &=&(7\pm 2)~{\rm MeV}/\cos 2\theta , \label{eqn:deltam}
\end{eqnarray}
where $\theta$ is the mixing angle.  Here, an unequal mixture of the two states
({\em i.e.}, $\theta\neq 45^0$) would generate isospin violation in $X_h$ and
$X_l$ decays.   
Tetraquarks made out of diquarks must be compact due to the confinement of color,
which naturally can explain the large $X(3872)$ prompt production rate at hadron
colliders.  
However, this model predicted the existence of three additional states: two $0^{++}$ states 
with lower mass values and a $2^{++}$ state at a higher mass.  
The production of two neutral $X(3872)$ states differing in mass by an amount consistent
with the Eq.~\ref{eqn:deltam} prediction was ruled for high energy $p\bar{p}$
production by CDF~\cite{Aaltonen:2009vj} and for production in $B$ meson decays by 
BaBar~\cite{Aubert:2008gu} and Belle~\cite{Choi:2011fc}.  Compelling candidates
for the predicted $0^{++}$ and $2^{++}$ states have yet to be identified.
Similar to the case for molecular models, 
the significant $X(3872)\to\gamma\psi(2S)$ rate relative to the $X(3872)\to\gamma \jpsi(1S)$ 
decays is not naturally explained by the tetraquark model, unless it also includes mixing with the
$\chi_{c1}(2{\rm P})$ charmonium state or dynamical effects \cite{Brodsky:2014xia} discussed below.    
Moreover, unlike molecular models, the tetraquark picture does not
offer natural explanations for the $X(3872)$ mass coincidence with the $D^0\bar{D}^{*0}$
threshold, or its very narrow width. 
The latter led to hybridized-tetraquark model, which explains the near-threshold states as an 
interplay between bound molecules and compact tetraquark states \cite{Esposito:2016itg}.    

Diquark interpretations have been proposed to explain most of the observed states.
For example, in this model, the $Y(4260)$ is a symmetric $[cq]_{S=0}[\bar{c}\bar{q}]_{S=1}$
diquark-diantiquark combination, like the $X(3872)$, but in a relative P-wave~\cite{Maiani:2014aja}.
The $\simeq 350$~MeV mass difference between the two states is consistent with the typical
mass penalty that is associated with the addition of one unit of orbital angular
momentum (see, {\em e.g.}, Table~\ref{tab:Dsj}) and the strong $Y(4260)\rt\gamma X(3872)$
transition reported by BESIII~\cite{Ablikim:2013dyn} would be an allowed electric dipole
transition between the two related states~\cite{Chen:2015dig}.  Likewise the $Z_c(3900)$
can be naturally accommodated as the {\em antisymmetric}, S-wave
$[cq]_{S=0}[\bar{c}\bar{q}^{\prime}]_{S=1}$ diquark-diantiquark combination, and the $Z(4430)$ as
its first radial excitation.  The diquark picture also can explain states that molecular
models have trouble with, such as the $X(3915)$~\cite{Lebed:2016yvr} and the four states 
seen decaying to $\jpsi\phi$~\cite{Chen:2016oma,Maiani:2016wlq}, and provide a
common origin for the $P_c(4380)$ and $P_c(4450)$ pentaquarks that can accommodate
their opposite parities~\cite{Maiani:2015vwa,Lebed:2015tna}.

These successes can be attributed to a larger number of degrees of freedom in strong color binding
as compared to weak, Yukawa-type exchange forces. Orbital excitations between colored objects 
can reach higher spin states at lower masses. Slight changes in free parameters that describe the 
strength of the color forces, can shift predicted masses to any value, while molecular models
can only accommodate structures near hadron-hadron mass thresholds. 
However, the free parameters of diquark models are usually readjusted when moving from the description of
one exotic hadron system to another, resulting in no single unified diquark model that can
describe all of the observed states at once.  The universal prediction of large prompt production rates for
tightly bound tetraquarks and pentaquarks at hadron colliders are in some conflict with the reality that
to date, only the $X(3872)$ and $X(4140)$ states have been seen in this production mode~\cite{Bondar:2016hva}.
 
One big problem with the diquark picture is that, since the strong radial color force
between the diquark and the diantiquark is universal, it has the same strength for
$q,q^{\prime}=u,d,s$ independently of their flavor or spin state. 
Therefore, every successful application
of the idea to an experimentally observed state carries with it predictions for a large number
of related states that typically are not seen, as was the case for the original
diquark-diantiquark interpretation of the $X(3872)$ discussed above. 
Another problem is that in this picture,  diquark separations are comparable to the diquark sizes,
and there is nothing in the model to prevent fast fall-apart widths to individually confined mesons
and baryons, thereby precluding the existence of narrow states. 

Brodsky, Hwang and Lebed~\cite{Brodsky:2014xia}
address these problems with a scheme that creates separation between diquarks, severely restricts the number
of observed states, and suppresses the superfluous ones. Their basic idea, which they call {\em dynamical diquarks},
is that at production, the diquark and diantiquark move rapidly apart, a motion that is opposed
by the increasingly stronger confinement force.  By the time the motion stops, the diquark
diantiquark separation is large and there is no substantial overlap between the quarks in
one and the antiquarks in the other.  Hadronization of diquarks separated dynamically by this
production mechanism depends on the overlap of the tails of the quark and antiquark wave-functions and only
occurs if this overlap is well matched to an accessible configuration of final-state hadrons. In this way
the process has a complex and intricate dependence on the production mechanism and decay final states
that suppresses all the unwanted states. But this comes at the cost of stripping the dynamical diquark picture
of virtually all of its ability to predict the existence of additional states. 

Baryons with heavy quarks are an interesting testing ground for diquark models. 
For example, in heavy-light-light quark baryons, spin couplings of the heavy quark are expected 
to be suppressed by its heavy mass creating favorable conditions for formation of a light diquark, $[qq]Q$. 
Since a good diquark would be in the same color state as $\bar{q}$, meson-like radial and orbital-angular-momentum excitation 
spectrum is expected for such a system. The known excitations of charmed and beauty baryons follow these predictions.  
For example, the LHCb has recently observed five narrow $\Omega_c$ excitations  \cite{Aaij:2017nav}, 
that lie in the mass range of 1P and 2S states predicted in diquark models \cite{Ebert:2011kk,Karliner:2017kfm}. 
Their masses extend to just below $m_\Xi+m_D$, which is the 
threshold for decays that leave the $ss$ diquark intact. 
Therefore, their strong decays to  $\Xi_c^+K^-$ requires the disintegration of the $ss$ diquark, 
which can explain their narrow widths (0.8-8.7 MeV) in spite of the fact that there is no OZI suppression.  

In doubly-heavy baryons, the heavy quark pair is likely to form a diquark, $[QQ]q$, acting like an effective heavy antiquark. 
Using an approach based on this idea, Karliner and Rosner \cite{Karliner:2014gca} predicted with good precision the mass of
the $\Xi_{cc}^{++}$ baryon, that was recently 
observed by LHCb \cite{Aaij:2017ueg}. 
They have extended this model to predict that the mass of the lightest 
$1^+$ $[bb][\bar{u}\bar{d}]$ tetraquark system would be well below the threshold 
for decays to $B^-\bar{B}^0\gamma$, and therefore be stable under strong and electromagnetic decays (see also refs.~\cite{Eichten:2017ffp}).  
The similar  $[bc][\bar{u}\bar{d}]$ tetraquark state is also likely to be stable, 
while  $[bc][\bar{u}\bar{d}]$ may be right below, or right above the $D^0D^{*+}$ threshold.  
Almost the same predictions for these states were obtained a year earlier 
by Francis {\em et al.} employing LQCD calculations with diquark interpolators \cite{Francis:2016hui}. 
Therefore, these are now perhaps the most firm theoretical predictions 
offering hope for unambiguously establishing the existence of diquark tetraquarks in not too distant future.

\subsubsection{QCD hybrids}

The existence of hybrid states is among the most intriguing predictions of QCD.
The most striking experimental signature for QCD hybrids is the possible existence of mesons
with exotic quantum numbers. However, since all the states discussed in this report have
non-exotic quantum numbers, there is no smoking gun candidate for a hybrid
state.  The state that has been most strongly promoted as a candidate for a $\ccbar$-gluon
charmonium hybrid is the $1^{--}$ $Y(4260)$~\cite{Zhu:2005hp,Close:2005iz,Kou:2005gt}.  However,
this assignment is not unique, and this state is also considered to be a good QCD tetraquark candidate. 
Charmonium hybrids are expected to have strong decays to S-wave $D^{(*)}\bar{D}^{**}$ final states, where $D^{**}$
designates the P-wave charmed mesons described in Table~\ref{tab:Dsj}.  For the $Y(4260)$, the
only possible accessible final state of this type would be $D^*\bar{D}_0^*(2400)$, where the
$D_0^*(2400)$ is a $0^{++}$ $D\pi$ resonance with mass and width $M=2318\pm 29$~MeV and
$\Gamma=267\pm 40$~MeV. In principle, a high-statistics study of $Y(4260)\rt D^*\bar{D}\pi$
might see evidence for a $D^*\bar{D}^*_0$ final state but, since it is a broad S-wave state, it would
probably be difficult to distinguish from a non-resonant background.  As mentioned in Section~\ref{sec:y4260},
the latest lattice QCD study of charmonium states has a lowest-mass $1^{--}$ hybrid candidate at
$M=4285\pm 14$~MeV~\cite{Liu:2012ze}.  This is $\simeq 65$~MeV above the recent BESIII measurement
of the $Y(4260)$ mass ($4220\pm4$~MeV). However, given that the LQCD calculations are missing 
simulations of couplings to the available decay channels, this difference should probably be considered small. 

\subsubsection{Hadrocharmonium}
In the hadrocharmonium model, the strong preference for the $Y(4260)$ ($Y(4360)$) meson to decay to
$\pipi\jpsi$ ($\pipi\psip$)  over the fall-apart $D^{(*)}\bar{D}^{(*)}$ modes is accommodated by
a meson structure that contains a color-singlet charmonium-state core surrounded by
a light quark excitation~\cite{Dubynskiy:2008mq}. In the case of the $Y(4260)$ ($Y(4360)$), this
core state was taken to be the $\jpsi$ ($\psip$). Since the $\jpsi$ ($\psip$) state is present
in its constituents, the $Y(4260)$ ($Y(4360)$) naturally prefers to decay to final states
that include it. However this model had trouble with
BESIII observations of substantial rates for $Y(4260)$ and $Y(4360)$ decays to $\pipi h_c$
final states~\cite{Ablikim:2013wzq}.  In the $\jpsi$ and $\psip$, the $c$- and
$\bar{c}$-quarks are in a spin-triplet state, while in the $h_c$ they are in a spin-singlet.
By themselves, the triplet and singlet cores cannot mix and a hadrocharmonium state should
have strong rates for decays to one of them, but not both.  In an attempt to fix this problem,
the model was revised to include two hadrocharmonium states, one with a spin-triplet core and
one with a spin-singlet core, and these two states mix, producing the observed $Y(4260)$ and
$Y(4360)$, which now can both decay to singlet and triplet states with relative
rates determined by the (unspecified) mixing angle~\cite{Li:2013ssa}. But, subsequent to the
appearance of this modification to the model, the BESIII group found that in the two resonance
$Y(4260)\rt\pipi\jpsi$ structure shown in Fig.~\ref{fig:bes3_pipijpsi-scan}~\cite{Ablikim:2016qzw},
both resonant components have strong decay rates to $\pipi\jpsi$, while the latest BESIII
$\pipi h_c$ cross section measurements~\cite{BESIII:2016adj}, shown in Fig.~\ref{fig:bes3_pipihc-scan},
find a strong $\pipi h_c$ decay rate for the lower mass resonant component, but no signal
for the higher mass component. Even with the above-described mixing, the hadrocharmonium model will
have a hard time reproducing this decay pattern.

\subsubsection{Kinematically induced resonance-like peaks}
As mentioned above in the discussion of molecular models, some authors have suggested that the
$Z_c$ and $Z_b$ peaks are due to kinematic effects caused by the nearby $D^{(*)}\bar{D}^*$ and
$B^{(*)}\bar{B}^*$ mass thresholds~\cite{Bugg:2011jr,Blitz:2015nra,Swanson:2015bsa,Chen:2013coa}.
Other authors claim that the same data show that the peaks are are associated with poles in the
corresponding $D^{(*)}\bar{D}^*$ or $B^{(*)}\bar{B}^*$ scattering t-matrix and, therefore,
qualify as genuine physical states~\cite{Guo:2014iya,Guo:2016bjq}.  This seems to be a
controversy that only more, and higher-precision data can resolve.  For example, although both
mechanisms produce strong phase motion that can be displayed in Argand plots, the detailed mass
dependence of this phase motion is distinctly different for the two models. Another suggestion requires the application
of sophisticated coupled-channel data analysis techniques~\cite{Pilloni:2016obd}. However, distinguishing between
the two scenarios with either method will require huge data samples, with increases over existing data sets
by factors that approach two orders of magnitude.  This would require at least a year-long dedicated BESIII
run at energies near the $Y(4260)$ peak and a long period of BelleII operation at the $\Upsilon(10860)$ peak.

\subsubsection{Comments}
At this point, the challenges posed by the observed heavy quarkonium-like states with 
unusual properties have not been answered well by any theory for hadronic states.    

In general the successful theoretical work in this area has been reactive rather than predictive. 
Different approaches have had some success for some states but fail or are not applicable to others.  
This could be because the non-standard hadrons that are observed are due to a variety of different, 
unrelated mechanisms, or that the actual underlying mechanism has not yet been discovered.  
The lack of good understanding of the nature of these states, sheds doubts into our understanding of
hadronic structures in general. Open-flavor thresholds in light hadron spectroscopy are 
usually right above the ground states of such systems and whatever mechanisms that are in play for
above-open-flavor-threshold heavy quarkonium-like hadrons are likely having a strong influence
on light hadron spectroscopy.

More theoretical and experimental work is required in all areas of hadron spectroscopy to 
overcome this crisis. 

\subsection{Experiment}
\label{sec:summ_expt}

To date, the progress in the field has mainly been driven by experiments. Although
T\"ornqvist proposed the existence of a $J^{PC}=1^{++}$ state with mass near the $D\bar{D}^*$
threshold ten years before its discovery~\cite{Tornqvist:1993ng}, the $X(3872)$'s narrow
width, close proximity to the $D^0\bar{D}^{*0}$, large isospin violation, and its
strong production in $B$-meson decays and high-energy $p\bar{p}$ collisions were big
surprises. Likewise, the discovery of the $Y(4260)$ as a huge peak in the $\ee\rt\pipi\jpsi$
cross section with no hints of a corresponding signal in the open charmed meson channel was
totally unexpected.  Subsequent discoveries, both in the $c$-quark and $b$-quark sectors, were
similarly unanticipated.
New developments in this field have closely tracked increases in the size and improvements in
the quality of the available data samples plus advances in the sophistication of analysis techniques.
The discovery of the $P_c(4380)$ and $P_c(4450)$ pentaquarks was possible because of the unprecedentedly
large sample of $\Lambda_b$ decays that was accumulated by the LHCb experiment. Likewise, the two-resonance
structure of the $Y(4260)$ was seen by BESIII only because they had a larger data sample and better mass
resolution than the earlier measurements. The clear demonstration of BW-amplitude-like phase motion for
the $Z(4430)$ meson and the $P_c(4450)$ pentaquark was the result of complex, multidimensional
amplitude analyses. 

There are huge differences in what we know about various candidate states. For some, like the $X(3872)$
and the $Y(4260)$, we know the quantum numbers and a number of decay modes, while for others we do not even
know their quantum numbers.  Even the ones with a lot of available information have important pieces of information
that are unknown.  For example, we still do not know the natural width of the $X(3872)$ or whether its mass is
above or below the $m_{D^0}+m_{D^{*0}}$ mass threshold.  We have not seen phase motion for the $Y(4260)$, or convincing
phase motion for the $P_c(4380)$.  While we learned a great deal about the nature of the $X(3872)$ from the fact
that it is produced promptly in high energy $p\bar{p}$ and $pp$ collisions, we only have limited information
about prompt hadro-production of other states.

For experimental reasons, most of the states seen so far were first discovered in decay modes that include a 
$\jpsi$ or a $\psip$ the final state.  Future experiments that access pairs of open-charm or beauty particles
may uncover a interesting dimensions of this spectroscopy that might give important clues about the
underlying dynamics.

In addition to finding new states, it would be very useful if our level of knowledge of all of the candidate
states could be brought to the same high level as the currently best known ones.  For example,
multidimensional amplitude analyses of $B\rt K\jpsi\omega$ and $K\chi_{c1}\pi$ are needed to establish the
$J^{PC}$ quantum numbers of the $X(3915)$, and the $Z(4050)^+$ and $Z(4250)^+$.

Fortunately there are powerful experiments that are currently running and producing important and unique
results. For example, $B\rt K\chi_{c1}\pi$ and $K\jpsi\omega$ are accessible to the LHCb even with their
existing data sample. Their recent discoveries of $\jpsi\phi$ mesons and $P_c$ pentaquarks were based on
analyses of their  3~fb$^{-1}$ Run-I data sample that was accumulated at $\ecmx{pp}=7~{\rm and}~8$~TeV. 
In Run-II, which is now underway and will finish in 2018, they will accumulate an additional 
8~fb$^{-1}$ at $\ecmx{pp}=13$~TeV. Since the
$b$-quark production cross section at 13~TeV is about twice that at 7~TeV~\cite{Aaij:2016avz}, the Run-II  
data set will be equivalent to approximately five times that for Run-I. Thus, we can anticipate the discovery
of additional states and significant improvements of existing results, including substantial improvements in the precision
of the $Z(4430)$ and $P_c$ Argand plots. A LHCb detector upgrade in 2021  will enable it to accumulate higher luminosities.
After that they expect that the total data sample accumulated  by 2030 will be 50~fb$^{-1}$ (collected at $\ecmx{pp}=14$~TeV). 
The LHCb collaboration is also discussing a major detector/luminosity upgrade after 2030 with an ultimate goal  
of 300 fb$^{-1}$.  Even larger integrated luminosities will be collected at the LHC by the CMS and ATLAS 
detectors. Therefore, it is expected that these experiments will also continue to contribute to advances in exotic
hadron spectroscopy in  spite of their lack of hadron identification and their more restrictive triggers.   

BESIII is planning long data runs in the Y(4260) peak region during the next few years, and these
should  provide sufficient data to support studies of the phase motion across the $Z_c$ peaks in 
$Y(4260)\rt \pipi\jpsi$ and $\pipi h_c$ decay channels, and enable amplitude analyses of
$Y(4260)\rt\pi D^{(*)}\bar{D}^*$  decays.  Long-term future BESIII running plans include high luminosity scans
over the $2m_D\le \ecmx{\ee}\le 2m_{\Lambda_c}+ 50$~MeV range to completely map out $c$-quark production in the
threshold region in a large assortment of decay channels.

The  BelleII experiment will start physics operation in late 2017.  The first year of running 
will be used to develop experience with machine operations without either the inner pixel or the silicon-strip
vertex detector in place. Since this configuration has limited capabilities for doing $B$-meson physics, which
is the main motivation for the project, operation at the $\Upsilon(10860)$ and higher energies for detailed studies
of the $Z_b$ mesons and searches for possible additional states is planned~\cite{Bondar:2016hva}.   Over the anticipated
ten-year operational lifetime of BelleII, a total data sample of 50~ab$^{-1}$ will be accumulated, mostly at
$\ecmx{\ee}=10.58$~GeV. Thus, the final BelleII data sample will be a factor of 50 larger than the Belle data
set.  The strengths of BelleII, {\em i.e.}~superior photon, $\pi^0$, $\omega$ and $\eta^{(')} $ detection capabilities and
higher absolute reconstruction efficiencies, are complementary to those of LHCb. High reconstruction efficiencies
are especially essential for studies of $D^{(*)}_{(s)}\bar{D}^{(*)}_{(s)}$ systems.  In addition, the $B$-factory environment
provides a unique opportunity for making precise measurements of inclusive branching fractions such as
$B^+\rt K^+\,X_{c\bar{c}}$; $X_{c\bar{c}}\rt\,{\rm anything}$ decays~\cite{Aubert:2005vi},
where $X_{c\bar{c}}$ designates relatively narrow charmonium-like states such as the $X(3872)$ or $X(3915)$.
Inclusive branching fractions are required input for converting product branching fraction and total
width measurements into decay partial width values, which are usually the theoretically relevant quantities.

There are several current and future medium-energy experiments designed to explore non-standard
hadrons and to provide information on the properties of the still enigmatic states
complementary to $e^+e^-$ and hadron collider experiments.

The $\rm {PANDA}$ experiment~\cite{Lutz:2009ff} at the Facility for Antiproton and Ion Research in Germany, expected
to start data taking in 2022, will study particles produced in collisions of an intense, nearly monoenergetic beam
of antiprotons with nucleons or nuclei in the c.m.\ energy range from 2.5 GeV to 5.5 GeV.
By performing a mass scan in several steps across the $X(3872)$ resonance,  $\rm {PANDA}$ will measure
the resonance line shape that is sensitive to the binding mechanism with an accuracy an order of magnitude better
than currently available.  To resolve the nature of   the  $D^*_{s0}(2317)$,  $\rm {PANDA}$  will do
precise measurements of its total width and branching fractions to different decay channels.
The $p\bar{p}$ initial state will provide access to possible states with exotic quantum numbers.
Lattice QCD indicates that the lightest  exotic $c\overline c$ hybrid with $J^{PC}=1^{-+}$ 
has a mass that is near 4.2 GeV~\cite{Liu:2012ze}, and will be accessible at  $\rm {PANDA}$.

Discovery and studies of properties of hybrid mesons are the primary goals of the 
GlueX~\cite{Ghoul:2015ifw}  and CLAS12~\cite{Burkert:2008rj} photo-production experiments
  at the Jefferson National Accelerator Facility. 
Soon after  the discovery of the LHCb pentaquark candidates, 
it was suggested~\cite{Wang:2015jsa,Kubarovsky:2015aaa,Karliner:2015voa,Blin:2016dlf}  that 
photo-production on a nucleon would be a promising way
to search for these states and to study their properties.
Recently, a proposal~\cite{Meziani:2016lhg} to  study photo-production of $J/\psi$ near threshold
in  search for the $P_c$ states was approved by JLab and designated as a ``high-impact'' activity. 

\subsection{Final remark}

The heavy-flavor, non-standard hadrons discussed in this report have severely challenged existing ideas
about the underlying structure of hadrons.  The puzzles that they pose have intrigued theorists in both
the particle and nuclear physics communities, and experimenters at all of the world's particle physics accelerator
facilities. 
While they do not challenge QCD as the exact theory of strong interactions, 
they expose the phenomenological disconnect between its Lagrangian  and types of structures it can produce at large distances, 
with possible implications to strongly coupled theories proposed as extensions of the Standard Model.  
Often in the history of physics, puzzles and their eventual resolution have produced important advances
in our understanding of nature.  We hope that this will ultimately be the case for the issues that are discussed here.  

\vskip0.5cm

\noindent{\bf Acknowledgements:}
This work was supported by the Institute for Basic Science (Korea) by project code IBS-R016-D1
and by the National Science Foundation (USA) Award Number 1507572.
We would like to thank Steven Gottlieb for reading
the manuscript and providing a number of useful comments.

\myclearpage

\bibliographystyle{LHCb}
\bibliography{rmp_master}

\ifx\mcitethebibliography\mciteundefinedmacro
\PackageError{LHCb.bst}{mciteplus.sty has not been loaded}
{This bibstyle requires the use of the mciteplus package.}\fi
\providecommand{\href}[2]{#2}
\begin{mcitethebibliography}{100}
\mciteSetBstSublistMode{n}
\mciteSetBstMaxWidthForm{subitem}{\alph{mcitesubitemcount})}
\mciteSetBstSublistLabelBeginEnd{\mcitemaxwidthsubitemform\space}
{\relax}{\relax}

\bibitem{Gellmann:1964nj}
M.~Gell-Mann, \ifthenelse{\boolean{articletitles}}{\emph{{A schematic model of
  baryons and mesons}},
  }{}\href{http://dx.doi.org/10.1016/S0031-9163(64)92001-3}{Phys.\ Lett.\
  \textbf{8} (1964) 214}\relax
\mciteBstWouldAddEndPuncttrue
\mciteSetBstMidEndSepPunct{\mcitedefaultmidpunct}
{\mcitedefaultendpunct}{\mcitedefaultseppunct}\relax
\EndOfBibitem
\bibitem{Zweig:1981pd}
G.~Zweig, \ifthenelse{\boolean{articletitles}}{\emph{{An SU$_3$ model for
  strong interaction symmetry and its breaking}}, }{}
\newblock
  {\href{http://cds.cern.ch/record/352337/files/CERN-TH-401.pdf}{CERN-TH-401,
  1964}}\relax
\mciteBstWouldAddEndPuncttrue
\mciteSetBstMidEndSepPunct{\mcitedefaultmidpunct}
{\mcitedefaultendpunct}{\mcitedefaultseppunct}\relax
\EndOfBibitem
\bibitem{Greenberg:1964pe}
O.~W. Greenberg, \ifthenelse{\boolean{articletitles}}{\emph{{Spin and Unitary
  Spin Independence in a Paraquark Model of Baryons and Mesons}},
  }{}\href{http://dx.doi.org/10.1103/PhysRevLett.13.598}{Phys.\ Rev.\ Lett.\
  \textbf{13} (1964) 598}\relax
\mciteBstWouldAddEndPuncttrue
\mciteSetBstMidEndSepPunct{\mcitedefaultmidpunct}
{\mcitedefaultendpunct}{\mcitedefaultseppunct}\relax
\EndOfBibitem
\bibitem{Han:1965pf}
M.~Y. Han and Y.~Nambu, \ifthenelse{\boolean{articletitles}}{\emph{{Three
  Triplet Model with Double SU(3) Symmetry}},
  }{}\href{http://dx.doi.org/10.1103/PhysRev.139.B1006}{Phys.\ Rev.\
  \textbf{139} (1965) B1006}\relax
\mciteBstWouldAddEndPuncttrue
\mciteSetBstMidEndSepPunct{\mcitedefaultmidpunct}
{\mcitedefaultendpunct}{\mcitedefaultseppunct}\relax
\EndOfBibitem
\bibitem{Litke:1973np}
Mark I, A.~Litke {\em et~al.},
  \ifthenelse{\boolean{articletitles}}{\emph{{Hadron Production by electron
  positron annihilation at 4-GeV center-of-mass energy}},
  }{}\href{http://dx.doi.org/10.1103/PhysRevLett.30.1189}{Phys.\ Rev.\ Lett.\
  \textbf{30} (1973) 1189}\relax
\mciteBstWouldAddEndPuncttrue
\mciteSetBstMidEndSepPunct{\mcitedefaultmidpunct}
{\mcitedefaultendpunct}{\mcitedefaultseppunct}\relax
\EndOfBibitem
\bibitem{Bardeen:1972xk}
W.~A. Bardeen, H.~Fritzsch, and M.~Gell-Mann,
  \ifthenelse{\boolean{articletitles}}{\emph{{Light cone current algebra,
  $\pi^0$ decay, and $e^+e^-$ annihilation}}, }{} in {\em {Topical Meeting on
  the Outlook for Broken Conformal Symmetry in Elementary Particle Physics
  Frascati, Italy, May 4-5, 1972}}, 1972.
\newblock \href{http://arxiv.org/abs/hep-ph/0211388}{{\normalfont\ttfamily
  arXiv:hep-ph/0211388}}\relax
\mciteBstWouldAddEndPuncttrue
\mciteSetBstMidEndSepPunct{\mcitedefaultmidpunct}
{\mcitedefaultendpunct}{\mcitedefaultseppunct}\relax
\EndOfBibitem
\bibitem{Gross:1973id}
D.~J. Gross and F.~Wilczek,
  \ifthenelse{\boolean{articletitles}}{\emph{{Ultraviolet Behavior of
  Nonabelian Gauge Theories}},
  }{}\href{http://dx.doi.org/10.1103/PhysRevLett.30.1343}{Phys.\ Rev.\ Lett.\
  \textbf{30} (1973) 1343}\relax
\mciteBstWouldAddEndPuncttrue
\mciteSetBstMidEndSepPunct{\mcitedefaultmidpunct}
{\mcitedefaultendpunct}{\mcitedefaultseppunct}\relax
\EndOfBibitem
\bibitem{Politzer:1973fx}
H.~D. Politzer, \ifthenelse{\boolean{articletitles}}{\emph{{Reliable
  Perturbative Results for Strong Interactions?}},
  }{}\href{http://dx.doi.org/10.1103/PhysRevLett.30.1346}{Phys.\ Rev.\ Lett.\
  \textbf{30} (1973) 1346}\relax
\mciteBstWouldAddEndPuncttrue
\mciteSetBstMidEndSepPunct{\mcitedefaultmidpunct}
{\mcitedefaultendpunct}{\mcitedefaultseppunct}\relax
\EndOfBibitem
\bibitem{Olive:2016xmw}
Particle Data Group, C.~Patrignani {\em et~al.},
  \ifthenelse{\boolean{articletitles}}{\emph{{Review of Particle Physics}},
  }{}\href{http://dx.doi.org/10.1088/1674-1137/40/10/100001}{Chin.\ Phys.\
  \textbf{C40} (2016) 100001}\relax
\mciteBstWouldAddEndPuncttrue
\mciteSetBstMidEndSepPunct{\mcitedefaultmidpunct}
{\mcitedefaultendpunct}{\mcitedefaultseppunct}\relax
\EndOfBibitem
\bibitem{Jaffe:1976ig}
R.~L. Jaffe, \ifthenelse{\boolean{articletitles}}{\emph{{Multi-Quark Hadrons.
  1. The Phenomenology of (2 Quark 2 anti-Quark) Mesons}},
  }{}\href{http://dx.doi.org/10.1103/PhysRevD.15.267}{Phys.\ Rev.\
  \textbf{D15} (1977) 267}\relax
\mciteBstWouldAddEndPuncttrue
\mciteSetBstMidEndSepPunct{\mcitedefaultmidpunct}
{\mcitedefaultendpunct}{\mcitedefaultseppunct}\relax
\EndOfBibitem
\bibitem{AlvarezRuso:2010xr}
L.~Alvarez-Ruso, \ifthenelse{\boolean{articletitles}}{\emph{{On the nature of
  the Roper resonance}}, }{} in {\em {Dressing hadrons. Proceedings,
  Mini-Workshop, Bled, Slovenia, July 4-11, 2010}}, pp.~1--8, 2010.
\newblock \href{http://arxiv.org/abs/1011.0609}{{\normalfont\ttfamily
  arXiv:1011.0609}}\relax
\mciteBstWouldAddEndPuncttrue
\mciteSetBstMidEndSepPunct{\mcitedefaultmidpunct}
{\mcitedefaultendpunct}{\mcitedefaultseppunct}\relax
\EndOfBibitem
\bibitem{Close:1980ab}
F.~E. Close and R.~H. Dalitz, \ifthenelse{\boolean{articletitles}}{\emph{{The
  Antisymmetric Spin Orbit Interaction Between Quarks}}, }{} in {\em {Workshop
  on Low-energy and Intermediate-energy Kaon-Nucleon Physics Rome, Italy, March
  24-28, 1980}}, p.~411, 1980\relax
\mciteBstWouldAddEndPuncttrue
\mciteSetBstMidEndSepPunct{\mcitedefaultmidpunct}
{\mcitedefaultendpunct}{\mcitedefaultseppunct}\relax
\EndOfBibitem
\bibitem{wilczek16}
F.~Wliczek, \ifthenelse{\boolean{articletitles}}{\emph{{Power Over Nature}},
  }{}
\newblock
  {\href{http://www.edge.org/conversation/frank\_wilczek-power-over-nature}{http://www.edge.org/conversation/frank\_wilczek-power-over-nature}}\relax
\mciteBstWouldAddEndPuncttrue
\mciteSetBstMidEndSepPunct{\mcitedefaultmidpunct}
{\mcitedefaultendpunct}{\mcitedefaultseppunct}\relax
\EndOfBibitem
\bibitem{Jegerlehner:2009ry}
F.~Jegerlehner and A.~Nyffeler, \ifthenelse{\boolean{articletitles}}{\emph{{The
  Muon g-2}}, }{}\href{http://dx.doi.org/10.1016/j.physrep.2009.04.003}{Phys.\
  Rept.\  \textbf{477} (2009) 1},
  \href{http://arxiv.org/abs/0902.3360}{{\normalfont\ttfamily
  arXiv:0902.3360}}\relax
\mciteBstWouldAddEndPuncttrue
\mciteSetBstMidEndSepPunct{\mcitedefaultmidpunct}
{\mcitedefaultendpunct}{\mcitedefaultseppunct}\relax
\EndOfBibitem
\bibitem{pdg76}
Particle Data Group, T.~G. Trippe {\em et~al.},
  \ifthenelse{\boolean{articletitles}}{\emph{{Review of Particle Properties.
  Particle Data Group}},
  }{}\href{http://dx.doi.org/10.1103/RevModPhys.48.S1}{Rev.\ Mod.\ Phys.\
  \textbf{48} (1976) S1}, [Erratum: Rev. Mod. Phys.48,497(1976)]\relax
\mciteBstWouldAddEndPuncttrue
\mciteSetBstMidEndSepPunct{\mcitedefaultmidpunct}
{\mcitedefaultendpunct}{\mcitedefaultseppunct}\relax
\EndOfBibitem
\bibitem{pdg94}
Particle Data Group, L.~Montanet {\em et~al.},
  \ifthenelse{\boolean{articletitles}}{\emph{{Review of particle properties.
  Particle Data Group}},
  }{}\href{http://dx.doi.org/10.1103/PhysRevD.50.1173}{Phys.\ Rev.\
  \textbf{D50} (1994) 1173}\relax
\mciteBstWouldAddEndPuncttrue
\mciteSetBstMidEndSepPunct{\mcitedefaultmidpunct}
{\mcitedefaultendpunct}{\mcitedefaultseppunct}\relax
\EndOfBibitem
\bibitem{Nakano:2003qx}
LEPS, T.~Nakano {\em et~al.},
  \ifthenelse{\boolean{articletitles}}{\emph{{Evidence for a narrow S = +1
  baryon resonance in photoproduction from the neutron}},
  }{}\href{http://dx.doi.org/10.1103/PhysRevLett.91.012002}{Phys.\ Rev.\ Lett.\
   \textbf{91} (2003) 012002},
  \href{http://arxiv.org/abs/hep-ex/0301020}{{\normalfont\ttfamily
  arXiv:hep-ex/0301020}}\relax
\mciteBstWouldAddEndPuncttrue
\mciteSetBstMidEndSepPunct{\mcitedefaultmidpunct}
{\mcitedefaultendpunct}{\mcitedefaultseppunct}\relax
\EndOfBibitem
\bibitem{Diakonov:1997mm}
D.~Diakonov, V.~Petrov, and M.~V. Polyakov,
  \ifthenelse{\boolean{articletitles}}{\emph{{Exotic anti-decuplet of baryons:
  Prediction from chiral solitons}},
  }{}\href{http://dx.doi.org/10.1007/s002180050406}{Z.\ Phys.\  \textbf{A359}
  (1997) 305}, \href{http://arxiv.org/abs/hep-ph/9703373}{{\normalfont\ttfamily
  arXiv:hep-ph/9703373}}\relax
\mciteBstWouldAddEndPuncttrue
\mciteSetBstMidEndSepPunct{\mcitedefaultmidpunct}
{\mcitedefaultendpunct}{\mcitedefaultseppunct}\relax
\EndOfBibitem
\bibitem{pdg06}
W.-M. Yao {\em et~al.}, \ifthenelse{\boolean{articletitles}}{\emph{{Review of
  Particle Physics}}, }{}{Journal of Physics G} \textbf{33} (2006) 1\relax
\mciteBstWouldAddEndPuncttrue
\mciteSetBstMidEndSepPunct{\mcitedefaultmidpunct}
{\mcitedefaultendpunct}{\mcitedefaultseppunct}\relax
\EndOfBibitem
\bibitem{Schumacher:2005wu}
R.~A. Schumacher, \ifthenelse{\boolean{articletitles}}{\emph{{The Rise and fall
  of pentaquarks in experiments}},
  }{}\href{http://dx.doi.org/10.1063/1.2220285}{AIP Conf.\ Proc.\  \textbf{842}
  (2006) 409},
  \href{http://arxiv.org/abs/nucl-ex/0512042}{{\normalfont\ttfamily
  arXiv:nucl-ex/0512042}}, [,409(2005)]\relax
\mciteBstWouldAddEndPuncttrue
\mciteSetBstMidEndSepPunct{\mcitedefaultmidpunct}
{\mcitedefaultendpunct}{\mcitedefaultseppunct}\relax
\EndOfBibitem
\bibitem{Dzierba:2004db}
A.~R. Dzierba, C.~A. Meyer, and A.~P. Szczepaniak,
  \ifthenelse{\boolean{articletitles}}{\emph{{Reviewing the evidence for
  pentaquarks}}, }{}\href{http://dx.doi.org/10.1088/1742-6596/9/1/036}{J.\
  Phys.\ Conf.\ Ser.\  \textbf{9} (2005) 192},
  \href{http://arxiv.org/abs/hep-ex/0412077}{{\normalfont\ttfamily
  arXiv:hep-ex/0412077}}\relax
\mciteBstWouldAddEndPuncttrue
\mciteSetBstMidEndSepPunct{\mcitedefaultmidpunct}
{\mcitedefaultendpunct}{\mcitedefaultseppunct}\relax
\EndOfBibitem
\bibitem{Hicks:2012zz}
K.~H. Hicks, \ifthenelse{\boolean{articletitles}}{\emph{{On the conundrum of
  the pentaquark}},
  }{}\href{http://dx.doi.org/10.1140/epjh/e2012-20032-0}{Eur.\ Phys.\ J.\
  \textbf{H37} (2012) 1}\relax
\mciteBstWouldAddEndPuncttrue
\mciteSetBstMidEndSepPunct{\mcitedefaultmidpunct}
{\mcitedefaultendpunct}{\mcitedefaultseppunct}\relax
\EndOfBibitem
\bibitem{Meyer:2010ku}
C.~A. Meyer and Y.~Van~Haarlem, \ifthenelse{\boolean{articletitles}}{\emph{{The
  Status of Exotic-quantum-number Mesons}},
  }{}\href{http://dx.doi.org/10.1103/PhysRevC.82.025208}{Phys.\ Rev.\
  \textbf{C82} (2010) 025208},
  \href{http://arxiv.org/abs/1004.5516}{{\normalfont\ttfamily
  arXiv:1004.5516}}\relax
\mciteBstWouldAddEndPuncttrue
\mciteSetBstMidEndSepPunct{\mcitedefaultmidpunct}
{\mcitedefaultendpunct}{\mcitedefaultseppunct}\relax
\EndOfBibitem
\bibitem{Meyer:2015eta}
C.~A. Meyer and E.~S. Swanson,
  \ifthenelse{\boolean{articletitles}}{\emph{{Hybrid Mesons}},
  }{}\href{http://dx.doi.org/10.1016/j.ppnp.2015.03.001}{Prog.\ Part.\ Nucl.\
  Phys.\  \textbf{82} (2015) 21},
  \href{http://arxiv.org/abs/1502.07276}{{\normalfont\ttfamily
  arXiv:1502.07276}}\relax
\mciteBstWouldAddEndPuncttrue
\mciteSetBstMidEndSepPunct{\mcitedefaultmidpunct}
{\mcitedefaultendpunct}{\mcitedefaultseppunct}\relax
\EndOfBibitem
\bibitem{Achasov:2008ut}
N.~N. Achasov, A.~V. Kiselev, and G.~N. Shestakov,
  \ifthenelse{\boolean{articletitles}}{\emph{{Theory of Scalars}},
  }{}\href{http://dx.doi.org/10.1016/j.nuclphysbps.2008.09.012}{Nucl.\ Phys.\
  Proc.\ Suppl.\  \textbf{181-182} (2008) 169},
  \href{http://arxiv.org/abs/0806.0521}{{\normalfont\ttfamily
  arXiv:0806.0521}}\relax
\mciteBstWouldAddEndPuncttrue
\mciteSetBstMidEndSepPunct{\mcitedefaultmidpunct}
{\mcitedefaultendpunct}{\mcitedefaultseppunct}\relax
\EndOfBibitem
\bibitem{Weinstein:1982gc}
J.~D. Weinstein and N.~Isgur, \ifthenelse{\boolean{articletitles}}{\emph{{Do
  Multi-Quark Hadrons Exist?}},
  }{}\href{http://dx.doi.org/10.1103/PhysRevLett.48.659}{Phys.\ Rev.\ Lett.\
  \textbf{48} (1982) 659}\relax
\mciteBstWouldAddEndPuncttrue
\mciteSetBstMidEndSepPunct{\mcitedefaultmidpunct}
{\mcitedefaultendpunct}{\mcitedefaultseppunct}\relax
\EndOfBibitem
\bibitem{Jaffe:2004ph}
R.~L. Jaffe, \ifthenelse{\boolean{articletitles}}{\emph{{Exotica}},
  }{}\href{http://dx.doi.org/10.1016/j.physrep.2004.11.005}{Phys.\ Rept.\
  \textbf{409} (2005) 1},
  \href{http://arxiv.org/abs/hep-ph/0409065}{{\normalfont\ttfamily
  arXiv:hep-ph/0409065}}\relax
\mciteBstWouldAddEndPuncttrue
\mciteSetBstMidEndSepPunct{\mcitedefaultmidpunct}
{\mcitedefaultendpunct}{\mcitedefaultseppunct}\relax
\EndOfBibitem
\bibitem{Jones:1976xy}
L.~W. Jones, \ifthenelse{\boolean{articletitles}}{\emph{{A Review of Quark
  Search Experiments}},
  }{}\href{http://dx.doi.org/10.1103/RevModPhys.49.717}{Rev.\ Mod.\ Phys.\
  \textbf{49} (1977) 717}\relax
\mciteBstWouldAddEndPuncttrue
\mciteSetBstMidEndSepPunct{\mcitedefaultmidpunct}
{\mcitedefaultendpunct}{\mcitedefaultseppunct}\relax
\EndOfBibitem
\bibitem{Lyons:1980ad}
L.~Lyons, \ifthenelse{\boolean{articletitles}}{\emph{{Current Status of Quark
  Search Experiments}},
  }{}\href{http://dx.doi.org/10.1016/0146-6410(81)90014-4}{Prog.\ Part.\ Nucl.\
  Phys.\  \textbf{7} (1981) 157}\relax
\mciteBstWouldAddEndPuncttrue
\mciteSetBstMidEndSepPunct{\mcitedefaultmidpunct}
{\mcitedefaultendpunct}{\mcitedefaultseppunct}\relax
\EndOfBibitem
\bibitem{Aubert:1974js}
E598, J.~J. Aubert {\em et~al.},
  \ifthenelse{\boolean{articletitles}}{\emph{{Experimental Observation of a
  Heavy Particle J}},
  }{}\href{http://dx.doi.org/10.1103/PhysRevLett.33.1404}{Phys.\ Rev.\ Lett.\
  \textbf{33} (1974) 1404}\relax
\mciteBstWouldAddEndPuncttrue
\mciteSetBstMidEndSepPunct{\mcitedefaultmidpunct}
{\mcitedefaultendpunct}{\mcitedefaultseppunct}\relax
\EndOfBibitem
\bibitem{Augustin:1974xw}
SLAC-SP-017, J.~E. Augustin {\em et~al.},
  \ifthenelse{\boolean{articletitles}}{\emph{{Discovery of a Narrow Resonance
  in $e^+e^-$ Annihilation}},
  }{}\href{http://dx.doi.org/10.1103/PhysRevLett.33.1406}{Phys.\ Rev.\ Lett.\
  \textbf{33} (1974) 1406}, [Adv. Exp. Phys.5,141(1976)]\relax
\mciteBstWouldAddEndPuncttrue
\mciteSetBstMidEndSepPunct{\mcitedefaultmidpunct}
{\mcitedefaultendpunct}{\mcitedefaultseppunct}\relax
\EndOfBibitem
\bibitem{Abrams:1974yy}
G.~S. Abrams {\em et~al.}, \ifthenelse{\boolean{articletitles}}{\emph{{The
  Discovery of a Second Narrow Resonance in $e^+e^-$ Annihilation}},
  }{}\href{http://dx.doi.org/10.1103/PhysRevLett.33.1453}{Phys.\ Rev.\ Lett.\
  \textbf{33} (1974) 1453}, [Adv. Exp. Phys.5,150(1976)]\relax
\mciteBstWouldAddEndPuncttrue
\mciteSetBstMidEndSepPunct{\mcitedefaultmidpunct}
{\mcitedefaultendpunct}{\mcitedefaultseppunct}\relax
\EndOfBibitem
\bibitem{Feldman:1975bq}
G.~J. Feldman {\em et~al.},
  \ifthenelse{\boolean{articletitles}}{\emph{{$\psi'(3684)$ Radiative Decays to
  High Mass States}},
  }{}\href{http://dx.doi.org/10.1103/PhysRevLett.35.821}{Phys.\ Rev.\ Lett.\
  \textbf{35} (1975) 821}, [Erratum: Phys. Rev. Lett.35,1184(1975)]\relax
\mciteBstWouldAddEndPuncttrue
\mciteSetBstMidEndSepPunct{\mcitedefaultmidpunct}
{\mcitedefaultendpunct}{\mcitedefaultseppunct}\relax
\EndOfBibitem
\bibitem{Appelquist:1974zd}
T.~Appelquist and H.~D. Politzer,
  \ifthenelse{\boolean{articletitles}}{\emph{{Orthocharmonium and $e^+e^-$
  Annihilation}}, }{}\href{http://dx.doi.org/10.1103/PhysRevLett.34.43}{Phys.\
  Rev.\ Lett.\  \textbf{34} (1975) 43}\relax
\mciteBstWouldAddEndPuncttrue
\mciteSetBstMidEndSepPunct{\mcitedefaultmidpunct}
{\mcitedefaultendpunct}{\mcitedefaultseppunct}\relax
\EndOfBibitem
\bibitem{Eichten:1978tg}
E.~Eichten {\em et~al.},
  \ifthenelse{\boolean{articletitles}}{\emph{{Charmonium: The Model}},
  }{}\href{http://dx.doi.org/10.1103/PhysRevD.17.3090,
  10.1103/PhysRevD.21.313}{Phys.\ Rev.\  \textbf{D17} (1978) 3090}, [Erratum:
  Phys. Rev.D21,313(1980)]\relax
\mciteBstWouldAddEndPuncttrue
\mciteSetBstMidEndSepPunct{\mcitedefaultmidpunct}
{\mcitedefaultendpunct}{\mcitedefaultseppunct}\relax
\EndOfBibitem
\bibitem{Necco:2001xg}
S.~Necco and R.~Sommer, \ifthenelse{\boolean{articletitles}}{\emph{{The N(f) =
  0 heavy quark potential from short to intermediate distances}},
  }{}\href{http://dx.doi.org/10.1016/S0550-3213(01)00582-X}{Nucl.\ Phys.\
  \textbf{B622} (2002) 328},
  \href{http://arxiv.org/abs/hep-lat/0108008}{{\normalfont\ttfamily
  arXiv:hep-lat/0108008}}\relax
\mciteBstWouldAddEndPuncttrue
\mciteSetBstMidEndSepPunct{\mcitedefaultmidpunct}
{\mcitedefaultendpunct}{\mcitedefaultseppunct}\relax
\EndOfBibitem
\bibitem{Brambilla:1999xf}
N.~Brambilla, A.~Pineda, J.~Soto, and A.~Vairo,
  \ifthenelse{\boolean{articletitles}}{\emph{{Potential NRQCD: An Effective
  theory for heavy quarkonium}},
  }{}\href{http://dx.doi.org/10.1016/S0550-3213(99)00693-8}{Nucl.\ Phys.\
  \textbf{B566} (2000) 275},
  \href{http://arxiv.org/abs/hep-ph/9907240}{{\normalfont\ttfamily
  arXiv:hep-ph/9907240}}\relax
\mciteBstWouldAddEndPuncttrue
\mciteSetBstMidEndSepPunct{\mcitedefaultmidpunct}
{\mcitedefaultendpunct}{\mcitedefaultseppunct}\relax
\EndOfBibitem
\bibitem{Brambilla:2014jmp}
N.~Brambilla {\em et~al.}, \ifthenelse{\boolean{articletitles}}{\emph{{QCD and
  Strongly Coupled Gauge Theories: Challenges and Perspectives}},
  }{}\href{http://dx.doi.org/10.1140/epjc/s10052-014-2981-5}{Eur.\ Phys.\ J.\
  \textbf{C74} (2014), no.~10 2981},
  \href{http://arxiv.org/abs/1404.3723}{{\normalfont\ttfamily
  arXiv:1404.3723}}\relax
\mciteBstWouldAddEndPuncttrue
\mciteSetBstMidEndSepPunct{\mcitedefaultmidpunct}
{\mcitedefaultendpunct}{\mcitedefaultseppunct}\relax
\EndOfBibitem
\bibitem{Barnes:2005pb}
T.~Barnes, S.~Godfrey, and E.~S. Swanson,
  \ifthenelse{\boolean{articletitles}}{\emph{{Higher charmonia}},
  }{}\href{http://dx.doi.org/10.1103/PhysRevD.72.054026}{Phys.\ Rev.\
  \textbf{D72} (2005) 054026},
  \href{http://arxiv.org/abs/hep-ph/0505002}{{\normalfont\ttfamily
  arXiv:hep-ph/0505002}}\relax
\mciteBstWouldAddEndPuncttrue
\mciteSetBstMidEndSepPunct{\mcitedefaultmidpunct}
{\mcitedefaultendpunct}{\mcitedefaultseppunct}\relax
\EndOfBibitem
\bibitem{Godfrey:1985xj}
S.~Godfrey and N.~Isgur, \ifthenelse{\boolean{articletitles}}{\emph{{Mesons in
  a Relativized Quark Model with Chromodynamics}},
  }{}\href{http://dx.doi.org/10.1103/PhysRevD.32.189}{Phys.\ Rev.\
  \textbf{D32} (1985) 189}\relax
\mciteBstWouldAddEndPuncttrue
\mciteSetBstMidEndSepPunct{\mcitedefaultmidpunct}
{\mcitedefaultendpunct}{\mcitedefaultseppunct}\relax
\EndOfBibitem
\bibitem{Godfrey:2015dia}
S.~Godfrey and K.~Moats,
  \ifthenelse{\boolean{articletitles}}{\emph{{Bottomonium Mesons and Strategies
  for their Observation}},
  }{}\href{http://dx.doi.org/10.1103/PhysRevD.92.054034}{Phys.\ Rev.\
  \textbf{D92} (2015) 054034},
  \href{http://arxiv.org/abs/1507.00024}{{\normalfont\ttfamily
  arXiv:1507.00024}}\relax
\mciteBstWouldAddEndPuncttrue
\mciteSetBstMidEndSepPunct{\mcitedefaultmidpunct}
{\mcitedefaultendpunct}{\mcitedefaultseppunct}\relax
\EndOfBibitem
\bibitem{Herb:1977ek}
S.~W. Herb {\em et~al.},
  \ifthenelse{\boolean{articletitles}}{\emph{{Observation of a Dimuon Resonance
  at 9.5-GeV in 400-GeV Proton-Nucleus Collisions}},
  }{}\href{http://dx.doi.org/10.1103/PhysRevLett.39.252}{Phys.\ Rev.\ Lett.\
  \textbf{39} (1977) 252}\relax
\mciteBstWouldAddEndPuncttrue
\mciteSetBstMidEndSepPunct{\mcitedefaultmidpunct}
{\mcitedefaultendpunct}{\mcitedefaultseppunct}\relax
\EndOfBibitem
\bibitem{Ueno:1978vr}
K.~Ueno {\em et~al.}, \ifthenelse{\boolean{articletitles}}{\emph{{Evidence for
  the $\Upsilon^{\prime\prime}$ and a search for new narrow resonances}},
  }{}\href{http://dx.doi.org/10.1103/PhysRevLett.42.486}{Phys.\ Rev.\ Lett.\
  \textbf{42} (1979) 486}\relax
\mciteBstWouldAddEndPuncttrue
\mciteSetBstMidEndSepPunct{\mcitedefaultmidpunct}
{\mcitedefaultendpunct}{\mcitedefaultseppunct}\relax
\EndOfBibitem
\bibitem{Andrews:1980ha}
CLEO, D.~Andrews {\em et~al.},
  \ifthenelse{\boolean{articletitles}}{\emph{{Observation of Three Upsilon
  States}}, }{}\href{http://dx.doi.org/10.1103/PhysRevLett.44.1108}{Phys.\
  Rev.\ Lett.\  \textbf{44} (1980) 1108}\relax
\mciteBstWouldAddEndPuncttrue
\mciteSetBstMidEndSepPunct{\mcitedefaultmidpunct}
{\mcitedefaultendpunct}{\mcitedefaultseppunct}\relax
\EndOfBibitem
\bibitem{Choi:2007wga}
Belle collaboraation, S.~K. Choi {\em et~al.},
  \ifthenelse{\boolean{articletitles}}{\emph{{Observation of a resonance-like
  structure in the $\pi^\pm\psi'$ mass distribution in exclusive $B\to
  K\pi^\pm\psi'$ decays}},
  }{}\href{http://dx.doi.org/10.1103/PhysRevLett.100.142001}{Phys.\ Rev.\
  Lett.\  \textbf{100} (2008) 142001},
  \href{http://arxiv.org/abs/0708.1790}{{\normalfont\ttfamily
  arXiv:0708.1790}}\relax
\mciteBstWouldAddEndPuncttrue
\mciteSetBstMidEndSepPunct{\mcitedefaultmidpunct}
{\mcitedefaultendpunct}{\mcitedefaultseppunct}\relax
\EndOfBibitem
\bibitem{Aaij:2014jqa}
LHCb, R.~Aaij {\em et~al.},
  \ifthenelse{\boolean{articletitles}}{\emph{{Observation of the resonant
  character of the $Z(4430)^-$ state}},
  }{}\href{http://dx.doi.org/10.1103/PhysRevLett.112.222002}{Phys.\ Rev.\
  Lett.\  \textbf{112} (2014) 222002},
  \href{http://arxiv.org/abs/1404.1903}{{\normalfont\ttfamily
  arXiv:1404.1903}}\relax
\mciteBstWouldAddEndPuncttrue
\mciteSetBstMidEndSepPunct{\mcitedefaultmidpunct}
{\mcitedefaultendpunct}{\mcitedefaultseppunct}\relax
\EndOfBibitem
\bibitem{Aaij:2015tga}
LHCb, R.~Aaij {\em et~al.},
  \ifthenelse{\boolean{articletitles}}{\emph{{Observation of $J/\psi p$
  Resonances Consistent with Pentaquark States in $\Lambda_b^0 \to J/\psi K^-
  p$ Decays}},
  }{}\href{http://dx.doi.org/10.1103/PhysRevLett.115.072001}{Phys.\ Rev.\
  Lett.\  \textbf{115} (2015) 072001},
  \href{http://arxiv.org/abs/1507.03414}{{\normalfont\ttfamily
  arXiv:1507.03414}}\relax
\mciteBstWouldAddEndPuncttrue
\mciteSetBstMidEndSepPunct{\mcitedefaultmidpunct}
{\mcitedefaultendpunct}{\mcitedefaultseppunct}\relax
\EndOfBibitem
\bibitem{D0:2016mwd}
D0, V.~M. Abazov {\em et~al.},
  \ifthenelse{\boolean{articletitles}}{\emph{{Evidence for a $B_s^0 \pi^\pm$
  state}}, }{}\href{http://dx.doi.org/10.1103/PhysRevLett.117.022003}{Phys.\
  Rev.\ Lett.\  \textbf{117} (2016) 022003},
  \href{http://arxiv.org/abs/1602.07588}{{\normalfont\ttfamily
  arXiv:1602.07588}}\relax
\mciteBstWouldAddEndPuncttrue
\mciteSetBstMidEndSepPunct{\mcitedefaultmidpunct}
{\mcitedefaultendpunct}{\mcitedefaultseppunct}\relax
\EndOfBibitem
\bibitem{Aaij:2016iza}
LHCb, R.~Aaij {\em et~al.},
  \ifthenelse{\boolean{articletitles}}{\emph{{Observation of $J/\psi\phi$
  structures consistent with exotic states from amplitude analysis of $B^+\to
  J/\psi \phi K^+$ decays}},
  }{}\href{http://dx.doi.org/10.1103/PhysRevLett.118.022003}{Phys.\ Rev.\
  Lett.\  \textbf{118} (2017) 022003},
  \href{http://arxiv.org/abs/1606.07895}{{\normalfont\ttfamily
  arXiv:1606.07895}}\relax
\mciteBstWouldAddEndPuncttrue
\mciteSetBstMidEndSepPunct{\mcitedefaultmidpunct}
{\mcitedefaultendpunct}{\mcitedefaultseppunct}\relax
\EndOfBibitem
\bibitem{Aaij:2016nsc}
LHCb, R.~Aaij {\em et~al.},
  \ifthenelse{\boolean{articletitles}}{\emph{{Amplitude analysis of $B^+\to
  J/\psi \phi K^+$ decays}},
  }{}\href{http://dx.doi.org/10.1103/PhysRevD.95.012002}{Phys.\ Rev.\
  \textbf{D95} (2017) 012002},
  \href{http://arxiv.org/abs/1606.07898}{{\normalfont\ttfamily
  arXiv:1606.07898}}\relax
\mciteBstWouldAddEndPuncttrue
\mciteSetBstMidEndSepPunct{\mcitedefaultmidpunct}
{\mcitedefaultendpunct}{\mcitedefaultseppunct}\relax
\EndOfBibitem
\bibitem{Chilikin:2014bkk}
Belle, K.~Chilikin {\em et~al.},
  \ifthenelse{\boolean{articletitles}}{\emph{{Observation of a new charged
  charmoniumlike state in $\overline{B^0}\to\jpsi K^-\pi^+$ decays}},
  }{}\href{http://dx.doi.org/10.1103/PhysRevD.90.112009}{Phys.\ Rev.\
  \textbf{D90} (2014) 112009},
  \href{http://arxiv.org/abs/1408.6457}{{\normalfont\ttfamily
  arXiv:1408.6457}}\relax
\mciteBstWouldAddEndPuncttrue
\mciteSetBstMidEndSepPunct{\mcitedefaultmidpunct}
{\mcitedefaultendpunct}{\mcitedefaultseppunct}\relax
\EndOfBibitem
\bibitem{Ablikim:2016qzw}
BESIII, M.~Ablikim {\em et~al.},
  \ifthenelse{\boolean{articletitles}}{\emph{{Precise measurement of the
  $e^+e^-\to \pi^+\pi^-J/\psi$ cross section at center-of-mass energies from
  3.77 to 4.60 GeV}},
  }{}\href{http://dx.doi.org/10.1103/PhysRevLett.118.092001}{Phys.\ Rev.\
  Lett.\  \textbf{118} (2017) 092001},
  \href{http://arxiv.org/abs/1611.01317}{{\normalfont\ttfamily
  arXiv:1611.01317}}\relax
\mciteBstWouldAddEndPuncttrue
\mciteSetBstMidEndSepPunct{\mcitedefaultmidpunct}
{\mcitedefaultendpunct}{\mcitedefaultseppunct}\relax
\EndOfBibitem
\bibitem{Choi:2003ue}
Belle, S.~K. Choi {\em et~al.},
  \ifthenelse{\boolean{articletitles}}{\emph{{Observation of a narrow
  charmonium-like state in exclusive $B^\pm\to K^\pm \pi^+ \pi^- \jpsi$
  decays}}, }{}\href{http://dx.doi.org/10.1103/PhysRevLett.91.262001}{Phys.\
  Rev.\ Lett.\  \textbf{91} (2003) 262001},
  \href{http://arxiv.org/abs/hep-ex/0309032}{{\normalfont\ttfamily
  arXiv:hep-ex/0309032}}\relax
\mciteBstWouldAddEndPuncttrue
\mciteSetBstMidEndSepPunct{\mcitedefaultmidpunct}
{\mcitedefaultendpunct}{\mcitedefaultseppunct}\relax
\EndOfBibitem
\bibitem{Choi:2011fc}
S.-K. Choi {\em et~al.}, \ifthenelse{\boolean{articletitles}}{\emph{{Bounds on
  the width, mass difference and other properties of
  $X(3872)\to\pi^+\pi^-\jpsi$ decays}},
  }{}\href{http://dx.doi.org/10.1103/PhysRevD.84.052004}{Phys.\ Rev.\
  \textbf{D84} (2011) 052004},
  \href{http://arxiv.org/abs/1107.0163}{{\normalfont\ttfamily
  arXiv:1107.0163}}\relax
\mciteBstWouldAddEndPuncttrue
\mciteSetBstMidEndSepPunct{\mcitedefaultmidpunct}
{\mcitedefaultendpunct}{\mcitedefaultseppunct}\relax
\EndOfBibitem
\bibitem{Aubert:2004ns}
BaBar, B.~Aubert {\em et~al.},
  \ifthenelse{\boolean{articletitles}}{\emph{{Study of the $B \to J/\psi K^-
  \pi^+ \pi^-$ decay and measurement of the $B \to X(3872) K^-$ branching
  fraction}}, }{}\href{http://dx.doi.org/10.1103/PhysRevD.71.071103}{Phys.\
  Rev.\  \textbf{D71} (2005) 071103},
  \href{http://arxiv.org/abs/hep-ex/0406022}{{\normalfont\ttfamily
  arXiv:hep-ex/0406022}}\relax
\mciteBstWouldAddEndPuncttrue
\mciteSetBstMidEndSepPunct{\mcitedefaultmidpunct}
{\mcitedefaultendpunct}{\mcitedefaultseppunct}\relax
\EndOfBibitem
\bibitem{Aaij:2013zoa}
LHCb, R.~Aaij {\em et~al.},
  \ifthenelse{\boolean{articletitles}}{\emph{{Determination of the $X(3872)$
  meson quantum numbers}},
  }{}\href{http://dx.doi.org/10.1103/PhysRevLett.110.222001}{Phys.\ Rev.\
  Lett.\  \textbf{110} (2013) 222001},
  \href{http://arxiv.org/abs/1302.6269}{{\normalfont\ttfamily
  arXiv:1302.6269}}\relax
\mciteBstWouldAddEndPuncttrue
\mciteSetBstMidEndSepPunct{\mcitedefaultmidpunct}
{\mcitedefaultendpunct}{\mcitedefaultseppunct}\relax
\EndOfBibitem
\bibitem{Aaij:2015eva}
LHCb, R.~Aaij {\em et~al.}, \ifthenelse{\boolean{articletitles}}{\emph{{Quantum
  numbers of the $X(3872)$ state and orbital angular momentum in its $\rho^0
  J\psi$ decay}}, }{}\href{http://dx.doi.org/10.1103/PhysRevD.92.011102}{Phys.\
  Rev.\  \textbf{D92} (2015) 011102},
  \href{http://arxiv.org/abs/1504.06339}{{\normalfont\ttfamily
  arXiv:1504.06339}}\relax
\mciteBstWouldAddEndPuncttrue
\mciteSetBstMidEndSepPunct{\mcitedefaultmidpunct}
{\mcitedefaultendpunct}{\mcitedefaultseppunct}\relax
\EndOfBibitem
\bibitem{Acosta:2003zx}
CDF, D.~Acosta {\em et~al.},
  \ifthenelse{\boolean{articletitles}}{\emph{{Observation of the narrow state
  $X(3872) \to J/\psi \pi^+ \pi^-$ in $\bar{p}p$ collisions at $\sqrt{s} =
  1.96$ TeV}}, }{}\href{http://dx.doi.org/10.1103/PhysRevLett.93.072001}{Phys.\
  Rev.\ Lett.\  \textbf{93} (2004) 072001},
  \href{http://arxiv.org/abs/hep-ex/0312021}{{\normalfont\ttfamily
  arXiv:hep-ex/0312021}}\relax
\mciteBstWouldAddEndPuncttrue
\mciteSetBstMidEndSepPunct{\mcitedefaultmidpunct}
{\mcitedefaultendpunct}{\mcitedefaultseppunct}\relax
\EndOfBibitem
\bibitem{Abulencia:2005zc}
CDF, A.~Abulencia {\em et~al.},
  \ifthenelse{\boolean{articletitles}}{\emph{{Measurement of the dipion mass
  spectrum in $X(3872) \to J/\psi \pi^+ \pi^-$ decays.}},
  }{}\href{http://dx.doi.org/10.1103/PhysRevLett.96.102002}{Phys.\ Rev.\ Lett.\
   \textbf{96} (2006) 102002},
  \href{http://arxiv.org/abs/hep-ex/0512074}{{\normalfont\ttfamily
  arXiv:hep-ex/0512074}}\relax
\mciteBstWouldAddEndPuncttrue
\mciteSetBstMidEndSepPunct{\mcitedefaultmidpunct}
{\mcitedefaultendpunct}{\mcitedefaultseppunct}\relax
\EndOfBibitem
\bibitem{Aaltonen:2009vj}
CDF, T.~Aaltonen {\em et~al.},
  \ifthenelse{\boolean{articletitles}}{\emph{{Precision Measurement of the
  $X(3872)$ Mass in $J/\psi \pi^+ \pi^-$ Decays}},
  }{}\href{http://dx.doi.org/10.1103/PhysRevLett.103.152001}{Phys.\ Rev.\
  Lett.\  \textbf{103} (2009) 152001},
  \href{http://arxiv.org/abs/0906.5218}{{\normalfont\ttfamily
  arXiv:0906.5218}}\relax
\mciteBstWouldAddEndPuncttrue
\mciteSetBstMidEndSepPunct{\mcitedefaultmidpunct}
{\mcitedefaultendpunct}{\mcitedefaultseppunct}\relax
\EndOfBibitem
\bibitem{Abazov:2004kp}
D0, V.~M. Abazov {\em et~al.},
  \ifthenelse{\boolean{articletitles}}{\emph{{Observation and properties of the
  $X(3872)$ decaying to $J/\psi \pi^+ \pi^-$ in $p\bar{p}$ collisions at
  $\sqrt{s} = 1.96$ TeV}},
  }{}\href{http://dx.doi.org/10.1103/PhysRevLett.93.162002}{Phys.\ Rev.\ Lett.\
   \textbf{93} (2004) 162002},
  \href{http://arxiv.org/abs/hep-ex/0405004}{{\normalfont\ttfamily
  arXiv:hep-ex/0405004}}\relax
\mciteBstWouldAddEndPuncttrue
\mciteSetBstMidEndSepPunct{\mcitedefaultmidpunct}
{\mcitedefaultendpunct}{\mcitedefaultseppunct}\relax
\EndOfBibitem
\bibitem{Abe:2005ax}
Belle, K.~Abe {\em et~al.},
  \ifthenelse{\boolean{articletitles}}{\emph{{Evidence for
  $X(3872)\to\gamma\jpsi$ and the sub-threshold decay
  $X(3872)\to\omega\jpsi$}}, }{} in {\em {Lepton and photon interactions at
  high energies. Proceedings, 22nd International Symposium, LP 2005, Uppsala,
  Sweden, June 30-July 5, 2005}}, 2005.
\newblock \href{http://arxiv.org/abs/hep-ex/0505037}{{\normalfont\ttfamily
  arXiv:hep-ex/0505037}}\relax
\mciteBstWouldAddEndPuncttrue
\mciteSetBstMidEndSepPunct{\mcitedefaultmidpunct}
{\mcitedefaultendpunct}{\mcitedefaultseppunct}\relax
\EndOfBibitem
\bibitem{delAmoSanchez:2010jr}
BaBar, P.~del Amo~Sanchez {\em et~al.},
  \ifthenelse{\boolean{articletitles}}{\emph{{Evidence for the decay
  $X(3872)\to\jpsi\omega$}},
  }{}\href{http://dx.doi.org/10.1103/PhysRevD.82.011101}{Phys.\ Rev.\
  \textbf{D82} (2010) 011101},
  \href{http://arxiv.org/abs/1005.5190}{{\normalfont\ttfamily
  arXiv:1005.5190}}\relax
\mciteBstWouldAddEndPuncttrue
\mciteSetBstMidEndSepPunct{\mcitedefaultmidpunct}
{\mcitedefaultendpunct}{\mcitedefaultseppunct}\relax
\EndOfBibitem
\bibitem{Gokhroo:2006bt}
Belle, G.~Gokhroo {\em et~al.},
  \ifthenelse{\boolean{articletitles}}{\emph{{Observation of a Near-threshold
  $D^0 \overline{D^0} \pi^0$ Enhancement in $B\to D^0 \overline{D^0} \pi^0 K$
  Decay}}, }{}\href{http://dx.doi.org/10.1103/PhysRevLett.97.162002}{Phys.\
  Rev.\ Lett.\  \textbf{97} (2006) 162002},
  \href{http://arxiv.org/abs/hep-ex/0606055}{{\normalfont\ttfamily
  arXiv:hep-ex/0606055}}\relax
\mciteBstWouldAddEndPuncttrue
\mciteSetBstMidEndSepPunct{\mcitedefaultmidpunct}
{\mcitedefaultendpunct}{\mcitedefaultseppunct}\relax
\EndOfBibitem
\bibitem{Adachi:2008sua}
Belle, T.~Aushev {\em et~al.},
  \ifthenelse{\boolean{articletitles}}{\emph{{Study of the $B\to X(3872)(
  D^{*0} \overline{D^0}) K$ decay}},
  }{}\href{http://dx.doi.org/10.1103/PhysRevD.81.031103}{Phys.\ Rev.\
  \textbf{D81} (2010) 031103},
  \href{http://arxiv.org/abs/0810.0358}{{\normalfont\ttfamily
  arXiv:0810.0358}}\relax
\mciteBstWouldAddEndPuncttrue
\mciteSetBstMidEndSepPunct{\mcitedefaultmidpunct}
{\mcitedefaultendpunct}{\mcitedefaultseppunct}\relax
\EndOfBibitem
\bibitem{Aubert:2007rva}
BaBar, B.~Aubert {\em et~al.},
  \ifthenelse{\boolean{articletitles}}{\emph{{Study of Resonances in Exclusive
  $B$ Decays to $\overline{D^{(*)}} D^{(*)} K$}},
  }{}\href{http://dx.doi.org/10.1103/PhysRevD.77.011102}{Phys.\ Rev.\
  \textbf{D77} (2008) 011102},
  \href{http://arxiv.org/abs/0708.1565}{{\normalfont\ttfamily
  arXiv:0708.1565}}\relax
\mciteBstWouldAddEndPuncttrue
\mciteSetBstMidEndSepPunct{\mcitedefaultmidpunct}
{\mcitedefaultendpunct}{\mcitedefaultseppunct}\relax
\EndOfBibitem
\bibitem{Bhardwaj:2011dj}
Belle, V.~Bhardwaj {\em et~al.},
  \ifthenelse{\boolean{articletitles}}{\emph{{Observation of $X(3872)\to J/\psi
  \gamma$ and search for $X(3872)\to\psi'\gamma$ in B decays}},
  }{}\href{http://dx.doi.org/10.1103/PhysRevLett.107.091803}{Phys.\ Rev.\
  Lett.\  \textbf{107} (2011) 091803},
  \href{http://arxiv.org/abs/1105.0177}{{\normalfont\ttfamily
  arXiv:1105.0177}}\relax
\mciteBstWouldAddEndPuncttrue
\mciteSetBstMidEndSepPunct{\mcitedefaultmidpunct}
{\mcitedefaultendpunct}{\mcitedefaultseppunct}\relax
\EndOfBibitem
\bibitem{Aaij:2011sn}
LHCb, R.~Aaij {\em et~al.},
  \ifthenelse{\boolean{articletitles}}{\emph{{Observation of $X(3872) $
  production in $pp$ collisions at $\sqrt{s}=7$ TeV}},
  }{}\href{http://dx.doi.org/10.1140/epjc/s10052-012-1972-7}{Eur.\ Phys.\ J.\
  \textbf{C72} (2012) 1972},
  \href{http://arxiv.org/abs/1112.5310}{{\normalfont\ttfamily
  arXiv:1112.5310}}\relax
\mciteBstWouldAddEndPuncttrue
\mciteSetBstMidEndSepPunct{\mcitedefaultmidpunct}
{\mcitedefaultendpunct}{\mcitedefaultseppunct}\relax
\EndOfBibitem
\bibitem{Aubert:2008ae}
BaBar, B.~Aubert {\em et~al.},
  \ifthenelse{\boolean{articletitles}}{\emph{{Evidence for $X(3872) \to
  \psi(2S) \gamma$ in $B^\pm \to X(3872) K^\pm$ decays, and a study of $B \to c
  \bar{c} \gamma K$}},
  }{}\href{http://dx.doi.org/10.1103/PhysRevLett.102.132001}{Phys.\ Rev.\
  Lett.\  \textbf{102} (2009) 132001},
  \href{http://arxiv.org/abs/0809.0042}{{\normalfont\ttfamily
  arXiv:0809.0042}}\relax
\mciteBstWouldAddEndPuncttrue
\mciteSetBstMidEndSepPunct{\mcitedefaultmidpunct}
{\mcitedefaultendpunct}{\mcitedefaultseppunct}\relax
\EndOfBibitem
\bibitem{Aaij:2014ala}
LHCb, R.~Aaij {\em et~al.},
  \ifthenelse{\boolean{articletitles}}{\emph{{Evidence for the decay
  $X(3872)\rightarrow\psi(2S)\gamma$}},
  }{}\href{http://dx.doi.org/10.1016/j.nuclphysb.2014.06.011}{Nucl.\ Phys.\
  \textbf{B886} (2014) 665},
  \href{http://arxiv.org/abs/1404.0275}{{\normalfont\ttfamily
  arXiv:1404.0275}}\relax
\mciteBstWouldAddEndPuncttrue
\mciteSetBstMidEndSepPunct{\mcitedefaultmidpunct}
{\mcitedefaultendpunct}{\mcitedefaultseppunct}\relax
\EndOfBibitem
\bibitem{Chatrchyan:2013cld}
CMS, S.~Chatrchyan {\em et~al.},
  \ifthenelse{\boolean{articletitles}}{\emph{{Measurement of the X(3872)
  production cross section via decays to $J/\psi \pi \pi$ in $pp$ collisions at
  $\sqrt{s} = 7$ TeV}},
  }{}\href{http://dx.doi.org/10.1007/JHEP04(2013)154}{JHEP \textbf{04} (2013)
  154}, \href{http://arxiv.org/abs/1302.3968}{{\normalfont\ttfamily
  arXiv:1302.3968}}\relax
\mciteBstWouldAddEndPuncttrue
\mciteSetBstMidEndSepPunct{\mcitedefaultmidpunct}
{\mcitedefaultendpunct}{\mcitedefaultseppunct}\relax
\EndOfBibitem
\bibitem{Aaboud:2016vzw}
ATLAS, M.~Aaboud {\em et~al.},
  \ifthenelse{\boolean{articletitles}}{\emph{{Measurements of $\psi(2S)$ and
  $X(3872) \to J/\psi\pi^+\pi^-$ production in $pp$ collisions at $\sqrt{s} =
  8$ TeV with the ATLAS detector}},
  }{}\href{http://dx.doi.org/10.1007/JHEP01(2017)117}{JHEP \textbf{01} (2017)
  117}, \href{http://arxiv.org/abs/1610.09303}{{\normalfont\ttfamily
  arXiv:1610.09303}}\relax
\mciteBstWouldAddEndPuncttrue
\mciteSetBstMidEndSepPunct{\mcitedefaultmidpunct}
{\mcitedefaultendpunct}{\mcitedefaultseppunct}\relax
\EndOfBibitem
\bibitem{Ablikim:2013dyn}
BESIII, M.~Ablikim {\em et~al.},
  \ifthenelse{\boolean{articletitles}}{\emph{{Observation of $e^+e^\to\gamma
  X(3872)$ at BESIII}},
  }{}\href{http://dx.doi.org/10.1103/PhysRevLett.112.092001}{Phys.\ Rev.\
  Lett.\  \textbf{112} (2014) 092001},
  \href{http://arxiv.org/abs/1310.4101}{{\normalfont\ttfamily
  arXiv:1310.4101}}\relax
\mciteBstWouldAddEndPuncttrue
\mciteSetBstMidEndSepPunct{\mcitedefaultmidpunct}
{\mcitedefaultendpunct}{\mcitedefaultseppunct}\relax
\EndOfBibitem
\bibitem{Abe:2004zs}
Belle, S.-K. Choi {\em et~al.},
  \ifthenelse{\boolean{articletitles}}{\emph{{Observation of a near-threshold
  $\omega J/\psi$ mass enhancement in exclusive $B \to K \omega J/\psi$
  decays}}, }{}\href{http://dx.doi.org/10.1103/PhysRevLett.94.182002}{Phys.\
  Rev.\ Lett.\  \textbf{94} (2005) 182002},
  \href{http://arxiv.org/abs/hep-ex/0408126}{{\normalfont\ttfamily
  arXiv:hep-ex/0408126}}\relax
\mciteBstWouldAddEndPuncttrue
\mciteSetBstMidEndSepPunct{\mcitedefaultmidpunct}
{\mcitedefaultendpunct}{\mcitedefaultseppunct}\relax
\EndOfBibitem
\bibitem{Aubert:2007vj}
BaBar, B.~Aubert {\em et~al.},
  \ifthenelse{\boolean{articletitles}}{\emph{{Observation of Y(3940) $\to
  J/\psi \omega$ in $B \to J/\psi \omega K$ at BABAR}},
  }{}\href{http://dx.doi.org/10.1103/PhysRevLett.101.082001}{Phys.\ Rev.\
  Lett.\  \textbf{101} (2008) 082001},
  \href{http://arxiv.org/abs/0711.2047}{{\normalfont\ttfamily
  arXiv:0711.2047}}\relax
\mciteBstWouldAddEndPuncttrue
\mciteSetBstMidEndSepPunct{\mcitedefaultmidpunct}
{\mcitedefaultendpunct}{\mcitedefaultseppunct}\relax
\EndOfBibitem
\bibitem{Uehara:2009tx}
Belle, S.~Uehara {\em et~al.},
  \ifthenelse{\boolean{articletitles}}{\emph{{Observation of a charmonium-like
  enhancement in the $\gamma\gamma\to\omega J/\psi$ process}},
  }{}\href{http://dx.doi.org/10.1103/PhysRevLett.104.092001}{Phys.\ Rev.\
  Lett.\  \textbf{104} (2010) 092001},
  \href{http://arxiv.org/abs/0912.4451}{{\normalfont\ttfamily
  arXiv:0912.4451}}\relax
\mciteBstWouldAddEndPuncttrue
\mciteSetBstMidEndSepPunct{\mcitedefaultmidpunct}
{\mcitedefaultendpunct}{\mcitedefaultseppunct}\relax
\EndOfBibitem
\bibitem{Lees:2012xs}
BaBar, J.~P. Lees {\em et~al.},
  \ifthenelse{\boolean{articletitles}}{\emph{{Study of $X(3915) \to J/\psi
  \omega$ in two-photon collisions}},
  }{}\href{http://dx.doi.org/10.1103/PhysRevD.86.072002}{Phys.\ Rev.\
  \textbf{D86} (2012) 072002},
  \href{http://arxiv.org/abs/1207.2651}{{\normalfont\ttfamily
  arXiv:1207.2651}}\relax
\mciteBstWouldAddEndPuncttrue
\mciteSetBstMidEndSepPunct{\mcitedefaultmidpunct}
{\mcitedefaultendpunct}{\mcitedefaultseppunct}\relax
\EndOfBibitem
\bibitem{Abe:2007sya}
Belle, P.~Pakhlov {\em et~al.},
  \ifthenelse{\boolean{articletitles}}{\emph{{Production of New Charmoniumlike
  States in $e^+ e^- \to J/\psi D^{(*)} \overline{D^{(*)}}$ at $\sqrt{s}\sim10$
  GeV}}, }{}\href{http://dx.doi.org/10.1103/PhysRevLett.100.202001}{Phys.\
  Rev.\ Lett.\  \textbf{100} (2008) 202001},
  \href{http://arxiv.org/abs/0708.3812}{{\normalfont\ttfamily
  arXiv:0708.3812}}\relax
\mciteBstWouldAddEndPuncttrue
\mciteSetBstMidEndSepPunct{\mcitedefaultmidpunct}
{\mcitedefaultendpunct}{\mcitedefaultseppunct}\relax
\EndOfBibitem
\bibitem{Abe:2007jna}
Belle, K.~Abe {\em et~al.},
  \ifthenelse{\boolean{articletitles}}{\emph{{Observation of a new charmonium
  state in double charmonium production in $e^+e^-$ annihilation at
  $\sqrt{s}\sim$ 10.6-GeV}},
  }{}\href{http://dx.doi.org/10.1103/PhysRevLett.98.082001}{Phys.\ Rev.\ Lett.\
   \textbf{98} (2007) 082001},
  \href{http://arxiv.org/abs/hep-ex/0507019}{{\normalfont\ttfamily
  arXiv:hep-ex/0507019}}\relax
\mciteBstWouldAddEndPuncttrue
\mciteSetBstMidEndSepPunct{\mcitedefaultmidpunct}
{\mcitedefaultendpunct}{\mcitedefaultseppunct}\relax
\EndOfBibitem
\bibitem{Aaltonen:2009tz}
CDF, T.~Aaltonen {\em et~al.},
  \ifthenelse{\boolean{articletitles}}{\emph{{Evidence for a Narrow
  Near-Threshold Structure in the $J/\psi\phi$ Mass Spectrum in $B^+\to
  J/\psi\phi K^+$ Decays}},
  }{}\href{http://dx.doi.org/10.1103/PhysRevLett.102.242002}{Phys.\ Rev.\
  Lett.\  \textbf{102} (2009) 242002},
  \href{http://arxiv.org/abs/0903.2229}{{\normalfont\ttfamily
  arXiv:0903.2229}}\relax
\mciteBstWouldAddEndPuncttrue
\mciteSetBstMidEndSepPunct{\mcitedefaultmidpunct}
{\mcitedefaultendpunct}{\mcitedefaultseppunct}\relax
\EndOfBibitem
\bibitem{Chatrchyan:2013dma}
CMS, S.~Chatrchyan {\em et~al.},
  \ifthenelse{\boolean{articletitles}}{\emph{{Observation of a peaking
  structure in the $J/\psi \phi$ mass spectrum from $B^{\pm} \to J/\psi \phi
  K^{\pm}$ decays}},
  }{}\href{http://dx.doi.org/10.1016/j.physletb.2014.05.055}{Phys.\ Lett.\
  \textbf{B734} (2014) 261},
  \href{http://arxiv.org/abs/1309.6920}{{\normalfont\ttfamily
  arXiv:1309.6920}}\relax
\mciteBstWouldAddEndPuncttrue
\mciteSetBstMidEndSepPunct{\mcitedefaultmidpunct}
{\mcitedefaultendpunct}{\mcitedefaultseppunct}\relax
\EndOfBibitem
\bibitem{Abazov:2013xda}
D0, V.~M. Abazov {\em et~al.},
  \ifthenelse{\boolean{articletitles}}{\emph{{Search for the $X$(4140) state in
  $B^+ \to J/\psi\phi K^+$ decays with the D0 Detector}},
  }{}\href{http://dx.doi.org/10.1103/PhysRevD.89.012004}{Phys.\ Rev.\
  \textbf{D89} (2014) 012004},
  \href{http://arxiv.org/abs/1309.6580}{{\normalfont\ttfamily
  arXiv:1309.6580}}\relax
\mciteBstWouldAddEndPuncttrue
\mciteSetBstMidEndSepPunct{\mcitedefaultmidpunct}
{\mcitedefaultendpunct}{\mcitedefaultseppunct}\relax
\EndOfBibitem
\bibitem{Abazov:2015sxa}
D0, V.~M. Abazov {\em et~al.},
  \ifthenelse{\boolean{articletitles}}{\emph{{Inclusive Production of the
  X(4140) State in $p \overline p$ Collisions at D0}},
  }{}\href{http://dx.doi.org/10.1103/PhysRevLett.115.232001}{Phys.\ Rev.\
  Lett.\  \textbf{115} (2015) 232001},
  \href{http://arxiv.org/abs/1508.07846}{{\normalfont\ttfamily
  arXiv:1508.07846}}\relax
\mciteBstWouldAddEndPuncttrue
\mciteSetBstMidEndSepPunct{\mcitedefaultmidpunct}
{\mcitedefaultendpunct}{\mcitedefaultseppunct}\relax
\EndOfBibitem
\bibitem{Aubert:2005rm}
BaBar, B.~Aubert {\em et~al.},
  \ifthenelse{\boolean{articletitles}}{\emph{{Observation of a broad structure
  in the $\pi^+ \pi^- J/\psi$ mass spectrum around 4.26-GeV/c$^2$}},
  }{}\href{http://dx.doi.org/10.1103/PhysRevLett.95.142001}{Phys.\ Rev.\ Lett.\
   \textbf{95} (2005) 142001},
  \href{http://arxiv.org/abs/hep-ex/0506081}{{\normalfont\ttfamily
  arXiv:hep-ex/0506081}}\relax
\mciteBstWouldAddEndPuncttrue
\mciteSetBstMidEndSepPunct{\mcitedefaultmidpunct}
{\mcitedefaultendpunct}{\mcitedefaultseppunct}\relax
\EndOfBibitem
\bibitem{Lees:2012cn}
BaBar, J.~P. Lees {\em et~al.},
  \ifthenelse{\boolean{articletitles}}{\emph{{Study of the reaction $e^{+}e^{-}
  \to J/\psi\pi^{+}\pi^{-}$ via initial-state radiation at BaBar}},
  }{}\href{http://dx.doi.org/10.1103/PhysRevD.86.051102}{Phys.\ Rev.\
  \textbf{D86} (2012) 051102},
  \href{http://arxiv.org/abs/1204.2158}{{\normalfont\ttfamily
  arXiv:1204.2158}}\relax
\mciteBstWouldAddEndPuncttrue
\mciteSetBstMidEndSepPunct{\mcitedefaultmidpunct}
{\mcitedefaultendpunct}{\mcitedefaultseppunct}\relax
\EndOfBibitem
\bibitem{He:2006kg}
CLEO, Q.~He {\em et~al.},
  \ifthenelse{\boolean{articletitles}}{\emph{{Confirmation of the Y(4260)
  resonance production in ISR}},
  }{}\href{http://dx.doi.org/10.1103/PhysRevD.74.091104}{Phys.\ Rev.\
  \textbf{D74} (2006) 091104},
  \href{http://arxiv.org/abs/hep-ex/0611021}{{\normalfont\ttfamily
  arXiv:hep-ex/0611021}}\relax
\mciteBstWouldAddEndPuncttrue
\mciteSetBstMidEndSepPunct{\mcitedefaultmidpunct}
{\mcitedefaultendpunct}{\mcitedefaultseppunct}\relax
\EndOfBibitem
\bibitem{Yuan:2007sj}
Belle, C.~Z. Yuan {\em et~al.},
  \ifthenelse{\boolean{articletitles}}{\emph{{Measurement of $e^+ e^- \to \pi^+
  \pi^- J/\psi$ cross-section via initial state radiation at Belle}},
  }{}\href{http://dx.doi.org/10.1103/PhysRevLett.99.182004}{Phys.\ Rev.\ Lett.\
   \textbf{99} (2007) 182004},
  \href{http://arxiv.org/abs/0707.2541}{{\normalfont\ttfamily
  arXiv:0707.2541}}\relax
\mciteBstWouldAddEndPuncttrue
\mciteSetBstMidEndSepPunct{\mcitedefaultmidpunct}
{\mcitedefaultendpunct}{\mcitedefaultseppunct}\relax
\EndOfBibitem
\bibitem{Liu:2013dau}
Belle, Z.~Q. Liu {\em et~al.},
  \ifthenelse{\boolean{articletitles}}{\emph{{Study of
  $e^+e^-\to\pi^+\pi^-J/\psi$ and Observation of a Charged Charmoniumlike State
  at Belle}}, }{}\href{http://dx.doi.org/10.1103/PhysRevLett.110.252002}{Phys.\
  Rev.\ Lett.\  \textbf{110} (2013) 252002},
  \href{http://arxiv.org/abs/1304.0121}{{\normalfont\ttfamily
  arXiv:1304.0121}}\relax
\mciteBstWouldAddEndPuncttrue
\mciteSetBstMidEndSepPunct{\mcitedefaultmidpunct}
{\mcitedefaultendpunct}{\mcitedefaultseppunct}\relax
\EndOfBibitem
\bibitem{BESIII:2016adj}
BESIII, M.~Ablikim {\em et~al.},
  \ifthenelse{\boolean{articletitles}}{\emph{{Evidence of Two Resonant
  Structures in $e^+ e^- \to \pi^+ \pi^- h_c$}},
  }{}\href{http://dx.doi.org/10.1103/PhysRevLett.118.092002}{Phys.\ Rev.\
  Lett.\  \textbf{118} (2017) 092002},
  \href{http://arxiv.org/abs/1610.07044}{{\normalfont\ttfamily
  arXiv:1610.07044}}\relax
\mciteBstWouldAddEndPuncttrue
\mciteSetBstMidEndSepPunct{\mcitedefaultmidpunct}
{\mcitedefaultendpunct}{\mcitedefaultseppunct}\relax
\EndOfBibitem
\bibitem{Ablikim:2014qwy}
BESIII, M.~Ablikim {\em et~al.},
  \ifthenelse{\boolean{articletitles}}{\emph{{Study of
  $e^+e^-\to\omega\chi_{cJ}$ at center-of-mass energies from 4.21 to 4.42
  GeV}}, }{}\href{http://dx.doi.org/10.1103/PhysRevLett.114.092003}{Phys.\
  Rev.\ Lett.\  \textbf{114} (2015) 092003},
  \href{http://arxiv.org/abs/1410.6538}{{\normalfont\ttfamily
  arXiv:1410.6538}}\relax
\mciteBstWouldAddEndPuncttrue
\mciteSetBstMidEndSepPunct{\mcitedefaultmidpunct}
{\mcitedefaultendpunct}{\mcitedefaultseppunct}\relax
\EndOfBibitem
\bibitem{Ablikim:2015xhk}
BESIII, M.~Ablikim {\em et~al.},
  \ifthenelse{\boolean{articletitles}}{\emph{{Measurement of the $e^{+}e^{-}
  \to \eta J/\psi$ cross section and search for $e^{+}e^{-} \to \pi^{0} J/\psi$
  at center-of-mass energies between 3.810 and 4.600 GeV}},
  }{}\href{http://dx.doi.org/10.1103/PhysRevD.91.112005}{Phys.\ Rev.\
  \textbf{D91} (2015) 112005},
  \href{http://arxiv.org/abs/1503.06644}{{\normalfont\ttfamily
  arXiv:1503.06644}}\relax
\mciteBstWouldAddEndPuncttrue
\mciteSetBstMidEndSepPunct{\mcitedefaultmidpunct}
{\mcitedefaultendpunct}{\mcitedefaultseppunct}\relax
\EndOfBibitem
\bibitem{Ablikim:2013mio}
BESIII, M.~Ablikim {\em et~al.},
  \ifthenelse{\boolean{articletitles}}{\emph{{Observation of a Charged
  Charmoniumlike Structure in $e^+e^-\to\pi^+\pi^-J/\psi$ at
  $\sqrt{s}$=4.26  GeV}},
  }{}\href{http://dx.doi.org/10.1103/PhysRevLett.110.252001}{Phys.\ Rev.\
  Lett.\  \textbf{110} (2013) 252001},
  \href{http://arxiv.org/abs/1303.5949}{{\normalfont\ttfamily
  arXiv:1303.5949}}\relax
\mciteBstWouldAddEndPuncttrue
\mciteSetBstMidEndSepPunct{\mcitedefaultmidpunct}
{\mcitedefaultendpunct}{\mcitedefaultseppunct}\relax
\EndOfBibitem
\bibitem{Ablikim:2013wzq}
BESIII, M.~Ablikim {\em et~al.},
  \ifthenelse{\boolean{articletitles}}{\emph{{Observation of a Charged
  Charmoniumlike Structure $Z_c$(4020) and Search for the $Z_c$(3900) in
  $e^+e^- \to\pi^+\pi^-h_c$}},
  }{}\href{http://dx.doi.org/10.1103/PhysRevLett.111.242001}{Phys.\ Rev.\
  Lett.\  \textbf{111} (2013) 242001},
  \href{http://arxiv.org/abs/1309.1896}{{\normalfont\ttfamily
  arXiv:1309.1896}}\relax
\mciteBstWouldAddEndPuncttrue
\mciteSetBstMidEndSepPunct{\mcitedefaultmidpunct}
{\mcitedefaultendpunct}{\mcitedefaultseppunct}\relax
\EndOfBibitem
\bibitem{Aaltonen:2011at}
CDF, T.~Aaltonen {\em et~al.},
  \ifthenelse{\boolean{articletitles}}{\emph{{Observation of the $Y(4140)$
  structure in the $J/\psi\,\phi$ Mass Spectrum in $B^\pm\to J/\psi\,\phi K$
  decays}}, }{}\href{http://dx.doi.org/10.1142/S0217732317501395}{Mod.\ Phys.\
  Lett.\  \textbf{A32} (2017), no.~26 1750139},
  \href{http://arxiv.org/abs/1101.6058}{{\normalfont\ttfamily
  arXiv:1101.6058}}\relax
\mciteBstWouldAddEndPuncttrue
\mciteSetBstMidEndSepPunct{\mcitedefaultmidpunct}
{\mcitedefaultendpunct}{\mcitedefaultseppunct}\relax
\EndOfBibitem
\bibitem{Shen:2009vs}
Belle, C.~P. Shen {\em et~al.},
  \ifthenelse{\boolean{articletitles}}{\emph{{Evidence for a new resonance and
  search for the $Y(4140)$ in the $\gamma\gamma\to\phi\jpsi$ process}},
  }{}\href{http://dx.doi.org/10.1103/PhysRevLett.104.112004}{Phys.\ Rev.\
  Lett.\  \textbf{104} (2010) 112004},
  \href{http://arxiv.org/abs/0912.2383}{{\normalfont\ttfamily
  arXiv:0912.2383}}\relax
\mciteBstWouldAddEndPuncttrue
\mciteSetBstMidEndSepPunct{\mcitedefaultmidpunct}
{\mcitedefaultendpunct}{\mcitedefaultseppunct}\relax
\EndOfBibitem
\bibitem{Aubert:2007zz}
BaBar, B.~Aubert {\em et~al.},
  \ifthenelse{\boolean{articletitles}}{\emph{{Evidence of a broad structure at
  an invariant mass of 4.32- $GeV/c^{2}$ in the reaction $e^{+} e^{-} \to
  \pi^{+} \pi^{-} \psi_{2S}$ measured at BaBar}},
  }{}\href{http://dx.doi.org/10.1103/PhysRevLett.98.212001}{Phys.\ Rev.\ Lett.\
   \textbf{98} (2007) 212001},
  \href{http://arxiv.org/abs/hep-ex/0610057}{{\normalfont\ttfamily
  arXiv:hep-ex/0610057}}\relax
\mciteBstWouldAddEndPuncttrue
\mciteSetBstMidEndSepPunct{\mcitedefaultmidpunct}
{\mcitedefaultendpunct}{\mcitedefaultseppunct}\relax
\EndOfBibitem
\bibitem{Lees:2012pv}
BaBar, J.~P. Lees {\em et~al.},
  \ifthenelse{\boolean{articletitles}}{\emph{{Study of the reaction
  $e^{+}e^{-}\to \psi(2S)\pi^{-}\pi^{-}$ via initial-state radiation at
  BaBar}}, }{}\href{http://dx.doi.org/10.1103/PhysRevD.89.111103}{Phys.\ Rev.\
  \textbf{D89} (2014) 111103},
  \href{http://arxiv.org/abs/1211.6271}{{\normalfont\ttfamily
  arXiv:1211.6271}}\relax
\mciteBstWouldAddEndPuncttrue
\mciteSetBstMidEndSepPunct{\mcitedefaultmidpunct}
{\mcitedefaultendpunct}{\mcitedefaultseppunct}\relax
\EndOfBibitem
\bibitem{Wang:2007ea}
Belle, X.~L. Wang {\em et~al.},
  \ifthenelse{\boolean{articletitles}}{\emph{{Observation of Two Resonant
  Structures in $e^+e^- \to \pi^+ \pi^- \psi(2S)$ via Initial State Radiation
  at Belle}}, }{}\href{http://dx.doi.org/10.1103/PhysRevLett.99.142002}{Phys.\
  Rev.\ Lett.\  \textbf{99} (2007) 142002},
  \href{http://arxiv.org/abs/0707.3699}{{\normalfont\ttfamily
  arXiv:0707.3699}}\relax
\mciteBstWouldAddEndPuncttrue
\mciteSetBstMidEndSepPunct{\mcitedefaultmidpunct}
{\mcitedefaultendpunct}{\mcitedefaultseppunct}\relax
\EndOfBibitem
\bibitem{Wang:2014hta}
Belle, X.~L. Wang {\em et~al.},
  \ifthenelse{\boolean{articletitles}}{\emph{{Measurement of $e^+e^- \to
  \pi^+\pi^-\psi(2S)$ via Initial State Radiation at Belle}},
  }{}\href{http://dx.doi.org/10.1103/PhysRevD.91.112007}{Phys.\ Rev.\
  \textbf{D91} (2015) 112007},
  \href{http://arxiv.org/abs/1410.7641}{{\normalfont\ttfamily
  arXiv:1410.7641}}\relax
\mciteBstWouldAddEndPuncttrue
\mciteSetBstMidEndSepPunct{\mcitedefaultmidpunct}
{\mcitedefaultendpunct}{\mcitedefaultseppunct}\relax
\EndOfBibitem
\bibitem{Pakhlova:2008vn}
Belle, G.~Pakhlova {\em et~al.},
  \ifthenelse{\boolean{articletitles}}{\emph{{Observation of a near-threshold
  enhancement in the $e^+e^- \to \Lambda^+_{c} \Lambda^-_{c}$ cross section
  using initial-state radiation}},
  }{}\href{http://dx.doi.org/10.1103/PhysRevLett.101.172001}{Phys.\ Rev.\
  Lett.\  \textbf{101} (2008) 172001},
  \href{http://arxiv.org/abs/0807.4458}{{\normalfont\ttfamily
  arXiv:0807.4458}}\relax
\mciteBstWouldAddEndPuncttrue
\mciteSetBstMidEndSepPunct{\mcitedefaultmidpunct}
{\mcitedefaultendpunct}{\mcitedefaultseppunct}\relax
\EndOfBibitem
\bibitem{Ablikim:2015tbp}
BESIII, M.~Ablikim {\em et~al.},
  \ifthenelse{\boolean{articletitles}}{\emph{{Observation of $Z_c(3900)^{0}$ in
  $e^+e^-\to\pi^0\pi^0 J/\psi$}},
  }{}\href{http://dx.doi.org/10.1103/PhysRevLett.115.112003}{Phys.\ Rev.\
  Lett.\  \textbf{115} (2015) 112003},
  \href{http://arxiv.org/abs/1506.06018}{{\normalfont\ttfamily
  arXiv:1506.06018}}\relax
\mciteBstWouldAddEndPuncttrue
\mciteSetBstMidEndSepPunct{\mcitedefaultmidpunct}
{\mcitedefaultendpunct}{\mcitedefaultseppunct}\relax
\EndOfBibitem
\bibitem{Ablikim:2013xfr}
BESIII, M.~Ablikim {\em et~al.},
  \ifthenelse{\boolean{articletitles}}{\emph{{Observation of a charged
  $(D\bar{D}^{*})^\pm$ mass peak in $e^{+}e^{-} \to \pi D\bar{D}^{*}$ at
  $\sqrt{s} =$ 4.26 GeV}},
  }{}\href{http://dx.doi.org/10.1103/PhysRevLett.112.022001}{Phys.\ Rev.\
  Lett.\  \textbf{112} (2014) 022001},
  \href{http://arxiv.org/abs/1310.1163}{{\normalfont\ttfamily
  arXiv:1310.1163}}\relax
\mciteBstWouldAddEndPuncttrue
\mciteSetBstMidEndSepPunct{\mcitedefaultmidpunct}
{\mcitedefaultendpunct}{\mcitedefaultseppunct}\relax
\EndOfBibitem
\bibitem{Ablikim:2015gda}
BESIII, M.~Ablikim {\em et~al.},
  \ifthenelse{\boolean{articletitles}}{\emph{{Observation of a Neutral
  Structure near the $D\bar{D}^{*}$ Mass Threshold in $e^{+}e^{-}\to (D
  \bar{D}^*)^0\pi^0$ at $\sqrt{s}$ = 4.226 and 4.257 GeV}},
  }{}\href{http://dx.doi.org/10.1103/PhysRevLett.115.222002}{Phys.\ Rev.\
  Lett.\  \textbf{115} (2015) 222002},
  \href{http://arxiv.org/abs/1509.05620}{{\normalfont\ttfamily
  arXiv:1509.05620}}\relax
\mciteBstWouldAddEndPuncttrue
\mciteSetBstMidEndSepPunct{\mcitedefaultmidpunct}
{\mcitedefaultendpunct}{\mcitedefaultseppunct}\relax
\EndOfBibitem
\bibitem{Ablikim:2014dxl}
BESIII, M.~Ablikim {\em et~al.},
  \ifthenelse{\boolean{articletitles}}{\emph{{Observation of
  $e^+e^-\to\pi^0\pi^0 h_c$ and a Neutral Charmoniumlike Structure
  $Z_c(4020)^0$}},
  }{}\href{http://dx.doi.org/10.1103/PhysRevLett.113.212002}{Phys.\ Rev.\
  Lett.\  \textbf{113} (2014) 212002},
  \href{http://arxiv.org/abs/1409.6577}{{\normalfont\ttfamily
  arXiv:1409.6577}}\relax
\mciteBstWouldAddEndPuncttrue
\mciteSetBstMidEndSepPunct{\mcitedefaultmidpunct}
{\mcitedefaultendpunct}{\mcitedefaultseppunct}\relax
\EndOfBibitem
\bibitem{Ablikim:2013emm}
BESIII, M.~Ablikim {\em et~al.},
  \ifthenelse{\boolean{articletitles}}{\emph{{Observation of a charged
  charmoniumlike structure in $e^+e^- \to (D^{*} \bar{D}^{*})^{\pm} \pi^\mp$ at
  $\sqrt{s}=4.26$GeV}},
  }{}\href{http://dx.doi.org/10.1103/PhysRevLett.112.132001}{Phys.\ Rev.\
  Lett.\  \textbf{112} (2014) 132001},
  \href{http://arxiv.org/abs/1308.2760}{{\normalfont\ttfamily
  arXiv:1308.2760}}\relax
\mciteBstWouldAddEndPuncttrue
\mciteSetBstMidEndSepPunct{\mcitedefaultmidpunct}
{\mcitedefaultendpunct}{\mcitedefaultseppunct}\relax
\EndOfBibitem
\bibitem{Ablikim:2015vvn}
BESIII, M.~Ablikim {\em et~al.},
  \ifthenelse{\boolean{articletitles}}{\emph{{Observation of a neutral
  charmoniumlike state $Z_c(4025)^0$ in $e^{+} e^{-} \to (D^{*}
  \bar{D}^{*})^{0} \pi^0$}},
  }{}\href{http://dx.doi.org/10.1103/PhysRevLett.115.182002}{Phys.\ Rev.\
  Lett.\  \textbf{115} (2015) 182002},
  \href{http://arxiv.org/abs/1507.02404}{{\normalfont\ttfamily
  arXiv:1507.02404}}\relax
\mciteBstWouldAddEndPuncttrue
\mciteSetBstMidEndSepPunct{\mcitedefaultmidpunct}
{\mcitedefaultendpunct}{\mcitedefaultseppunct}\relax
\EndOfBibitem
\bibitem{Mizuk:2008me}
Belle, R.~Mizuk {\em et~al.},
  \ifthenelse{\boolean{articletitles}}{\emph{{Observation of two resonance-like
  structures in the $\pi^+ \chi_{c1}$ mass distribution in exclusive
  $\bar{B}^0\to K^- \pi^+ \chi_{c1}$ decays}},
  }{}\href{http://dx.doi.org/10.1103/PhysRevD.78.072004}{Phys.\ Rev.\
  \textbf{D78} (2008) 072004},
  \href{http://arxiv.org/abs/0806.4098}{{\normalfont\ttfamily
  arXiv:0806.4098}}\relax
\mciteBstWouldAddEndPuncttrue
\mciteSetBstMidEndSepPunct{\mcitedefaultmidpunct}
{\mcitedefaultendpunct}{\mcitedefaultseppunct}\relax
\EndOfBibitem
\bibitem{Lees:2011ik}
BaBar, J.~P. Lees {\em et~al.},
  \ifthenelse{\boolean{articletitles}}{\emph{{Search for the $Z_1(4050)^+$ and
  $Z_2(4250)^+$ states in $\overline{B^0} \to \chi_{c1} K^- \pi^+$ and $B^+ \to
  \chi_{c1} K^0_S \pi^+$}},
  }{}\href{http://dx.doi.org/10.1103/PhysRevD.85.052003}{Phys.\ Rev.\
  \textbf{D85} (2012) 052003},
  \href{http://arxiv.org/abs/1111.5919}{{\normalfont\ttfamily
  arXiv:1111.5919}}\relax
\mciteBstWouldAddEndPuncttrue
\mciteSetBstMidEndSepPunct{\mcitedefaultmidpunct}
{\mcitedefaultendpunct}{\mcitedefaultseppunct}\relax
\EndOfBibitem
\bibitem{Mizuk:2009da}
Belle, R.~Mizuk {\em et~al.},
  \ifthenelse{\boolean{articletitles}}{\emph{{Dalitz analysis of $B\to K \pi^+
  \psi'$ decays and the $Z(4430)^+$}},
  }{}\href{http://dx.doi.org/10.1103/PhysRevD.80.031104}{Phys.\ Rev.\
  \textbf{D80} (2009) 031104},
  \href{http://arxiv.org/abs/0905.2869}{{\normalfont\ttfamily
  arXiv:0905.2869}}\relax
\mciteBstWouldAddEndPuncttrue
\mciteSetBstMidEndSepPunct{\mcitedefaultmidpunct}
{\mcitedefaultendpunct}{\mcitedefaultseppunct}\relax
\EndOfBibitem
\bibitem{Chilikin:2013tch}
Belle, K.~Chilikin {\em et~al.},
  \ifthenelse{\boolean{articletitles}}{\emph{{Experimental constraints on the
  spin and parity of the $Z(4430)^+$}},
  }{}\href{http://dx.doi.org/10.1103/PhysRevD.88.074026}{Phys.\ Rev.\
  \textbf{D88} (2013) 074026},
  \href{http://arxiv.org/abs/1306.4894}{{\normalfont\ttfamily
  arXiv:1306.4894}}\relax
\mciteBstWouldAddEndPuncttrue
\mciteSetBstMidEndSepPunct{\mcitedefaultmidpunct}
{\mcitedefaultendpunct}{\mcitedefaultseppunct}\relax
\EndOfBibitem
\bibitem{Aaij:2015zxa}
LHCb, R.~Aaij {\em et~al.},
  \ifthenelse{\boolean{articletitles}}{\emph{{Model-independent confirmation of
  the $Z(4430)^-$ state}},
  }{}\href{http://dx.doi.org/10.1103/PhysRevD.92.112009}{Phys.\ Rev.\
  \textbf{D92} (2015) 112009},
  \href{http://arxiv.org/abs/1510.01951}{{\normalfont\ttfamily
  arXiv:1510.01951}}\relax
\mciteBstWouldAddEndPuncttrue
\mciteSetBstMidEndSepPunct{\mcitedefaultmidpunct}
{\mcitedefaultendpunct}{\mcitedefaultseppunct}\relax
\EndOfBibitem
\bibitem{Abe:2007tk}
Belle, K.~F. Chen {\em et~al.},
  \ifthenelse{\boolean{articletitles}}{\emph{{Observation of anomalous
  $\Upsilon(1S) \pi^+ \pi^-$ and $\Upsilon(2S) \pi^+ \pi^-$ production near the
  $\Upsilon(5S)$ resonance}},
  }{}\href{http://dx.doi.org/10.1103/PhysRevLett.100.112001}{Phys.\ Rev.\
  Lett.\  \textbf{100} (2008) 112001},
  \href{http://arxiv.org/abs/0710.2577}{{\normalfont\ttfamily
  arXiv:0710.2577}}\relax
\mciteBstWouldAddEndPuncttrue
\mciteSetBstMidEndSepPunct{\mcitedefaultmidpunct}
{\mcitedefaultendpunct}{\mcitedefaultseppunct}\relax
\EndOfBibitem
\bibitem{Santel:2015qga}
Belle, D.~Santel {\em et~al.},
  \ifthenelse{\boolean{articletitles}}{\emph{{Measurements of the
  $\Upsilon$(10860) and $\Upsilon$(11020) resonances via $\sigma(e^+e^-\to
  \Upsilon(nS)\pi^+ \pi^-)$}},
  }{}\href{http://dx.doi.org/10.1103/PhysRevD.93.011101}{Phys.\ Rev.\
  \textbf{D93} (2016) 011101},
  \href{http://arxiv.org/abs/1501.01137}{{\normalfont\ttfamily
  arXiv:1501.01137}}\relax
\mciteBstWouldAddEndPuncttrue
\mciteSetBstMidEndSepPunct{\mcitedefaultmidpunct}
{\mcitedefaultendpunct}{\mcitedefaultseppunct}\relax
\EndOfBibitem
\bibitem{Belle:2011aa}
Belle, A.~Bondar {\em et~al.},
  \ifthenelse{\boolean{articletitles}}{\emph{{Observation of two charged
  bottomonium-like resonances in $\Upsilon(5S)$ decays}},
  }{}\href{http://dx.doi.org/10.1103/PhysRevLett.108.122001}{Phys.\ Rev.\
  Lett.\  \textbf{108} (2012) 122001},
  \href{http://arxiv.org/abs/1110.2251}{{\normalfont\ttfamily
  arXiv:1110.2251}}\relax
\mciteBstWouldAddEndPuncttrue
\mciteSetBstMidEndSepPunct{\mcitedefaultmidpunct}
{\mcitedefaultendpunct}{\mcitedefaultseppunct}\relax
\EndOfBibitem
\bibitem{Garmash:2014dhx}
Belle, A.~Garmash {\em et~al.},
  \ifthenelse{\boolean{articletitles}}{\emph{{Amplitude analysis of $e^+e^- \to
  \Upsilon(nS) \pi^+\pi^-$ at $\sqrt{s}=10.865$~GeV}},
  }{}\href{http://dx.doi.org/10.1103/PhysRevD.91.072003}{Phys.\ Rev.\
  \textbf{D91} (2015) 072003},
  \href{http://arxiv.org/abs/1403.0992}{{\normalfont\ttfamily
  arXiv:1403.0992}}\relax
\mciteBstWouldAddEndPuncttrue
\mciteSetBstMidEndSepPunct{\mcitedefaultmidpunct}
{\mcitedefaultendpunct}{\mcitedefaultseppunct}\relax
\EndOfBibitem
\bibitem{Krokovny:2013mgx}
Belle, P.~Krokovny {\em et~al.},
  \ifthenelse{\boolean{articletitles}}{\emph{{First observation of the $Z
  \frac{0}{b}$(10610) in a Dalitz analysis of $\Upsilon$(10860) $\to
  \Upsilon$(nS)$\pi^0 \pi^0$}},
  }{}\href{http://dx.doi.org/10.1103/PhysRevD.88.052016}{Phys.\ Rev.\
  \textbf{D88} (2013) 052016},
  \href{http://arxiv.org/abs/1308.2646}{{\normalfont\ttfamily
  arXiv:1308.2646}}\relax
\mciteBstWouldAddEndPuncttrue
\mciteSetBstMidEndSepPunct{\mcitedefaultmidpunct}
{\mcitedefaultendpunct}{\mcitedefaultseppunct}\relax
\EndOfBibitem
\bibitem{Garmash:2015rfd}
Belle, A.~Garmash {\em et~al.},
  \ifthenelse{\boolean{articletitles}}{\emph{{Observation of $Z_b(10610)$ and
  $Z_b(10650)$ Decaying to $B$ Mesons}},
  }{}\href{http://dx.doi.org/10.1103/PhysRevLett.116.212001}{Phys.\ Rev.\
  Lett.\  \textbf{116} (2016) 212001},
  \href{http://arxiv.org/abs/1512.07419}{{\normalfont\ttfamily
  arXiv:1512.07419}}\relax
\mciteBstWouldAddEndPuncttrue
\mciteSetBstMidEndSepPunct{\mcitedefaultmidpunct}
{\mcitedefaultendpunct}{\mcitedefaultseppunct}\relax
\EndOfBibitem
\bibitem{Lebed:2016hpi}
R.~F. Lebed, R.~E. Mitchell, and E.~S. Swanson,
  \ifthenelse{\boolean{articletitles}}{\emph{{Heavy-Quark QCD Exotica}},
  }{}\href{http://dx.doi.org/10.1016/j.ppnp.2016.11.003}{Prog.\ Part.\ Nucl.\
  Phys.\  \textbf{93} (2017) 143},
  \href{http://arxiv.org/abs/1610.04528}{{\normalfont\ttfamily
  arXiv:1610.04528}}\relax
\mciteBstWouldAddEndPuncttrue
\mciteSetBstMidEndSepPunct{\mcitedefaultmidpunct}
{\mcitedefaultendpunct}{\mcitedefaultseppunct}\relax
\EndOfBibitem
\bibitem{Esposito:2016noz}
A.~Esposito, A.~Pilloni, and A.~D. Polosa,
  \ifthenelse{\boolean{articletitles}}{\emph{{Multiquark Resonances}},
  }{}\href{http://dx.doi.org/10.1016/j.physrep.2016.11.002}{Phys.\ Rept.\
  \textbf{668} (2016) 1},
  \href{http://arxiv.org/abs/1611.07920}{{\normalfont\ttfamily
  arXiv:1611.07920}}\relax
\mciteBstWouldAddEndPuncttrue
\mciteSetBstMidEndSepPunct{\mcitedefaultmidpunct}
{\mcitedefaultendpunct}{\mcitedefaultseppunct}\relax
\EndOfBibitem
\bibitem{Guo:2017jvc}
F.-K. Guo {\em et~al.}, \ifthenelse{\boolean{articletitles}}{\emph{{Hadronic
  molecules}}, }{}\href{http://arxiv.org/abs/1705.00141}{{\normalfont\ttfamily
  arXiv:1705.00141}}\relax
\mciteBstWouldAddEndPuncttrue
\mciteSetBstMidEndSepPunct{\mcitedefaultmidpunct}
{\mcitedefaultendpunct}{\mcitedefaultseppunct}\relax
\EndOfBibitem
\bibitem{Ali:2017jda}
A.~Ali, J.~S. Lange, and S.~Stone,
  \ifthenelse{\boolean{articletitles}}{\emph{{Exotics: Heavy Pentaquarks and
  Tetraquarks}},
  }{}\href{http://arxiv.org/abs/1706.00610}{{\normalfont\ttfamily
  arXiv:1706.00610}}\relax
\mciteBstWouldAddEndPuncttrue
\mciteSetBstMidEndSepPunct{\mcitedefaultmidpunct}
{\mcitedefaultendpunct}{\mcitedefaultseppunct}\relax
\EndOfBibitem
\bibitem{Afach:2015sja}
J.~M. Pendlebury {\em et~al.},
  \ifthenelse{\boolean{articletitles}}{\emph{{Revised experimental upper limit
  on the electric dipole moment of the neutron}},
  }{}\href{http://dx.doi.org/10.1103/PhysRevD.92.092003}{Phys.\ Rev.\
  \textbf{D92} (2015) 092003},
  \href{http://arxiv.org/abs/1509.04411}{{\normalfont\ttfamily
  arXiv:1509.04411}}\relax
\mciteBstWouldAddEndPuncttrue
\mciteSetBstMidEndSepPunct{\mcitedefaultmidpunct}
{\mcitedefaultendpunct}{\mcitedefaultseppunct}\relax
\EndOfBibitem
\bibitem{Jaffe:1976ih}
R.~L. Jaffe, \ifthenelse{\boolean{articletitles}}{\emph{{Multi-Quark Hadrons.
  2. Methods}}, }{}\href{http://dx.doi.org/10.1103/PhysRevD.15.281}{Phys.\
  Rev.\  \textbf{D15} (1977) 281}\relax
\mciteBstWouldAddEndPuncttrue
\mciteSetBstMidEndSepPunct{\mcitedefaultmidpunct}
{\mcitedefaultendpunct}{\mcitedefaultseppunct}\relax
\EndOfBibitem
\bibitem{Maiani:2004uc}
L.~Maiani, F.~Piccinini, A.~D. Polosa, and V.~Riquer,
  \ifthenelse{\boolean{articletitles}}{\emph{{A New look at scalar mesons}},
  }{}\href{http://dx.doi.org/10.1103/PhysRevLett.93.212002}{Phys.\ Rev.\ Lett.\
   \textbf{93} (2004) 212002},
  \href{http://arxiv.org/abs/hep-ph/0407017}{{\normalfont\ttfamily
  arXiv:hep-ph/0407017}}\relax
\mciteBstWouldAddEndPuncttrue
\mciteSetBstMidEndSepPunct{\mcitedefaultmidpunct}
{\mcitedefaultendpunct}{\mcitedefaultseppunct}\relax
\EndOfBibitem
\bibitem{Hooft:2008we}
G.~'t~Hooft {\em et~al.}, \ifthenelse{\boolean{articletitles}}{\emph{{A Theory
  of Scalar Mesons}},
  }{}\href{http://dx.doi.org/10.1016/j.physletb.2008.03.036}{Phys.\ Lett.\
  \textbf{B662} (2008) 424},
  \href{http://arxiv.org/abs/0801.2288}{{\normalfont\ttfamily
  arXiv:0801.2288}}\relax
\mciteBstWouldAddEndPuncttrue
\mciteSetBstMidEndSepPunct{\mcitedefaultmidpunct}
{\mcitedefaultendpunct}{\mcitedefaultseppunct}\relax
\EndOfBibitem
\bibitem{Maiani:2004vq}
L.~Maiani, F.~Piccinini, A.~D. Polosa, and V.~Riquer,
  \ifthenelse{\boolean{articletitles}}{\emph{{Diquark-antidiquarks with hidden
  or open charm and the nature of X(3872)}},
  }{}\href{http://dx.doi.org/10.1103/PhysRevD.71.014028}{Phys.\ Rev.\
  \textbf{D71} (2005) 014028},
  \href{http://arxiv.org/abs/hep-ph/0412098}{{\normalfont\ttfamily
  arXiv:hep-ph/0412098}}\relax
\mciteBstWouldAddEndPuncttrue
\mciteSetBstMidEndSepPunct{\mcitedefaultmidpunct}
{\mcitedefaultendpunct}{\mcitedefaultseppunct}\relax
\EndOfBibitem
\bibitem{Terasaki:2004yx}
K.~Terasaki, \ifthenelse{\boolean{articletitles}}{\emph{{Charmed scalar mesons
  and related}}, }{} 2004\relax
\mciteBstWouldAddEndPuncttrue
\mciteSetBstMidEndSepPunct{\mcitedefaultmidpunct}
{\mcitedefaultendpunct}{\mcitedefaultseppunct}\relax
\EndOfBibitem
\bibitem{Manohar:2000dt}
A.~V. Manohar and M.~B. Wise, \ifthenelse{\boolean{articletitles}}{\emph{{Heavy
  quark physics}}, }{}Camb.\ Monogr.\ Part.\ Phys.\ Nucl.\ Phys.\ Cosmol.\
  \textbf{10} (2000) 1\relax
\mciteBstWouldAddEndPuncttrue
\mciteSetBstMidEndSepPunct{\mcitedefaultmidpunct}
{\mcitedefaultendpunct}{\mcitedefaultseppunct}\relax
\EndOfBibitem
\bibitem{Esposito:2014rxa}
A.~Esposito {\em et~al.},
  \ifthenelse{\boolean{articletitles}}{\emph{{Four-Quark Hadrons: an Updated
  Review}}, }{}\href{http://dx.doi.org/10.1142/S0217751X15300021}{Int.\ J.\
  Mod.\ Phys.\  \textbf{A30} (2015) 1530002},
  \href{http://arxiv.org/abs/1411.5997}{{\normalfont\ttfamily
  arXiv:1411.5997}}\relax
\mciteBstWouldAddEndPuncttrue
\mciteSetBstMidEndSepPunct{\mcitedefaultmidpunct}
{\mcitedefaultendpunct}{\mcitedefaultseppunct}\relax
\EndOfBibitem
\bibitem{Isgur:1983wj}
N.~Isgur and J.~E. Paton, \ifthenelse{\boolean{articletitles}}{\emph{{A Flux
  Tube Model for Hadrons}},
  }{}\href{http://dx.doi.org/10.1016/0370-2693(83)91445-4}{Phys.\ Lett.\
  \textbf{B124} (1983) 247}\relax
\mciteBstWouldAddEndPuncttrue
\mciteSetBstMidEndSepPunct{\mcitedefaultmidpunct}
{\mcitedefaultendpunct}{\mcitedefaultseppunct}\relax
\EndOfBibitem
\bibitem{Horn:1977rq}
D.~Horn and J.~Mandula, \ifthenelse{\boolean{articletitles}}{\emph{{A Model of
  Mesons with Constituent Gluons}},
  }{}\href{http://dx.doi.org/10.1103/PhysRevD.17.898}{Phys.\ Rev.\
  \textbf{D17} (1978) 898}\relax
\mciteBstWouldAddEndPuncttrue
\mciteSetBstMidEndSepPunct{\mcitedefaultmidpunct}
{\mcitedefaultendpunct}{\mcitedefaultseppunct}\relax
\EndOfBibitem
\bibitem{Isgur:1985vy}
N.~Isgur, R.~Kokoski, and J.~Paton,
  \ifthenelse{\boolean{articletitles}}{\emph{{Gluonic Excitations of Mesons:
  Why They Are Missing and Where to Find Them}},
  }{}\href{http://dx.doi.org/10.1103/PhysRevLett.54.869,
  10.1063/1.35357}{Phys.\ Rev.\ Lett.\  \textbf{54} (1985) 869}, [AIP Conf.
  Proc.132,242(1985)]\relax
\mciteBstWouldAddEndPuncttrue
\mciteSetBstMidEndSepPunct{\mcitedefaultmidpunct}
{\mcitedefaultendpunct}{\mcitedefaultseppunct}\relax
\EndOfBibitem
\bibitem{Page:1998gz}
P.~R. Page, E.~S. Swanson, and A.~P. Szczepaniak,
  \ifthenelse{\boolean{articletitles}}{\emph{{Hybrid meson decay
  phenomenology}},
  }{}\href{http://dx.doi.org/10.1103/PhysRevD.59.034016}{Phys.\ Rev.\
  \textbf{D59} (1999) 034016},
  \href{http://arxiv.org/abs/hep-ph/9808346}{{\normalfont\ttfamily
  arXiv:hep-ph/9808346}}\relax
\mciteBstWouldAddEndPuncttrue
\mciteSetBstMidEndSepPunct{\mcitedefaultmidpunct}
{\mcitedefaultendpunct}{\mcitedefaultseppunct}\relax
\EndOfBibitem
\bibitem{Voloshin:1976ap}
M.~B. Voloshin and L.~B. Okun,
  \ifthenelse{\boolean{articletitles}}{\emph{{Hadron Molecules and Charmonium
  Atom}}, }{}JETP Lett.\  \textbf{23} (1976) 333, [Pisma Zh. Eksp. Teor.
  Fiz.23,369(1976)]\relax
\mciteBstWouldAddEndPuncttrue
\mciteSetBstMidEndSepPunct{\mcitedefaultmidpunct}
{\mcitedefaultendpunct}{\mcitedefaultseppunct}\relax
\EndOfBibitem
\bibitem{Bander:1975fb}
M.~Bander, G.~L. Shaw, P.~Thomas, and S.~Meshkov,
  \ifthenelse{\boolean{articletitles}}{\emph{{Exotic Mesons and $e^+e^-$
  Annihilation}}, }{}\href{http://dx.doi.org/10.1103/PhysRevLett.36.695}{Phys.\
  Rev.\ Lett.\  \textbf{36} (1976) 695}\relax
\mciteBstWouldAddEndPuncttrue
\mciteSetBstMidEndSepPunct{\mcitedefaultmidpunct}
{\mcitedefaultendpunct}{\mcitedefaultseppunct}\relax
\EndOfBibitem
\bibitem{DeRujula:1976zlg}
A.~De~Rujula, H.~Georgi, and S.~L. Glashow,
  \ifthenelse{\boolean{articletitles}}{\emph{{Molecular Charmonium: A New
  Spectroscopy?}},
  }{}\href{http://dx.doi.org/10.1103/PhysRevLett.38.317}{Phys.\ Rev.\ Lett.\
  \textbf{38} (1977) 317}\relax
\mciteBstWouldAddEndPuncttrue
\mciteSetBstMidEndSepPunct{\mcitedefaultmidpunct}
{\mcitedefaultendpunct}{\mcitedefaultseppunct}\relax
\EndOfBibitem
\bibitem{Manohar:1992nd}
A.~V. Manohar and M.~B. Wise,
  \ifthenelse{\boolean{articletitles}}{\emph{{Exotic $QQ\bar{q}\bar{q}$ states
  in QCD}}, }{}\href{http://dx.doi.org/10.1016/0550-3213(93)90614-U}{Nucl.\
  Phys.\  \textbf{B399} (1993) 17},
  \href{http://arxiv.org/abs/hep-ph/9212236}{{\normalfont\ttfamily
  arXiv:hep-ph/9212236}}\relax
\mciteBstWouldAddEndPuncttrue
\mciteSetBstMidEndSepPunct{\mcitedefaultmidpunct}
{\mcitedefaultendpunct}{\mcitedefaultseppunct}\relax
\EndOfBibitem
\bibitem{Tornqvist:1993ng}
N.~A. Tornqvist, \ifthenelse{\boolean{articletitles}}{\emph{{From the deuteron
  to deusons, an analysis of deuteron - like meson meson bound states}},
  }{}\href{http://dx.doi.org/10.1007/BF01413192}{Z.\ Phys.\  \textbf{C61}
  (1994) 525}, \href{http://arxiv.org/abs/hep-ph/9310247}{{\normalfont\ttfamily
  arXiv:hep-ph/9310247}}\relax
\mciteBstWouldAddEndPuncttrue
\mciteSetBstMidEndSepPunct{\mcitedefaultmidpunct}
{\mcitedefaultendpunct}{\mcitedefaultseppunct}\relax
\EndOfBibitem
\bibitem{Swanson:2006st}
E.~S. Swanson, \ifthenelse{\boolean{articletitles}}{\emph{{The New heavy
  mesons: A Status report}},
  }{}\href{http://dx.doi.org/10.1016/j.physrep.2006.04.003}{Phys.\ Rept.\
  \textbf{429} (2006) 243},
  \href{http://arxiv.org/abs/hep-ph/0601110}{{\normalfont\ttfamily
  arXiv:hep-ph/0601110}}\relax
\mciteBstWouldAddEndPuncttrue
\mciteSetBstMidEndSepPunct{\mcitedefaultmidpunct}
{\mcitedefaultendpunct}{\mcitedefaultseppunct}\relax
\EndOfBibitem
\bibitem{Polosa:2015tra}
A.~D. Polosa, \ifthenelse{\boolean{articletitles}}{\emph{{Constraints from
  precision measurements on the hadron-molecule interpretation of X , Y , Z
  resonances}},
  }{}\href{http://dx.doi.org/10.1016/j.physletb.2015.05.017}{Phys.\ Lett.\
  \textbf{B746} (2015) 248},
  \href{http://arxiv.org/abs/1505.03083}{{\normalfont\ttfamily
  arXiv:1505.03083}}\relax
\mciteBstWouldAddEndPuncttrue
\mciteSetBstMidEndSepPunct{\mcitedefaultmidpunct}
{\mcitedefaultendpunct}{\mcitedefaultseppunct}\relax
\EndOfBibitem
\bibitem{Dubynskiy:2008mq}
S.~Dubynskiy and M.~B. Voloshin,
  \ifthenelse{\boolean{articletitles}}{\emph{{Hadro-Charmonium}},
  }{}\href{http://dx.doi.org/10.1016/j.physletb.2008.07.086}{Phys.\ Lett.\
  \textbf{B666} (2008) 344},
  \href{http://arxiv.org/abs/0803.2224}{{\normalfont\ttfamily
  arXiv:0803.2224}}\relax
\mciteBstWouldAddEndPuncttrue
\mciteSetBstMidEndSepPunct{\mcitedefaultmidpunct}
{\mcitedefaultendpunct}{\mcitedefaultseppunct}\relax
\EndOfBibitem
\bibitem{Dubynskiy:2008di}
S.~Dubynskiy, A.~Gorsky, and M.~B. Voloshin,
  \ifthenelse{\boolean{articletitles}}{\emph{{Holographic Hadro-Quarkonium}},
  }{}\href{http://dx.doi.org/10.1016/j.physletb.2008.11.040}{Phys.\ Lett.\
  \textbf{B671} (2009) 82},
  \href{http://arxiv.org/abs/0804.2244}{{\normalfont\ttfamily
  arXiv:0804.2244}}\relax
\mciteBstWouldAddEndPuncttrue
\mciteSetBstMidEndSepPunct{\mcitedefaultmidpunct}
{\mcitedefaultendpunct}{\mcitedefaultseppunct}\relax
\EndOfBibitem
\bibitem{Braaten:2013boa}
E.~Braaten, \ifthenelse{\boolean{articletitles}}{\emph{{How the $Z_c(3900)$
  Reveals the Spectra of Quarkonium Hybrid and Tetraquark Mesons}},
  }{}\href{http://dx.doi.org/10.1103/PhysRevLett.111.162003}{Phys.\ Rev.\
  Lett.\  \textbf{111} (2013) 162003},
  \href{http://arxiv.org/abs/1305.6905}{{\normalfont\ttfamily
  arXiv:1305.6905}}\relax
\mciteBstWouldAddEndPuncttrue
\mciteSetBstMidEndSepPunct{\mcitedefaultmidpunct}
{\mcitedefaultendpunct}{\mcitedefaultseppunct}\relax
\EndOfBibitem
\bibitem{Braaten:2014qka}
E.~Braaten, C.~Langmack, and D.~H. Smith,
  \ifthenelse{\boolean{articletitles}}{\emph{{Born-Oppenheimer Approximation
  for the XYZ Mesons}},
  }{}\href{http://dx.doi.org/10.1103/PhysRevD.90.014044}{Phys.\ Rev.\
  \textbf{D90} (2014) 014044},
  \href{http://arxiv.org/abs/1402.0438}{{\normalfont\ttfamily
  arXiv:1402.0438}}\relax
\mciteBstWouldAddEndPuncttrue
\mciteSetBstMidEndSepPunct{\mcitedefaultmidpunct}
{\mcitedefaultendpunct}{\mcitedefaultseppunct}\relax
\EndOfBibitem
\bibitem{Juge:1999ie}
K.~J. Juge, J.~Kuti, and C.~J. Morningstar,
  \ifthenelse{\boolean{articletitles}}{\emph{{Ab initio study of hybrid
  $\bar{b} g b$ mesons}},
  }{}\href{http://dx.doi.org/10.1103/PhysRevLett.82.4400}{Phys.\ Rev.\ Lett.\
  \textbf{82} (1999) 4400},
  \href{http://arxiv.org/abs/hep-ph/9902336}{{\normalfont\ttfamily
  arXiv:hep-ph/9902336}}\relax
\mciteBstWouldAddEndPuncttrue
\mciteSetBstMidEndSepPunct{\mcitedefaultmidpunct}
{\mcitedefaultendpunct}{\mcitedefaultseppunct}\relax
\EndOfBibitem
\bibitem{Bugg:2011jr}
D.~V. Bugg, \ifthenelse{\boolean{articletitles}}{\emph{{An Explanation of Belle
  states $Z_b(10610)$ and $Z_b(10650)$}},
  }{}\href{http://dx.doi.org/10.1209/0295-5075/96/11002}{Europhys.\ Lett.\
  \textbf{96} (2011) 11002},
  \href{http://arxiv.org/abs/1105.5492}{{\normalfont\ttfamily
  arXiv:1105.5492}}\relax
\mciteBstWouldAddEndPuncttrue
\mciteSetBstMidEndSepPunct{\mcitedefaultmidpunct}
{\mcitedefaultendpunct}{\mcitedefaultseppunct}\relax
\EndOfBibitem
\bibitem{Blitz:2015nra}
S.~H. Blitz and R.~F. Lebed,
  \ifthenelse{\boolean{articletitles}}{\emph{{Tetraquark Cusp Effects from
  Diquark Pair Production}},
  }{}\href{http://dx.doi.org/10.1103/PhysRevD.91.094025}{Phys.\ Rev.\
  \textbf{D91} (2015) 094025},
  \href{http://arxiv.org/abs/1503.04802}{{\normalfont\ttfamily
  arXiv:1503.04802}}\relax
\mciteBstWouldAddEndPuncttrue
\mciteSetBstMidEndSepPunct{\mcitedefaultmidpunct}
{\mcitedefaultendpunct}{\mcitedefaultseppunct}\relax
\EndOfBibitem
\bibitem{Swanson:2015bsa}
E.~S. Swanson, \ifthenelse{\boolean{articletitles}}{\emph{{Cusps and Exotic
  Charmonia}}, }{}\href{http://dx.doi.org/10.1142/S0218301316420106}{Int.\ J.\
  Mod.\ Phys.\  \textbf{E25} (2016) 1642010},
  \href{http://arxiv.org/abs/1504.07952}{{\normalfont\ttfamily
  arXiv:1504.07952}}\relax
\mciteBstWouldAddEndPuncttrue
\mciteSetBstMidEndSepPunct{\mcitedefaultmidpunct}
{\mcitedefaultendpunct}{\mcitedefaultseppunct}\relax
\EndOfBibitem
\bibitem{Guo:2014iya}
F.-K. Guo, C.~Hanhart, Q.~Wang, and Q.~Zhao,
  \ifthenelse{\boolean{articletitles}}{\emph{{Could the near-threshold $XYZ$
  states be simply kinematic effects?}},
  }{}\href{http://dx.doi.org/10.1103/PhysRevD.91.051504}{Phys.\ Rev.\
  \textbf{D91} (2015) 051504},
  \href{http://arxiv.org/abs/1411.5584}{{\normalfont\ttfamily
  arXiv:1411.5584}}\relax
\mciteBstWouldAddEndPuncttrue
\mciteSetBstMidEndSepPunct{\mcitedefaultmidpunct}
{\mcitedefaultendpunct}{\mcitedefaultseppunct}\relax
\EndOfBibitem
\bibitem{Swanson:2014tra}
E.~S. Swanson, \ifthenelse{\boolean{articletitles}}{\emph{{$Z_b$ and $Z_c$
  Exotic States as Coupled Channel Cusps}},
  }{}\href{http://dx.doi.org/10.1103/PhysRevD.91.034009}{Phys.\ Rev.\
  \textbf{D91} (2015) 034009},
  \href{http://arxiv.org/abs/1409.3291}{{\normalfont\ttfamily
  arXiv:1409.3291}}\relax
\mciteBstWouldAddEndPuncttrue
\mciteSetBstMidEndSepPunct{\mcitedefaultmidpunct}
{\mcitedefaultendpunct}{\mcitedefaultseppunct}\relax
\EndOfBibitem
\bibitem{Landau:1959fi}
L.~D. Landau, \ifthenelse{\boolean{articletitles}}{\emph{{On analytic
  properties of vertex parts in quantum field theory}},
  }{}\href{http://dx.doi.org/10.1016/0029-5582(59)90154-3}{Nucl.\ Phys.\
  \textbf{13} (1959) 181}\relax
\mciteBstWouldAddEndPuncttrue
\mciteSetBstMidEndSepPunct{\mcitedefaultmidpunct}
{\mcitedefaultendpunct}{\mcitedefaultseppunct}\relax
\EndOfBibitem
\bibitem{Coleman:1965xm}
S.~Coleman and R.~E. Norton,
  \ifthenelse{\boolean{articletitles}}{\emph{{Singularities in the physical
  region}}, }{}\href{http://dx.doi.org/10.1007/BF02750472}{Nuovo Cim.\
  \textbf{38} (1965) 438}\relax
\mciteBstWouldAddEndPuncttrue
\mciteSetBstMidEndSepPunct{\mcitedefaultmidpunct}
{\mcitedefaultendpunct}{\mcitedefaultseppunct}\relax
\EndOfBibitem
\bibitem{Schmid:1967ojm}
C.~Schmid, \ifthenelse{\boolean{articletitles}}{\emph{{Final-State Interactions
  and the Simulation of Resonances}},
  }{}\href{http://dx.doi.org/10.1103/PhysRev.154.1363}{Phys.\ Rev.\
  \textbf{154} (1967), no.~5 1363}\relax
\mciteBstWouldAddEndPuncttrue
\mciteSetBstMidEndSepPunct{\mcitedefaultmidpunct}
{\mcitedefaultendpunct}{\mcitedefaultseppunct}\relax
\EndOfBibitem
\bibitem{Szczepaniak:2015hya}
A.~P. Szczepaniak, \ifthenelse{\boolean{articletitles}}{\emph{{Dalitz plot
  distributions in presence of triangle singularities}},
  }{}\href{http://dx.doi.org/10.1016/j.physletb.2016.03.064}{Phys.\ Lett.\
  \textbf{B757} (2016) 61},
  \href{http://arxiv.org/abs/1510.01789}{{\normalfont\ttfamily
  arXiv:1510.01789}}\relax
\mciteBstWouldAddEndPuncttrue
\mciteSetBstMidEndSepPunct{\mcitedefaultmidpunct}
{\mcitedefaultendpunct}{\mcitedefaultseppunct}\relax
\EndOfBibitem
\bibitem{Liu:2015taa}
X.-H. Liu, M.~Oka, and Q.~Zhao,
  \ifthenelse{\boolean{articletitles}}{\emph{{Searching for observable effects
  induced by anomalous triangle singularities}},
  }{}\href{http://dx.doi.org/10.1016/j.physletb.2015.12.027}{Phys.\ Lett.\
  \textbf{B753} (2016) 297},
  \href{http://arxiv.org/abs/1507.01674}{{\normalfont\ttfamily
  arXiv:1507.01674}}\relax
\mciteBstWouldAddEndPuncttrue
\mciteSetBstMidEndSepPunct{\mcitedefaultmidpunct}
{\mcitedefaultendpunct}{\mcitedefaultseppunct}\relax
\EndOfBibitem
\bibitem{Szczepaniak:2015eza}
A.~P. Szczepaniak, \ifthenelse{\boolean{articletitles}}{\emph{{Triangle
  Singularities and XYZ Quarkonium Peaks}},
  }{}\href{http://dx.doi.org/10.1016/j.physletb.2015.06.029}{Phys.\ Lett.\
  \textbf{B747} (2015) 410},
  \href{http://arxiv.org/abs/1501.01691}{{\normalfont\ttfamily
  arXiv:1501.01691}}\relax
\mciteBstWouldAddEndPuncttrue
\mciteSetBstMidEndSepPunct{\mcitedefaultmidpunct}
{\mcitedefaultendpunct}{\mcitedefaultseppunct}\relax
\EndOfBibitem
\bibitem{Pilloni:2016obd}
JPAC, A.~Pilloni {\em et~al.},
  \ifthenelse{\boolean{articletitles}}{\emph{{Amplitude analysis and the nature
  of the Z$_c$(3900)}},
  }{}\href{http://dx.doi.org/10.1016/j.physletb.2017.06.030}{Phys.\ Lett.\
  \textbf{B772} (2017) 200},
  \href{http://arxiv.org/abs/1612.06490}{{\normalfont\ttfamily
  arXiv:1612.06490}}\relax
\mciteBstWouldAddEndPuncttrue
\mciteSetBstMidEndSepPunct{\mcitedefaultmidpunct}
{\mcitedefaultendpunct}{\mcitedefaultseppunct}\relax
\EndOfBibitem
\bibitem{Weber:2013eba}
M.~Wagner, S.~Diehl, T.~Kuske, and J.~Weber,
  \ifthenelse{\boolean{articletitles}}{\emph{{An introduction to lattice hadron
  spectroscopy for students without quantum field theoretical background}}, }{}
  2013\relax
\mciteBstWouldAddEndPuncttrue
\mciteSetBstMidEndSepPunct{\mcitedefaultmidpunct}
{\mcitedefaultendpunct}{\mcitedefaultseppunct}\relax
\EndOfBibitem
\bibitem{Wilson:1974sk}
K.~G. Wilson, \ifthenelse{\boolean{articletitles}}{\emph{{Confinement of
  Quarks}}, }{}\href{http://dx.doi.org/10.1103/PhysRevD.10.2445}{Phys.\ Rev.\
  \textbf{D10} (1974) 2445}, [,45(1974)]\relax
\mciteBstWouldAddEndPuncttrue
\mciteSetBstMidEndSepPunct{\mcitedefaultmidpunct}
{\mcitedefaultendpunct}{\mcitedefaultseppunct}\relax
\EndOfBibitem
\bibitem{Blum:2013mhx}
T.~Blum {\em et~al.}, \ifthenelse{\boolean{articletitles}}{\emph{{Working Group
  Report: Lattice Field Theory}}, }{} in {\em {Community Summer Study 2013:
  Snowmass on the Mississippi (CSS2013) Minneapolis, MN, USA, July 29-August 6,
  2013}}, 2013.
\newblock \href{http://arxiv.org/abs/1310.6087}{{\normalfont\ttfamily
  arXiv:1310.6087}}\relax
\mciteBstWouldAddEndPuncttrue
\mciteSetBstMidEndSepPunct{\mcitedefaultmidpunct}
{\mcitedefaultendpunct}{\mcitedefaultseppunct}\relax
\EndOfBibitem
\bibitem{Orginos:2015tha}
K.~Orginos and D.~Richards,
  \ifthenelse{\boolean{articletitles}}{\emph{{Improved methods for the study of
  hadronic physics from lattice QCD}},
  }{}\href{http://dx.doi.org/10.1088/0954-3899/42/3/034011}{J.\ Phys.\
  \textbf{G42} (2015) 034011}\relax
\mciteBstWouldAddEndPuncttrue
\mciteSetBstMidEndSepPunct{\mcitedefaultmidpunct}
{\mcitedefaultendpunct}{\mcitedefaultseppunct}\relax
\EndOfBibitem
\bibitem{Bietenholz:2011qq}
W.~Bietenholz {\em et~al.}, \ifthenelse{\boolean{articletitles}}{\emph{{Flavour
  blindness and patterns of flavour symmetry breaking in lattice simulations of
  up, down and strange quarks}},
  }{}\href{http://dx.doi.org/10.1103/PhysRevD.84.054509}{Phys.\ Rev.\
  \textbf{D84} (2011) 054509},
  \href{http://arxiv.org/abs/1102.5300}{{\normalfont\ttfamily
  arXiv:1102.5300}}\relax
\mciteBstWouldAddEndPuncttrue
\mciteSetBstMidEndSepPunct{\mcitedefaultmidpunct}
{\mcitedefaultendpunct}{\mcitedefaultseppunct}\relax
\EndOfBibitem
\bibitem{Aubin:2004wf}
C.~Aubin {\em et~al.}, \ifthenelse{\boolean{articletitles}}{\emph{{Light
  hadrons with improved staggered quarks: Approaching the continuum limit}},
  }{}\href{http://dx.doi.org/10.1103/PhysRevD.70.094505}{Phys.\ Rev.\
  \textbf{D70} (2004) 094505},
  \href{http://arxiv.org/abs/hep-lat/0402030}{{\normalfont\ttfamily
  arXiv:hep-lat/0402030}}\relax
\mciteBstWouldAddEndPuncttrue
\mciteSetBstMidEndSepPunct{\mcitedefaultmidpunct}
{\mcitedefaultendpunct}{\mcitedefaultseppunct}\relax
\EndOfBibitem
\bibitem{Bazavov:2009bb}
MILC, A.~Bazavov {\em et~al.},
  \ifthenelse{\boolean{articletitles}}{\emph{{Nonperturbative QCD Simulations
  with 2+1 Flavors of Improved Staggered Quarks}},
  }{}\href{http://dx.doi.org/10.1103/RevModPhys.82.1349}{Rev.\ Mod.\ Phys.\
  \textbf{82} (2010) 1349},
  \href{http://arxiv.org/abs/0903.3598}{{\normalfont\ttfamily
  arXiv:0903.3598}}\relax
\mciteBstWouldAddEndPuncttrue
\mciteSetBstMidEndSepPunct{\mcitedefaultmidpunct}
{\mcitedefaultendpunct}{\mcitedefaultseppunct}\relax
\EndOfBibitem
\bibitem{Aoki:2008sm}
PACS-CS, S.~Aoki {\em et~al.}, \ifthenelse{\boolean{articletitles}}{\emph{{2+1
  Flavor Lattice QCD toward the Physical Point}},
  }{}\href{http://dx.doi.org/10.1103/PhysRevD.79.034503}{Phys.\ Rev.\
  \textbf{D79} (2009) 034503},
  \href{http://arxiv.org/abs/0807.1661}{{\normalfont\ttfamily
  arXiv:0807.1661}}\relax
\mciteBstWouldAddEndPuncttrue
\mciteSetBstMidEndSepPunct{\mcitedefaultmidpunct}
{\mcitedefaultendpunct}{\mcitedefaultseppunct}\relax
\EndOfBibitem
\bibitem{Durr:2008zz}
S.~Durr {\em et~al.}, \ifthenelse{\boolean{articletitles}}{\emph{{Ab-Initio
  Determination of Light Hadron Masses}},
  }{}\href{http://dx.doi.org/10.1126/science.1163233}{Science \textbf{322}
  (2008) 1224}, \href{http://arxiv.org/abs/0906.3599}{{\normalfont\ttfamily
  arXiv:0906.3599}}\relax
\mciteBstWouldAddEndPuncttrue
\mciteSetBstMidEndSepPunct{\mcitedefaultmidpunct}
{\mcitedefaultendpunct}{\mcitedefaultseppunct}\relax
\EndOfBibitem
\bibitem{Christ:2010dd}
N.~H. Christ {\em et~al.}, \ifthenelse{\boolean{articletitles}}{\emph{{The
  $\eta$ and $\eta^\prime$ mesons from Lattice QCD}},
  }{}\href{http://dx.doi.org/10.1103/PhysRevLett.105.241601}{Phys.\ Rev.\
  Lett.\  \textbf{105} (2010) 241601},
  \href{http://arxiv.org/abs/1002.2999}{{\normalfont\ttfamily
  arXiv:1002.2999}}\relax
\mciteBstWouldAddEndPuncttrue
\mciteSetBstMidEndSepPunct{\mcitedefaultmidpunct}
{\mcitedefaultendpunct}{\mcitedefaultseppunct}\relax
\EndOfBibitem
\bibitem{Dudek:2011tt}
J.~J. Dudek {\em et~al.}, \ifthenelse{\boolean{articletitles}}{\emph{{Isoscalar
  meson spectroscopy from lattice QCD}},
  }{}\href{http://dx.doi.org/10.1103/PhysRevD.83.111502}{Phys.\ Rev.\
  \textbf{D83} (2011) 111502},
  \href{http://arxiv.org/abs/1102.4299}{{\normalfont\ttfamily
  arXiv:1102.4299}}\relax
\mciteBstWouldAddEndPuncttrue
\mciteSetBstMidEndSepPunct{\mcitedefaultmidpunct}
{\mcitedefaultendpunct}{\mcitedefaultseppunct}\relax
\EndOfBibitem
\bibitem{Gregory:2011sg}
UKQCD, E.~B. Gregory, A.~C. Irving, C.~M. Richards, and C.~McNeile,
  \ifthenelse{\boolean{articletitles}}{\emph{{A study of the eta and eta'
  mesons with improved staggered fermions}},
  }{}\href{http://dx.doi.org/10.1103/PhysRevD.86.014504}{Phys.\ Rev.\
  \textbf{D86} (2012) 014504},
  \href{http://arxiv.org/abs/1112.4384}{{\normalfont\ttfamily
  arXiv:1112.4384}}\relax
\mciteBstWouldAddEndPuncttrue
\mciteSetBstMidEndSepPunct{\mcitedefaultmidpunct}
{\mcitedefaultendpunct}{\mcitedefaultseppunct}\relax
\EndOfBibitem
\bibitem{Bernard:2010fr}
Fermilab Lattice, MILC, C.~Bernard {\em et~al.},
  \ifthenelse{\boolean{articletitles}}{\emph{{Tuning Fermilab Heavy Quarks in
  2+1 Flavor Lattice QCD with Application to Hyperfine Splittings}},
  }{}\href{http://dx.doi.org/10.1103/PhysRevD.83.034503}{Phys.\ Rev.\
  \textbf{D83} (2011) 034503},
  \href{http://arxiv.org/abs/1003.1937}{{\normalfont\ttfamily
  arXiv:1003.1937}}\relax
\mciteBstWouldAddEndPuncttrue
\mciteSetBstMidEndSepPunct{\mcitedefaultmidpunct}
{\mcitedefaultendpunct}{\mcitedefaultseppunct}\relax
\EndOfBibitem
\bibitem{Gregory:2010gm}
E.~B. Gregory {\em et~al.}, \ifthenelse{\boolean{articletitles}}{\emph{{Precise
  B, $B_s$ and $B_c$ meson spectroscopy from full lattice QCD}},
  }{}\href{http://dx.doi.org/10.1103/PhysRevD.83.014506}{Phys.\ Rev.\
  \textbf{D83} (2011) 014506},
  \href{http://arxiv.org/abs/1010.3848}{{\normalfont\ttfamily
  arXiv:1010.3848}}\relax
\mciteBstWouldAddEndPuncttrue
\mciteSetBstMidEndSepPunct{\mcitedefaultmidpunct}
{\mcitedefaultendpunct}{\mcitedefaultseppunct}\relax
\EndOfBibitem
\bibitem{Mohler:2011ke}
D.~Mohler and R.~M. Woloshyn, \ifthenelse{\boolean{articletitles}}{\emph{{$D$
  and $D_s$ meson spectroscopy}},
  }{}\href{http://dx.doi.org/10.1103/PhysRevD.84.054505}{Phys.\ Rev.\
  \textbf{D84} (2011) 054505},
  \href{http://arxiv.org/abs/1103.5506}{{\normalfont\ttfamily
  arXiv:1103.5506}}\relax
\mciteBstWouldAddEndPuncttrue
\mciteSetBstMidEndSepPunct{\mcitedefaultmidpunct}
{\mcitedefaultendpunct}{\mcitedefaultseppunct}\relax
\EndOfBibitem
\bibitem{Kronfeld:2012uk}
A.~S. Kronfeld, \ifthenelse{\boolean{articletitles}}{\emph{{Twenty-first
  Century Lattice Gauge Theory: Results from the QCD Lagrangian}},
  }{}\href{http://dx.doi.org/10.1146/annurev-nucl-102711-094942}{Ann.\ Rev.\
  Nucl.\ Part.\ Sci.\  \textbf{62} (2012) 265},
  \href{http://arxiv.org/abs/1203.1204}{{\normalfont\ttfamily
  arXiv:1203.1204}}\relax
\mciteBstWouldAddEndPuncttrue
\mciteSetBstMidEndSepPunct{\mcitedefaultmidpunct}
{\mcitedefaultendpunct}{\mcitedefaultseppunct}\relax
\EndOfBibitem
\bibitem{Borsanyi:2014jba}
S.~Borsanyi {\em et~al.}, \ifthenelse{\boolean{articletitles}}{\emph{{Ab initio
  calculation of the neutron-proton mass difference}},
  }{}\href{http://dx.doi.org/10.1126/science.1257050}{Science \textbf{347}
  (2015) 1452}, \href{http://arxiv.org/abs/1406.4088}{{\normalfont\ttfamily
  arXiv:1406.4088}}\relax
\mciteBstWouldAddEndPuncttrue
\mciteSetBstMidEndSepPunct{\mcitedefaultmidpunct}
{\mcitedefaultendpunct}{\mcitedefaultseppunct}\relax
\EndOfBibitem
\bibitem{Wilczek:2015exa}
F.~Wilczek, \ifthenelse{\boolean{articletitles}}{\emph{{Particle physics: A
  weighty mass difference}},
  }{}\href{http://dx.doi.org/10.1038/nature14381}{Nature \textbf{520} (2015)
  303}\relax
\mciteBstWouldAddEndPuncttrue
\mciteSetBstMidEndSepPunct{\mcitedefaultmidpunct}
{\mcitedefaultendpunct}{\mcitedefaultseppunct}\relax
\EndOfBibitem
\bibitem{Mattson:2002vu}
SELEX, M.~Mattson {\em et~al.},
  \ifthenelse{\boolean{articletitles}}{\emph{{First observation of the doubly
  charmed baryon $\Xi^+_{cc}$}},
  }{}\href{http://dx.doi.org/10.1103/PhysRevLett.89.112001}{Phys.\ Rev.\ Lett.\
   \textbf{89} (2002) 112001},
  \href{http://arxiv.org/abs/hep-ex/0208014}{{\normalfont\ttfamily
  arXiv:hep-ex/0208014}}\relax
\mciteBstWouldAddEndPuncttrue
\mciteSetBstMidEndSepPunct{\mcitedefaultmidpunct}
{\mcitedefaultendpunct}{\mcitedefaultseppunct}\relax
\EndOfBibitem
\bibitem{Ocherashvili:2004hi}
SELEX, A.~Ocherashvili {\em et~al.},
  \ifthenelse{\boolean{articletitles}}{\emph{{Confirmation of the double charm
  baryon $\Xi^+_{cc}(3520)$ via its decay to $p D^+ K^-$}},
  }{}\href{http://dx.doi.org/10.1016/j.physletb.2005.09.043}{Phys.\ Lett.\
  \textbf{B628} (2005) 18},
  \href{http://arxiv.org/abs/hep-ex/0406033}{{\normalfont\ttfamily
  arXiv:hep-ex/0406033}}\relax
\mciteBstWouldAddEndPuncttrue
\mciteSetBstMidEndSepPunct{\mcitedefaultmidpunct}
{\mcitedefaultendpunct}{\mcitedefaultseppunct}\relax
\EndOfBibitem
\bibitem{Chistov:2006zj}
Belle, R.~Chistov {\em et~al.},
  \ifthenelse{\boolean{articletitles}}{\emph{{Observation of new states
  decaying into $\Lambda_c^+ K^- \pi^+$ and $\Lambda_c^+ K^0_s \pi^-$}},
  }{}\href{http://dx.doi.org/10.1103/PhysRevLett.97.162001}{Phys.\ Rev.\ Lett.\
   \textbf{97} (2006) 162001},
  \href{http://arxiv.org/abs/hep-ex/0606051}{{\normalfont\ttfamily
  arXiv:hep-ex/0606051}}\relax
\mciteBstWouldAddEndPuncttrue
\mciteSetBstMidEndSepPunct{\mcitedefaultmidpunct}
{\mcitedefaultendpunct}{\mcitedefaultseppunct}\relax
\EndOfBibitem
\bibitem{Aubert:2006qw}
BaBar, B.~Aubert {\em et~al.},
  \ifthenelse{\boolean{articletitles}}{\emph{{Search for doubly charmed baryons
  $\Xi_{cc}^+$ and $\Xi_{cc}^{++}$ in BABAR}},
  }{}\href{http://dx.doi.org/10.1103/PhysRevD.74.011103}{Phys.\ Rev.\
  \textbf{D74} (2006) 011103},
  \href{http://arxiv.org/abs/hep-ex/0605075}{{\normalfont\ttfamily
  arXiv:hep-ex/0605075}}\relax
\mciteBstWouldAddEndPuncttrue
\mciteSetBstMidEndSepPunct{\mcitedefaultmidpunct}
{\mcitedefaultendpunct}{\mcitedefaultseppunct}\relax
\EndOfBibitem
\bibitem{Aaij:2013voa}
LHCb, R.~Aaij {\em et~al.}, \ifthenelse{\boolean{articletitles}}{\emph{{Search
  for the doubly charmed baryon $\Xi_{cc}^+$}},
  }{}\href{http://dx.doi.org/10.1007/JHEP12(2013)090}{JHEP \textbf{12} (2013)
  090}, \href{http://arxiv.org/abs/1310.2538}{{\normalfont\ttfamily
  arXiv:1310.2538}}\relax
\mciteBstWouldAddEndPuncttrue
\mciteSetBstMidEndSepPunct{\mcitedefaultmidpunct}
{\mcitedefaultendpunct}{\mcitedefaultseppunct}\relax
\EndOfBibitem
\bibitem{Aaij:2017ueg}
LHCb, R.~Aaij {\em et~al.},
  \ifthenelse{\boolean{articletitles}}{\emph{{Observation of the doubly charmed
  baryon $\Xi_{cc}^{++}$}},
  }{}\href{http://arxiv.org/abs/1707.01621}{{\normalfont\ttfamily
  arXiv:1707.01621}}\relax
\mciteBstWouldAddEndPuncttrue
\mciteSetBstMidEndSepPunct{\mcitedefaultmidpunct}
{\mcitedefaultendpunct}{\mcitedefaultseppunct}\relax
\EndOfBibitem
\bibitem{Coleman:1961jn}
S.~R. Coleman and S.~L. Glashow,
  \ifthenelse{\boolean{articletitles}}{\emph{{Electrodynamic properties of
  baryons in the unitary symmetry scheme}},
  }{}\href{http://dx.doi.org/10.1103/PhysRevLett.6.423}{Phys.\ Rev.\ Lett.\
  \textbf{6} (1961) 423}\relax
\mciteBstWouldAddEndPuncttrue
\mciteSetBstMidEndSepPunct{\mcitedefaultmidpunct}
{\mcitedefaultendpunct}{\mcitedefaultseppunct}\relax
\EndOfBibitem
\bibitem{Cichy:2016bci}
K.~Cichy, M.~Kalinowski, and M.~Wagner,
  \ifthenelse{\boolean{articletitles}}{\emph{{Continuum limit of the $D$ meson,
  $D_s$ meson and charmonium spectrum from $N_f=2+1+1$ twisted mass lattice
  QCD}}, }{}\href{http://dx.doi.org/10.1103/PhysRevD.94.094503}{Phys.\ Rev.\
  \textbf{D94} (2016) 094503},
  \href{http://arxiv.org/abs/1603.06467}{{\normalfont\ttfamily
  arXiv:1603.06467}}\relax
\mciteBstWouldAddEndPuncttrue
\mciteSetBstMidEndSepPunct{\mcitedefaultmidpunct}
{\mcitedefaultendpunct}{\mcitedefaultseppunct}\relax
\EndOfBibitem
\bibitem{Leskovec:2015naf}
L.~Leskovec {\em et~al.}, \ifthenelse{\boolean{articletitles}}{\emph{{Positive
  Parity $D_s$ Mesons}}, }{} in {\em {27th International Symposium on Lepton
  Photon Interactions at High Energy (LP15) Ljubljana, Slovenia, August 17-22,
  2015}}, 2015.
\newblock \href{http://arxiv.org/abs/1511.04140}{{\normalfont\ttfamily
  arXiv:1511.04140}}\relax
\mciteBstWouldAddEndPuncttrue
\mciteSetBstMidEndSepPunct{\mcitedefaultmidpunct}
{\mcitedefaultendpunct}{\mcitedefaultseppunct}\relax
\EndOfBibitem
\bibitem{Luscher:1990ux}
M.~Luscher, \ifthenelse{\boolean{articletitles}}{\emph{{Two particle states on
  a torus and their relation to the scattering matrix}},
  }{}\href{http://dx.doi.org/10.1016/0550-3213(91)90366-6}{Nucl.\ Phys.\
  \textbf{B354} (1991) 531}\relax
\mciteBstWouldAddEndPuncttrue
\mciteSetBstMidEndSepPunct{\mcitedefaultmidpunct}
{\mcitedefaultendpunct}{\mcitedefaultseppunct}\relax
\EndOfBibitem
\bibitem{Liu:2012ze}
Hadron Spectrum, L.~Liu {\em et~al.},
  \ifthenelse{\boolean{articletitles}}{\emph{{Excited and exotic charmonium
  spectroscopy from lattice QCD}},
  }{}\href{http://dx.doi.org/10.1007/JHEP07(2012)126}{JHEP \textbf{07} (2012)
  126}, \href{http://arxiv.org/abs/1204.5425}{{\normalfont\ttfamily
  arXiv:1204.5425}}\relax
\mciteBstWouldAddEndPuncttrue
\mciteSetBstMidEndSepPunct{\mcitedefaultmidpunct}
{\mcitedefaultendpunct}{\mcitedefaultseppunct}\relax
\EndOfBibitem
\bibitem{Hansen:2014eka}
M.~T. Hansen and S.~R. Sharpe,
  \ifthenelse{\boolean{articletitles}}{\emph{{Relativistic, model-independent,
  three-particle quantization condition}},
  }{}\href{http://dx.doi.org/10.1103/PhysRevD.90.116003}{Phys.\ Rev.\
  \textbf{D90} (2014), no.~11 116003},
  \href{http://arxiv.org/abs/1408.5933}{{\normalfont\ttfamily
  arXiv:1408.5933}}\relax
\mciteBstWouldAddEndPuncttrue
\mciteSetBstMidEndSepPunct{\mcitedefaultmidpunct}
{\mcitedefaultendpunct}{\mcitedefaultseppunct}\relax
\EndOfBibitem
\bibitem{Hansen:2015zga}
M.~T. Hansen and S.~R. Sharpe,
  \ifthenelse{\boolean{articletitles}}{\emph{{Expressing the three-particle
  finite-volume spectrum in terms of the three-to-three scattering amplitude}},
  }{}\href{http://dx.doi.org/10.1103/PhysRevD.92.114509}{Phys.\ Rev.\
  \textbf{D92} (2015), no.~11 114509},
  \href{http://arxiv.org/abs/1504.04248}{{\normalfont\ttfamily
  arXiv:1504.04248}}\relax
\mciteBstWouldAddEndPuncttrue
\mciteSetBstMidEndSepPunct{\mcitedefaultmidpunct}
{\mcitedefaultendpunct}{\mcitedefaultseppunct}\relax
\EndOfBibitem
\bibitem{Briceno:2016ffu}
R.~A. Briceno, M.~T. Hansen, and S.~R. Sharpe,
  \ifthenelse{\boolean{articletitles}}{\emph{{Progress on the three-particle
  quantization condition}}, }{}PoS \textbf{LATTICE2016} (2016) 115,
  \href{http://arxiv.org/abs/1609.09805}{{\normalfont\ttfamily
  arXiv:1609.09805}}\relax
\mciteBstWouldAddEndPuncttrue
\mciteSetBstMidEndSepPunct{\mcitedefaultmidpunct}
{\mcitedefaultendpunct}{\mcitedefaultseppunct}\relax
\EndOfBibitem
\bibitem{Padmanath:2015era}
M.~Padmanath, C.~B. Lang, and S.~Prelovsek,
  \ifthenelse{\boolean{articletitles}}{\emph{{X(3872) and Y(4140) using
  diquark-antidiquark operators with lattice QCD}},
  }{}\href{http://dx.doi.org/10.1103/PhysRevD.92.034501}{Phys.\ Rev.\
  \textbf{D92} (2015) 034501},
  \href{http://arxiv.org/abs/1503.03257}{{\normalfont\ttfamily
  arXiv:1503.03257}}\relax
\mciteBstWouldAddEndPuncttrue
\mciteSetBstMidEndSepPunct{\mcitedefaultmidpunct}
{\mcitedefaultendpunct}{\mcitedefaultseppunct}\relax
\EndOfBibitem
\bibitem{Bicudo:2015kna}
P.~Bicudo, K.~Cichy, A.~Peters, and M.~Wagner,
  \ifthenelse{\boolean{articletitles}}{\emph{{BB interactions with static
  bottom quarks from Lattice QCD}},
  }{}\href{http://dx.doi.org/10.1103/PhysRevD.93.034501}{Phys.\ Rev.\
  \textbf{D93} (2016) 034501},
  \href{http://arxiv.org/abs/1510.03441}{{\normalfont\ttfamily
  arXiv:1510.03441}}\relax
\mciteBstWouldAddEndPuncttrue
\mciteSetBstMidEndSepPunct{\mcitedefaultmidpunct}
{\mcitedefaultendpunct}{\mcitedefaultseppunct}\relax
\EndOfBibitem
\bibitem{Francis:2016hui}
A.~Francis, R.~J. Hudspith, R.~Lewis, and K.~Maltman,
  \ifthenelse{\boolean{articletitles}}{\emph{{Lattice Prediction for Deeply
  Bound Doubly Heavy Tetraquarks}},
  }{}\href{http://dx.doi.org/10.1103/PhysRevLett.118.142001}{Phys.\ Rev.\
  Lett.\  \textbf{118} (2017), no.~14 142001},
  \href{http://arxiv.org/abs/1607.05214}{{\normalfont\ttfamily
  arXiv:1607.05214}}\relax
\mciteBstWouldAddEndPuncttrue
\mciteSetBstMidEndSepPunct{\mcitedefaultmidpunct}
{\mcitedefaultendpunct}{\mcitedefaultseppunct}\relax
\EndOfBibitem
\bibitem{Briceno:2017max}
R.~A. Briceno, J.~J. Dudek, and R.~D. Young,
  \ifthenelse{\boolean{articletitles}}{\emph{{Scattering processes and
  resonances from lattice QCD}},
  }{}\href{http://arxiv.org/abs/1706.06223}{{\normalfont\ttfamily
  arXiv:1706.06223}}\relax
\mciteBstWouldAddEndPuncttrue
\mciteSetBstMidEndSepPunct{\mcitedefaultmidpunct}
{\mcitedefaultendpunct}{\mcitedefaultseppunct}\relax
\EndOfBibitem
\bibitem{Aubert:2001tu}
BaBar, B.~Aubert {\em et~al.}, \ifthenelse{\boolean{articletitles}}{\emph{{The
  BaBar detector}},
  }{}\href{http://dx.doi.org/10.1016/S0168-9002(01)02012-5}{Nucl.\ Instrum.\
  Meth.\  \textbf{A479} (2002) 1},
  \href{http://arxiv.org/abs/hep-ex/0105044}{{\normalfont\ttfamily
  arXiv:hep-ex/0105044}}\relax
\mciteBstWouldAddEndPuncttrue
\mciteSetBstMidEndSepPunct{\mcitedefaultmidpunct}
{\mcitedefaultendpunct}{\mcitedefaultseppunct}\relax
\EndOfBibitem
\bibitem{Abashian:2000cg}
A.~Abashian {\em et~al.}, \ifthenelse{\boolean{articletitles}}{\emph{{The Belle
  Detector}}, }{}\href{http://dx.doi.org/10.1016/S0168-9002(01)02013-7}{Nucl.\
  Instrum.\ Meth.\  \textbf{A479} (2002) 117}\relax
\mciteBstWouldAddEndPuncttrue
\mciteSetBstMidEndSepPunct{\mcitedefaultmidpunct}
{\mcitedefaultendpunct}{\mcitedefaultseppunct}\relax
\EndOfBibitem
\bibitem{pepii:1994}
\ifthenelse{\boolean{articletitles}}{\emph{{PEP-II: An Asymmetric B Factory.
  Conceptual Design Report. June 1993}}, }{},
  \href{http://www-public.slac.stanford.edu/sciDoc/docMeta.aspx?slacPubNumber=slac-r-418}{SLAC-418,
  LBL-PUB-5379, CALT-68-1869, UCRL-ID-114055, UCIIRPA-93-01}\relax
\mciteBstWouldAddEndPuncttrue
\mciteSetBstMidEndSepPunct{\mcitedefaultmidpunct}
{\mcitedefaultendpunct}{\mcitedefaultseppunct}\relax
\EndOfBibitem
\bibitem{Kurokawa:2001nw}
S.~Kurokawa and E.~Kikutani,
  \ifthenelse{\boolean{articletitles}}{\emph{{Overview of the KEKB
  accelerators}},
  }{}\href{http://dx.doi.org/10.1016/S0168-9002(02)01771-0}{Nucl.\ Instrum.\
  Meth.\  \textbf{A499} (2003) 1}\relax
\mciteBstWouldAddEndPuncttrue
\mciteSetBstMidEndSepPunct{\mcitedefaultmidpunct}
{\mcitedefaultendpunct}{\mcitedefaultseppunct}\relax
\EndOfBibitem
\bibitem{Kobayashi:1973fv}
M.~Kobayashi and T.~Maskawa, \ifthenelse{\boolean{articletitles}}{\emph{{CP
  Violation in the Renormalizable Theory of Weak Interaction}},
  }{}\href{http://dx.doi.org/10.1143/PTP.49.652}{Prog.\ Theor.\ Phys.\
  \textbf{49} (1973) 652}\relax
\mciteBstWouldAddEndPuncttrue
\mciteSetBstMidEndSepPunct{\mcitedefaultmidpunct}
{\mcitedefaultendpunct}{\mcitedefaultseppunct}\relax
\EndOfBibitem
\bibitem{Okubo:1963fa}
S.~Okubo, \ifthenelse{\boolean{articletitles}}{\emph{{Phi meson and unitary
  symmetry model}},
  }{}\href{http://dx.doi.org/10.1016/S0375-9601(63)92548-9}{Phys.\ Lett.\
  \textbf{5} (1963) 165}\relax
\mciteBstWouldAddEndPuncttrue
\mciteSetBstMidEndSepPunct{\mcitedefaultmidpunct}
{\mcitedefaultendpunct}{\mcitedefaultseppunct}\relax
\EndOfBibitem
\bibitem{Iizuka:1966fk}
J.~Iizuka, \ifthenelse{\boolean{articletitles}}{\emph{{Systematics and
  phenomenology of meson family}},
  }{}\href{http://dx.doi.org/10.1143/PTPS.37.21}{Prog.\ Theor.\ Phys.\ Suppl.\
  \textbf{37} (1966) 21}\relax
\mciteBstWouldAddEndPuncttrue
\mciteSetBstMidEndSepPunct{\mcitedefaultmidpunct}
{\mcitedefaultendpunct}{\mcitedefaultseppunct}\relax
\EndOfBibitem
\bibitem{Gu:2003ck}
P.~D. Gu {\em et~al.}, \ifthenelse{\boolean{articletitles}}{\emph{{Accelerator
  physics design of BEPCII}}, }{}ICFA Beam Dyn.\ Newslett.\  \textbf{31} (2003)
  32\relax
\mciteBstWouldAddEndPuncttrue
\mciteSetBstMidEndSepPunct{\mcitedefaultmidpunct}
{\mcitedefaultendpunct}{\mcitedefaultseppunct}\relax
\EndOfBibitem
\bibitem{Ablikim:2009aa}
BESIII, M.~Ablikim {\em et~al.},
  \ifthenelse{\boolean{articletitles}}{\emph{{Design and Construction of the
  BESIII Detector}},
  }{}\href{http://dx.doi.org/10.1016/j.nima.2009.12.050}{Nucl.\ Instrum.\
  Meth.\  \textbf{A614} (2010) 345},
  \href{http://arxiv.org/abs/0911.4960}{{\normalfont\ttfamily
  arXiv:0911.4960}}\relax
\mciteBstWouldAddEndPuncttrue
\mciteSetBstMidEndSepPunct{\mcitedefaultmidpunct}
{\mcitedefaultendpunct}{\mcitedefaultseppunct}\relax
\EndOfBibitem
\bibitem{Criegee:1981qx}
L.~Criegee and G.~Knies, \ifthenelse{\boolean{articletitles}}{\emph{{Review of
  $e^+ e^-$ Experiments With Pluto From 3 GeV to 31 GeV}},
  }{}\href{http://dx.doi.org/10.1016/0370-1573(82)90012-6}{Phys.\ Rept.\
  \textbf{83} (1982) 151}\relax
\mciteBstWouldAddEndPuncttrue
\mciteSetBstMidEndSepPunct{\mcitedefaultmidpunct}
{\mcitedefaultendpunct}{\mcitedefaultseppunct}\relax
\EndOfBibitem
\bibitem{Augustin:1975yq}
J.~E. Augustin {\em et~al.}, \ifthenelse{\boolean{articletitles}}{\emph{{Total
  Cross-Section for Hadron Production by electron-Positron Annihilation Between
  2.4 GeV and 5.0 GeV Center-Of-Mass Energy}},
  }{}\href{http://dx.doi.org/10.1103/PhysRevLett.34.764}{Phys.\ Rev.\ Lett.\
  \textbf{34} (1975) 764}\relax
\mciteBstWouldAddEndPuncttrue
\mciteSetBstMidEndSepPunct{\mcitedefaultmidpunct}
{\mcitedefaultendpunct}{\mcitedefaultseppunct}\relax
\EndOfBibitem
\bibitem{Bacci:1980zs}
C.~Bacci {\em et~al.}, \ifthenelse{\boolean{articletitles}}{\emph{{Measurement
  of Hadronic Exclusive Cross-sections in $e^+ e^-$ Annihilation From 1.42
  {GeV} to 2.20 {GeV}}},
  }{}\href{http://dx.doi.org/10.1016/0550-3213(81)90208-X}{Nucl.\ Phys.\
  \textbf{B184} (1981) 31}\relax
\mciteBstWouldAddEndPuncttrue
\mciteSetBstMidEndSepPunct{\mcitedefaultmidpunct}
{\mcitedefaultendpunct}{\mcitedefaultseppunct}\relax
\EndOfBibitem
\bibitem{Bai:1999pk}
BES, J.~Z. Bai {\em et~al.},
  \ifthenelse{\boolean{articletitles}}{\emph{{Measurement of the total
  cross-section for hadronic production by $e^+ e^-$ annihilation at energies
  between 2.6-GeV - 5-GeV}},
  }{}\href{http://dx.doi.org/10.1103/PhysRevLett.84.594}{Phys.\ Rev.\ Lett.\
  \textbf{84} (2000) 594},
  \href{http://arxiv.org/abs/hep-ex/9908046}{{\normalfont\ttfamily
  arXiv:hep-ex/9908046}}\relax
\mciteBstWouldAddEndPuncttrue
\mciteSetBstMidEndSepPunct{\mcitedefaultmidpunct}
{\mcitedefaultendpunct}{\mcitedefaultseppunct}\relax
\EndOfBibitem
\bibitem{Bai:2001ct}
BES, J.~Z. Bai {\em et~al.},
  \ifthenelse{\boolean{articletitles}}{\emph{{Measurements of the cross-section
  for $e^+ e^-$ to hadrons at center-of-mass energies from 2-GeV to 5-GeV}},
  }{}\href{http://dx.doi.org/10.1103/PhysRevLett.88.101802}{Phys.\ Rev.\ Lett.\
   \textbf{88} (2002) 101802},
  \href{http://arxiv.org/abs/hep-ex/0102003}{{\normalfont\ttfamily
  arXiv:hep-ex/0102003}}\relax
\mciteBstWouldAddEndPuncttrue
\mciteSetBstMidEndSepPunct{\mcitedefaultmidpunct}
{\mcitedefaultendpunct}{\mcitedefaultseppunct}\relax
\EndOfBibitem
\bibitem{Ablikim:2007gd}
BES, M.~Ablikim {\em et~al.},
  \ifthenelse{\boolean{articletitles}}{\emph{{Determination of the
  $\psi(3770)$, $\psi(4040)$, $\psi(4160)$ and $\psi(4415)$ resonance
  parameters}},
  }{}\href{http://dx.doi.org/10.1016/j.physletb.2007.11.100}{eConf
  \textbf{C070805} (2007) 02},
  \href{http://arxiv.org/abs/0705.4500}{{\normalfont\ttfamily
  arXiv:0705.4500}}, [Phys. Lett.B660,315(2008)]\relax
\mciteBstWouldAddEndPuncttrue
\mciteSetBstMidEndSepPunct{\mcitedefaultmidpunct}
{\mcitedefaultendpunct}{\mcitedefaultseppunct}\relax
\EndOfBibitem
\bibitem{Rice:1982br}
E.~Rice {\em et~al.}, \ifthenelse{\boolean{articletitles}}{\emph{{Search for
  structure in $\sigma(e^+e^-\to{\rm hadrons})$ between $\sqrt{s}=10.24$ GeV
  and 11.5 GeV}}, }{}\href{http://dx.doi.org/10.1103/PhysRevLett.48.906}{Phys.\
  Rev.\ Lett.\  \textbf{48} (1982) 906}\relax
\mciteBstWouldAddEndPuncttrue
\mciteSetBstMidEndSepPunct{\mcitedefaultmidpunct}
{\mcitedefaultendpunct}{\mcitedefaultseppunct}\relax
\EndOfBibitem
\bibitem{Besson:1984bd}
CLEO, D.~Besson {\em et~al.},
  \ifthenelse{\boolean{articletitles}}{\emph{{Observation of New Structure in
  the $e^+ e^-$ Annihilation Cross-Section Above $B\bar B$ Threshold}},
  }{}\href{http://dx.doi.org/10.1103/PhysRevLett.54.381}{Phys.\ Rev.\ Lett.\
  \textbf{54} (1985) 381}\relax
\mciteBstWouldAddEndPuncttrue
\mciteSetBstMidEndSepPunct{\mcitedefaultmidpunct}
{\mcitedefaultendpunct}{\mcitedefaultseppunct}\relax
\EndOfBibitem
\bibitem{Kubota:1991ww}
CLEO, Y.~Kubota {\em et~al.}, \ifthenelse{\boolean{articletitles}}{\emph{{The
  CLEO-II detector}},
  }{}\href{http://dx.doi.org/10.1016/0168-9002(92)90770-5}{Nucl.\ Instrum.\
  Meth.\  \textbf{A320} (1992) 66}\relax
\mciteBstWouldAddEndPuncttrue
\mciteSetBstMidEndSepPunct{\mcitedefaultmidpunct}
{\mcitedefaultendpunct}{\mcitedefaultseppunct}\relax
\EndOfBibitem
\bibitem{Beneke:1999br}
M.~Beneke, G.~Buchalla, M.~Neubert, and C.~T. Sachrajda,
  \ifthenelse{\boolean{articletitles}}{\emph{{QCD factorization for
  $B\to\pi\pi$ decays: Strong phases and CP violation in the heavy quark
  limit}}, }{}\href{http://dx.doi.org/10.1103/PhysRevLett.83.1914}{Phys.\ Rev.\
  Lett.\  \textbf{83} (1999) 1914},
  \href{http://arxiv.org/abs/hep-ph/9905312}{{\normalfont\ttfamily
  arXiv:hep-ph/9905312}}\relax
\mciteBstWouldAddEndPuncttrue
\mciteSetBstMidEndSepPunct{\mcitedefaultmidpunct}
{\mcitedefaultendpunct}{\mcitedefaultseppunct}\relax
\EndOfBibitem
\bibitem{Aubert:2005vi}
BaBar, B.~Aubert {\em et~al.},
  \ifthenelse{\boolean{articletitles}}{\emph{{Measurements of the absolute
  branching fractions of $B^\pm \to K^\pm$ X($c \bar{c}$)}},
  }{}\href{http://dx.doi.org/10.1103/PhysRevLett.96.052002}{Phys.\ Rev.\ Lett.\
   \textbf{96} (2006) 052002},
  \href{http://arxiv.org/abs/hep-ex/0510070}{{\normalfont\ttfamily
  arXiv:hep-ex/0510070}}\relax
\mciteBstWouldAddEndPuncttrue
\mciteSetBstMidEndSepPunct{\mcitedefaultmidpunct}
{\mcitedefaultendpunct}{\mcitedefaultseppunct}\relax
\EndOfBibitem
\bibitem{Bevan:2014iga}
Belle, BaBar, A.~J. Bevan {\em et~al.},
  \ifthenelse{\boolean{articletitles}}{\emph{{The Physics of the B Factories}},
  }{}\href{http://dx.doi.org/10.1140/epjc/s10052-014-3026-9}{Eur.\ Phys.\ J.\
  \textbf{C74} (2014) 3026},
  \href{http://arxiv.org/abs/1406.6311}{{\normalfont\ttfamily
  arXiv:1406.6311}}\relax
\mciteBstWouldAddEndPuncttrue
\mciteSetBstMidEndSepPunct{\mcitedefaultmidpunct}
{\mcitedefaultendpunct}{\mcitedefaultseppunct}\relax
\EndOfBibitem
\bibitem{Uehara:2005qd}
Belle, S.~Uehara {\em et~al.},
  \ifthenelse{\boolean{articletitles}}{\emph{{Observation of a $\chi_{c2}'$
  candidate in $\gamma\gamma \to D \bar{D}$ production at BELLE}},
  }{}\href{http://dx.doi.org/10.1103/PhysRevLett.96.082003}{Phys.\ Rev.\ Lett.\
   \textbf{96} (2006) 082003},
  \href{http://arxiv.org/abs/hep-ex/0512035}{{\normalfont\ttfamily
  arXiv:hep-ex/0512035}}\relax
\mciteBstWouldAddEndPuncttrue
\mciteSetBstMidEndSepPunct{\mcitedefaultmidpunct}
{\mcitedefaultendpunct}{\mcitedefaultseppunct}\relax
\EndOfBibitem
\bibitem{Berezhnoy:2003hz}
A.~V. Berezhnoy and A.~K. Likhoded,
  \ifthenelse{\boolean{articletitles}}{\emph{{Processes $e^+e^-\to c\bar{c}$
  and $e^+e^-\to\jpsi$+gg at $\sqrt{s}$ = 10.59-GeV}},
  }{}\href{http://dx.doi.org/10.1134/1.1707137}{Phys.\ Atom.\ Nucl.\
  \textbf{67} (2004) 757},
  \href{http://arxiv.org/abs/hep-ph/0303145}{{\normalfont\ttfamily
  arXiv:hep-ph/0303145}}, [Yad. Fiz.67,778(2004)]\relax
\mciteBstWouldAddEndPuncttrue
\mciteSetBstMidEndSepPunct{\mcitedefaultmidpunct}
{\mcitedefaultendpunct}{\mcitedefaultseppunct}\relax
\EndOfBibitem
\bibitem{Pakhlov:2009nj}
Belle, P.~Pakhlov {\em et~al.},
  \ifthenelse{\boolean{articletitles}}{\emph{{Measurement of the
  $e^+e^-\to\jpsi c\bar{c}$ cross section at $\sqrt{s}\sim$10.6-GeV}},
  }{}\href{http://dx.doi.org/10.1103/PhysRevD.79.071101}{Phys.\ Rev.\
  \textbf{D79} (2009) 071101},
  \href{http://arxiv.org/abs/0901.2775}{{\normalfont\ttfamily
  arXiv:0901.2775}}\relax
\mciteBstWouldAddEndPuncttrue
\mciteSetBstMidEndSepPunct{\mcitedefaultmidpunct}
{\mcitedefaultendpunct}{\mcitedefaultseppunct}\relax
\EndOfBibitem
\bibitem{Bodwin:2002fk}
G.~T. Bodwin, J.~Lee, and E.~Braaten,
  \ifthenelse{\boolean{articletitles}}{\emph{{$e^+e^-$ annihilation into
  $\jpsi\jpsi$}},
  }{}\href{http://dx.doi.org/10.1103/PhysRevLett.90.162001}{Phys.\ Rev.\ Lett.\
   \textbf{90} (2003) 162001},
  \href{http://arxiv.org/abs/hep-ph/0212181}{{\normalfont\ttfamily
  arXiv:hep-ph/0212181}}\relax
\mciteBstWouldAddEndPuncttrue
\mciteSetBstMidEndSepPunct{\mcitedefaultmidpunct}
{\mcitedefaultendpunct}{\mcitedefaultseppunct}\relax
\EndOfBibitem
\bibitem{Rosner:2005ry}
CLEO, J.~L. Rosner {\em et~al.},
  \ifthenelse{\boolean{articletitles}}{\emph{{Observation of $h_{c}(1P)$ state
  of charmonium}},
  }{}\href{http://dx.doi.org/10.1103/PhysRevLett.95.102003}{Phys.\ Rev.\ Lett.\
   \textbf{95} (2005) 102003},
  \href{http://arxiv.org/abs/hep-ex/0505073}{{\normalfont\ttfamily
  arXiv:hep-ex/0505073}}\relax
\mciteBstWouldAddEndPuncttrue
\mciteSetBstMidEndSepPunct{\mcitedefaultmidpunct}
{\mcitedefaultendpunct}{\mcitedefaultseppunct}\relax
\EndOfBibitem
\bibitem{zhao:2016zg}
Z.~Zhao, L.~Luo, B.~Zhang, and W.~Xu,
  \ifthenelse{\boolean{articletitles}}{\emph{{Preliminary Concept and Key
  Technologies of HIEPA Accelerator}}, }{}
\newblock
  \href{http://inspirehep.net/record/1470642/files/thpor047.pdf}{Proceedings of
  the 7th International Particle Accelerator Conference, (IPAC 2016): Busan,
  Korea, May 8-13, 2016.}\relax
\mciteBstWouldAddEndPunctfalse
\mciteSetBstMidEndSepPunct{\mcitedefaultmidpunct}
{}{\mcitedefaultseppunct}\relax
\EndOfBibitem
\bibitem{Bondar:2013cja}
Charm-Tau Factory, A.~E. Bondar {\em et~al.},
  \ifthenelse{\boolean{articletitles}}{\emph{{Project of a Super Charm-Tau
  factory at the Budker Institute of Nuclear Physics in Novosibirsk}},
  }{}\href{http://dx.doi.org/10.1134/S1063778813090032}{Phys.\ Atom.\ Nucl.\
  \textbf{76} (2013) 1072}, [Yad. Fiz.76,no.9,1132(2013)]\relax
\mciteBstWouldAddEndPuncttrue
\mciteSetBstMidEndSepPunct{\mcitedefaultmidpunct}
{\mcitedefaultendpunct}{\mcitedefaultseppunct}\relax
\EndOfBibitem
\bibitem{Ohnishi:2013fma}
Y.~Ohnishi {\em et~al.},
  \ifthenelse{\boolean{articletitles}}{\emph{{Accelerator design at
  SuperKEKB}}, }{}\href{http://dx.doi.org/10.1093/ptep/pts083}{PTEP
  \textbf{2013} (2013) 03A011}\relax
\mciteBstWouldAddEndPuncttrue
\mciteSetBstMidEndSepPunct{\mcitedefaultmidpunct}
{\mcitedefaultendpunct}{\mcitedefaultseppunct}\relax
\EndOfBibitem
\bibitem{Aushev:2010bq}
T.~Aushev {\em et~al.}, \ifthenelse{\boolean{articletitles}}{\emph{{Physics at
  Super B Factory}},
  }{}\href{http://arxiv.org/abs/1002.5012}{{\normalfont\ttfamily
  arXiv:1002.5012}}\relax
\mciteBstWouldAddEndPuncttrue
\mciteSetBstMidEndSepPunct{\mcitedefaultmidpunct}
{\mcitedefaultendpunct}{\mcitedefaultseppunct}\relax
\EndOfBibitem
\bibitem{Bondar:2016hva}
A.~E. Bondar, R.~V. Mizuk, and M.~B. Voloshin,
  \ifthenelse{\boolean{articletitles}}{\emph{{Bottomonium-like states: Physics
  case for energy scan above the $B\bar{B}$ threshold at Belle-II}},
  }{}\href{http://dx.doi.org/10.1142/S0217732317500250}{Mod.\ Phys.\ Lett.\
  \textbf{A32} (2017) 1750025},
  \href{http://arxiv.org/abs/1610.01102}{{\normalfont\ttfamily
  arXiv:1610.01102}}\relax
\mciteBstWouldAddEndPuncttrue
\mciteSetBstMidEndSepPunct{\mcitedefaultmidpunct}
{\mcitedefaultendpunct}{\mcitedefaultseppunct}\relax
\EndOfBibitem
\bibitem{Abe:1995hr}
CDF, F.~Abe {\em et~al.},
  \ifthenelse{\boolean{articletitles}}{\emph{{Observation of top quark
  production in $\bar{p}p$ collisions}},
  }{}\href{http://dx.doi.org/10.1103/PhysRevLett.74.2626}{Phys.\ Rev.\ Lett.\
  \textbf{74} (1995) 2626},
  \href{http://arxiv.org/abs/hep-ex/9503002}{{\normalfont\ttfamily
  arXiv:hep-ex/9503002}}\relax
\mciteBstWouldAddEndPuncttrue
\mciteSetBstMidEndSepPunct{\mcitedefaultmidpunct}
{\mcitedefaultendpunct}{\mcitedefaultseppunct}\relax
\EndOfBibitem
\bibitem{Abachi:1995iq}
D0, S.~Abachi {\em et~al.},
  \ifthenelse{\boolean{articletitles}}{\emph{{Observation of the top quark}},
  }{}\href{http://dx.doi.org/10.1103/PhysRevLett.74.2632}{Phys.\ Rev.\ Lett.\
  \textbf{74} (1995) 2632},
  \href{http://arxiv.org/abs/hep-ex/9503003}{{\normalfont\ttfamily
  arXiv:hep-ex/9503003}}\relax
\mciteBstWouldAddEndPuncttrue
\mciteSetBstMidEndSepPunct{\mcitedefaultmidpunct}
{\mcitedefaultendpunct}{\mcitedefaultseppunct}\relax
\EndOfBibitem
\bibitem{Bauer:2004bc}
CDF, G.~Bauer, \ifthenelse{\boolean{articletitles}}{\emph{{The $X(3872) $ at
  CDF II}}, }{}\href{http://dx.doi.org/10.1142/S0217751X05027552}{Int.\ J.\
  Mod.\ Phys.\  \textbf{A20} (2005) 3765},
  \href{http://arxiv.org/abs/hep-ex/0409052}{{\normalfont\ttfamily
  arXiv:hep-ex/0409052}}\relax
\mciteBstWouldAddEndPuncttrue
\mciteSetBstMidEndSepPunct{\mcitedefaultmidpunct}
{\mcitedefaultendpunct}{\mcitedefaultseppunct}\relax
\EndOfBibitem
\bibitem{Aad:2008zzm}
ATLAS, G.~Aad {\em et~al.}, \ifthenelse{\boolean{articletitles}}{\emph{{The
  ATLAS Experiment at the CERN Large Hadron Collider}},
  }{}\href{http://dx.doi.org/10.1088/1748-0221/3/08/S08003}{JINST \textbf{3}
  (2008) S08003}\relax
\mciteBstWouldAddEndPuncttrue
\mciteSetBstMidEndSepPunct{\mcitedefaultmidpunct}
{\mcitedefaultendpunct}{\mcitedefaultseppunct}\relax
\EndOfBibitem
\bibitem{Chatrchyan:2008aa}
CMS, S.~Chatrchyan {\em et~al.},
  \ifthenelse{\boolean{articletitles}}{\emph{{The CMS experiment at the CERN
  LHC}}, }{}\href{http://dx.doi.org/10.1088/1748-0221/3/08/S08004}{JINST
  \textbf{3} (2008) S08004}\relax
\mciteBstWouldAddEndPuncttrue
\mciteSetBstMidEndSepPunct{\mcitedefaultmidpunct}
{\mcitedefaultendpunct}{\mcitedefaultseppunct}\relax
\EndOfBibitem
\bibitem{Artoisenet:2009wk}
P.~Artoisenet and E.~Braaten,
  \ifthenelse{\boolean{articletitles}}{\emph{{Production of the X(3872) at the
  Tevatron and the LHC}},
  }{}\href{http://dx.doi.org/10.1103/PhysRevD.81.114018}{Phys.\ Rev.\
  \textbf{D81} (2010) 114018},
  \href{http://arxiv.org/abs/0911.2016}{{\normalfont\ttfamily
  arXiv:0911.2016}}\relax
\mciteBstWouldAddEndPuncttrue
\mciteSetBstMidEndSepPunct{\mcitedefaultmidpunct}
{\mcitedefaultendpunct}{\mcitedefaultseppunct}\relax
\EndOfBibitem
\bibitem{Chatrchyan:2013mea}
CMS, S.~Chatrchyan {\em et~al.},
  \ifthenelse{\boolean{articletitles}}{\emph{{Search for a new bottomonium
  state decaying to $\Upsilon(1S)\pi^+\pi^-$ in $pp$ collisions at $\sqrt{s}$ =
  8 TeV}}, }{}\href{http://dx.doi.org/10.1016/j.physletb.2013.10.016}{Phys.\
  Lett.\  \textbf{B727} (2013) 57},
  \href{http://arxiv.org/abs/1309.0250}{{\normalfont\ttfamily
  arXiv:1309.0250}}\relax
\mciteBstWouldAddEndPuncttrue
\mciteSetBstMidEndSepPunct{\mcitedefaultmidpunct}
{\mcitedefaultendpunct}{\mcitedefaultseppunct}\relax
\EndOfBibitem
\bibitem{Aad:2014ama}
ATLAS, G.~Aad {\em et~al.}, \ifthenelse{\boolean{articletitles}}{\emph{{Search
  for the $X_b$ and other hidden-beauty states in the $\pi^+ \pi^- \Upsilon(1
  \rm S)$ channel at ATLAS}},
  }{}\href{http://dx.doi.org/10.1016/j.physletb.2014.11.055}{Phys.\ Lett.\
  \textbf{B740} (2015) 199},
  \href{http://arxiv.org/abs/1410.4409}{{\normalfont\ttfamily
  arXiv:1410.4409}}\relax
\mciteBstWouldAddEndPuncttrue
\mciteSetBstMidEndSepPunct{\mcitedefaultmidpunct}
{\mcitedefaultendpunct}{\mcitedefaultseppunct}\relax
\EndOfBibitem
\bibitem{Alves:2008zz}
LHCb, A.~A. Alves~Jr.\ {\em et~al.},
  \ifthenelse{\boolean{articletitles}}{\emph{{The LHCb detector at the LHC}},
  }{}\href{http://dx.doi.org/10.1088/1748-0221/3/08/S08005}{JINST \textbf{3}
  (2008) S08005}\relax
\mciteBstWouldAddEndPuncttrue
\mciteSetBstMidEndSepPunct{\mcitedefaultmidpunct}
{\mcitedefaultendpunct}{\mcitedefaultseppunct}\relax
\EndOfBibitem
\bibitem{LHCb-DP-2014-002}
LHCb, R.~Aaij {\em et~al.}, \ifthenelse{\boolean{articletitles}}{\emph{{LHCb
  detector performance}},
  }{}\href{http://dx.doi.org/10.1142/S0217751X15300227}{Int.\ J.\ Mod.\ Phys.\
  \textbf{A30} (2015) 1530022},
  \href{http://arxiv.org/abs/1412.6352}{{\normalfont\ttfamily
  arXiv:1412.6352}}\relax
\mciteBstWouldAddEndPuncttrue
\mciteSetBstMidEndSepPunct{\mcitedefaultmidpunct}
{\mcitedefaultendpunct}{\mcitedefaultseppunct}\relax
\EndOfBibitem
\bibitem{Aaij:2015bpa}
LHCb, R.~Aaij {\em et~al.},
  \ifthenelse{\boolean{articletitles}}{\emph{{Measurements of prompt charm
  production cross-sections in $pp$ collisions at $ \sqrt{s}=13 $ TeV}},
  }{}\href{http://dx.doi.org/10.1007/JHEP03(2016)159,
  10.1007/JHEP09(2016)013}{JHEP \textbf{03} (2016) 159},
  \href{http://arxiv.org/abs/1510.01707}{{\normalfont\ttfamily
  arXiv:1510.01707}}, [Erratum: JHEP09,013(2016)]\relax
\mciteBstWouldAddEndPuncttrue
\mciteSetBstMidEndSepPunct{\mcitedefaultmidpunct}
{\mcitedefaultendpunct}{\mcitedefaultseppunct}\relax
\EndOfBibitem
\bibitem{Bediaga:1443882}
LHCb, I.~Bediaga {\em et~al.},
  \ifthenelse{\boolean{articletitles}}{\emph{{Framework TDR for the LHCb
  Upgrade: Technical Design Report}}, }{} 2012.
\newblock \href{https://cds.cern.ch/record/1443882}{CERN-LHCC-2012-007.
  LHCb-TDR-12}\relax
\mciteBstWouldAddEndPuncttrue
\mciteSetBstMidEndSepPunct{\mcitedefaultmidpunct}
{\mcitedefaultendpunct}{\mcitedefaultseppunct}\relax
\EndOfBibitem
\bibitem{Godfrey:2015dva}
S.~Godfrey and K.~Moats, \ifthenelse{\boolean{articletitles}}{\emph{{Properties
  of Excited Charm and Charm-Strange Mesons}},
  }{}\href{http://dx.doi.org/10.1103/PhysRevD.93.034035}{Phys.\ Rev.\
  \textbf{D93} (2016) 034035},
  \href{http://arxiv.org/abs/1510.08305}{{\normalfont\ttfamily
  arXiv:1510.08305}}\relax
\mciteBstWouldAddEndPuncttrue
\mciteSetBstMidEndSepPunct{\mcitedefaultmidpunct}
{\mcitedefaultendpunct}{\mcitedefaultseppunct}\relax
\EndOfBibitem
\bibitem{Link:2003bd}
FOCUS, J.~M. Link {\em et~al.},
  \ifthenelse{\boolean{articletitles}}{\emph{{Measurement of masses and widths
  of excited charm mesons $D_2^*$ and evidence for broad states}},
  }{}\href{http://dx.doi.org/10.1016/j.physletb.2004.02.017}{Phys.\ Lett.\
  \textbf{B586} (2004) 11},
  \href{http://arxiv.org/abs/hep-ex/0312060}{{\normalfont\ttfamily
  arXiv:hep-ex/0312060}}\relax
\mciteBstWouldAddEndPuncttrue
\mciteSetBstMidEndSepPunct{\mcitedefaultmidpunct}
{\mcitedefaultendpunct}{\mcitedefaultseppunct}\relax
\EndOfBibitem
\bibitem{Abe:2003zm}
Belle, K.~Abe {\em et~al.}, \ifthenelse{\boolean{articletitles}}{\emph{{Study
  of $B^- \to D^{**0} \pi^- (D^{**0}\to D^{(*)+} \pi^-$) decays}},
  }{}\href{http://dx.doi.org/10.1103/PhysRevD.69.112002}{Phys.\ Rev.\
  \textbf{D69} (2004) 112002},
  \href{http://arxiv.org/abs/hep-ex/0307021}{{\normalfont\ttfamily
  arXiv:hep-ex/0307021}}\relax
\mciteBstWouldAddEndPuncttrue
\mciteSetBstMidEndSepPunct{\mcitedefaultmidpunct}
{\mcitedefaultendpunct}{\mcitedefaultseppunct}\relax
\EndOfBibitem
\bibitem{Aubert:2009wg}
BaBar, B.~Aubert {\em et~al.},
  \ifthenelse{\boolean{articletitles}}{\emph{{Dalitz Plot Analysis of $B^-\to
  D^+ \pi^- \pi^-$}},
  }{}\href{http://dx.doi.org/10.1103/PhysRevD.79.112004}{Phys.\ Rev.\
  \textbf{D79} (2009) 112004},
  \href{http://arxiv.org/abs/0901.1291}{{\normalfont\ttfamily
  arXiv:0901.1291}}\relax
\mciteBstWouldAddEndPuncttrue
\mciteSetBstMidEndSepPunct{\mcitedefaultmidpunct}
{\mcitedefaultendpunct}{\mcitedefaultseppunct}\relax
\EndOfBibitem
\bibitem{Besson:2003cp}
CLEO, D.~Besson {\em et~al.},
  \ifthenelse{\boolean{articletitles}}{\emph{{Observation of a narrow resonance
  of mass 2.46-GeV/c$^2$ decaying to $D^{*+}_{s}\pi^0$ and confirmation of the
  $D^{*}_{sJ}(2317)$ state}},
  }{}\href{http://dx.doi.org/10.1103/PhysRevD.68.032002,
  10.1103/PhysRevD.75.119908}{Phys.\ Rev.\  \textbf{D68} (2003) 032002},
  \href{http://arxiv.org/abs/hep-ex/0305100}{{\normalfont\ttfamily
  arXiv:hep-ex/0305100}}, [Erratum: Phys. Rev.D75,119908(2007)]\relax
\mciteBstWouldAddEndPuncttrue
\mciteSetBstMidEndSepPunct{\mcitedefaultmidpunct}
{\mcitedefaultendpunct}{\mcitedefaultseppunct}\relax
\EndOfBibitem
\bibitem{Krokovny:2003zq}
Belle, P.~Krokovny {\em et~al.},
  \ifthenelse{\boolean{articletitles}}{\emph{{Observation of the $D_{sJ}(2317)$
  and $D_{sJ}(2457)$ in $B$ decays}},
  }{}\href{http://dx.doi.org/10.1103/PhysRevLett.91.262002}{Phys.\ Rev.\ Lett.\
   \textbf{91} (2003) 262002},
  \href{http://arxiv.org/abs/hep-ex/0308019}{{\normalfont\ttfamily
  arXiv:hep-ex/0308019}}\relax
\mciteBstWouldAddEndPuncttrue
\mciteSetBstMidEndSepPunct{\mcitedefaultmidpunct}
{\mcitedefaultendpunct}{\mcitedefaultseppunct}\relax
\EndOfBibitem
\bibitem{Cheng:2003kg}
H.-Y. Cheng and W.-S. Hou, \ifthenelse{\boolean{articletitles}}{\emph{{B decays
  as spectroscope for charmed four quark states}},
  }{}\href{http://dx.doi.org/10.1016/S0370-2693(03)00834-7}{Phys.\ Lett.\
  \textbf{B566} (2003) 193},
  \href{http://arxiv.org/abs/hep-ph/0305038}{{\normalfont\ttfamily
  arXiv:hep-ph/0305038}}\relax
\mciteBstWouldAddEndPuncttrue
\mciteSetBstMidEndSepPunct{\mcitedefaultmidpunct}
{\mcitedefaultendpunct}{\mcitedefaultseppunct}\relax
\EndOfBibitem
\bibitem{Terasaki:2003qa}
K.~Terasaki, \ifthenelse{\boolean{articletitles}}{\emph{{BABAR resonance as a
  new window of hadron physics}},
  }{}\href{http://dx.doi.org/10.1103/PhysRevD.68.011501}{Phys.\ Rev.\
  \textbf{D68} (2003) 011501},
  \href{http://arxiv.org/abs/hep-ph/0305213}{{\normalfont\ttfamily
  arXiv:hep-ph/0305213}}\relax
\mciteBstWouldAddEndPuncttrue
\mciteSetBstMidEndSepPunct{\mcitedefaultmidpunct}
{\mcitedefaultendpunct}{\mcitedefaultseppunct}\relax
\EndOfBibitem
\bibitem{Bracco:2005kt}
M.~E. Bracco {\em et~al.},
  \ifthenelse{\boolean{articletitles}}{\emph{{Disentangling two- and four-quark
  state pictures of the charmed scalar mesons}},
  }{}\href{http://dx.doi.org/10.1016/j.physletb.2005.08.037}{Phys.\ Lett.\
  \textbf{B624} (2005) 217},
  \href{http://arxiv.org/abs/hep-ph/0503137}{{\normalfont\ttfamily
  arXiv:hep-ph/0503137}}\relax
\mciteBstWouldAddEndPuncttrue
\mciteSetBstMidEndSepPunct{\mcitedefaultmidpunct}
{\mcitedefaultendpunct}{\mcitedefaultseppunct}\relax
\EndOfBibitem
\bibitem{Aubert:2006bk}
BaBar, B.~Aubert {\em et~al.}, \ifthenelse{\boolean{articletitles}}{\emph{{A
  Study of the $D^*_{sJ}(2317)$ and $D_{sJ}(2460)$ Mesons in Inclusive $c\bar
  c$ Production Near $\sqrt{s} = 10.6$ GeV}},
  }{}\href{http://dx.doi.org/10.1103/PhysRevD.74.032007}{Phys.\ Rev.\
  \textbf{D74} (2006) 032007},
  \href{http://arxiv.org/abs/hep-ex/0604030}{{\normalfont\ttfamily
  arXiv:hep-ex/0604030}}\relax
\mciteBstWouldAddEndPuncttrue
\mciteSetBstMidEndSepPunct{\mcitedefaultmidpunct}
{\mcitedefaultendpunct}{\mcitedefaultseppunct}\relax
\EndOfBibitem
\bibitem{Choi:2015lpc}
Belle, S.-K. Choi {\em et~al.},
  \ifthenelse{\boolean{articletitles}}{\emph{{Measurements of $B\rightarrow
  \bar{D} D_{s0}^{*+}(2317)$ decay rates and a search for isospin partners of
  the $D_{s0}^{*+} (2317)$}},
  }{}\href{http://dx.doi.org/10.1103/PhysRevD.91.092011,
  10.1103/PhysRevD.92.039905}{Phys.\ Rev.\  \textbf{D91} (2015) 092011},
  \href{http://arxiv.org/abs/1504.02637}{{\normalfont\ttfamily
  arXiv:1504.02637}}, [Addendum: Phys. Rev.D92,no.3,039905(2015)]\relax
\mciteBstWouldAddEndPuncttrue
\mciteSetBstMidEndSepPunct{\mcitedefaultmidpunct}
{\mcitedefaultendpunct}{\mcitedefaultseppunct}\relax
\EndOfBibitem
\bibitem{Barnes:2003dj}
T.~Barnes, F.~E. Close, and H.~J. Lipkin,
  \ifthenelse{\boolean{articletitles}}{\emph{{Implications of a DK molecule at
  2.32-GeV}}, }{}\href{http://dx.doi.org/10.1103/PhysRevD.68.054006}{Phys.\
  Rev.\  \textbf{D68} (2003) 054006},
  \href{http://arxiv.org/abs/hep-ph/0305025}{{\normalfont\ttfamily
  arXiv:hep-ph/0305025}}\relax
\mciteBstWouldAddEndPuncttrue
\mciteSetBstMidEndSepPunct{\mcitedefaultmidpunct}
{\mcitedefaultendpunct}{\mcitedefaultseppunct}\relax
\EndOfBibitem
\bibitem{Browder:2003fk}
T.~E. Browder, S.~Pakvasa, and A.~A. Petrov,
  \ifthenelse{\boolean{articletitles}}{\emph{{Comment on the new
  $D_s^{*+}\pi^0$ resonances}},
  }{}\href{http://dx.doi.org/10.1016/j.physletb.2003.10.067}{Phys.\ Lett.\
  \textbf{B578} (2004) 365},
  \href{http://arxiv.org/abs/hep-ph/0307054}{{\normalfont\ttfamily
  arXiv:hep-ph/0307054}}\relax
\mciteBstWouldAddEndPuncttrue
\mciteSetBstMidEndSepPunct{\mcitedefaultmidpunct}
{\mcitedefaultendpunct}{\mcitedefaultseppunct}\relax
\EndOfBibitem
\bibitem{bspisl}
D0, \ifthenelse{\boolean{articletitles}}{\emph{{Confirmation of the $X(5568)$
  with the semileptonic decay of $B_s^0$}}, }{}
\newblock
  {\href{http://www-d0.fnal.gov/Run2Physics/WWW/results/prelim/B/B68/B68.pdf}{D0
  Note 6496-CONF}}\relax
\mciteBstWouldAddEndPuncttrue
\mciteSetBstMidEndSepPunct{\mcitedefaultmidpunct}
{\mcitedefaultendpunct}{\mcitedefaultseppunct}\relax
\EndOfBibitem
\bibitem{Aaij:2016iev}
LHCb, R.~Aaij {\em et~al.}, \ifthenelse{\boolean{articletitles}}{\emph{{Search
  for Structure in the $B_s^0\pi^\pm$ Invariant Mass Spectrum}},
  }{}\href{http://dx.doi.org/10.1103/PhysRevLett.117.152003}{Phys.\ Rev.\
  Lett.\  \textbf{117} (2016) 152003},
  \href{http://arxiv.org/abs/1608.00435}{{\normalfont\ttfamily
  arXiv:1608.00435}}\relax
\mciteBstWouldAddEndPuncttrue
\mciteSetBstMidEndSepPunct{\mcitedefaultmidpunct}
{\mcitedefaultendpunct}{\mcitedefaultseppunct}\relax
\EndOfBibitem
\bibitem{CMS:2016fvl}
CMS, CMS, \ifthenelse{\boolean{articletitles}}{\emph{{Search for the X(5568)
  state in $B^0_s\pi^{\pm}$ decays}}, }{},
  \href{http://cds.cern.ch/record/2204918/}{CMS-PAS-BPH-16-002}\relax
\mciteBstWouldAddEndPuncttrue
\mciteSetBstMidEndSepPunct{\mcitedefaultmidpunct}
{\mcitedefaultendpunct}{\mcitedefaultseppunct}\relax
\EndOfBibitem
\bibitem{Burns:2016gvy}
T.~J. Burns and E.~S. Swanson,
  \ifthenelse{\boolean{articletitles}}{\emph{{Interpreting the X(5568)}},
  }{}\href{http://dx.doi.org/10.1016/j.physletb.2016.07.049}{Phys.\ Lett.\
  \textbf{B760} (2016) 627},
  \href{http://arxiv.org/abs/1603.04366}{{\normalfont\ttfamily
  arXiv:1603.04366}}\relax
\mciteBstWouldAddEndPuncttrue
\mciteSetBstMidEndSepPunct{\mcitedefaultmidpunct}
{\mcitedefaultendpunct}{\mcitedefaultseppunct}\relax
\EndOfBibitem
\bibitem{Guo:2016nhb}
F.-K. Guo, U.-G. Meißner, and B.-S. Zou,
  \ifthenelse{\boolean{articletitles}}{\emph{{How the X(5568) challenges our
  understanding of QCD}},
  }{}\href{http://dx.doi.org/10.1088/0253-6102/65/5/593}{Commun.\ Theor.\
  Phys.\  \textbf{65} (2016) 593},
  \href{http://arxiv.org/abs/1603.06316}{{\normalfont\ttfamily
  arXiv:1603.06316}}\relax
\mciteBstWouldAddEndPuncttrue
\mciteSetBstMidEndSepPunct{\mcitedefaultmidpunct}
{\mcitedefaultendpunct}{\mcitedefaultseppunct}\relax
\EndOfBibitem
\bibitem{Esposito:2016itg}
A.~Esposito, A.~Pilloni, and A.~D. Polosa,
  \ifthenelse{\boolean{articletitles}}{\emph{{Hybridized Tetraquarks}},
  }{}\href{http://dx.doi.org/10.1016/j.physletb.2016.05.028}{Phys.\ Lett.\
  \textbf{B758} (2016) 292},
  \href{http://arxiv.org/abs/1603.07667}{{\normalfont\ttfamily
  arXiv:1603.07667}}\relax
\mciteBstWouldAddEndPuncttrue
\mciteSetBstMidEndSepPunct{\mcitedefaultmidpunct}
{\mcitedefaultendpunct}{\mcitedefaultseppunct}\relax
\EndOfBibitem
\bibitem{Yang:2016sws}
Z.~Yang, Q.~Wang, and U.-G. Meissner,
  \ifthenelse{\boolean{articletitles}}{\emph{{Where does the X(5568) structure
  come from?}},
  }{}\href{http://dx.doi.org/10.1016/j.physletb.2017.01.023}{Phys.\ Lett.\
  \textbf{B767} (2017) 470},
  \href{http://arxiv.org/abs/1609.08807}{{\normalfont\ttfamily
  arXiv:1609.08807}}\relax
\mciteBstWouldAddEndPuncttrue
\mciteSetBstMidEndSepPunct{\mcitedefaultmidpunct}
{\mcitedefaultendpunct}{\mcitedefaultseppunct}\relax
\EndOfBibitem
\bibitem{Lang:2016jpk}
C.~B. Lang, D.~Mohler, and S.~Prelovsek,
  \ifthenelse{\boolean{articletitles}}{\emph{{$B_s\pi^+$ scattering and search
  for X(5568) with lattice QCD}},
  }{}\href{http://dx.doi.org/10.1103/PhysRevD.94.074509}{Phys.\ Rev.\
  \textbf{D94} (2016) 074509},
  \href{http://arxiv.org/abs/1607.03185}{{\normalfont\ttfamily
  arXiv:1607.03185}}\relax
\mciteBstWouldAddEndPuncttrue
\mciteSetBstMidEndSepPunct{\mcitedefaultmidpunct}
{\mcitedefaultendpunct}{\mcitedefaultseppunct}\relax
\EndOfBibitem
\bibitem{Bala:2015wep}
Belle, A.~Bala {\em et~al.},
  \ifthenelse{\boolean{articletitles}}{\emph{{Observation of X(3872) in $B\to
  X(3872)K\pi$ decays}},
  }{}\href{http://dx.doi.org/10.1103/PhysRevD.91.051101}{Phys.\ Rev.\
  \textbf{D91} (2015), no.~5 051101},
  \href{http://arxiv.org/abs/1501.06867}{{\normalfont\ttfamily
  arXiv:1501.06867}}\relax
\mciteBstWouldAddEndPuncttrue
\mciteSetBstMidEndSepPunct{\mcitedefaultmidpunct}
{\mcitedefaultendpunct}{\mcitedefaultseppunct}\relax
\EndOfBibitem
\bibitem{Aubert:2006aj}
BaBar, B.~Aubert {\em et~al.},
  \ifthenelse{\boolean{articletitles}}{\emph{{Search for $B^{+} \to X(3872)
  K^{+}$, $X(3872) \to J/\psi \gamma$}},
  }{}\href{http://dx.doi.org/10.1103/PhysRevD.74.071101}{Phys.\ Rev.\
  \textbf{D74} (2006) 071101},
  \href{http://arxiv.org/abs/hep-ex/0607050}{{\normalfont\ttfamily
  arXiv:hep-ex/0607050}}\relax
\mciteBstWouldAddEndPuncttrue
\mciteSetBstMidEndSepPunct{\mcitedefaultmidpunct}
{\mcitedefaultendpunct}{\mcitedefaultseppunct}\relax
\EndOfBibitem
\bibitem{Eichten:2005ga}
E.~J. Eichten, K.~Lane, and C.~Quigg,
  \ifthenelse{\boolean{articletitles}}{\emph{{New states above charm
  threshold}}, }{}\href{http://dx.doi.org/10.1103/PhysRevD.73.014014,
  10.1103/PhysRevD.73.079903}{Phys.\ Rev.\  \textbf{D73} (2006) 014014},
  \href{http://arxiv.org/abs/hep-ph/0511179}{{\normalfont\ttfamily
  arXiv:hep-ph/0511179}}, [Erratum: Phys. Rev.D73,079903(2006)]\relax
\mciteBstWouldAddEndPuncttrue
\mciteSetBstMidEndSepPunct{\mcitedefaultmidpunct}
{\mcitedefaultendpunct}{\mcitedefaultseppunct}\relax
\EndOfBibitem
\bibitem{Aubert:2010ab}
BaBar, B.~Aubert {\em et~al.},
  \ifthenelse{\boolean{articletitles}}{\emph{{Observation of the
  $\chi_{c2}(2p)$ Meson in the Reaction $\gamma \gamma \to D \bar{D}$ at
  {BaBar}}}, }{}\href{http://dx.doi.org/10.1103/PhysRevD.81.092003}{Phys.\
  Rev.\  \textbf{D81} (2010) 092003},
  \href{http://arxiv.org/abs/1002.0281}{{\normalfont\ttfamily
  arXiv:1002.0281}}\relax
\mciteBstWouldAddEndPuncttrue
\mciteSetBstMidEndSepPunct{\mcitedefaultmidpunct}
{\mcitedefaultendpunct}{\mcitedefaultseppunct}\relax
\EndOfBibitem
\bibitem{Brambilla:2010cs}
N.~Brambilla {\em et~al.}, \ifthenelse{\boolean{articletitles}}{\emph{{Heavy
  quarkonium: progress, puzzles, and opportunities}},
  }{}\href{http://dx.doi.org/10.1140/epjc/s10052-010-1534-9}{Eur.\ Phys.\ J.\
  \textbf{C71} (2011) 1534},
  \href{http://arxiv.org/abs/1010.5827}{{\normalfont\ttfamily
  arXiv:1010.5827}}\relax
\mciteBstWouldAddEndPuncttrue
\mciteSetBstMidEndSepPunct{\mcitedefaultmidpunct}
{\mcitedefaultendpunct}{\mcitedefaultseppunct}\relax
\EndOfBibitem
\bibitem{Eichten:2004uh}
E.~J. Eichten, K.~Lane, and C.~Quigg,
  \ifthenelse{\boolean{articletitles}}{\emph{{Charmonium levels near threshold
  and the narrow state $X(3872) \to \pi^{+}\pi^{-}J/\psi$}},
  }{}\href{http://dx.doi.org/10.1103/PhysRevD.69.094019}{Phys.\ Rev.\
  \textbf{D69} (2004) 094019},
  \href{http://arxiv.org/abs/hep-ph/0401210}{{\normalfont\ttfamily
  arXiv:hep-ph/0401210}}\relax
\mciteBstWouldAddEndPuncttrue
\mciteSetBstMidEndSepPunct{\mcitedefaultmidpunct}
{\mcitedefaultendpunct}{\mcitedefaultseppunct}\relax
\EndOfBibitem
\bibitem{Li:2009zu}
B.-Q. Li and K.-T. Chao, \ifthenelse{\boolean{articletitles}}{\emph{{Higher
  Charmonia and X,Y,Z states with Screened Potential}},
  }{}\href{http://dx.doi.org/10.1103/PhysRevD.79.094004}{Phys.\ Rev.\
  \textbf{D79} (2009) 094004},
  \href{http://arxiv.org/abs/0903.5506}{{\normalfont\ttfamily
  arXiv:0903.5506}}\relax
\mciteBstWouldAddEndPuncttrue
\mciteSetBstMidEndSepPunct{\mcitedefaultmidpunct}
{\mcitedefaultendpunct}{\mcitedefaultseppunct}\relax
\EndOfBibitem
\bibitem{Wang:2014voa}
H.~Wang, Y.~Yang, and J.~Ping,
  \ifthenelse{\boolean{articletitles}}{\emph{{Strong decays of $\chi_{cJ}(2P)$
  and $\chi_{cJ}(3P)$}},
  }{}\href{http://dx.doi.org/10.1140/epja/i2014-14076-y}{Eur.\ Phys.\ J.\
  \textbf{A50} (2014) 76}\relax
\mciteBstWouldAddEndPuncttrue
\mciteSetBstMidEndSepPunct{\mcitedefaultmidpunct}
{\mcitedefaultendpunct}{\mcitedefaultseppunct}\relax
\EndOfBibitem
\bibitem{Tornqvist:2003na}
N.~A. Tornqvist, \ifthenelse{\boolean{articletitles}}{\emph{{Comment on the
  narrow charmonium state of Belle at 3871.8-MeV as a deuson}},
  }{}\href{http://arxiv.org/abs/hep-ph/0308277}{{\normalfont\ttfamily
  arXiv:hep-ph/0308277}}\relax
\mciteBstWouldAddEndPuncttrue
\mciteSetBstMidEndSepPunct{\mcitedefaultmidpunct}
{\mcitedefaultendpunct}{\mcitedefaultseppunct}\relax
\EndOfBibitem
\bibitem{Suzuki:2005ha}
M.~Suzuki, \ifthenelse{\boolean{articletitles}}{\emph{{The X(3872) boson:
  Molecule or charmonium}},
  }{}\href{http://dx.doi.org/10.1103/PhysRevD.72.114013}{Phys.\ Rev.\
  \textbf{D72} (2005) 114013},
  \href{http://arxiv.org/abs/hep-ph/0508258}{{\normalfont\ttfamily
  arXiv:hep-ph/0508258}}\relax
\mciteBstWouldAddEndPuncttrue
\mciteSetBstMidEndSepPunct{\mcitedefaultmidpunct}
{\mcitedefaultendpunct}{\mcitedefaultseppunct}\relax
\EndOfBibitem
\bibitem{Aubert:2004zr}
BaBar, B.~Aubert {\em et~al.},
  \ifthenelse{\boolean{articletitles}}{\emph{{Search for a charged partner of
  the X(3872) in the $B$ meson decay $B \to X^- K$, $X^- \to J/\psi \pi^-
  \pi^0$}}, }{}\href{http://dx.doi.org/10.1103/PhysRevD.71.031501}{Phys.\ Rev.\
   \textbf{D71} (2005) 031501},
  \href{http://arxiv.org/abs/hep-ex/0412051}{{\normalfont\ttfamily
  arXiv:hep-ex/0412051}}\relax
\mciteBstWouldAddEndPuncttrue
\mciteSetBstMidEndSepPunct{\mcitedefaultmidpunct}
{\mcitedefaultendpunct}{\mcitedefaultseppunct}\relax
\EndOfBibitem
\bibitem{Aushev:2008sua}
Belle, T.~Aushev {\em et~al.},
  \ifthenelse{\boolean{articletitles}}{\emph{{Study of the $B\to
  X(3872)(D^{*0}\overline{D^0})K$ decay}},
  }{}\href{http://dx.doi.org/10.1103/PhysRevD.81.031103}{Phys.\ Rev.\
  \textbf{D81} (2010) 031103},
  \href{http://arxiv.org/abs/0810.0358}{{\normalfont\ttfamily
  arXiv:0810.0358}}\relax
\mciteBstWouldAddEndPuncttrue
\mciteSetBstMidEndSepPunct{\mcitedefaultmidpunct}
{\mcitedefaultendpunct}{\mcitedefaultseppunct}\relax
\EndOfBibitem
\bibitem{Braaten:2007dw}
E.~Braaten and M.~Lu, \ifthenelse{\boolean{articletitles}}{\emph{{Line shapes
  of the $X(3872)$}},
  }{}\href{http://dx.doi.org/10.1103/PhysRevD.76.094028}{Phys.\ Rev.\
  \textbf{D76} (2007) 094028},
  \href{http://arxiv.org/abs/0709.2697}{{\normalfont\ttfamily
  arXiv:0709.2697}}\relax
\mciteBstWouldAddEndPuncttrue
\mciteSetBstMidEndSepPunct{\mcitedefaultmidpunct}
{\mcitedefaultendpunct}{\mcitedefaultseppunct}\relax
\EndOfBibitem
\bibitem{Braaten:2004rn}
E.~Braaten and H.-W. Hammer,
  \ifthenelse{\boolean{articletitles}}{\emph{{Universality in few-body systems
  with large scattering length}},
  }{}\href{http://dx.doi.org/10.1016/j.physrep.2006.03.001}{Phys.\ Rept.\
  \textbf{428} (2006) 259},
  \href{http://arxiv.org/abs/cond-mat/0410417}{{\normalfont\ttfamily
  arXiv:cond-mat/0410417}}\relax
\mciteBstWouldAddEndPuncttrue
\mciteSetBstMidEndSepPunct{\mcitedefaultmidpunct}
{\mcitedefaultendpunct}{\mcitedefaultseppunct}\relax
\EndOfBibitem
\bibitem{Coito:2012vf}
S.~Coito, G.~Rupp, and E.~van Beveren,
  \ifthenelse{\boolean{articletitles}}{\emph{{X(3872) is not a true molecule}},
  }{}\href{http://dx.doi.org/10.1140/epjc/s10052-013-2351-8}{Eur.\ Phys.\ J.\
  \textbf{C73} (2013) 2351},
  \href{http://arxiv.org/abs/1212.0648}{{\normalfont\ttfamily
  arXiv:1212.0648}}\relax
\mciteBstWouldAddEndPuncttrue
\mciteSetBstMidEndSepPunct{\mcitedefaultmidpunct}
{\mcitedefaultendpunct}{\mcitedefaultseppunct}\relax
\EndOfBibitem
\bibitem{Tornqvist:2004qy}
N.~A. Tornqvist, \ifthenelse{\boolean{articletitles}}{\emph{{Isospin breaking
  of the narrow charmonium state of Belle at 3872-MeV as a deuson}},
  }{}\href{http://dx.doi.org/10.1016/j.physletb.2004.03.077}{Phys.\ Lett.\
  \textbf{B590} (2004) 209},
  \href{http://arxiv.org/abs/hep-ph/0402237}{{\normalfont\ttfamily
  arXiv:hep-ph/0402237}}\relax
\mciteBstWouldAddEndPuncttrue
\mciteSetBstMidEndSepPunct{\mcitedefaultmidpunct}
{\mcitedefaultendpunct}{\mcitedefaultseppunct}\relax
\EndOfBibitem
\bibitem{Swanson:2004pp}
E.~S. Swanson, \ifthenelse{\boolean{articletitles}}{\emph{{Diagnostic decays of
  the X(3872)}},
  }{}\href{http://dx.doi.org/10.1016/j.physletb.2004.07.059}{Phys.\ Lett.\
  \textbf{B598} (2004) 197},
  \href{http://arxiv.org/abs/hep-ph/0406080}{{\normalfont\ttfamily
  arXiv:hep-ph/0406080}}\relax
\mciteBstWouldAddEndPuncttrue
\mciteSetBstMidEndSepPunct{\mcitedefaultmidpunct}
{\mcitedefaultendpunct}{\mcitedefaultseppunct}\relax
\EndOfBibitem
\bibitem{Swanson:2003tb}
E.~S. Swanson, \ifthenelse{\boolean{articletitles}}{\emph{{Short range
  structure in the X(3872)}},
  }{}\href{http://dx.doi.org/10.1016/j.physletb.2004.03.033}{Phys.\ Lett.\
  \textbf{B588} (2004) 189},
  \href{http://arxiv.org/abs/hep-ph/0311229}{{\normalfont\ttfamily
  arXiv:hep-ph/0311229}}\relax
\mciteBstWouldAddEndPuncttrue
\mciteSetBstMidEndSepPunct{\mcitedefaultmidpunct}
{\mcitedefaultendpunct}{\mcitedefaultseppunct}\relax
\EndOfBibitem
\bibitem{Dong:2009uf}
Y.~Dong, A.~Faessler, T.~Gutsche, and V.~E. Lyubovitskij,
  \ifthenelse{\boolean{articletitles}}{\emph{{$J/\psi\gamma$ and $\psi(2S)
  \gamma$ decay modes of the X(3872)}},
  }{}\href{http://dx.doi.org/10.1088/0954-3899/38/1/015001}{J.\ Phys.\
  \textbf{G38} (2011) 015001},
  \href{http://arxiv.org/abs/0909.0380}{{\normalfont\ttfamily
  arXiv:0909.0380}}\relax
\mciteBstWouldAddEndPuncttrue
\mciteSetBstMidEndSepPunct{\mcitedefaultmidpunct}
{\mcitedefaultendpunct}{\mcitedefaultseppunct}\relax
\EndOfBibitem
\bibitem{Bignamini:2009sk}
C.~Bignamini {\em et~al.}, \ifthenelse{\boolean{articletitles}}{\emph{{Is the
  X(3872) Production Cross Section at Tevatron Compatible with a Hadron
  Molecule Interpretation?}},
  }{}\href{http://dx.doi.org/10.1103/PhysRevLett.103.162001}{Phys.\ Rev.\
  Lett.\  \textbf{103} (2009) 162001},
  \href{http://arxiv.org/abs/0906.0882}{{\normalfont\ttfamily
  arXiv:0906.0882}}\relax
\mciteBstWouldAddEndPuncttrue
\mciteSetBstMidEndSepPunct{\mcitedefaultmidpunct}
{\mcitedefaultendpunct}{\mcitedefaultseppunct}\relax
\EndOfBibitem
\bibitem{CDFprompt:2004}
CDFII, \ifthenelse{\boolean{articletitles}}{\emph{{The lifetime distribution of
  $X(3872)$ mesons produced in ppbar collisions at CDF}}, }{} 2004.
\newblock
  {\href{http://www-cdf.fnal.gov/physics/new/bottom/051020.blessed-x3872/XLife/xlonglivedWWW.ps}{CDF
  note 7159}}\relax
\mciteBstWouldAddEndPuncttrue
\mciteSetBstMidEndSepPunct{\mcitedefaultmidpunct}
{\mcitedefaultendpunct}{\mcitedefaultseppunct}\relax
\EndOfBibitem
\bibitem{Esposito:2015fsa}
A.~Esposito {\em et~al.},
  \ifthenelse{\boolean{articletitles}}{\emph{{Observation of light nuclei at
  ALICE and the $X(3872)$ conundrum}},
  }{}\href{http://dx.doi.org/10.1103/PhysRevD.92.034028}{Phys.\ Rev.\
  \textbf{D92} (2015) 034028},
  \href{http://arxiv.org/abs/1508.00295}{{\normalfont\ttfamily
  arXiv:1508.00295}}\relax
\mciteBstWouldAddEndPuncttrue
\mciteSetBstMidEndSepPunct{\mcitedefaultmidpunct}
{\mcitedefaultendpunct}{\mcitedefaultseppunct}\relax
\EndOfBibitem
\bibitem{Adam:2015yta}
ALICE, J.~Adam {\em et~al.},
  \ifthenelse{\boolean{articletitles}}{\emph{{$^{3}_{\Lambda}\mathrm H$ and
  $^{3}_{\bar{\Lambda}} \overline{\mathrm H}$ production in Pb-Pb collisions at
  $\sqrt{s_{\rm NN}} =$ 2.76 TeV}},
  }{}\href{http://dx.doi.org/10.1016/j.physletb.2016.01.040}{Phys.\ Lett.\
  \textbf{B754} (2016) 360},
  \href{http://arxiv.org/abs/1506.08453}{{\normalfont\ttfamily
  arXiv:1506.08453}}\relax
\mciteBstWouldAddEndPuncttrue
\mciteSetBstMidEndSepPunct{\mcitedefaultmidpunct}
{\mcitedefaultendpunct}{\mcitedefaultseppunct}\relax
\EndOfBibitem
\bibitem{Adam:2015vda}
ALICE, J.~Adam {\em et~al.},
  \ifthenelse{\boolean{articletitles}}{\emph{{Production of light nuclei and
  anti-nuclei in $pp$ and Pb-Pb collisions at energies available at the CERN
  Large Hadron Collider}},
  }{}\href{http://dx.doi.org/10.1103/PhysRevC.93.024917}{Phys.\ Rev.\
  \textbf{C93} (2016) 024917},
  \href{http://arxiv.org/abs/1506.08951}{{\normalfont\ttfamily
  arXiv:1506.08951}}\relax
\mciteBstWouldAddEndPuncttrue
\mciteSetBstMidEndSepPunct{\mcitedefaultmidpunct}
{\mcitedefaultendpunct}{\mcitedefaultseppunct}\relax
\EndOfBibitem
\bibitem{Schnedermann:1993ws}
E.~Schnedermann, J.~Sollfrank, and U.~W. Heinz,
  \ifthenelse{\boolean{articletitles}}{\emph{{Thermal phenomenology of hadrons
  from 200-A/GeV S+S collisions}},
  }{}\href{http://dx.doi.org/10.1103/PhysRevC.48.2462}{Phys.\ Rev.\
  \textbf{C48} (1993) 2462},
  \href{http://arxiv.org/abs/nucl-th/9307020}{{\normalfont\ttfamily
  arXiv:nucl-th/9307020}}\relax
\mciteBstWouldAddEndPuncttrue
\mciteSetBstMidEndSepPunct{\mcitedefaultmidpunct}
{\mcitedefaultendpunct}{\mcitedefaultseppunct}\relax
\EndOfBibitem
\bibitem{pdg14}
Particle Data Group, K.~A. Olive {\em et~al.},
  \ifthenelse{\boolean{articletitles}}{\emph{{Review of particle physics}},
  }{}\href{http://dx.doi.org/10.1088/1674-1137/38/9/090001}{Chin.\ Phys.\
  \textbf{C38} (2014) 090001}\relax
\mciteBstWouldAddEndPuncttrue
\mciteSetBstMidEndSepPunct{\mcitedefaultmidpunct}
{\mcitedefaultendpunct}{\mcitedefaultseppunct}\relax
\EndOfBibitem
\bibitem{Guo:2012tv}
F.-K. Guo and U.-G. Meissner, \ifthenelse{\boolean{articletitles}}{\emph{{Where
  is the $\chi_{c0}(2P)$?}},
  }{}\href{http://dx.doi.org/10.1103/PhysRevD.86.091501}{Phys.\ Rev.\
  \textbf{D86} (2012) 091501},
  \href{http://arxiv.org/abs/1208.1134}{{\normalfont\ttfamily
  arXiv:1208.1134}}\relax
\mciteBstWouldAddEndPuncttrue
\mciteSetBstMidEndSepPunct{\mcitedefaultmidpunct}
{\mcitedefaultendpunct}{\mcitedefaultseppunct}\relax
\EndOfBibitem
\bibitem{Olsen:2014maa}
S.~L. Olsen, \ifthenelse{\boolean{articletitles}}{\emph{{Is the X(3915) the
  $\chi_{c0}(2P)$?}},
  }{}\href{http://dx.doi.org/10.1103/PhysRevD.91.057501}{Phys.\ Rev.\
  \textbf{D91} (2015) 057501},
  \href{http://arxiv.org/abs/1410.6534}{{\normalfont\ttfamily
  arXiv:1410.6534}}\relax
\mciteBstWouldAddEndPuncttrue
\mciteSetBstMidEndSepPunct{\mcitedefaultmidpunct}
{\mcitedefaultendpunct}{\mcitedefaultseppunct}\relax
\EndOfBibitem
\bibitem{Zhou:2015uva}
Z.-Y. Zhou, Z.~Xiao, and H.-Q. Zhou,
  \ifthenelse{\boolean{articletitles}}{\emph{{Could the $X(3915)$ and the
  $X(3930)$ Be the Same Tensor State?}},
  }{}\href{http://dx.doi.org/10.1103/PhysRevLett.115.022001}{Phys.\ Rev.\
  Lett.\  \textbf{115} (2015) 022001},
  \href{http://arxiv.org/abs/1501.00879}{{\normalfont\ttfamily
  arXiv:1501.00879}}\relax
\mciteBstWouldAddEndPuncttrue
\mciteSetBstMidEndSepPunct{\mcitedefaultmidpunct}
{\mcitedefaultendpunct}{\mcitedefaultseppunct}\relax
\EndOfBibitem
\bibitem{Bodwin:1992qr}
G.~T. Bodwin, E.~Braaten, T.~C. Yuan, and G.~P. Lepage,
  \ifthenelse{\boolean{articletitles}}{\emph{{P wave charmonium production in B
  meson decays}}, }{}\href{http://dx.doi.org/10.1103/PhysRevD.46.R3703}{Phys.\
  Rev.\  \textbf{D46} (1992) R3703},
  \href{http://arxiv.org/abs/hep-ph/9208254}{{\normalfont\ttfamily
  arXiv:hep-ph/9208254}}\relax
\mciteBstWouldAddEndPuncttrue
\mciteSetBstMidEndSepPunct{\mcitedefaultmidpunct}
{\mcitedefaultendpunct}{\mcitedefaultseppunct}\relax
\EndOfBibitem
\bibitem{Chilikin:2017evr}
Belle, K.~Chilikin {\em et~al.},
  \ifthenelse{\boolean{articletitles}}{\emph{{Observation of an alternative
  $\chi_{c0}(2P)$ candidate in $e^+ e^- \rightarrow J/\psi D \bar{D}$}},
  }{}\href{http://dx.doi.org/10.1103/PhysRevD.95.112003}{Phys.\ Rev.\
  \textbf{D95} (2017) 112003},
  \href{http://arxiv.org/abs/1704.01872}{{\normalfont\ttfamily
  arXiv:1704.01872}}\relax
\mciteBstWouldAddEndPuncttrue
\mciteSetBstMidEndSepPunct{\mcitedefaultmidpunct}
{\mcitedefaultendpunct}{\mcitedefaultseppunct}\relax
\EndOfBibitem
\bibitem{Li:2015iga}
X.~Li and M.~B. Voloshin, \ifthenelse{\boolean{articletitles}}{\emph{{$X(3915)$
  as a $D_s \bar D_s$ bound state}},
  }{}\href{http://dx.doi.org/10.1103/PhysRevD.91.114014}{Phys.\ Rev.\
  \textbf{D91} (2015) 114014},
  \href{http://arxiv.org/abs/1503.04431}{{\normalfont\ttfamily
  arXiv:1503.04431}}\relax
\mciteBstWouldAddEndPuncttrue
\mciteSetBstMidEndSepPunct{\mcitedefaultmidpunct}
{\mcitedefaultendpunct}{\mcitedefaultseppunct}\relax
\EndOfBibitem
\bibitem{Lebed:2016yvr}
R.~F. Lebed and A.~D. Polosa,
  \ifthenelse{\boolean{articletitles}}{\emph{{$\chi^{\vphantom\dagger}_{c0}(3915)$
  as the lightest $c\bar c s \bar s$ state}},
  }{}\href{http://dx.doi.org/10.1103/PhysRevD.93.094024}{Phys.\ Rev.\
  \textbf{D93} (2016) 094024},
  \href{http://arxiv.org/abs/1602.08421}{{\normalfont\ttfamily
  arXiv:1602.08421}}\relax
\mciteBstWouldAddEndPuncttrue
\mciteSetBstMidEndSepPunct{\mcitedefaultmidpunct}
{\mcitedefaultendpunct}{\mcitedefaultseppunct}\relax
\EndOfBibitem
\bibitem{Benayoun:1999fv}
M.~Benayoun {\em et~al.}, \ifthenelse{\boolean{articletitles}}{\emph{{Radiative
  decays, nonet symmetry and SU(3) breaking}},
  }{}\href{http://dx.doi.org/10.1103/PhysRevD.59.114027}{Phys.\ Rev.\
  \textbf{D59} (1999) 114027},
  \href{http://arxiv.org/abs/hep-ph/9902326}{{\normalfont\ttfamily
  arXiv:hep-ph/9902326}}\relax
\mciteBstWouldAddEndPuncttrue
\mciteSetBstMidEndSepPunct{\mcitedefaultmidpunct}
{\mcitedefaultendpunct}{\mcitedefaultseppunct}\relax
\EndOfBibitem
\bibitem{Vinokurova:2015txd}
Belle, A.~Vinokurova {\em et~al.},
  \ifthenelse{\boolean{articletitles}}{\emph{{Search for B decays to final
  states with the $\eta_c$ meson}},
  }{}\href{http://dx.doi.org/10.1007/JHEP06(2015)132}{JHEP \textbf{06} (2015)
  132}, \href{http://arxiv.org/abs/1501.06351}{{\normalfont\ttfamily
  arXiv:1501.06351}}\relax
\mciteBstWouldAddEndPuncttrue
\mciteSetBstMidEndSepPunct{\mcitedefaultmidpunct}
{\mcitedefaultendpunct}{\mcitedefaultseppunct}\relax
\EndOfBibitem
\bibitem{Abe:2010gxa}
Belle-II, T.~Abe {\em et~al.},
  \ifthenelse{\boolean{articletitles}}{\emph{{Belle II Technical Design
  Report}}, }{}\href{http://arxiv.org/abs/1011.0352}{{\normalfont\ttfamily
  arXiv:1011.0352}}\relax
\mciteBstWouldAddEndPuncttrue
\mciteSetBstMidEndSepPunct{\mcitedefaultmidpunct}
{\mcitedefaultendpunct}{\mcitedefaultseppunct}\relax
\EndOfBibitem
\bibitem{LHCb:2012cw}
LHCb, R.~Aaij {\em et~al.},
  \ifthenelse{\boolean{articletitles}}{\emph{{Evidence for the decay $B^0\to
  J/\psi \omega$ and measurement of the relative branching fractions of $B^0_s$
  meson decays to $J/\psi\eta$ and $J/\psi\eta^{'}$}},
  }{}\href{http://dx.doi.org/10.1016/j.nuclphysb.2012.10.021}{Nucl.\ Phys.\
  \textbf{B867} (2013) 547},
  \href{http://arxiv.org/abs/1210.2631}{{\normalfont\ttfamily
  arXiv:1210.2631}}\relax
\mciteBstWouldAddEndPuncttrue
\mciteSetBstMidEndSepPunct{\mcitedefaultmidpunct}
{\mcitedefaultendpunct}{\mcitedefaultseppunct}\relax
\EndOfBibitem
\bibitem{Aubert:2005eg}
BaBar, B.~Aubert {\em et~al.}, \ifthenelse{\boolean{articletitles}}{\emph{{The
  $e^+e^- \to \pi^+ \pi^- \pi^+ \pi^-$, $K^+ K^- \pi^+ \pi^-$, and $K^+ K^- K^+
  K^-$ cross sections at center-of-mass energies 0.5-GeV - 4.5-GeV measured
  with initial-state radiation}},
  }{}\href{http://dx.doi.org/10.1103/PhysRevD.71.052001}{Phys.\ Rev.\
  \textbf{D71} (2005) 052001},
  \href{http://arxiv.org/abs/hep-ex/0502025}{{\normalfont\ttfamily
  arXiv:hep-ex/0502025}}\relax
\mciteBstWouldAddEndPuncttrue
\mciteSetBstMidEndSepPunct{\mcitedefaultmidpunct}
{\mcitedefaultendpunct}{\mcitedefaultseppunct}\relax
\EndOfBibitem
\bibitem{Coan:2006rv}
CLEO, T.~E. Coan {\em et~al.},
  \ifthenelse{\boolean{articletitles}}{\emph{{Charmonium decays of Y(4260),
  $\psi(4160)$ and $\psi(4040)$}},
  }{}\href{http://dx.doi.org/10.1103/PhysRevLett.96.162003}{Phys.\ Rev.\ Lett.\
   \textbf{96} (2006) 162003},
  \href{http://arxiv.org/abs/hep-ex/0602034}{{\normalfont\ttfamily
  arXiv:hep-ex/0602034}}\relax
\mciteBstWouldAddEndPuncttrue
\mciteSetBstMidEndSepPunct{\mcitedefaultmidpunct}
{\mcitedefaultendpunct}{\mcitedefaultseppunct}\relax
\EndOfBibitem
\bibitem{Maiani:2014aja}
L.~Maiani, F.~Piccinini, A.~D. Polosa, and V.~Riquer,
  \ifthenelse{\boolean{articletitles}}{\emph{{The Z(4430) and a New Paradigm
  for Spin Interactions in Tetraquarks}},
  }{}\href{http://dx.doi.org/10.1103/PhysRevD.89.114010}{Phys.\ Rev.\
  \textbf{D89} (2014) 114010},
  \href{http://arxiv.org/abs/1405.1551}{{\normalfont\ttfamily
  arXiv:1405.1551}}\relax
\mciteBstWouldAddEndPuncttrue
\mciteSetBstMidEndSepPunct{\mcitedefaultmidpunct}
{\mcitedefaultendpunct}{\mcitedefaultseppunct}\relax
\EndOfBibitem
\bibitem{Guo:2014zva}
P.~Guo, T.~Yépez-Martínez, and A.~P. Szczepaniak,
  \ifthenelse{\boolean{articletitles}}{\emph{{Charmonium meson and hybrid
  radiative transitions}},
  }{}\href{http://dx.doi.org/10.1103/PhysRevD.89.116005}{Phys.\ Rev.\
  \textbf{D89} (2014), no.~11 116005},
  \href{http://arxiv.org/abs/1402.5863}{{\normalfont\ttfamily
  arXiv:1402.5863}}\relax
\mciteBstWouldAddEndPuncttrue
\mciteSetBstMidEndSepPunct{\mcitedefaultmidpunct}
{\mcitedefaultendpunct}{\mcitedefaultseppunct}\relax
\EndOfBibitem
\bibitem{Ablikim:2012ht}
BESIII, M.~Ablikim {\em et~al.},
  \ifthenelse{\boolean{articletitles}}{\emph{{Observation of $e^{+}e^{-} \to
  \eta J/\psi$ at center-of-mass energy $\sqrt{s}=4.009$ GeV}},
  }{}\href{http://dx.doi.org/10.1103/PhysRevD.86.071101}{Phys.\ Rev.\
  \textbf{D86} (2012) 071101},
  \href{http://arxiv.org/abs/1208.1857}{{\normalfont\ttfamily
  arXiv:1208.1857}}\relax
\mciteBstWouldAddEndPuncttrue
\mciteSetBstMidEndSepPunct{\mcitedefaultmidpunct}
{\mcitedefaultendpunct}{\mcitedefaultseppunct}\relax
\EndOfBibitem
\bibitem{Ding:2008gr}
G.-J. Ding, \ifthenelse{\boolean{articletitles}}{\emph{{Are Y(4260) and
  $Z^+_2(4250)$ $D_1D$ or $D_0D^*$ Hadronic Molecules?}},
  }{}\href{http://dx.doi.org/10.1103/PhysRevD.79.014001}{Phys.\ Rev.\
  \textbf{D79} (2009) 014001},
  \href{http://arxiv.org/abs/0809.4818}{{\normalfont\ttfamily
  arXiv:0809.4818}}\relax
\mciteBstWouldAddEndPuncttrue
\mciteSetBstMidEndSepPunct{\mcitedefaultmidpunct}
{\mcitedefaultendpunct}{\mcitedefaultseppunct}\relax
\EndOfBibitem
\bibitem{Wang:2013cya}
Q.~Wang, C.~Hanhart, and Q.~Zhao,
  \ifthenelse{\boolean{articletitles}}{\emph{{Decoding the riddle of $Y(4260)$
  and $Z_c(3900)$}},
  }{}\href{http://dx.doi.org/10.1103/PhysRevLett.111.132003}{Phys.\ Rev.\
  Lett.\  \textbf{111} (2013) 132003},
  \href{http://arxiv.org/abs/1303.6355}{{\normalfont\ttfamily
  arXiv:1303.6355}}\relax
\mciteBstWouldAddEndPuncttrue
\mciteSetBstMidEndSepPunct{\mcitedefaultmidpunct}
{\mcitedefaultendpunct}{\mcitedefaultseppunct}\relax
\EndOfBibitem
\bibitem{Zhu:2005hp}
S.-L. Zhu, \ifthenelse{\boolean{articletitles}}{\emph{{The Possible
  interpretations of Y(4260)}},
  }{}\href{http://dx.doi.org/10.1016/j.physletb.2005.08.068}{Phys.\ Lett.\
  \textbf{B625} (2005) 212},
  \href{http://arxiv.org/abs/hep-ph/0507025}{{\normalfont\ttfamily
  arXiv:hep-ph/0507025}}\relax
\mciteBstWouldAddEndPuncttrue
\mciteSetBstMidEndSepPunct{\mcitedefaultmidpunct}
{\mcitedefaultendpunct}{\mcitedefaultseppunct}\relax
\EndOfBibitem
\bibitem{Close:2005iz}
F.~E. Close and P.~R. Page, \ifthenelse{\boolean{articletitles}}{\emph{{Gluonic
  charmonium resonances at BaBar and Belle?}},
  }{}\href{http://dx.doi.org/10.1016/j.physletb.2005.09.016}{Phys.\ Lett.\
  \textbf{B628} (2005) 215},
  \href{http://arxiv.org/abs/hep-ph/0507199}{{\normalfont\ttfamily
  arXiv:hep-ph/0507199}}\relax
\mciteBstWouldAddEndPuncttrue
\mciteSetBstMidEndSepPunct{\mcitedefaultmidpunct}
{\mcitedefaultendpunct}{\mcitedefaultseppunct}\relax
\EndOfBibitem
\bibitem{Kou:2005gt}
E.~Kou and O.~Pene, \ifthenelse{\boolean{articletitles}}{\emph{{Suppressed
  decay into open charm for the Y(4260) being an hybrid}},
  }{}\href{http://dx.doi.org/10.1016/j.physletb.2005.09.013}{Phys.\ Lett.\
  \textbf{B631} (2005) 164},
  \href{http://arxiv.org/abs/hep-ph/0507119}{{\normalfont\ttfamily
  arXiv:hep-ph/0507119}}\relax
\mciteBstWouldAddEndPuncttrue
\mciteSetBstMidEndSepPunct{\mcitedefaultmidpunct}
{\mcitedefaultendpunct}{\mcitedefaultseppunct}\relax
\EndOfBibitem
\bibitem{Liu:2009ei}
X.~Liu and S.-L. Zhu, \ifthenelse{\boolean{articletitles}}{\emph{{$Y(4143)$ is
  probably a molecular partner of $Y(3930)$}},
  }{}\href{http://dx.doi.org/10.1103/PhysRevD.80.017502}{Phys.\ Rev.\
  \textbf{D80} (2009) 017502},
  \href{http://arxiv.org/abs/0903.2529}{{\normalfont\ttfamily
  arXiv:0903.2529}}\relax
\mciteBstWouldAddEndPuncttrue
\mciteSetBstMidEndSepPunct{\mcitedefaultmidpunct}
{\mcitedefaultendpunct}{\mcitedefaultseppunct}\relax
\EndOfBibitem
\bibitem{Branz:2009yt}
T.~Branz, T.~Gutsche, and V.~E. Lyubovitskij,
  \ifthenelse{\boolean{articletitles}}{\emph{{Hadronic molecule structure of
  the $Y(3940)$ and $Y(4140)$}},
  }{}\href{http://dx.doi.org/10.1103/PhysRevD.80.054019}{Phys.\ Rev.\
  \textbf{D80} (2009) 054019},
  \href{http://arxiv.org/abs/0903.5424}{{\normalfont\ttfamily
  arXiv:0903.5424}}\relax
\mciteBstWouldAddEndPuncttrue
\mciteSetBstMidEndSepPunct{\mcitedefaultmidpunct}
{\mcitedefaultendpunct}{\mcitedefaultseppunct}\relax
\EndOfBibitem
\bibitem{Albuquerque:2009ak}
R.~M. Albuquerque, M.~E. Bracco, and M.~Nielsen,
  \ifthenelse{\boolean{articletitles}}{\emph{{A QCD sum rule calculation for
  the $Y(4140)$ narrow structure}},
  }{}\href{http://dx.doi.org/10.1016/j.physletb.2009.06.022}{Phys.\ Lett.\
  \textbf{B678} (2009) 186},
  \href{http://arxiv.org/abs/0903.5540}{{\normalfont\ttfamily
  arXiv:0903.5540}}\relax
\mciteBstWouldAddEndPuncttrue
\mciteSetBstMidEndSepPunct{\mcitedefaultmidpunct}
{\mcitedefaultendpunct}{\mcitedefaultseppunct}\relax
\EndOfBibitem
\bibitem{Ding:2009vd}
G.-J. Ding, \ifthenelse{\boolean{articletitles}}{\emph{{Possible molecular
  states of $D^{*}_s\overline{D}^*_s$ system and $Y(4140)$}},
  }{}\href{http://dx.doi.org/10.1140/epjc/s10052-009-1146-4}{Eur.\ Phys.\ J.\
  \textbf{C64} (2009) 297},
  \href{http://arxiv.org/abs/0904.1782}{{\normalfont\ttfamily
  arXiv:0904.1782}}\relax
\mciteBstWouldAddEndPuncttrue
\mciteSetBstMidEndSepPunct{\mcitedefaultmidpunct}
{\mcitedefaultendpunct}{\mcitedefaultseppunct}\relax
\EndOfBibitem
\bibitem{Zhang:2009st}
J.-R. Zhang and M.-Q. Huang,
  \ifthenelse{\boolean{articletitles}}{\emph{{$(Q\bar{s})^{(*)}(\overline{Q}s)^{(*)}$
  molecular states from QCD sum rules: a view on $Y(4140)$}},
  }{}\href{http://dx.doi.org/10.1088/0954-3899/37/2/025005}{J.\ Phys.\
  \textbf{G37} (2010) 025005},
  \href{http://arxiv.org/abs/0905.4178}{{\normalfont\ttfamily
  arXiv:0905.4178}}\relax
\mciteBstWouldAddEndPuncttrue
\mciteSetBstMidEndSepPunct{\mcitedefaultmidpunct}
{\mcitedefaultendpunct}{\mcitedefaultseppunct}\relax
\EndOfBibitem
\bibitem{Liu:2009pu}
X.~Liu and H.-W. Ke, \ifthenelse{\boolean{articletitles}}{\emph{{The line shape
  of the radiative open-charm decay of $Y(4140)$ and $Y(3930)$}},
  }{}\href{http://dx.doi.org/10.1103/PhysRevD.80.034009}{Phys.\ Rev.\
  \textbf{D80} (2009) 034009},
  \href{http://arxiv.org/abs/0907.1349}{{\normalfont\ttfamily
  arXiv:0907.1349}}\relax
\mciteBstWouldAddEndPuncttrue
\mciteSetBstMidEndSepPunct{\mcitedefaultmidpunct}
{\mcitedefaultendpunct}{\mcitedefaultseppunct}\relax
\EndOfBibitem
\bibitem{Wang:2009ry}
Z.-G. Wang, Z.-C. Liu, and X.-H. Zhang,
  \ifthenelse{\boolean{articletitles}}{\emph{{Analysis of the $Y(4140)$ and
  related molecular states with QCD sum rules}},
  }{}\href{http://dx.doi.org/10.1140/epjc/s10052-009-1156-2}{Eur.\ Phys.\ J.\
  \textbf{C64} (2009) 373},
  \href{http://arxiv.org/abs/0907.1467}{{\normalfont\ttfamily
  arXiv:0907.1467}}\relax
\mciteBstWouldAddEndPuncttrue
\mciteSetBstMidEndSepPunct{\mcitedefaultmidpunct}
{\mcitedefaultendpunct}{\mcitedefaultseppunct}\relax
\EndOfBibitem
\bibitem{Molina:2009ct}
R.~Molina and E.~Oset, \ifthenelse{\boolean{articletitles}}{\emph{{The Y(3940),
  Z(3930) and the X(4160) as dynamically generated resonances from the
  vector-vector interaction}},
  }{}\href{http://dx.doi.org/10.1103/PhysRevD.80.114013}{Phys.\ Rev.\
  \textbf{D80} (2009) 114013},
  \href{http://arxiv.org/abs/0907.3043}{{\normalfont\ttfamily
  arXiv:0907.3043}}\relax
\mciteBstWouldAddEndPuncttrue
\mciteSetBstMidEndSepPunct{\mcitedefaultmidpunct}
{\mcitedefaultendpunct}{\mcitedefaultseppunct}\relax
\EndOfBibitem
\bibitem{Chen:2015fdn}
X.~Chen, X.~Lu, R.~Shi, and X.~Guo,
  \ifthenelse{\boolean{articletitles}}{\emph{{Mass of $Y(4140)$ in
  Bethe--Salpeter equation for quarks}},
  }{}\href{http://arxiv.org/abs/1512.06483}{{\normalfont\ttfamily
  arXiv:1512.06483}}\relax
\mciteBstWouldAddEndPuncttrue
\mciteSetBstMidEndSepPunct{\mcitedefaultmidpunct}
{\mcitedefaultendpunct}{\mcitedefaultseppunct}\relax
\EndOfBibitem
\bibitem{Karliner:2016ith}
M.~Karliner and J.~L. Rosner,
  \ifthenelse{\boolean{articletitles}}{\emph{{Exotic resonances due to $\eta$
  exchange}},
  }{}\href{http://dx.doi.org/10.1016/j.nuclphysa.2016.03.057}{Nucl.\ Phys.\
  \textbf{A954} (2016) 365},
  \href{http://arxiv.org/abs/1601.00565}{{\normalfont\ttfamily
  arXiv:1601.00565}}\relax
\mciteBstWouldAddEndPuncttrue
\mciteSetBstMidEndSepPunct{\mcitedefaultmidpunct}
{\mcitedefaultendpunct}{\mcitedefaultseppunct}\relax
\EndOfBibitem
\bibitem{Stancu:2009ka}
F.~Stancu, \ifthenelse{\boolean{articletitles}}{\emph{{Can $Y(4140)$ be a $c
  \bar c s \bar s$ tetraquark?}},
  }{}\href{http://dx.doi.org/10.1088/0954-3899/37/7/075017}{J.\ Phys.\
  \textbf{G37} (2010) 075017},
  \href{http://arxiv.org/abs/0906.2485}{{\normalfont\ttfamily
  arXiv:0906.2485}}\relax
\mciteBstWouldAddEndPuncttrue
\mciteSetBstMidEndSepPunct{\mcitedefaultmidpunct}
{\mcitedefaultendpunct}{\mcitedefaultseppunct}\relax
\EndOfBibitem
\bibitem{Drenska:2009cd}
N.~V. Drenska, R.~Faccini, and A.~D. Polosa,
  \ifthenelse{\boolean{articletitles}}{\emph{{Exotic hadrons with hidden charm
  and strangeness}},
  }{}\href{http://dx.doi.org/10.1103/PhysRevD.79.077502}{Phys.\ Rev.\
  \textbf{D79} (2009) 077502},
  \href{http://arxiv.org/abs/0902.2803}{{\normalfont\ttfamily
  arXiv:0902.2803}}\relax
\mciteBstWouldAddEndPuncttrue
\mciteSetBstMidEndSepPunct{\mcitedefaultmidpunct}
{\mcitedefaultendpunct}{\mcitedefaultseppunct}\relax
\EndOfBibitem
\bibitem{Wang:2015pea}
Z.-G. Wang and Y.-F. Tian,
  \ifthenelse{\boolean{articletitles}}{\emph{{Tetraquark state candidates:
  $Y(4140)$, $Y(4274)$ and $X(4350)$}},
  }{}\href{http://dx.doi.org/10.1142/S0217751X15500049}{Int.\ J.\ Mod.\ Phys.\
  \textbf{A30} (2015) 1550004},
  \href{http://arxiv.org/abs/1502.04619}{{\normalfont\ttfamily
  arXiv:1502.04619}}\relax
\mciteBstWouldAddEndPuncttrue
\mciteSetBstMidEndSepPunct{\mcitedefaultmidpunct}
{\mcitedefaultendpunct}{\mcitedefaultseppunct}\relax
\EndOfBibitem
\bibitem{Anisovich:2015caa}
V.~V. Anisovich, M.~A. Matveev, A.~V. Sarantsev, and A.~N. Semenova,
  \ifthenelse{\boolean{articletitles}}{\emph{{Exotic mesons with hidden charm
  as diquark-antidiquark states}},
  }{}\href{http://dx.doi.org/10.1142/S0217751X15501869}{Int.\ J.\ Mod.\ Phys.\
  \textbf{A30} (2015) 1550186},
  \href{http://arxiv.org/abs/1507.07232}{{\normalfont\ttfamily
  arXiv:1507.07232}}\relax
\mciteBstWouldAddEndPuncttrue
\mciteSetBstMidEndSepPunct{\mcitedefaultmidpunct}
{\mcitedefaultendpunct}{\mcitedefaultseppunct}\relax
\EndOfBibitem
\bibitem{Mahajan:2009pj}
N.~Mahajan, \ifthenelse{\boolean{articletitles}}{\emph{{$Y(4140)$: possible
  options}}, }{}\href{http://dx.doi.org/10.1016/j.physletb.2009.07.043}{Phys.\
  Lett.\  \textbf{B679} (2009) 228},
  \href{http://arxiv.org/abs/0903.3107}{{\normalfont\ttfamily
  arXiv:0903.3107}}\relax
\mciteBstWouldAddEndPuncttrue
\mciteSetBstMidEndSepPunct{\mcitedefaultmidpunct}
{\mcitedefaultendpunct}{\mcitedefaultseppunct}\relax
\EndOfBibitem
\bibitem{Wang:2009ue}
Z.-G. Wang, \ifthenelse{\boolean{articletitles}}{\emph{{Analysis of the
  $Y(4140)$ with QCD sum rules}},
  }{}\href{http://dx.doi.org/10.1140/epjc/s10052-009-1097-9}{Eur.\ Phys.\ J.\
  \textbf{C63} (2009) 115},
  \href{http://arxiv.org/abs/0903.5200}{{\normalfont\ttfamily
  arXiv:0903.5200}}\relax
\mciteBstWouldAddEndPuncttrue
\mciteSetBstMidEndSepPunct{\mcitedefaultmidpunct}
{\mcitedefaultendpunct}{\mcitedefaultseppunct}\relax
\EndOfBibitem
\bibitem{Liu:2009iw}
X.~Liu, \ifthenelse{\boolean{articletitles}}{\emph{{The hidden charm decay of
  $Y(4140)$ by the rescattering mechanism}},
  }{}\href{http://dx.doi.org/10.1016/j.physletb.2009.08.049}{Phys.\ Lett.\
  \textbf{B680} (2009) 137},
  \href{http://arxiv.org/abs/0904.0136}{{\normalfont\ttfamily
  arXiv:0904.0136}}\relax
\mciteBstWouldAddEndPuncttrue
\mciteSetBstMidEndSepPunct{\mcitedefaultmidpunct}
{\mcitedefaultendpunct}{\mcitedefaultseppunct}\relax
\EndOfBibitem
\bibitem{Aaij:2012pz}
LHCb, R.~Aaij {\em et~al.}, \ifthenelse{\boolean{articletitles}}{\emph{{Search
  for the $X(4140)$ state in $B^+ \to J/\psi \phi K^+$ decays}},
  }{}\href{http://dx.doi.org/10.1103/PhysRevD.85.091103}{Phys.\ Rev.\
  \textbf{D85} (2012) 091103},
  \href{http://arxiv.org/abs/1202.5087}{{\normalfont\ttfamily
  arXiv:1202.5087}}\relax
\mciteBstWouldAddEndPuncttrue
\mciteSetBstMidEndSepPunct{\mcitedefaultmidpunct}
{\mcitedefaultendpunct}{\mcitedefaultseppunct}\relax
\EndOfBibitem
\bibitem{Brodzicka:2010zz}
J.~Brodzicka, \ifthenelse{\boolean{articletitles}}{\emph{{Heavy flavour
  spectroscopy}},
  }{}\href{http://dx.doi.org/10.3204/DESY-PROC-2010-04/38}{Conf.\ Proc.\
  \textbf{C0908171} (2009) 299}, Proceedings, 24th International Symposium on
  Lepton-Photon Interactions at High Energy (LP09), 2009.\relax
\mciteBstWouldAddEndPunctfalse
\mciteSetBstMidEndSepPunct{\mcitedefaultmidpunct}
{}{\mcitedefaultseppunct}\relax
\EndOfBibitem
\bibitem{ChengPing:2009vu}
C.-P. Shen, \ifthenelse{\boolean{articletitles}}{\emph{{XYZ particles at
  Belle}}, }{}\href{http://dx.doi.org/10.1088/1674-1137/34/6/001}{Chin.\ Phys.\
   \textbf{C34} (2010) 615},
  \href{http://arxiv.org/abs/0912.2386}{{\normalfont\ttfamily
  arXiv:0912.2386}}, {Proceedings, 6th International Workshop on $e^+e^-$
  Collisions from Phi to Psi (PHIPSI09), 2009.}\relax
\mciteBstWouldAddEndPunctfalse
\mciteSetBstMidEndSepPunct{\mcitedefaultmidpunct}
{}{\mcitedefaultseppunct}\relax
\EndOfBibitem
\bibitem{Lees:2014lra}
BaBar, J.~P. Lees {\em et~al.},
  \ifthenelse{\boolean{articletitles}}{\emph{{Study of $B^{\pm,0} \to \jpsi K^+
  K^- K^{\pm,0}$ and search for $B^0 \to \jpsi\phi$ at BABAR}},
  }{}\href{http://dx.doi.org/10.1103/PhysRevD.91.012003}{Phys.\ Rev.\
  \textbf{D91} (2015) 012003},
  \href{http://arxiv.org/abs/1407.7244}{{\normalfont\ttfamily
  arXiv:1407.7244}}\relax
\mciteBstWouldAddEndPuncttrue
\mciteSetBstMidEndSepPunct{\mcitedefaultmidpunct}
{\mcitedefaultendpunct}{\mcitedefaultseppunct}\relax
\EndOfBibitem
\bibitem{Ablikim:2014atq}
BES-III, M.~Ablikim {\em et~al.},
  \ifthenelse{\boolean{articletitles}}{\emph{{Search for the $Y(4140)$ via
  $e^+e^-\to\gamma\jpsi\phi$ at $\sqrt{s}$=4.23 , 4.26 and 4.36 GeV}},
  }{}\href{http://dx.doi.org/10.1103/PhysRevD.91.032002}{Phys.\ Rev.\
  \textbf{D91} (2015) 032002},
  \href{http://arxiv.org/abs/1412.1867}{{\normalfont\ttfamily
  arXiv:1412.1867}}\relax
\mciteBstWouldAddEndPuncttrue
\mciteSetBstMidEndSepPunct{\mcitedefaultmidpunct}
{\mcitedefaultendpunct}{\mcitedefaultseppunct}\relax
\EndOfBibitem
\bibitem{He:2011ed}
J.~He and X.~Liu, \ifthenelse{\boolean{articletitles}}{\emph{{The line shape of
  the open-charm radiative and pionic decays of $Y(4274)$ as a molecular
  charmonium}}, }{}\href{http://arxiv.org/abs/1102.1127}{{\normalfont\ttfamily
  arXiv:1102.1127}}\relax
\mciteBstWouldAddEndPuncttrue
\mciteSetBstMidEndSepPunct{\mcitedefaultmidpunct}
{\mcitedefaultendpunct}{\mcitedefaultseppunct}\relax
\EndOfBibitem
\bibitem{Finazzo:2011he}
S.~I. Finazzo, M.~Nielsen, and X.~Liu,
  \ifthenelse{\boolean{articletitles}}{\emph{{QCD sum rule calculation for the
  charmonium-like structures in the $\jpsi \phi$ and $\jpsi \omega$ invariant
  mass spectra}},
  }{}\href{http://dx.doi.org/10.1016/j.physletb.2011.05.042}{Phys.\ Lett.\
  \textbf{B701} (2011) 101},
  \href{http://arxiv.org/abs/1102.2347}{{\normalfont\ttfamily
  arXiv:1102.2347}}\relax
\mciteBstWouldAddEndPuncttrue
\mciteSetBstMidEndSepPunct{\mcitedefaultmidpunct}
{\mcitedefaultendpunct}{\mcitedefaultseppunct}\relax
\EndOfBibitem
\bibitem{Maiani:2016wlq}
L.~Maiani, A.~D. Polosa, and V.~Riquer,
  \ifthenelse{\boolean{articletitles}}{\emph{{Interpretation of Axial
  Resonances in $J/\psi\phi$ at LHCb}},
  }{}\href{http://dx.doi.org/10.1103/PhysRevD.94.054026}{Phys.\ Rev.\
  \textbf{D94} (2016) 054026},
  \href{http://arxiv.org/abs/1607.02405}{{\normalfont\ttfamily
  arXiv:1607.02405}}\relax
\mciteBstWouldAddEndPuncttrue
\mciteSetBstMidEndSepPunct{\mcitedefaultmidpunct}
{\mcitedefaultendpunct}{\mcitedefaultseppunct}\relax
\EndOfBibitem
\bibitem{Chen:2016iua}
D.-Y. Chen, \ifthenelse{\boolean{articletitles}}{\emph{{Where are $\chi
  _{cJ}(3P)$ ?}},
  }{}\href{http://dx.doi.org/10.1140/epjc/s10052-016-4531-9}{Eur.\ Phys.\ J.\
  \textbf{C76} (2016) 671},
  \href{http://arxiv.org/abs/1611.00109}{{\normalfont\ttfamily
  arXiv:1611.00109}}\relax
\mciteBstWouldAddEndPuncttrue
\mciteSetBstMidEndSepPunct{\mcitedefaultmidpunct}
{\mcitedefaultendpunct}{\mcitedefaultseppunct}\relax
\EndOfBibitem
\bibitem{Li:2009ad}
B.-Q. Li, C.~Meng, and K.-T. Chao,
  \ifthenelse{\boolean{articletitles}}{\emph{{Coupled-Channel and Screening
  Effects in Charmonium Spectrum}},
  }{}\href{http://dx.doi.org/10.1103/PhysRevD.80.014012}{Phys.\ Rev.\
  \textbf{D80} (2009) 014012},
  \href{http://arxiv.org/abs/0904.4068}{{\normalfont\ttfamily
  arXiv:0904.4068}}\relax
\mciteBstWouldAddEndPuncttrue
\mciteSetBstMidEndSepPunct{\mcitedefaultmidpunct}
{\mcitedefaultendpunct}{\mcitedefaultseppunct}\relax
\EndOfBibitem
\bibitem{Lu:2016cwr}
Q.-F. Lu and Y.-B. Dong, \ifthenelse{\boolean{articletitles}}{\emph{{X(4140) ,
  X(4274) , X(4500) , and X(4700) in the relativized quark model}},
  }{}\href{http://dx.doi.org/10.1103/PhysRevD.94.074007}{Phys.\ Rev.\
  \textbf{D94} (2016) 074007},
  \href{http://arxiv.org/abs/1607.05570}{{\normalfont\ttfamily
  arXiv:1607.05570}}\relax
\mciteBstWouldAddEndPuncttrue
\mciteSetBstMidEndSepPunct{\mcitedefaultmidpunct}
{\mcitedefaultendpunct}{\mcitedefaultseppunct}\relax
\EndOfBibitem
\bibitem{Ortega:2016hde}
P.~G. Ortega, J.~Segovia, D.~R. Entem, and F.~Fernández,
  \ifthenelse{\boolean{articletitles}}{\emph{{Canonical description of the new
  LHCb resonances}},
  }{}\href{http://dx.doi.org/10.1103/PhysRevD.94.114018}{Phys.\ Rev.\
  \textbf{D94} (2016) 114018},
  \href{http://arxiv.org/abs/1608.01325}{{\normalfont\ttfamily
  arXiv:1608.01325}}\relax
\mciteBstWouldAddEndPuncttrue
\mciteSetBstMidEndSepPunct{\mcitedefaultmidpunct}
{\mcitedefaultendpunct}{\mcitedefaultseppunct}\relax
\EndOfBibitem
\bibitem{Chao:2007it}
K.-T. Chao, \ifthenelse{\boolean{articletitles}}{\emph{{Interpretations for the
  $X(4160)$ observed in the double charm production at B factories}},
  }{}\href{http://dx.doi.org/10.1016/j.physletb.2008.02.039}{Phys.\ Lett.\
  \textbf{B661} (2008) 348},
  \href{http://arxiv.org/abs/0707.3982}{{\normalfont\ttfamily
  arXiv:0707.3982}}\relax
\mciteBstWouldAddEndPuncttrue
\mciteSetBstMidEndSepPunct{\mcitedefaultmidpunct}
{\mcitedefaultendpunct}{\mcitedefaultseppunct}\relax
\EndOfBibitem
\bibitem{Aubert:2008aa}
BaBar, B.~Aubert {\em et~al.},
  \ifthenelse{\boolean{articletitles}}{\emph{{Search for the $Z(4430)^-$ at
  BaBar}}, }{}\href{http://dx.doi.org/10.1103/PhysRevD.79.112001}{Phys.\ Rev.\
  \textbf{D79} (2009) 112001},
  \href{http://arxiv.org/abs/0811.0564}{{\normalfont\ttfamily
  arXiv:0811.0564}}\relax
\mciteBstWouldAddEndPuncttrue
\mciteSetBstMidEndSepPunct{\mcitedefaultmidpunct}
{\mcitedefaultendpunct}{\mcitedefaultseppunct}\relax
\EndOfBibitem
\bibitem{Pakhlov:2014qva}
P.~Pakhlov and T.~Uglov, \ifthenelse{\boolean{articletitles}}{\emph{{Charged
  charmonium-like Z$^+$(4430) from rescattering in conventional $B$ decays}},
  }{}\href{http://dx.doi.org/10.1016/j.physletb.2015.06.074}{Phys.\ Lett.\
  \textbf{B748} (2015) 183},
  \href{http://arxiv.org/abs/1408.5295}{{\normalfont\ttfamily
  arXiv:1408.5295}}\relax
\mciteBstWouldAddEndPuncttrue
\mciteSetBstMidEndSepPunct{\mcitedefaultmidpunct}
{\mcitedefaultendpunct}{\mcitedefaultseppunct}\relax
\EndOfBibitem
\bibitem{Hou:2006it}
W.-S. Hou, \ifthenelse{\boolean{articletitles}}{\emph{{Searching for the bottom
  counterparts of $X(3872) $ and $Y(4260)$ via $\pi^{+} \pi^{-} \Upsilon$}},
  }{}\href{http://dx.doi.org/10.1103/PhysRevD.74.017504}{Phys.\ Rev.\
  \textbf{D74} (2006) 017504},
  \href{http://arxiv.org/abs/hep-ph/0606016}{{\normalfont\ttfamily
  arXiv:hep-ph/0606016}}\relax
\mciteBstWouldAddEndPuncttrue
\mciteSetBstMidEndSepPunct{\mcitedefaultmidpunct}
{\mcitedefaultendpunct}{\mcitedefaultseppunct}\relax
\EndOfBibitem
\bibitem{Chen:2008xia}
Belle, K.-F. Chen {\em et~al.},
  \ifthenelse{\boolean{articletitles}}{\emph{{Observation of an enhancement in
  $e^+e^- \to \Upsilon(1S)\pi^+ \pi^-$, $\Upsilon(2S)\pi^+ \pi^-$, and
  $\Upsilon(3S)\pi^+ \pi^-$ production around $\sqrt{s}=10.89$ GeV at Belle}},
  }{}\href{http://dx.doi.org/10.1103/PhysRevD.82.091106}{Phys.\ Rev.\
  \textbf{D82} (2010) 091106},
  \href{http://arxiv.org/abs/0810.3829}{{\normalfont\ttfamily
  arXiv:0810.3829}}\relax
\mciteBstWouldAddEndPuncttrue
\mciteSetBstMidEndSepPunct{\mcitedefaultmidpunct}
{\mcitedefaultendpunct}{\mcitedefaultseppunct}\relax
\EndOfBibitem
\bibitem{Ali:2009pi}
A.~Ali, C.~Hambrock, I.~Ahmed, and M.~J. Aslam,
  \ifthenelse{\boolean{articletitles}}{\emph{{A case for hidden $b\bar{b}$
  tetraquarks based on $e^+e^- \to b\bar{b}$ cross section between
  $\sqrt{s}=10.54$ and 11.20 GeV}},
  }{}\href{http://dx.doi.org/10.1016/j.physletb.2009.12.053}{Phys.\ Lett.\
  \textbf{B684} (2010) 28},
  \href{http://arxiv.org/abs/0911.2787}{{\normalfont\ttfamily
  arXiv:0911.2787}}\relax
\mciteBstWouldAddEndPuncttrue
\mciteSetBstMidEndSepPunct{\mcitedefaultmidpunct}
{\mcitedefaultendpunct}{\mcitedefaultseppunct}\relax
\EndOfBibitem
\bibitem{Adachi:2011ji}
Belle, I.~Adachi {\em et~al.},
  \ifthenelse{\boolean{articletitles}}{\emph{{First observation of the $P$-wave
  spin-singlet bottomonium states $h_b(1P)$ and $h_b(2P)$}},
  }{}\href{http://dx.doi.org/10.1103/PhysRevLett.108.032001}{Phys.\ Rev.\
  Lett.\  \textbf{108} (2012) 032001},
  \href{http://arxiv.org/abs/1103.3419}{{\normalfont\ttfamily
  arXiv:1103.3419}}\relax
\mciteBstWouldAddEndPuncttrue
\mciteSetBstMidEndSepPunct{\mcitedefaultmidpunct}
{\mcitedefaultendpunct}{\mcitedefaultseppunct}\relax
\EndOfBibitem
\bibitem{Bondar:2011ev}
A.~E. Bondar {\em et~al.}, \ifthenelse{\boolean{articletitles}}{\emph{{Heavy
  quark spin structure in $Z_b$ resonances}},
  }{}\href{http://dx.doi.org/10.1103/PhysRevD.84.054010}{Phys.\ Rev.\
  \textbf{D84} (2011) 054010},
  \href{http://arxiv.org/abs/1105.4473}{{\normalfont\ttfamily
  arXiv:1105.4473}}\relax
\mciteBstWouldAddEndPuncttrue
\mciteSetBstMidEndSepPunct{\mcitedefaultmidpunct}
{\mcitedefaultendpunct}{\mcitedefaultseppunct}\relax
\EndOfBibitem
\bibitem{Karliner:2015ina}
M.~Karliner and J.~L. Rosner, \ifthenelse{\boolean{articletitles}}{\emph{{New
  Exotic Meson and Baryon Resonances from Doubly-Heavy Hadronic Molecules}},
  }{}\href{http://dx.doi.org/10.1103/PhysRevLett.115.122001}{Phys.\ Rev.\
  Lett.\  \textbf{115} (2015) 122001},
  \href{http://arxiv.org/abs/1506.06386}{{\normalfont\ttfamily
  arXiv:1506.06386}}\relax
\mciteBstWouldAddEndPuncttrue
\mciteSetBstMidEndSepPunct{\mcitedefaultmidpunct}
{\mcitedefaultendpunct}{\mcitedefaultseppunct}\relax
\EndOfBibitem
\bibitem{Liu:2008fh}
Y.-R. Liu, X.~Liu, W.-Z. Deng, and S.-L. Zhu,
  \ifthenelse{\boolean{articletitles}}{\emph{{Is $X(3872) $ Really a Molecular
  State?}}, }{}\href{http://dx.doi.org/10.1140/epjc/s10052-008-0640-4}{Eur.\
  Phys.\ J.\  \textbf{C56} (2008) 63},
  \href{http://arxiv.org/abs/0801.3540}{{\normalfont\ttfamily
  arXiv:0801.3540}}\relax
\mciteBstWouldAddEndPuncttrue
\mciteSetBstMidEndSepPunct{\mcitedefaultmidpunct}
{\mcitedefaultendpunct}{\mcitedefaultseppunct}\relax
\EndOfBibitem
\bibitem{Liu:2008tn}
X.~Liu, Z.-G. Luo, Y.-R. Liu, and S.-L. Zhu,
  \ifthenelse{\boolean{articletitles}}{\emph{{X(3872) and Other Possible Heavy
  Molecular States}},
  }{}\href{http://dx.doi.org/10.1140/epjc/s10052-009-1020-4}{Eur.\ Phys.\ J.\
  \textbf{C61} (2009) 411},
  \href{http://arxiv.org/abs/0808.0073}{{\normalfont\ttfamily
  arXiv:0808.0073}}\relax
\mciteBstWouldAddEndPuncttrue
\mciteSetBstMidEndSepPunct{\mcitedefaultmidpunct}
{\mcitedefaultendpunct}{\mcitedefaultseppunct}\relax
\EndOfBibitem
\bibitem{Ali:2014dva}
A.~Ali, L.~Maiani, A.~D. Polosa, and V.~Riquer,
  \ifthenelse{\boolean{articletitles}}{\emph{{Hidden-Beauty Charged Tetraquarks
  and Heavy Quark Spin Conservation}},
  }{}\href{http://dx.doi.org/10.1103/PhysRevD.91.017502}{Phys.\ Rev.\
  \textbf{D91} (2015), no.~1 017502},
  \href{http://arxiv.org/abs/1412.2049}{{\normalfont\ttfamily
  arXiv:1412.2049}}\relax
\mciteBstWouldAddEndPuncttrue
\mciteSetBstMidEndSepPunct{\mcitedefaultmidpunct}
{\mcitedefaultendpunct}{\mcitedefaultseppunct}\relax
\EndOfBibitem
\bibitem{Karliner:2008rc}
M.~Karliner and H.~J. Lipkin,
  \ifthenelse{\boolean{articletitles}}{\emph{{Possibility of Exotic States in
  the Upsilon system}},
  }{}\href{http://arxiv.org/abs/0802.0649}{{\normalfont\ttfamily
  arXiv:0802.0649}}\relax
\mciteBstWouldAddEndPuncttrue
\mciteSetBstMidEndSepPunct{\mcitedefaultmidpunct}
{\mcitedefaultendpunct}{\mcitedefaultseppunct}\relax
\EndOfBibitem
\bibitem{Ablikim:2015swa}
BESIII, M.~Ablikim {\em et~al.},
  \ifthenelse{\boolean{articletitles}}{\emph{{Confirmation of a charged
  charmoniumlike state $Z_c(3885)^{\mp}$ in
  $e^+e^-\to\pi^{\pm}(D\bar{D}^*)^\mp$ with double $D$ tag}},
  }{}\href{http://dx.doi.org/10.1103/PhysRevD.92.092006}{Phys.\ Rev.\
  \textbf{D92} (2015) 092006},
  \href{http://arxiv.org/abs/1509.01398}{{\normalfont\ttfamily
  arXiv:1509.01398}}\relax
\mciteBstWouldAddEndPuncttrue
\mciteSetBstMidEndSepPunct{\mcitedefaultmidpunct}
{\mcitedefaultendpunct}{\mcitedefaultseppunct}\relax
\EndOfBibitem
\bibitem{Ablikim:2017oaf}
BESIII, M.~Ablikim {\em et~al.},
  \ifthenelse{\boolean{articletitles}}{\emph{{Measurement of
  $e^{+}e^{-}\rightarrow \pi^{+}\pi^{-}\psi(3686)$ from 4.008 to 4.600~GeV and
  observation of a charged structure in the $\pi^{\pm}\psi(3686)$ mass
  spectrum}}, }{}\href{http://arxiv.org/abs/1703.08787}{{\normalfont\ttfamily
  arXiv:1703.08787}}\relax
\mciteBstWouldAddEndPuncttrue
\mciteSetBstMidEndSepPunct{\mcitedefaultmidpunct}
{\mcitedefaultendpunct}{\mcitedefaultseppunct}\relax
\EndOfBibitem
\bibitem{Esposito:2014hsa}
A.~Esposito, A.~L. Guerrieri, and A.~Pilloni,
  \ifthenelse{\boolean{articletitles}}{\emph{{Probing the nature of $Z_c^{(')}$
  states via the $?_c?$ decay}},
  }{}\href{http://dx.doi.org/10.1016/j.physletb.2015.04.057}{Phys.\ Lett.\
  \textbf{B746} (2015) 194},
  \href{http://arxiv.org/abs/1409.3551}{{\normalfont\ttfamily
  arXiv:1409.3551}}\relax
\mciteBstWouldAddEndPuncttrue
\mciteSetBstMidEndSepPunct{\mcitedefaultmidpunct}
{\mcitedefaultendpunct}{\mcitedefaultseppunct}\relax
\EndOfBibitem
\bibitem{Jurik:2016bdm}
N.~P. Jurik, {\em {Observation of $J/\psi$ p resonances consistent with
  pentaquark states in $\Lambda_ b^0\to J/\psi K^-p$ decays}}, PhD thesis,
  Syracuse U., 2016-08-08\relax
\mciteBstWouldAddEndPuncttrue
\mciteSetBstMidEndSepPunct{\mcitedefaultmidpunct}
{\mcitedefaultendpunct}{\mcitedefaultseppunct}\relax
\EndOfBibitem
\bibitem{Fernandez-Ramirez:2015tfa}
C.~Fernandez-Ramirez {\em et~al.},
  \ifthenelse{\boolean{articletitles}}{\emph{{Coupled-channel model for
  $\bar{K}N$ scattering in the resonant region}},
  }{}\href{http://dx.doi.org/10.1103/PhysRevD.93.034029}{Phys.\ Rev.\
  \textbf{D93} (2016) 034029},
  \href{http://arxiv.org/abs/1510.07065}{{\normalfont\ttfamily
  arXiv:1510.07065}}\relax
\mciteBstWouldAddEndPuncttrue
\mciteSetBstMidEndSepPunct{\mcitedefaultmidpunct}
{\mcitedefaultendpunct}{\mcitedefaultseppunct}\relax
\EndOfBibitem
\bibitem{SPECFG}
R.~N. Faustov and V.~O. Galkin, \ifthenelse{\boolean{articletitles}}{\emph{{
  Strange baryon spectroscopy in the relativistic quark model}},
  }{}\href{http://dx.doi.org/10.1103/PhysRevD.92.054005}{Phys.\ Rev.\
  \textbf{D92} (2015) 054005},
  \href{http://arxiv.org/abs/1507.04530}{{\normalfont\ttfamily
  arXiv:1507.04530}}\relax
\mciteBstWouldAddEndPuncttrue
\mciteSetBstMidEndSepPunct{\mcitedefaultmidpunct}
{\mcitedefaultendpunct}{\mcitedefaultseppunct}\relax
\EndOfBibitem
\bibitem{SPECCI}
S.~Capstick and N.~Isgur, \ifthenelse{\boolean{articletitles}}{\emph{{Baryons
  in a relativized quark model with chromodynamics}},
  }{}\href{http://dx.doi.org/10.1103/PhysRevD.34.2809}{Phys.\ Rev.\
  \textbf{D34} (1986) 2809}\relax
\mciteBstWouldAddEndPuncttrue
\mciteSetBstMidEndSepPunct{\mcitedefaultmidpunct}
{\mcitedefaultendpunct}{\mcitedefaultseppunct}\relax
\EndOfBibitem
\bibitem{SPECLMP}
U.~Loring, B.~C. Metsch, and H.~R. Petry,
  \ifthenelse{\boolean{articletitles}}{\emph{{The Light baryon spectrum in a
  relativistic quark model with instanton induced quark forces: The Strange
  baryon spectrum}}, }{}\href{http://dx.doi.org/10.1007/s100500170106}{Eur.\
  Phys.\ J.\  \textbf{A10} (2001) 447},
  \href{http://arxiv.org/abs/hep-ph/0103290}{{\normalfont\ttfamily
  arXiv:hep-ph/0103290}}\relax
\mciteBstWouldAddEndPuncttrue
\mciteSetBstMidEndSepPunct{\mcitedefaultmidpunct}
{\mcitedefaultendpunct}{\mcitedefaultseppunct}\relax
\EndOfBibitem
\bibitem{SPECMPS}
T.~Melde, W.~Plessas, and B.~Sengl,
  \ifthenelse{\boolean{articletitles}}{\emph{{Quark-model identification of
  baryon ground and resonant states}},
  }{}\href{http://dx.doi.org/10.1103/PhysRevD.77.114002}{Phys.\ Rev.\
  \textbf{D77} (2008) 114002},
  \href{http://arxiv.org/abs/0806.1454}{{\normalfont\ttfamily
  arXiv:0806.1454}}\relax
\mciteBstWouldAddEndPuncttrue
\mciteSetBstMidEndSepPunct{\mcitedefaultmidpunct}
{\mcitedefaultendpunct}{\mcitedefaultseppunct}\relax
\EndOfBibitem
\bibitem{SPECSF}
E.~Santopinto and J.~Ferretti,
  \ifthenelse{\boolean{articletitles}}{\emph{{Strange and nonstrange baryon
  spectra in the relativistic interacting quark-diquark model with a G\"ursey
  and Radicati-inspired exchange interaction}},
  }{}\href{http://dx.doi.org/10.1103/PhysRevC.92.025202}{Phys.\ Rev.\
  \textbf{C92} (2015) 025202},
  \href{http://arxiv.org/abs/1412.7571}{{\normalfont\ttfamily
  arXiv:1412.7571}}\relax
\mciteBstWouldAddEndPuncttrue
\mciteSetBstMidEndSepPunct{\mcitedefaultmidpunct}
{\mcitedefaultendpunct}{\mcitedefaultseppunct}\relax
\EndOfBibitem
\bibitem{SPECELMS}
G.~P. Engel, C.~B. Lang, D.~Mohler, and A.~Schaefer,
  \ifthenelse{\boolean{articletitles}}{\emph{{ QCD with two light dynamical
  chirally improved quarks: baryons}},
  }{}\href{http://dx.doi.org/10.1103/PhysRevD.87.074504}{Phys.\ Rev.\
  \textbf{D87} (2013) 074504},
  \href{http://arxiv.org/abs/1301.4318}{{\normalfont\ttfamily
  arXiv:1301.4318}}\relax
\mciteBstWouldAddEndPuncttrue
\mciteSetBstMidEndSepPunct{\mcitedefaultmidpunct}
{\mcitedefaultendpunct}{\mcitedefaultseppunct}\relax
\EndOfBibitem
\bibitem{Aaij:2016phn}
LHCb, R.~Aaij {\em et~al.},
  \ifthenelse{\boolean{articletitles}}{\emph{{Model-independent evidence for
  $J/\psi p$ contributions to $\Lambda_b^0\to J/\psi p K^-$ decays}},
  }{}\href{http://dx.doi.org/10.1103/PhysRevLett.117.082002}{Phys.\ Rev.\
  Lett.\  \textbf{117} (2016) 082002},
  \href{http://arxiv.org/abs/1604.05708}{{\normalfont\ttfamily
  arXiv:1604.05708}}\relax
\mciteBstWouldAddEndPuncttrue
\mciteSetBstMidEndSepPunct{\mcitedefaultmidpunct}
{\mcitedefaultendpunct}{\mcitedefaultseppunct}\relax
\EndOfBibitem
\bibitem{Aaij:2016ymb}
LHCb, R.~Aaij {\em et~al.},
  \ifthenelse{\boolean{articletitles}}{\emph{{Evidence for exotic hadron
  contributions to $\Lambda_b^0 \to J/\psi p \pi^-$ decays}},
  }{}\href{http://dx.doi.org/10.1103/PhysRevLett.117.082003,
  10.1103/PhysRevLett.117.109902}{Phys.\ Rev.\ Lett.\  \textbf{117} (2016)
  082003}, \href{http://arxiv.org/abs/1606.06999}{{\normalfont\ttfamily
  arXiv:1606.06999}}, [Addendum: Phys. Rev. Lett.117,no.10,109902(2016)]\relax
\mciteBstWouldAddEndPuncttrue
\mciteSetBstMidEndSepPunct{\mcitedefaultmidpunct}
{\mcitedefaultendpunct}{\mcitedefaultseppunct}\relax
\EndOfBibitem
\bibitem{Chen:2016qju}
H.-X. Chen, W.~Chen, X.~Liu, and S.-L. Zhu,
  \ifthenelse{\boolean{articletitles}}{\emph{{The hidden-charm pentaquark and
  tetraquark states}},
  }{}\href{http://dx.doi.org/10.1016/j.physrep.2016.05.004}{Phys.\ Rept.\
  \textbf{639} (2016) 1},
  \href{http://arxiv.org/abs/1601.02092}{{\normalfont\ttfamily
  arXiv:1601.02092}}\relax
\mciteBstWouldAddEndPuncttrue
\mciteSetBstMidEndSepPunct{\mcitedefaultmidpunct}
{\mcitedefaultendpunct}{\mcitedefaultseppunct}\relax
\EndOfBibitem
\bibitem{Hagiwara:2002fs}
Particle Data Group, K.~Hagiwara {\em et~al.},
  \ifthenelse{\boolean{articletitles}}{\emph{{Review of particle physics.
  Particle Data Group}},
  }{}\href{http://dx.doi.org/10.1103/PhysRevD.66.010001}{Phys.\ Rev.\
  \textbf{D66} (2002) 010001}\relax
\mciteBstWouldAddEndPuncttrue
\mciteSetBstMidEndSepPunct{\mcitedefaultmidpunct}
{\mcitedefaultendpunct}{\mcitedefaultseppunct}\relax
\EndOfBibitem
\bibitem{Close:2003sg}
F.~E. Close and P.~R. Page, \ifthenelse{\boolean{articletitles}}{\emph{{The
  $D^{*0}\bar{D}^0$ threshold resonance}},
  }{}\href{http://dx.doi.org/10.1016/j.physletb.2003.10.032}{Phys.\ Lett.\
  \textbf{B578} (2004) 119},
  \href{http://arxiv.org/abs/hep-ph/0309253}{{\normalfont\ttfamily
  arXiv:hep-ph/0309253}}\relax
\mciteBstWouldAddEndPuncttrue
\mciteSetBstMidEndSepPunct{\mcitedefaultmidpunct}
{\mcitedefaultendpunct}{\mcitedefaultseppunct}\relax
\EndOfBibitem
\bibitem{Takeuchi:2014rsa}
S.~Takeuchi, K.~Shimizu, and M.~Takizawa,
  \ifthenelse{\boolean{articletitles}}{\emph{{On the origin of the narrow peak
  and the isospin symmetry breaking of the $X$(3872)}},
  }{}\href{http://dx.doi.org/10.1093/ptep/ptu160, 10.1093/ptep/ptv104}{PTEP
  \textbf{2014} (2014) 123D01},
  \href{http://arxiv.org/abs/1408.0973}{{\normalfont\ttfamily
  arXiv:1408.0973}}, [Erratum: PTEP2015,no.7,079203(2015)]\relax
\mciteBstWouldAddEndPuncttrue
\mciteSetBstMidEndSepPunct{\mcitedefaultmidpunct}
{\mcitedefaultendpunct}{\mcitedefaultseppunct}\relax
\EndOfBibitem
\bibitem{Pakvasa:2003ea}
S.~Pakvasa and M.~Suzuki, \ifthenelse{\boolean{articletitles}}{\emph{{On the
  hidden charm state at 3872-MeV}},
  }{}\href{http://dx.doi.org/10.1016/j.physletb.2003.11.005}{Phys.\ Lett.\
  \textbf{B579} (2004) 67},
  \href{http://arxiv.org/abs/hep-ph/0309294}{{\normalfont\ttfamily
  arXiv:hep-ph/0309294}}\relax
\mciteBstWouldAddEndPuncttrue
\mciteSetBstMidEndSepPunct{\mcitedefaultmidpunct}
{\mcitedefaultendpunct}{\mcitedefaultseppunct}\relax
\EndOfBibitem
\bibitem{Braaten:2004rw}
E.~Braaten and M.~Kusunoki,
  \ifthenelse{\boolean{articletitles}}{\emph{{Production of the X(3870) at the
  $\Upsilon(4S)$ by the coalescence of charm mesons from B decays}},
  }{}\href{http://dx.doi.org/10.1103/PhysRevD.69.114012}{Phys.\ Rev.\
  \textbf{D69} (2004) 114012},
  \href{http://arxiv.org/abs/hep-ph/0402177}{{\normalfont\ttfamily
  arXiv:hep-ph/0402177}}\relax
\mciteBstWouldAddEndPuncttrue
\mciteSetBstMidEndSepPunct{\mcitedefaultmidpunct}
{\mcitedefaultendpunct}{\mcitedefaultseppunct}\relax
\EndOfBibitem
\bibitem{Voloshin:2003nt}
M.~B. Voloshin, \ifthenelse{\boolean{articletitles}}{\emph{{Interference and
  binding effects in decays of possible molecular component of X(3872)}},
  }{}\href{http://dx.doi.org/10.1016/j.physletb.2003.11.014}{Phys.\ Lett.\
  \textbf{B579} (2004) 316},
  \href{http://arxiv.org/abs/hep-ph/0309307}{{\normalfont\ttfamily
  arXiv:hep-ph/0309307}}\relax
\mciteBstWouldAddEndPuncttrue
\mciteSetBstMidEndSepPunct{\mcitedefaultmidpunct}
{\mcitedefaultendpunct}{\mcitedefaultseppunct}\relax
\EndOfBibitem
\bibitem{Fleming:2007rp}
S.~Fleming, M.~Kusunoki, T.~Mehen, and U.~van Kolck,
  \ifthenelse{\boolean{articletitles}}{\emph{{Pion interactions in the
  $X(3872)$}}, }{}\href{http://dx.doi.org/10.1103/PhysRevD.76.034006}{Phys.\
  Rev.\  \textbf{D76} (2007) 034006},
  \href{http://arxiv.org/abs/hep-ph/0703168}{{\normalfont\ttfamily
  arXiv:hep-ph/0703168}}\relax
\mciteBstWouldAddEndPuncttrue
\mciteSetBstMidEndSepPunct{\mcitedefaultmidpunct}
{\mcitedefaultendpunct}{\mcitedefaultseppunct}\relax
\EndOfBibitem
\bibitem{Zhu:2007wz}
S.-L. Zhu, \ifthenelse{\boolean{articletitles}}{\emph{{New hadron states}},
  }{}\href{http://dx.doi.org/10.1142/S0218301308009446}{Int.\ J.\ Mod.\ Phys.\
  \textbf{E17} (2008) 283},
  \href{http://arxiv.org/abs/hep-ph/0703225}{{\normalfont\ttfamily
  arXiv:hep-ph/0703225}}\relax
\mciteBstWouldAddEndPuncttrue
\mciteSetBstMidEndSepPunct{\mcitedefaultmidpunct}
{\mcitedefaultendpunct}{\mcitedefaultseppunct}\relax
\EndOfBibitem
\bibitem{Gamermann:2009fv}
D.~Gamermann and E.~Oset, \ifthenelse{\boolean{articletitles}}{\emph{{Isospin
  breaking effects in the X(3872) resonance}},
  }{}\href{http://dx.doi.org/10.1103/PhysRevD.80.014003}{Phys.\ Rev.\
  \textbf{D80} (2009) 014003},
  \href{http://arxiv.org/abs/0905.0402}{{\normalfont\ttfamily
  arXiv:0905.0402}}\relax
\mciteBstWouldAddEndPuncttrue
\mciteSetBstMidEndSepPunct{\mcitedefaultmidpunct}
{\mcitedefaultendpunct}{\mcitedefaultseppunct}\relax
\EndOfBibitem
\bibitem{Zhang:2009bv}
O.~Zhang, C.~Meng, and H.~Q. Zheng,
  \ifthenelse{\boolean{articletitles}}{\emph{{Ambiversion of X(3872)}},
  }{}\href{http://dx.doi.org/10.1016/j.physletb.2009.09.033}{Phys.\ Lett.\
  \textbf{B680} (2009) 453},
  \href{http://arxiv.org/abs/0901.1553}{{\normalfont\ttfamily
  arXiv:0901.1553}}\relax
\mciteBstWouldAddEndPuncttrue
\mciteSetBstMidEndSepPunct{\mcitedefaultmidpunct}
{\mcitedefaultendpunct}{\mcitedefaultseppunct}\relax
\EndOfBibitem
\bibitem{Sun:2012sy}
Z.-F. Sun, X.~Liu, M.~Nielsen, and S.-L. Zhu,
  \ifthenelse{\boolean{articletitles}}{\emph{{Hadronic molecules with both open
  charm and bottom}},
  }{}\href{http://dx.doi.org/10.1103/PhysRevD.85.094008}{Phys.\ Rev.\
  \textbf{D85} (2012) 094008},
  \href{http://arxiv.org/abs/1203.1090}{{\normalfont\ttfamily
  arXiv:1203.1090}}\relax
\mciteBstWouldAddEndPuncttrue
\mciteSetBstMidEndSepPunct{\mcitedefaultmidpunct}
{\mcitedefaultendpunct}{\mcitedefaultseppunct}\relax
\EndOfBibitem
\bibitem{Chen:2013coa}
D.-Y. Chen, X.~Liu, and T.~Matsuki,
  \ifthenelse{\boolean{articletitles}}{\emph{{Reproducing the $Z_c(3900)$
  structure through the initial-single-pion-emission mechanism}},
  }{}\href{http://dx.doi.org/10.1103/PhysRevD.88.036008}{Phys.\ Rev.\
  \textbf{D88} (2013) 036008},
  \href{http://arxiv.org/abs/1304.5845}{{\normalfont\ttfamily
  arXiv:1304.5845}}\relax
\mciteBstWouldAddEndPuncttrue
\mciteSetBstMidEndSepPunct{\mcitedefaultmidpunct}
{\mcitedefaultendpunct}{\mcitedefaultseppunct}\relax
\EndOfBibitem
\bibitem{Prelovsek:2013xba}
S.~Prelovsek and L.~Leskovec,
  \ifthenelse{\boolean{articletitles}}{\emph{{Search for $Z^{+}_{c}$(3900) in
  the $1^{+-}$ Channel on the Lattice}},
  }{}\href{http://dx.doi.org/10.1016/j.physletb.2013.10.009}{Phys.\ Lett.\
  \textbf{B727} (2013) 172},
  \href{http://arxiv.org/abs/1308.2097}{{\normalfont\ttfamily
  arXiv:1308.2097}}\relax
\mciteBstWouldAddEndPuncttrue
\mciteSetBstMidEndSepPunct{\mcitedefaultmidpunct}
{\mcitedefaultendpunct}{\mcitedefaultseppunct}\relax
\EndOfBibitem
\bibitem{Prelovsek:2014swa}
S.~Prelovsek, C.~B. Lang, L.~Leskovec, and D.~Mohler,
  \ifthenelse{\boolean{articletitles}}{\emph{{Study of the $Z_c^+$ channel
  using lattice QCD}},
  }{}\href{http://dx.doi.org/10.1103/PhysRevD.91.014504}{Phys.\ Rev.\
  \textbf{D91} (2015) 014504},
  \href{http://arxiv.org/abs/1405.7623}{{\normalfont\ttfamily
  arXiv:1405.7623}}\relax
\mciteBstWouldAddEndPuncttrue
\mciteSetBstMidEndSepPunct{\mcitedefaultmidpunct}
{\mcitedefaultendpunct}{\mcitedefaultseppunct}\relax
\EndOfBibitem
\bibitem{Lee:2014bea}
Fermilab Lattice, MILC, S.-H. Lee, C.~Detar, D.~Mohler, and H.~Na,
  \ifthenelse{\boolean{articletitles}}{\emph{{Searching for the $X(3872)$ and
  $Z_c^+(3900)$ on HISQ lattices}}, }{}PoS \textbf{LATTICE2014} (2014)
  125\relax
\mciteBstWouldAddEndPuncttrue
\mciteSetBstMidEndSepPunct{\mcitedefaultmidpunct}
{\mcitedefaultendpunct}{\mcitedefaultseppunct}\relax
\EndOfBibitem
\bibitem{Ikeda:2016zwx}
HAL QCD, Y.~Ikeda {\em et~al.},
  \ifthenelse{\boolean{articletitles}}{\emph{{Fate of the Tetraquark Candidate
  $Z_c$(3900) from Lattice QCD}},
  }{}\href{http://dx.doi.org/10.1103/PhysRevLett.117.242001}{Phys.\ Rev.\
  Lett.\  \textbf{117} (2016) 242001},
  \href{http://arxiv.org/abs/1602.03465}{{\normalfont\ttfamily
  arXiv:1602.03465}}\relax
\mciteBstWouldAddEndPuncttrue
\mciteSetBstMidEndSepPunct{\mcitedefaultmidpunct}
{\mcitedefaultendpunct}{\mcitedefaultseppunct}\relax
\EndOfBibitem
\bibitem{Guo:2016bjq}
F.-K. Guo {\em et~al.}, \ifthenelse{\boolean{articletitles}}{\emph{{Interplay
  of quark and meson degrees of freedom in near-threshold states: A practical
  parametrization for line shapes}},
  }{}\href{http://dx.doi.org/10.1103/PhysRevD.93.074031}{Phys.\ Rev.\
  \textbf{D93} (2016) 074031},
  \href{http://arxiv.org/abs/1602.00940}{{\normalfont\ttfamily
  arXiv:1602.00940}}\relax
\mciteBstWouldAddEndPuncttrue
\mciteSetBstMidEndSepPunct{\mcitedefaultmidpunct}
{\mcitedefaultendpunct}{\mcitedefaultseppunct}\relax
\EndOfBibitem
\bibitem{Albaladejo:2015lob}
M.~Albaladejo, F.-K. Guo, C.~Hidalgo-Duque, and J.~Nieves,
  \ifthenelse{\boolean{articletitles}}{\emph{{$Z_c(3900)$: What has been really
  seen?}}, }{}\href{http://dx.doi.org/10.1016/j.physletb.2016.02.025}{Phys.\
  Lett.\  \textbf{B755} (2016) 337},
  \href{http://arxiv.org/abs/1512.03638}{{\normalfont\ttfamily
  arXiv:1512.03638}}\relax
\mciteBstWouldAddEndPuncttrue
\mciteSetBstMidEndSepPunct{\mcitedefaultmidpunct}
{\mcitedefaultendpunct}{\mcitedefaultseppunct}\relax
\EndOfBibitem
\bibitem{Barnes:2014csa}
T.~Barnes, F.~E. Close, and E.~S. Swanson,
  \ifthenelse{\boolean{articletitles}}{\emph{{Molecular Interpretation of the
  Supercharmonium State Z(4475)}},
  }{}\href{http://dx.doi.org/10.1103/PhysRevD.91.014004}{Phys.\ Rev.\
  \textbf{D91} (2015) 014004},
  \href{http://arxiv.org/abs/1409.6651}{{\normalfont\ttfamily
  arXiv:1409.6651}}\relax
\mciteBstWouldAddEndPuncttrue
\mciteSetBstMidEndSepPunct{\mcitedefaultmidpunct}
{\mcitedefaultendpunct}{\mcitedefaultseppunct}\relax
\EndOfBibitem
\bibitem{Ma:2014zua}
L.~Ma, X.-H. Liu, X.~Liu, and S.-L. Zhu,
  \ifthenelse{\boolean{articletitles}}{\emph{{Exotic Four Quark Matter:
  $Z_1(4475)$}}, }{}\href{http://dx.doi.org/10.1103/PhysRevD.90.037502}{Phys.\
  Rev.\  \textbf{D90} (2014) 037502},
  \href{http://arxiv.org/abs/1404.3450}{{\normalfont\ttfamily
  arXiv:1404.3450}}\relax
\mciteBstWouldAddEndPuncttrue
\mciteSetBstMidEndSepPunct{\mcitedefaultmidpunct}
{\mcitedefaultendpunct}{\mcitedefaultseppunct}\relax
\EndOfBibitem
\bibitem{delAmoSanchez:2010vq}
BaBar, P.~del Amo~Sanchez {\em et~al.},
  \ifthenelse{\boolean{articletitles}}{\emph{{Observation of new resonances
  decaying to $D\pi$ and $D^*\pi$ in inclusive $e^+e^-$ collisions near
  $\sqrt{s}=$10.58 GeV}},
  }{}\href{http://dx.doi.org/10.1103/PhysRevD.82.111101}{Phys.\ Rev.\
  \textbf{D82} (2010) 111101},
  \href{http://arxiv.org/abs/1009.2076}{{\normalfont\ttfamily
  arXiv:1009.2076}}\relax
\mciteBstWouldAddEndPuncttrue
\mciteSetBstMidEndSepPunct{\mcitedefaultmidpunct}
{\mcitedefaultendpunct}{\mcitedefaultseppunct}\relax
\EndOfBibitem
\bibitem{Aaij:2013sza}
LHCb, R.~Aaij {\em et~al.}, \ifthenelse{\boolean{articletitles}}{\emph{{Study
  of $D_J$ meson decays to $D^+\pi^-$, $D^0 \pi^+$ and $D^{*+}\pi^-$ final
  states in $pp$ collision}},
  }{}\href{http://dx.doi.org/10.1007/JHEP09(2013)145}{JHEP \textbf{09} (2013)
  145}, \href{http://arxiv.org/abs/1307.4556}{{\normalfont\ttfamily
  arXiv:1307.4556}}\relax
\mciteBstWouldAddEndPuncttrue
\mciteSetBstMidEndSepPunct{\mcitedefaultmidpunct}
{\mcitedefaultendpunct}{\mcitedefaultseppunct}\relax
\EndOfBibitem
\bibitem{Roca:2015dva}
L.~Roca, J.~Nieves, and E.~Oset,
  \ifthenelse{\boolean{articletitles}}{\emph{{LHCb pentaquark as a
  $\bar{D}^*\Sigma_c-\bar{D}^*\Sigma_c^*$ molecular state}},
  }{}\href{http://dx.doi.org/10.1103/PhysRevD.92.094003}{Phys.\ Rev.\
  \textbf{D92} (2015) 094003},
  \href{http://arxiv.org/abs/1507.04249}{{\normalfont\ttfamily
  arXiv:1507.04249}}\relax
\mciteBstWouldAddEndPuncttrue
\mciteSetBstMidEndSepPunct{\mcitedefaultmidpunct}
{\mcitedefaultendpunct}{\mcitedefaultseppunct}\relax
\EndOfBibitem
\bibitem{Karliner:2016via}
M.~Karliner, \ifthenelse{\boolean{articletitles}}{\emph{{Doubly Heavy Exotic
  Mesons and Baryons and How to Look for Them}},
  }{}\href{http://dx.doi.org/10.5506/APhysPolB.47.117}{Acta Phys.\ Polon.\
  \textbf{B47} (2016) 117}\relax
\mciteBstWouldAddEndPuncttrue
\mciteSetBstMidEndSepPunct{\mcitedefaultmidpunct}
{\mcitedefaultendpunct}{\mcitedefaultseppunct}\relax
\EndOfBibitem
\bibitem{Bayar:2016ftu}
M.~Bayar, F.~Aceti, F.-K. Guo, and E.~Oset,
  \ifthenelse{\boolean{articletitles}}{\emph{{A Discussion on Triangle
  Singularities in the $\Lambda_b \to J/\psi K^{-} p$ Reaction}},
  }{}\href{http://dx.doi.org/10.1103/PhysRevD.94.074039}{Phys.\ Rev.\
  \textbf{D94} (2016) 074039},
  \href{http://arxiv.org/abs/1609.04133}{{\normalfont\ttfamily
  arXiv:1609.04133}}\relax
\mciteBstWouldAddEndPuncttrue
\mciteSetBstMidEndSepPunct{\mcitedefaultmidpunct}
{\mcitedefaultendpunct}{\mcitedefaultseppunct}\relax
\EndOfBibitem
\bibitem{Aubert:2008gu}
BaBar, B.~Aubert {\em et~al.}, \ifthenelse{\boolean{articletitles}}{\emph{{A
  Study of $B \to X(3872) K$, with $X(3872) \to J/\Psi \pi^{+} \pi^{-}$}},
  }{}\href{http://dx.doi.org/10.1103/PhysRevD.77.111101}{Phys.\ Rev.\
  \textbf{D77} (2008) 111101},
  \href{http://arxiv.org/abs/0803.2838}{{\normalfont\ttfamily
  arXiv:0803.2838}}\relax
\mciteBstWouldAddEndPuncttrue
\mciteSetBstMidEndSepPunct{\mcitedefaultmidpunct}
{\mcitedefaultendpunct}{\mcitedefaultseppunct}\relax
\EndOfBibitem
\bibitem{Brodsky:2014xia}
S.~J. Brodsky, D.~S. Hwang, and R.~F. Lebed,
  \ifthenelse{\boolean{articletitles}}{\emph{{Dynamical Picture for the
  Formation and Decay of the Exotic XYZ Mesons}},
  }{}\href{http://dx.doi.org/10.1103/PhysRevLett.113.112001}{Phys.\ Rev.\
  Lett.\  \textbf{113} (2014) 112001},
  \href{http://arxiv.org/abs/1406.7281}{{\normalfont\ttfamily
  arXiv:1406.7281}}\relax
\mciteBstWouldAddEndPuncttrue
\mciteSetBstMidEndSepPunct{\mcitedefaultmidpunct}
{\mcitedefaultendpunct}{\mcitedefaultseppunct}\relax
\EndOfBibitem
\bibitem{Chen:2015dig}
H.-X. Chen, L.~Maiani, A.~D. Polosa, and V.~Riquer,
  \ifthenelse{\boolean{articletitles}}{\emph{{$Y(4260)\rightarrow \gamma +
  X(3872)$ in the diquarkonium picture}},
  }{}\href{http://dx.doi.org/10.1140/epjc/s10052-015-3781-2}{Eur.\ Phys.\ J.\
  \textbf{C75} (2015) 550},
  \href{http://arxiv.org/abs/1510.03626}{{\normalfont\ttfamily
  arXiv:1510.03626}}\relax
\mciteBstWouldAddEndPuncttrue
\mciteSetBstMidEndSepPunct{\mcitedefaultmidpunct}
{\mcitedefaultendpunct}{\mcitedefaultseppunct}\relax
\EndOfBibitem
\bibitem{Chen:2016oma}
H.-X. Chen {\em et~al.},
  \ifthenelse{\boolean{articletitles}}{\emph{{Understanding the internal
  structures of the $X(4140)$, $X(4274)$, $X(4500)$ and $X(4700)$}},
  }{}\href{http://dx.doi.org/10.1140/epjc/s10052-017-4737-5}{Eur.\ Phys.\ J.\
  \textbf{C77} (2017) 160},
  \href{http://arxiv.org/abs/1606.03179}{{\normalfont\ttfamily
  arXiv:1606.03179}}\relax
\mciteBstWouldAddEndPuncttrue
\mciteSetBstMidEndSepPunct{\mcitedefaultmidpunct}
{\mcitedefaultendpunct}{\mcitedefaultseppunct}\relax
\EndOfBibitem
\bibitem{Maiani:2015vwa}
L.~Maiani, A.~D. Polosa, and V.~Riquer,
  \ifthenelse{\boolean{articletitles}}{\emph{{The New Pentaquarks in the
  Diquark Model}},
  }{}\href{http://dx.doi.org/10.1016/j.physletb.2015.08.008}{Phys.\ Lett.\
  \textbf{B749} (2015) 289},
  \href{http://arxiv.org/abs/1507.04980}{{\normalfont\ttfamily
  arXiv:1507.04980}}\relax
\mciteBstWouldAddEndPuncttrue
\mciteSetBstMidEndSepPunct{\mcitedefaultmidpunct}
{\mcitedefaultendpunct}{\mcitedefaultseppunct}\relax
\EndOfBibitem
\bibitem{Lebed:2015tna}
R.~F. Lebed, \ifthenelse{\boolean{articletitles}}{\emph{{The Pentaquark
  Candidates in the Dynamical Diquark Picture}},
  }{}\href{http://dx.doi.org/10.1016/j.physletb.2015.08.032}{Phys.\ Lett.\
  \textbf{B749} (2015) 454},
  \href{http://arxiv.org/abs/1507.05867}{{\normalfont\ttfamily
  arXiv:1507.05867}}\relax
\mciteBstWouldAddEndPuncttrue
\mciteSetBstMidEndSepPunct{\mcitedefaultmidpunct}
{\mcitedefaultendpunct}{\mcitedefaultseppunct}\relax
\EndOfBibitem
\bibitem{Aaij:2017nav}
LHCb, R.~Aaij {\em et~al.},
  \ifthenelse{\boolean{articletitles}}{\emph{{Observation of five new narrow
  $\Omega_c^0$ states decaying to $\Xi_c^+ K^-$}},
  }{}\href{http://dx.doi.org/10.1103/PhysRevLett.118.182001}{Phys.\ Rev.\
  Lett.\  \textbf{118} (2017), no.~18 182001},
  \href{http://arxiv.org/abs/1703.04639}{{\normalfont\ttfamily
  arXiv:1703.04639}}\relax
\mciteBstWouldAddEndPuncttrue
\mciteSetBstMidEndSepPunct{\mcitedefaultmidpunct}
{\mcitedefaultendpunct}{\mcitedefaultseppunct}\relax
\EndOfBibitem
\bibitem{Ebert:2011kk}
D.~Ebert, R.~N. Faustov, and V.~O. Galkin,
  \ifthenelse{\boolean{articletitles}}{\emph{{Spectroscopy and Regge
  trajectories of heavy baryons in the relativistic quark-diquark picture}},
  }{}\href{http://dx.doi.org/10.1103/PhysRevD.84.014025}{Phys.\ Rev.\
  \textbf{D84} (2011) 014025},
  \href{http://arxiv.org/abs/1105.0583}{{\normalfont\ttfamily
  arXiv:1105.0583}}\relax
\mciteBstWouldAddEndPuncttrue
\mciteSetBstMidEndSepPunct{\mcitedefaultmidpunct}
{\mcitedefaultendpunct}{\mcitedefaultseppunct}\relax
\EndOfBibitem
\bibitem{Karliner:2017kfm}
M.~Karliner and J.~L. Rosner, \ifthenelse{\boolean{articletitles}}{\emph{{Very
  narrow excited $\Omega_c$ baryons}},
  }{}\href{http://dx.doi.org/10.1103/PhysRevD.95.114012}{Phys.\ Rev.\
  \textbf{D95} (2017), no.~11 114012},
  \href{http://arxiv.org/abs/1703.07774}{{\normalfont\ttfamily
  arXiv:1703.07774}}\relax
\mciteBstWouldAddEndPuncttrue
\mciteSetBstMidEndSepPunct{\mcitedefaultmidpunct}
{\mcitedefaultendpunct}{\mcitedefaultseppunct}\relax
\EndOfBibitem
\bibitem{Karliner:2014gca}
M.~Karliner and J.~L. Rosner,
  \ifthenelse{\boolean{articletitles}}{\emph{{Baryons with two heavy quarks:
  Masses, production, decays, and detection}},
  }{}\href{http://dx.doi.org/10.1103/PhysRevD.90.094007}{Phys.\ Rev.\
  \textbf{D90} (2014), no.~9 094007},
  \href{http://arxiv.org/abs/1408.5877}{{\normalfont\ttfamily
  arXiv:1408.5877}}\relax
\mciteBstWouldAddEndPuncttrue
\mciteSetBstMidEndSepPunct{\mcitedefaultmidpunct}
{\mcitedefaultendpunct}{\mcitedefaultseppunct}\relax
\EndOfBibitem
\bibitem{Eichten:2017ffp}
E.~J. Eichten and C.~Quigg,
  \ifthenelse{\boolean{articletitles}}{\emph{{Heavy-quark symmetry implies
  stable heavy tetraquark mesons $Q_iQ_j \bar q_k \bar q_l$}},
  }{}\href{http://arxiv.org/abs/1707.09575}{{\normalfont\ttfamily
  arXiv:1707.09575}}\relax
\mciteBstWouldAddEndPuncttrue
\mciteSetBstMidEndSepPunct{\mcitedefaultmidpunct}
{\mcitedefaultendpunct}{\mcitedefaultseppunct}\relax
\EndOfBibitem
\bibitem{Li:2013ssa}
X.~Li and M.~B. Voloshin, \ifthenelse{\boolean{articletitles}}{\emph{{$Y$(4260)
  and $Y$(4360) as mixed hadrocharmonium}},
  }{}\href{http://dx.doi.org/10.1142/S0217732314500606}{Mod.\ Phys.\ Lett.\
  \textbf{A29} (2014) 1450060},
  \href{http://arxiv.org/abs/1309.1681}{{\normalfont\ttfamily
  arXiv:1309.1681}}\relax
\mciteBstWouldAddEndPuncttrue
\mciteSetBstMidEndSepPunct{\mcitedefaultmidpunct}
{\mcitedefaultendpunct}{\mcitedefaultseppunct}\relax
\EndOfBibitem
\bibitem{Aaij:2016avz}
LHCb, R.~Aaij {\em et~al.},
  \ifthenelse{\boolean{articletitles}}{\emph{{Measurement of the $b$-quark
  production cross-section in 7 and 13 TeV $pp$ collisions}},
  }{}\href{http://dx.doi.org/10.1103/PhysRevLett.118.052002}{Phys.\ Rev.\
  Lett.\  \textbf{118} (2017) 052002},
  \href{http://arxiv.org/abs/1612.05140}{{\normalfont\ttfamily
  arXiv:1612.05140}}\relax
\mciteBstWouldAddEndPuncttrue
\mciteSetBstMidEndSepPunct{\mcitedefaultmidpunct}
{\mcitedefaultendpunct}{\mcitedefaultseppunct}\relax
\EndOfBibitem
\bibitem{Lutz:2009ff}
PANDA, M.~F.~M. Lutz {\em et~al.},
  \ifthenelse{\boolean{articletitles}}{\emph{{Physics Performance Report for
  PANDA: Strong Interaction Studies with Antiprotons}},
  }{}\href{http://arxiv.org/abs/0903.3905}{{\normalfont\ttfamily
  arXiv:0903.3905}}\relax
\mciteBstWouldAddEndPuncttrue
\mciteSetBstMidEndSepPunct{\mcitedefaultmidpunct}
{\mcitedefaultendpunct}{\mcitedefaultseppunct}\relax
\EndOfBibitem
\bibitem{Ghoul:2015ifw}
GlueX, H.~Al~Ghoul {\em et~al.},
  \ifthenelse{\boolean{articletitles}}{\emph{{First Results from The GlueX
  Experiment}}, }{}\href{http://dx.doi.org/10.1063/1.4949369}{AIP Conf.\ Proc.\
   \textbf{1735} (2016) 020001},
  \href{http://arxiv.org/abs/1512.03699}{{\normalfont\ttfamily
  arXiv:1512.03699}}\relax
\mciteBstWouldAddEndPuncttrue
\mciteSetBstMidEndSepPunct{\mcitedefaultmidpunct}
{\mcitedefaultendpunct}{\mcitedefaultseppunct}\relax
\EndOfBibitem
\bibitem{Burkert:2008rj}
V.~D. Burkert, \ifthenelse{\boolean{articletitles}}{\emph{{CLAS12 and its
  initial Science Program at the Jefferson Lab Upgrade}}, }{} in {\em {CLAS 12
  RICH Detector Workshop Jefferson Lab, Newport News, Virginia, January 28-29,
  2008}}, 2008.
\newblock \href{http://arxiv.org/abs/0810.4718}{{\normalfont\ttfamily
  arXiv:0810.4718}}\relax
\mciteBstWouldAddEndPuncttrue
\mciteSetBstMidEndSepPunct{\mcitedefaultmidpunct}
{\mcitedefaultendpunct}{\mcitedefaultseppunct}\relax
\EndOfBibitem
\bibitem{Wang:2015jsa}
Q.~Wang, X.-H. Liu, and Q.~Zhao,
  \ifthenelse{\boolean{articletitles}}{\emph{{Photoproduction of hidden charm
  pentaquark states $P_c^+(4380)$ and $P_c^+(4450)$}},
  }{}\href{http://dx.doi.org/10.1103/PhysRevD.92.034022}{Phys.\ Rev.\
  \textbf{D92} (2015) 034022},
  \href{http://arxiv.org/abs/1508.00339}{{\normalfont\ttfamily
  arXiv:1508.00339}}\relax
\mciteBstWouldAddEndPuncttrue
\mciteSetBstMidEndSepPunct{\mcitedefaultmidpunct}
{\mcitedefaultendpunct}{\mcitedefaultseppunct}\relax
\EndOfBibitem
\bibitem{Kubarovsky:2015aaa}
V.~Kubarovsky and M.~B. Voloshin,
  \ifthenelse{\boolean{articletitles}}{\emph{{Formation of hidden-charm
  pentaquarks in photon-nucleon collisions}},
  }{}\href{http://dx.doi.org/10.1103/PhysRevD.92.031502}{Phys.\ Rev.\
  \textbf{D92} (2015) 031502},
  \href{http://arxiv.org/abs/1508.00888}{{\normalfont\ttfamily
  arXiv:1508.00888}}\relax
\mciteBstWouldAddEndPuncttrue
\mciteSetBstMidEndSepPunct{\mcitedefaultmidpunct}
{\mcitedefaultendpunct}{\mcitedefaultseppunct}\relax
\EndOfBibitem
\bibitem{Karliner:2015voa}
M.~Karliner and J.~L. Rosner,
  \ifthenelse{\boolean{articletitles}}{\emph{{Photoproduction of Exotic Baryon
  Resonances}},
  }{}\href{http://dx.doi.org/10.1016/j.physletb.2015.11.068}{Phys.\ Lett.\
  \textbf{B752} (2016) 329},
  \href{http://arxiv.org/abs/1508.01496}{{\normalfont\ttfamily
  arXiv:1508.01496}}\relax
\mciteBstWouldAddEndPuncttrue
\mciteSetBstMidEndSepPunct{\mcitedefaultmidpunct}
{\mcitedefaultendpunct}{\mcitedefaultseppunct}\relax
\EndOfBibitem
\bibitem{Blin:2016dlf}
A.~N. Hiller~Blin {\em et~al.},
  \ifthenelse{\boolean{articletitles}}{\emph{{Studying the P$_c$(4450)
  resonance in J/$\psi$ photoproduction off protons}},
  }{}\href{http://dx.doi.org/10.1103/PhysRevD.94.034002}{Phys.\ Rev.\
  \textbf{D94} (2016) 034002},
  \href{http://arxiv.org/abs/1606.08912}{{\normalfont\ttfamily
  arXiv:1606.08912}}\relax
\mciteBstWouldAddEndPuncttrue
\mciteSetBstMidEndSepPunct{\mcitedefaultmidpunct}
{\mcitedefaultendpunct}{\mcitedefaultseppunct}\relax
\EndOfBibitem
\bibitem{Meziani:2016lhg}
Z.~E. Meziani {\em et~al.}, \ifthenelse{\boolean{articletitles}}{\emph{{A
  Search for the LHCb Charmed 'Pentaquark' using Photo-Production of $J/{\psi}$
  at Threshold in Hall C at Jefferson Lab}},
  }{}\href{http://arxiv.org/abs/1609.00676}{{\normalfont\ttfamily
  arXiv:1609.00676}}\relax
\mciteBstWouldAddEndPuncttrue
\mciteSetBstMidEndSepPunct{\mcitedefaultmidpunct}
{\mcitedefaultendpunct}{\mcitedefaultseppunct}\relax
\EndOfBibitem
\end{mcitethebibliography}

\end{document}